\documentclass[acmsmall]{acmart}

\AtBeginDocument{%
  \providecommand\BibTeX{{%
    \normalfont B\kern-0.5em{\scshape i\kern-0.25em b}\kern-0.8em\TeX}}}

\setcopyright{acmcopyright}
\copyrightyear{2018}
\acmYear{2018}
\acmDOI{10.1145/1122445.1122456}

\acmJournal{JACM}
\acmVolume{37}
\acmNumber{4}
\acmArticle{111}
\acmMonth{8}

\usepackage{amsmath}
\usepackage{graphicx}
\usepackage{multirow}
\usepackage{soul}
\usepackage{booktabs}
\usepackage{colortbl}
\usepackage{wrapfig}
\usepackage{pifont}

\usepackage{etoolbox}
\usepackage{tikz}

\newrobustcmd*{\mysquare}[1]{\tikz{\filldraw[draw=#1,fill=#1] (0,0)
rectangle (0.2cm,0.2cm);}}

\newrobustcmd*{\mycircle}[1]{\tikz{\filldraw[draw=#1,fill=#1] (0,0) circle [radius=0.1cm];}}

\newrobustcmd*{\mytriangle}[1]{\tikz{\filldraw[draw=#1,fill=#1] (0,0) --
(0.2cm,0) -- (0.1cm,0.2cm);}}

\newcommand{\etal}{\textit{et al. }}

\begin{document}

\title{Wavelength-based Attributed Deep Neural Network for Underwater Image Restoration}
\thanks{This work is supported by IITG Technology Innovation and Development Foundation (IITGTI \& DF), which has been set up at IIT Guwahati as a part of the National Mission on Interdisciplinary Cyber Physical Systems (NMICPS). IITGTI\& DF is undertaking research, development and training activities on Technologies for Under Water Exploration with financial assistance from the Department of Science and Technology, India through grant number DST/NMICPS/TIH12/IITG/2020. Authors gratefully acknowledge the support provided for the present work. \\We also acknowledge the Department of Biotechnology,
Govt. of India for the financial support for the project BT/COE/34/SP28408/2018 (for computing resources).}
\author{Prasen Kumar Sharma}
\orcid{0000-0003-4847-8866}


\author{Ira Bisht}
%

\author{Arijit Sur}
\affiliation{%
  \streetaddress{ Multimedia Lab, Dept. of Computer Science
and Eng.}
  \institution{\\Indian Institute of Technology Guwahati}
 \city{ Assam}
 \country{India}
  }
\email{emails:{kumar176101005, ibisht, arijit}@iitg.ac.in}
%
%
%
%
%
\begin{abstract}
\textbf{Background:} Underwater images, in general, suffer from low contrast and high color distortions due to the non-uniform attenuation of the light as it propagates through the water. In addition, the degree of attenuation varies with the wavelength resulting in the asymmetric traversing of colors. Despite the prolific works for \textit{underwater image restoration} (UIR) using deep learning, the above asymmetricity has not been addressed in the respective network engineering.
  
\noindent  \textbf{Contributions:} As the first novelty, this paper shows that attributing the right receptive field size (\textit{context}) based on the traversing range of the color channel may lead to a substantial performance gain for the task of UIR. Further, it is important to suppress the irrelevant multi-contextual features and increase the representational power of the model. Therefore, as a second novelty, we have incorporated an attentive skip mechanism to adaptively refine the learned multi-contextual features. The proposed framework, called \textit{Deep WaveNet}, is optimized using the traditional pixel-wise and feature-based cost functions. An extensive set of experiments have been carried out to show the efficacy of the proposed scheme over existing best-published literature on benchmark datasets. More importantly, we have demonstrated a comprehensive validation of enhanced images across various high-level vision tasks, \textit{e.g.}, underwater image semantic segmentation, and diver's 2D pose estimation. A sample video to exhibit our real-world performance is available at \url{https://tinyurl.com/yzcrup9n}. Also, we have open-sourced our framework at \url{https://github.com/pksvision/Deep-WaveNet-Underwater-Image-Restoration}.  
\end{abstract}



\maketitle

\section{Introduction}
\begin{wrapfigure}{r}{0.5\textwidth}
\begin{center}
\includegraphics[width = 0.4\textwidth]{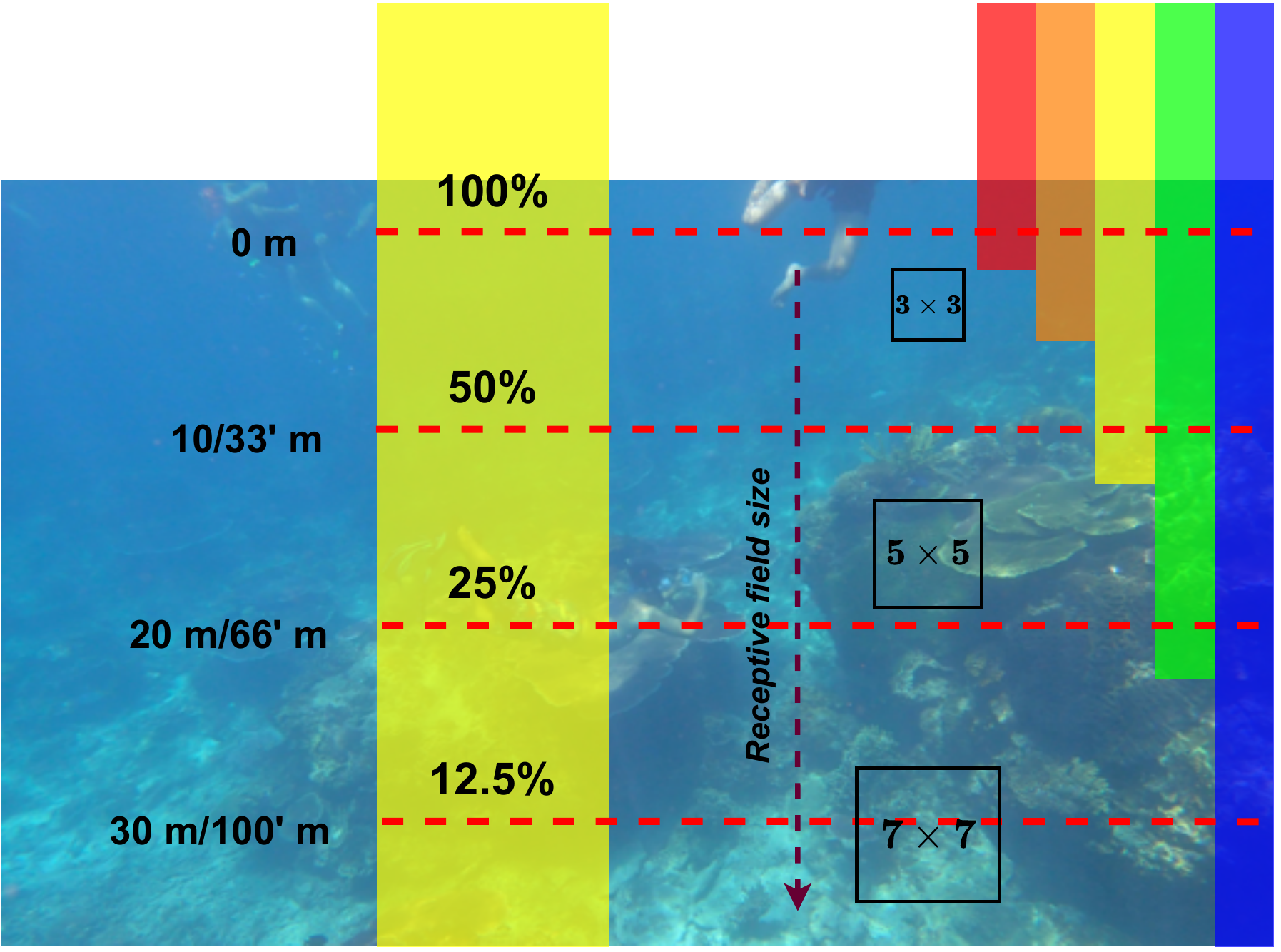}
\end{center}
\caption{\textbf{Wavelength vs. Receptive field size: }Graphic demonstration of attenuation rates corresponding to different wavelengths of light as it propagates through the water. The blue color traverses the longest because of its shortest wavelength.
It is one of the main reasons why underwater images are prevailed by the blue color \cite{chiang}. }
\label{fig:intro_wavelength}
\end{wrapfigure}
Marine ecosystem exploration, over the years, undoubtedly has been more adverse than the terrestrial due to the lack of survival endurance. The need for exploration includes oceanography, marine warfare, information navigation, and the analysis of marine life. Over the past few years, underwater exploration has gained significant attention from the machine vision research community. The study on the needs mentioned above has been conducted by performing high-level vision tasks, \textit{e.g.}, semantic segmentation, classification \textit{etc.}, on underwater images and videos. Deep \textbf{convolutional neural networks (CNNs)} have shown exceptional performance gain over the prior-based methods on these high-level vision tasks \cite{ilsvrc}. However, they, in general, yield undesirable performances when input with seen/unseen noisy data \cite{ijcai18}. Also, a few studies have shown that the deep CNNs can be deluded by an imperceivable noise perturbation for high-level vision tasks \cite{perturb1}. One straightforward solution to these problems has been the pre-processing of noisy data using low-level vision tasks such as image de-noising and restoration \cite{low-level-1,low-level-2}. 

Unlike outdoor, underwater images inhere complicated lighting conditions and environment, color casts,  making restoration a more challenging task. One of the main reasons behind such visual distortions is the \textit{non-uniform} attenuation of the light that varies with the wavelength \cite{berman_pami_20}. Further, the visibility of the underwater ecosystem through the lens heavily depends on the marine snow that increases the light scattering effect \cite{uieb}. \textcolor{black}{A detailed analysis exhibiting the differences between ordinary vs. underwater image restoration has been presented in Section 1 of the supplementary material. We encourage the readers to refer it.}

To solve these problems, recent efforts \cite{euvp,funiegan,sesr,uieb, dudhane, wang_ca_gan,wang_cnn_icip,liu_gan}  have been afoot towards deep learning-based methods that have shown remarkable results in other low-level vision tasks. It has been observed that most of the best-published UIR schemes (to name a few, \cite{euvp,funiegan,sesr, uieb, dudhane}) process color channels of the degraded images with equal receptive field sizes (\textit{alias context}). However, similar receptive fields for different color channels may not be a beneficial setting typically for underwater scenarios. To elaborate, consider Fig. \ref{fig:intro_wavelength}, wherein it has been shown that the blue color, due to its shortest wavelength, traverses longer in the water compared to red color \cite{chiang}. This type of traverse phenomenon results in most of the underwater images influenced by the blue color. It may also be concluded that the dominance of the blue color may have a direct effect on the efficiency of the high-level vision tasks, such as semantic segmentation.

It has been widely established that the larger receptive field plays a vital role in high-level vision tasks, especially classification and segmentation that includes dense per-pixel predictions \cite{large_kernel}. Moreover, it has been mentioned earlier that the efficiency of such high-level vision tasks depends on how well the noisy input has been pre-processed by the low-level vision frameworks. Although, usage of the larger receptive field may be directly related to the efficiency of the low-level vision tasks \cite{large_kernel_issue_intro}. However, this work dedicates the larger context to a specific color channel considering its vital role in the efficiency of high-level vision tasks. Given that a larger receptive field may be computationally expensive, one may also utilize the dilated filtering \cite{dilated_filtering} to add the element of multi-contextual essence. However, due to its sparse sampling, the dilated filtering may induce the gridding effect \cite{gridding_effect} in the restored images, leading to a significant reduction in performance \cite{large_kernel_issue_intro}. For a deep CNN (\textit{without pooling}), the spatial context can also be enlarged by increasing the depth of the network. However, such engineering may sometimes be inefficient, resulting in a higher computational cost. Another way of adding multi-contextual flavor in the network engineering is to process the whole RGB image, in parallel, with multiple convolution layers having different context sizes. Even though such a formulation may address the local and global coherence spatially, but may fail to address the same channel-wise. We have considered the designs mentioned above in Section \ref{sec:ablation} to show the potential of the proposed Deep WaveNet's engineering.  
 \begin{wrapfigure}{l}{0.5\textwidth}
\begin{center}
\includegraphics[width = 0.5\textwidth]{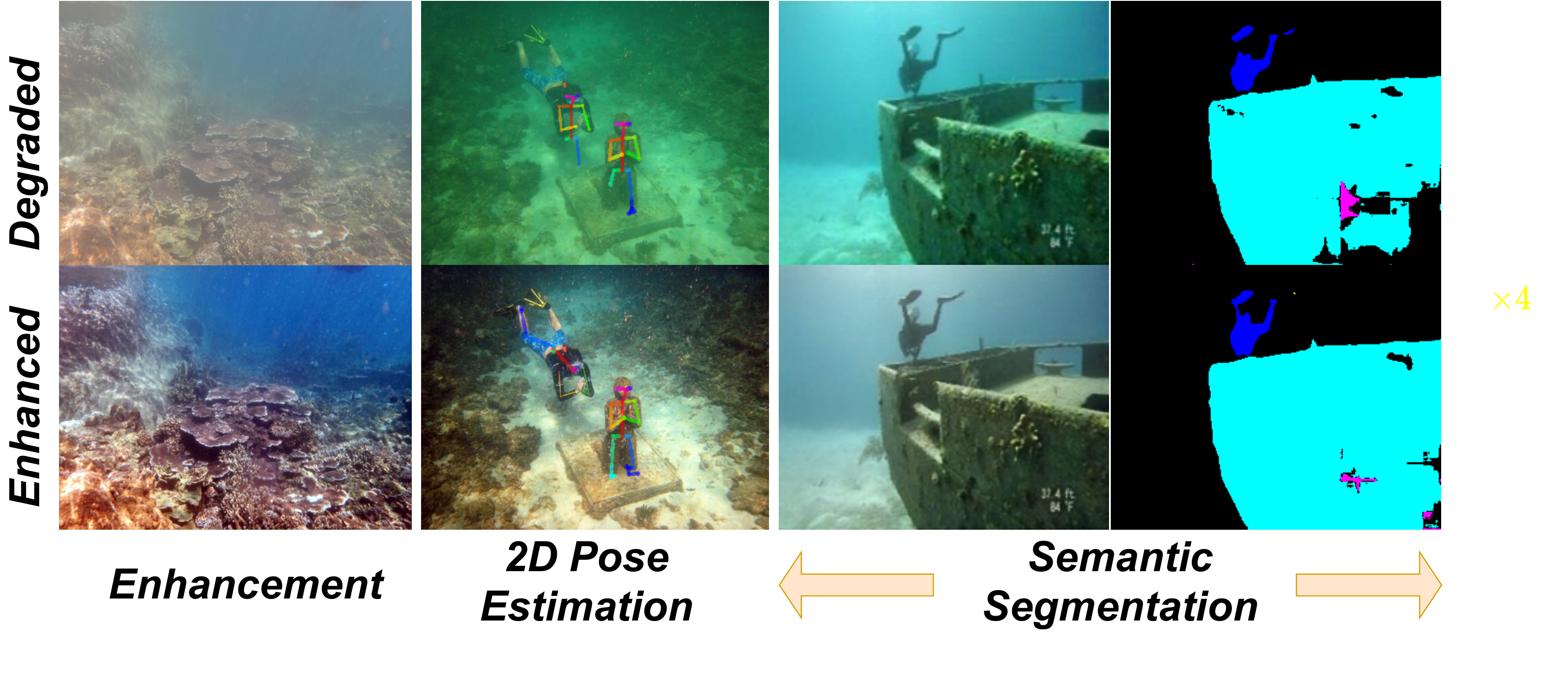}\\
\end{center}
\caption{\textbf{Impact of low-level efficiency on high-level tasks: }Sample results to show the effectiveness of the enhanced images generated by the Deep WaveNet across various high-level vision tasks.}
\label{fig:intro:super-resolution}
\end{wrapfigure}
To summarize, in an underwater scenario, efficient network engineering for UIR may help in improving the performance of underwater high-level vision tasks. Knowing that the blue color may directly influence the efficiency of high-level vision tasks, and the dominance of the blue color is because of its shortest wavelength, the high-level vision tasks may require a larger receptive field for better performance. Therefore, one may assign a larger receptive field to the blue color channel of the degraded underwater image for the task of UIR.

To the best of our knowledge, this is the first work that presents a wavelength-driven multi-contextual design of the deep CNNs for the task of UIR. For this, we have considered two of the main low-level vision tasks for the UIR, namely, \textbf{\textit{image enhancement}} and \textbf{\textit{super-resolution (SR)}}. We have shown that the Deep WaveNet outperforms the best-published works and outputs visually pleasant enhanced underwater images, which further boosts the efficiency on high-level vision tasks, as shown in Fig. \ref{fig:intro:super-resolution}. Additionally, we have incorporated the attention mechanism \cite{cbam} to improve the efficiency of the proposed Deep WaveNet. However, instead of its traditional usage just after the convolution layers \cite{cbam}, we propose to utilize it for adaptive residual learning to filter out the irrelevant features from the previous layers intelligently. It should be mentioned that both wavelength-driven multi-contextual design and attentive residual learning have not been proposed for UIR. We have presented a detailed study on how the performances of a few high-level vision tasks, such as diver's 2D pose estimation and underwater semantic segmentation, have been improved when presented with the enhanced images produced by Deep WaveNet.

\section{Related Research: An Overview}
This section briefly introduces the existing literature on UIR, categorized into underwater (a) image enhancement and (b) single image super-resolution. Further, we have classified the recent research based on (i) \textit{Non-Physical and Physical-model based} methods, and (ii) \textit{Data-driven} based approaches.

\textbf{Underwater Image Enhancement:} \textbf{\textit{Non-physical model-based}} approaches address the pixels to enhance the degraded perceptual quality of the image. In this line, Ancuti \textit{et al.} \cite{ancuti1} proposed a fusion-based scheme for enhancing the degraded underwater videos using temporal coherence. In \cite{ancuti_tip_18}, the authors presented an efficient technique built upon the fusion of two images distilled from the color compensated and white-balanced variants of the degraded underwater images. Later, Li \etal \cite{li_tip_retinex} utilized the robust retinex framework and optimized it using the augmented Lagrange multiplier-based Alternating Direction Minimization. \textcolor{black}{Ma \textit{et al.} \cite{jma_model} proposed to enhance the degraded underwater images in decorrelated color spaces such as YIQ and HSI, and then utilized Sobel edge detection algorithm to fuse them. }  

On the other hand, \textit{\textbf{physical-model based}} methods follow the inverse approach by assuming the optical model formulation and estimating its latent parameters from a given degraded underwater image. In this line, Chiang \textit{et al.} \cite{chiang} proposed a systematic approach towards UIR by first incorporating the dehazing module to predict the depth map. The produced depth map is then utilized to perform the color change compensation to estimate the enhanced image. Wen \etal \cite{wen} proposed to utilize the dark channel in underwater images for predicting the scattering rate and background light in the optical model for image enhancement. Subsequently, Peng \etal \cite{peng} suggested incorporating the image blurriness for estimating the depth map for color correction. Li \etal \cite{li_tip} proposed a contrast correction scheme by utilizing the histogram distribution as a prior for UIR. The authors of \cite{peng_tip} presented a depth estimation-based framework by utilizing blurriness and light absorption to enhance the degraded underwater images. Later, in \cite{wang_iscas}, the authors proposed to utilize the attenuation-curve prior for underwater single image enhancement. Li \etal \cite{li} proposed one of the earliest works that separately handle the color channels of degraded underwater images for UIR. The proposed model considers the blue and green channels for de-hazing, whereas the red channel for color enhancement for estimating the artifacts-free underwater images. Ultimately proved to be one of our earliest motivations of separately localizing the attentive residuals of color channels. The authors of \cite{wang_access} proposed to recover the depth maps using the maximum-attenuation identification. Yang \etal \cite{yang_icip} proposed a structure-texture decomposition-based framework that splits the degraded images into the structure and texture layers. The texture layer is denoised using gradient residual minimization and appended to the structure layer to produce the enhanced image. In \cite{wang_access_2019}, the authors proposed an $\ell_p$ norm-based minimization for single underwater image restoration. Yang \etal \cite{yang_unified} proposed to utilize the dark channel prior \cite{dcp} for an effective background light estimation. Huo \etal \cite{huo_access_20} designed a method that estimates the transmission map across each channel of the degraded image to obtain the enhanced image. Further, Bai \etal \cite{bai} proposed to utilize the local as well as global equalization of the histogram and fusion-based strategies for UIR. Berman \etal \cite{berman_pami_20} proposed to utilize the concept of haze lines for the task of UIR and presented a new quantitative dataset as well. 

\textit{\textbf{Data-driven Approaches:}}
Coming to learning-based schemes, Xu \textit{et al.} \cite{wnn} presented one of the earliest yet effective methods based on the Wavelet Neural Networks \cite{WCNN} for UIR. In \cite{wang_cnn_icip}, the authors proposed a deep CNN for the color-correction and haze removal in underwater images. Li \etal \cite{li_spl} adopted the {weakly supervised learning paradigm} for UIR.  Liu \etal \cite{residual_access} proposed a {CycleGAN} \cite{cyclegan} and VDSR \cite{VDSR} based joint approach for the task of UIR. Later, Guo \etal \cite{guo_oe} utilized the \textbf{generative adversarial network (GAN)} \cite{gan} based framework for single image restoration. Subsequently, in \cite{liu_gan}, the authors proposed a multi-scale deep CNN, which has been built upon the conditional GAN framework. Dudhane \etal \cite{dudhane} proposed to utilize the concepts of channel-wise features extraction for UIR. Wang \etal \cite{wang_ca_gan} introduced an attention \cite{attention} mechanism-based conditional GAN framework for the task of UIR. The authors of \cite{han_access_2020} proposed the spiral GAN-based framework for the task of underwater image enhancement. Recently, Chen \etal \cite{chen_csvt} proposed a deep learning-based scheme that utilizes the physical priors for a perceptual enhancement of the degraded underwater images. Fabbri \textit{et al.} \cite{ugan} utilized the GANs for the task of UIR. In \cite{uieb}, authors proposed to utilize the \textbf{white-balanced (WB)}, \textbf{gamma-corrected (GC)}, and \textbf{histogram equalized (HE)} versions of degraded underwater images using a gated-fusion deep CNN.

\textbf{Underwater Single Image Super-resolution:}
\textbf{Single image super-resolution (SISR)} aims to restore the \textbf{higher resolution (HR)} counterpart of the \textbf{lower resolution (LR)} input \cite{sisr6, sisr5, sisr4, sisr3}.
The authors of \cite{mlp_access} proposed a multi-layer perceptron and color features-based method, namely SRCNN, for underwater image super-resolution. Chen \etal \cite{chen_access_2020} utilized the Wavelet-based deep CNN for underwater single image super-resolution and reconstruction.
In \cite{uwsisr}, the LR version of the underwater image is first processed with spatial and frequency domain constraints. The final HR version is later restored by employing a convex fusion rule. The authors of \cite{srdrm_srdrmgan} utilized the generative residual learning-based approach for underwater image super-resolution. Recently, in \cite{sesr}, the authors proposed a multi-modal objective function that considers the chrominance-specific degradations, image sharpness, and global contrast enhancement for underwater SISR.

\textbf{Major Observations:}
Despite the prolific literature of prior and deep learning-based methods, existing works suffer from visual artifacts, such as color distortion, in the enhanced underwater images. One of the primary reasons may be the direct plug and play of deep CNNs without properly supervising the contextual formulation of receptive fields across the channels based on their traversing ranges. \textcolor{black}{As we have described, in Section 1 of the supplementary material, unlike outdoor images, underwater images require special considerations due to different attenuation ranges across the channels. Hence, the direct plug and play of outdoor models may not be suitable in underwater scenario. Moreover,} it has been observed in one \cite{li} of the earlier works that a few channels contribute more towards de-hazing while the other helps in spatial color enhancement. Due to the semantic difference between the sub-tasks mentioned above, the channels may require different receptive field sizes. Further, a majority of the methods \cite{residual_access, guo_oe, liu_gan, wang_ca_gan, han_access_2020, ugan, srdrm_srdrmgan, funiegan} rely on adversarial training \cite{gan}, which may lead to instability \textcolor{black}{during training. Particularly,} due to adversarial learning, if not carefully crafted, the generator model may quickly start to generate the images to confuse the discriminator rather than producing the enhanced or de-noised underwater images \cite{divakar_cvpr}. Also, unlike \cite{uieb, tip21}, the proposed approach does not utilize any priors, \textit{e.g.}, WB, GC, HE, for estimating the enhanced images. \textcolor{black}{Such priors may help avoid learning a model that may not
perform well on unseen data due to a lack of semantic guidance and a large-scale dataset. An
analogy can be borrowed from the case of predicting alpha matte from a single image only without
using any prior, which is quite difficult to achieve \cite{mgmatting}. Whereas the Deep WaveNet is relatively self-sufficient in learning those semantic features without using any prior and simultaneously estimates
both the enhanced and super-resolved versions of the degraded underwater image.} In addition, Deep WaveNet does not comprises multiple residual dense blocks \cite{sisr3} and multi-modal loss functions, as in \cite{sesr}. We show that the right formulation of contextual sizes and proper use of attention modules are sufficient to generate notable performance gains in both enhancement and underwater SISR.   

\textbf{Our Contributions:} 
Considering the drawbacks mentioned above, we propose a multi-stage CNN, namely Deep WaveNet, that simultaneously enhances and improves the spatial resolution of the degraded underwater images. For this, we have supervised the receptive field size based on the attenuation-guided local and global coherence of the color channels.  It is a well-known fact that a receptive field with a larger (smaller) size may learn the global (local) features in an image better.  The global coherence in a generic underwater image mostly aligns with blue color. Whereas marine life closely aligns with green color. Therefore, based on our earlier observations of the wavelength-driven contextual size relationship, we decide to assign a larger receptive field to the blue color channel and whereas a smaller one to green and further reduced the size for red. In addition, we have also adopted a block attention-based \cite{cbam} skip refinement mechanism to adaptively regulate the channel-specific information flow across the proposed Deep WaveNet. Our key contributions are summarized as follows:
\begin{itemize}
\item We propose a multi-stage deep CNN framework for underwater image restoration (see Section \ref{proposed_approach}). The first stage supervises the color channels of the degraded input image with different contextual sizes considering its local and global semantics based on its attenuation range.  The intermediate stages aggregate the learned multi-contextual features and suppress the irrelevant color-localized skip information from the previous layers using an attention mechanism. The final stage focuses on the reconstruction of the enhanced image.

\item \textcolor{black}{Although the magnitude of such innovations may look simplistic, sometimes this is what one may all
need to achieve the best performance, \textit{e.g.,} \cite{anonymous2022patches}.}

\item For training, we have utilized the feature reconstruction loss \cite{perceptual} along with the traditional Mean Squared Error (see Section \ref{proposed_approach}).

\item To show the efficacy, we have presented a comprehensive set of experiments against nearly 20 existing best-published works on underwater image enhancement and super-resolution on over 12 image quality metrics (see Sections \ref{sec:exp_setup},\ref{sec:results}). Also, an ablation study has been presented at the end to demonstrate the effect of various cost functions and network modules (see Section \ref{sec:ablation}). 

\item To demonstrate the robustness of the proposed model across a variety of tasks, we have shown the comparative results on underwater image semantic segmentation and 2D divers pose estimation (see Section \ref{sec:more_results}).
\end{itemize}
\section{The Approach Description}
\label{proposed_approach}

\begin{figure}
\includegraphics[width=0.8\textwidth, height=2.5cm]{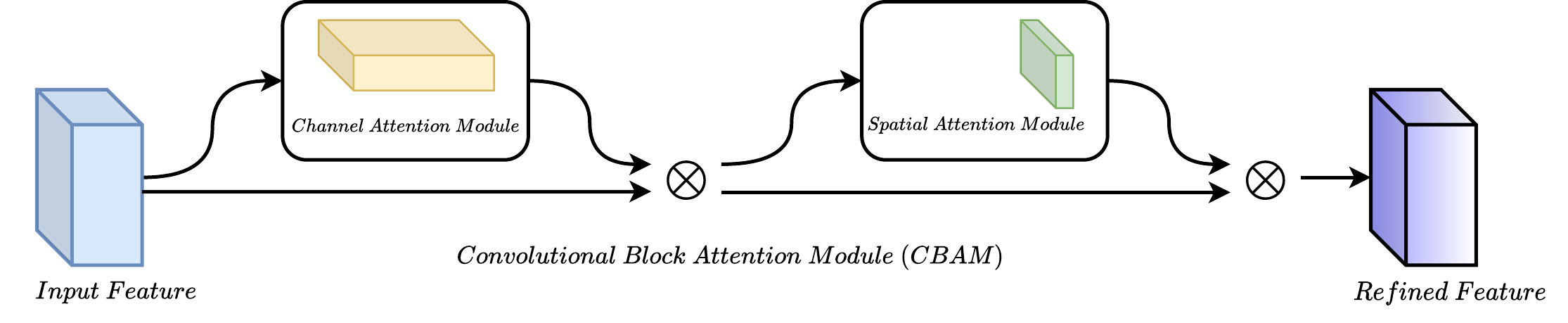}
\caption{\textbf{Overview of the CBAM module.} It has been used in the Deep WaveNet to adaptively refine the learned multi-contextual features.}
\label{fig:cbam_overview}
\end{figure}

\subsection{Convolutional Block Attention Module (CBAM)}
Woo \textit{et al.} \cite{cbam} proposed the CBAM module, which has been incorporated to extract the channel and spatial attention features in the proposed architecture for the given intermediate feature map as the input. The obtained attention maps are then multiplied with the given input features for the adaptive refinement, as shown in Fig. \ref{fig:cbam_overview}. Instead of its conventional usage just after the convolution layer, in Deep WaveNet, we leverage the CBAM's potential by utilizing it after color-localized skip connections of stages 2 and 4. 

To formally describe its working principle, let ${M}_s({J})$ and ${M}_c({J})$ be the spatial and channel attention maps for the intermediate feature map ${J}$, which can be computed as follows:
\begin{equation}
\begin{split}
    {M}_s({J})&=\sigma (p^{7\times 7}([\mathcal{A}({J});\mathcal{M}({J})]))\\
    					&= \sigma(p^{7\times 7}([{J}^s_{\mathbf{avg}};{J}^s_{\mathbf{max}}]))\\
    					{M}_c({J})&=\sigma (\mathbf{F}^{c}({J}^c_{\mathbf{avg}}) 
    					+ \mathbf{F}^{c}({J}^c_{\mathbf{max}})),
    \end{split}
\end{equation}
where $p^{7X7}$, $\sigma$, $\mathbf{F}^c$ refer to a convolution operation with a kernel of size $7X7$, Sigmoidal function, and a multi-layer perceptron with 1 hidden layer, respectively. $\mathcal{M}$, $\mathcal{A}$ denote the max and average pooling, respectively. Then the final refined feature ${J}^{r}$ can be written as
\begin{equation}
\begin{split}
{H} &= {M}_c ({J}) \otimes {J}\\
{J}^r &= {M}_s ({H}) \otimes {H},
\end{split}
\end{equation}
We encourage the readers to refer \cite{cbam} for more details. We have shown that the refined features have been beneficial in constructing the visually pleasant enhanced underwater images in Section \ref{sec:ablation_wave_cbam}.

\begin{figure*}[t]
\includegraphics[width=1.07\textwidth]{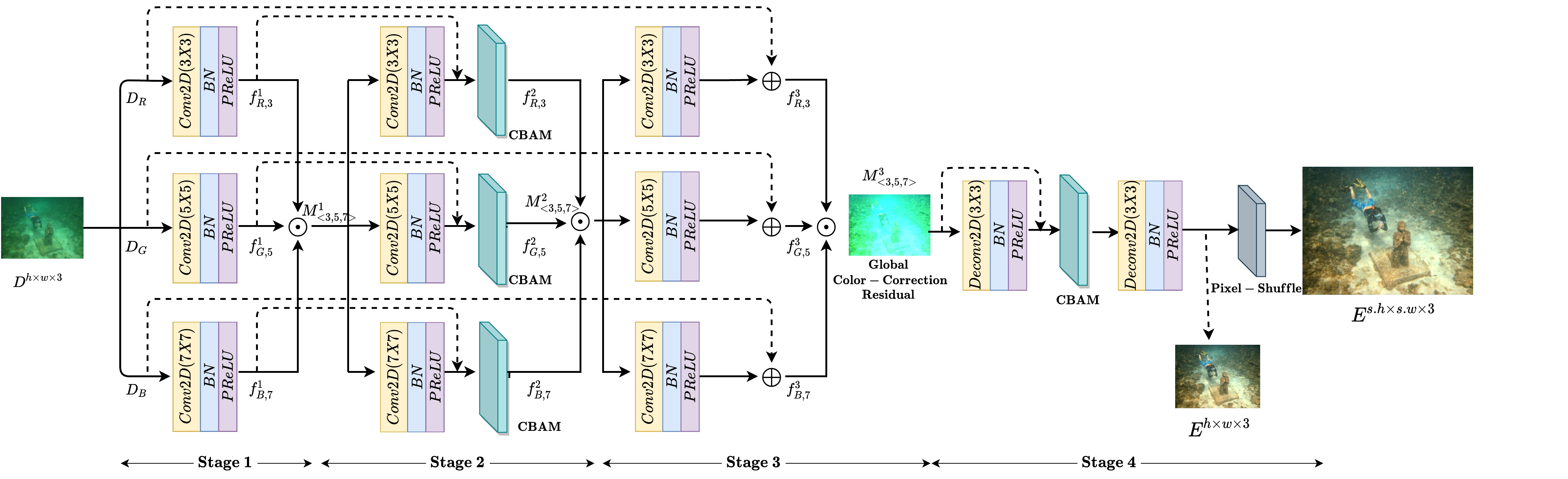}
\caption{\textbf{Dataflow of the proposed model for the simultaneous underwater image enhancement and super-resolution.} CBAM and Pixel-shuffle operations are described in Section \ref{proposed_approach}. The proposed model takes the degraded underwater image as input and outputs the visually and spatially enhanced image.}
\label{fig:main_model}
\end{figure*}
\subsection{Network Architecture}
\textbf{Goal:} Learn a unified deep learning-based model ($\mathcal{W}(\theta)$) for simultaneous enhancement and super-resolution of degraded underwater images, as shown in Fig. \ref{fig:main_model}. 

\noindent \textbf{Notations:} Let $D$, $E$ be the degraded and enhanced underwater images, respectively. We also denote the red, blue, and green color channels of $D$ as $D_R$, $D_G$, and $D_B$, respectively. We denote the contextual features channel $c$ at stage $i$ with receptive field size $s$ as $f^i_{c,s}$. Below, we describe the regimes of operations of each stage of the proposed Deep WaveNet. While $\odot$ denotes channel-wise concatenation, $\oplus$ refers to pixel-wise addition of the features.

\subsubsection{Stage 1}
The first stage aims to generate the channel-specific features with wavelength-driven contextual sizes. For this, we input the degraded underwater image $D$, channel-wise, and obtain the multi-contextual features as
\begin{equation}
M^1_{<3,5,7>} = f^1_{R,3} \odot f^1_{G,5} \odot f^1_{B,7},
\end{equation}
where, the above color-specific features can be estimated as 
\begin{equation}
\begin{split}
f^1_{R,3} &= g(bn(p^{3\times 3} (D_R)))\\
f^1_{G,5} &= g(bn(p^{5\times 5} (D_G)))\\
f^1_{B,7} &= g(bn(p^{7\times 7} (D_B))),
\end{split}
\end{equation} 
with $bn$, $g$ denote batch normalization \cite{bn} and parametric ReLU layers, respectively. 

\subsubsection{Stage 2}
The first layer of stage 2, similar to stage 1, consists of a stack of convolutional layers with different receptive fields. More specifically, stage 2 aims to generate the color-specific distortion residuals from the learned stage 1 multi-contextual features as  
\begin{equation}
\begin{split}
f^2_{R,3} &= g(bn(p^{3\times 3} (M^1_{<3,5,7>} ))) \odot f^1_{R,3}\\
f^2_{G,5} &= g(bn(p^{5\times 5} (M^1_{<3,5,7>} ))) \odot f^1_{G,5} \\
f^2_{B,7} &= g(bn(p^{7 \times 7} (M^1_{<3,5,7>} ))) \odot f^1_{B,7}.
\end{split}
\end{equation} 
The obtained residuals are further adaptively refined using CBAM \cite{cbam} modules as 
\begin{equation}
\begin{split}
f^2_{R,3} &= \textit{CBAM} (f^2_{R,3}) \\
f^2_{G,5} &= \textit{CBAM} (f^2_{G,5}) \\
f^2_{B,7} &= \textit{CBAM} (f^2_{B,7}).
\end{split}
\end{equation} 
\textcolor{black}{It has been done to ensure the two-fold goals: (a) the color-specific noisy features from stage 1
should not be propagated further to subsequent stages, and (b) towards predicting “global color correction residual”,
the model should not ignore the color-specific details, which it might while processing the whole input
image at one go.} The output of stage 2 can be defined as
\begin{equation}
M^2_{<3,5,7>} = f^2_{R,3} \odot f^2_{G,5} \odot f^2_{B,7}.
\end{equation} 

\subsubsection{Stage 3}
Stage 3 takes the intermediate multi-contextual attentive residuals as input and outputs the global color-correction residual map as 
\begin{equation}
M^3_{<3,5,7>} = f^3_{R,3} \odot f^3_{G,5} \odot f^3_{B,7},
\end{equation}  
where 
\begin{equation}
\begin{split}
f^3_{R,3} &= g(bn(p^{3\times 3} (M^2_{<3,5,7>} ))) \oplus D_R\\
f^3_{G,5} &= g(bn(p^{5\times 5} (M^2_{<3,5,7>} ))) \oplus D_G \\
f^3_{B,7} &= g(bn(p^{7 \times 7} (M^2_{<3,5,7>} ))) \oplus D_B.
\end{split}
\end{equation}

\subsubsection{Stage 4}
The final stage acts as the reconstruction module, consisting of a deconvolution layer, followed by an attentive residual block and final deconvolution layer. Stage 4 takes global color-correction residual as input and outputs the enhanced underwater image. The regimes of operations of stage 4 are as follows:
\begin{equation}
\begin{split}
f^4 &= g(bn(w^{3\times 3} (M^3_{<3,5,7>} ))) \odot M^3_{<3,5,7>}\\
f^4 &= \textit{CBAM}(f^4)\\
E &= g(bn(w^{3\times 3} (f^4 ))), 
\end{split}
\label{eq:stage_4}
\end{equation}
where $w^{3\times 3}$ denotes the 2D transpose convolution operation. 

Before deep-diving into the super-resolution sub-stage, note that the proposed work is dedicated to underwater image restoration, which includes either image enhancement, spatial super-resolution, or both. Given a paired dataset for underwater image enhancement, the proposed network can be trained till stage 4 without improving the spatial resolution of the enhanced image. In that case, the number of output channels in Eq. \ref{eq:stage_4} must be set to 3.

However, if the degraded input images of the dataset mentioned above can be transformed into the lower-resolution images using interpolation techniques, then it can be used for both underwater image enhancement and spatial super-resolution. In that case, to add the super-resolution sub-stage in the proposed model after stage 4 of the enhancement module, the output channels in Eq. \ref{eq:stage_4} must be set to $3s^2$, where $s$ denotes the scale factor by which the spatial resolution is to be enhanced. The detailed regimes of operations of the super-resolution sub-stage are explained below.  

\subsubsection{The Super-Resolution Sub-stage}
Once stage 4 outputs the enhanced image $E$, the post-processing layers consist of 2D convolution, and pixel-shuffle \cite{ps} operations aim to improve the spatial resolution of the image. The other way to improve the resolution of the input features from the downsampling or max-pooling layers is to utilize the transpose convolution, widely known as deconvolution \cite{deconv} or backward convolution \cite{back_conv}. The deconvolution operator with a functional stride $st = 1$ behaves identically to convolution \cite{deconv1, deconv2}. However, with the $st > 1$,  the computational cost increases, and the deconvolution operation induces the visual artifacts in the spatially super-resolved features \cite{ps}. Therefore, instead of employing deconvolution, we leverage the pixel-shuffle operation for spatial upscaling of the input features. The in-depth assessment of the efficiency of the sub-pixel convolution and pixel-shuffle operations over deconvolution is beyond the scope of this work. Hence, we encourage the readers to refer \cite{ps} for more details.

To formally define the super-resolution stage, given the \textit{stage 4} output, let $E \in \mathbb{R}^{h\times w\times 3.s^2}$ be the lower-resolution image or feature maps. With a spatial scale factor $s$, this stage aims to produce the higher-resolution image or features $\hat{E} \in \mathbb{R}^{s.h\times s.w\times 3}$. For simplicity of the notations, let $\hat{E}$ be denoted as $E$. Then, the higher-resolution $E^{s.h\times s.w\times 3}$ can be written as 
\begin{equation}
E^{s.h\times s.w\times 3} = \mathcal{PS} (E^{h\times w\times 3.s^2}),
\end{equation}
where $\mathcal{PS}$ denotes the pixel-shuffle operation. It periodically re-arranges the features of shape $h \times w \times k.s^2$ into $s.h \times s.w \times k$ where $k=3$ (an RGB image). It can be mathematically defined as 
\begin{equation}
\mathcal{PS}(Z)_{u,v,k} = Z_{\lfloor u/s \rfloor, \lfloor v/s \rfloor, k.s.\text{mod}(v,s)+ k.\text{mod}(u,s)},
\end{equation}
where $u,v$ are output pixel coordinates in higher-resolution space. We have designed the post-processing layers to incorporate $2\times$, $3\times$ and $4\times$ spatial scale resolution. It should be mentioned that the flexibility of setting number of output channels in Eq. \ref{eq:stage_4} shows the conceptual modularity of the proposed work across different UIR tasks.

\subsection{Model Learning}
Following the existing image restoration works \cite{ours_wacv, ours_icip}, to train the Deep WaveNet, we first have incorporated the traditional Mean Squared Error ($\ell_2$) as,
\begin{equation}
\mathcal{L}_2 (\theta) = \frac{1}{b} \sum_{j=1}^{b} \| \mathcal{W}(\theta; D_j) - O_j \|_2^2,
\end{equation}  
where $O$ is the original clean and higher-resolution underwater image. It is a widely known fact that the $\ell_2$-norm based minimization may suffer from blurry artefacts in the restored images. 

Therefore, to overcome this drawback, we have also incorporated the Perceptual loss \cite{perceptual} function that helps in retaining the high-frequency details of the image. For this, we have utilized VGG16 ($\mathcal{V}(\Theta)$) \cite{vgg} model pre-trained on ImageNet \cite{imagenet} dataset. We define the $\ell_2$ norm between the \texttt{relu2\_2} features obtained by using predicted enhanced image and ground truth underwater image, as cost function, and can be written as, 
\begin{equation}
\mathcal{L}_P (\theta) = \frac{1}{b} \sum_{j=1}^{b} \| \mathcal{V}(\mathcal{W}(\theta;D_j); \Theta) - \mathcal{V}(O_j;\Theta)  \|_2^2.
\end{equation}
In addition to $\mathcal{L}_2$ and $\mathcal{L}_P$ losses, we have also utilized the Structural Similarity Index Measure (SSIM) \cite{ssim} as a loss function to minimize the structural differences between $E$ and $O$. The SSIM reflects the similarity between two images and can be written as
\begin{equation}
{SSIM}(r) = \frac{2.\mu_m.\mu_n + Z_1}{\mu^2_m + \mu^2_n + Z_1}. \frac{2.\sigma_{mn} + Z_2}{\sigma^2_m + \sigma^2_n + Z_2},
\end{equation} 
where $m$, $n$ are patches from $E$, and $O$, respectively, $\mu$, $\sigma$, and $\sigma_{mn}$ denote mean, std-dev and covariance, given $r$ the center of patches $m$, $n$. $Z_1$, $Z_2$ are fixed parameters. Then, the incorporated SSIM loss can be written as
\begin{equation}
\mathcal{L}_{SSIM}(\theta) = \frac{1}{2b} \sum_{j=1}^{b} 1 - SSIM(\mathcal{W}(\theta; D_j), O_j)
\end{equation}

Finally, the proposed scheme is optimized using the following objective function
\begin{equation}
\arg \min_{\theta \in \mathbb{R}} \mathcal{L}_2 (.) + \lambda_P. \mathcal{L}_P (.) + \lambda_S. \mathcal{L}_{SSIM} (.),
\end{equation}
where $\lambda_P$, $\lambda_S$ have been empirically set to \texttt{0.02}, \texttt{0.5}, respectively. 

During our experimentation, it has been observed that the $\mathcal{L}_2$ and $\mathcal{L}_P$ losses are sufficient for the task of underwater image enhancement alone. However, for simultaneous image enhancement and super-resolution, $\mathcal{L}_{SSIM}$ loss has to be incorporated for better performance. Therefore, for better utilization of computing resources, $\lambda_S$ is set to 0 when trained for image enhancement. Also, intuitively, the addition of $\mathcal{L}_{SSIM}$ loss for the task of image enhancement, which mostly relates to color-correction, may not be much useful in the underwater scenario.
  
\section{Experimental Setup}
\label{sec:exp_setup}

\subsection{Datasets and Training Setup}
We have utilized the publicly available underwater image enhancement and super-resolution benchmarks, namely \texttt{UIEB} \cite{uieb}, \texttt{EUVP} \cite{euvp}, and \texttt{UFO-120} \cite{sesr}. For the enhancement task, we have adopted the \texttt{EUVP} training dataset that comprises of 11435 paired underwater images, each of size $256 \times 256$. The \texttt{EUVP} test-set consists of 515 image pairs of the same size. To show the results on the \texttt{UIEB} dataset, we have finetuned the pre-trained model learned using the \texttt{EUVP} dataset.

The \texttt{UIEB} dataset consists of 890 paired images. Following the \cite{uieb}, a random subset of 800 images is used as the training set, and the rest 90 images are utilized as the test-set. We have resized the training images to a size of $512\times 512$ given the memory constraint. Note that the chosen test-set may not align with the author's original test-set. Therefore, we have followed 5-fold cross-validation for a fair comparison and reported the mean results. We have also utilized the \texttt{Challenge} set of the \texttt{UIEB} dataset that consists of 60 degraded underwater images without ground truth references.

For simultaneous enhancement and super-resolution, we have utilized the \texttt{UFO-120} dataset that comprises 1500 paired images for training and 120 images for testing. The ground truth images are of shape $640\times 480$. Furthermore, we have modularized the proposed Deep WaveNet to assist $2\times$, $3\times$, and $4\times$ SR configurations. 

We have trained the proposed model using Pytorch \cite{pytorch} framework with Adam \cite{adam} optimization and an initial learning rate of $2e-4$ on Nvidia Tesla V100 GPU. The training progressed for about 2.3K iterations. The model takes $\sim 5$ GB of memory with a batch size of $5$. \textcolor{black}{Our proposed Deep WaveNet is a lightweight model of size just 3.23 MB, and can process an image of shape $640 \times 480$ within 0.38 seconds. }.
\subsection{Competing Methods}
\subsubsection{Underwater Image Enhancement:} For this task, we have compared the proposed scheme with  following existing best-published works: (i) Fusion-based \cite{ancuti1} (CVPR'12), (ii) Retinex-based \cite{retinex-based} (ICIP'14), (iii) Histogram Prior \cite{li_tip} (TIP'16), (iv) Blurriness-based \cite{peng_tip} (TIP'17), (v) GDCP \cite{gdcp} (TIP'18), (vi) Water CycleGAN \cite{li_spl} (SPL'18), (vii) Dense GAN \cite{guo_oe} (OE'19), (viii) Water-Net \cite{uieb} (TIP'19), (ix) Haze Lines \cite{berman_pami_20} (TPAMI'20), (x) UGAN \cite{ugan} (ICRA'18), (xi) Funie-GAN \cite{funiegan} (RAL'2020), (xii) Deep SESR \cite{sesr} (RSS'2020), (xiii) Ucolor \cite{tip21} (TIP'21).
\subsubsection{Underwater Image Super-Resolution:} For the task of single image super-resolution, we have compared the proposed work with the following state-of-the-art baselines: (i) SRCNN \cite{srcnn} (TPAMI'16), (ii) SRResNet \cite{srresnet_srgan} (CVPR'17), (iii) SRGAN \cite{srresnet_srgan} (CVPR'17), (iv) SRDRM \cite{srdrm_srdrmgan} (ICRA'20), (v) SRDRM-GAN \cite{srdrm_srdrmgan} (ICRA'20), (vi) Deep SESR \cite{sesr} (RSS'20).
\subsection{Evaluation Metrics}
We have incorporated both reference and reference-less image quality metrics for a robust comparison of the proposed scheme, as follows: Mean-Squared Error (MSE), Peak Signal-to-Noise Ratio (PSNR),  SSIM, Underwater Image Quality Measure (UIQM) \cite{uiqm}, Natural Image Quality Evaluator (NIQE) \cite{niqe}, Patch-based Contrast Quality Index (PCQI) \cite{pcqi}, Underwater Image Sharpness Measure (UISM) \cite{uiqm}, Visual Information Fidelity (VIF) \cite{vif}, Average Entropy (E), Average Gradient, Underwater Image Contrast Measure (UIConM) \cite{uiqm}, Underwater Color Image Quality Evaluation (UCIQE) \cite{uciqe}. 

\section{Results}
\label{sec:results}

\begin{table*}[t]
\begin{center}
\caption{Comparison against the best-published works on \texttt{EUVP} dataset for the task of image enhancement. $\bigtriangledown$ denotes lower is better. Best and second-best results are shown in \textcolor{red}{red} and \textcolor{blue}{blue} colors, respectively.}
\label{tab:euvp}
\resizebox{\textwidth}{!}{
\begin{tabular}{lcccccccccccc}
\toprule
\midrule
\text{Methods} & \text{MSE}$\bigtriangledown$ & \text{PSNR} & \text{SSIM} & \text{UIQM} & \text{NIQE}$\bigtriangledown$  & \text{PCQI} & \text{UISM} & \text{VIF} & \text{$\mathbf{E}$}$\bigtriangledown$  & \text{AG} & \text{UIConM} & \text{UCIQE}\\
\midrule
\midrule

UGAN \cite{ugan} & $.36$&$26.55$&{$.80$}&$2.89$&{\color{blue}$\mathbf{49.90}$}&$.700$&$6.84$&{\color{blue}$\mathbf{.402}$}&$7.52$&$7.48$&$.79$&$.581$\\

UGAN-P \cite{ugan} & $.36$&$26.54$&$.80$&$2.93$&$50.17$&{\color{blue}$\mathbf{.704}$}&$6.83$&$.400$&$7.54$&$7.58$&$.79$&{\color{red}$\mathbf{.590}$}\\

Funie-GAN \cite{funiegan} & $.39$&$26.22$&$.79$&{\color{black}${2.97}$}&$50.51$&{\color{red}$\mathbf{.706}$}&{\color{black}$6.90$}&$.384$&$7.55$&{\color{red}$\mathbf{8.58}$}&{\color{red}$\mathbf{.84}$}&{\color{red}$\mathbf{.590}$}\\

Funie-GAN-UP \cite{funiegan} &$.60$ &$25.22$&$.78$&{\color{black}${2.93}$}&$52.87$&$.702$&$6.86$&$.394$&$7.50$&{\color{blue}$\mathbf{7.80}$}&{\color{blue}$\mathbf{.79}$}&{\color{black}$.588$}\\

Deep SESR \cite{sesr} & {\color{blue}$\mathbf{.34}$}&{\color{blue}$\mathbf{27.08}$}&{\color{blue}$\mathbf{.80}$}&{\color{red}$\mathbf{3.09}$}&$55.68$&$.679$&{\color{red}$\mathbf{7.06}$}&$.384$&{\color{blue}$\mathbf{7.40}$}&$7.57$&$.78$&$.572$\\
\midrule
Deep WaveNet & {\textcolor{red}{$\mathbf{.29}$}}&{\color{red}$\mathbf{28.62}$}&{\color{red}$\mathbf{.83}$}&{\color{blue}$\mathbf{3.04}$}&{\color{red}$\mathbf{44.89}$}&$.694$&{\color{red}$\mathbf{7.06}$}&{\color{red}$\mathbf{.438}$}&{\color{red}$\mathbf{7.38}$}&$7.00$&$.77$&$.559$\\
\bottomrule
\bottomrule
\end{tabular}}

\end{center}
\end{table*}
\begin{table}
\caption{Comparison against the best-published works on \texttt{UIEB} dataset for the task of image enhancement. Best and second-best results are shown in \textcolor{red}{red} and \textcolor{blue}{blue} colors, respectively.}
\label{tab:uieb}
\centering
\resizebox{0.4\textwidth}{!}{
\begin{tabular}{lccc}
\toprule
\toprule
Methods & MSE & PSNR & SSIM\\
\midrule
\midrule
Input & $1.29$ & $17.11$ & $.61$ \\
Fusion-based \cite{ancuti1} & $0.91$ & \textcolor{blue}{$\mathbf{21.23}$} & $.78$ \\
Histogram Prior \cite{li_tip} & $1.70$ & $15.85$ & $.53$ \\
Retinex-based \cite{retinex-based} & $1.34$ & $17.66$ & $.61$ \\
GDCP \cite{gdcp} & $3.33$ & $13.86$ & $.55$\\
Blurriness-based \cite{peng_tip} & $1.91$ & $15.31$ & $.60$\\
Water CycleGAN \cite{li_spl} & $1.72$ & $15.75$ & $.52$\\
DenseGAN \cite{guo_oe} & $1.21$ & $17.28$ & $.44$\\
WaterNet \cite{uieb} & \textcolor{blue}{$\mathbf{0.79}$} & $19.11$ & \textcolor{blue}{$\mathbf{.79}$} \\
Hazelines \cite{berman_pami_20} & $2.44$ & $15.17$ & $.57$\\
Deep SESR \cite{sesr} & $1.70$ & $16.65$ & $.57$ \\
\midrule
Deep WaveNet & {\textcolor{red}{$\mathbf{0.60}$}} & {\textcolor{red}{$\mathbf{21.57}$}} & {\textcolor{red}{$\mathbf{.80}$}}\\
\bottomrule
\bottomrule
\end{tabular}}
\end{table}

\begin{table}[t]
\caption{Comparison against the best-published works on \texttt{Challenge} set for the task of image enhancement.
Best and second-best results are shown in \textcolor{red}{red} and \textcolor{blue}{blue} colors, respectively.}
\label{tab:tc60}
\resizebox{\linewidth}{!}{
\begin{tabular}{ccccccccccc}
\hline
\hline
Measure & Input & Fusion-based \cite{ancuti1} & Histogram Prior \cite{li_tip} & Blurriness-based \cite{peng_tip} & GDCP \cite{gdcp} & DenseGAN \cite{guo_oe} & Water CycleGAN \cite{li_spl} & WaterNet \cite{uieb} & Ucolor \cite{tip21} & Deep WaveNet \\
\hline
\hline
UIQM & 0.84 & 1.22 & \textcolor{blue}{1.27} & 1.13 & 1.07 & 1.11 & 0.91 & 0.97 & 0.88 & \textcolor{red}{2.14} \\
NIQE & 7.14 & \textcolor{blue}{4.94} & 5.32 & 6.01 & 5.92 & 5.71 & 7.67 & 6.04 & 6.21 & \textcolor{red}{3.98} \\
\hline
\hline
\end{tabular}}
\end{table}

\begin{table*}[t]
\centering
\caption{Comparison against the best-published works on \texttt{UFO-120} dataset for the task of underwater image super-resolution. Best and second-best results are shown in \textcolor{red}{red} and \textcolor{blue}{blue} colors, respectively.}
\label{tab:sr}
\resizebox{\textwidth}{!}{
\begin{tabular}{lccccccccc}
  \toprule
 \midrule
\text{Methods} & \multicolumn{3}{c}{\text{PSNR}} & \multicolumn{3}{c}{\text{SSIM}} & \multicolumn{3}{c}{\text{UIQM}}\\  
  \cmidrule{2-10}

& $2\times$ & $3\times$ & $4\times$ & $2\times$ & $3\times$ & $4\times$ & $2\times$ & $3\times$ & $4\times$\\

\midrule  
\midrule
  SRCNN \cite{srcnn} & $24.75\pm3.7$ & $22.22\pm3.9$ & $19.05\pm2.3$ & $.72\pm.07$ & $.65\pm.09$ & $.56\pm.12$ & $2.39\pm0.35$ & $2.24\pm0.17$ & $2.02\pm0.47$ \\ 
  SRResNet \cite{srresnet_srgan} & $25.23\pm4.1$ &$23.85\pm2.8$ &$19.13\pm2.4$ &$.74\pm.08$ & $.68\pm.07$ &$.56\pm.05$ &$2.42\pm0.37$ &$2.18\pm0.26$ &$2.09\pm0.30$ \\  
  SRGAN \cite{srresnet_srgan} &\color{red}$\mathbf{26.11\pm3.9}$ &$23.87\pm4.2$ &$21.08\pm2.3$ &{\color{blue}$\mathbf{.75\pm.06}$} &$.70\pm.05$ & $.58\pm.09$&$2.44\pm0.28$ &$2.39\pm0.25$ &$2.26\pm0.17$ \\ 
  SRDRM \cite{srdrm_srdrmgan}$^\ddagger$ & $24.62\pm2.8$ & $-$ & $23.15\pm2.9$ & $.72\pm.17$ & $-$ & {\color{blue}$\mathbf{.67\pm.19}$} & $2.59\pm0.64$ & $-$  & $2.56\pm0.63$ \\ 
  SRDRM-GAN \cite{srdrm_srdrmgan}$^\ddagger$ & $24.61\pm2.8$ & $-$ & $23.26\pm2.8$  &  $.72\pm.17$ &  $-$ & $.67\pm.19$ & $2.59\pm0.64$ & $-$  & {\color{blue}$\mathbf{2.57\pm0.63}$}     \\ 
  Deep SESR \cite{sesr}& {$25.70 \pm 3.2$}  & {\color{red}$\mathbf{26.86 \pm 4.1}$}  & {\color{blue}$\mathbf{24.75 \pm 2.8}$}  & {$.75 \pm .08$}  & {\color{blue}$\mathbf{.75 \pm .06}$}  & {$.66 \pm .05$}  & {\color{red}$\mathbf{3.15\pm0.48}$}  & {\color{blue}$\mathbf{2.87 \pm 0.39}$}  & {$2.55 \pm 0.35$}  \\ 
  \midrule
  Deep WaveNet & \textcolor{blue}{$\mathbf{25.71 \pm 3.0}$}& {\color{blue}$\mathbf{25.23 \pm 2.7}$} & {\color{red}$\mathbf{25.08 \pm 2.9}$}& \color{red}$\mathbf{.77 \pm .07}$& {\color{red}$\mathbf{.76 \pm .07}$} &{\color{red}$\mathbf{.74 \pm .07}$}& {\color{blue}$\mathbf{2.99\pm0.57}$}& {\color{red}$\mathbf{2.96\pm0.60}$}& {\color{red}$\mathbf{2.97\pm0.59}$} \\
  
  \bottomrule
  \bottomrule

\end{tabular}}
\end{table*}

\begin{figure*}
\resizebox{\textwidth}{!}{
\setlength{\tabcolsep}{1pt}
\begin{tabular}{cccccccc}

\includegraphics[width=3cm, height=2cm]{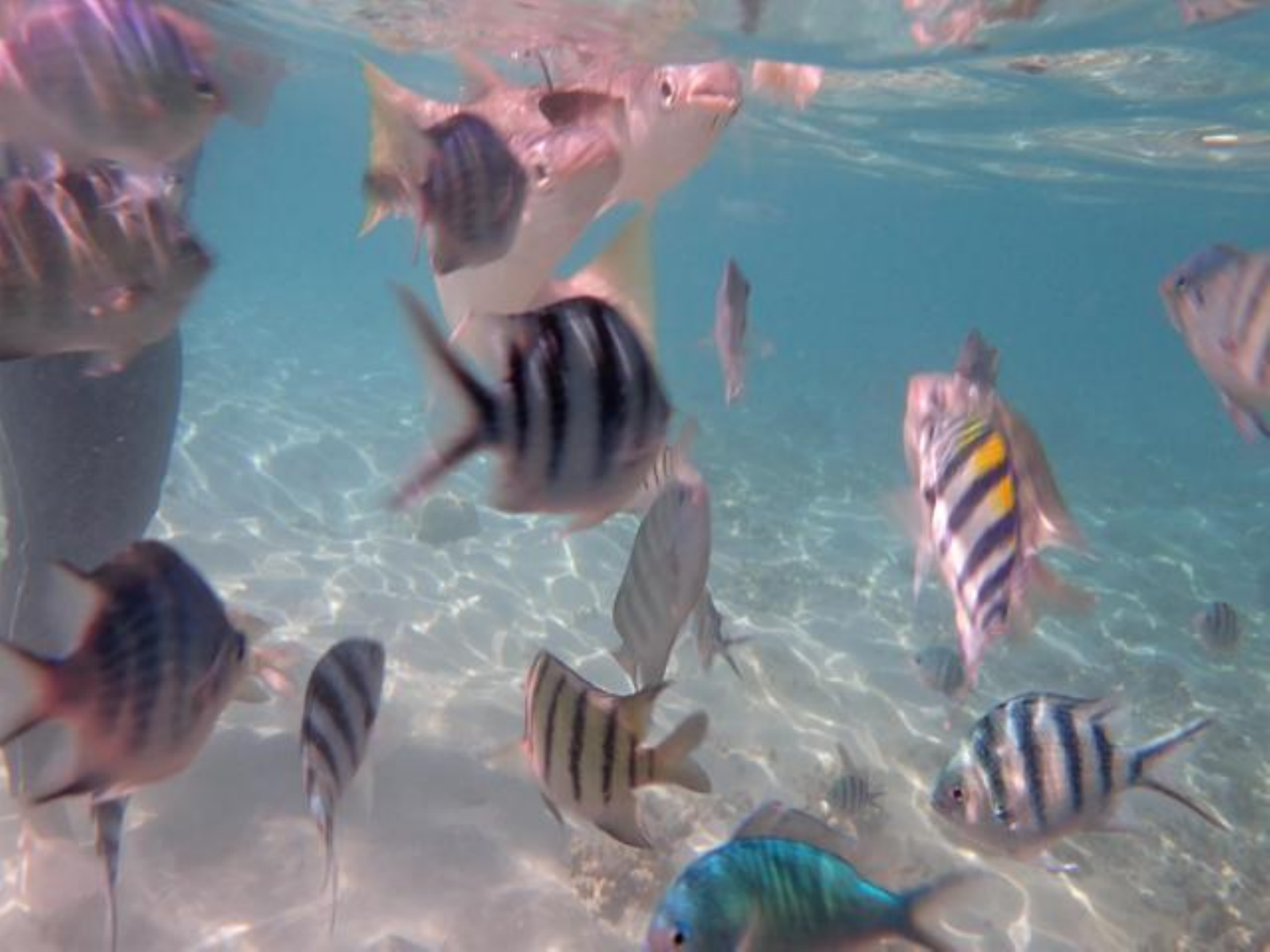}&
\includegraphics[width=3cm, height=2cm]{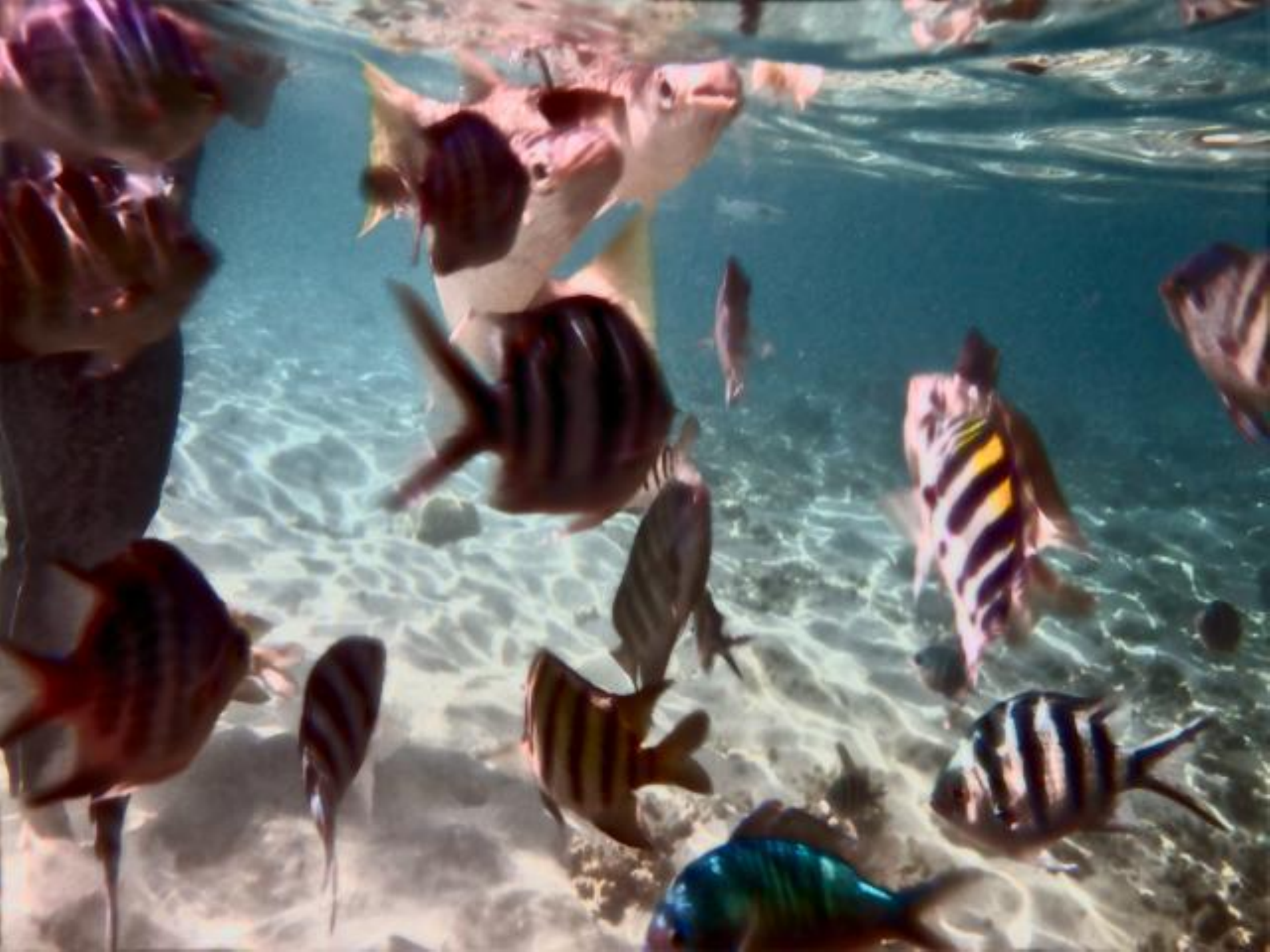}&
\includegraphics[width=3cm, height=2cm]{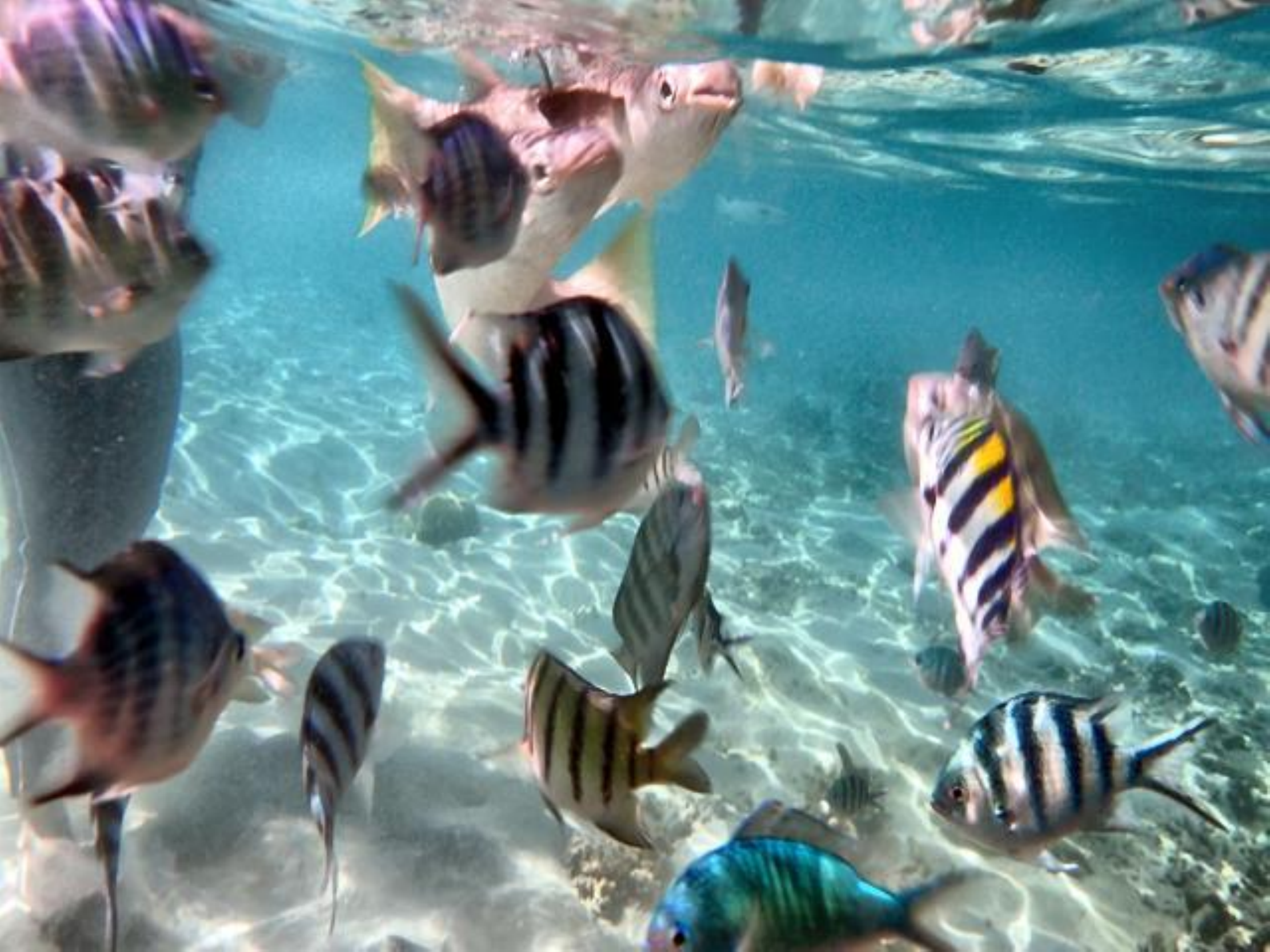}&
\includegraphics[width=3cm, height=2cm]{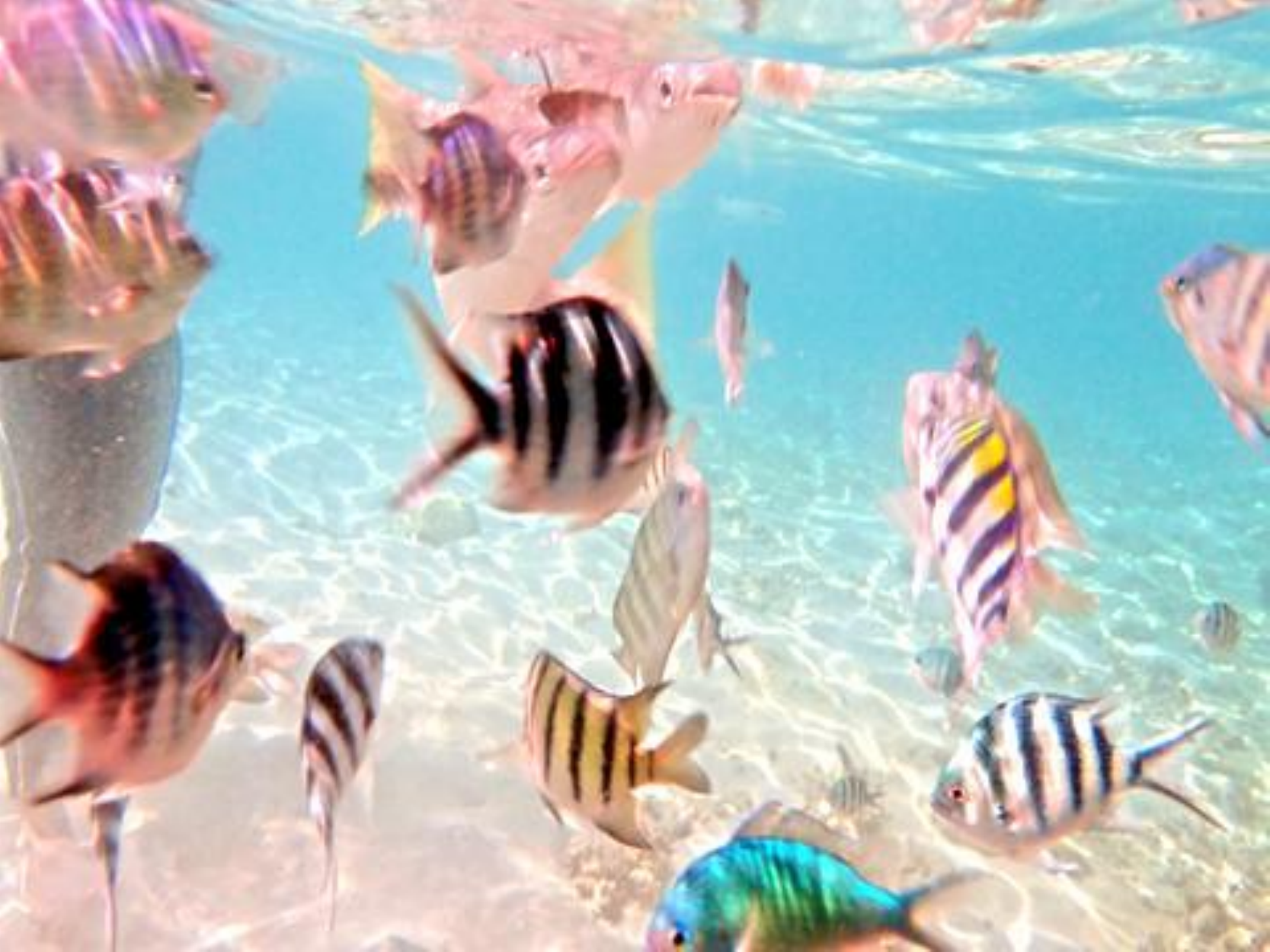}&
\includegraphics[width=3cm, height=2cm]{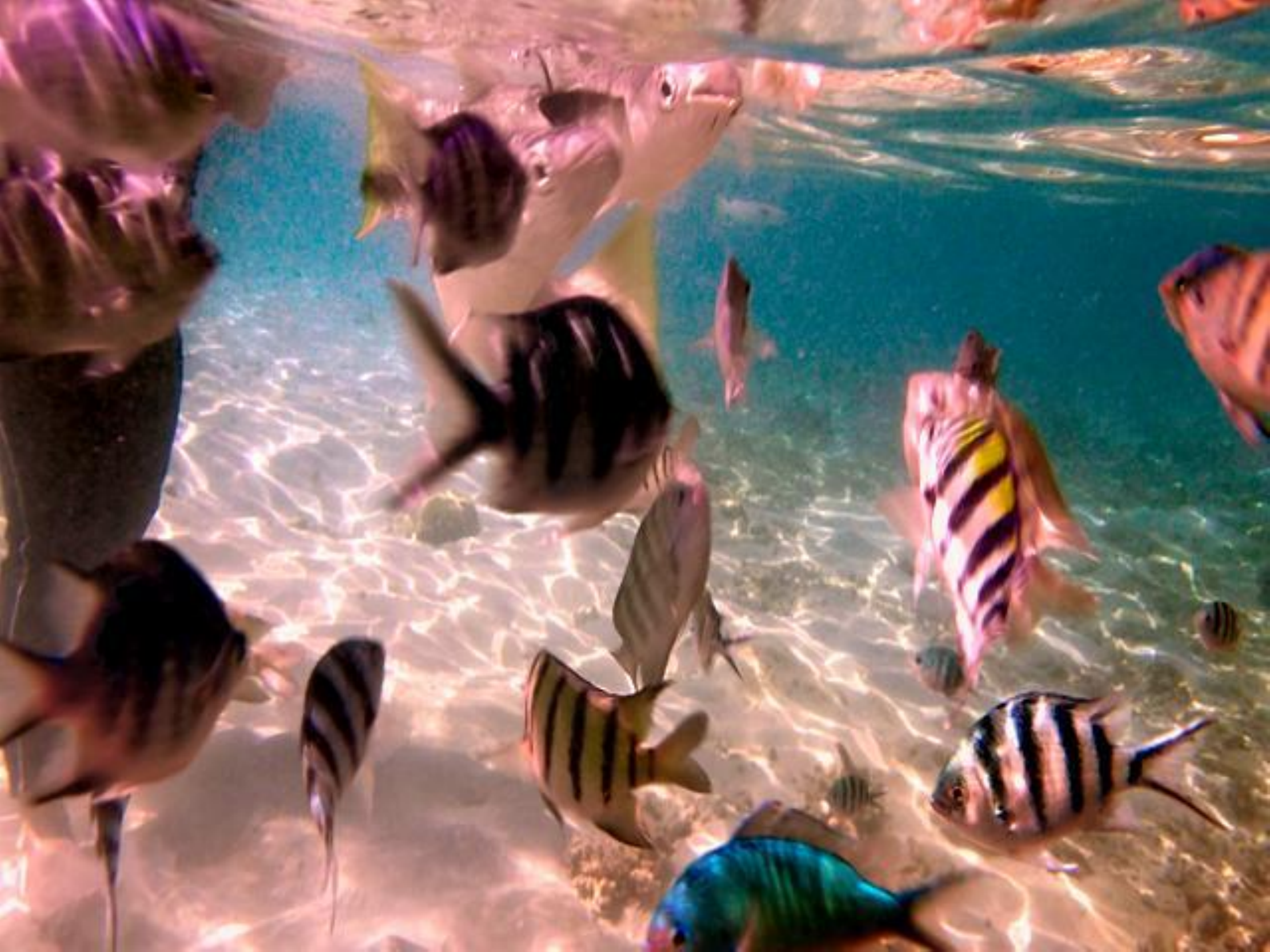}&
\includegraphics[width=3cm, height=2cm]{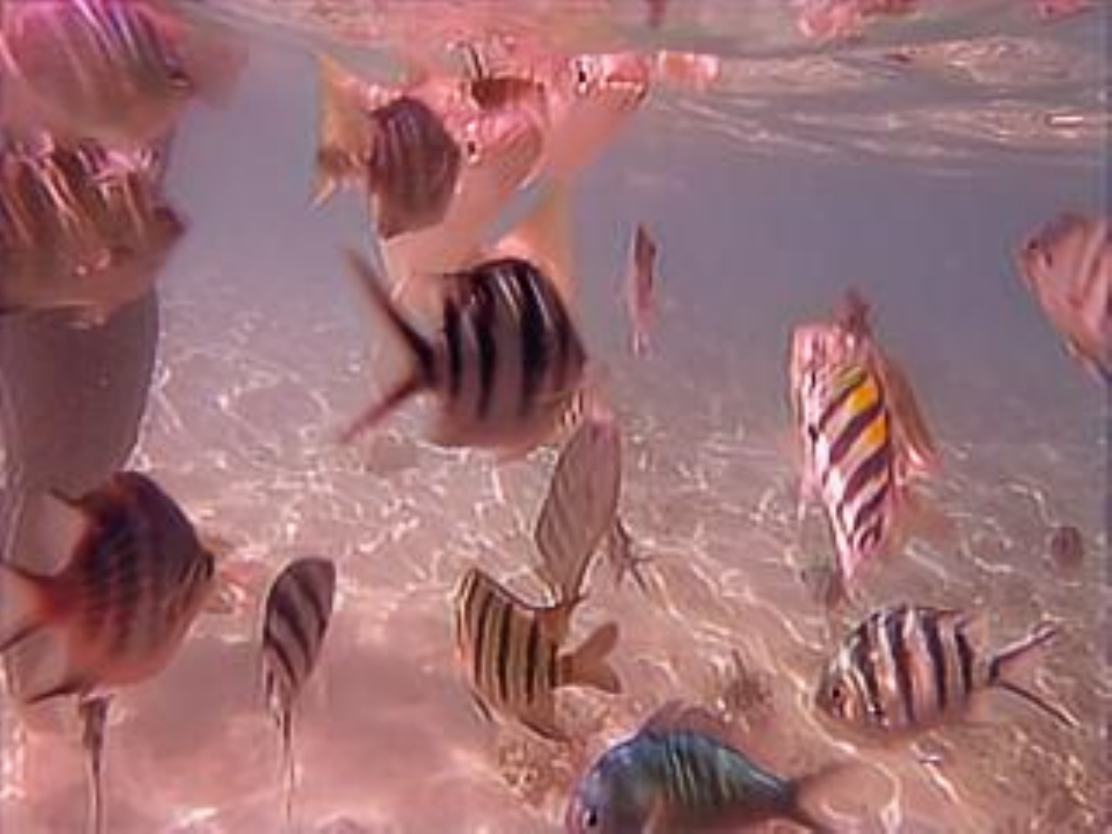}&
\includegraphics[width=3cm, height=2cm]{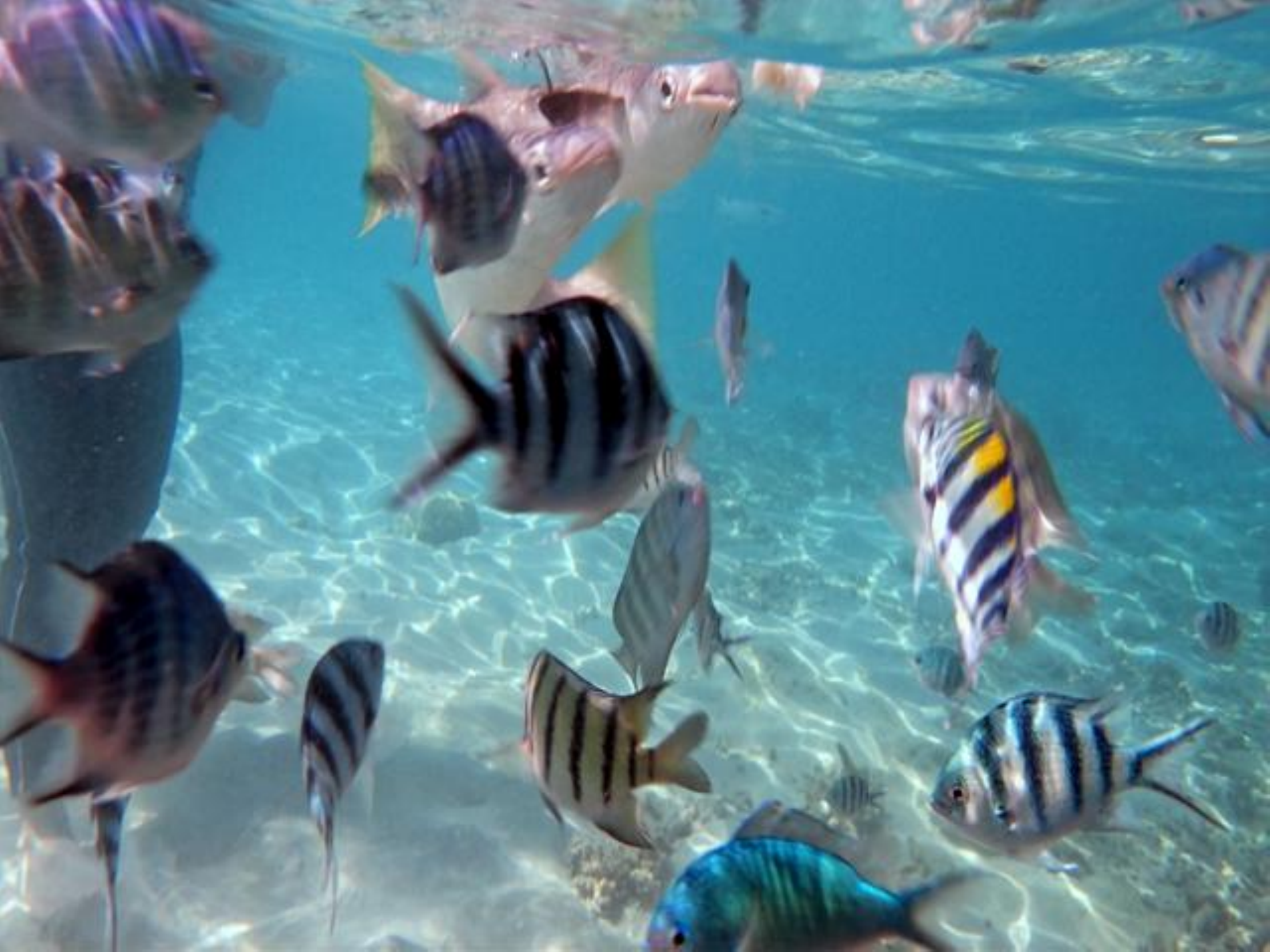}&
\includegraphics[width=3cm, height=2cm]{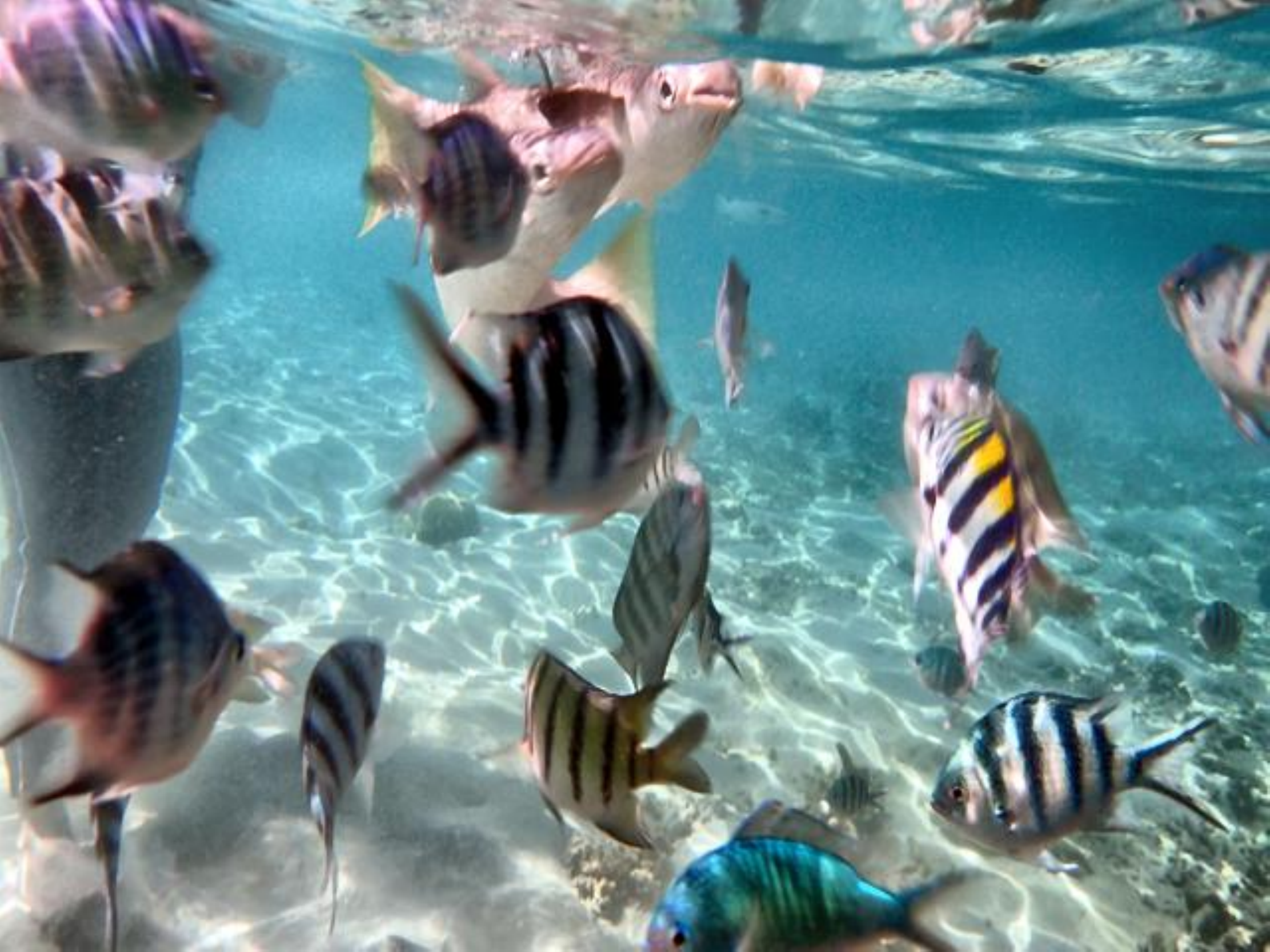}\\

\includegraphics[width=3cm, height=2cm]{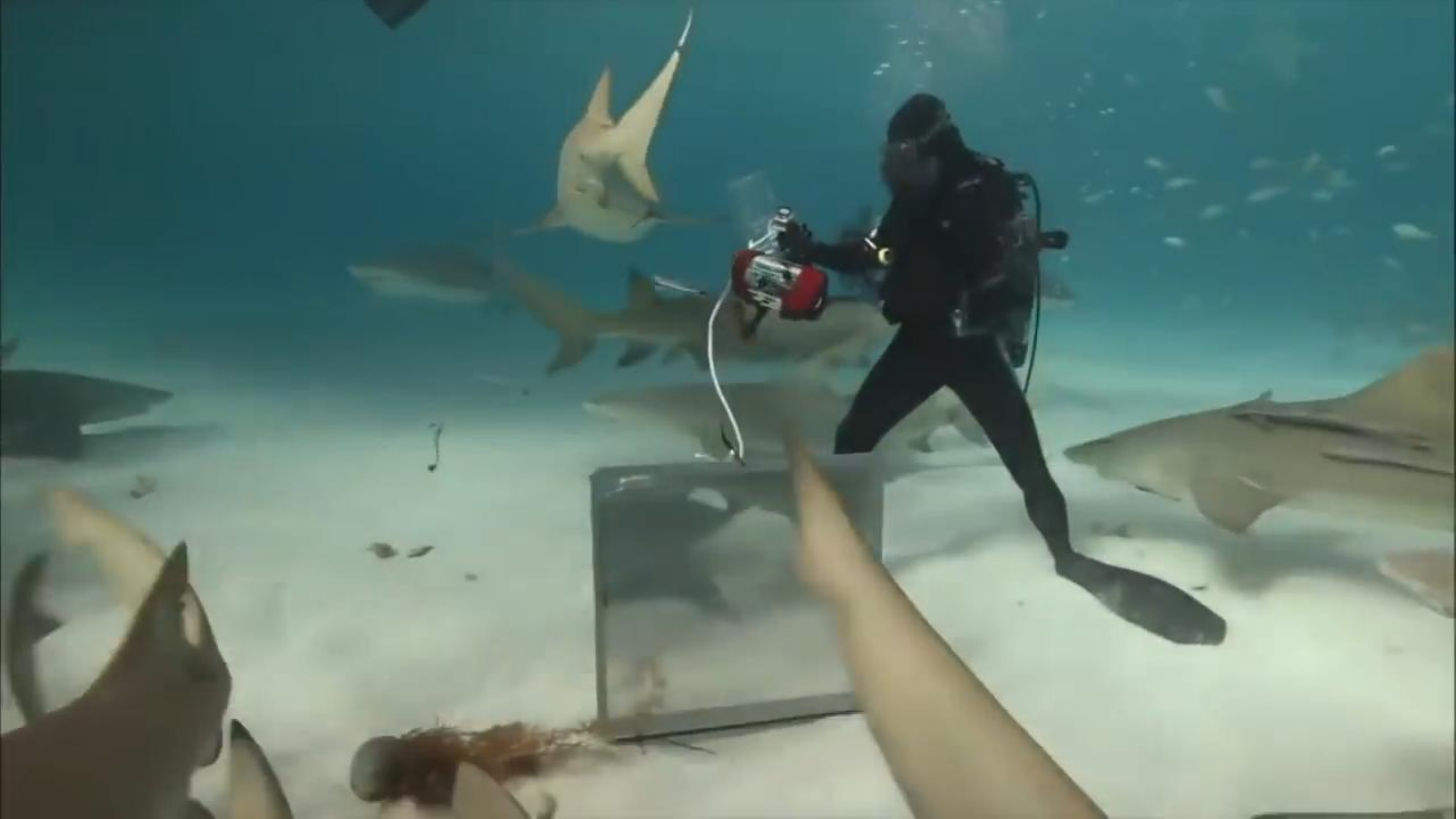}&
\includegraphics[width=3cm, height=2cm]{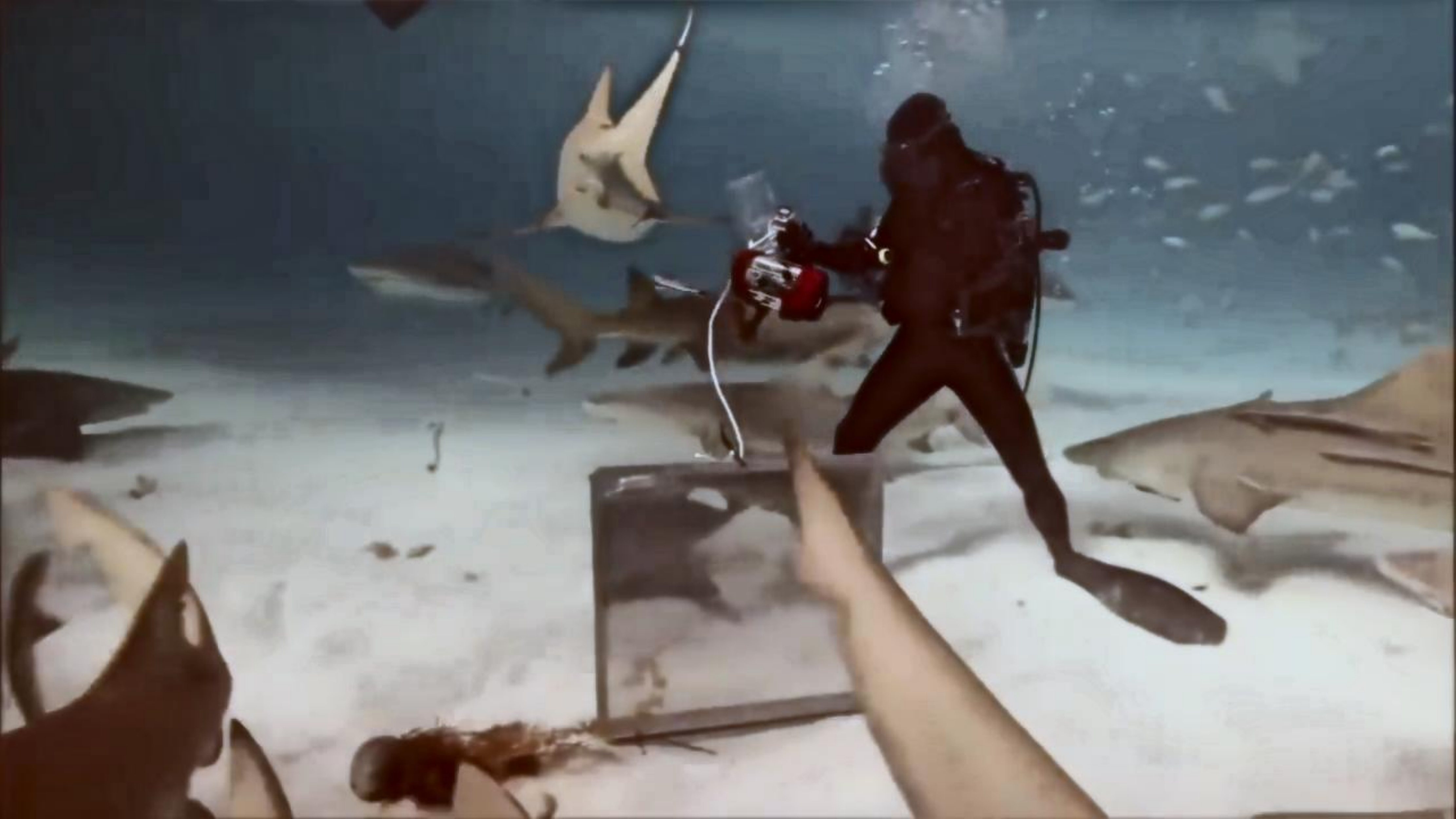}&
\includegraphics[width=3cm, height=2cm]{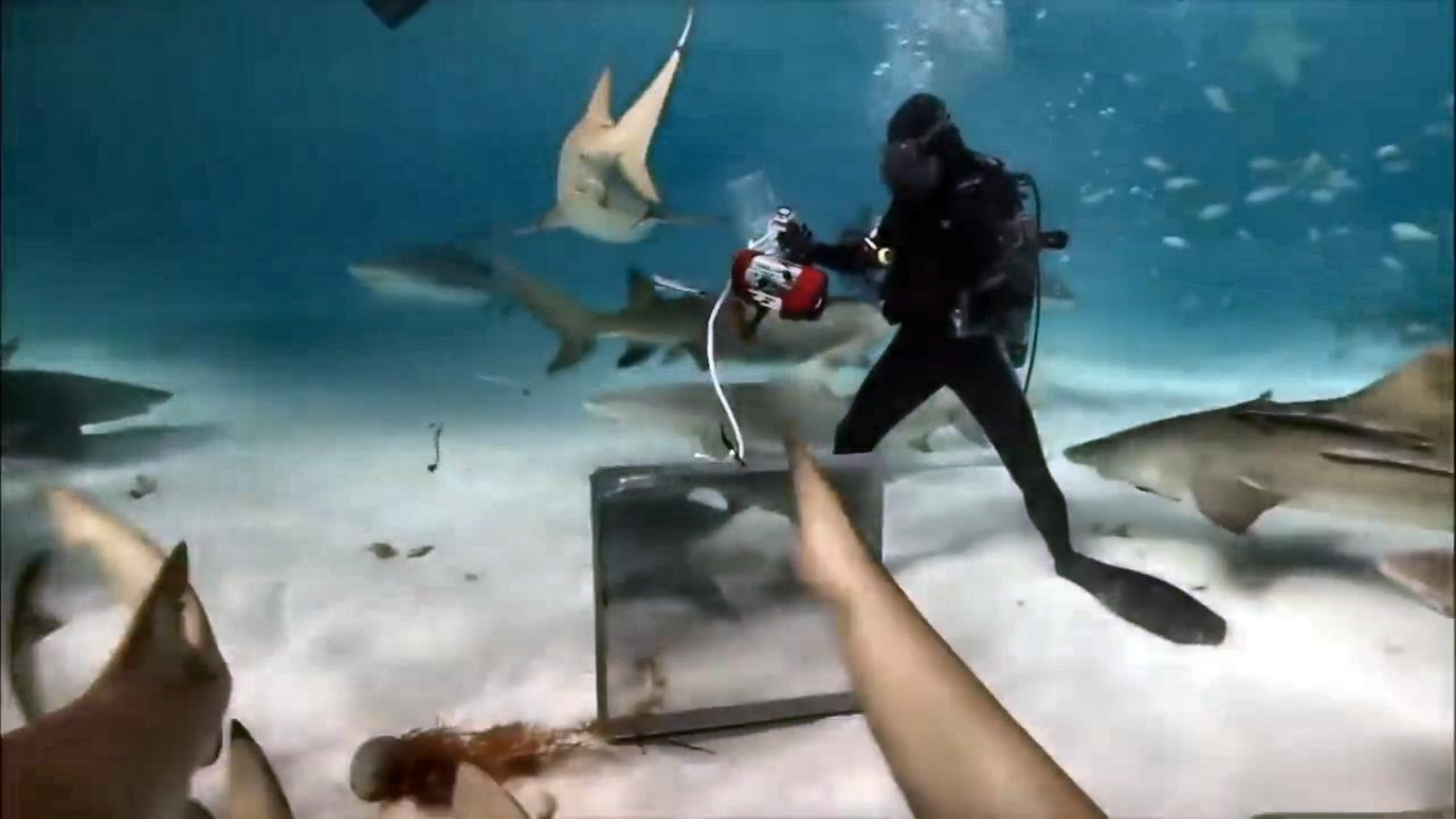}&
\includegraphics[width=3cm, height=2cm]{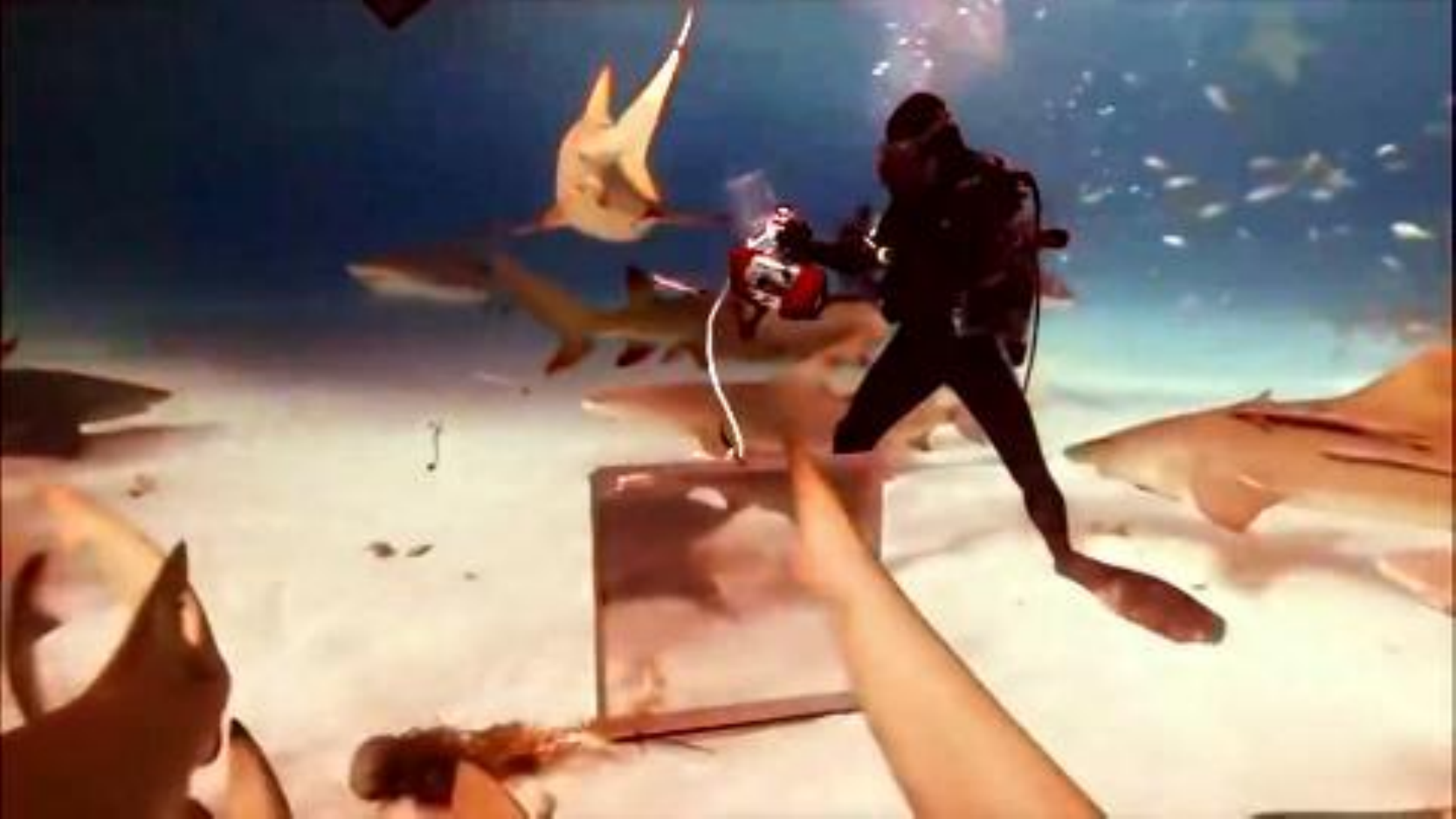}&
\includegraphics[width=3cm, height=2cm]{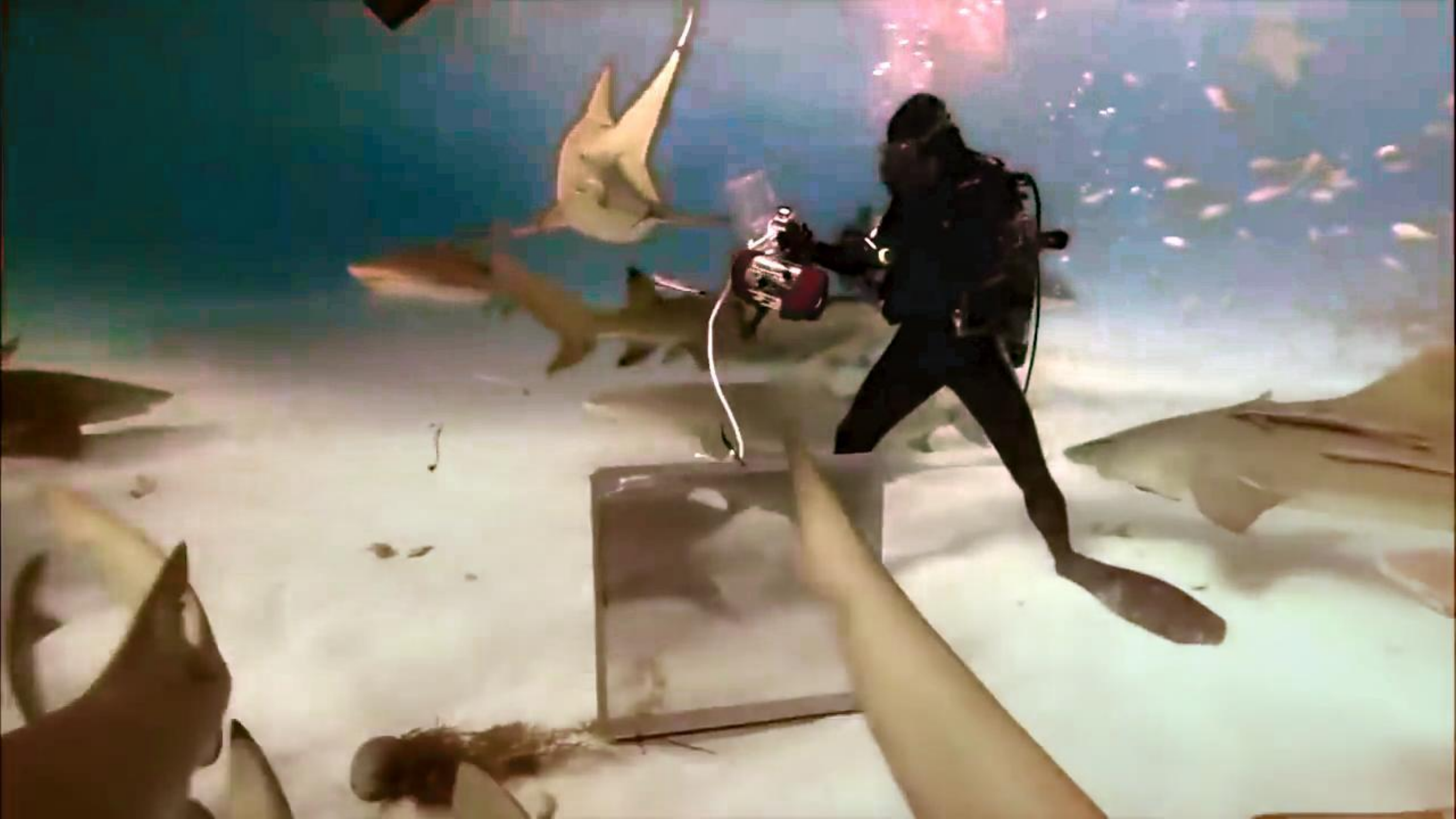}&
\includegraphics[width=3cm, height=2cm]{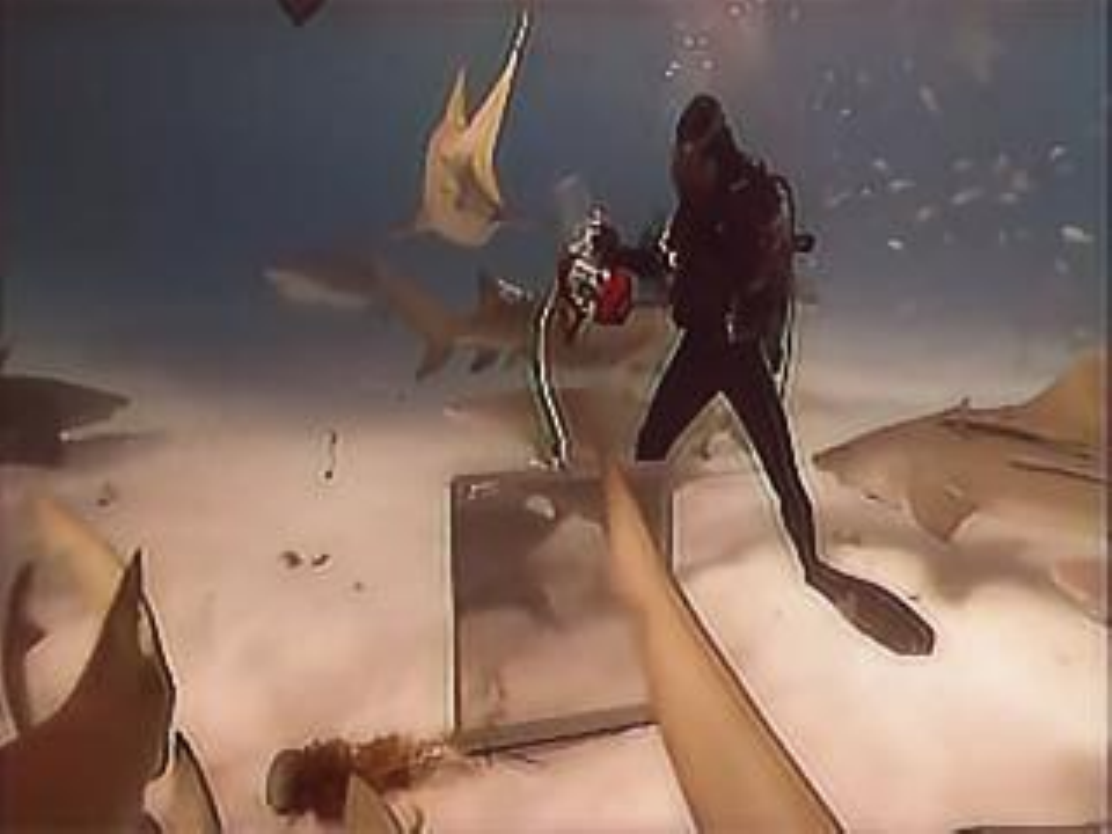}&
\includegraphics[width=3cm, height=2cm]{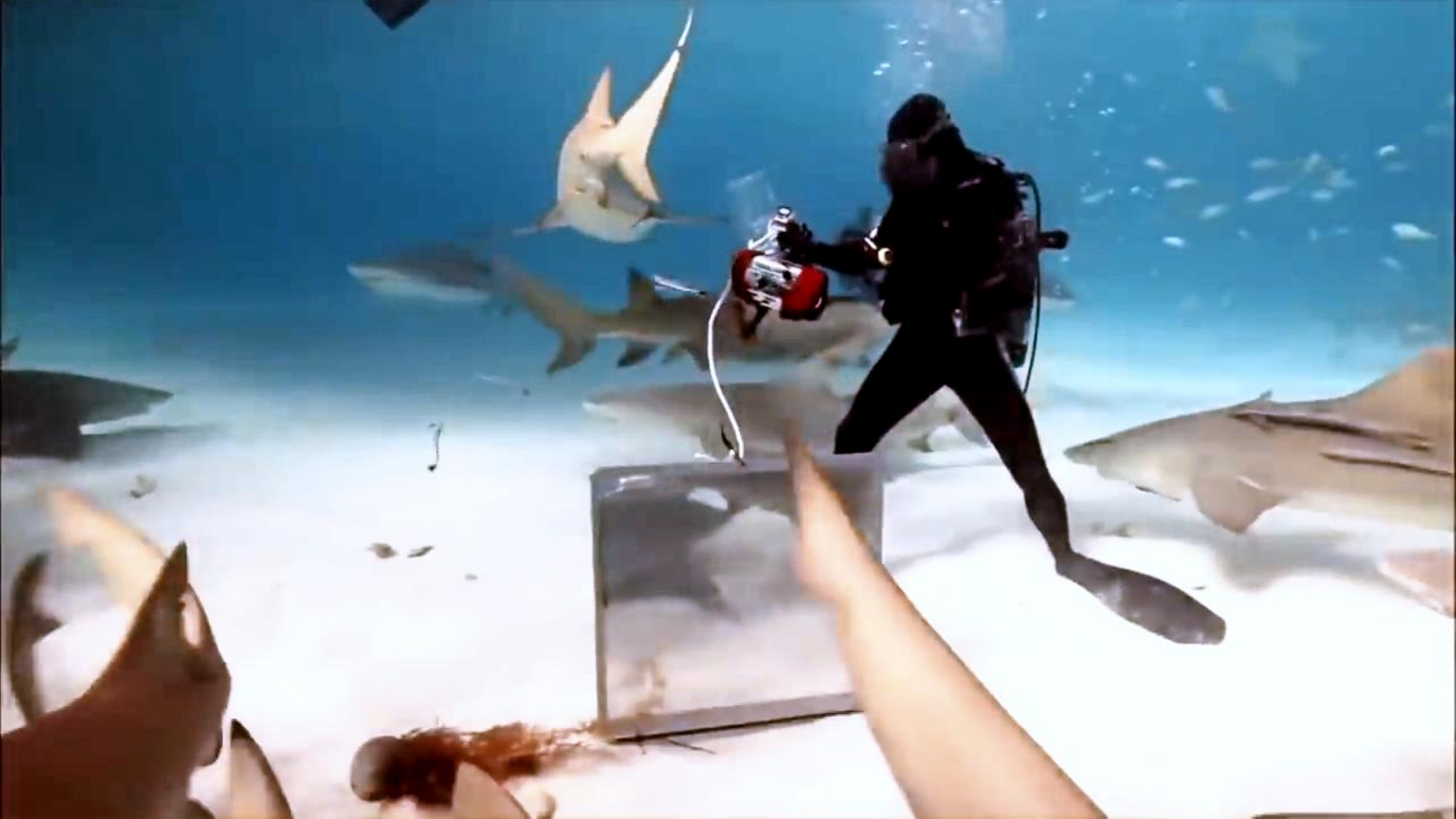}&
\includegraphics[width=3cm, height=2cm]{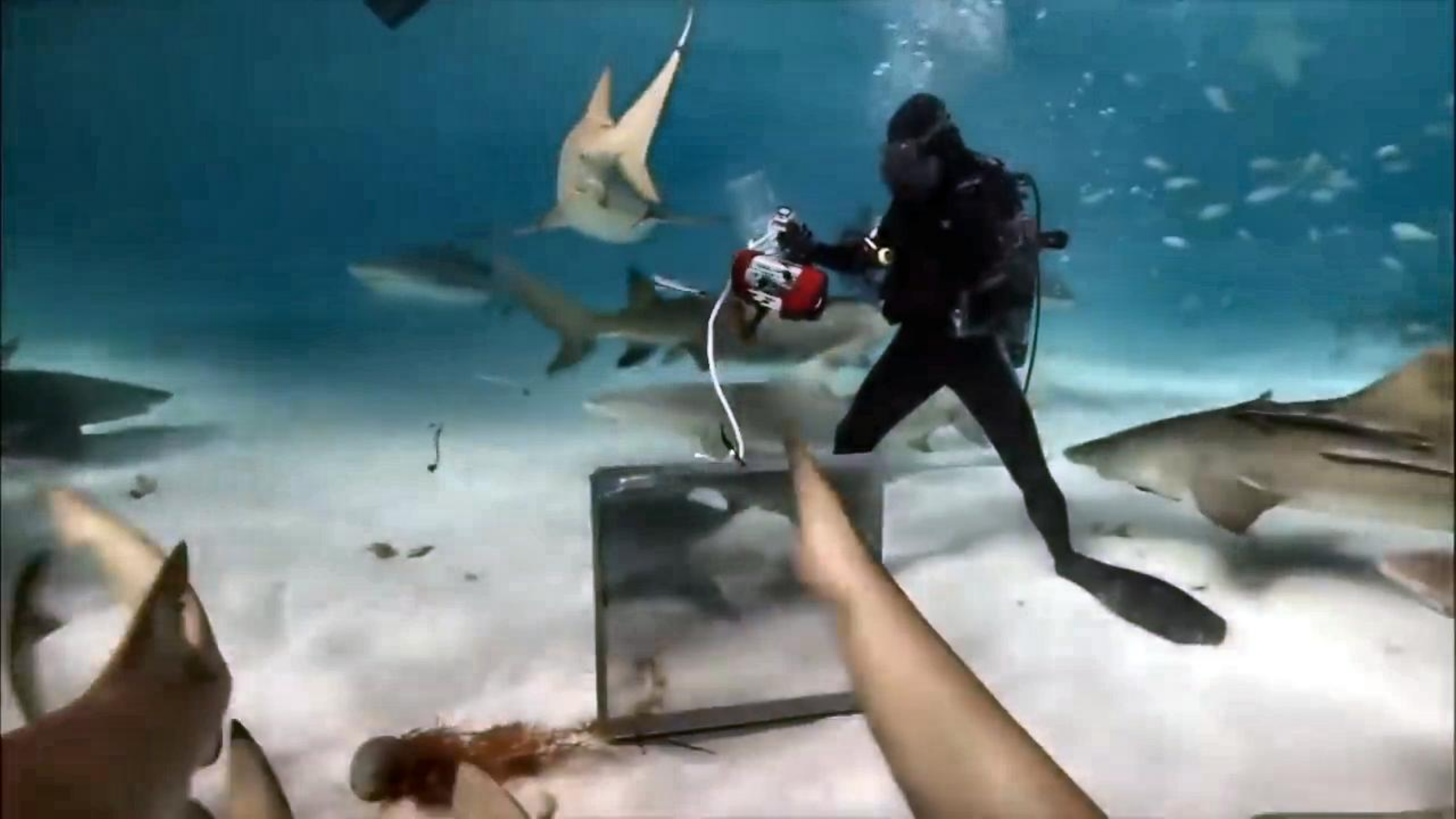}\\

\includegraphics[width=3cm, height=2cm]{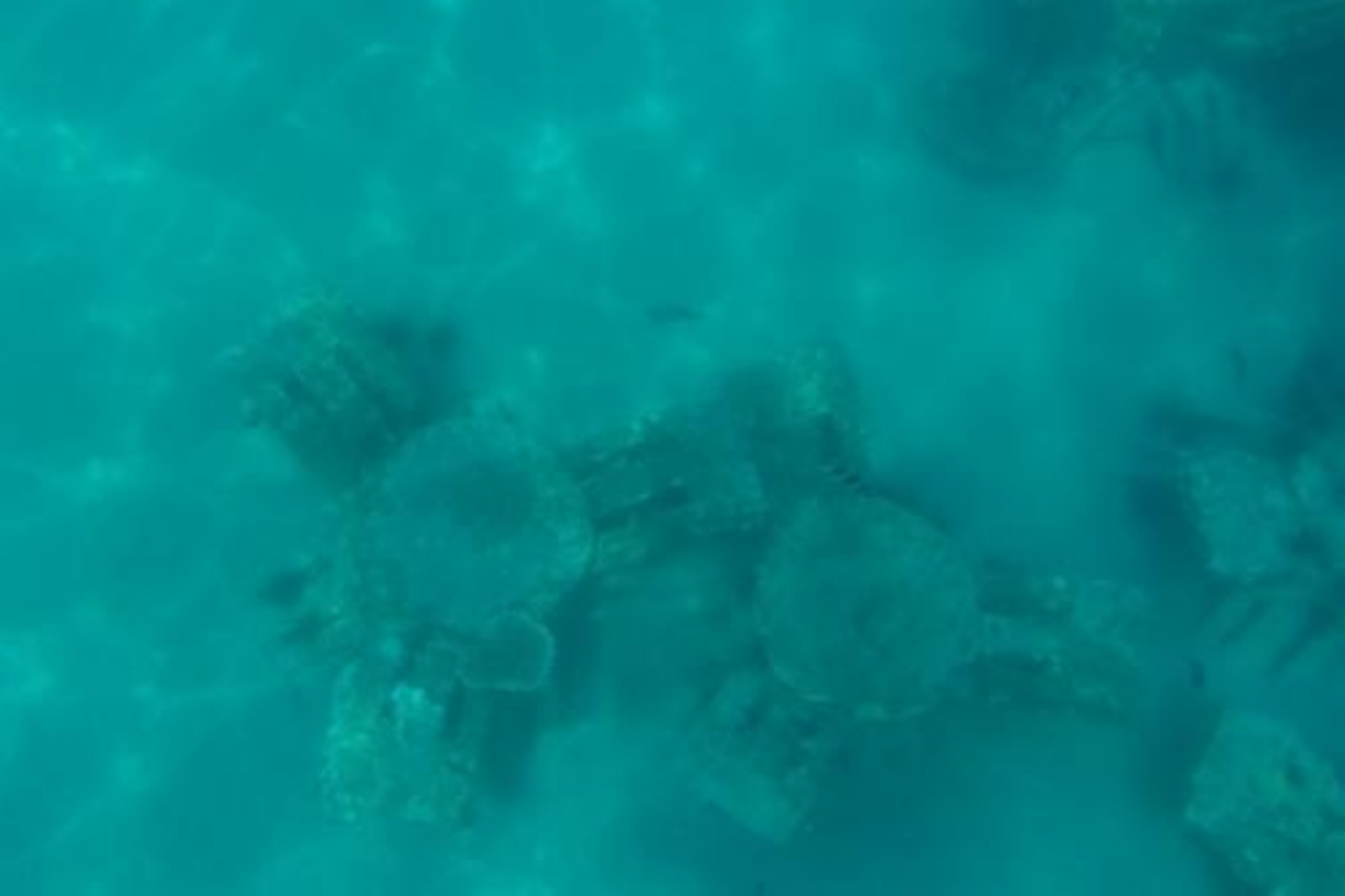}&
\includegraphics[width=3cm, height=2cm]{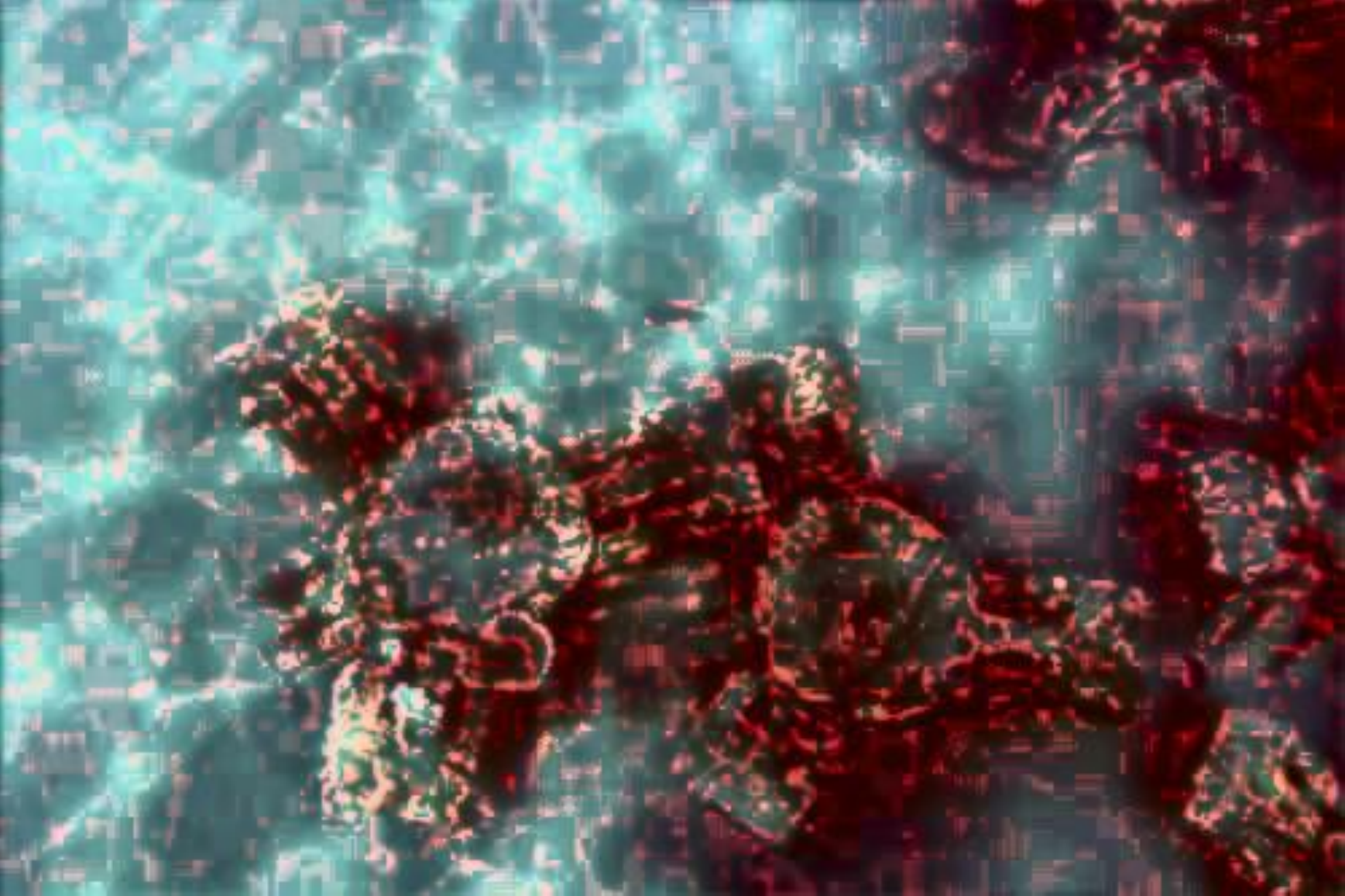}&
\includegraphics[width=3cm, height=2cm]{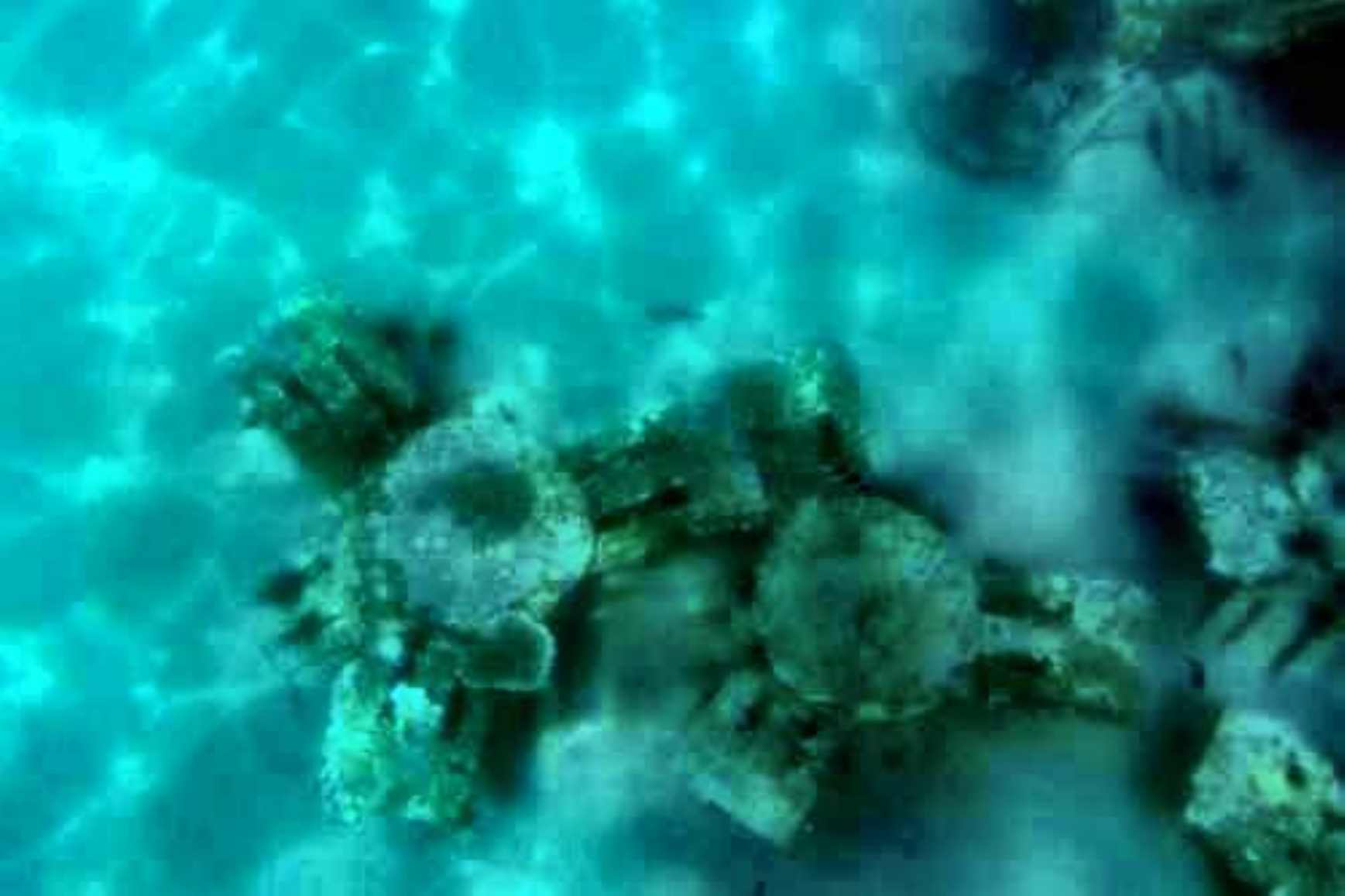}&
\includegraphics[width=3cm, height=2cm]{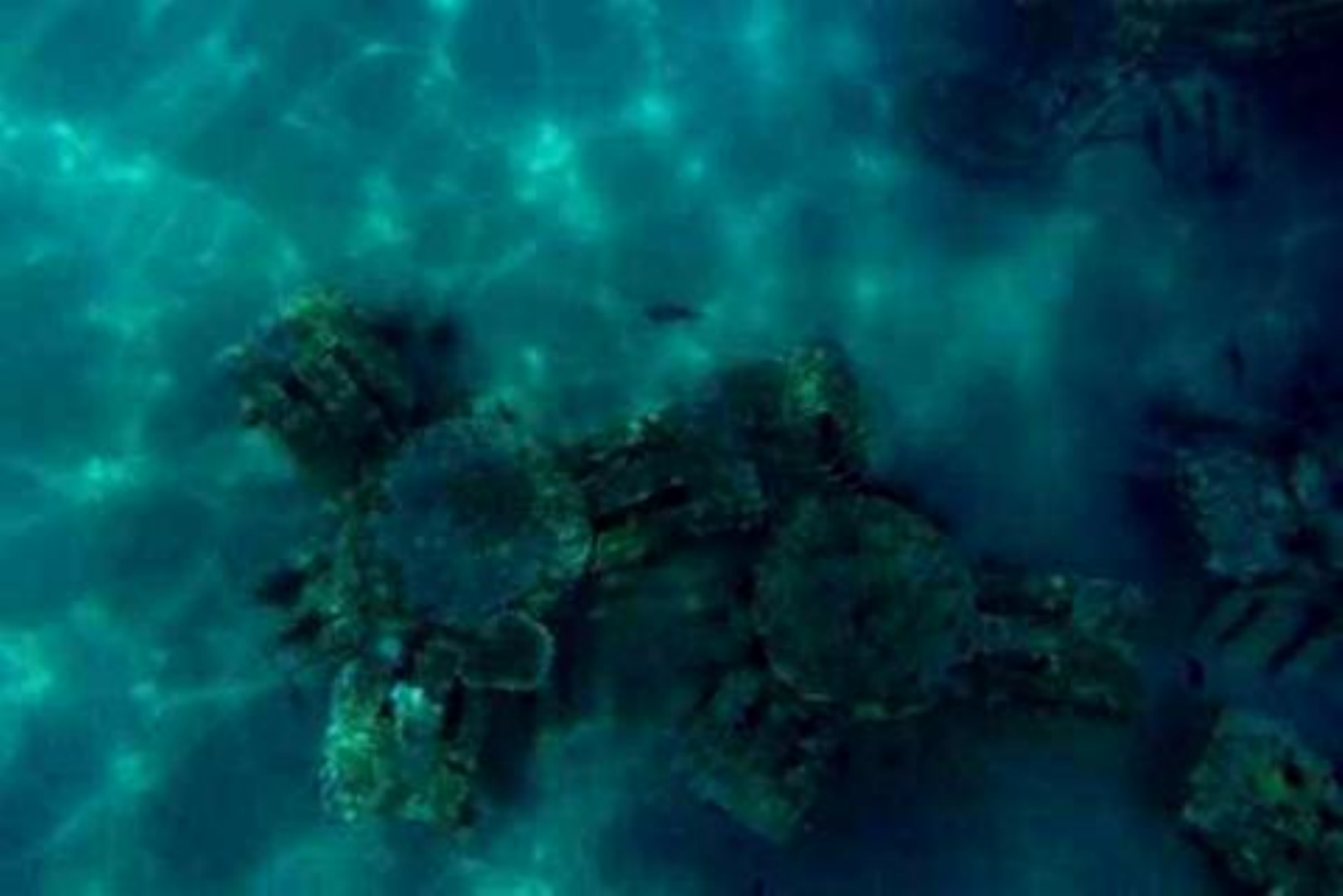}&
\includegraphics[width=3cm, height=2cm]{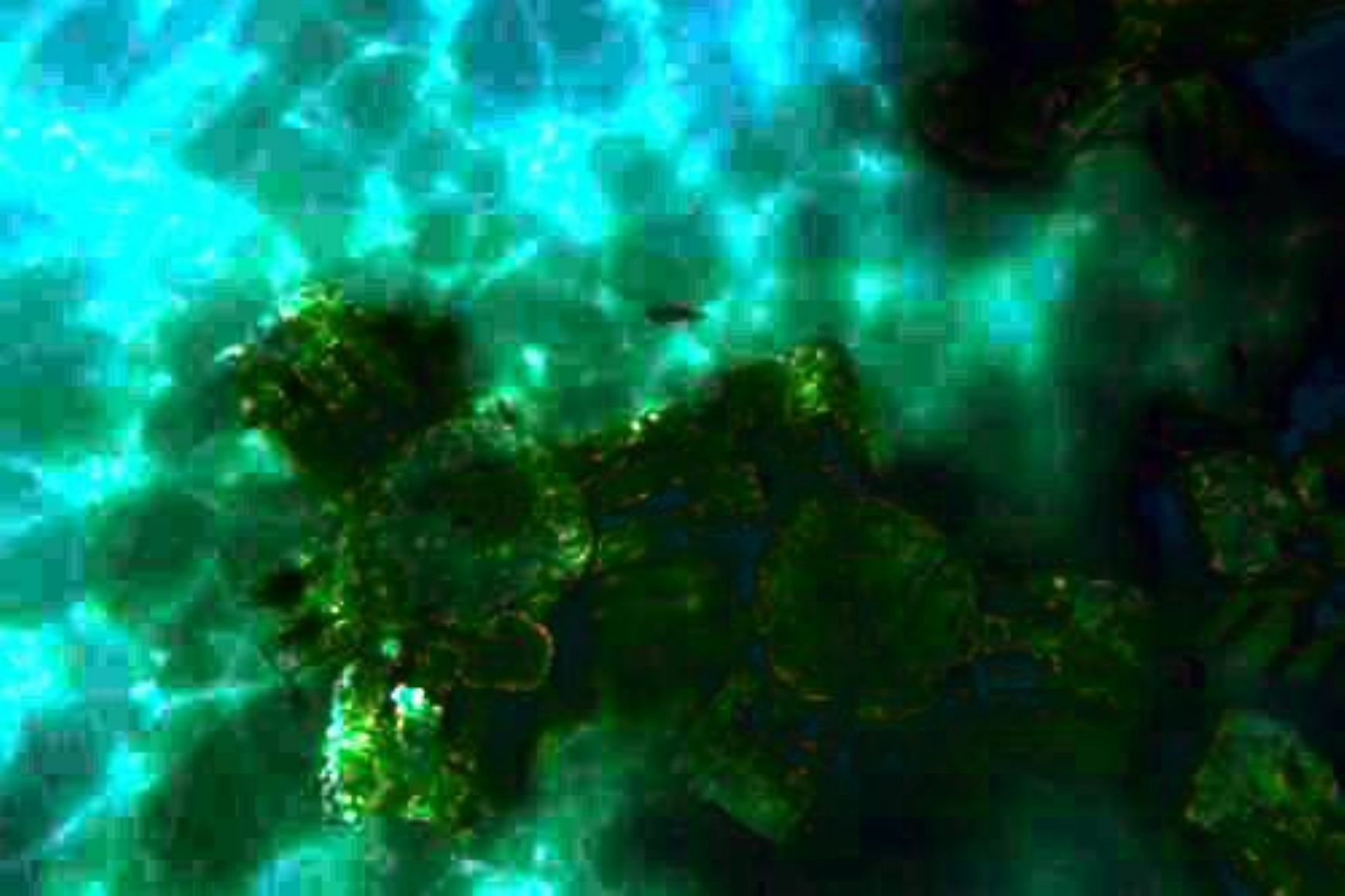}&
\includegraphics[width=3cm, height=2cm]{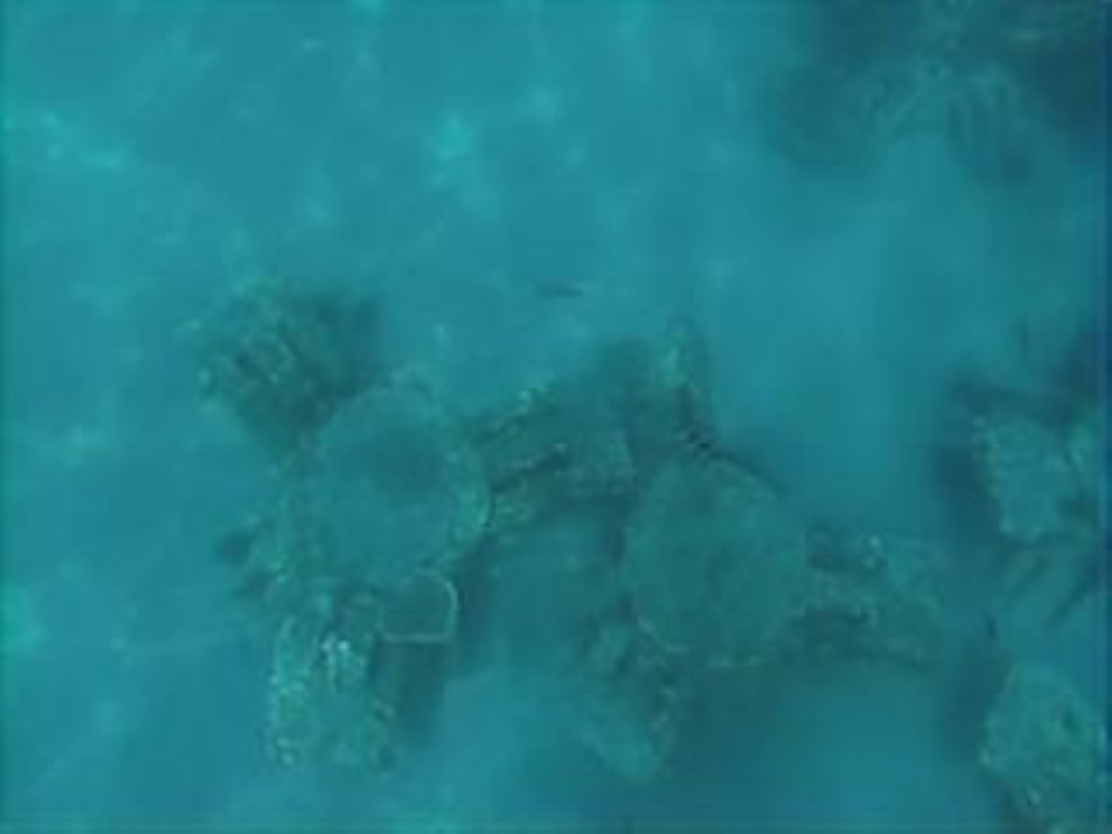}&
\includegraphics[width=3cm, height=2cm]{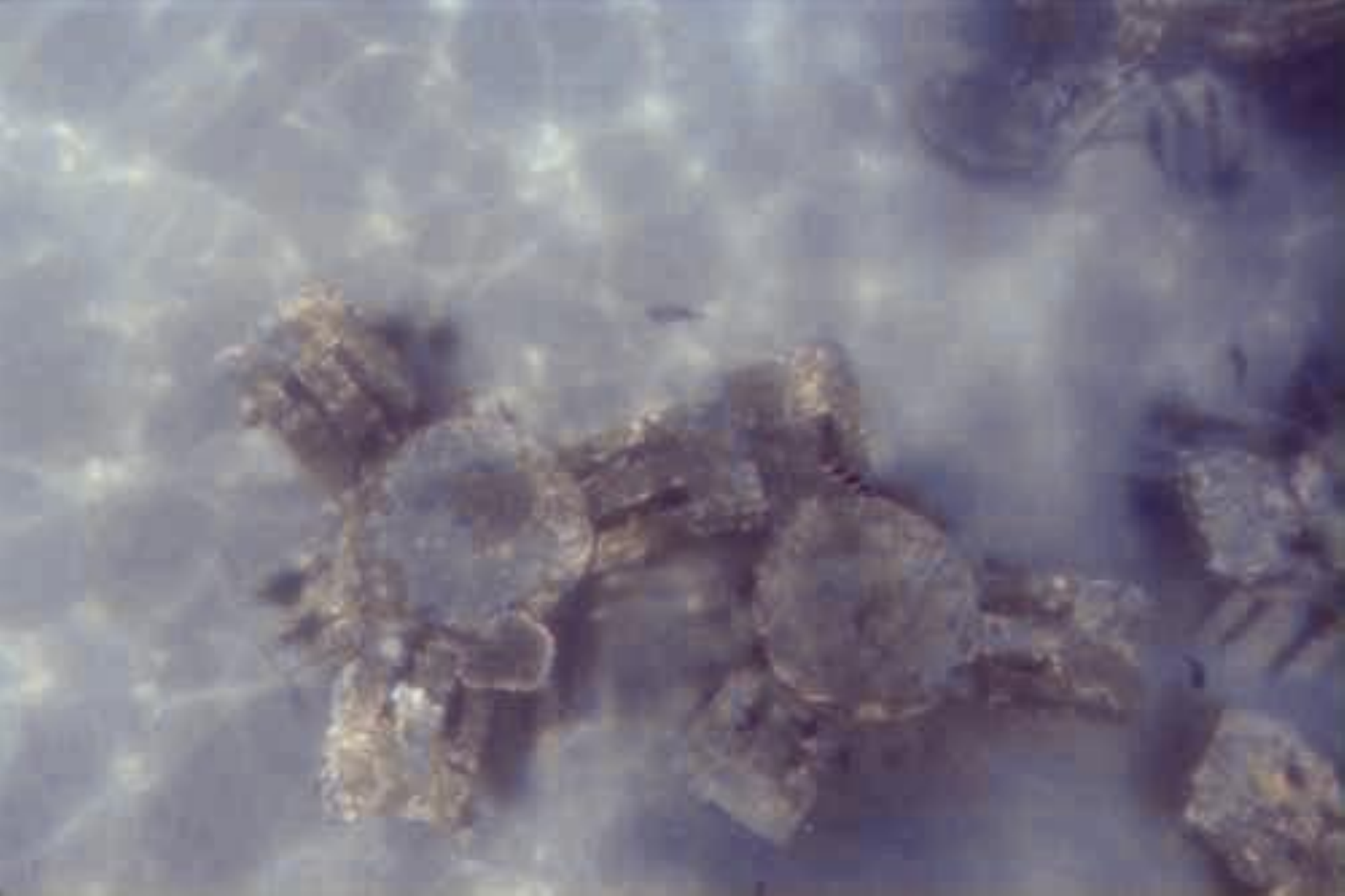}&
\includegraphics[width=3cm, height=2cm]{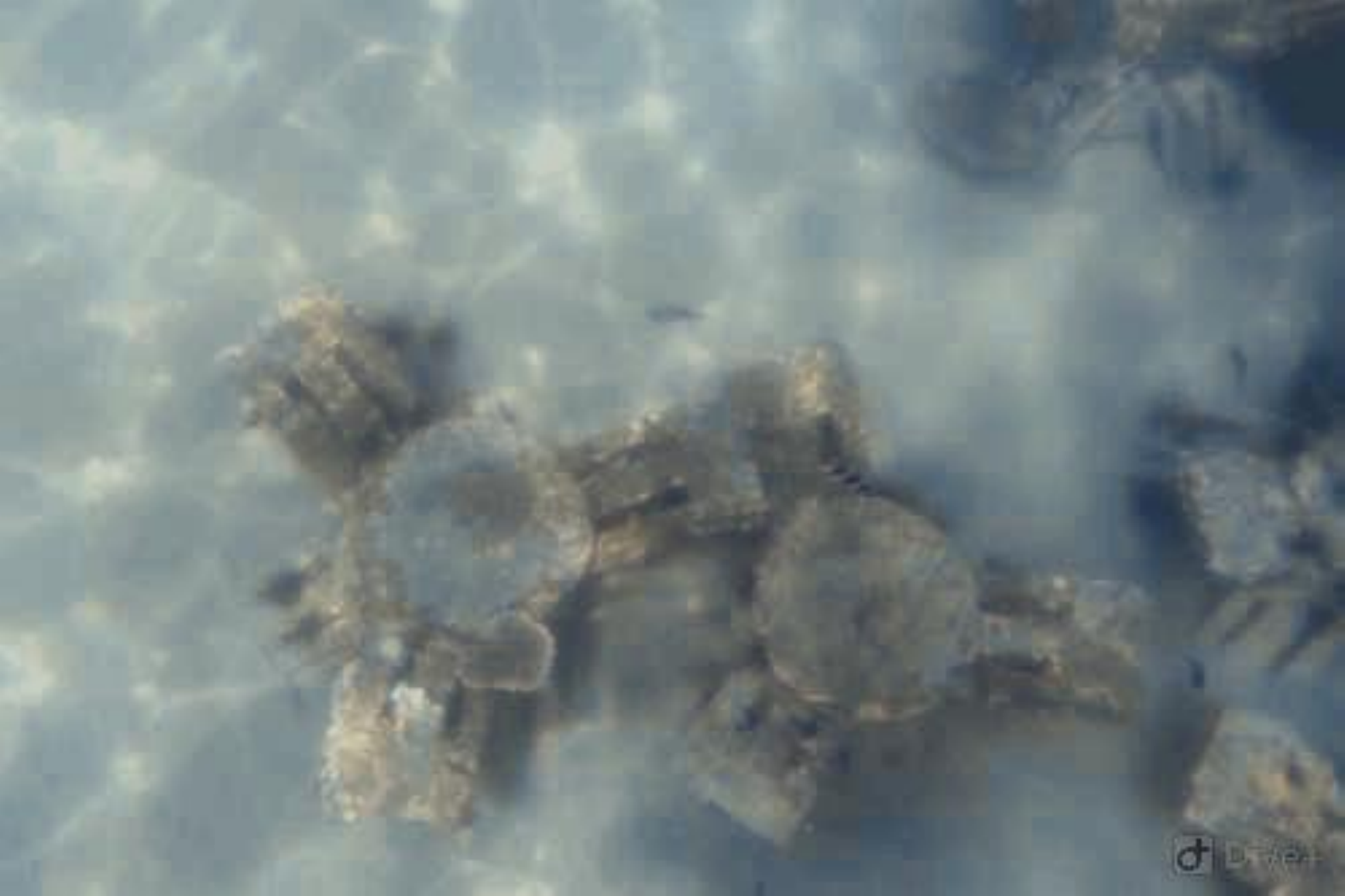}\\

\includegraphics[width=3cm, height=2cm]{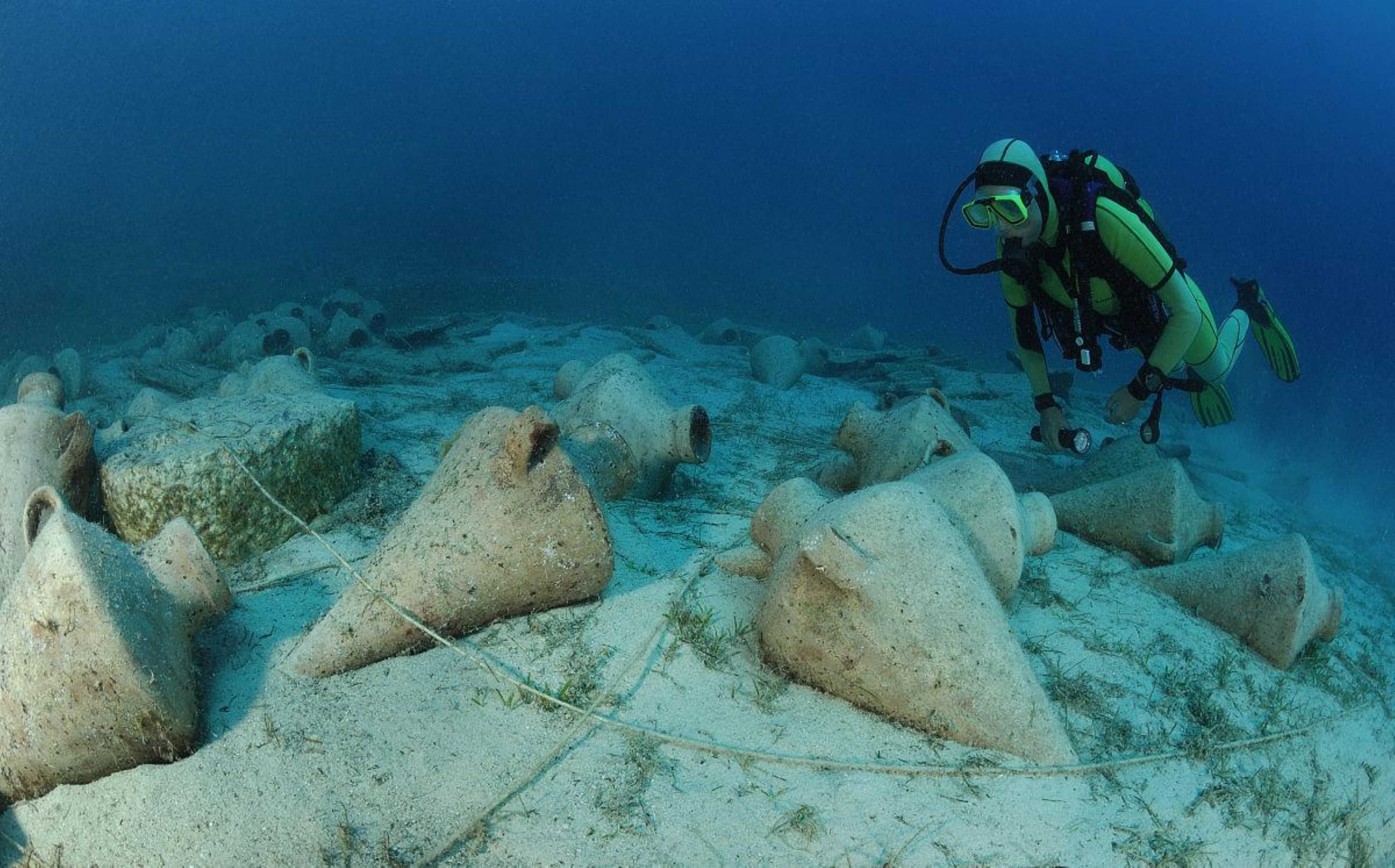}&
\includegraphics[width=3cm, height=2cm]{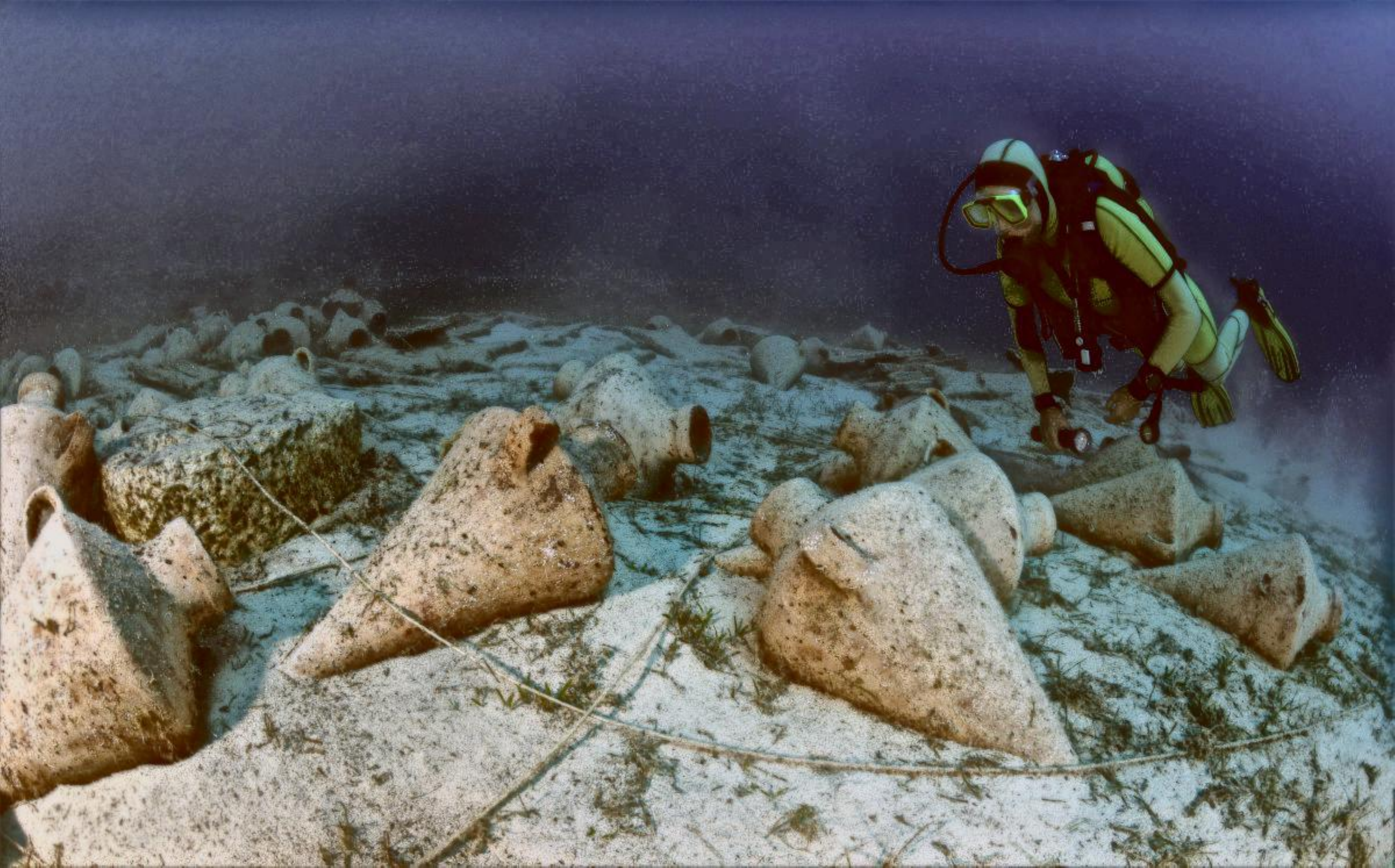}&
\includegraphics[width=3cm, height=2cm]{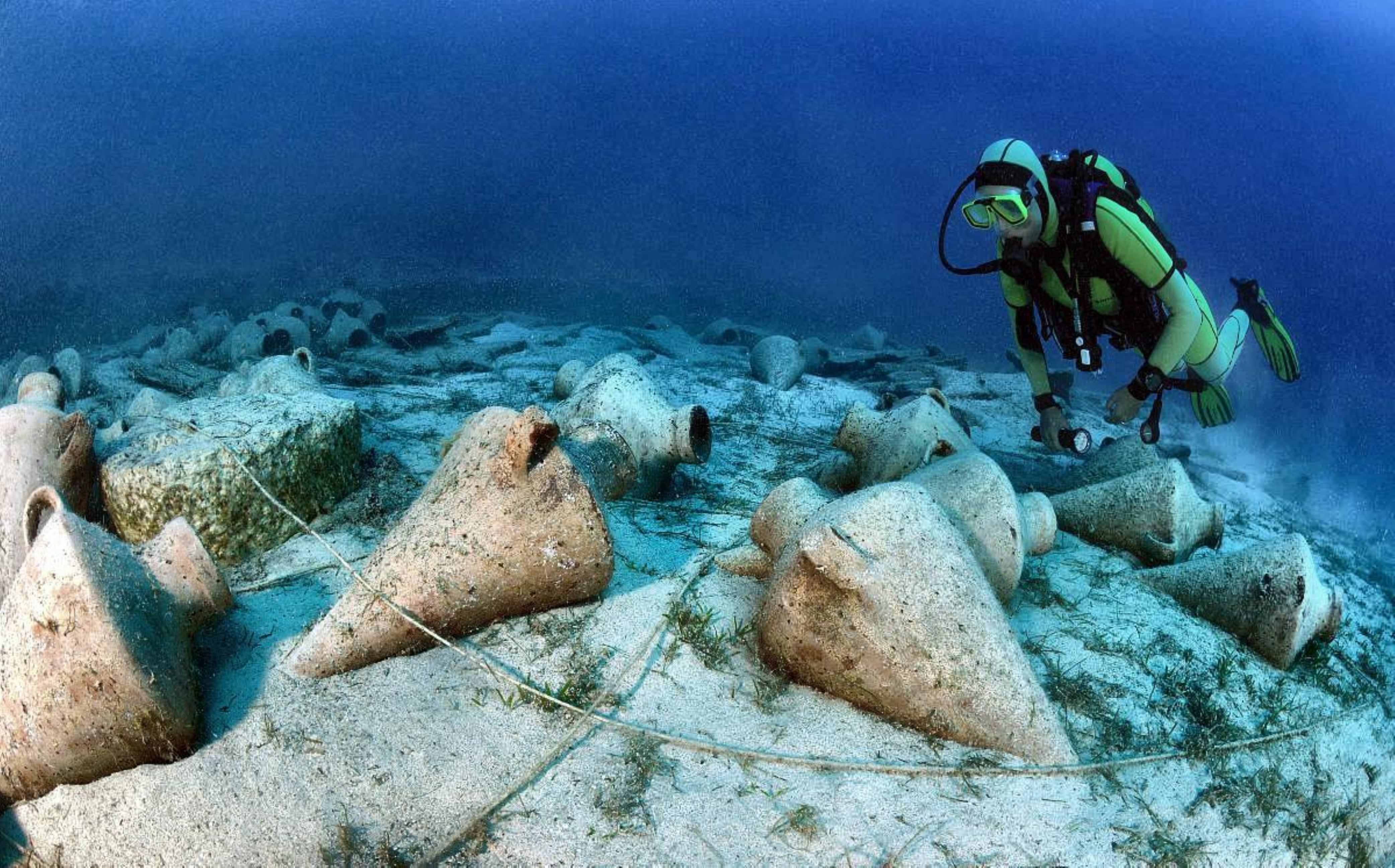}&
\includegraphics[width=3cm, height=2cm]{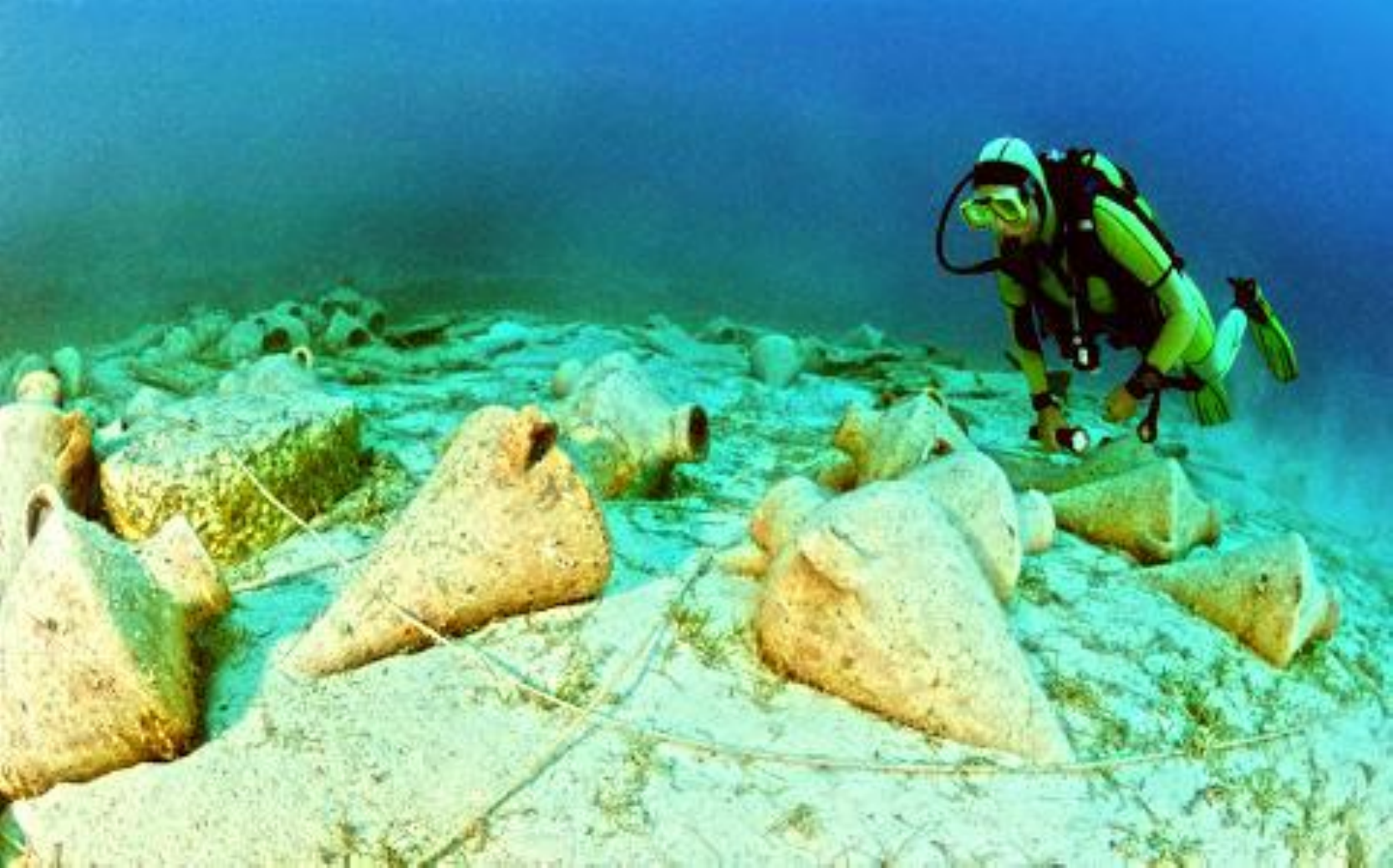}&
\includegraphics[width=3cm, height=2cm]{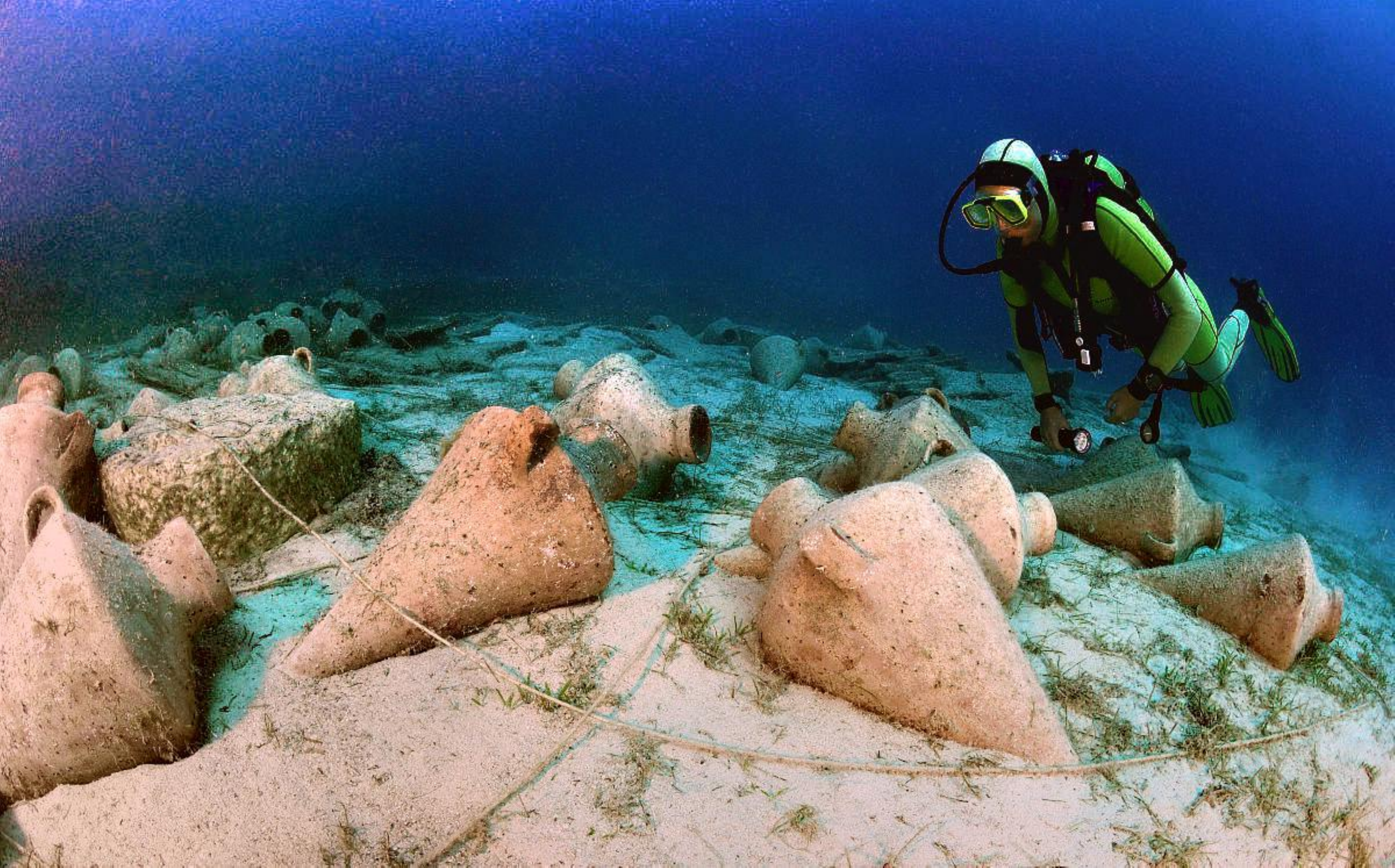}&
\includegraphics[width=3cm, height=2cm]{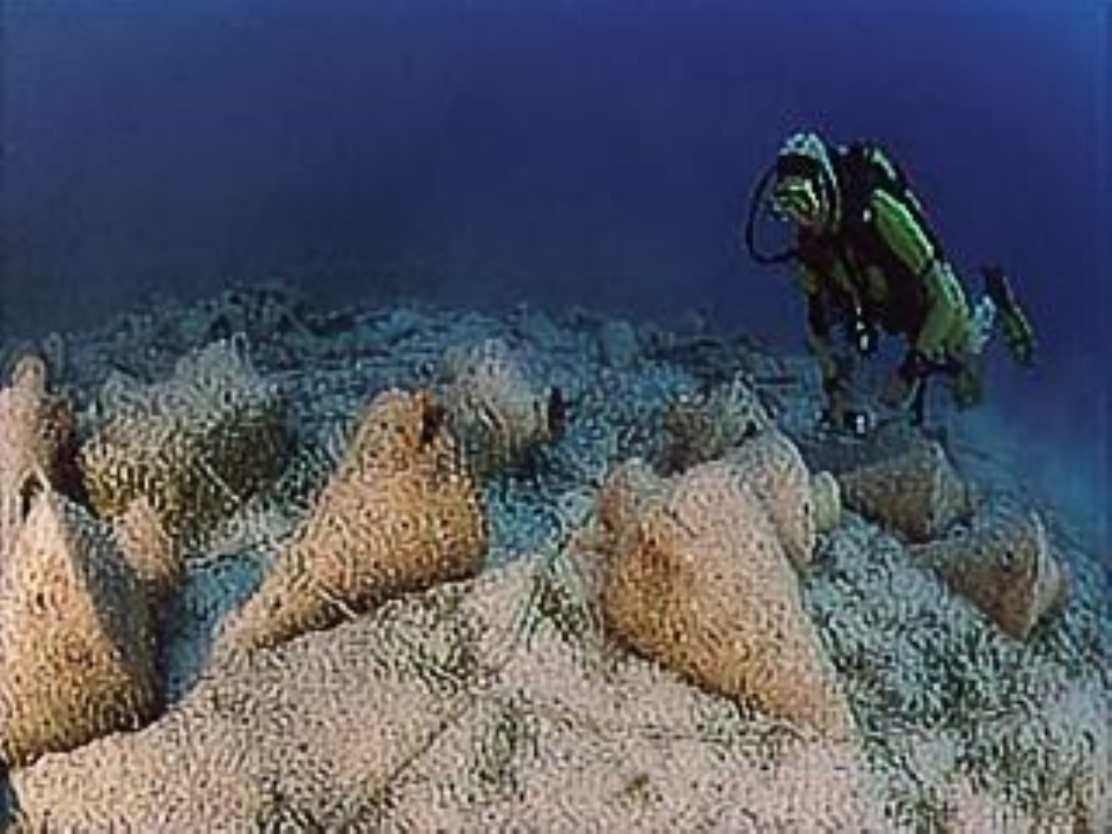}&
\includegraphics[width=3cm, height=2cm]{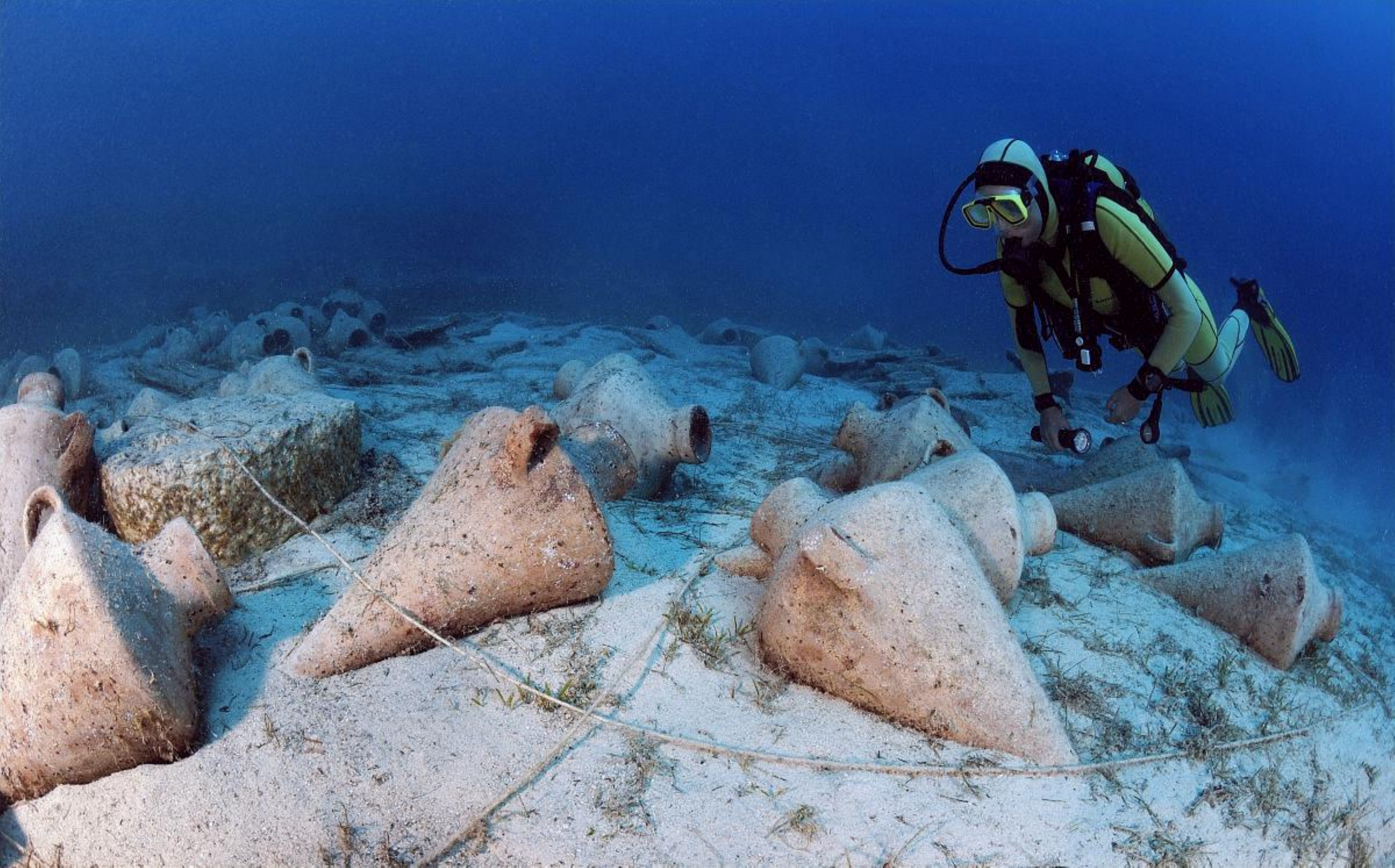}&
\includegraphics[width=3cm, height=2cm]{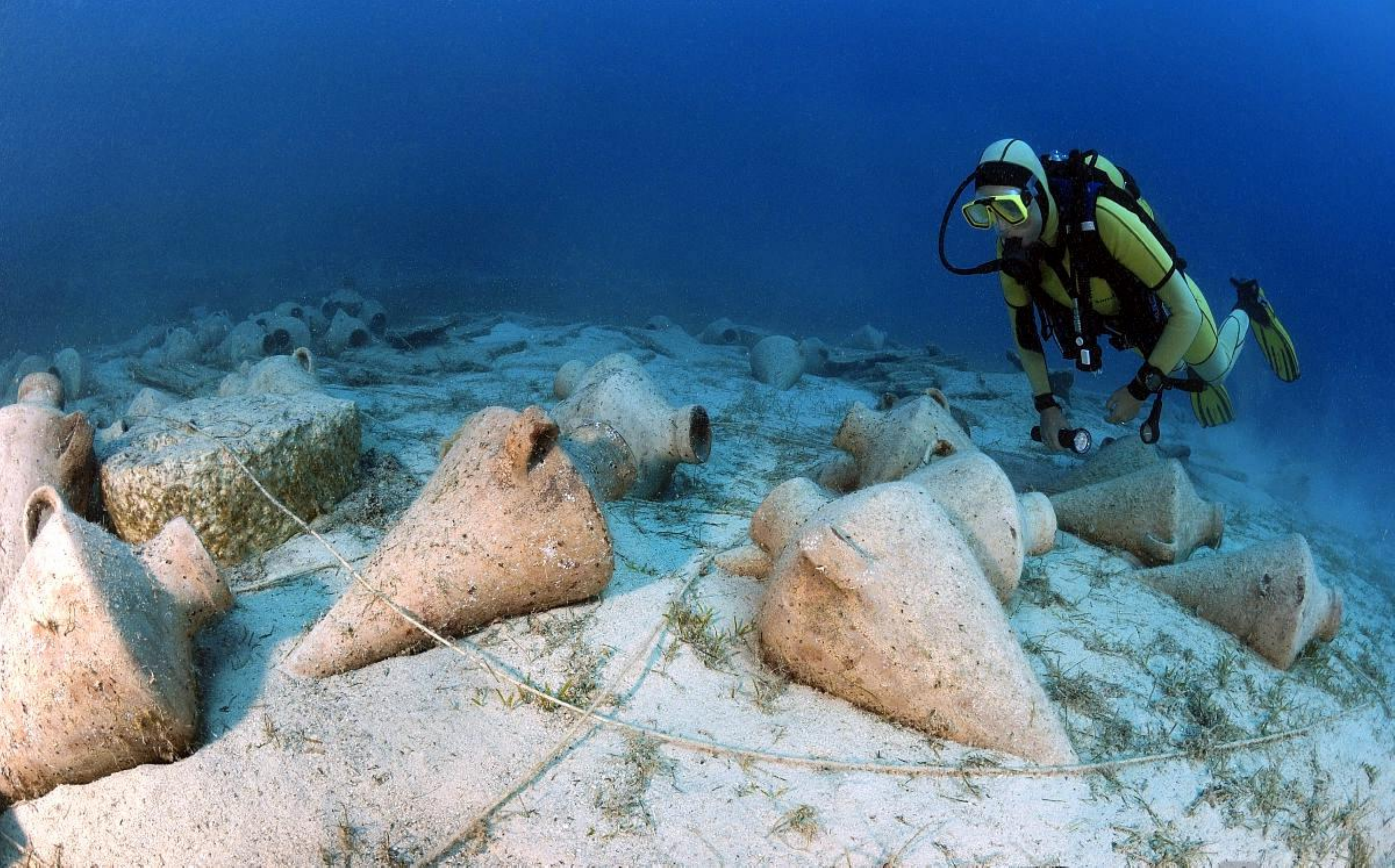}\\

\includegraphics[width=3cm, height=2cm]{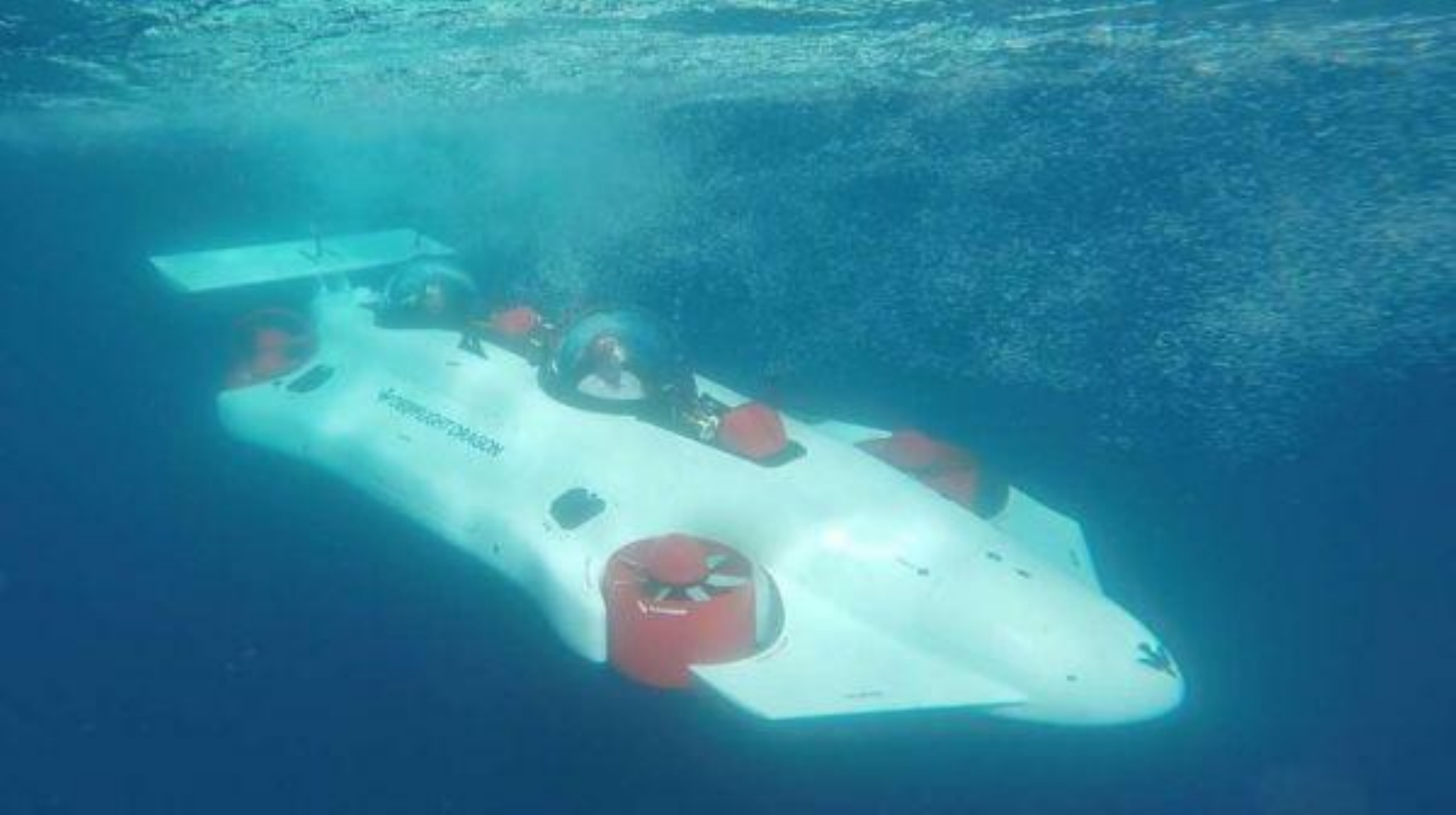}&
\includegraphics[width=3cm, height=2cm]{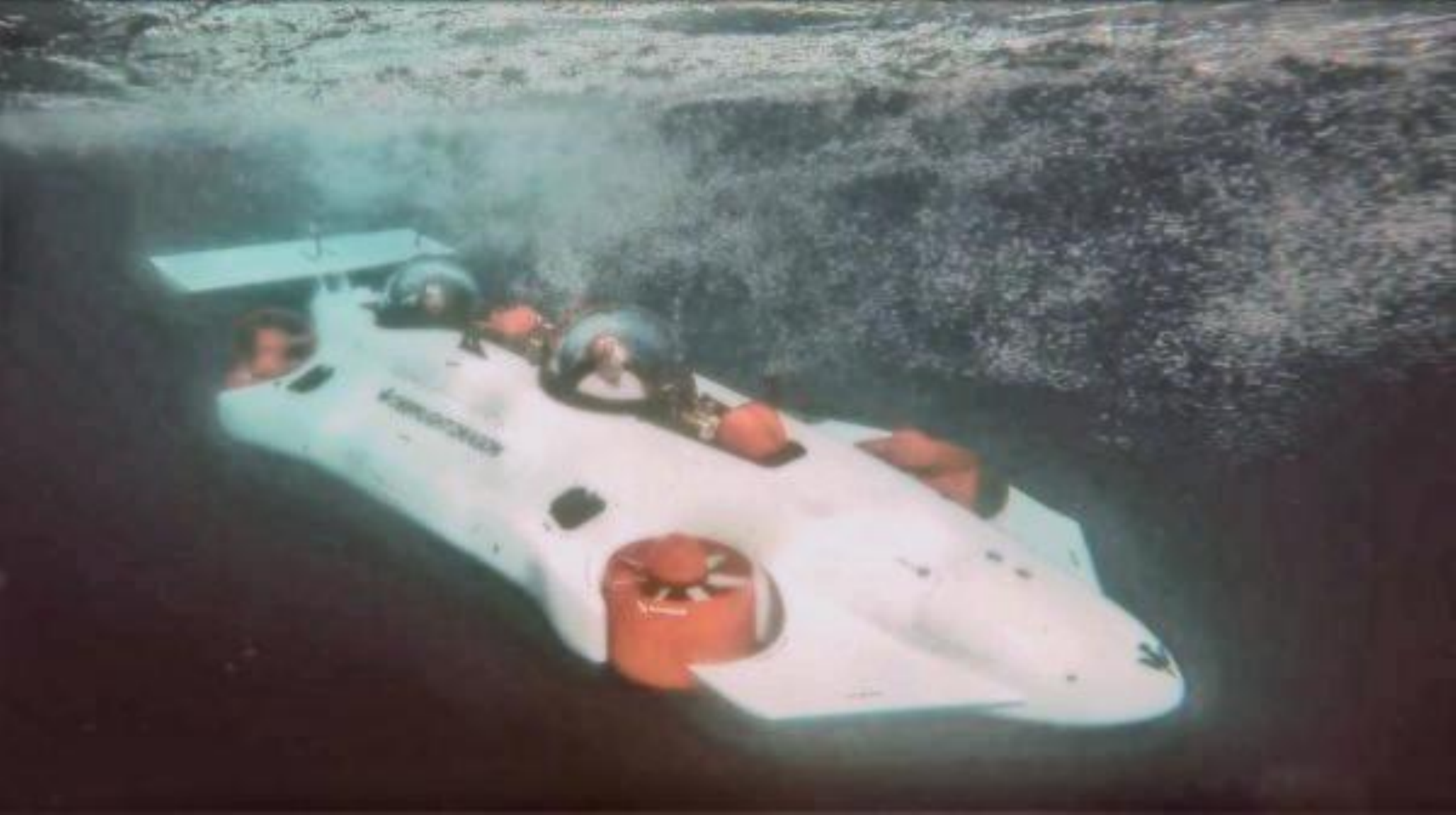}&
\includegraphics[width=3cm, height=2cm]{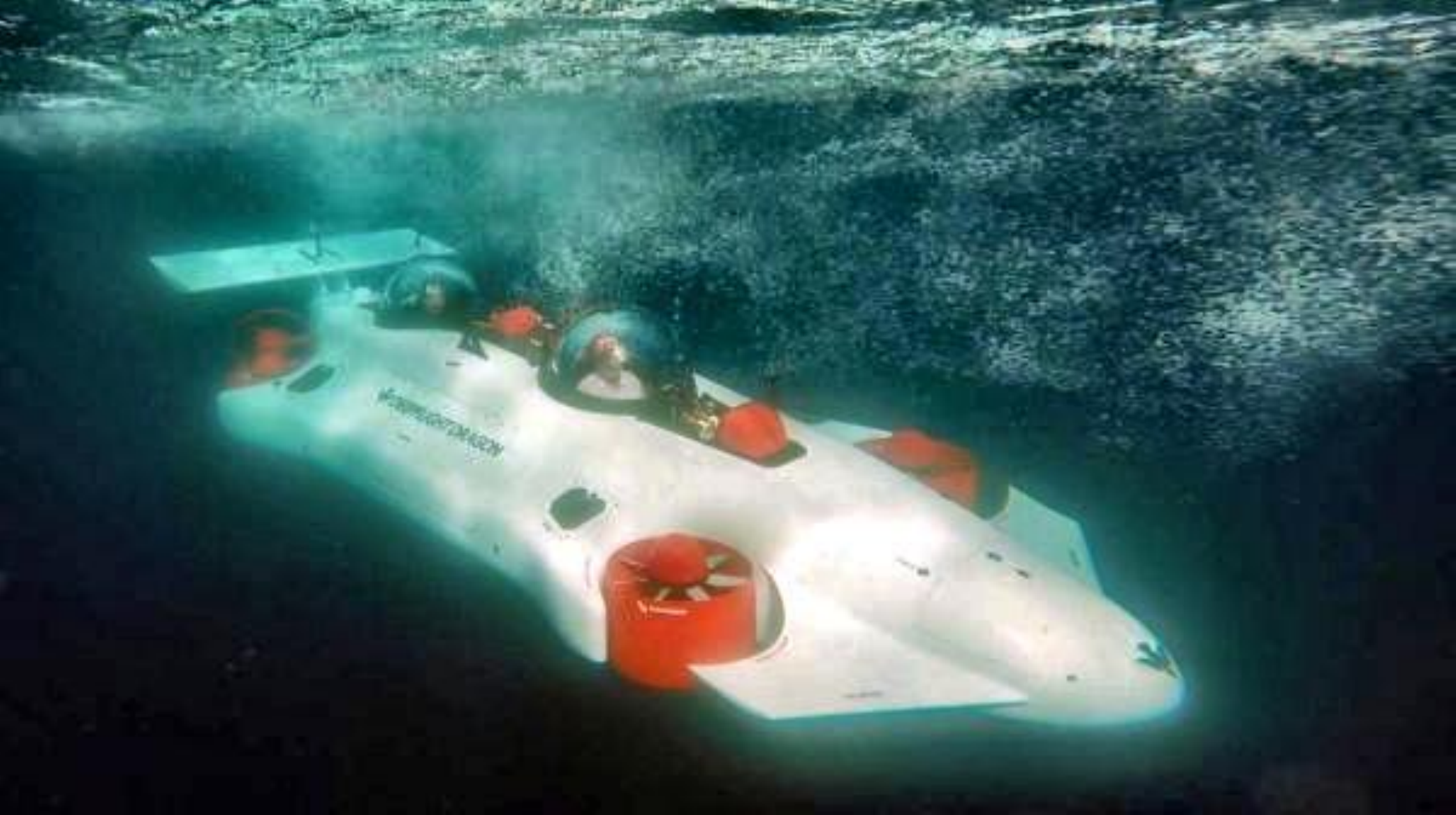}&
\includegraphics[width=3cm, height=2cm]{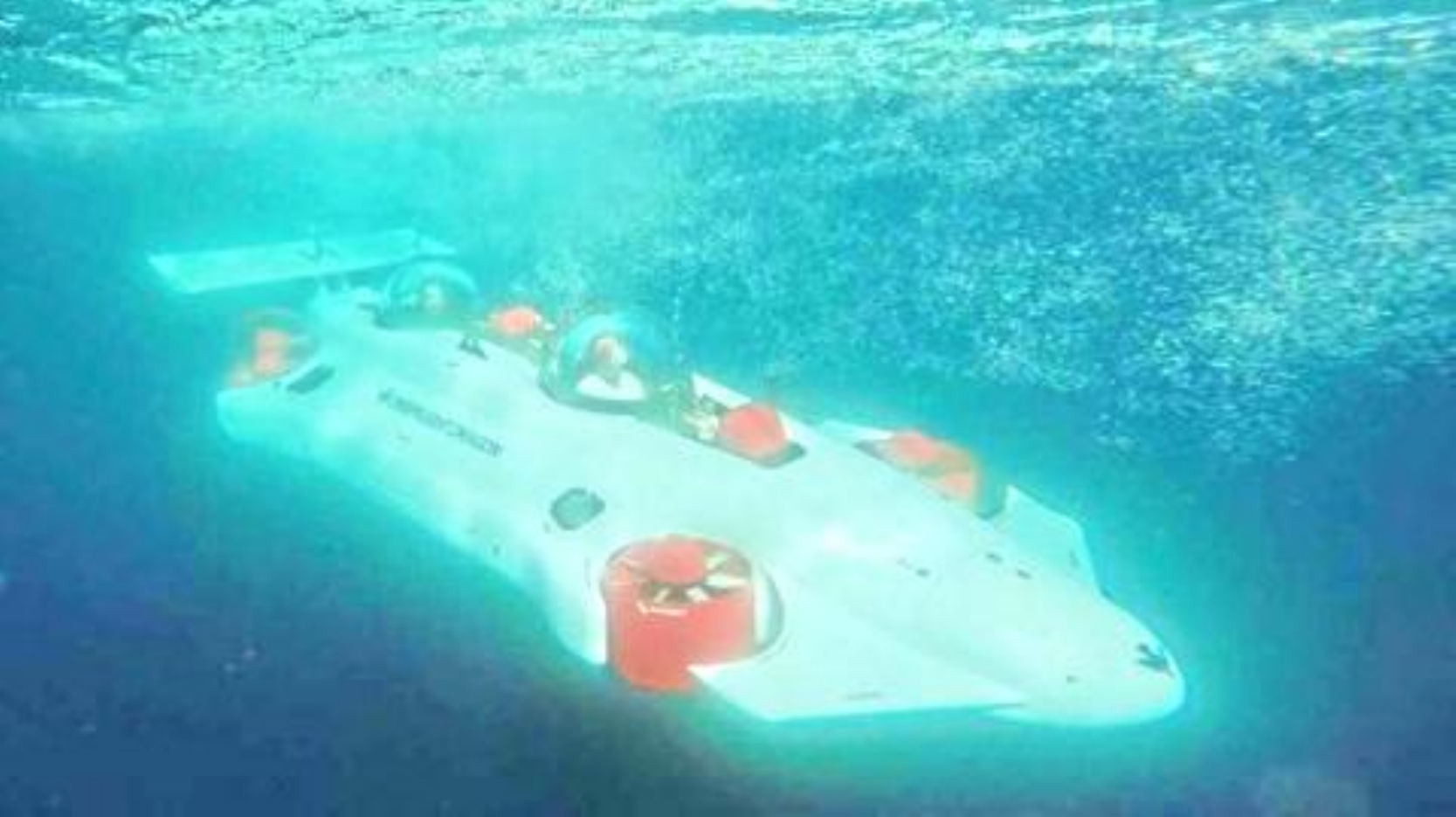}&
\includegraphics[width=3cm, height=2cm]{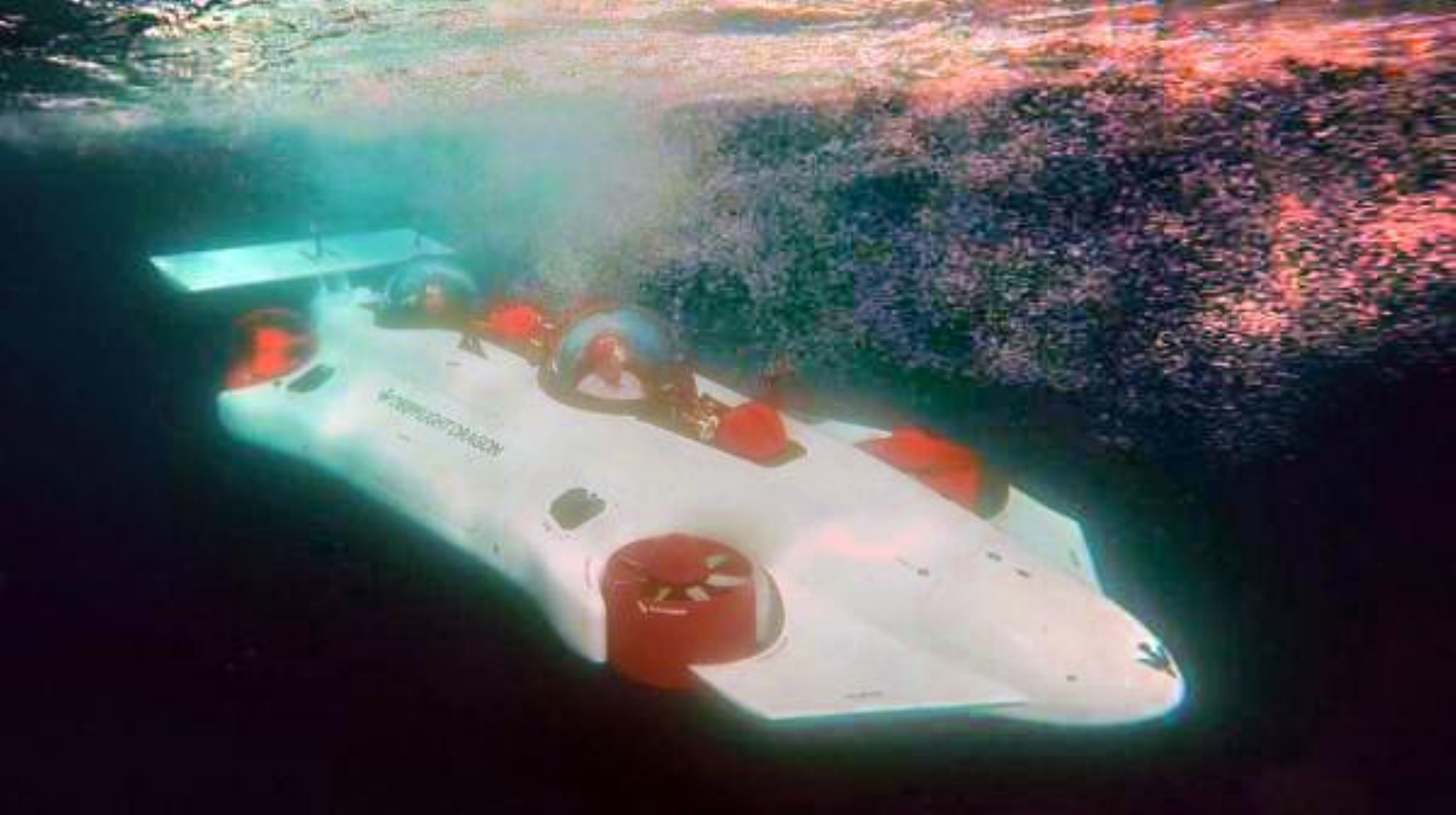}&
\includegraphics[width=3cm, height=2cm]{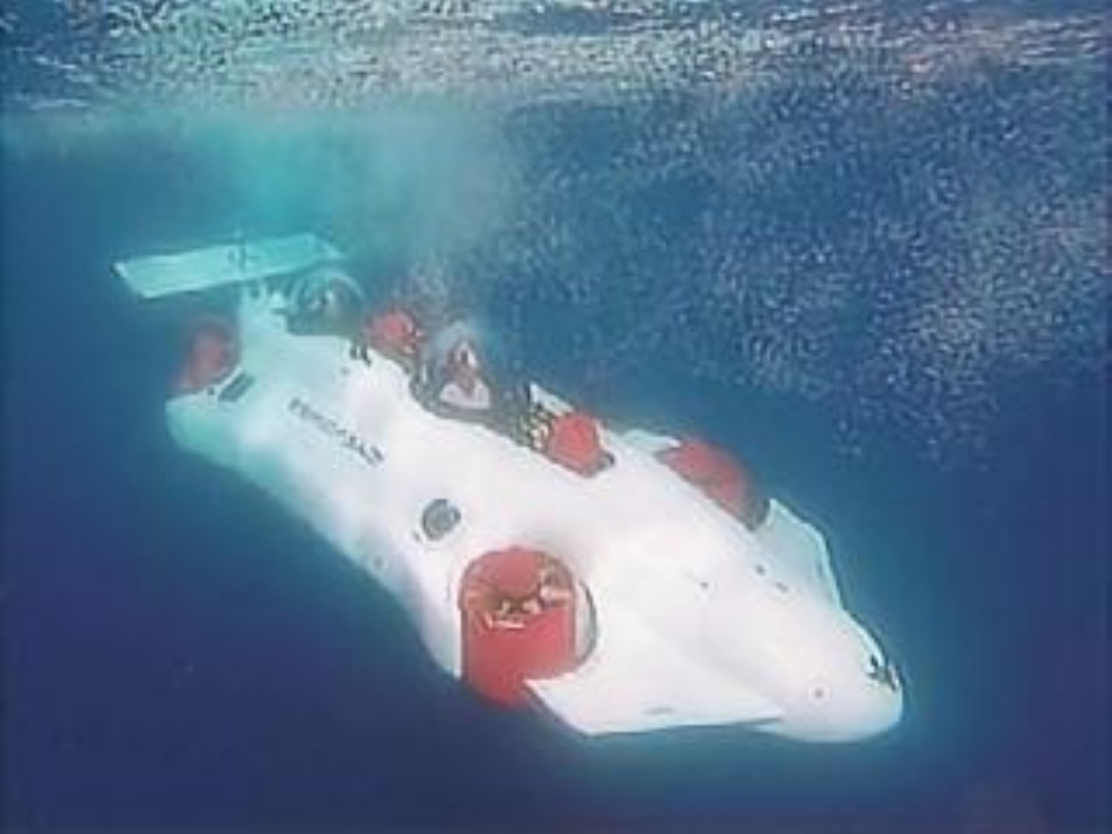}&
\includegraphics[width=3cm, height=2cm]{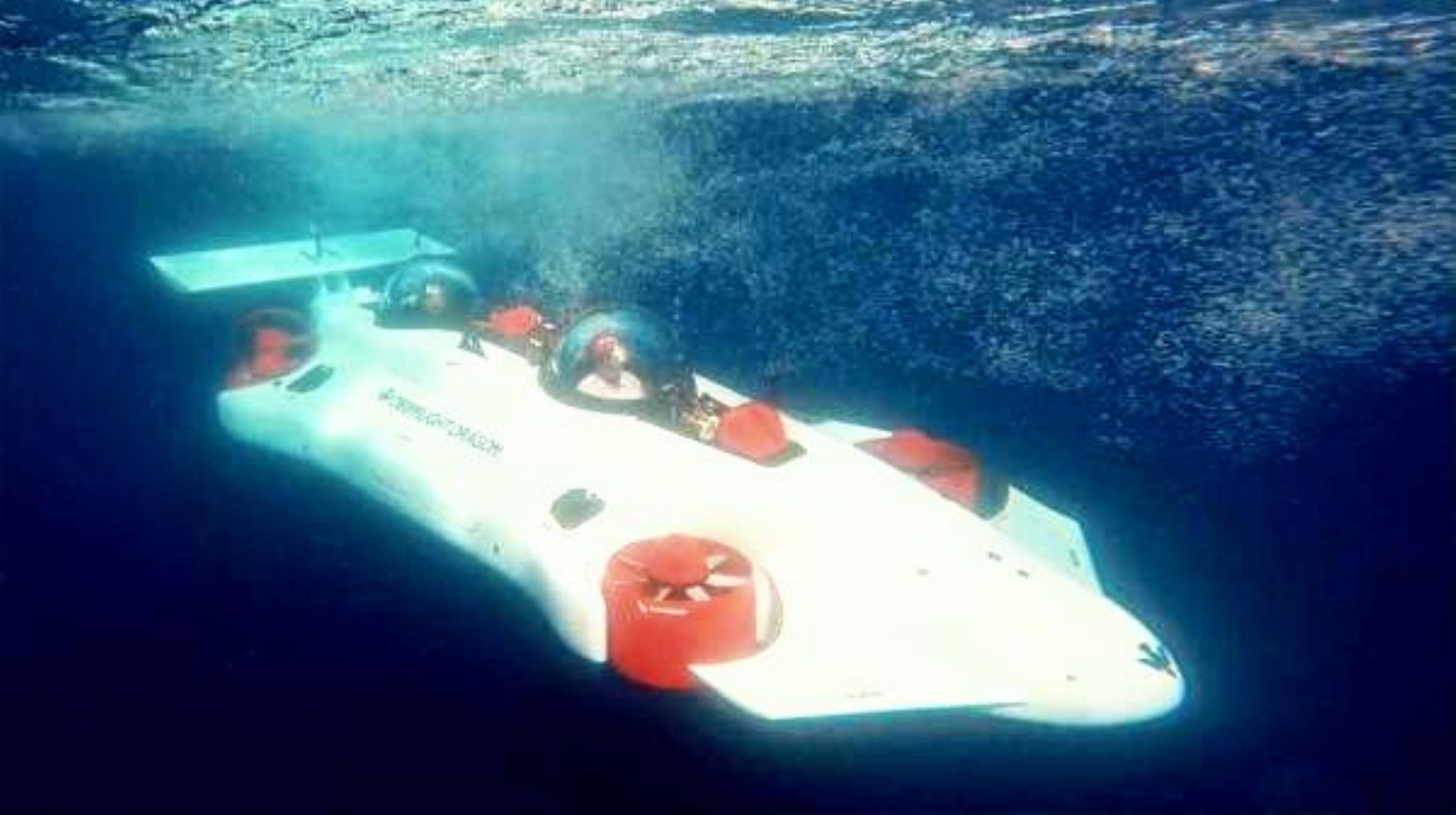}&
\includegraphics[width=3cm, height=2cm]{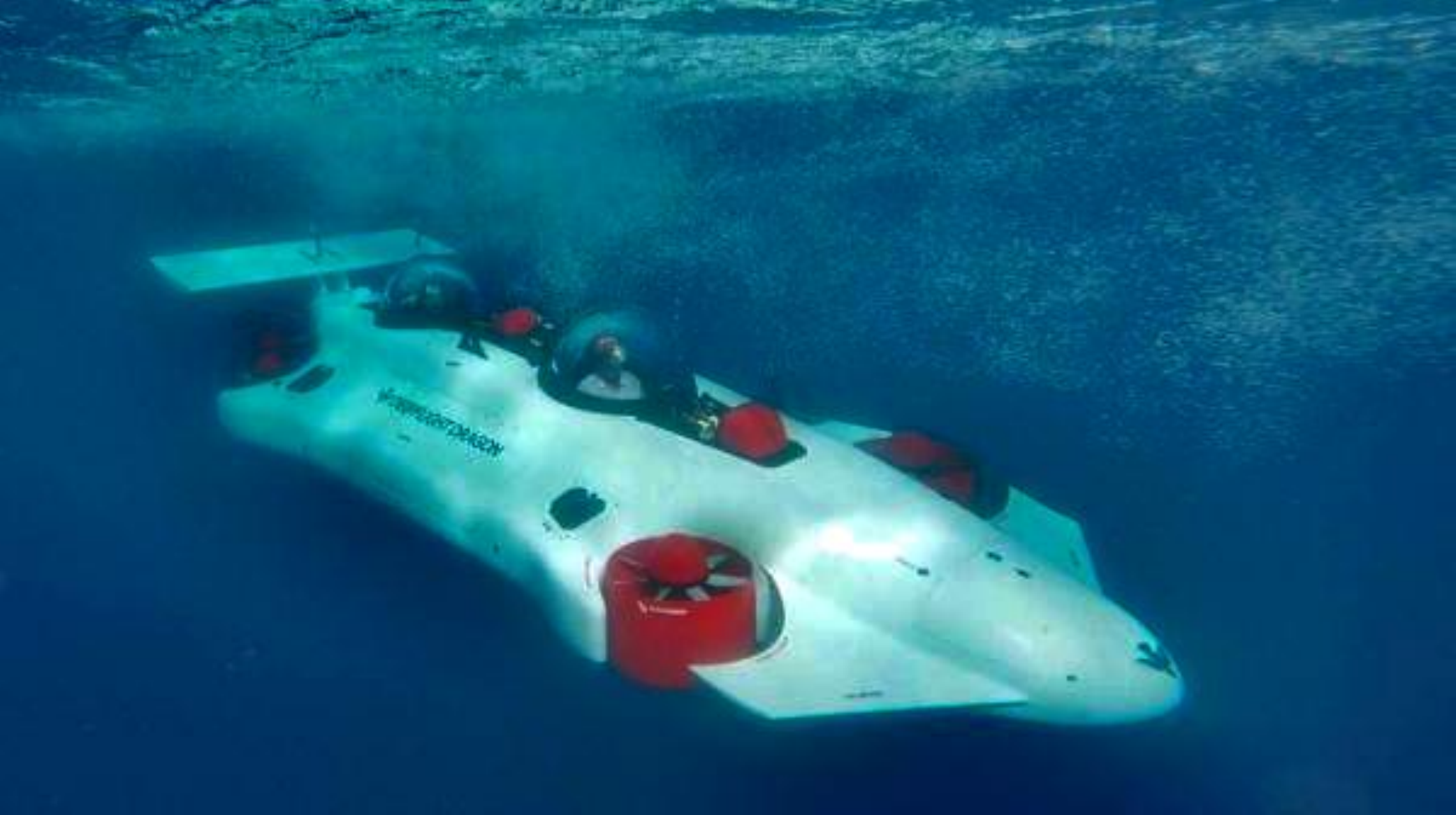}\\

\includegraphics[width=3cm, height=2cm]{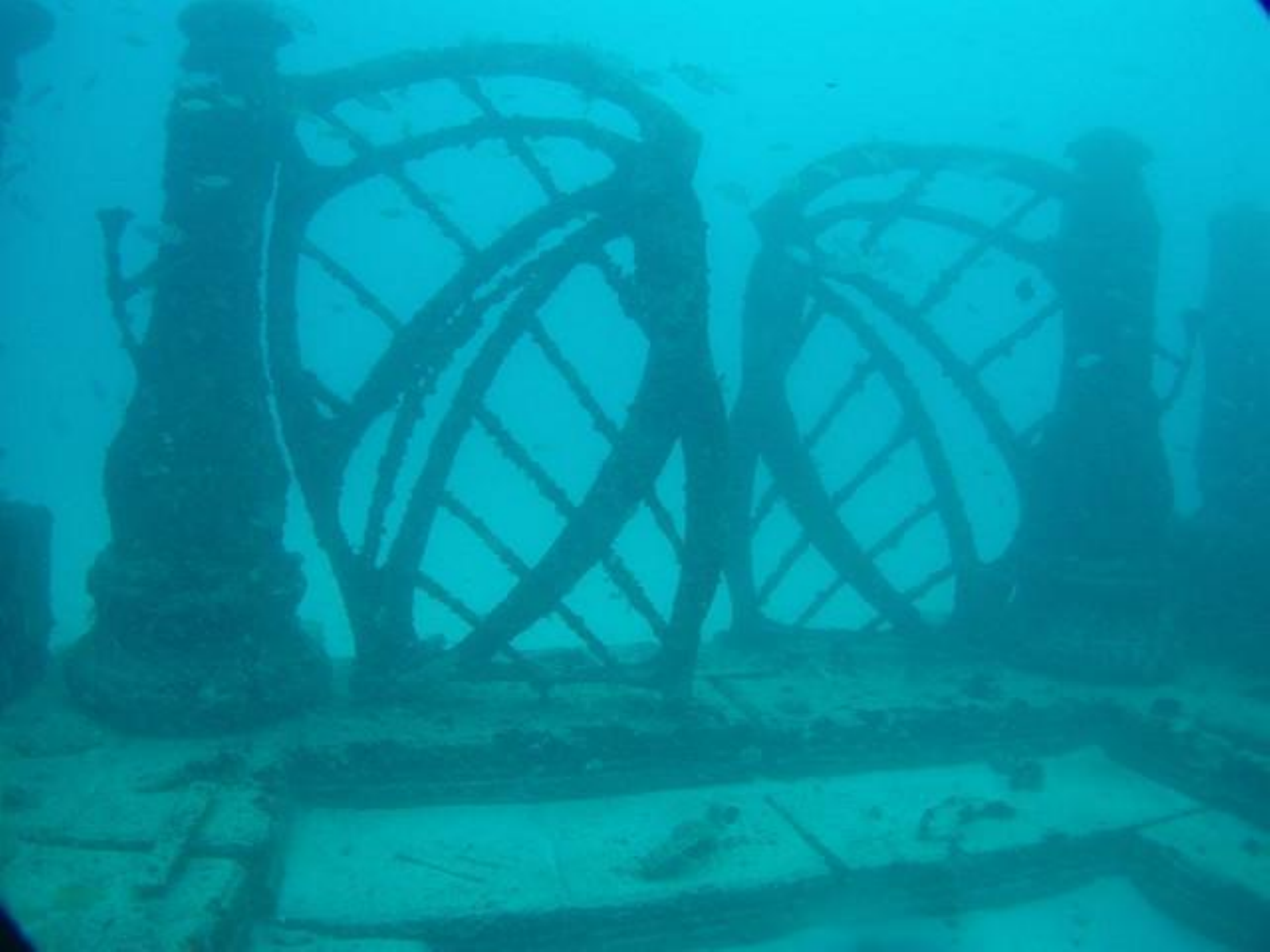}&
\includegraphics[width=3cm, height=2cm]{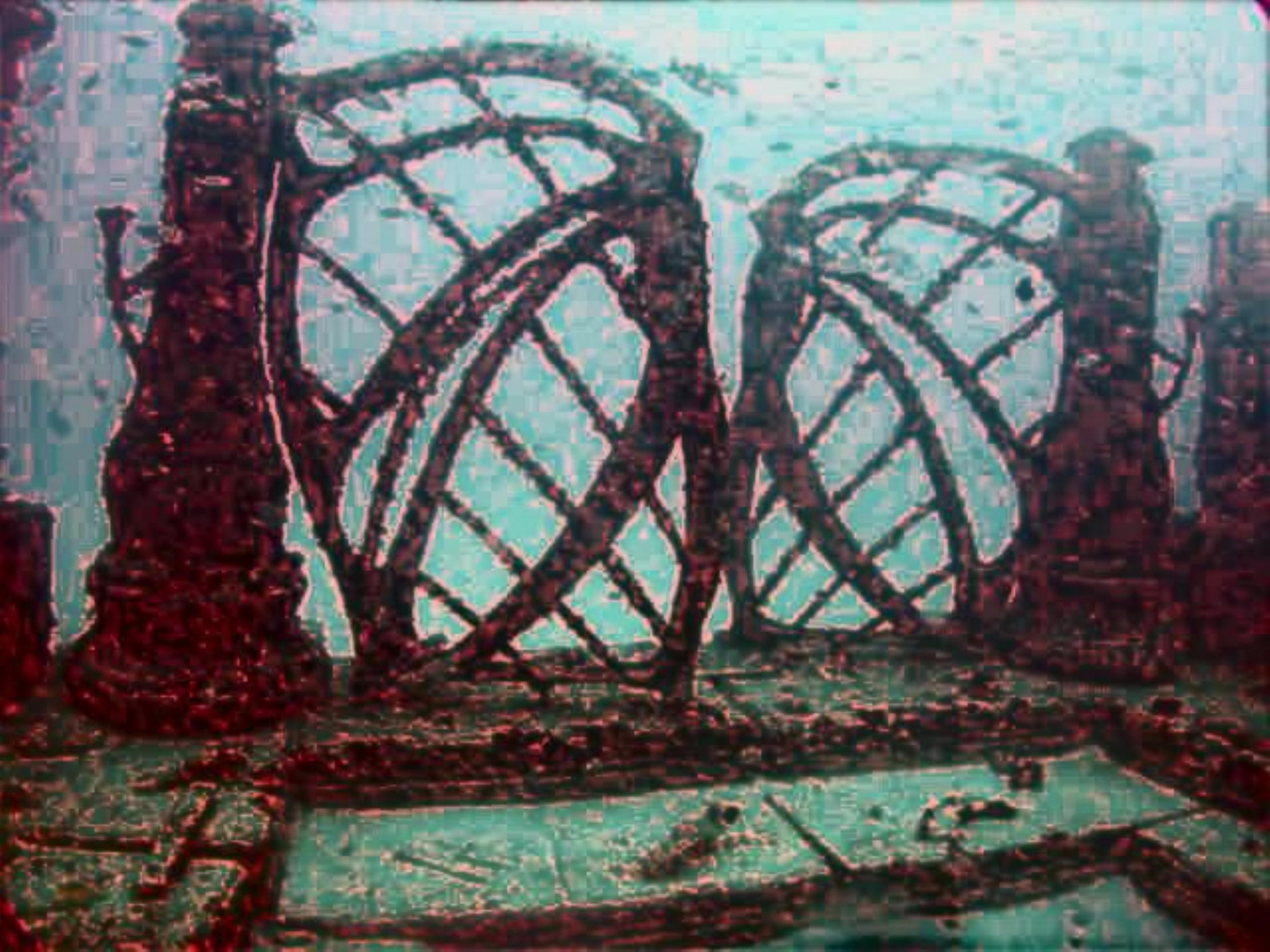}&
\includegraphics[width=3cm, height=2cm]{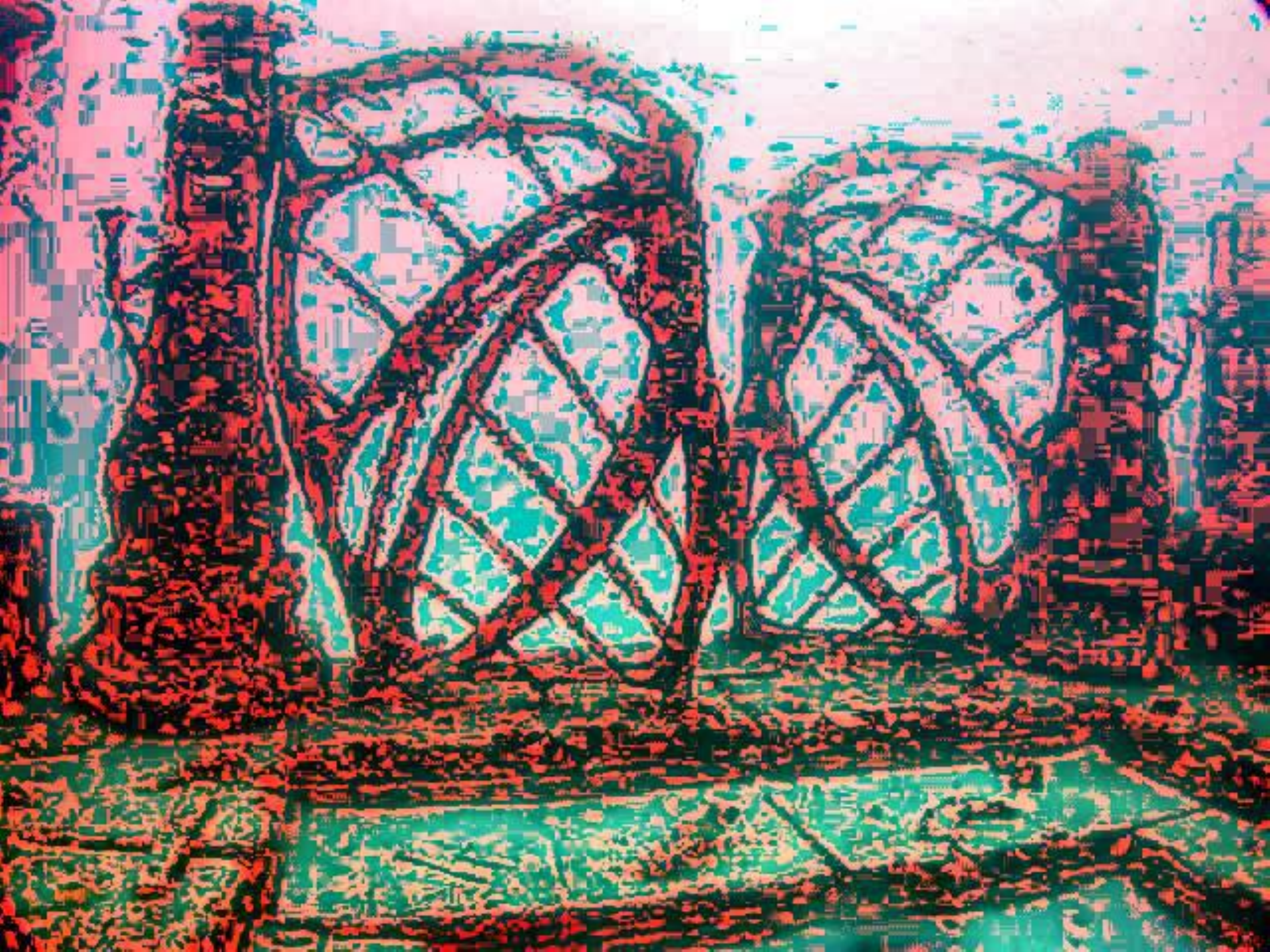}&
\includegraphics[width=3cm, height=2cm]{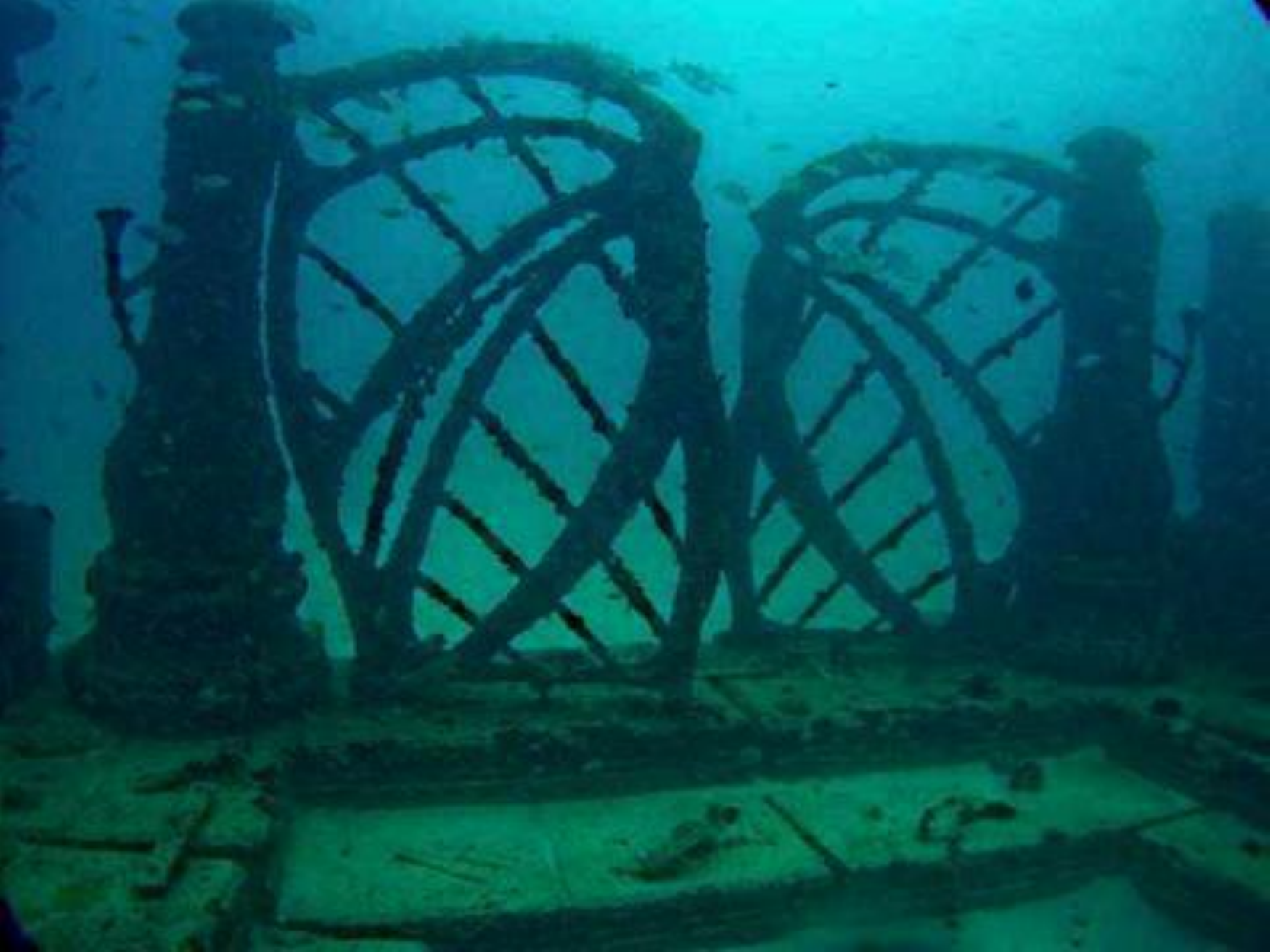}&
\includegraphics[width=3cm, height=2cm]{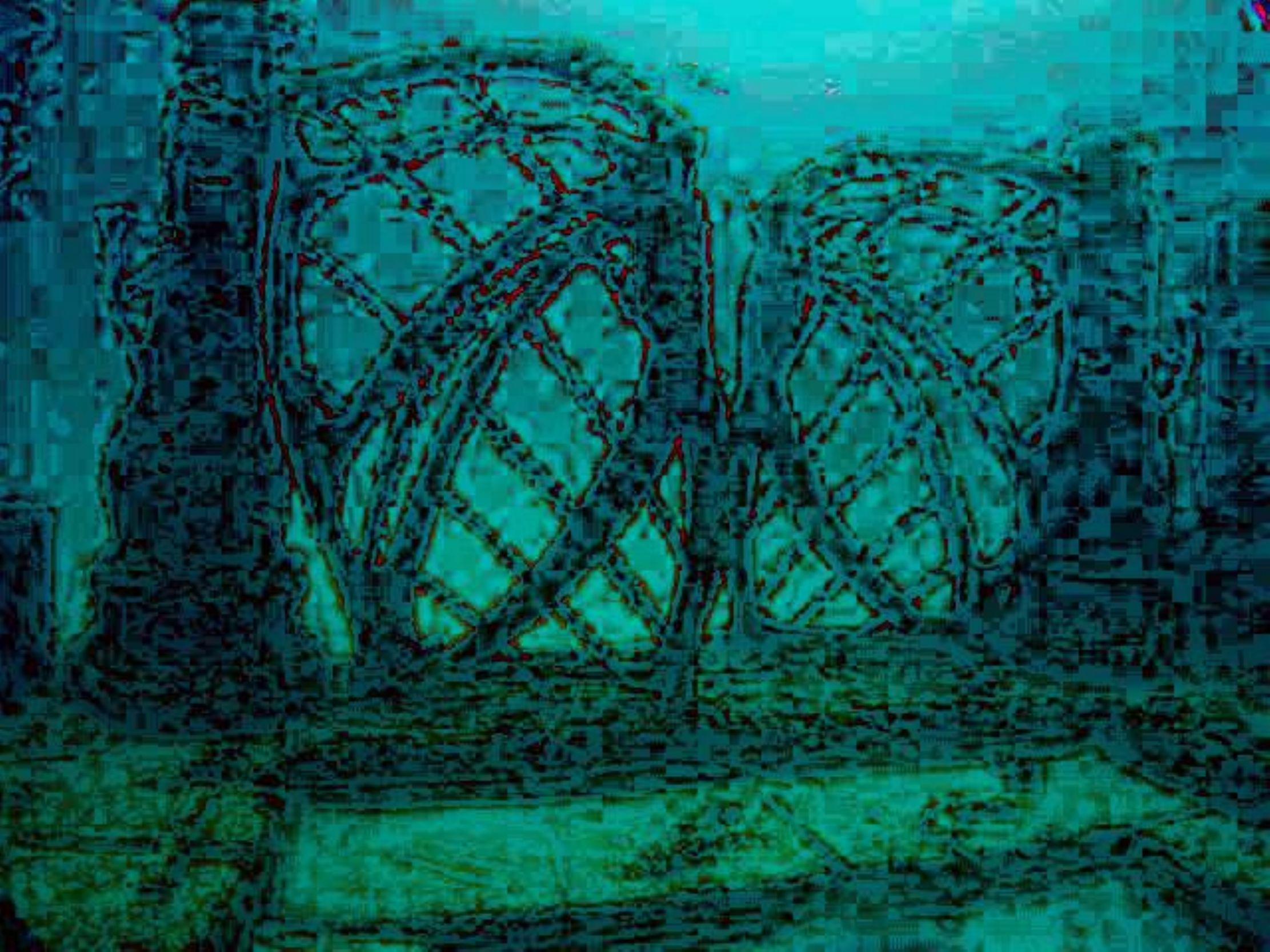}&
\includegraphics[width=3cm, height=2cm]{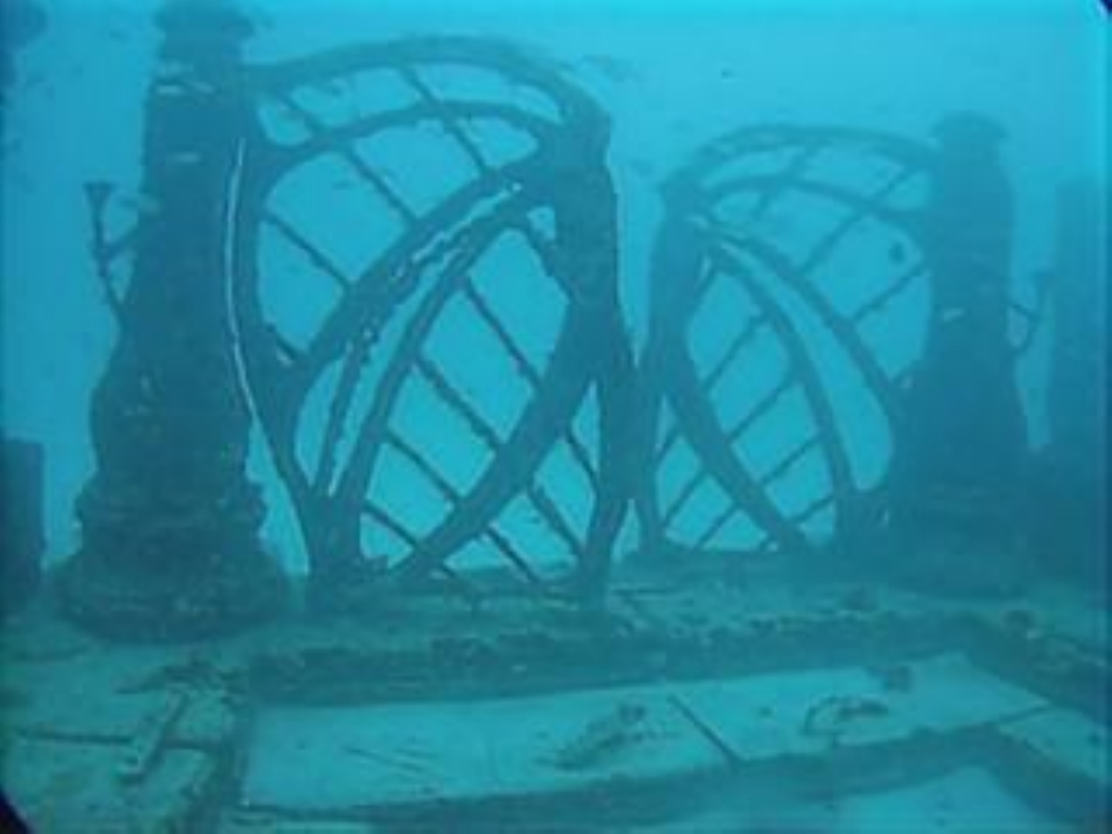}&
\includegraphics[width=3cm, height=2cm]{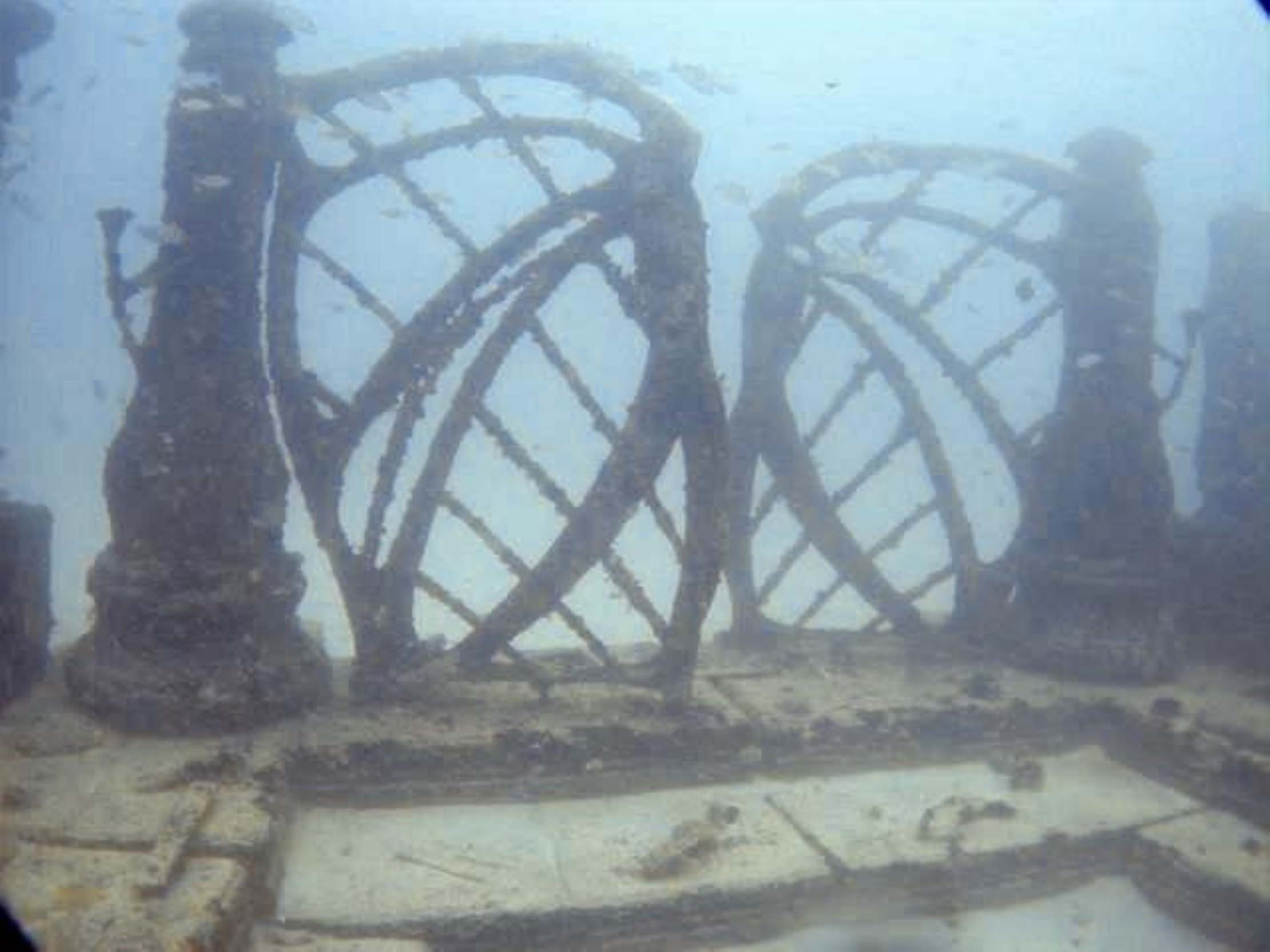}&
\includegraphics[width=3cm, height=2cm]{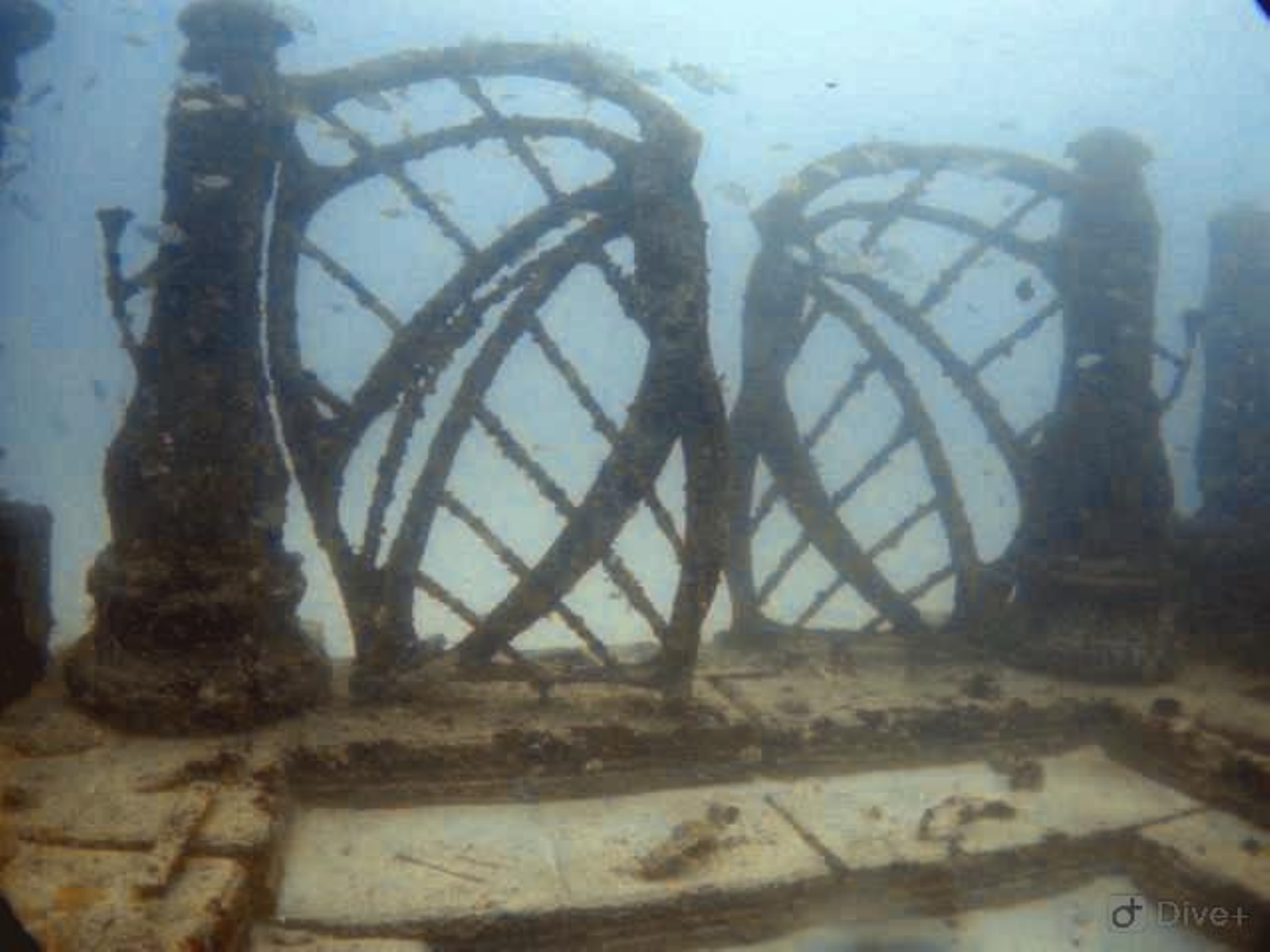}\\

\includegraphics[width=3cm, height=2cm]{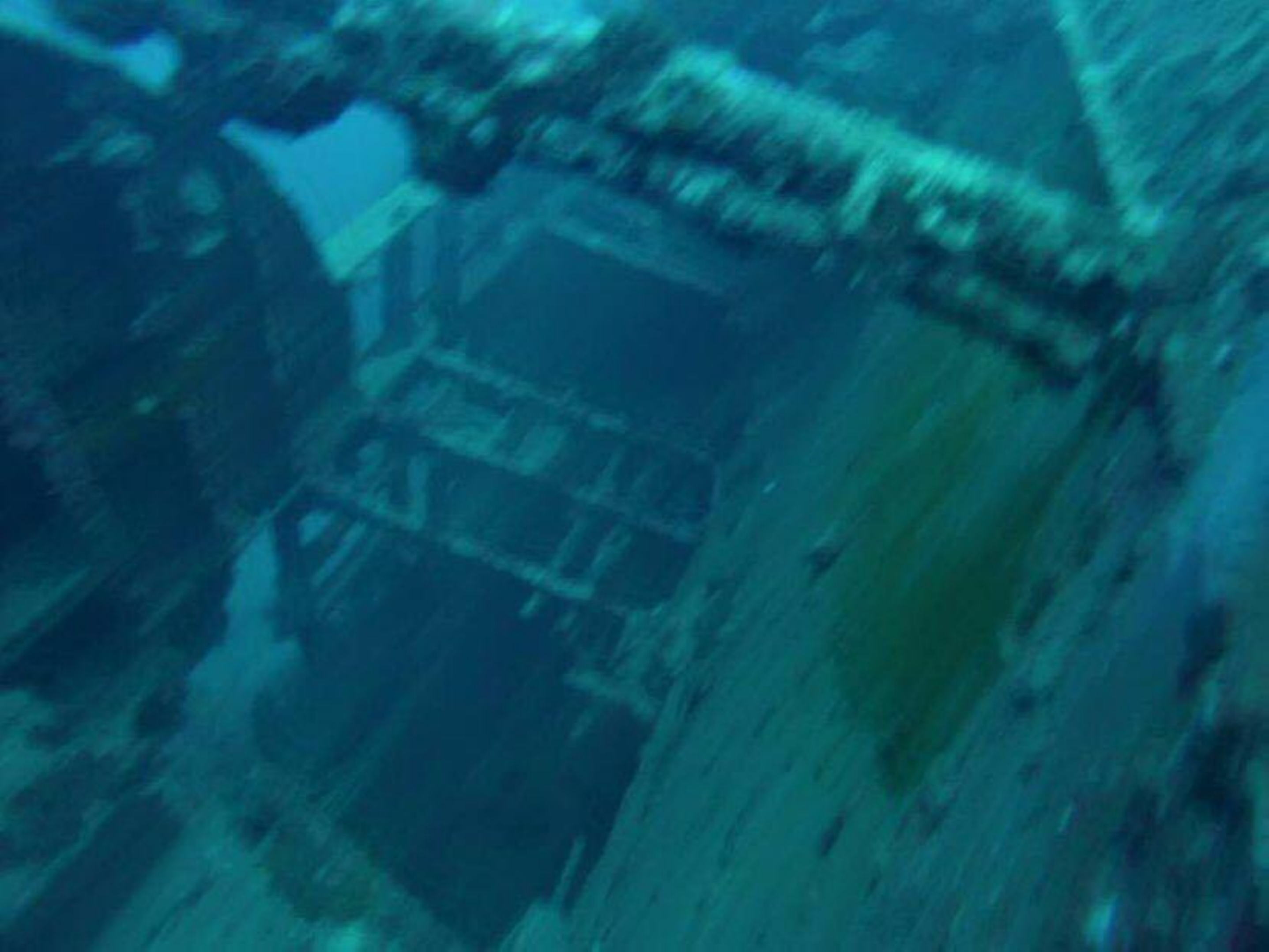}&
\includegraphics[width=3cm, height=2cm]{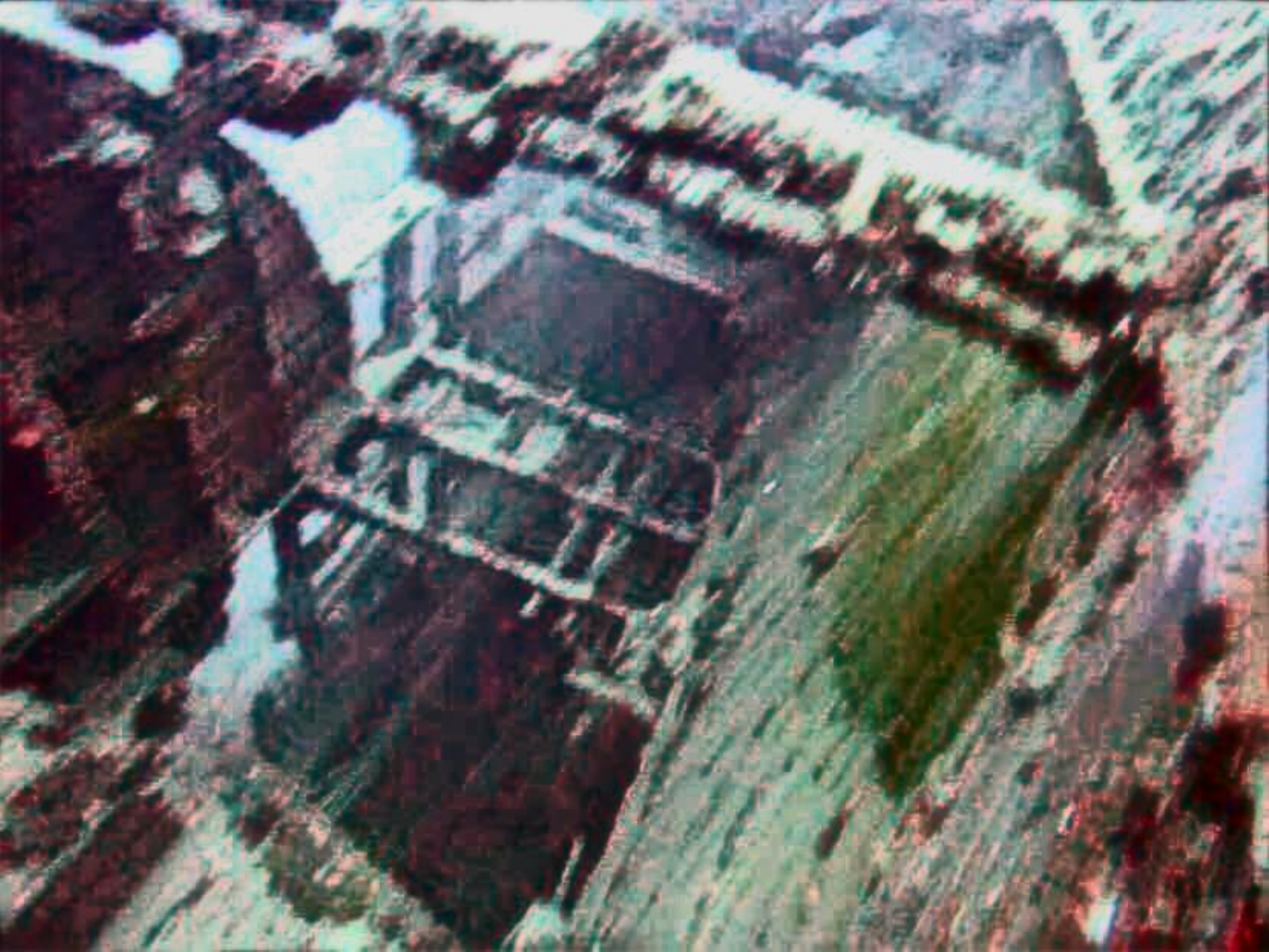}&
\includegraphics[width=3cm, height=2cm]{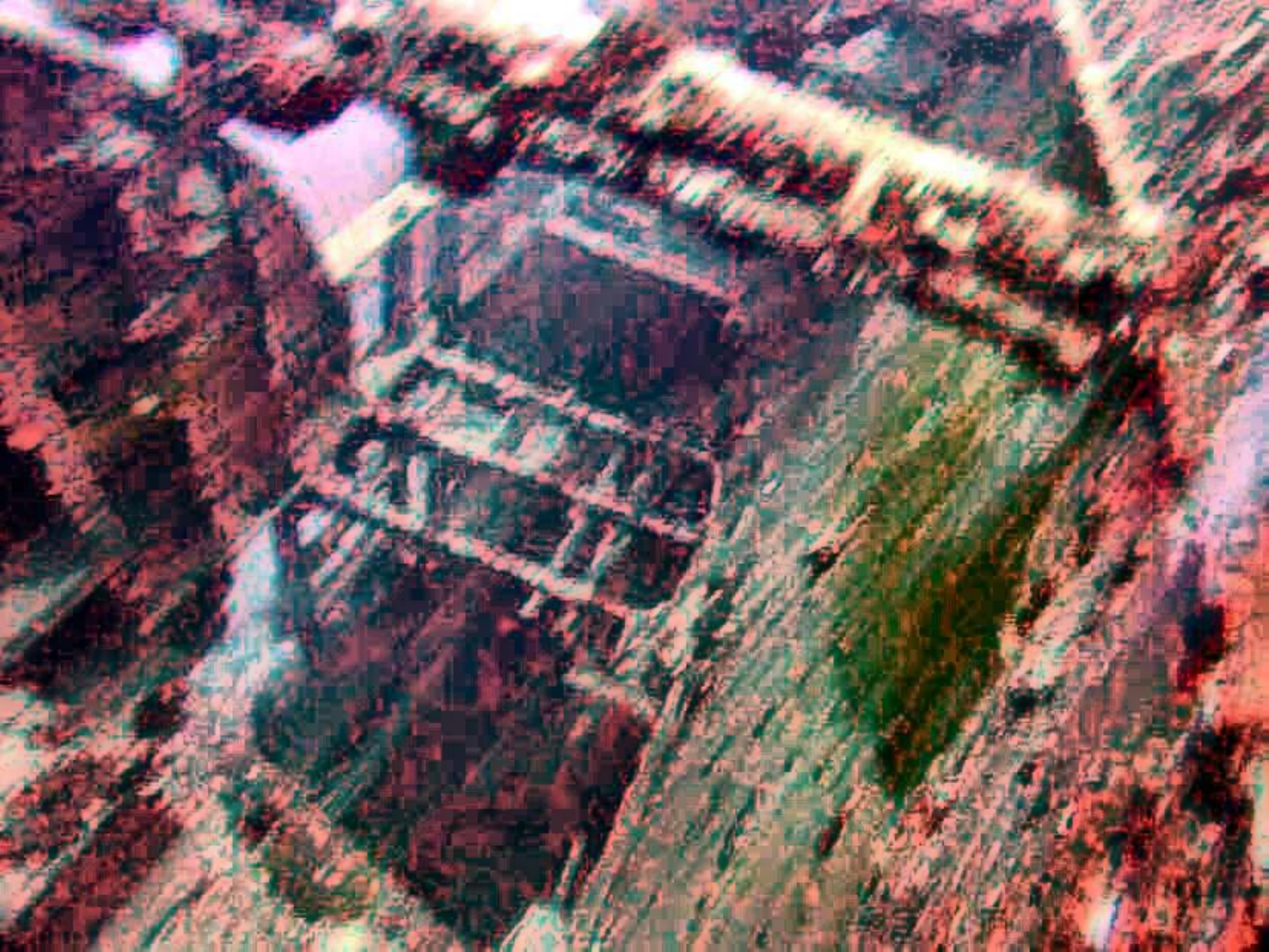}&
\includegraphics[width=3cm, height=2cm]{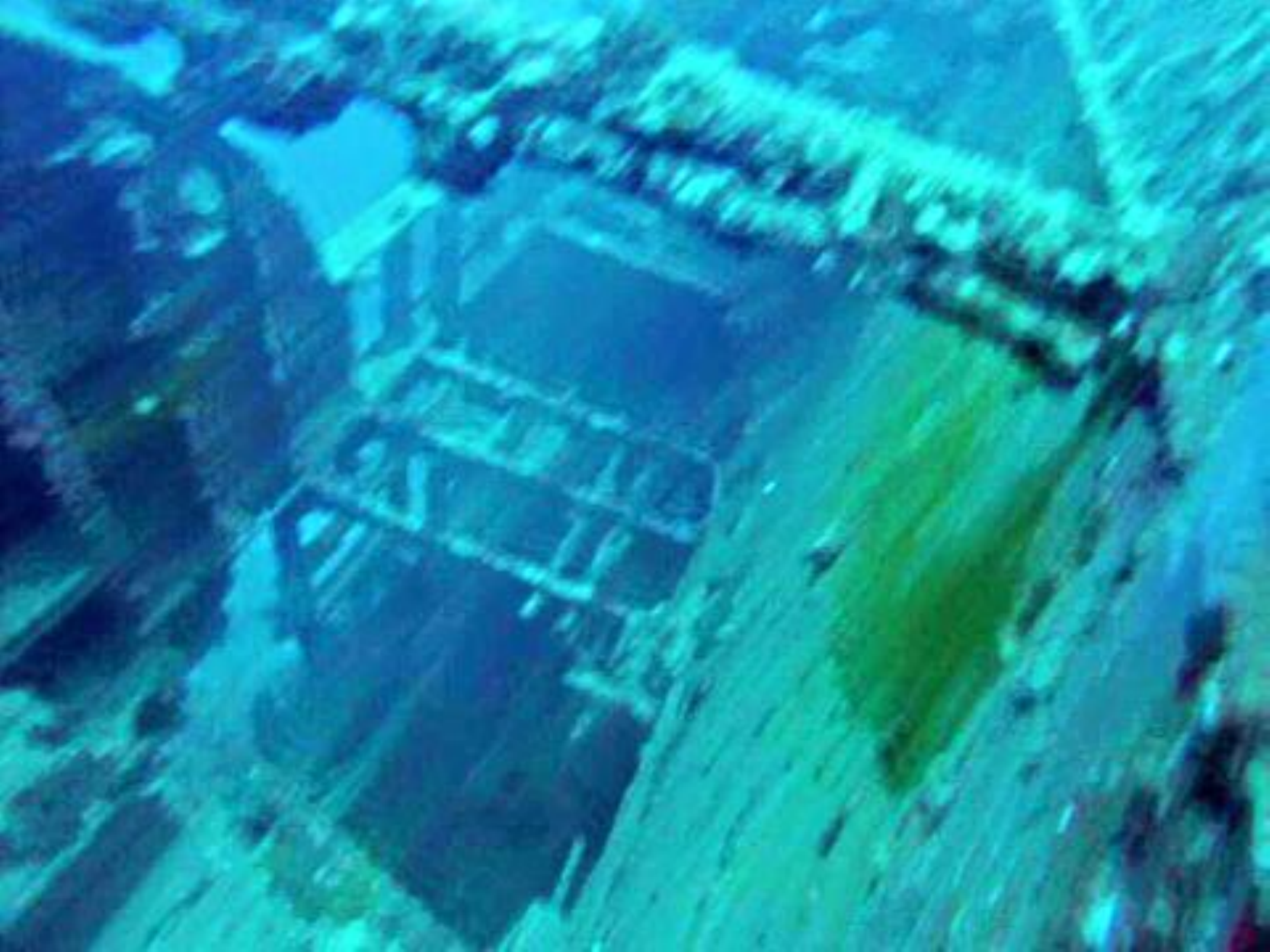}&
\includegraphics[width=3cm, height=2cm]{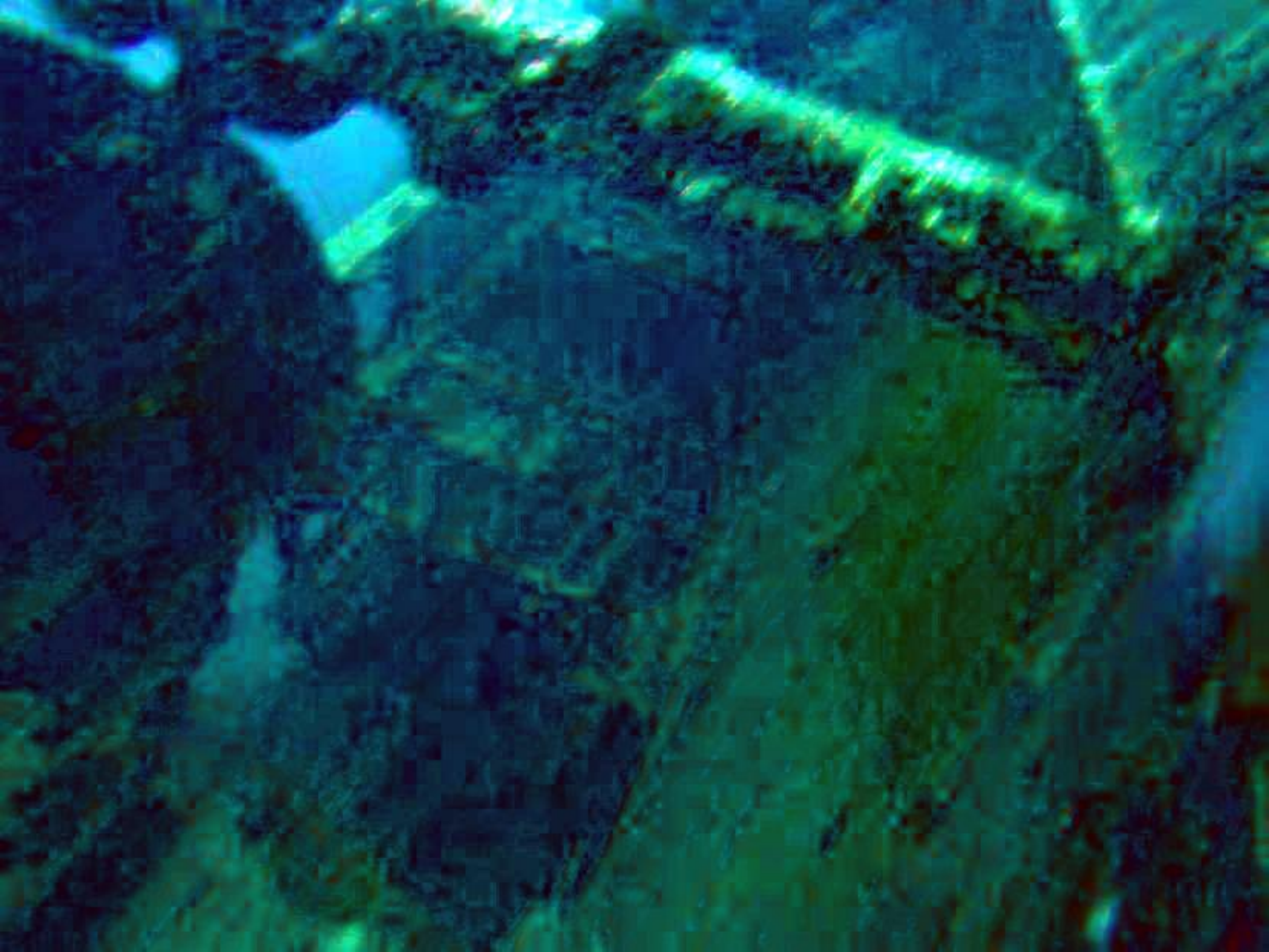}&
\includegraphics[width=3cm, height=2cm]{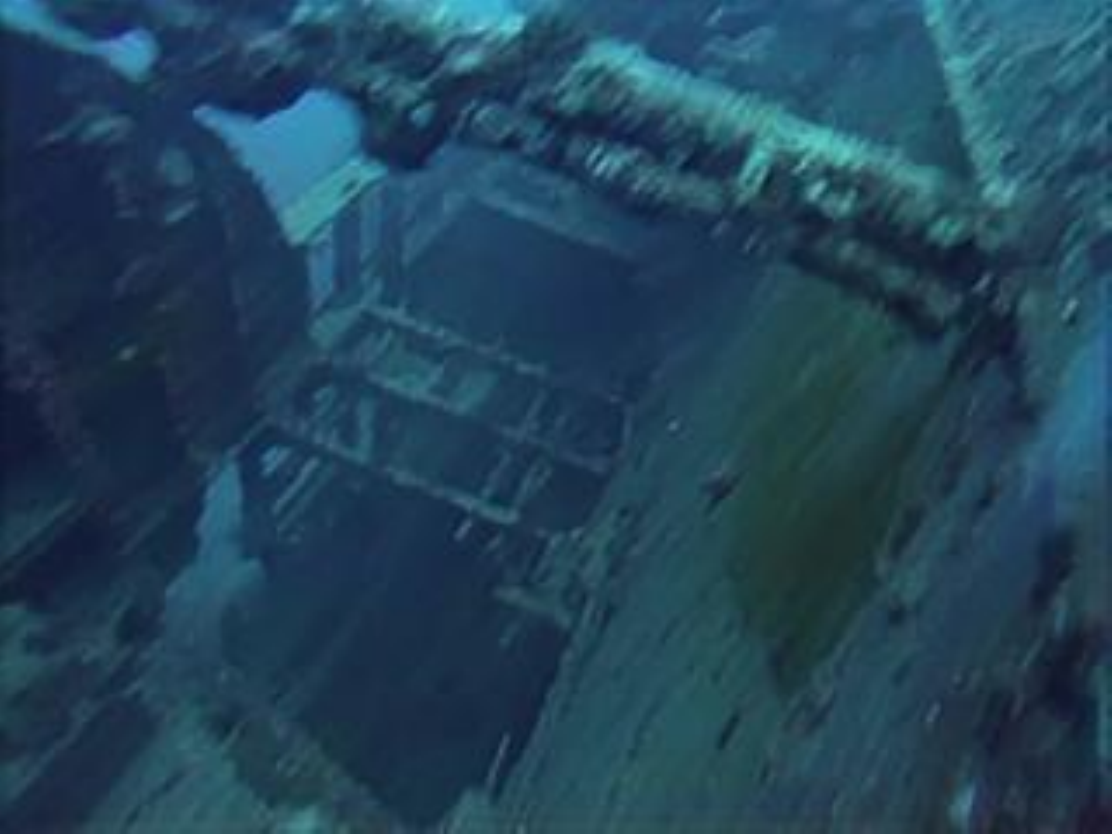}&
\includegraphics[width=3cm, height=2cm]{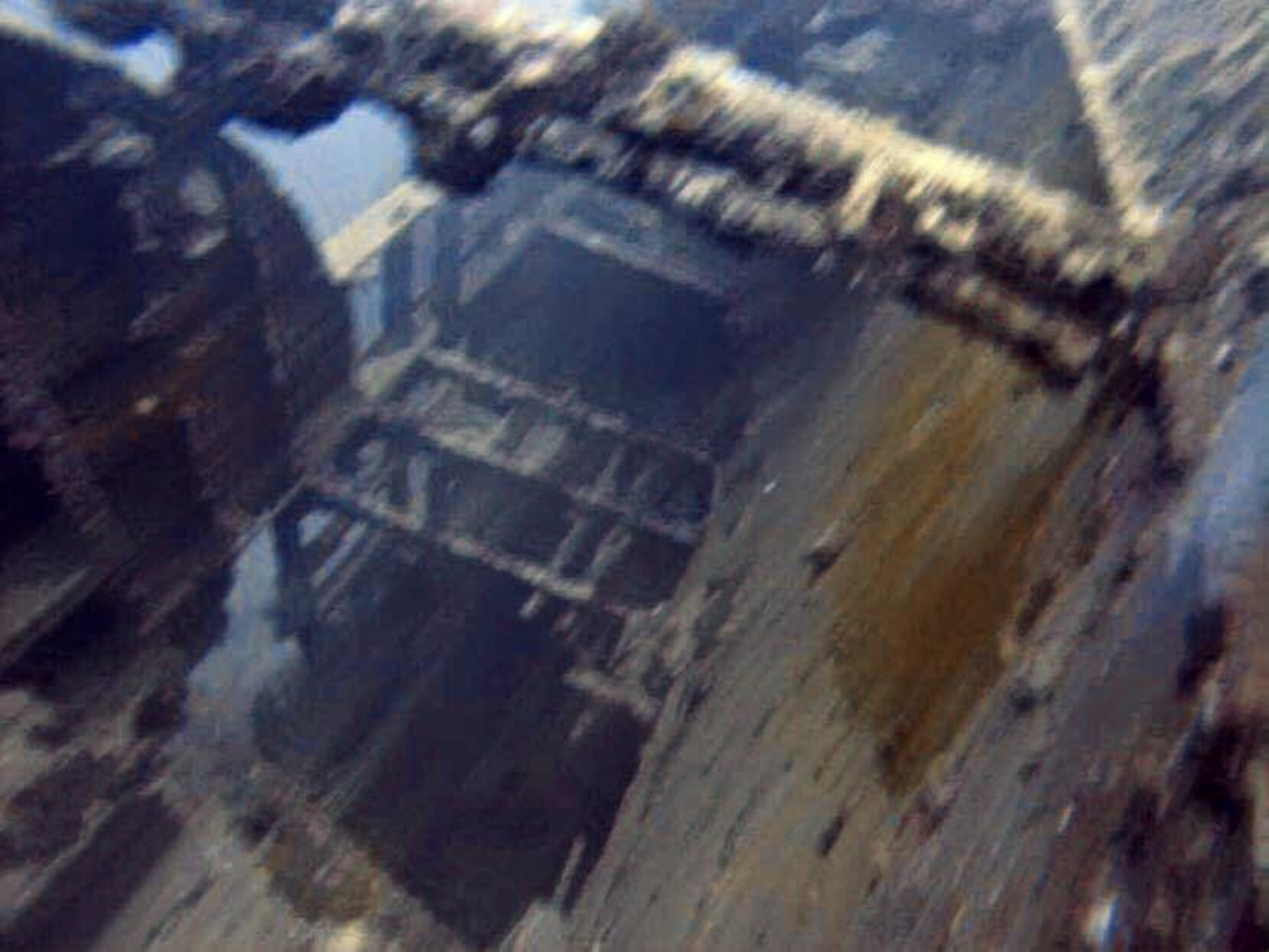}&
\includegraphics[width=3cm, height=2cm]{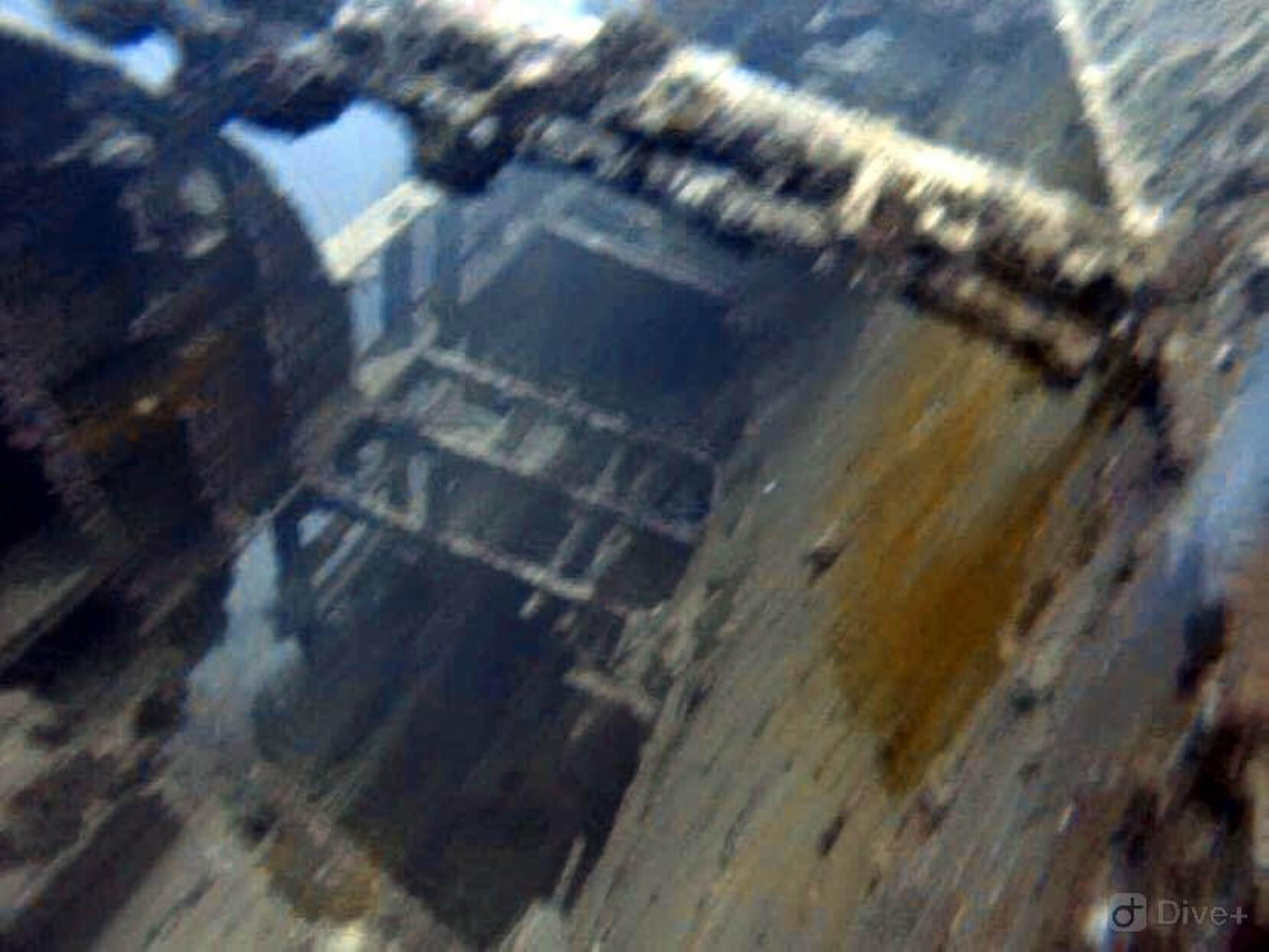}\\

\includegraphics[width=3cm, height=2cm]{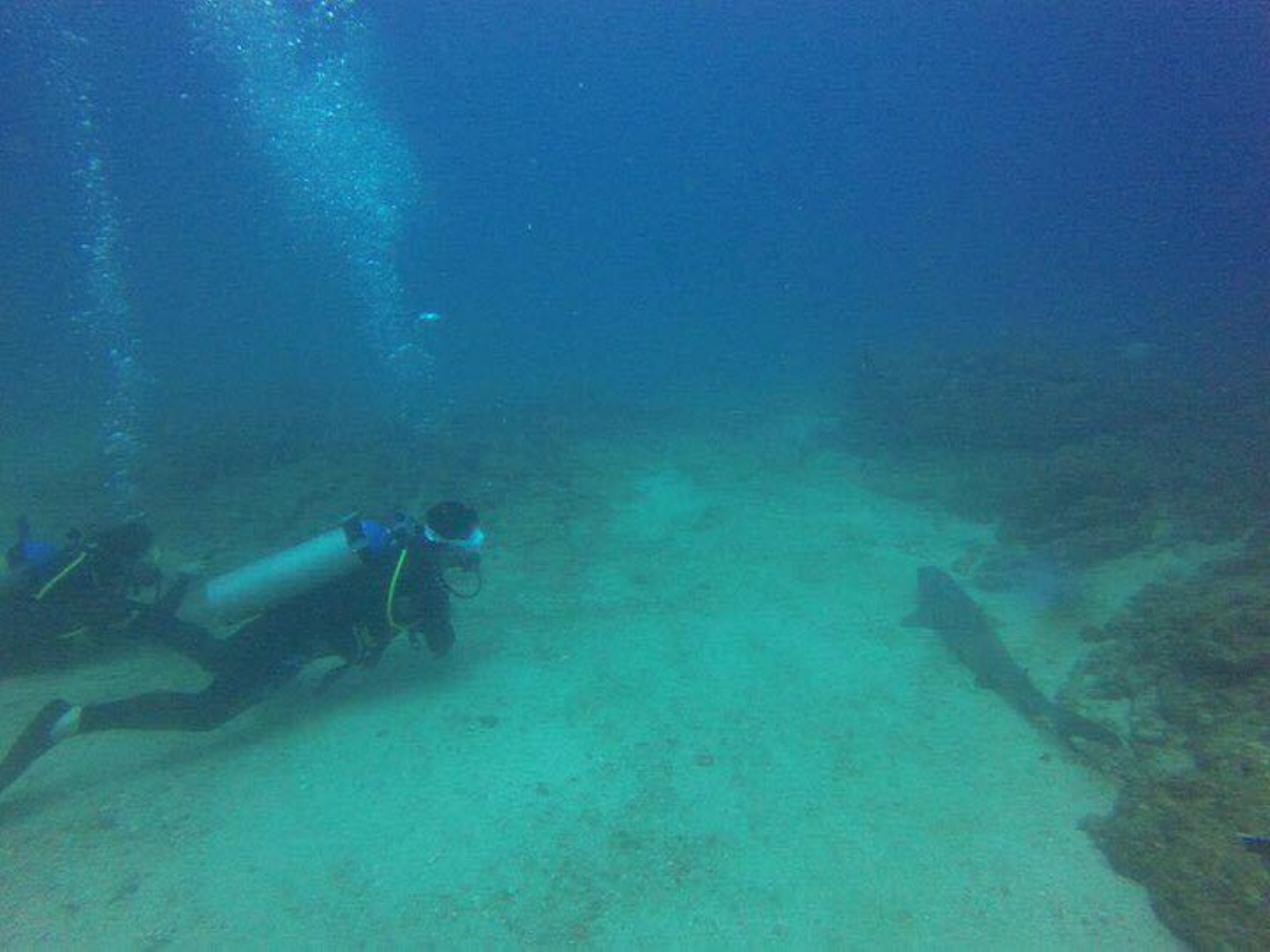}&
\includegraphics[width=3cm, height=2cm]{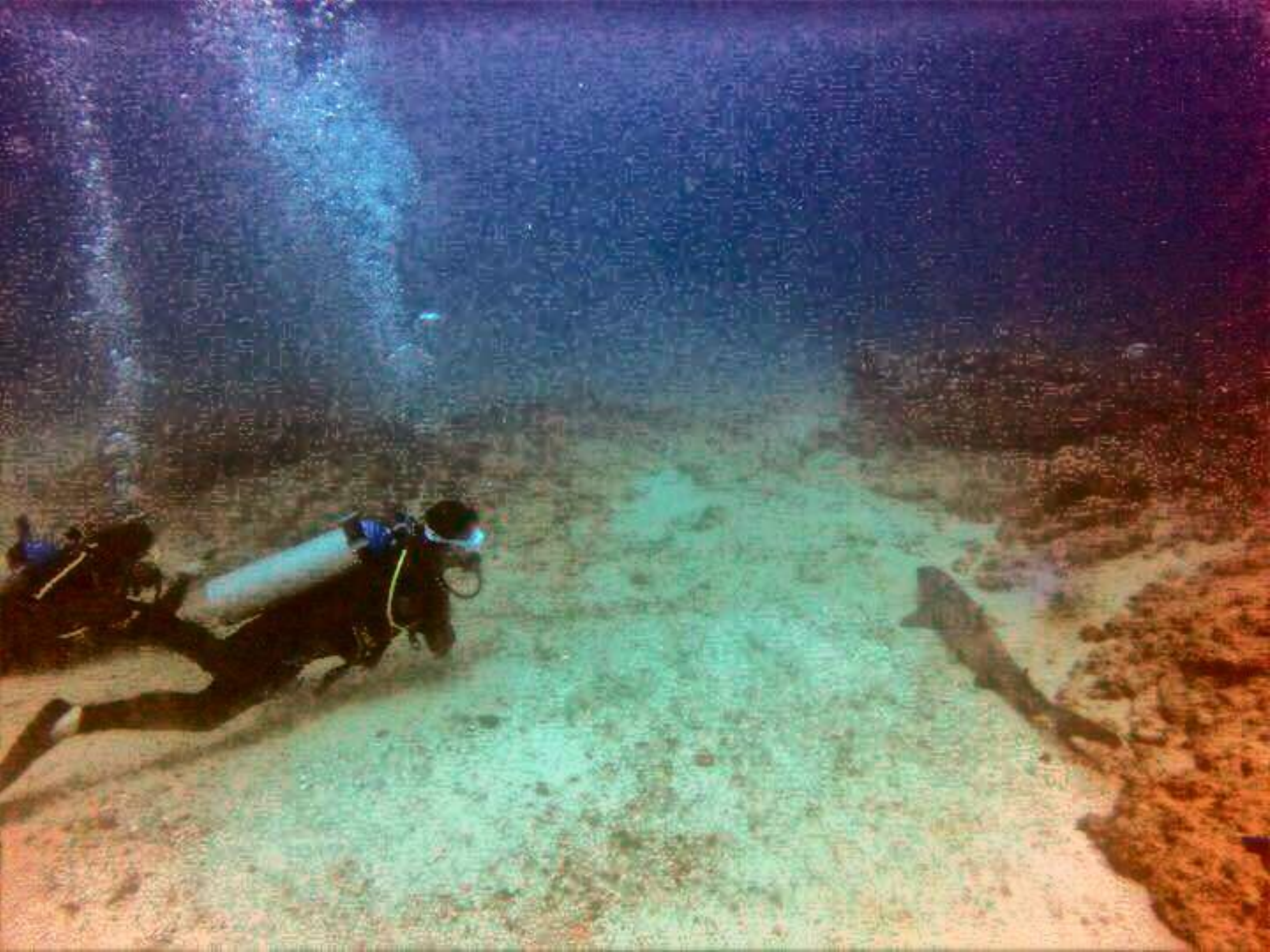}&
\includegraphics[width=3cm, height=2cm]{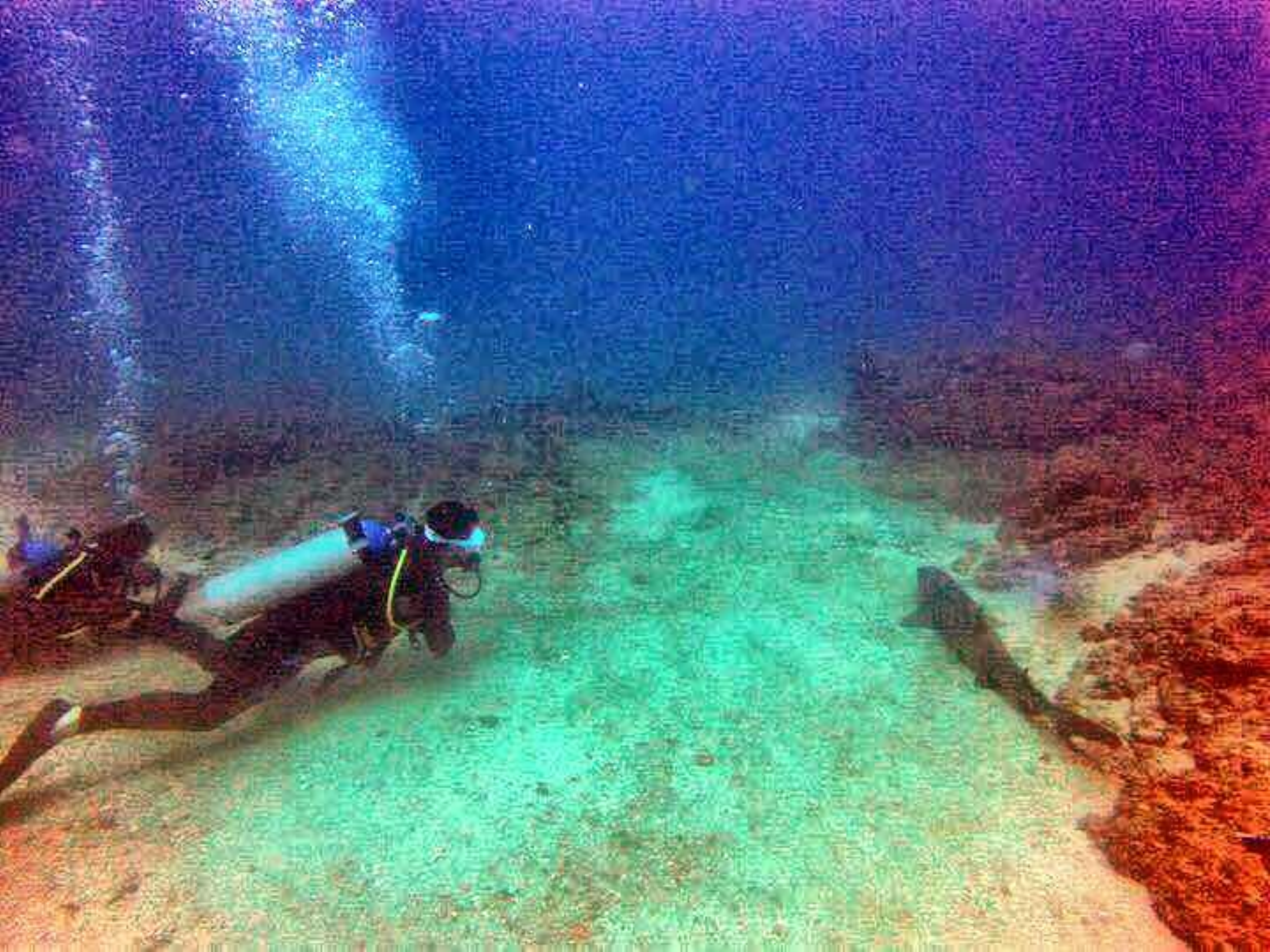}&
\includegraphics[width=3cm, height=2cm]{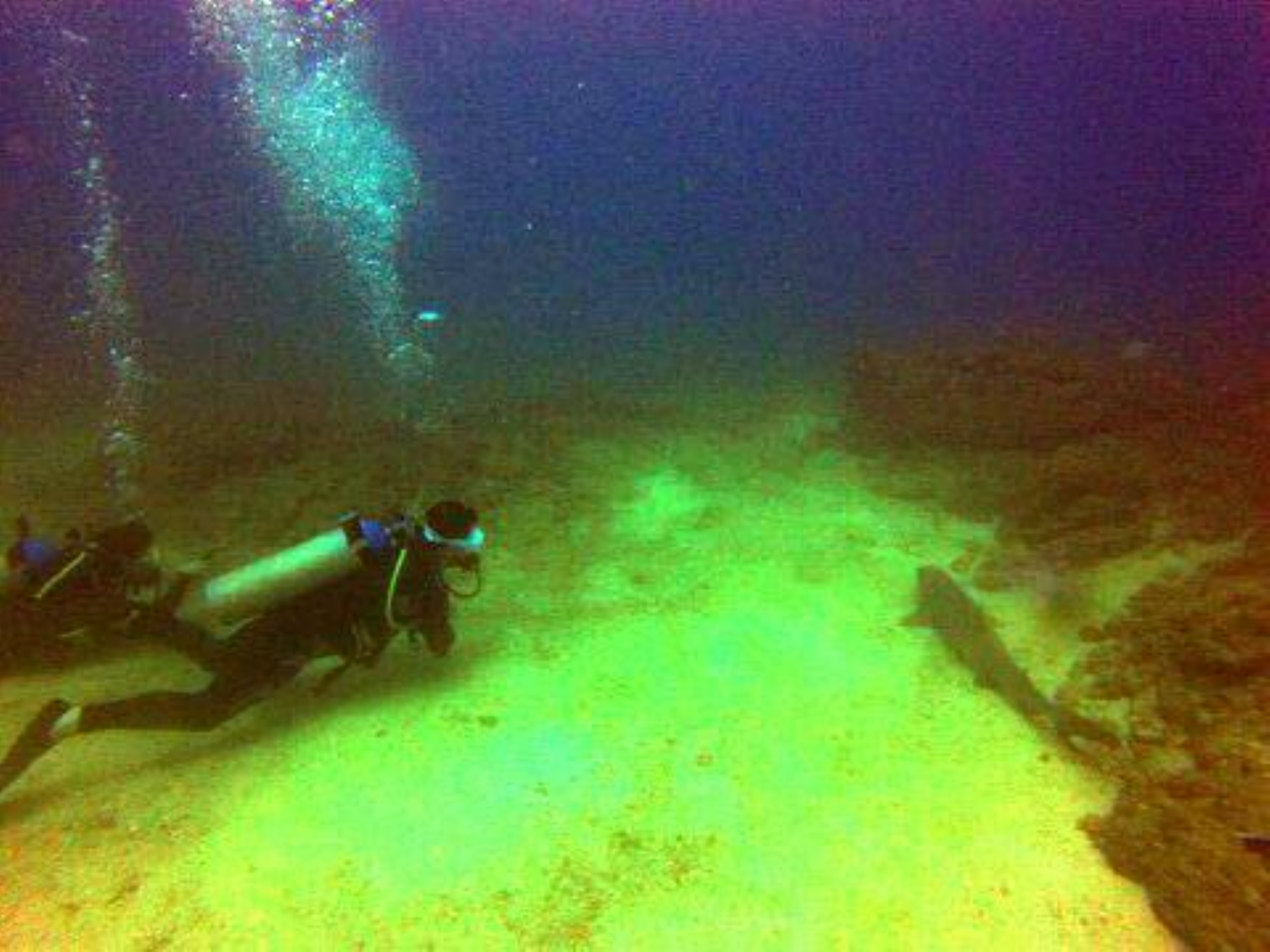}&
\includegraphics[width=3cm, height=2cm]{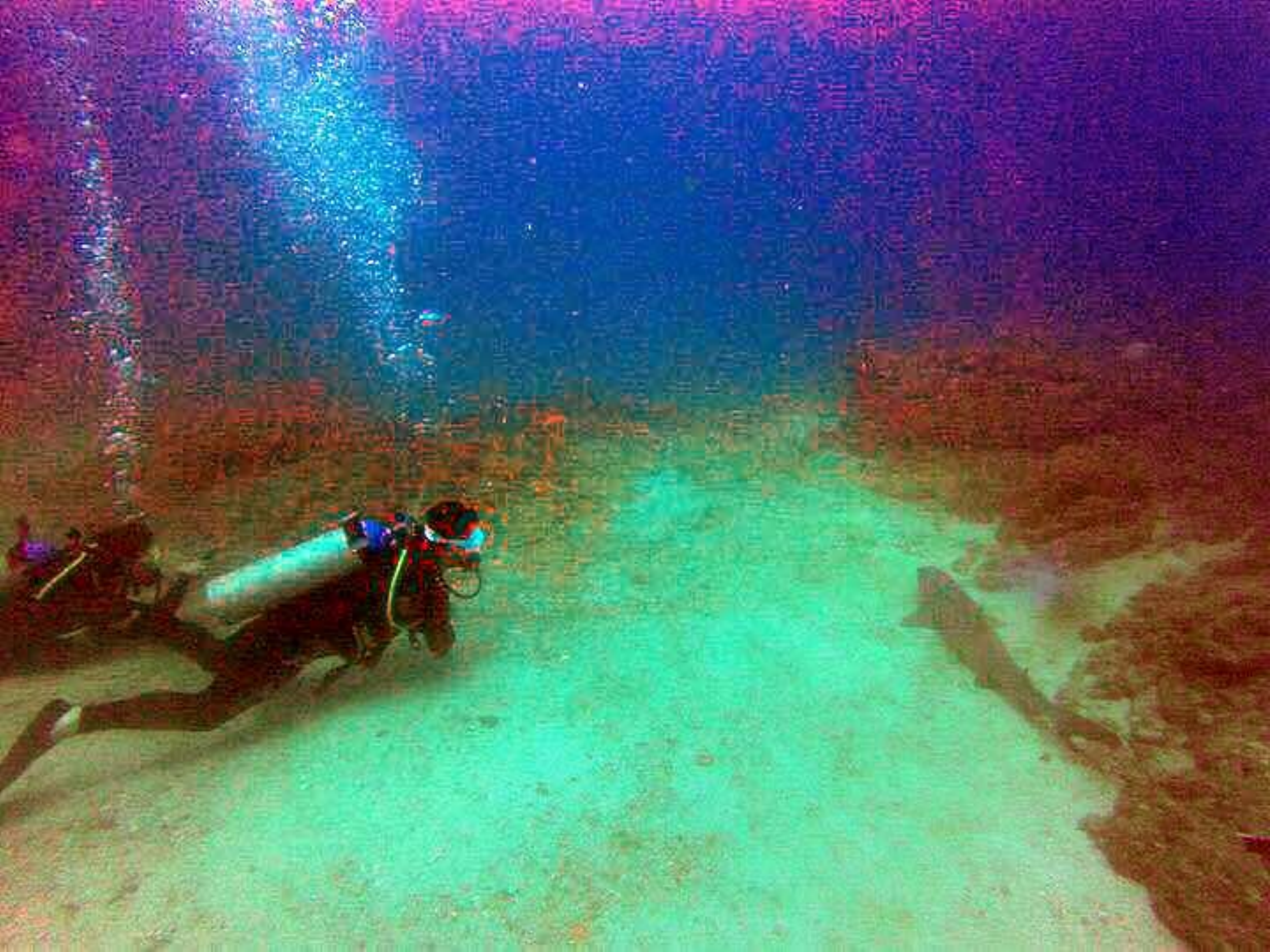}&
\includegraphics[width=3cm, height=2cm]{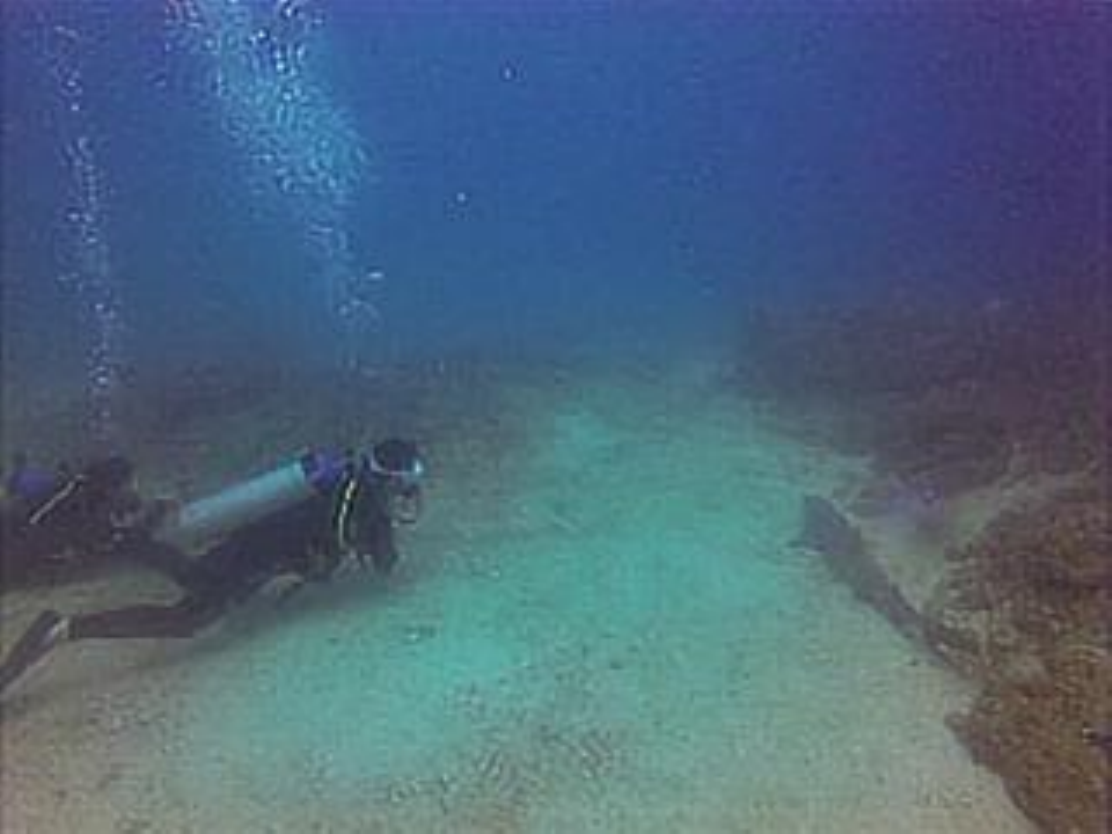}&
\includegraphics[width=3cm, height=2cm]{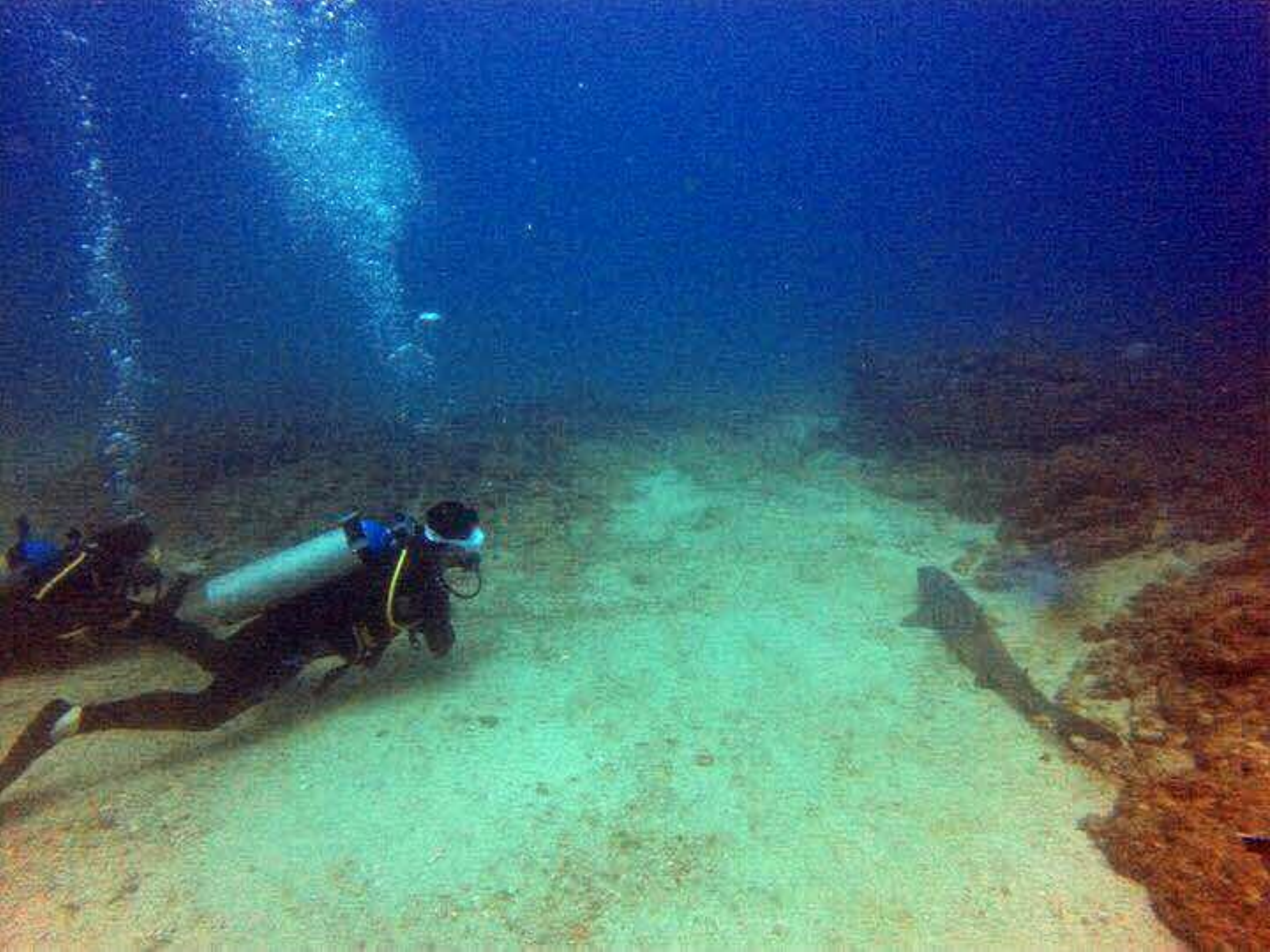}&
\includegraphics[width=3cm, height=2cm]{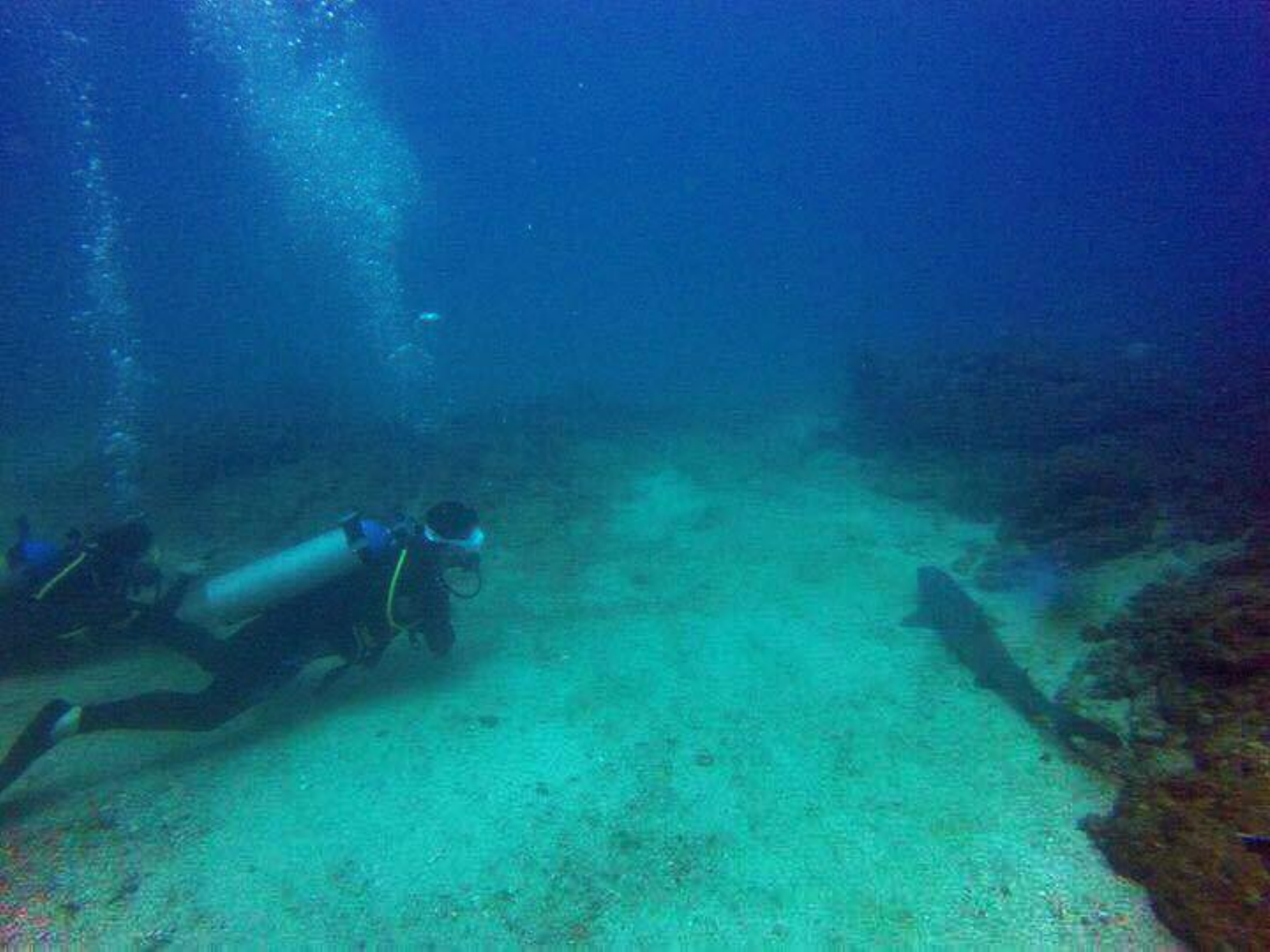}\\

{ Degraded} & { Retinex-based \cite{retinex-based}} & { Fusion-based \cite{ancuti1} } & { GDCP \cite{gdcp} } & { Haze Lines \cite{berman_pami_20} } & { Deep SESR \cite{sesr} } & { Deep WaveNet} & { Ground Truth}\\

\includegraphics[width=3cm, height=2cm]{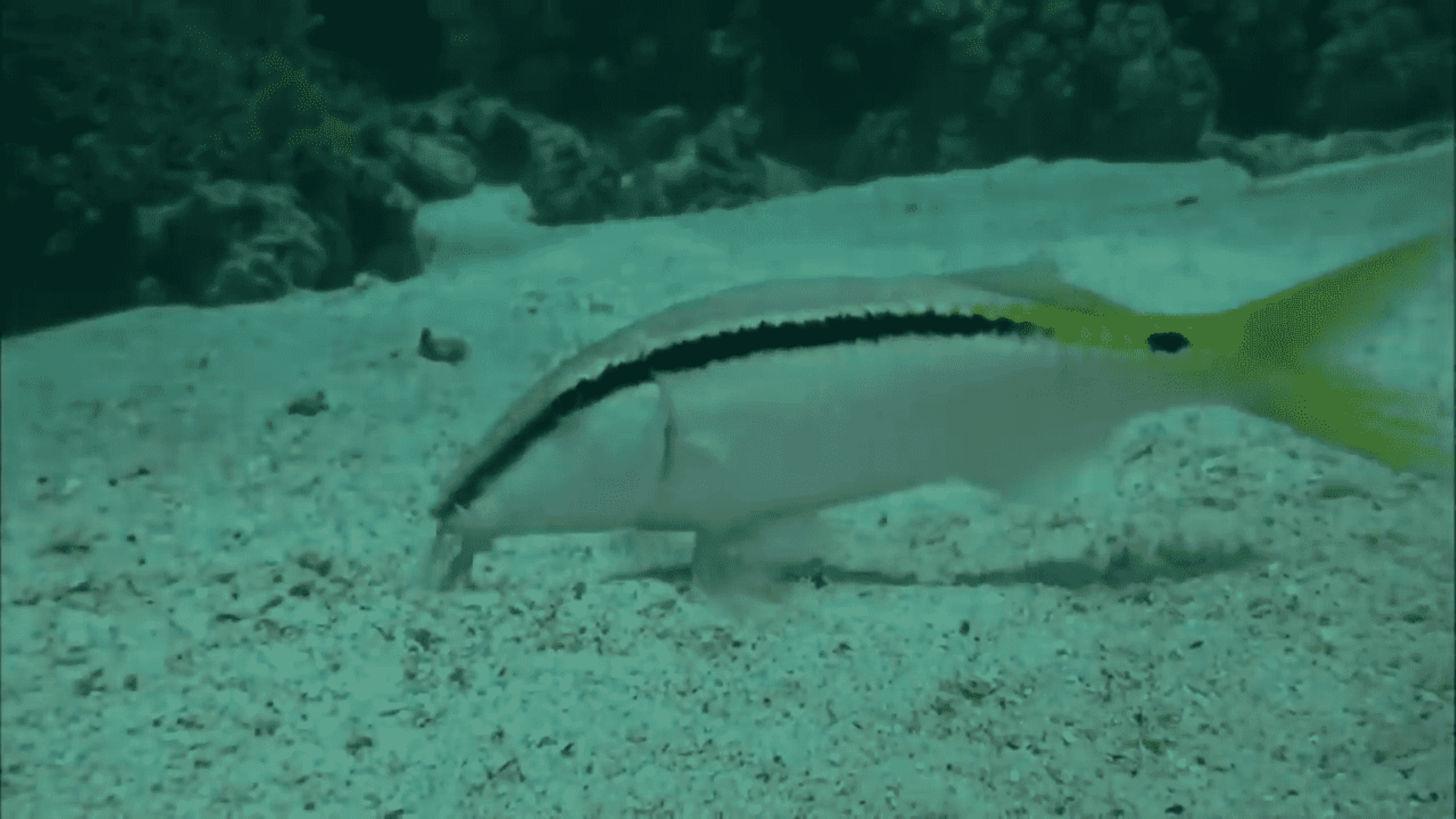}&
\includegraphics[width=3cm, height=2cm]{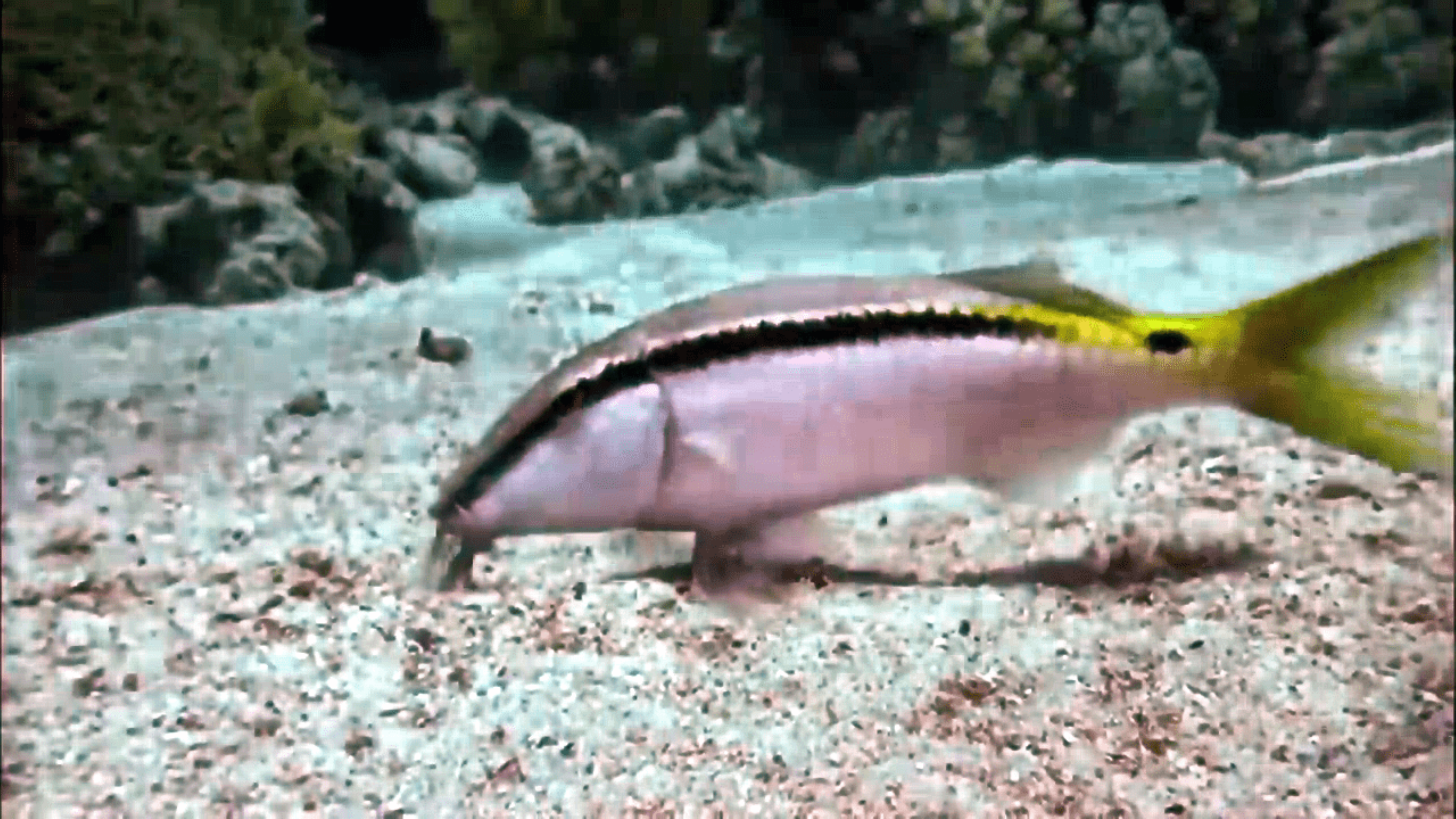}&
\includegraphics[width=3cm, height=2cm]{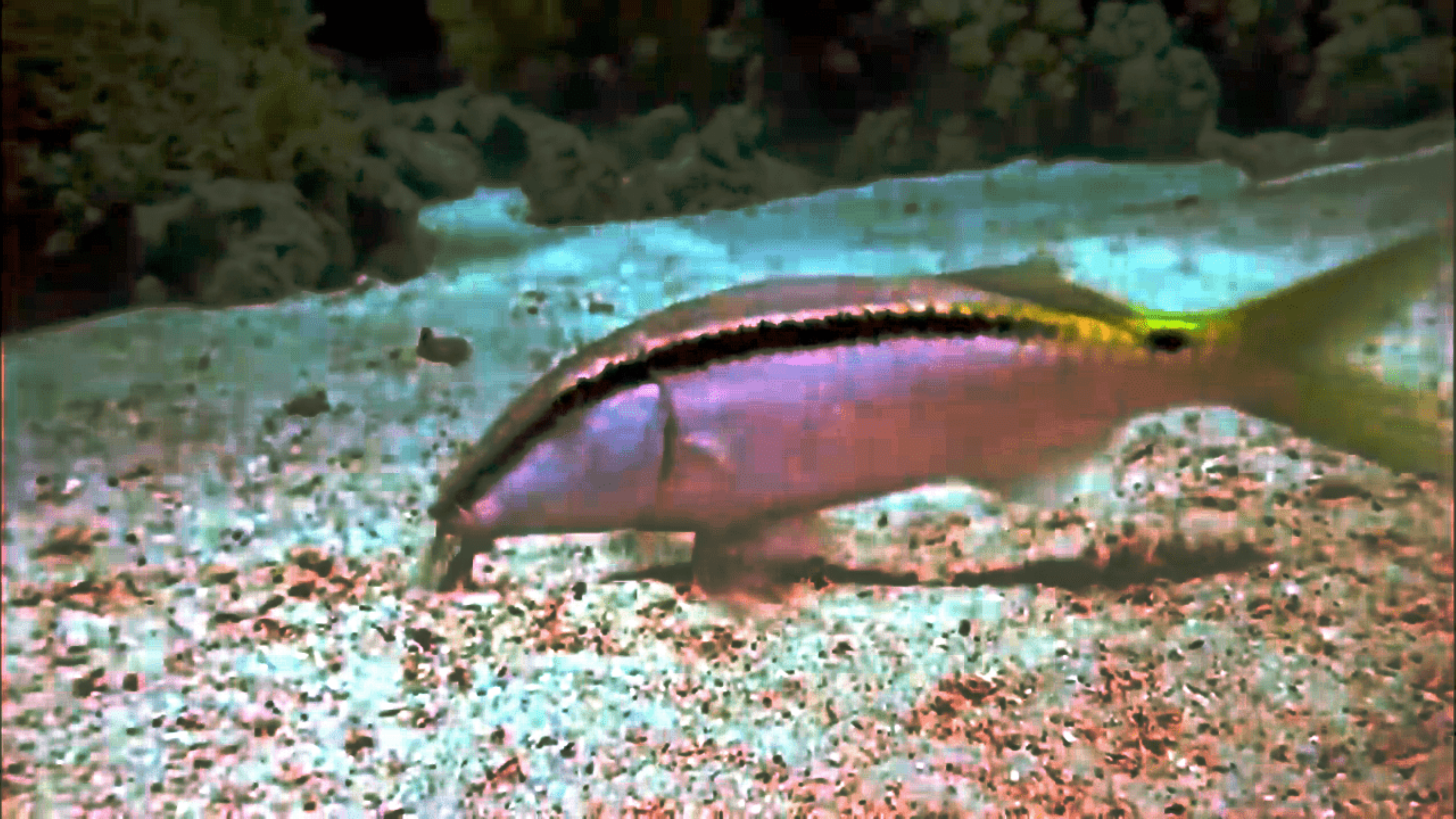}&
\includegraphics[width=3cm, height=2cm]{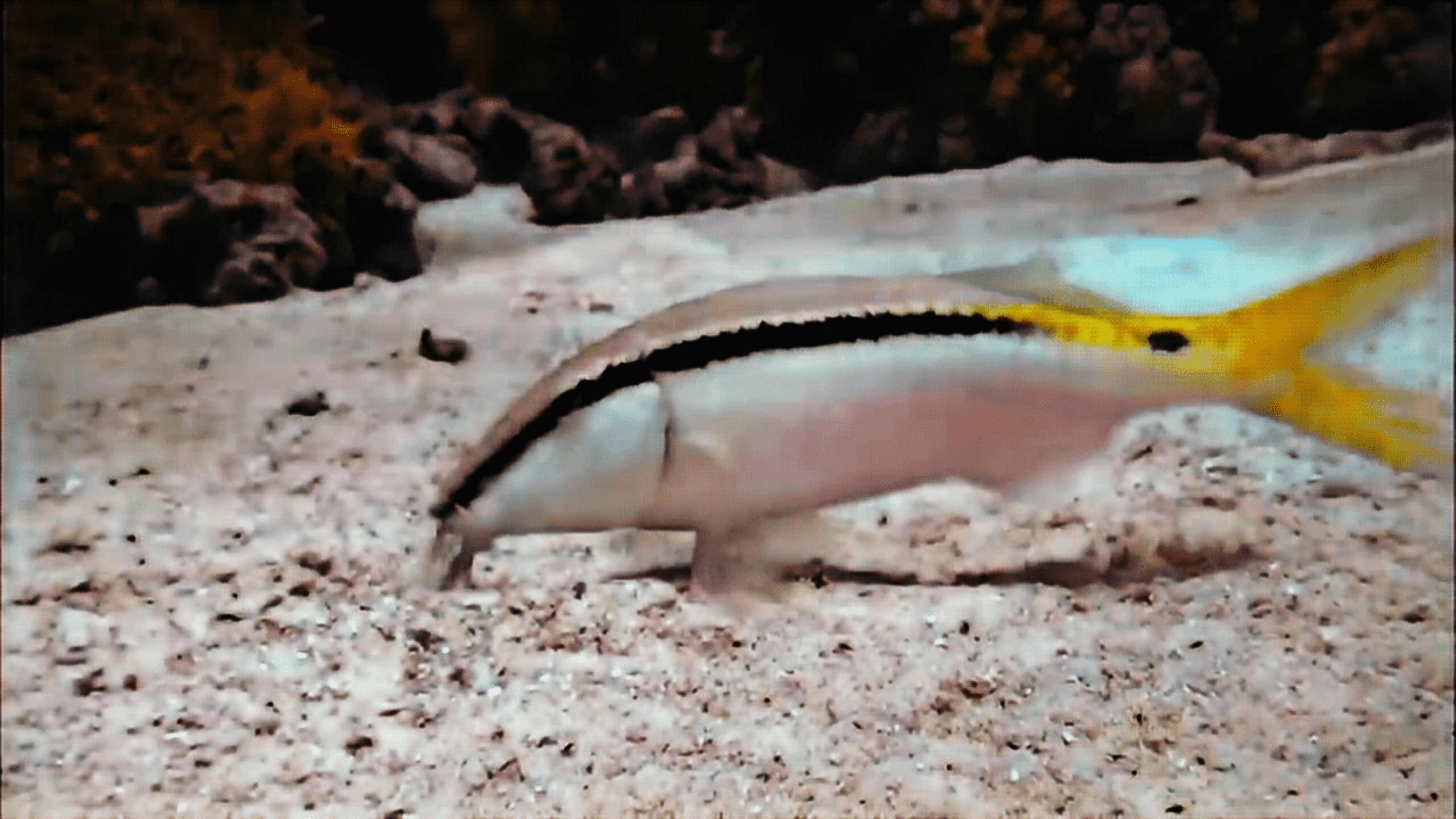}&
\includegraphics[width=3cm, height=2cm]{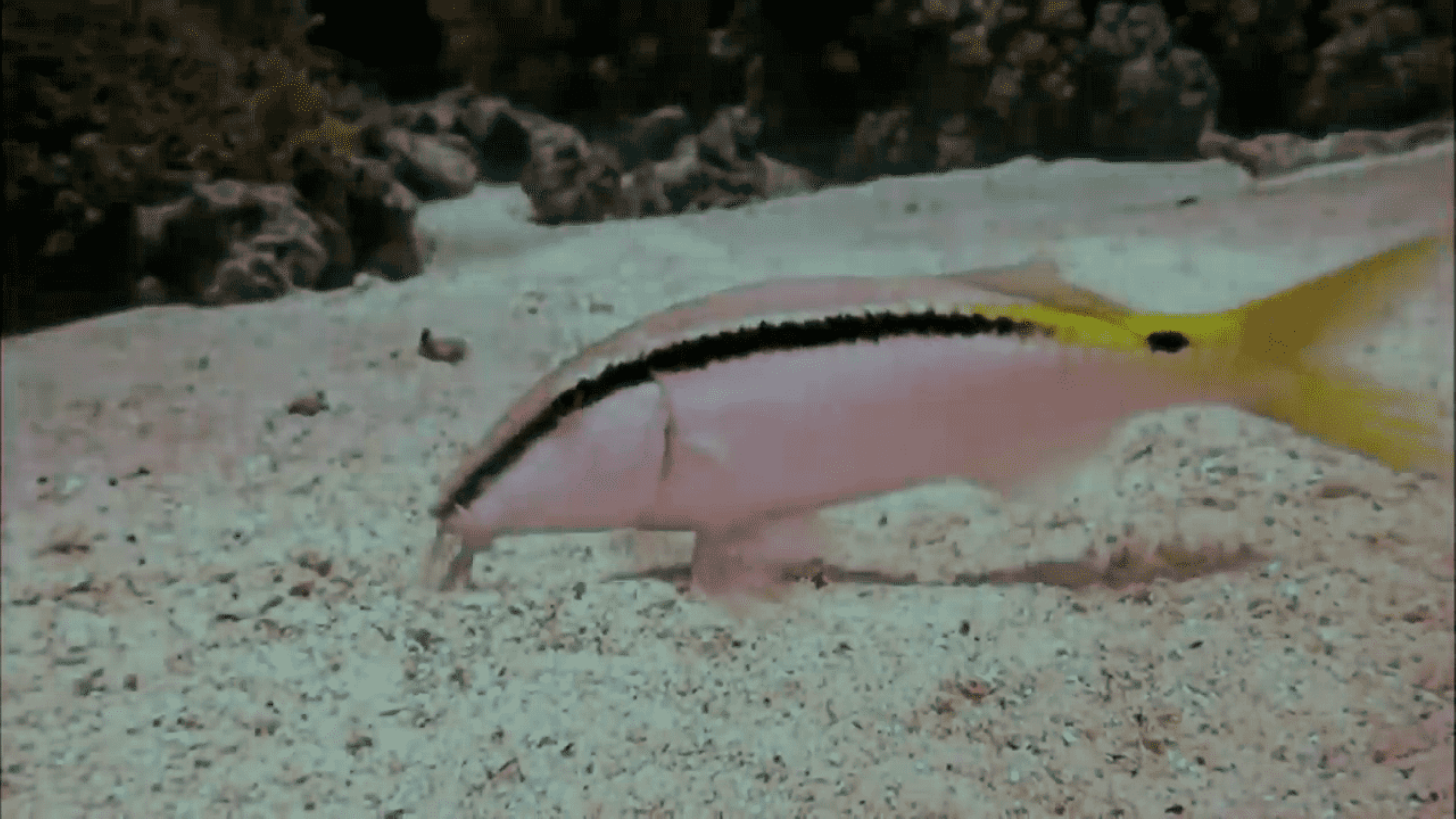}&
\includegraphics[width=3cm, height=2cm]{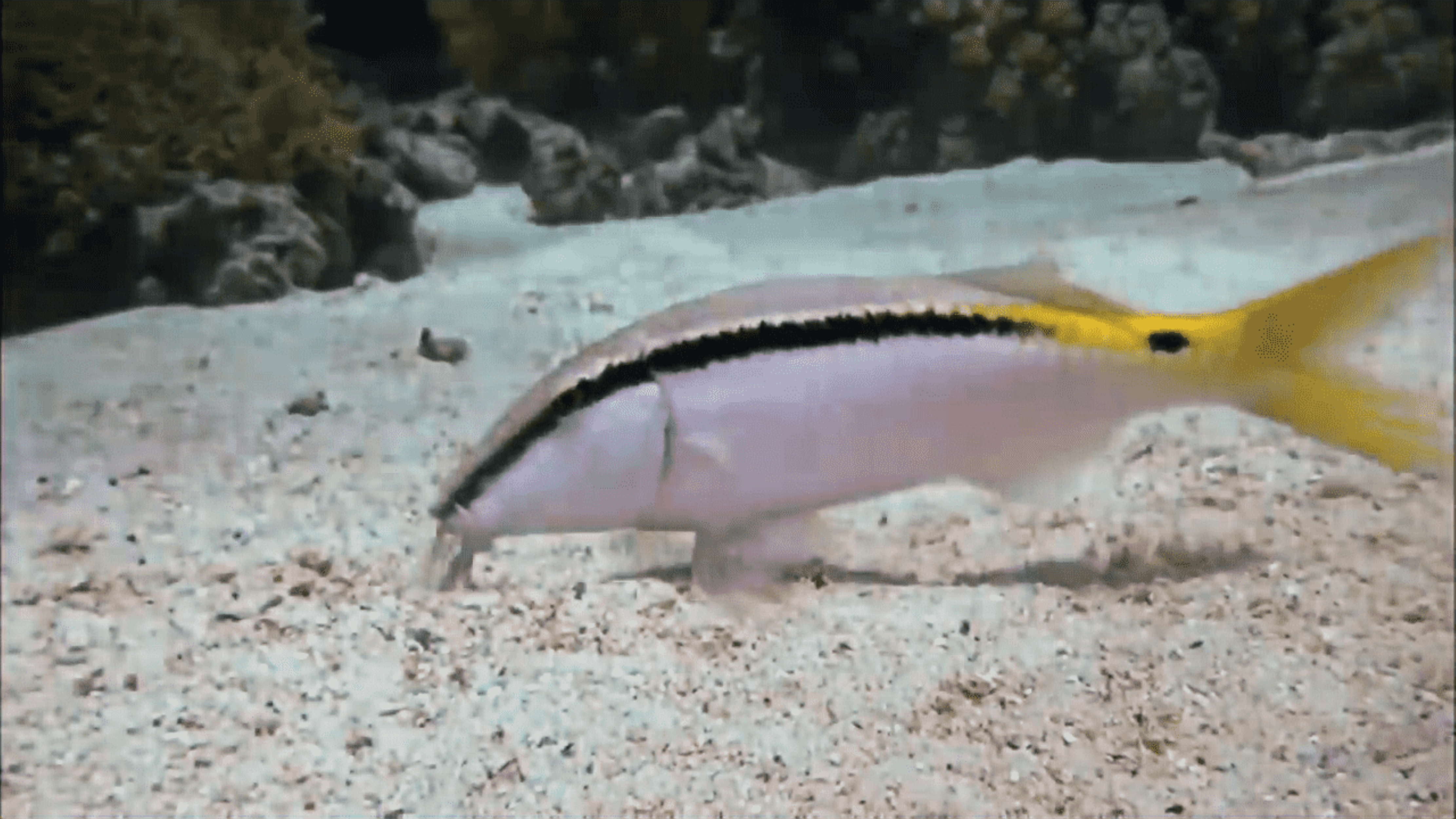}&
\includegraphics[width=3cm, height=2cm]{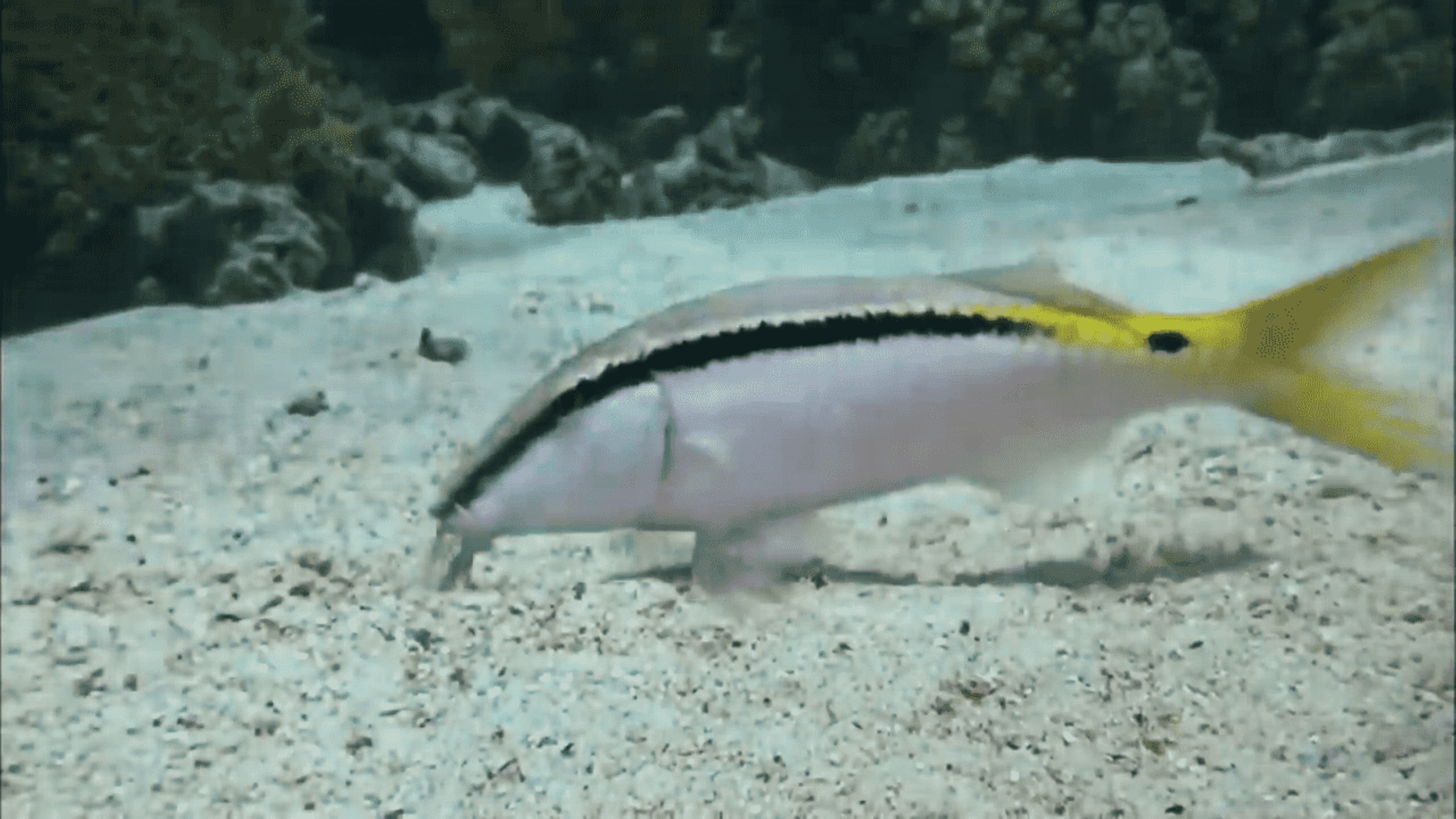}&
\includegraphics[width=3cm, height=2cm]{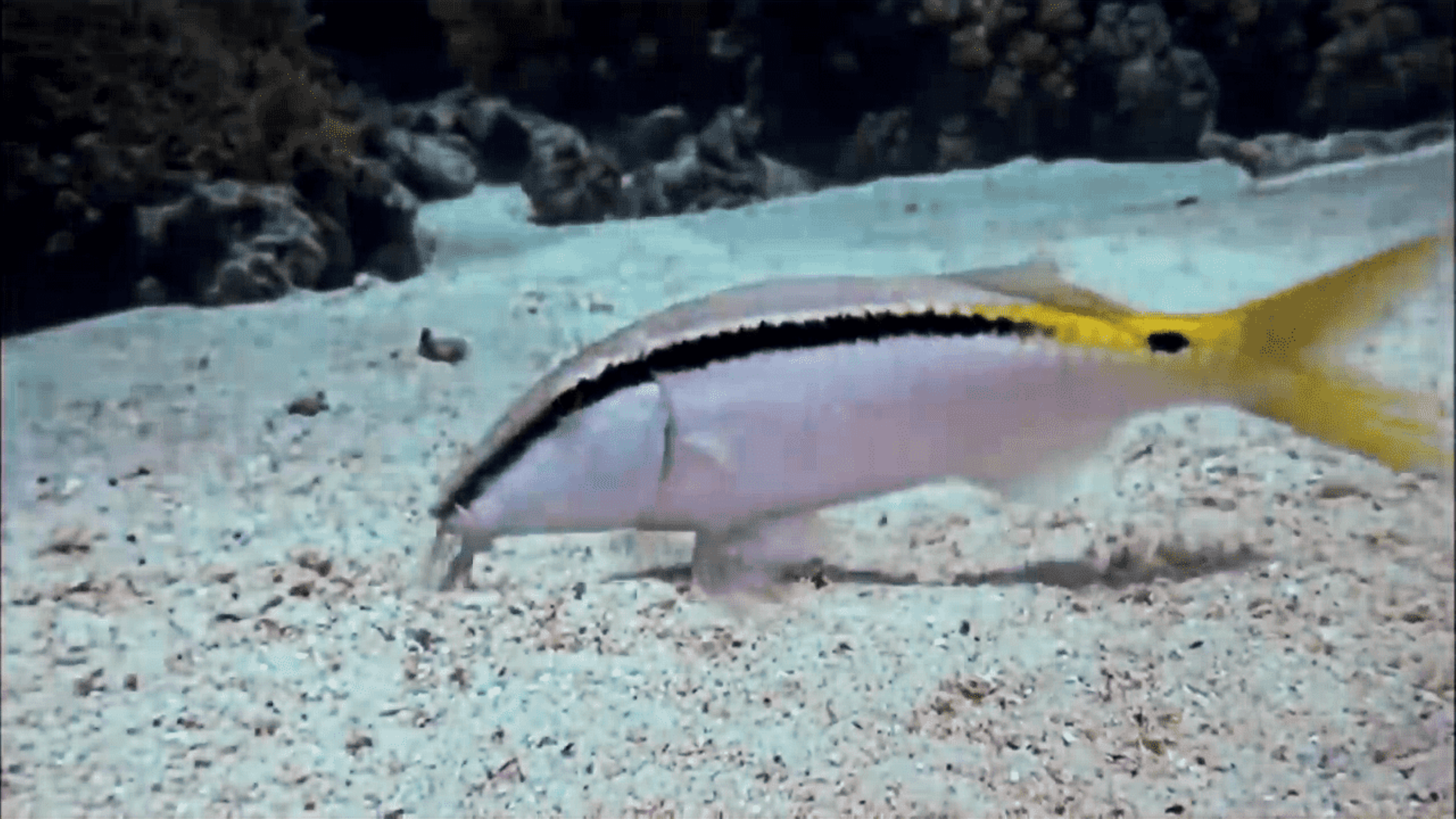}\\

\includegraphics[width=3cm, height=2cm]{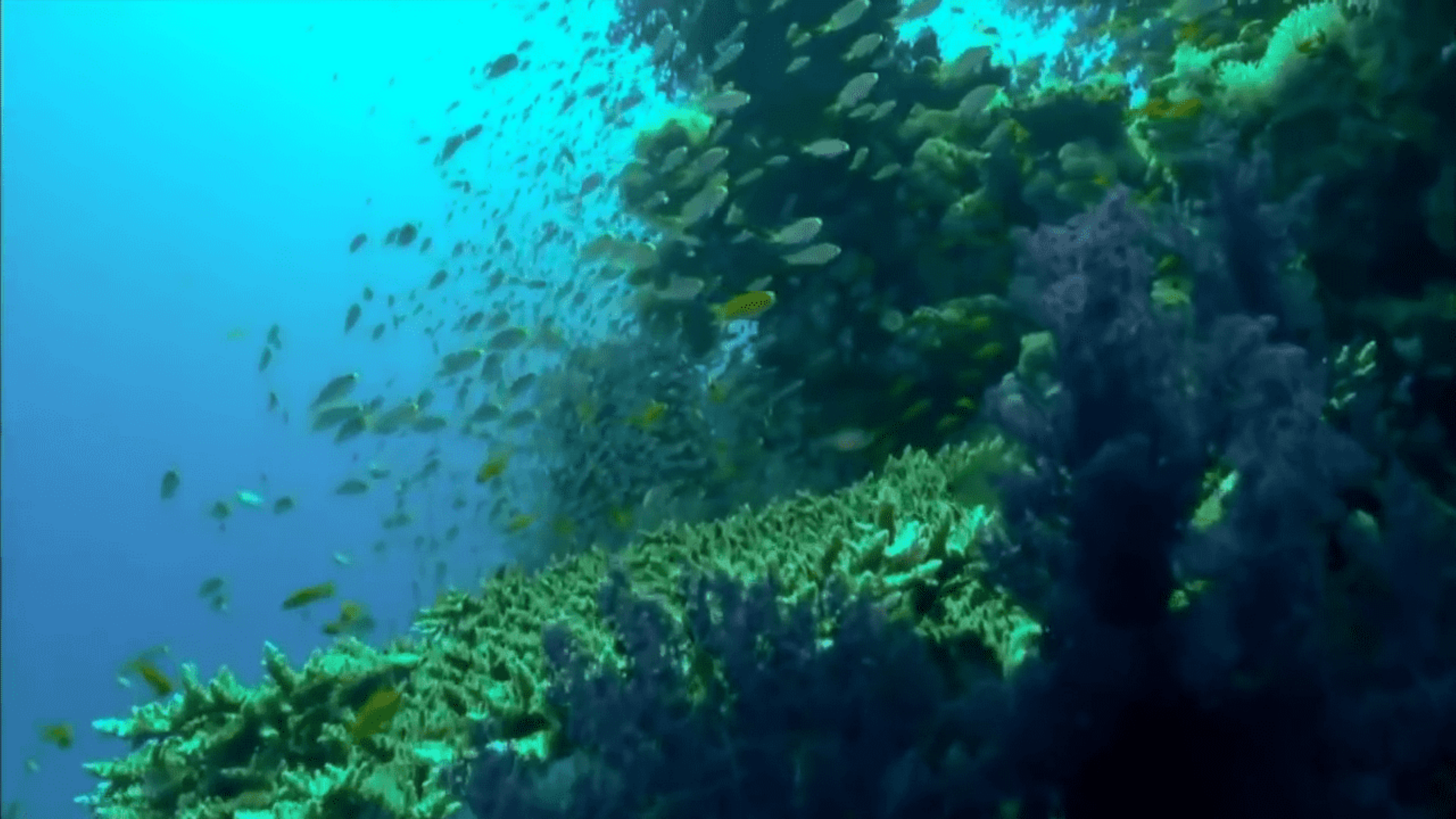}&
\includegraphics[width=3cm, height=2cm]{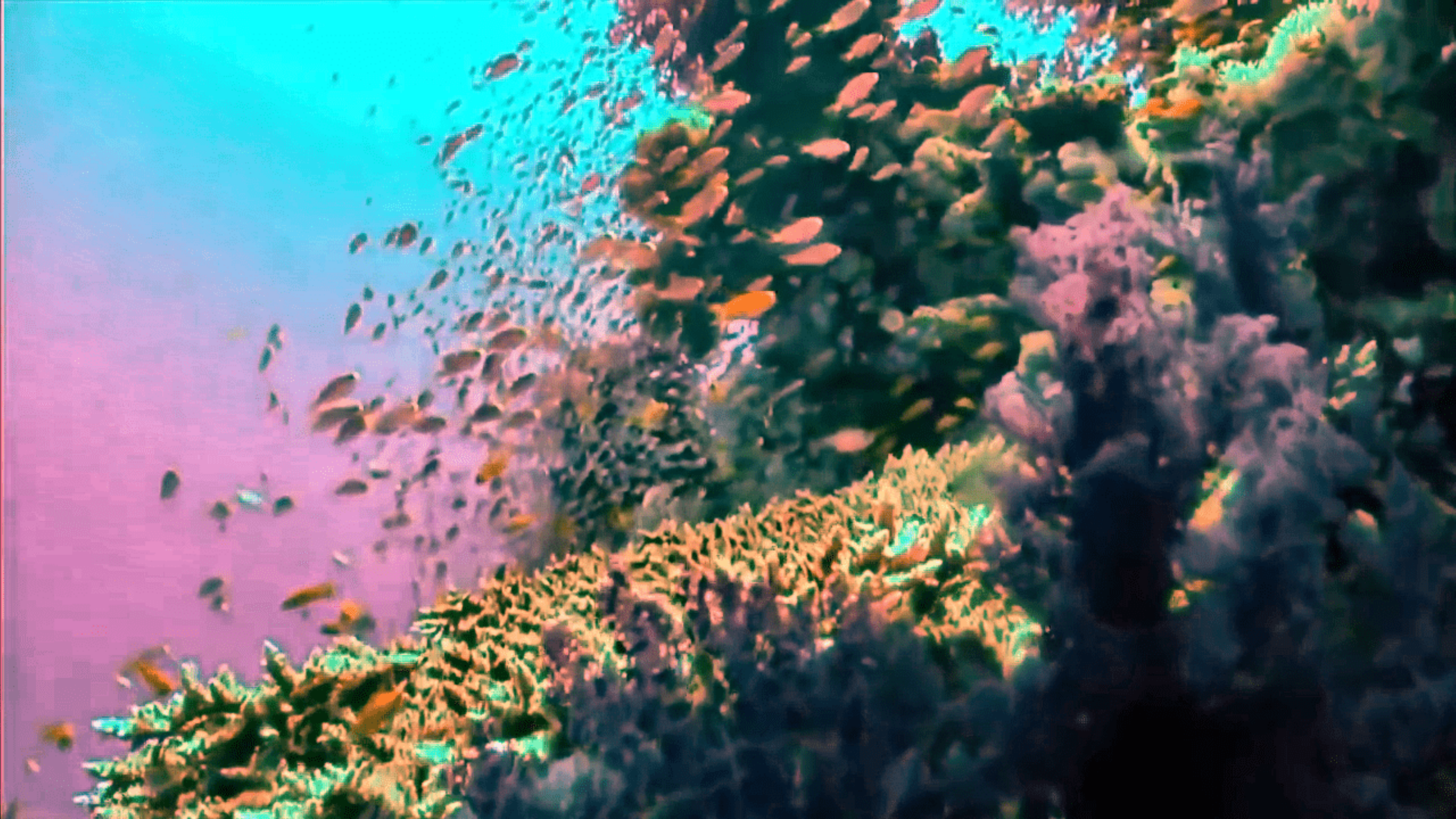}&
\includegraphics[width=3cm, height=2cm]{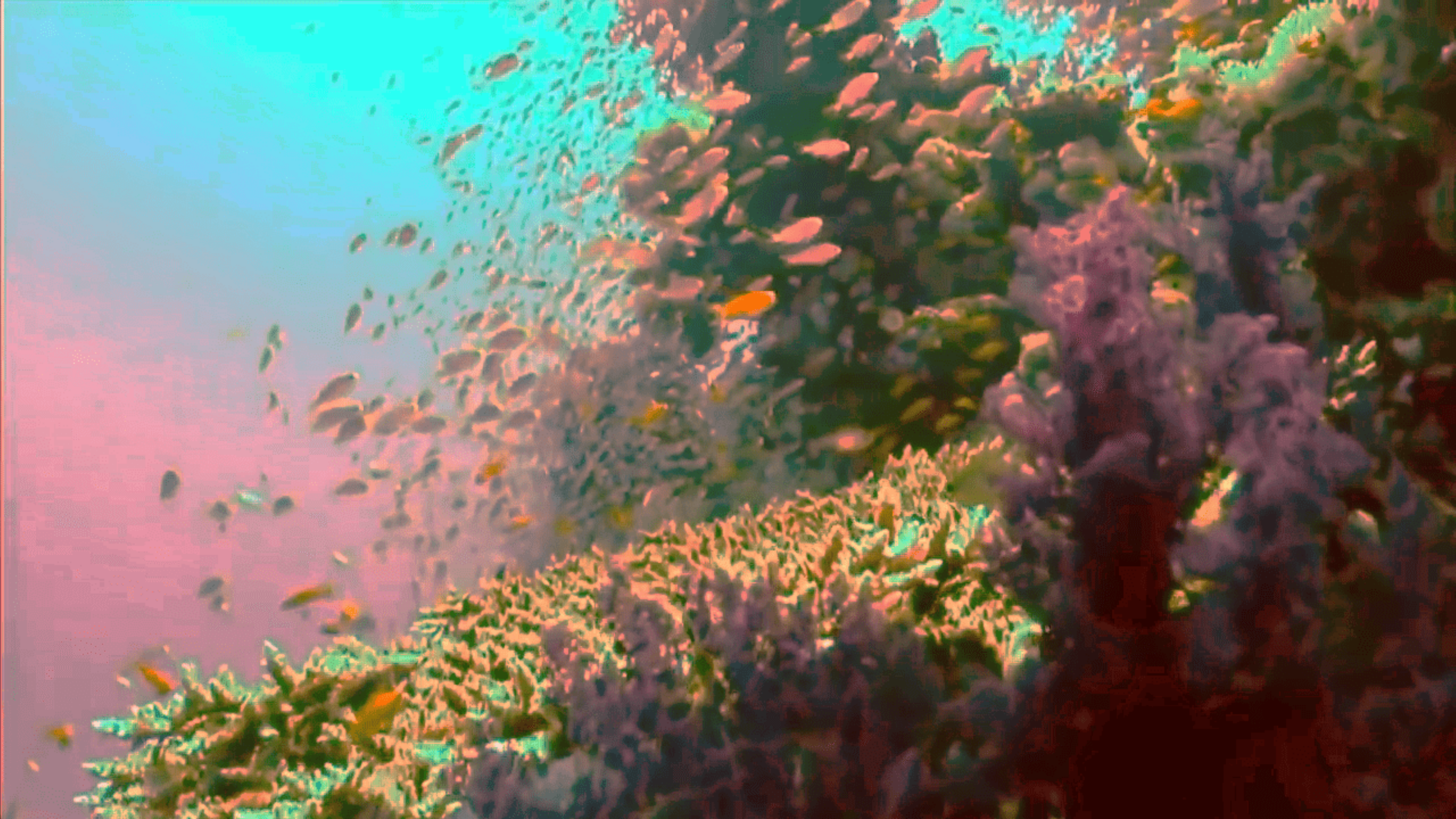}&
\includegraphics[width=3cm, height=2cm]{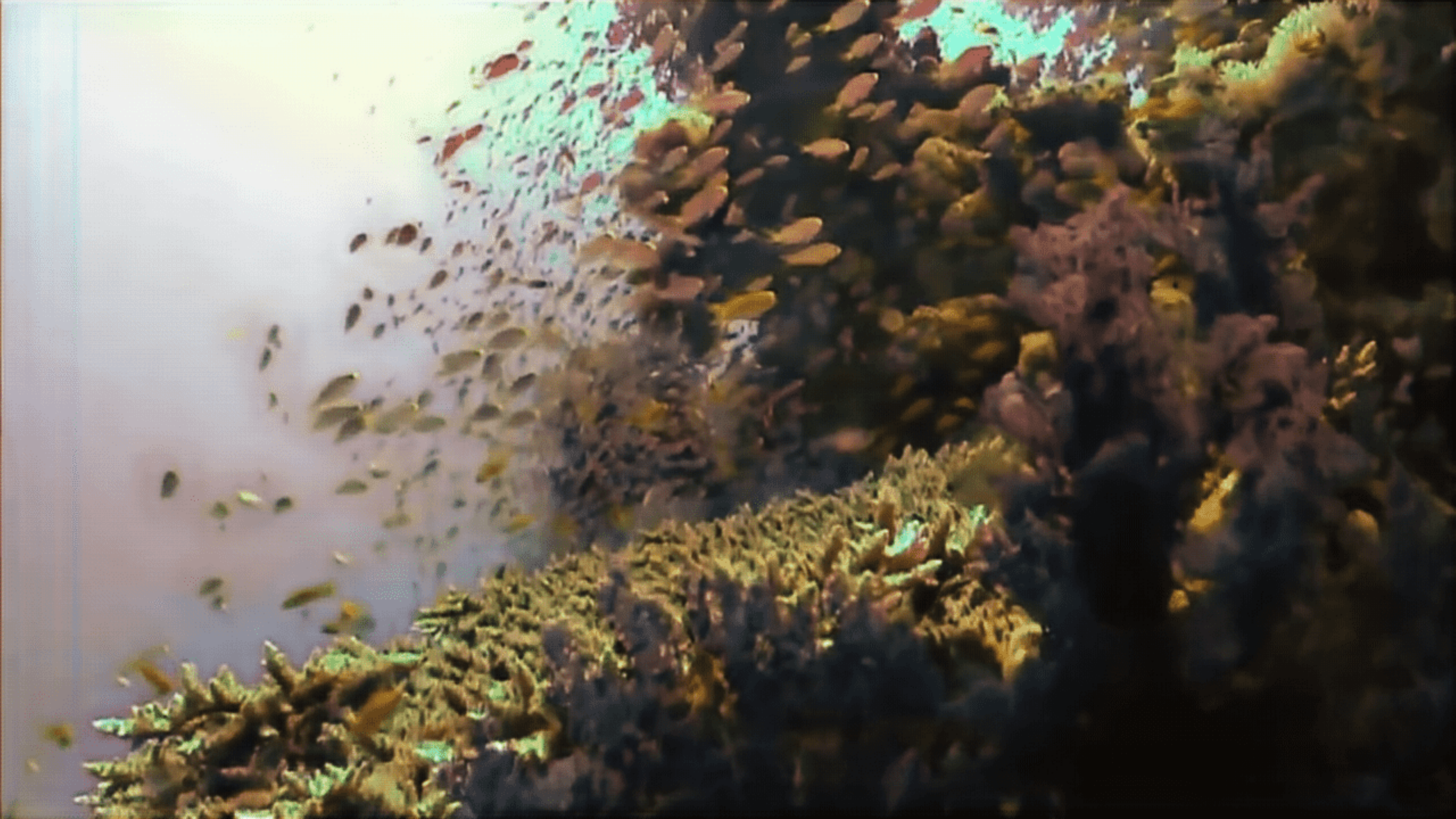}&
\includegraphics[width=3cm, height=2cm]{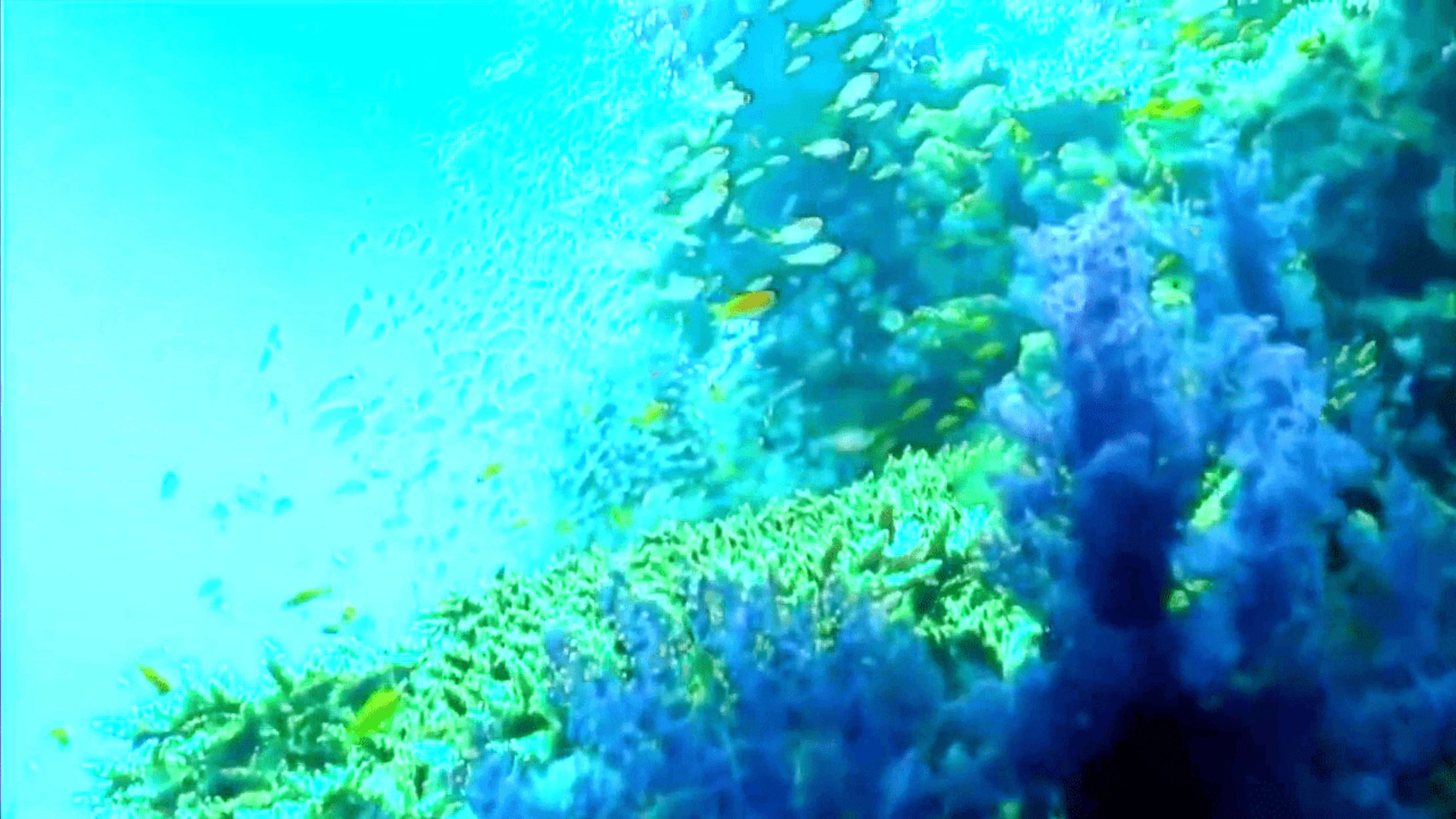}&
\includegraphics[width=3cm, height=2cm]{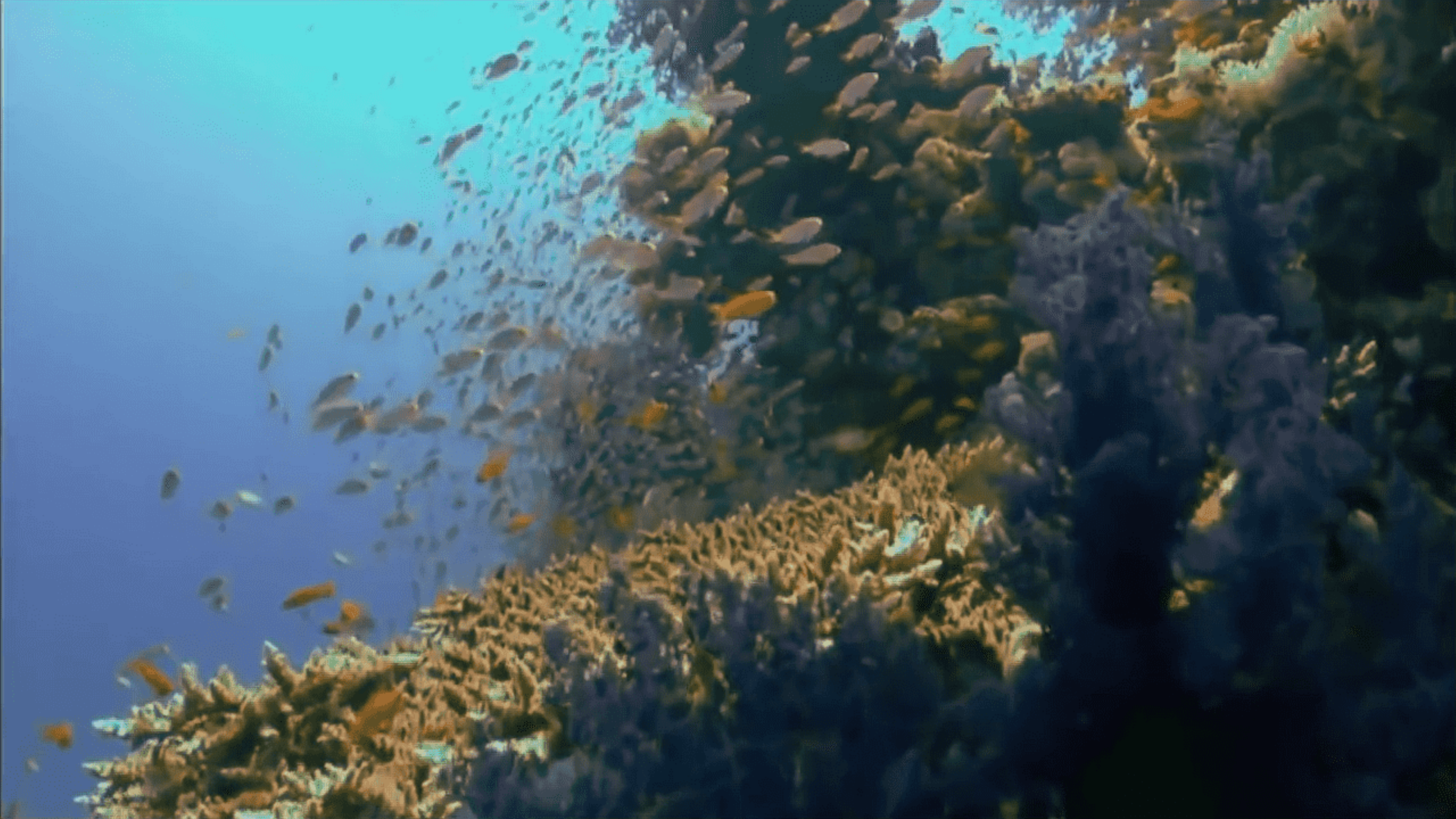}&
\includegraphics[width=3cm, height=2cm]{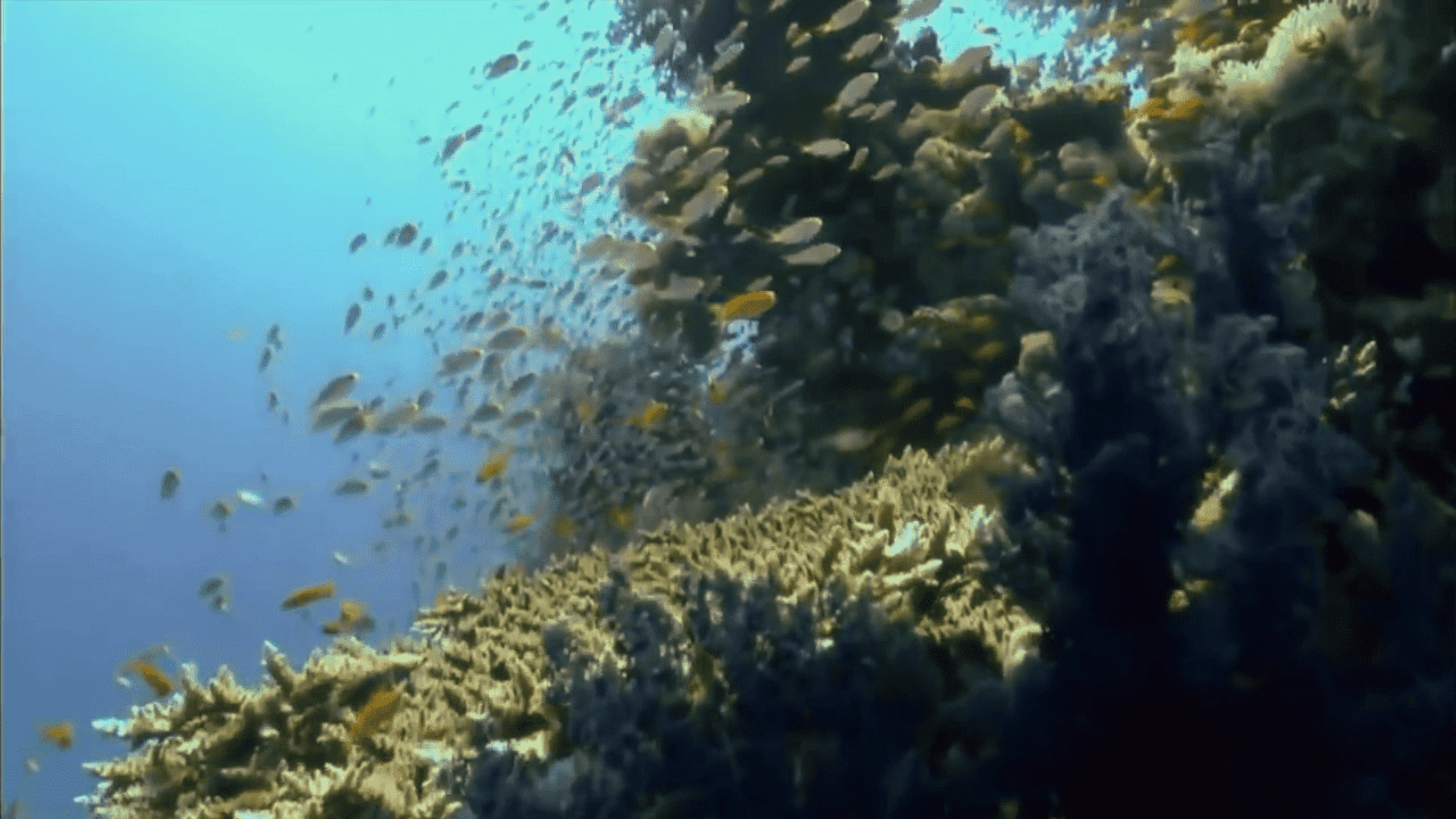}&
\includegraphics[width=3cm, height=2cm]{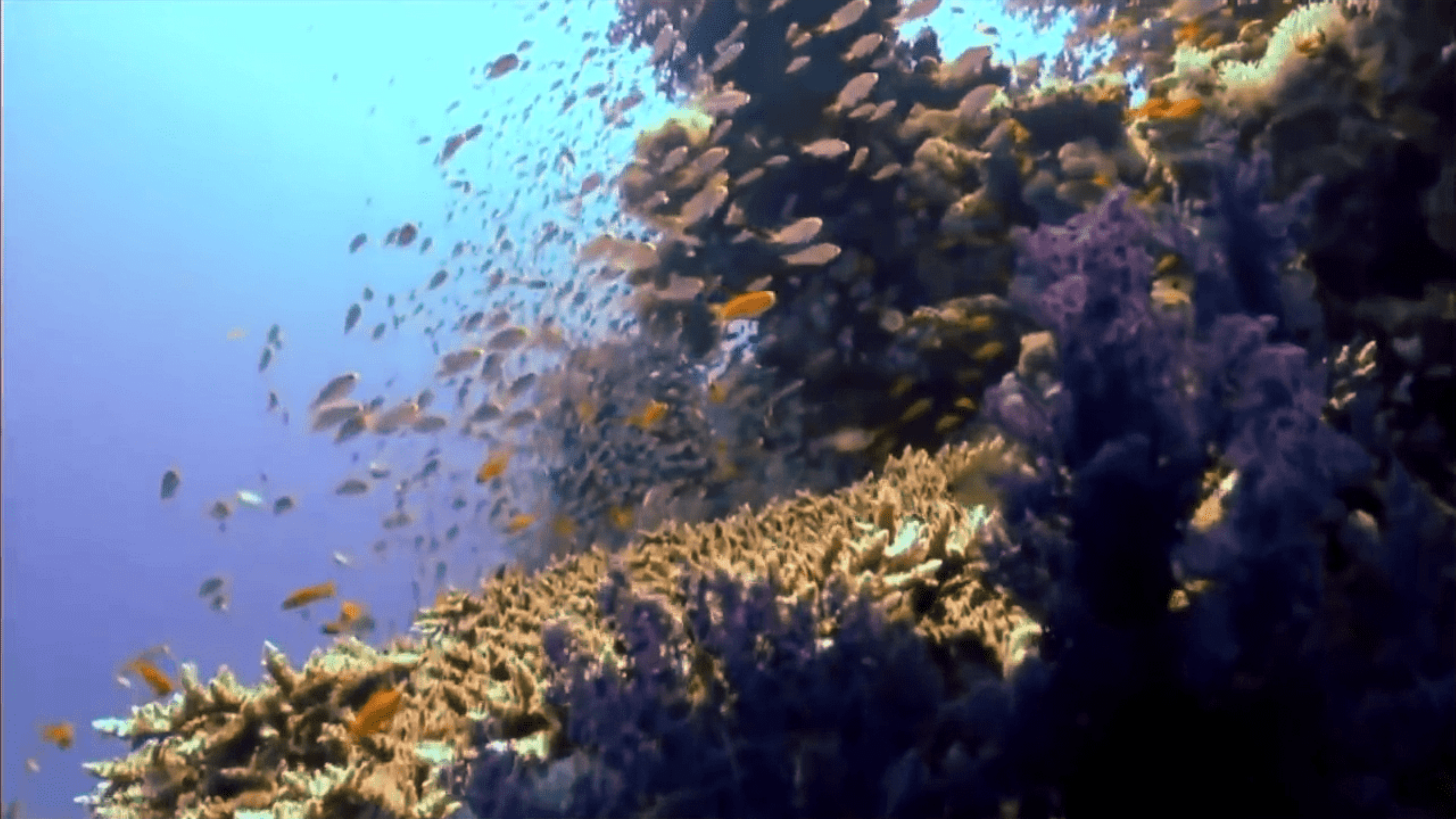}\\

{ Degraded} & { Fusion-based \cite{ancuti1} } & { Histogram Prior \cite{li_tip} } & { DenseGAN \cite{guo_oe} } & { Blurriness-based \cite{peng_tip} } & { WaterNet \cite{uieb}} & { Ucolor \cite{tip21}} & { Deep WaveNet}\\

\end{tabular}}
\caption{{Qualitative comparison} against the best-published works for the task of underwater image enhancement on \texttt{UIEB} dataset. Results, shown in last two rows, are obtained using \texttt{UIEB Challenge} test-set.}
\label{fig:uieb2}
\end{figure*}

\begin{figure*}[h]
\centering
\resizebox{\textwidth}{!}{
\setlength{\tabcolsep}{1pt}
\begin{tabular}{ccccccc}

&  & { \textit{SSIM: $0.4016$}} & {\textit{SSIM: $0.3604$}} & {\textit{SSIM: $0.3695$}} & {\textit{SSIM: $0.6540$}} & {\textit{SSIM: $1$}}\\

&  & {\textit{PSNR: $18.68$}} & {\textit{PSNR: $18.41$}} & {\textit{PSNR: $18.40$}} & {\textit{PSNR: $23.24$}} & {\textit{PSNR: $inf$}}\\

&
\multicolumn{1}{r}{\includegraphics[width=0.75cm, height=0.5cm]{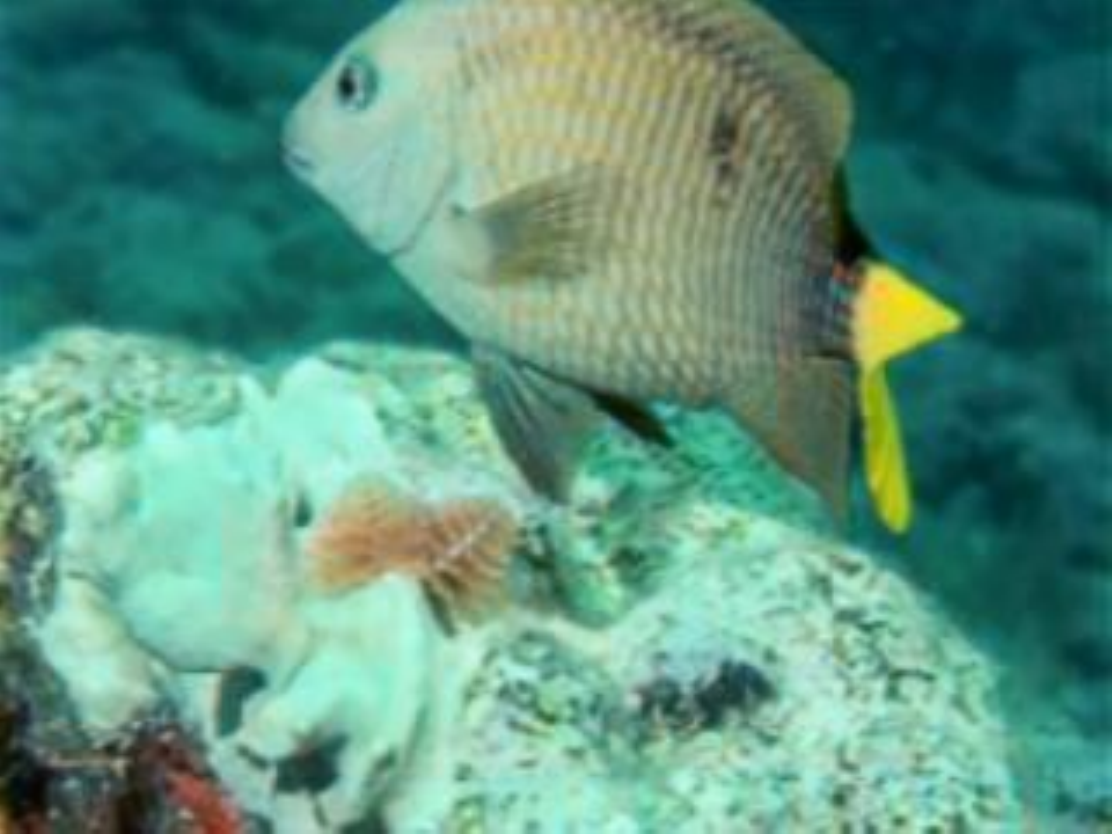}}
&
\includegraphics[width=3cm, height=2cm]{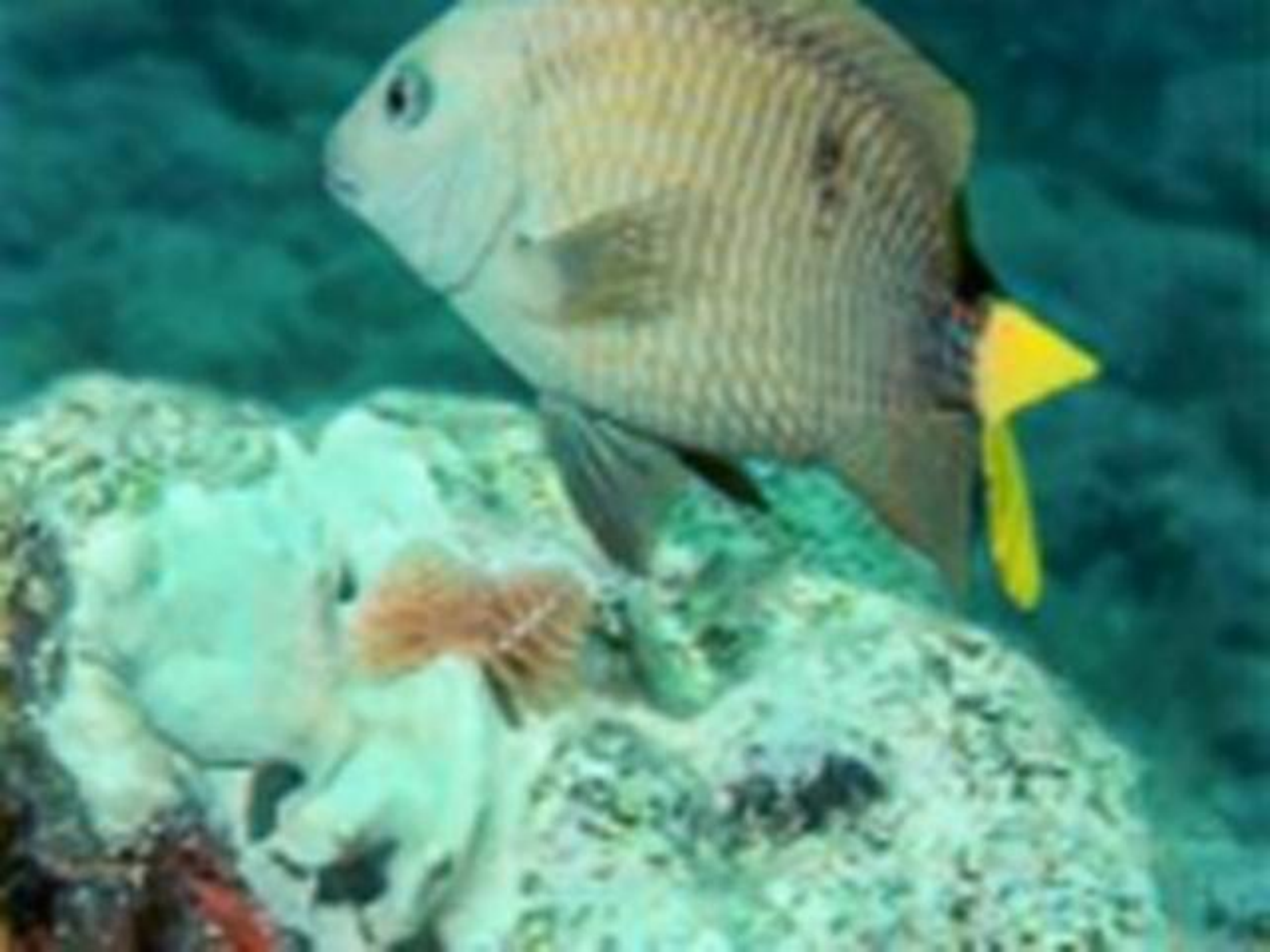}
&
\includegraphics[width=3cm, height=2cm]{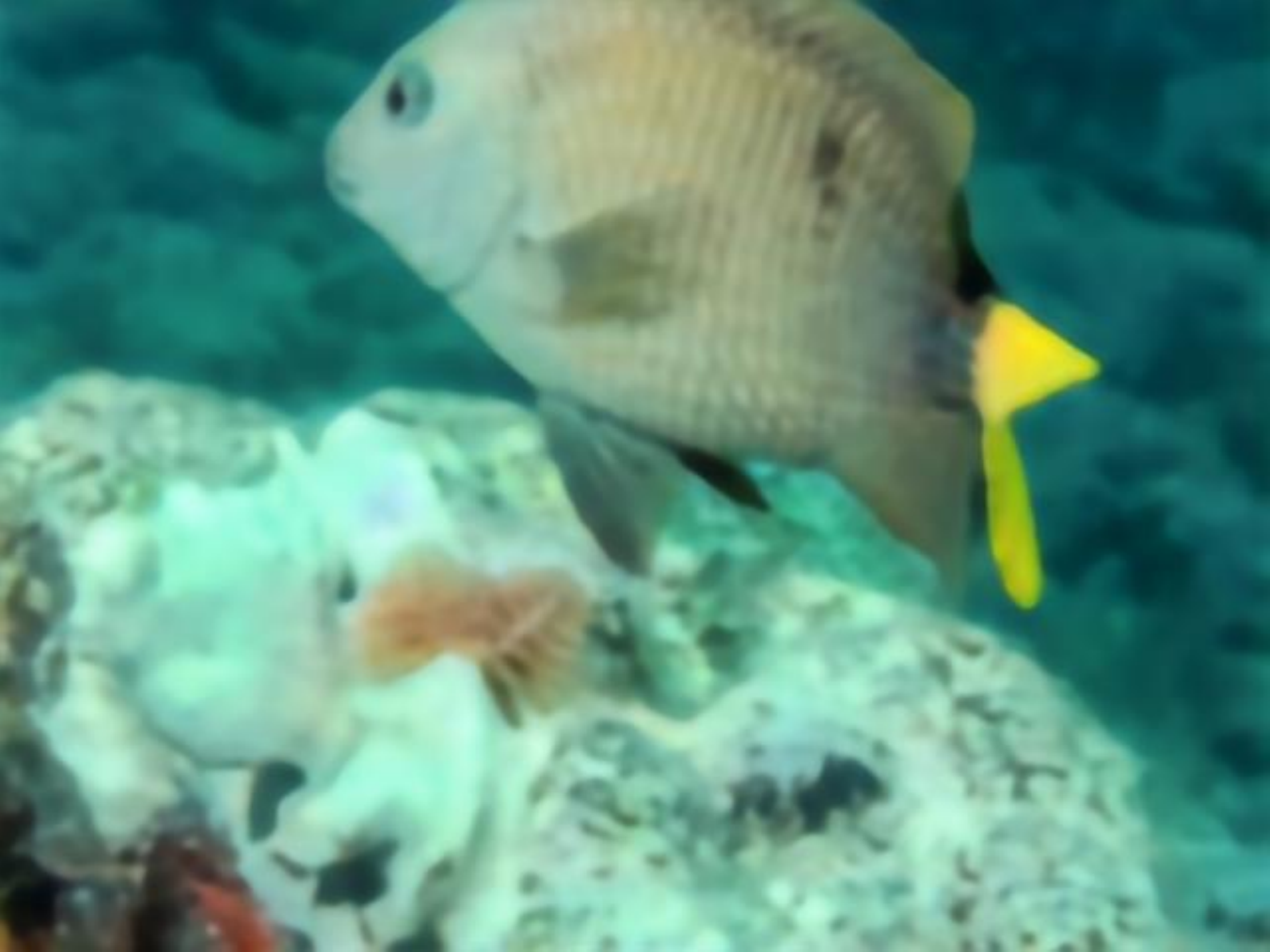}
&
\includegraphics[width=3cm, height=2cm]{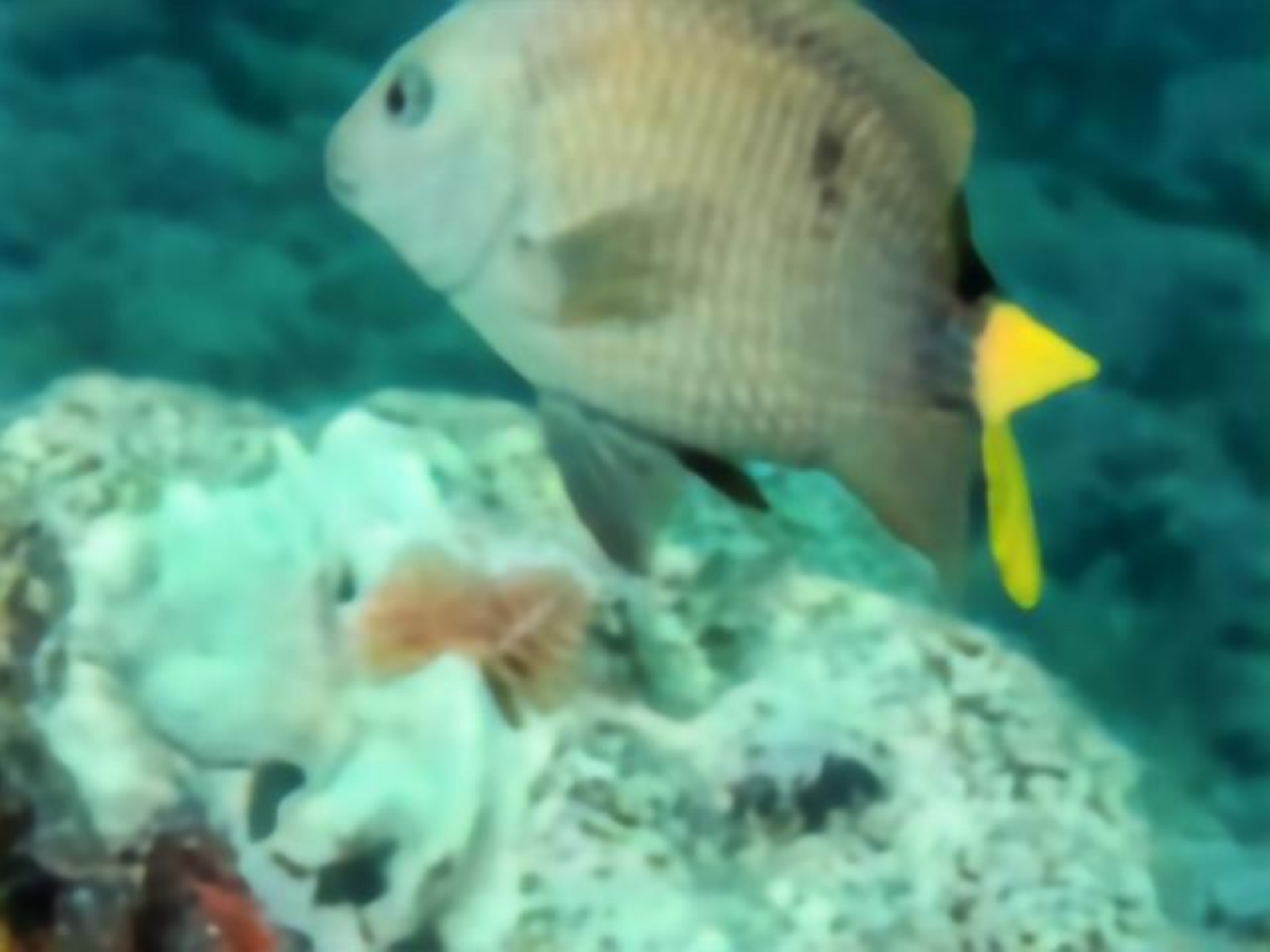}
&
\includegraphics[width=3cm, height=2cm]{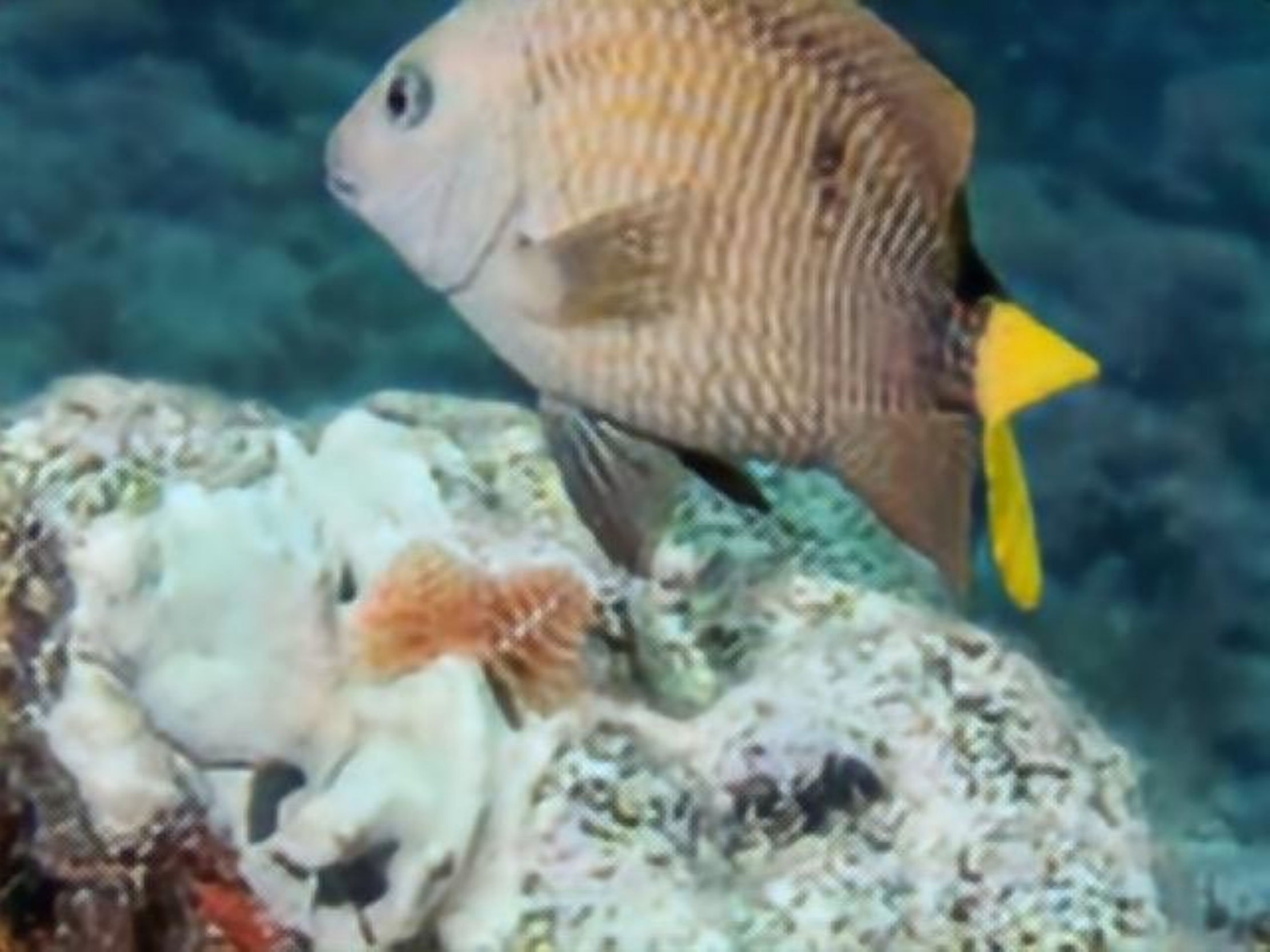}
&
\includegraphics[width=3cm, height=2cm]{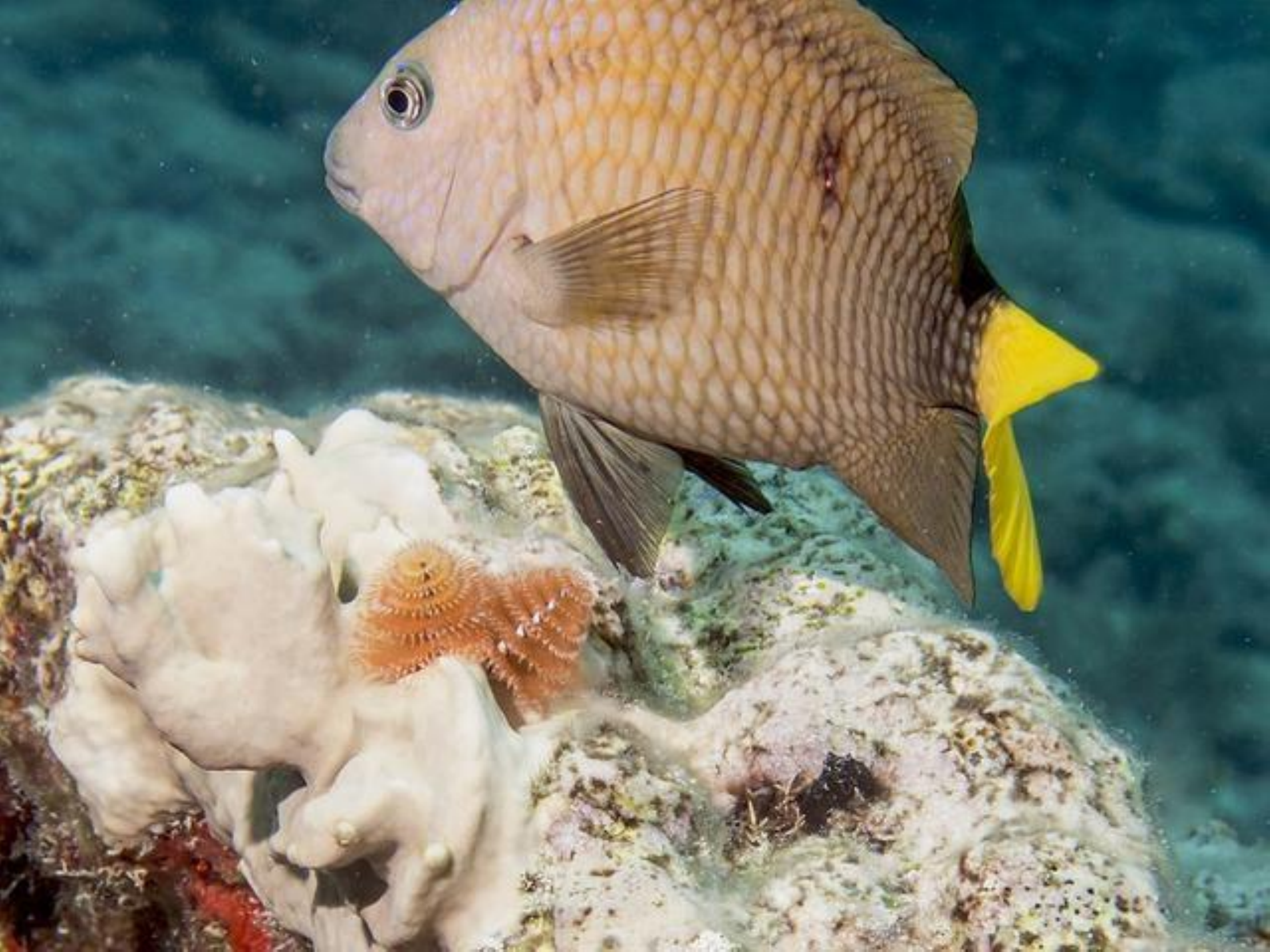}
\\

& \multicolumn{1}{r}{ $4\times$} & { Bicubic} & {SRDRM \cite{srdrm_srdrmgan}} & { SRDRM-GAN \cite{srdrm_srdrmgan}} & { Deep WaveNet }& {  GT}\\

 & {\textit{SSIM: $0.0932$}} & \textit{ SSIM: $0.0921$} & \textit{  SSIM: $0.0763$} & \textit{ SSIM: $0.6948$} & \textit{ SSIM: $0.8440$} & \textit{ SSIM: $1$}\\

& \textit{ PSNR: $14.84$} & \textit{ PSNR: $14.70$} & \textit{  PSNR: $14.64$} & \textit{ PSNR: $21.72$} & \textit{ PSNR: $20.81$} & \textit{ PSNR: $inf$}\\

\includegraphics[width=1.5cm, height=1cm]{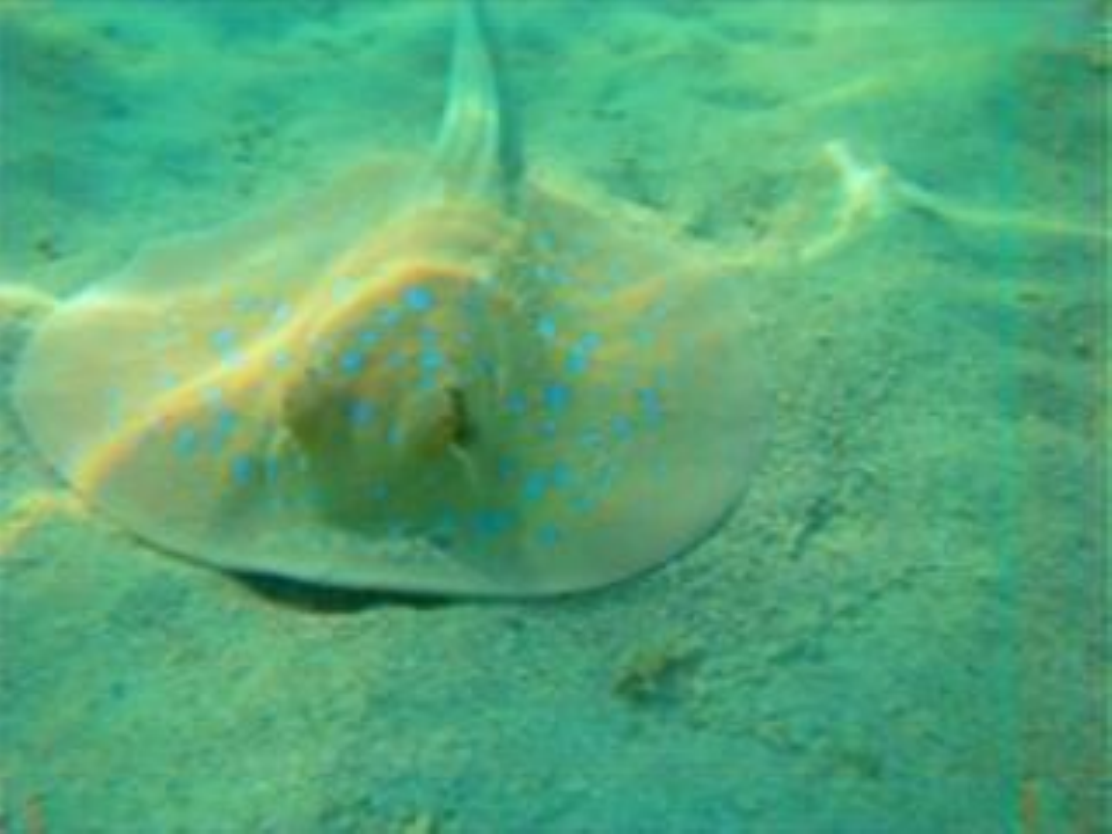}
&
\includegraphics[width=3cm, height=2cm]{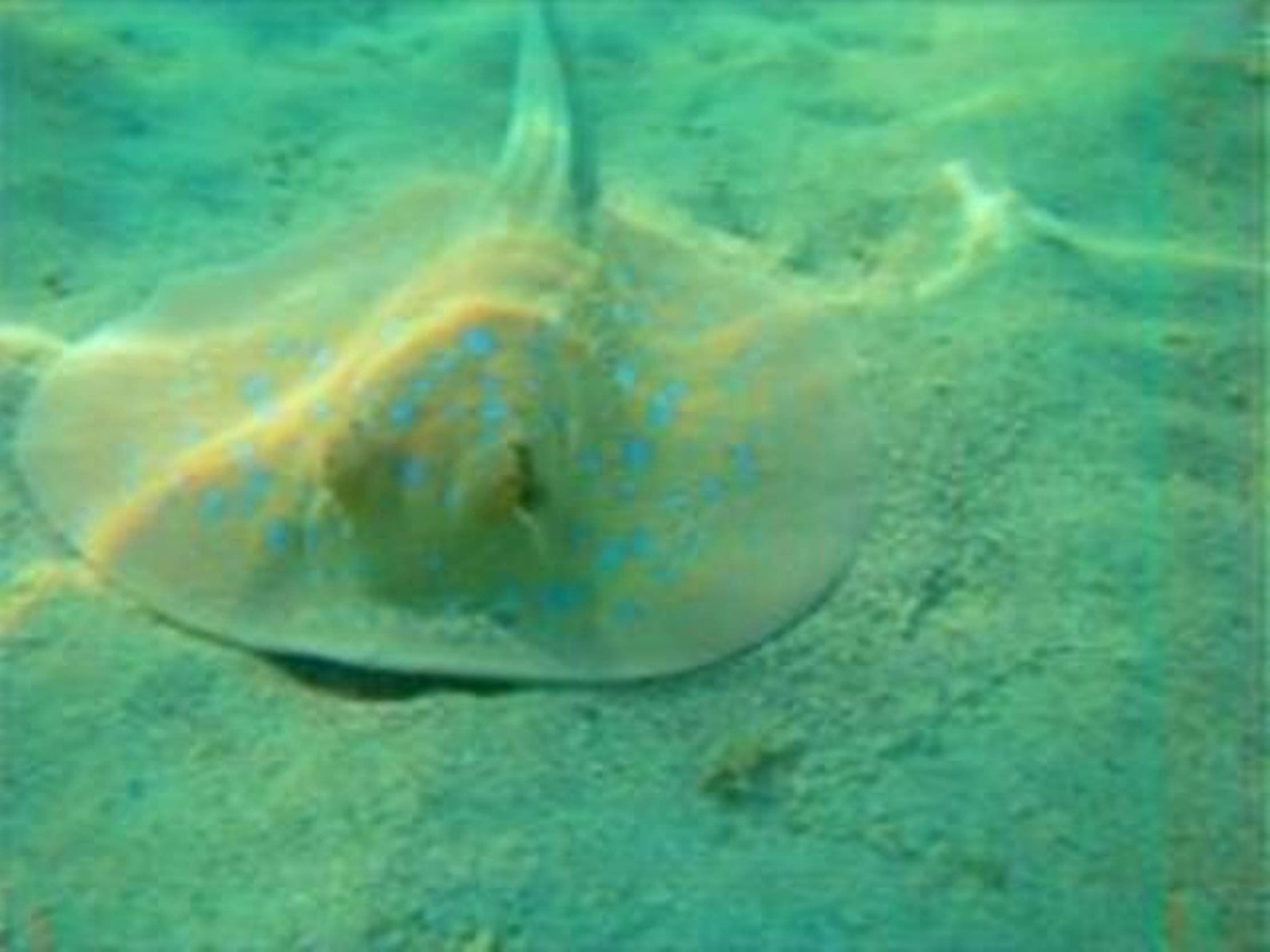}
&
\includegraphics[width=3cm, height=2cm]{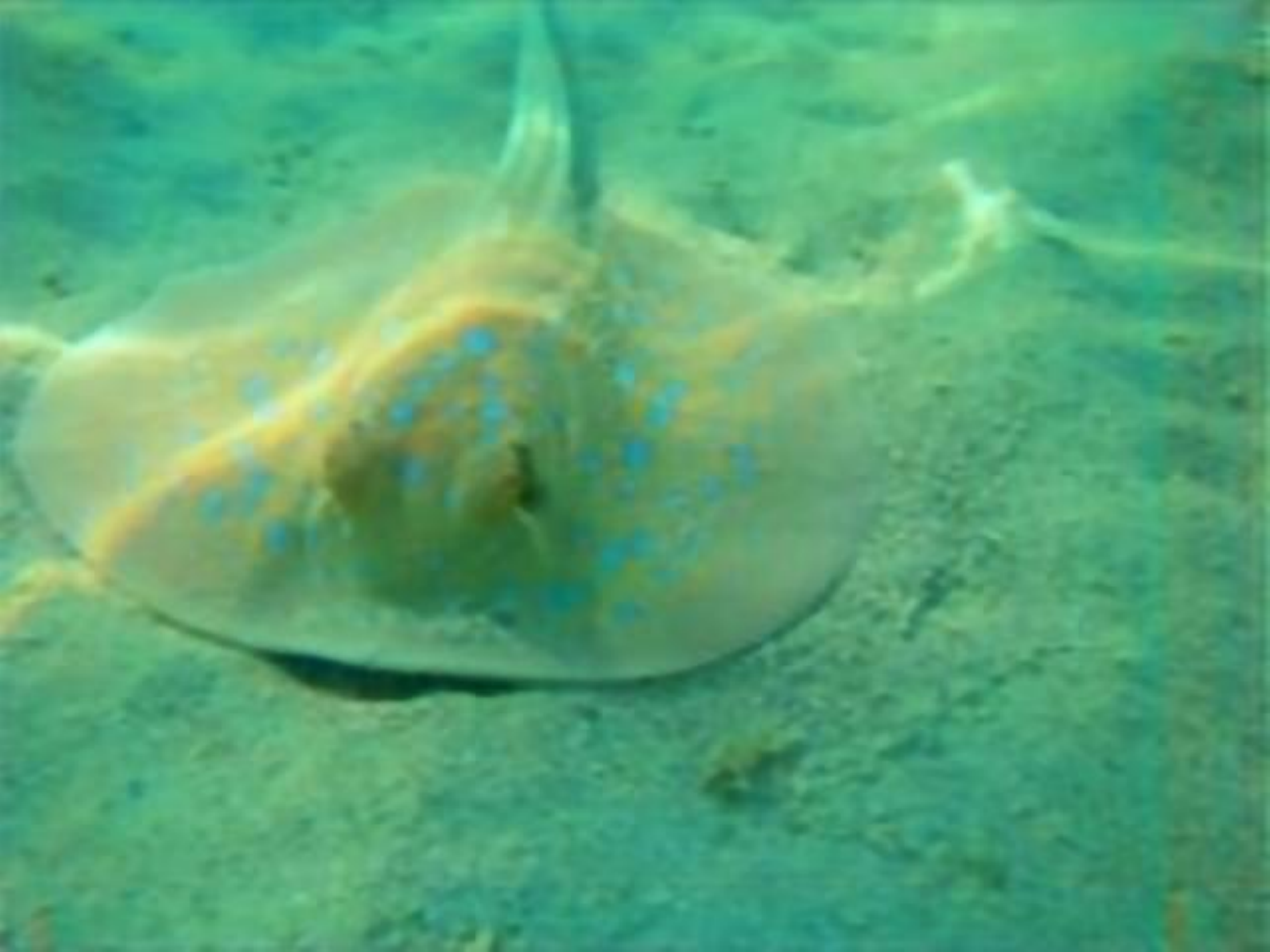}
&
\includegraphics[width=3cm, height=2cm]{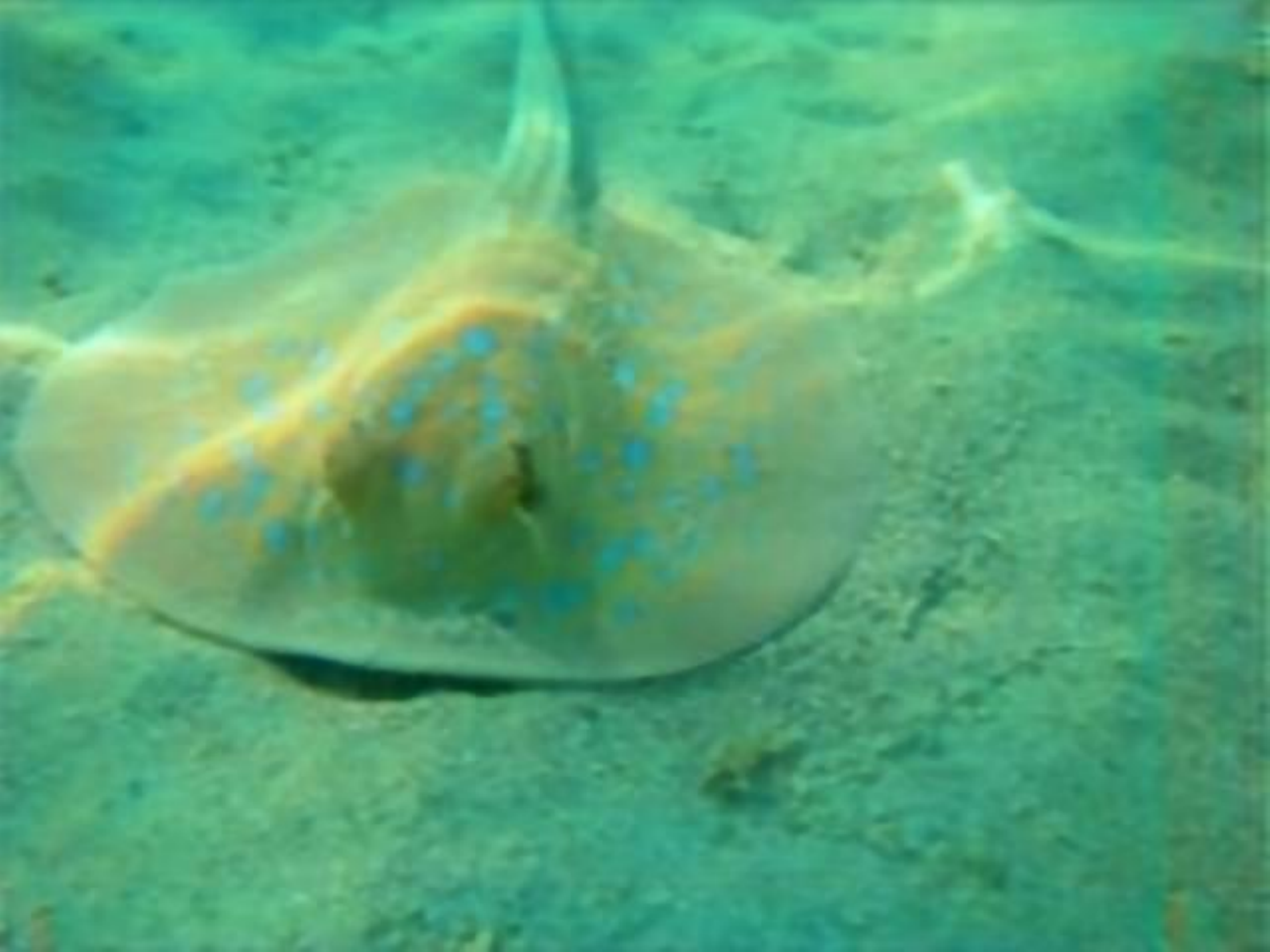}
&
\includegraphics[width=3cm, height=2cm]{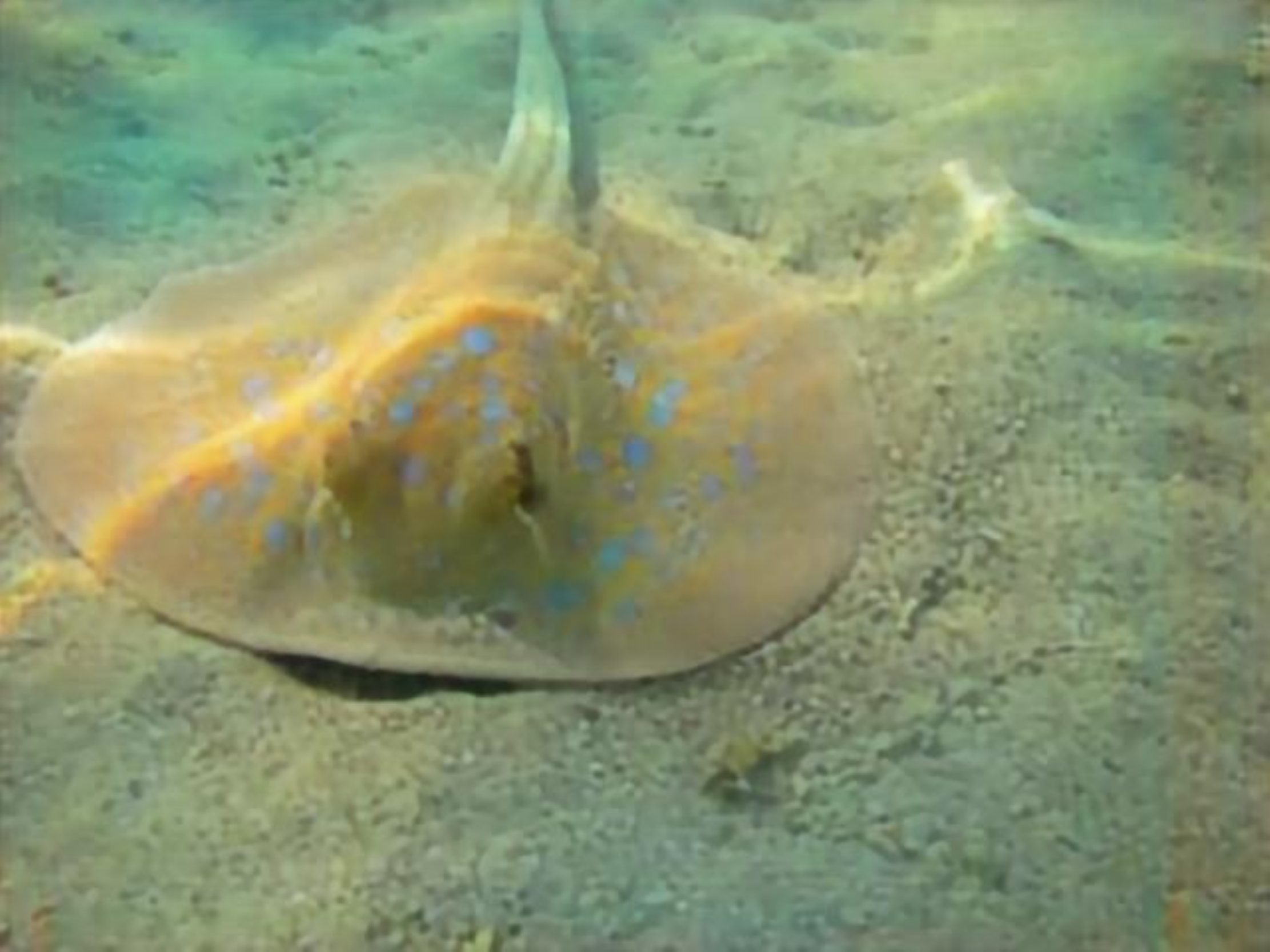}
&
\includegraphics[width=3cm, height=2cm]{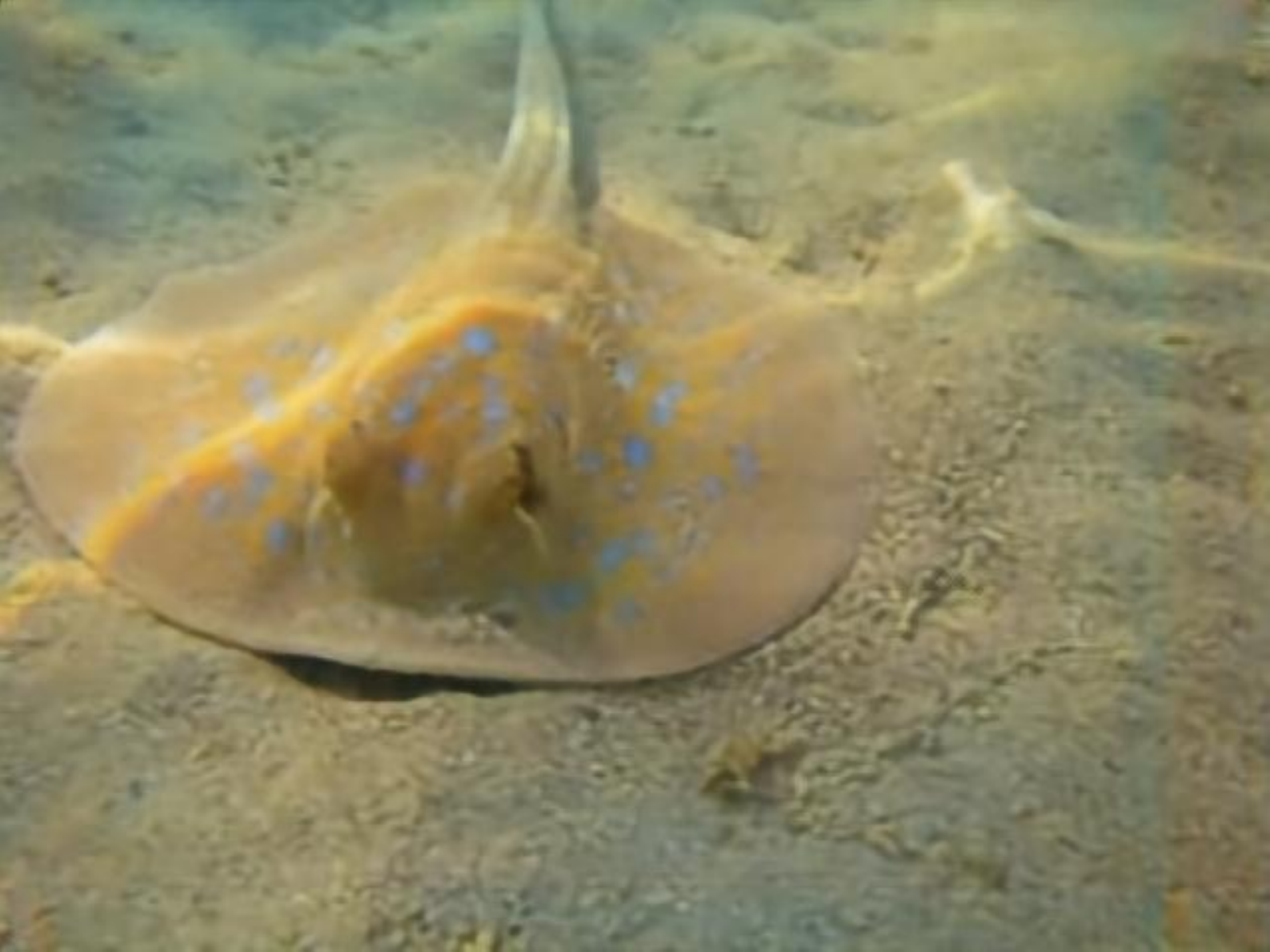}
&
\includegraphics[width=3cm, height=2cm]{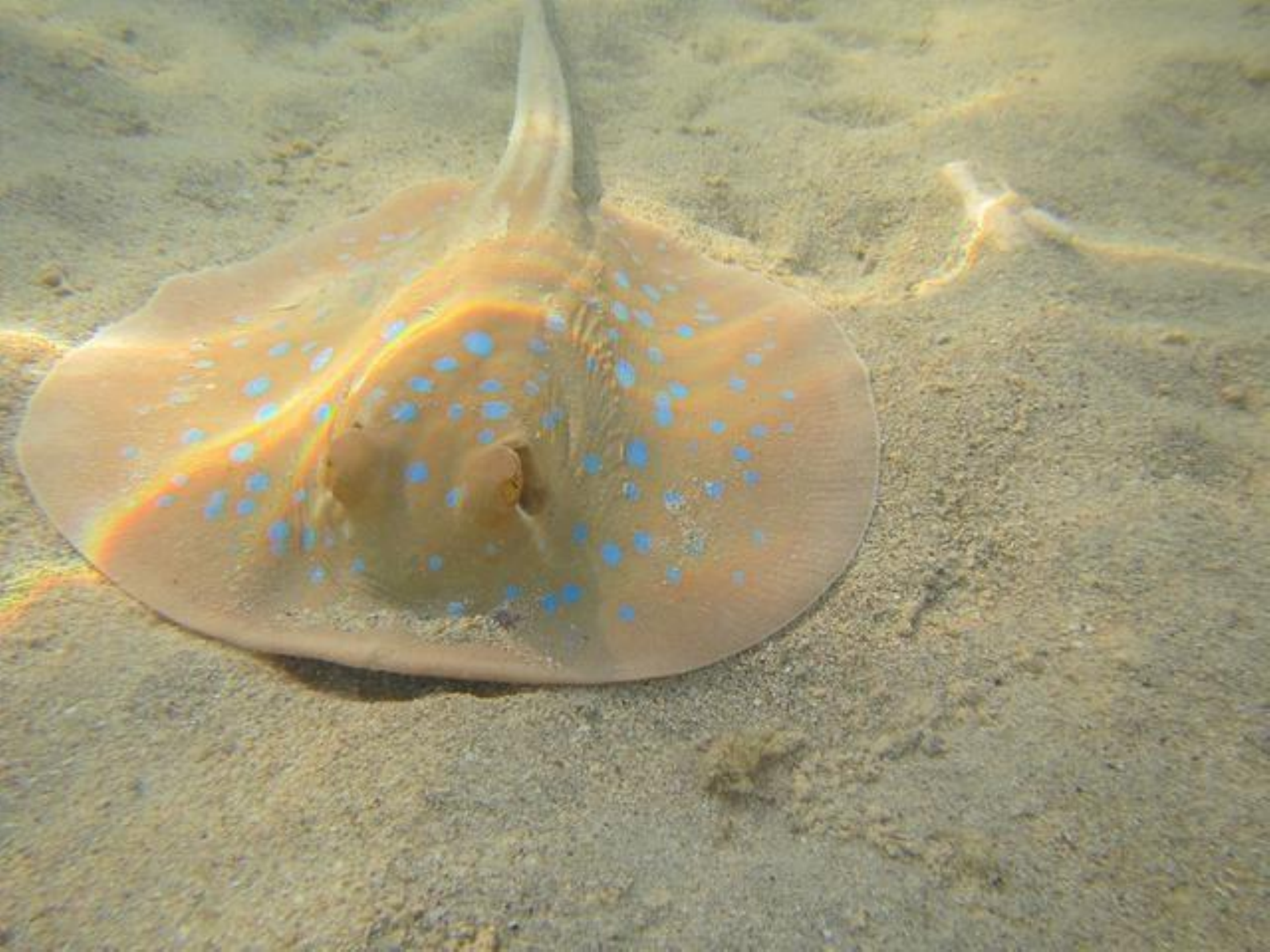}
\\
{  $2\times$} & { Bicubic} & {  SRDRM \cite{srdrm_srdrmgan}} & { SRDRM-GAN \cite{srdrm_srdrmgan}} & { Deep SESR \cite{sesr}} & {  Deep WaveNet} & { GT}\\

\end{tabular}
}
\caption{Qualitative comparison of the proposed method against existing works on underwater single image super-resolution. More results are shown in supplementary material.}
\label{fig:results_sr}
\end{figure*}

\begin{figure*}
\resizebox{\textwidth}{!}{
\setlength{\tabcolsep}{1pt}
\begin{tabular}{ccccccccc}

\includegraphics[width=3cm, height=2cm]{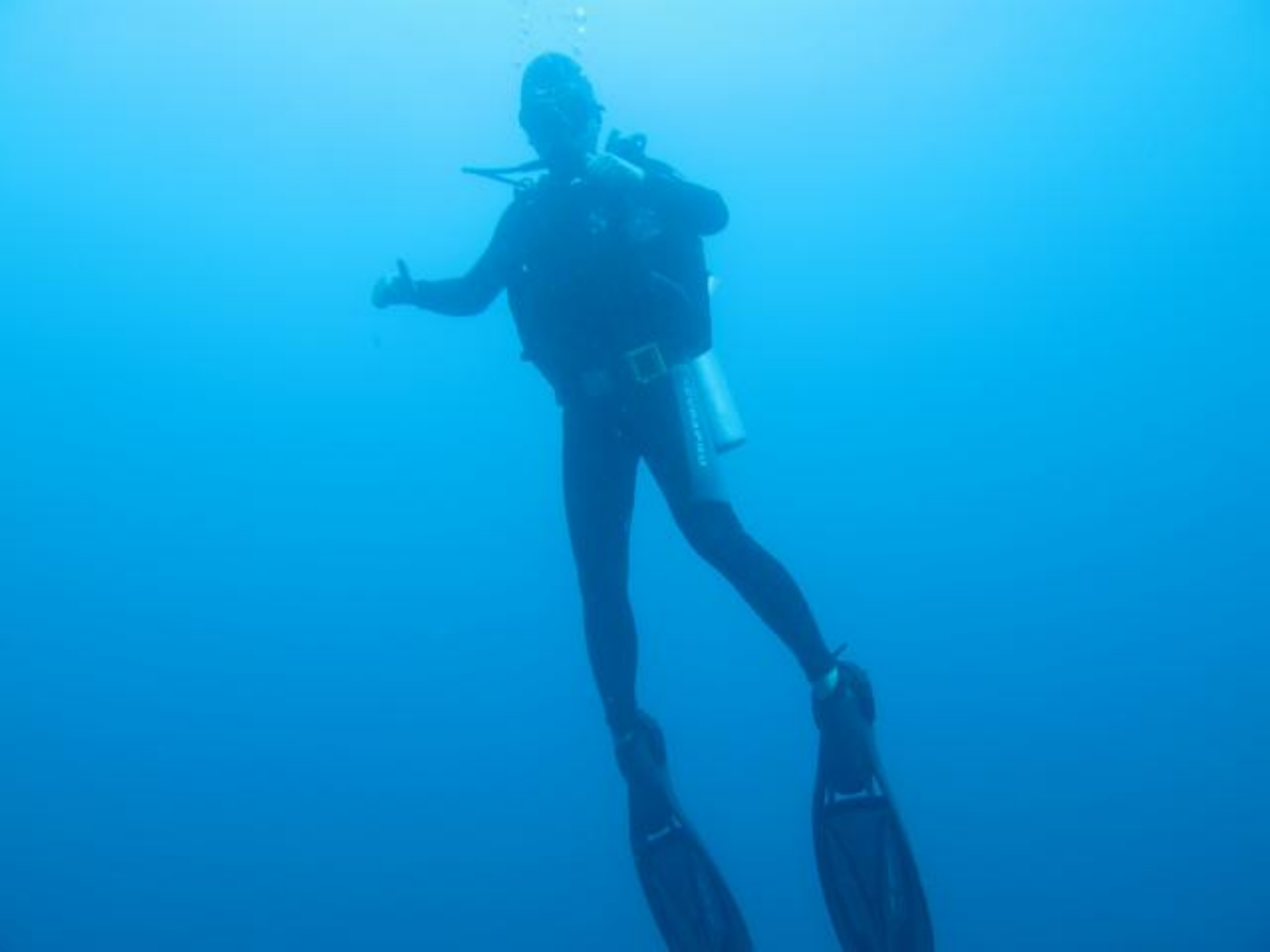}&
\includegraphics[width=3cm, height=2cm]{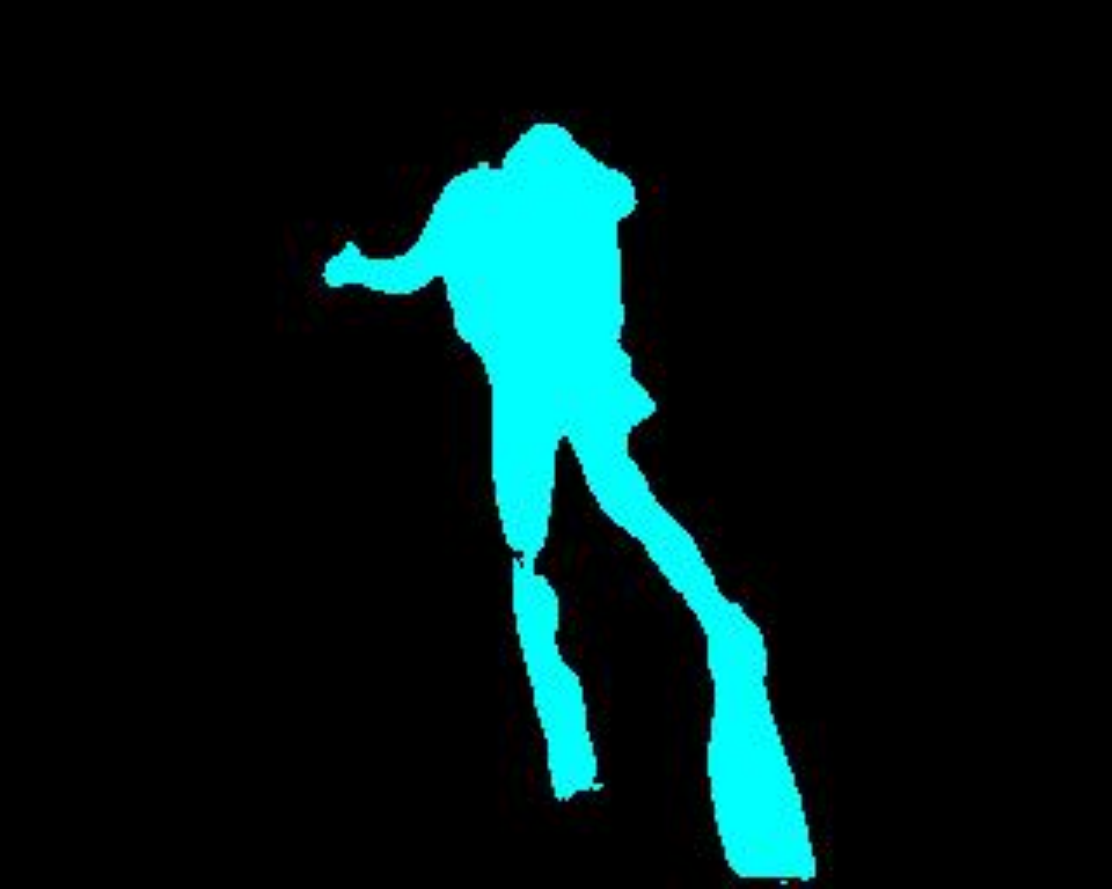}&
\includegraphics[width=3cm, height=2cm]{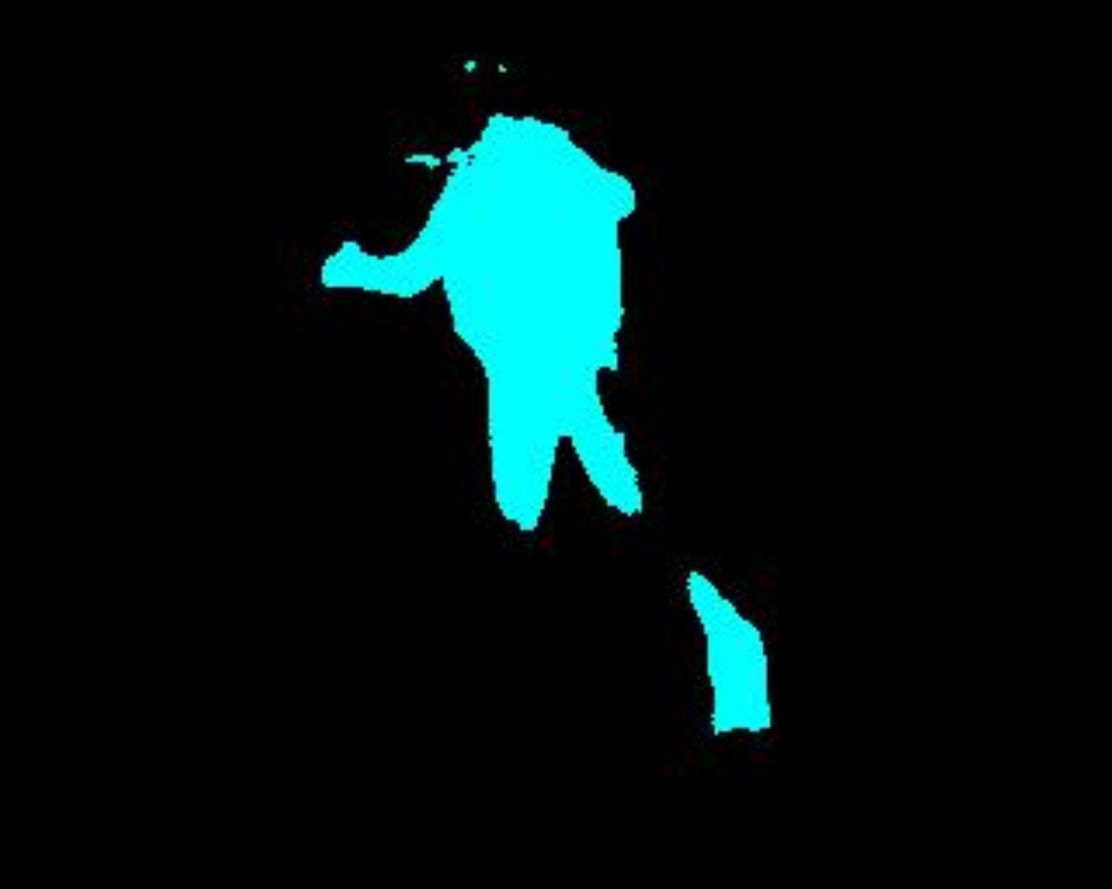}&
\includegraphics[width=3cm, height=2cm]{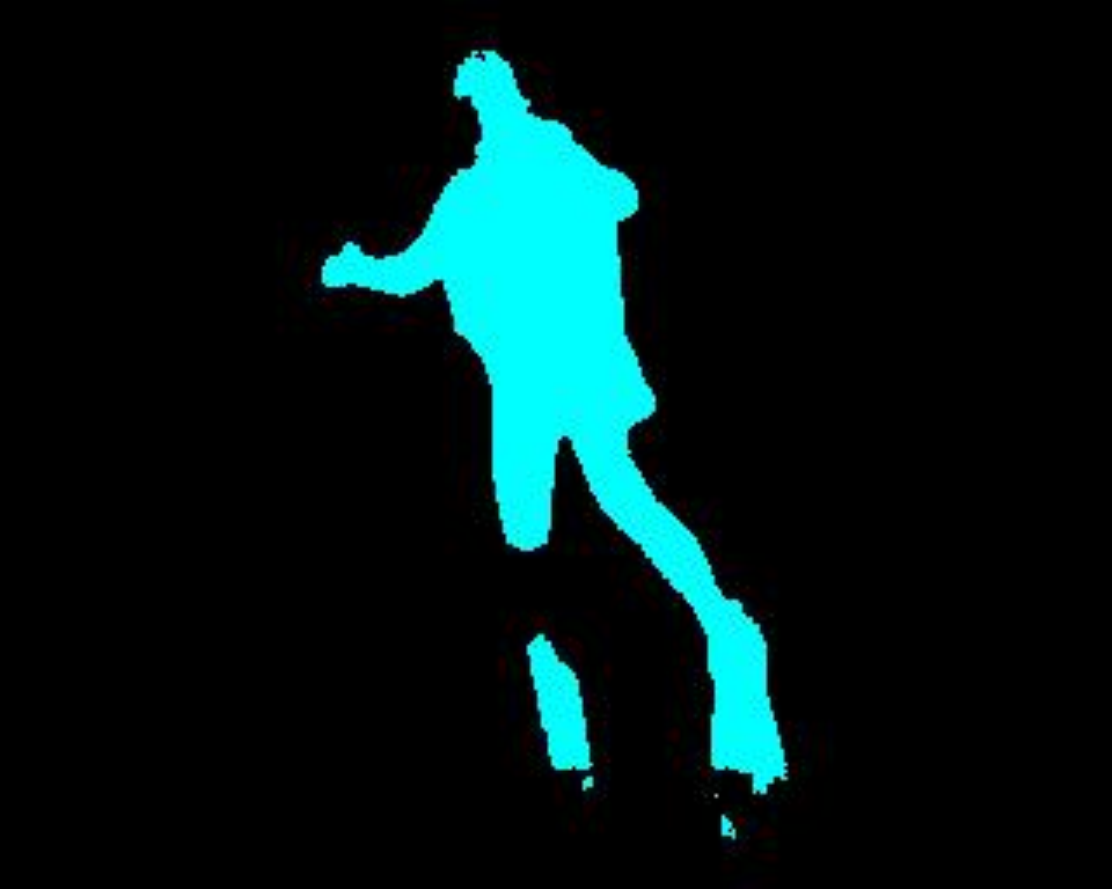}&
\includegraphics[width=3cm, height=2cm]{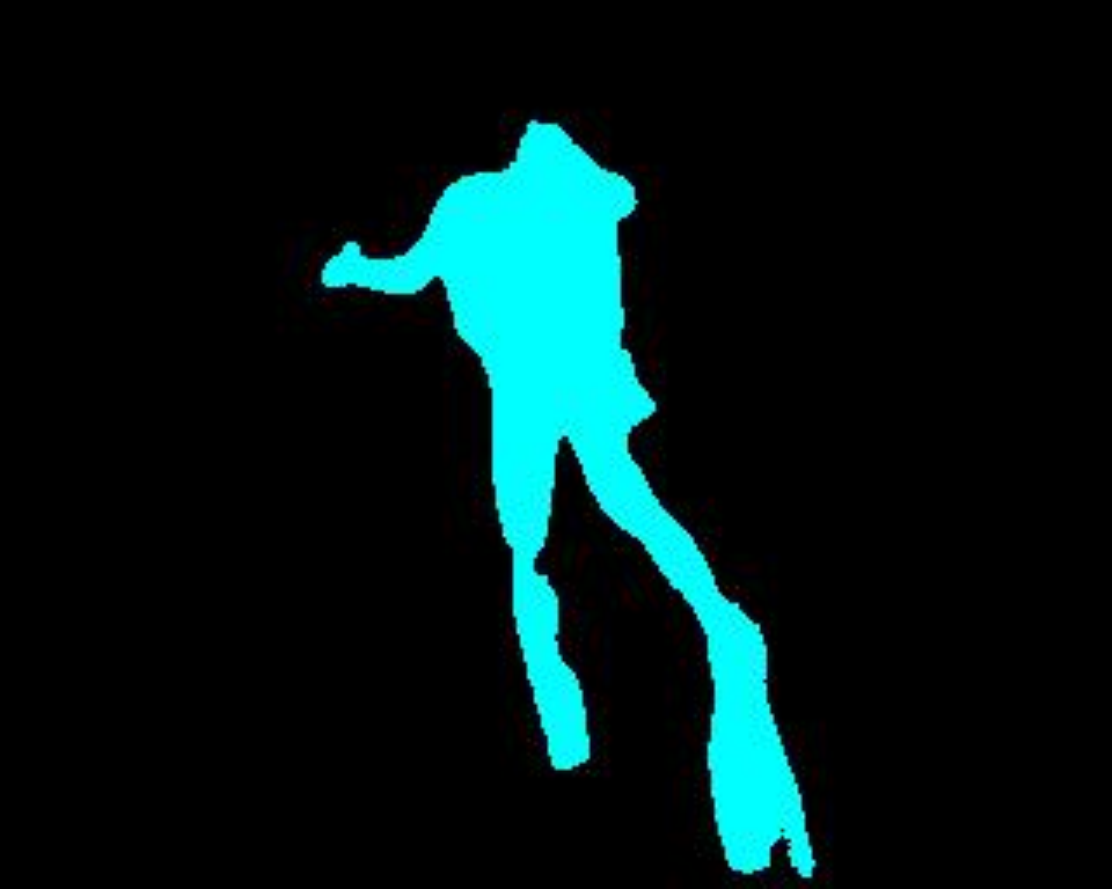}&
\includegraphics[width=3cm, height=2cm]{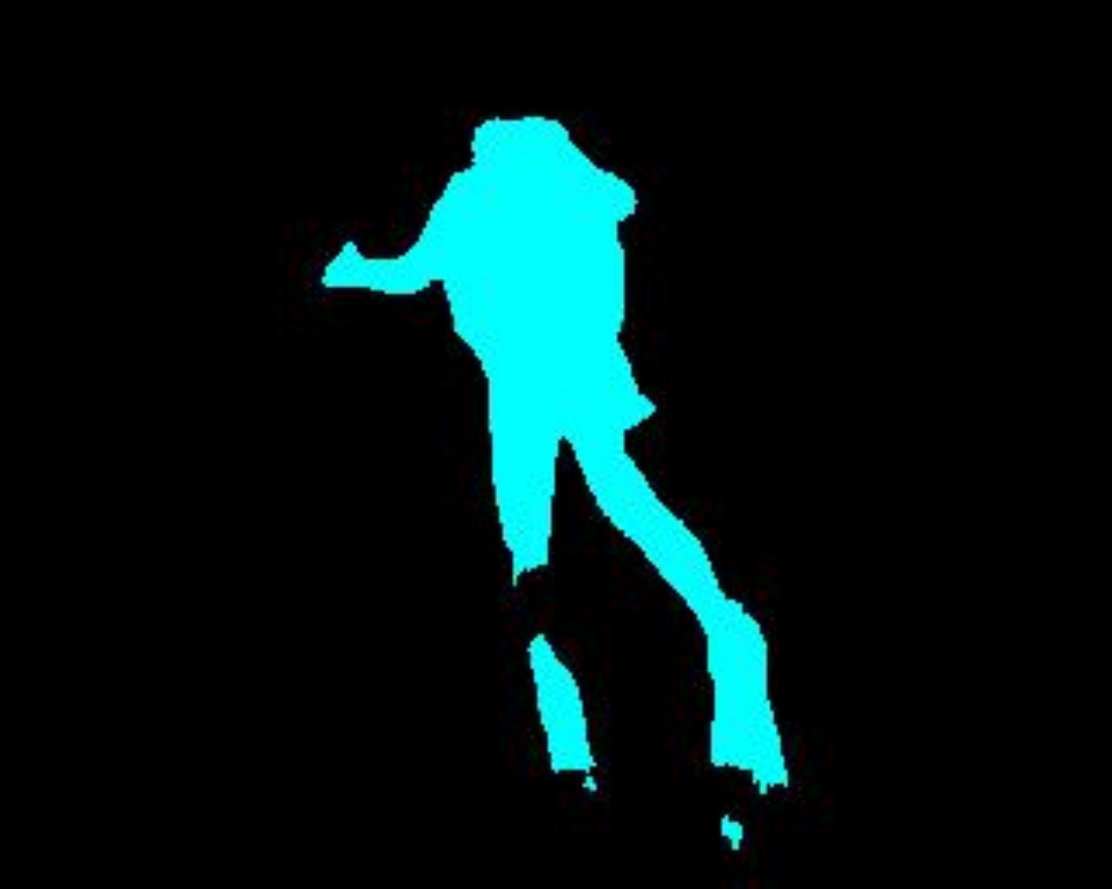}&
\includegraphics[width=3cm, height=2cm]{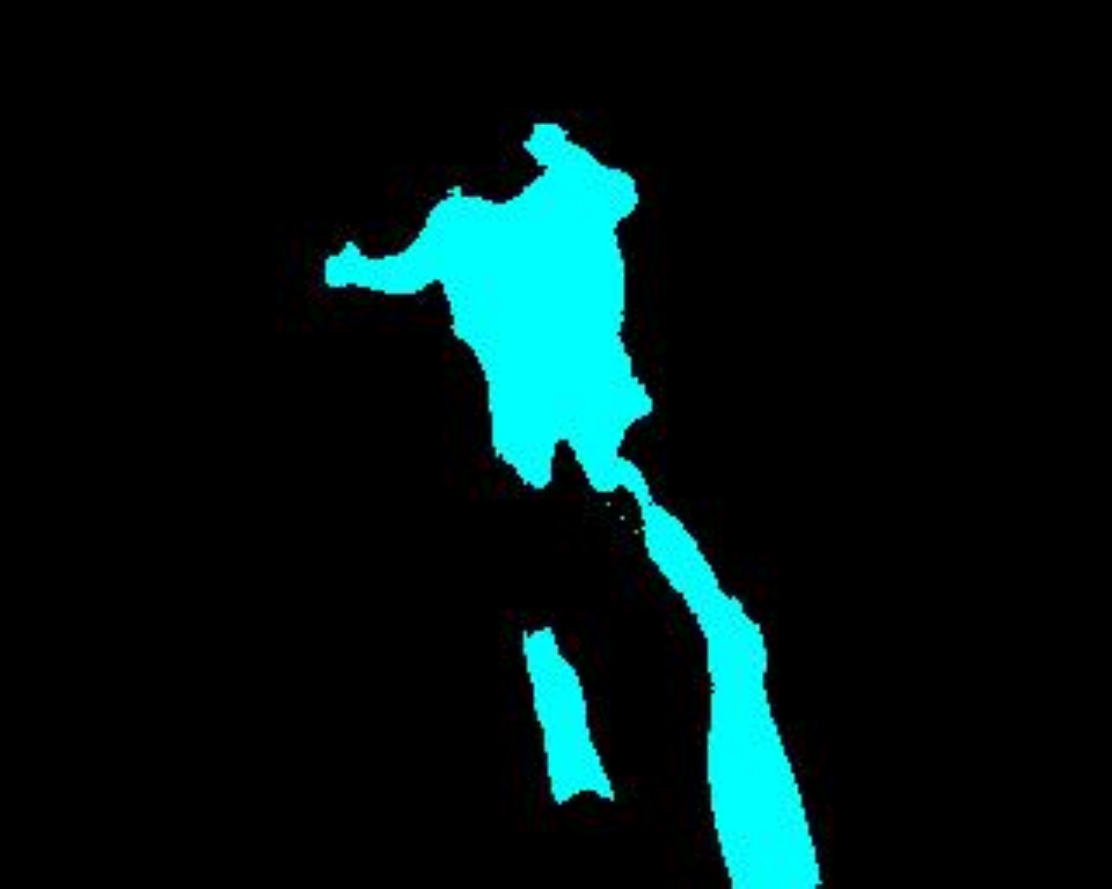}&
\includegraphics[width=3cm, height=2cm]{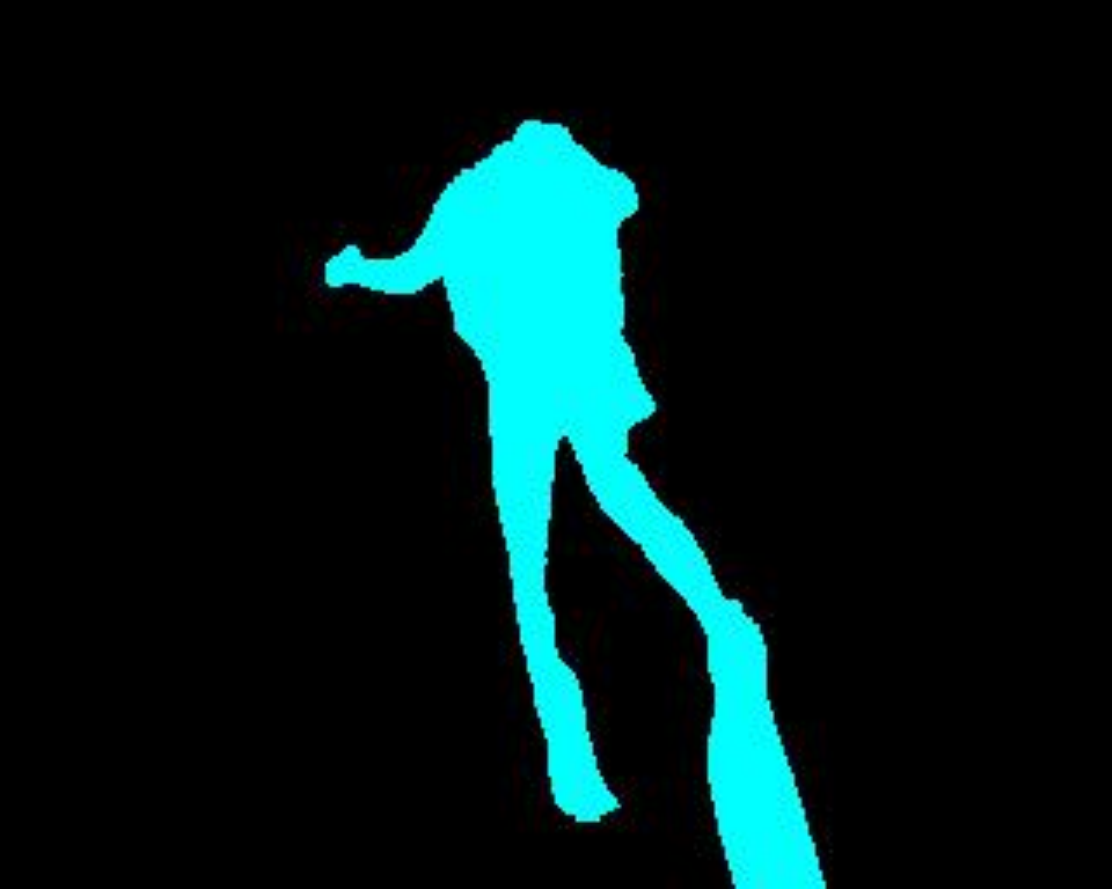}&
\includegraphics[width=3cm, height=2cm]{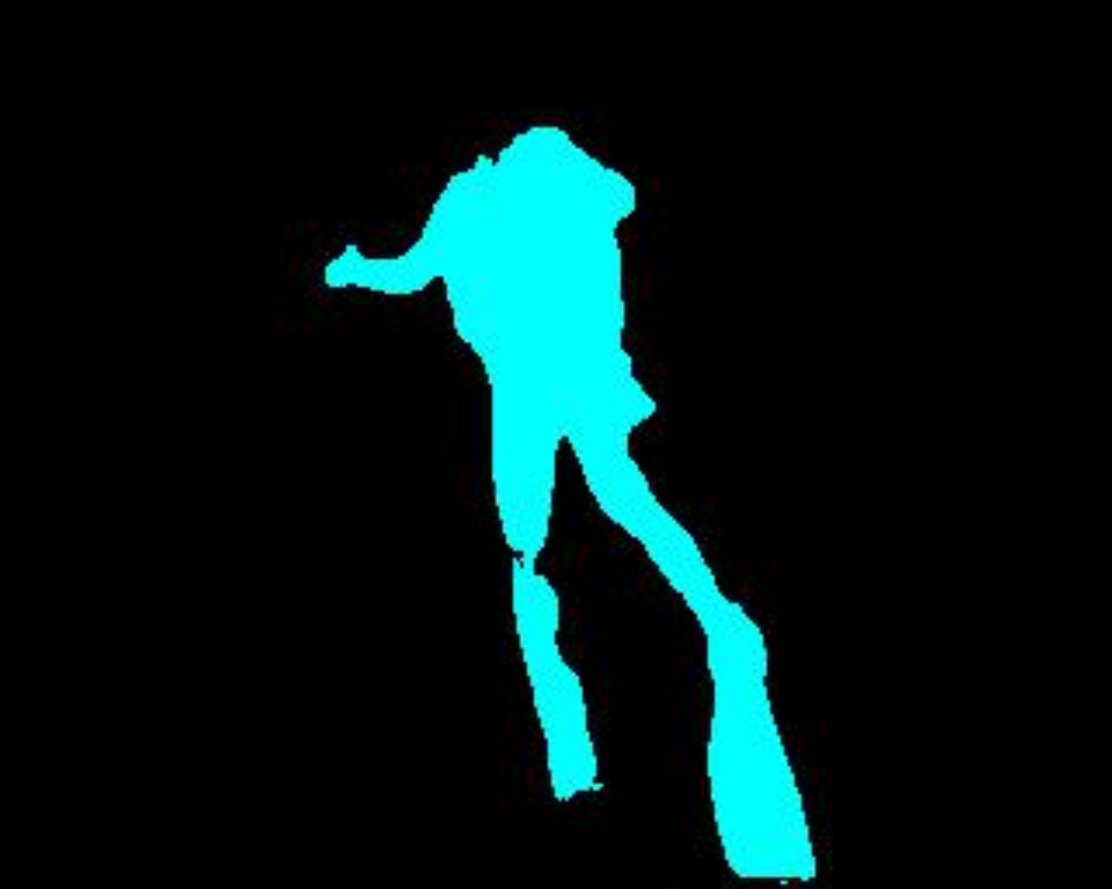}\\

\includegraphics[width=3cm, height=2cm]{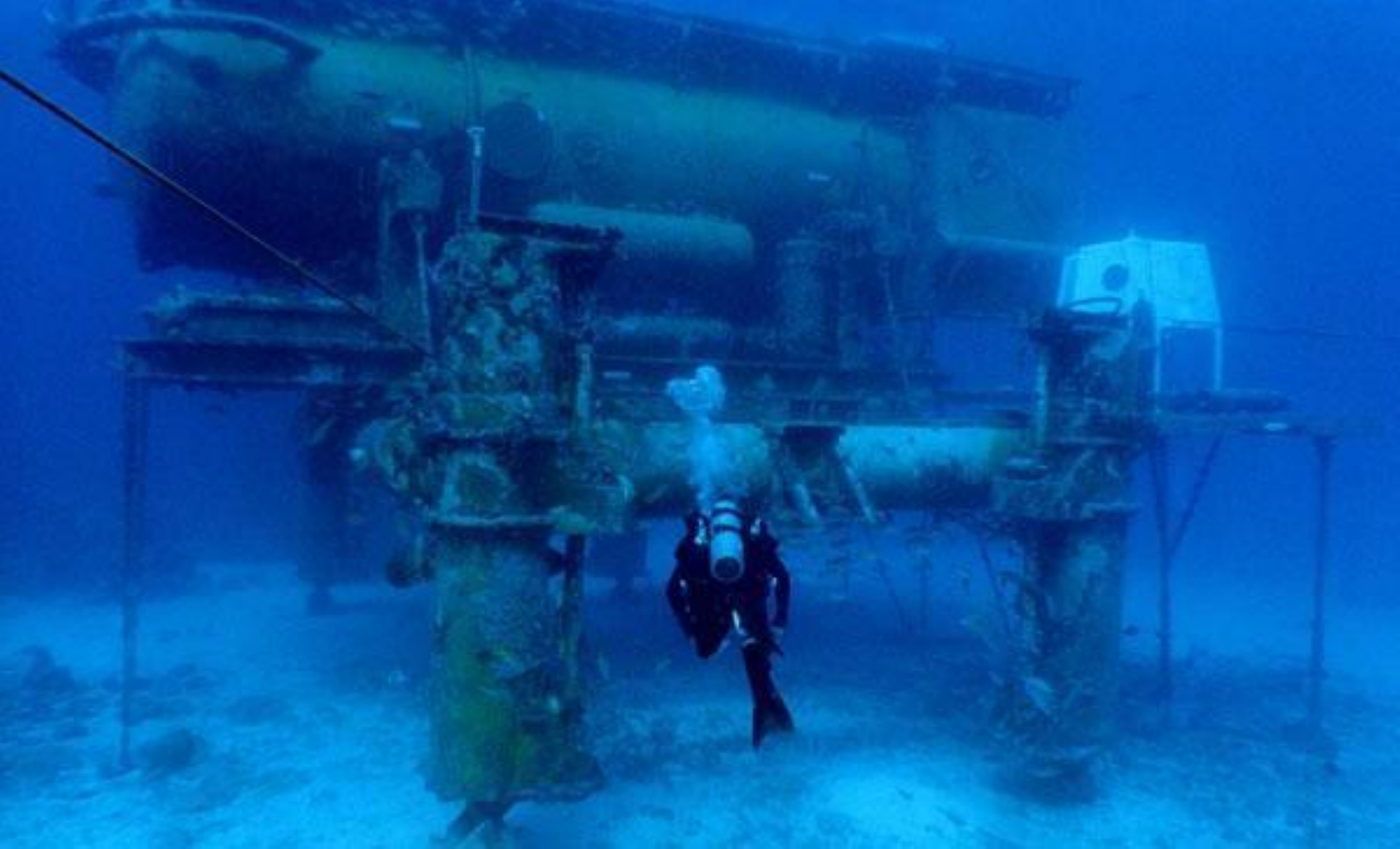}&
\includegraphics[width=3cm, height=2cm]{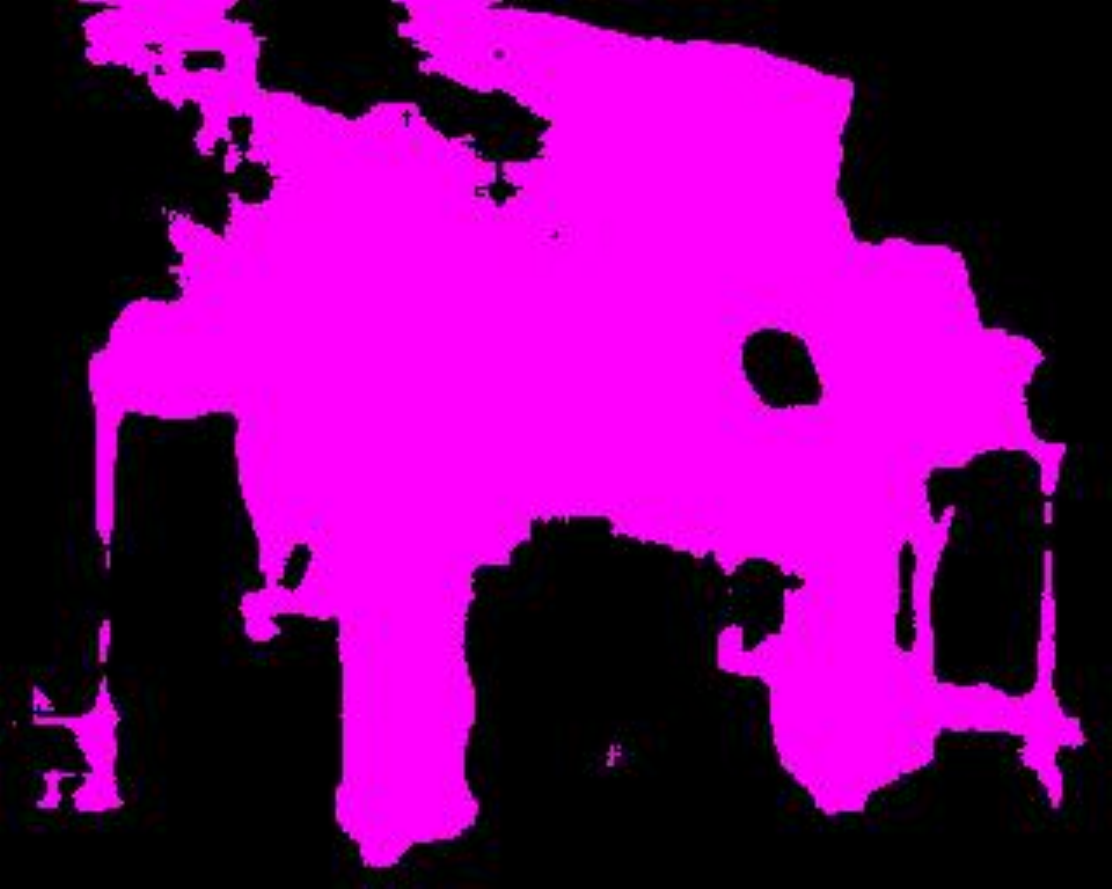}&
\includegraphics[width=3cm, height=2cm]{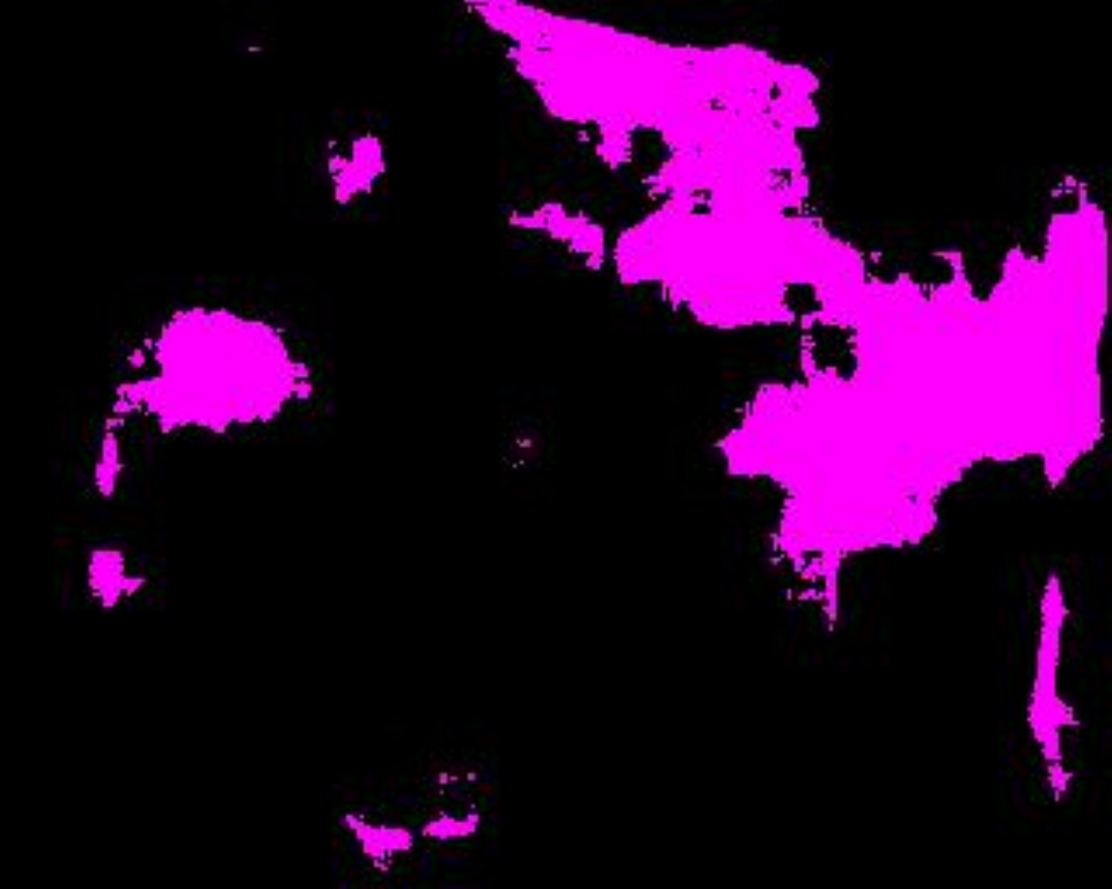}&
\includegraphics[width=3cm, height=2cm]{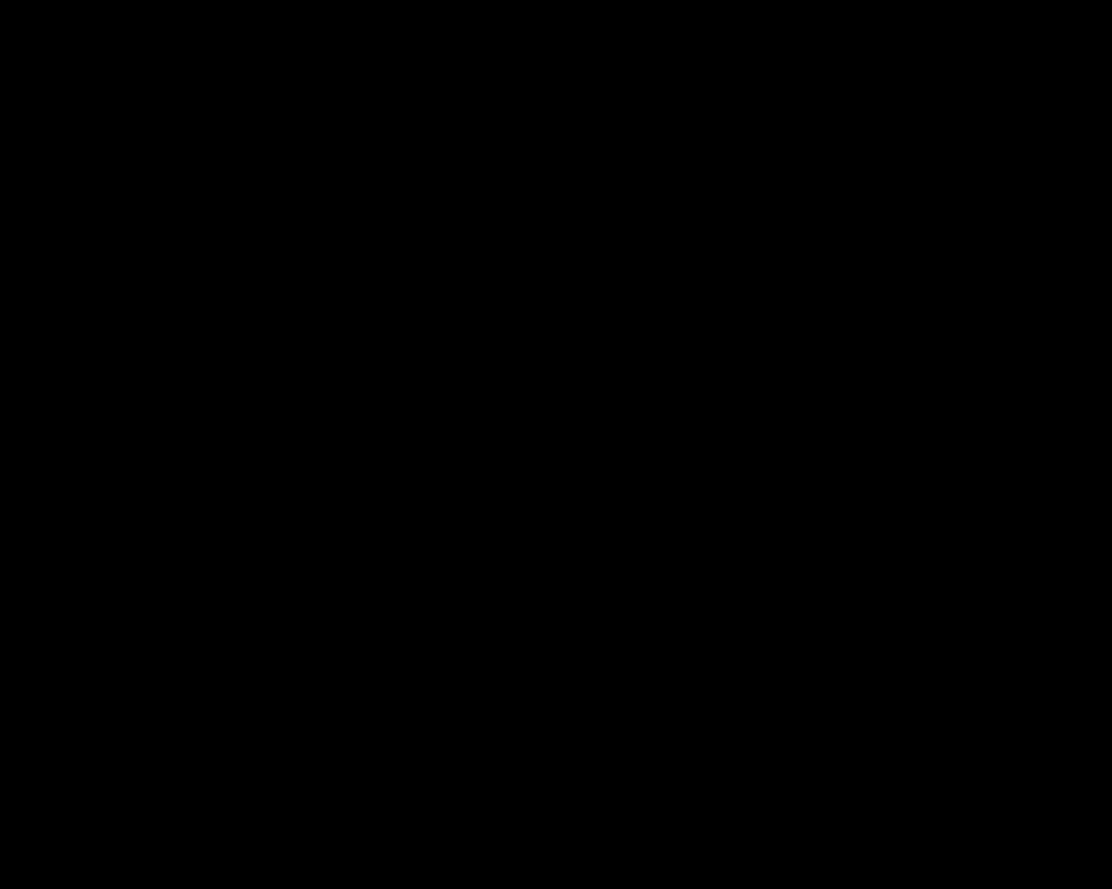}&
\includegraphics[width=3cm, height=2cm]{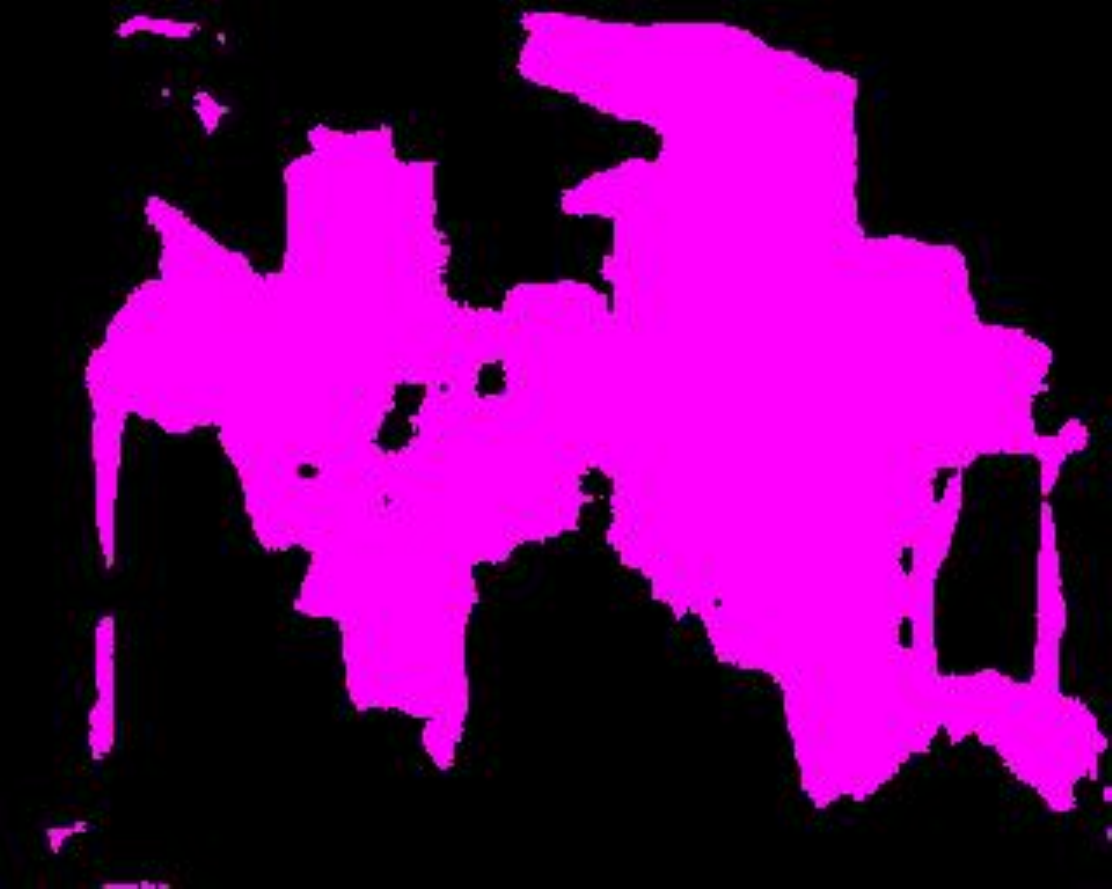}&
\includegraphics[width=3cm, height=2cm]{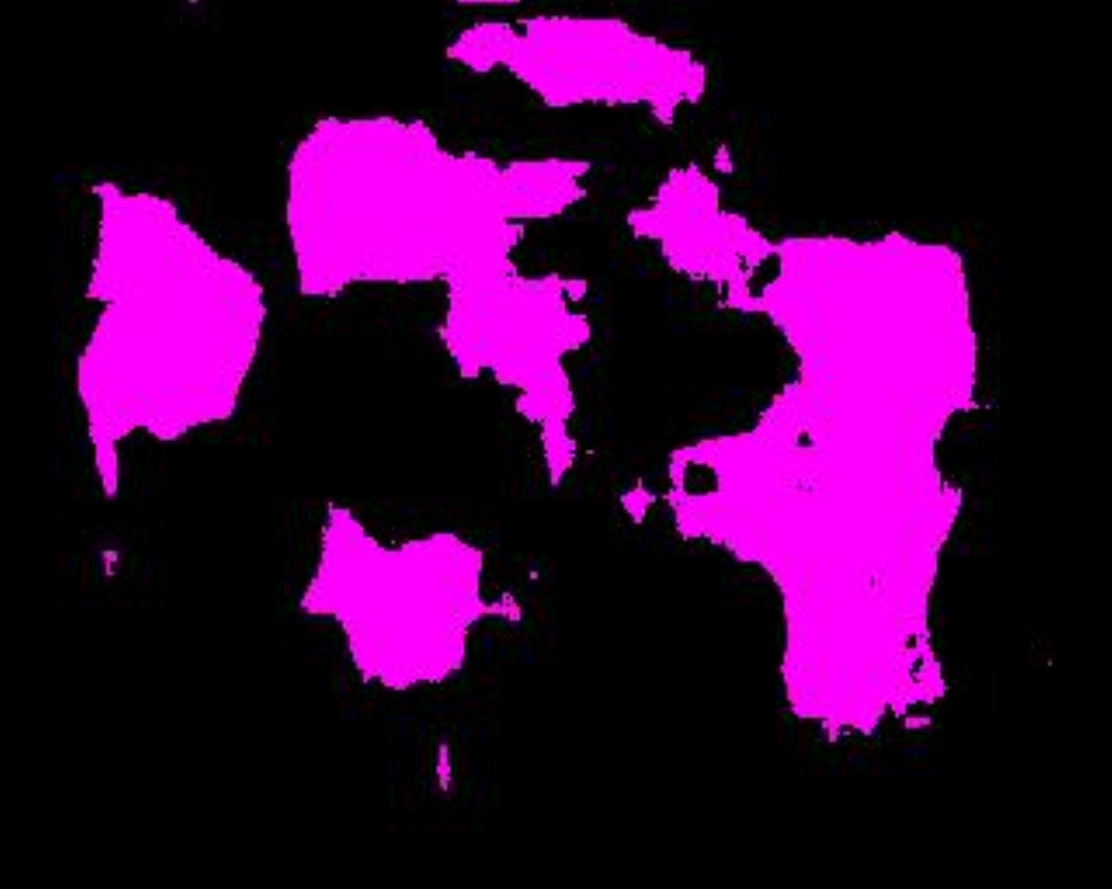}&
\includegraphics[width=3cm, height=2cm]{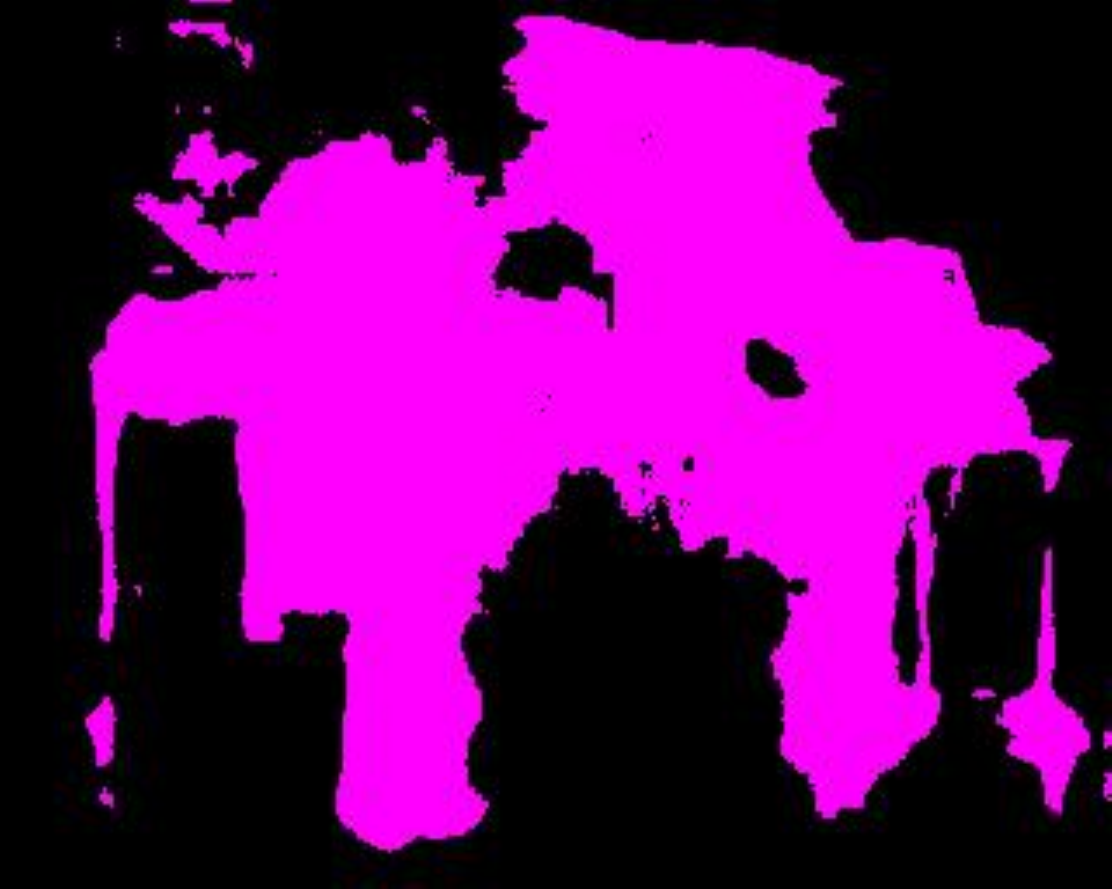}&
\includegraphics[width=3cm, height=2cm]{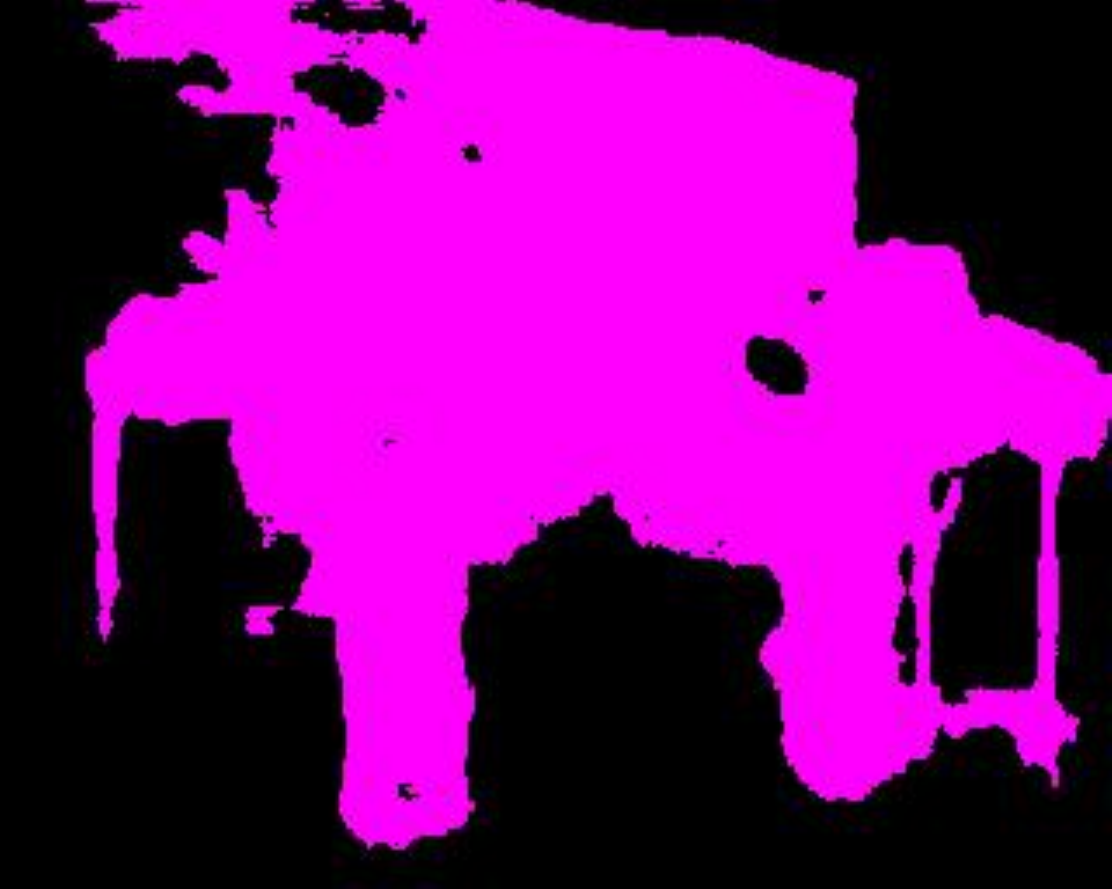}&
\includegraphics[width=3cm, height=2cm]{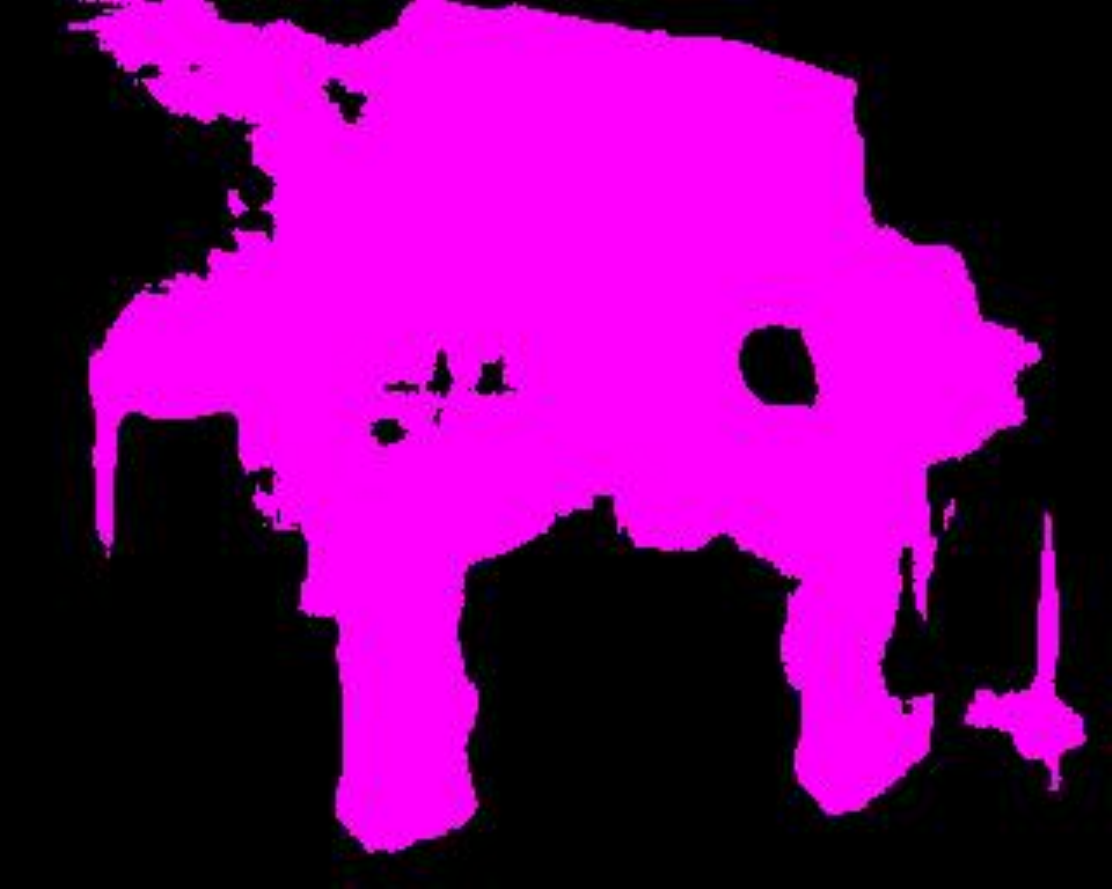}\\

\includegraphics[width=3cm, height=2cm]{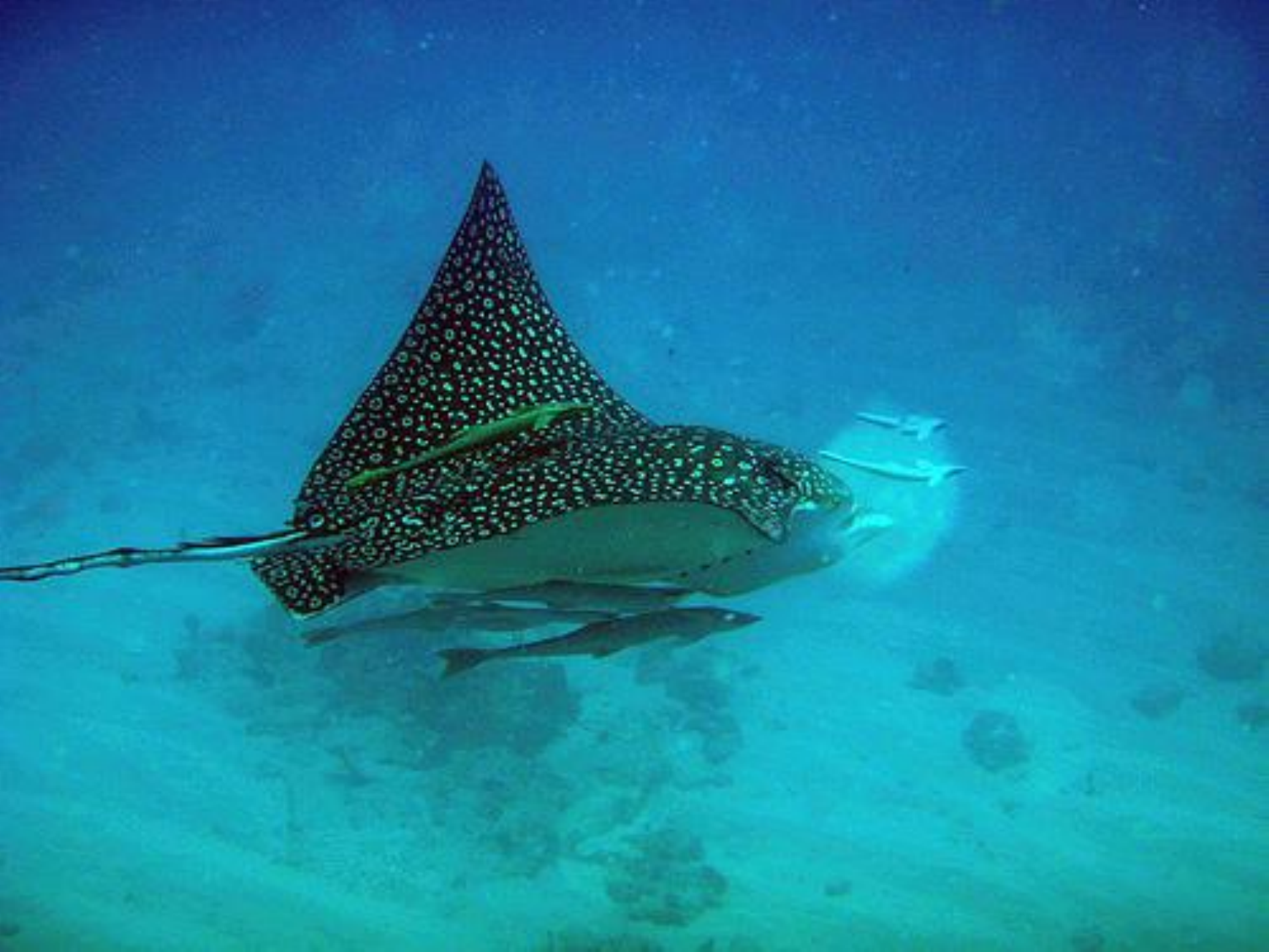}&
\includegraphics[width=3cm, height=2cm]{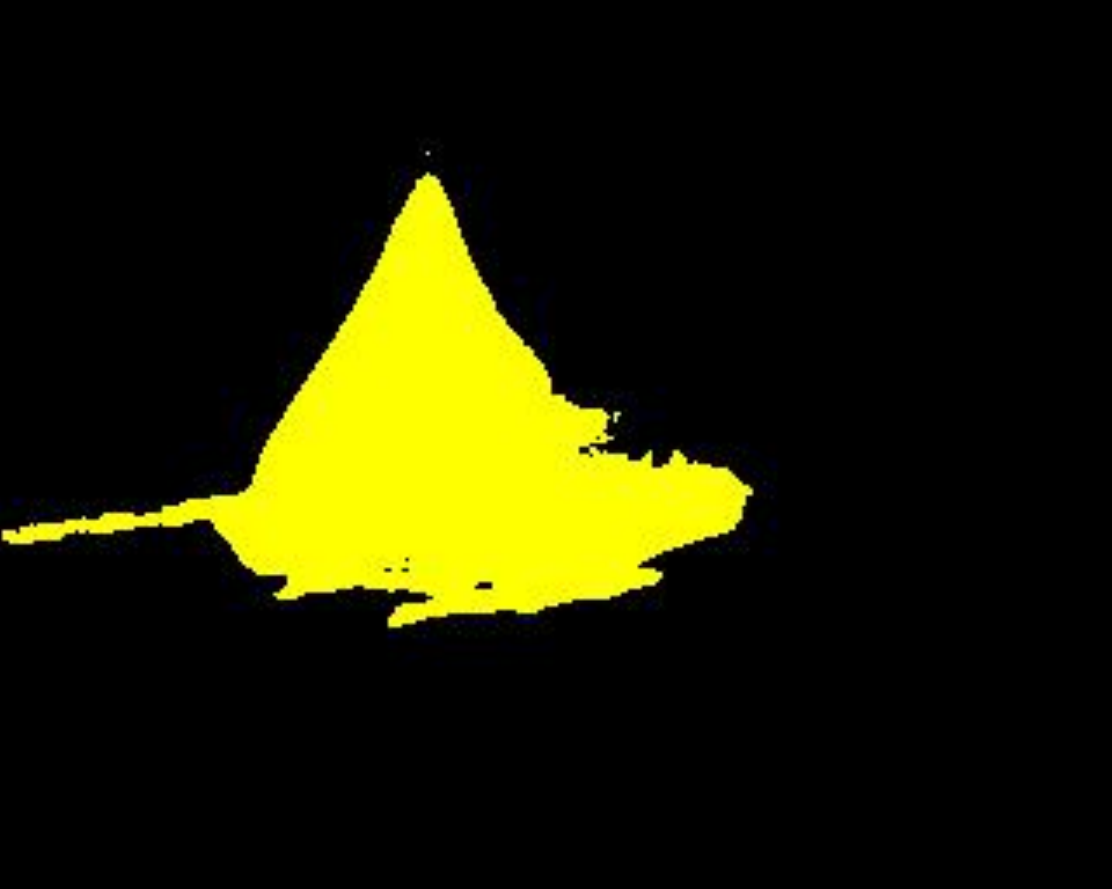}&
\includegraphics[width=3cm, height=2cm]{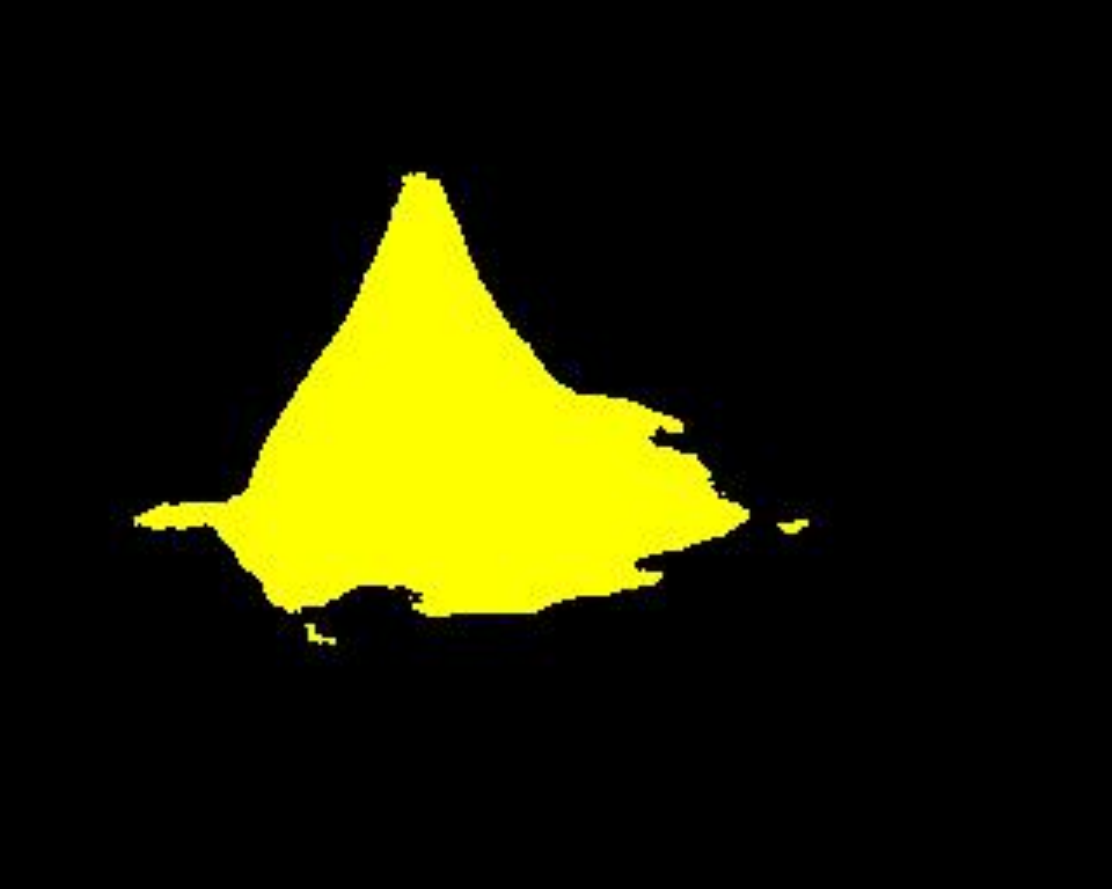}&
\includegraphics[width=3cm, height=2cm]{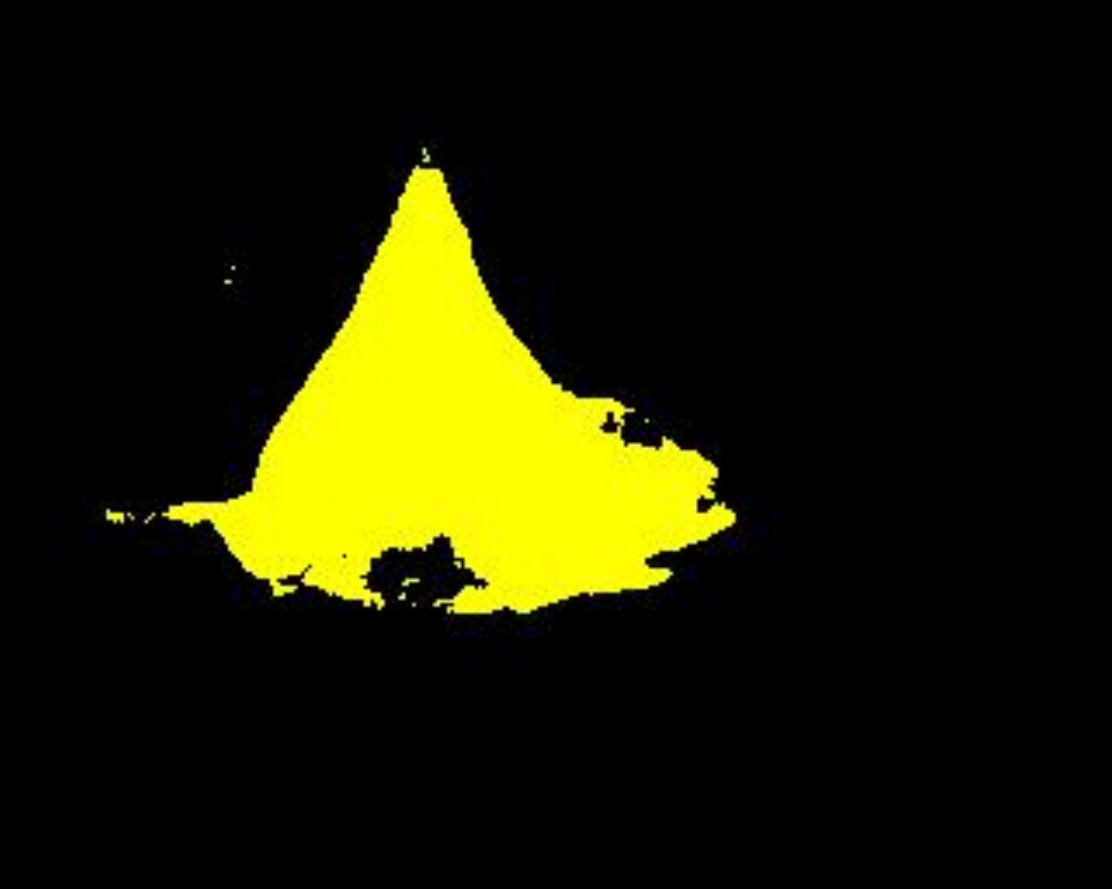}&
\includegraphics[width=3cm, height=2cm]{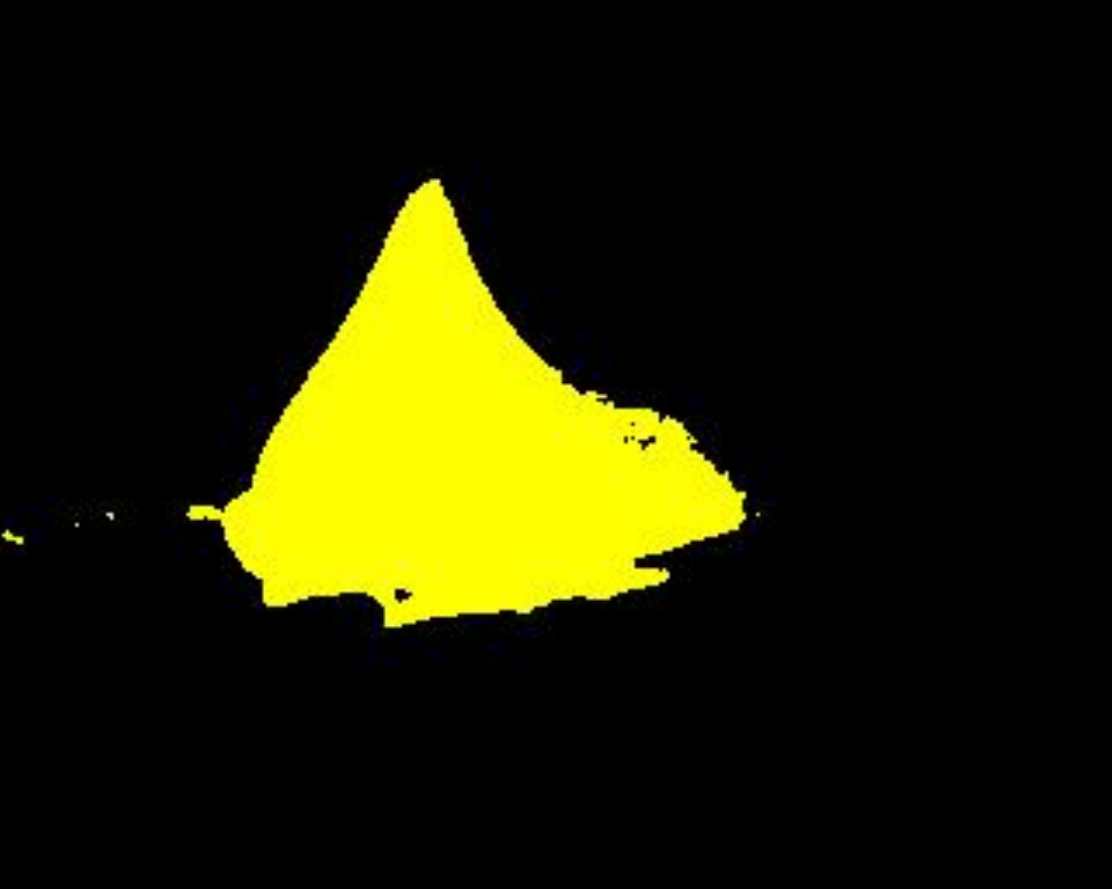}&
\includegraphics[width=3cm, height=2cm]{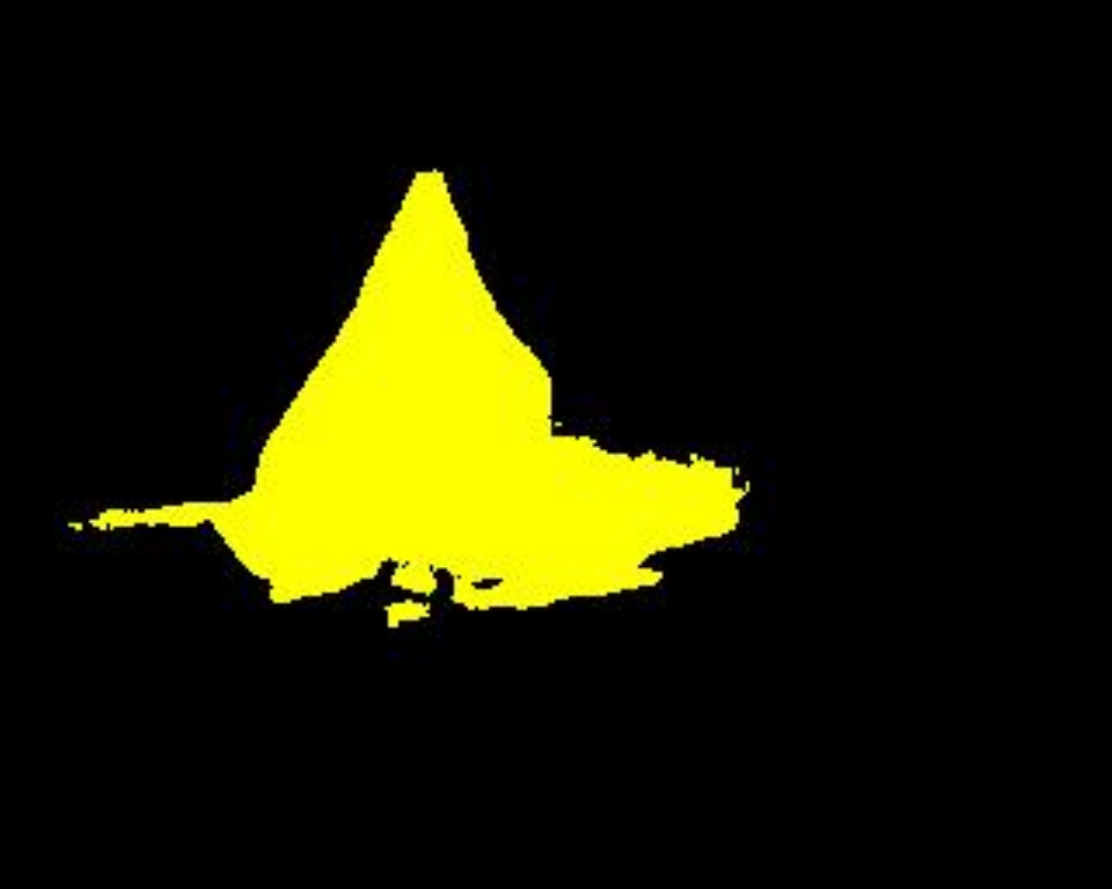}&
\includegraphics[width=3cm, height=2cm]{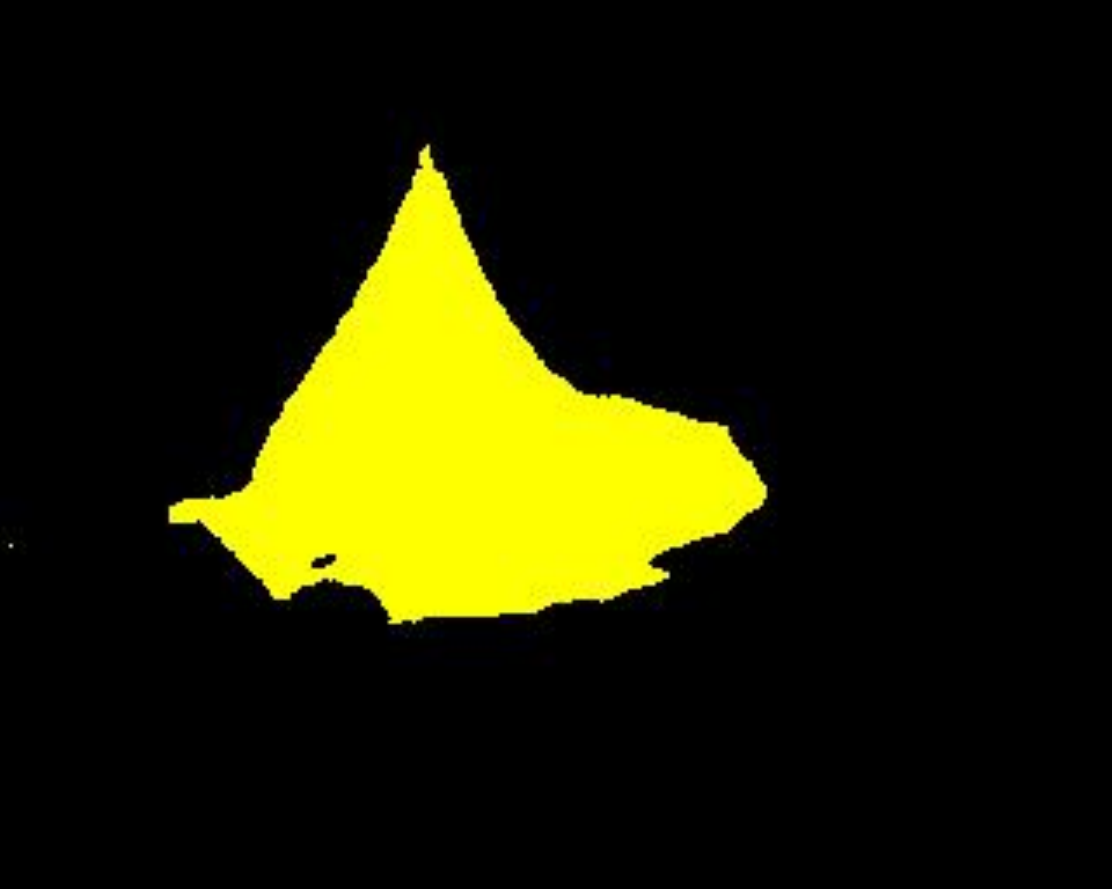}&
\includegraphics[width=3cm, height=2cm]{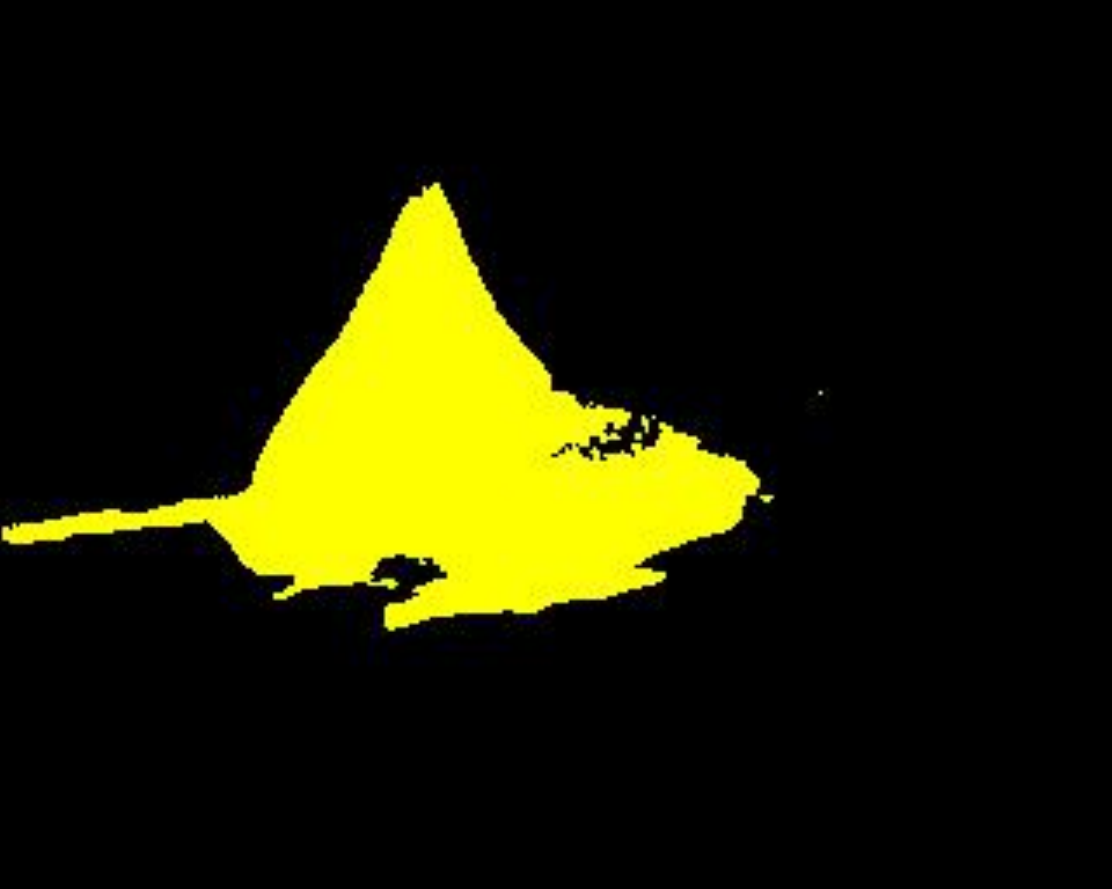}&
\includegraphics[width=3cm, height=2cm]{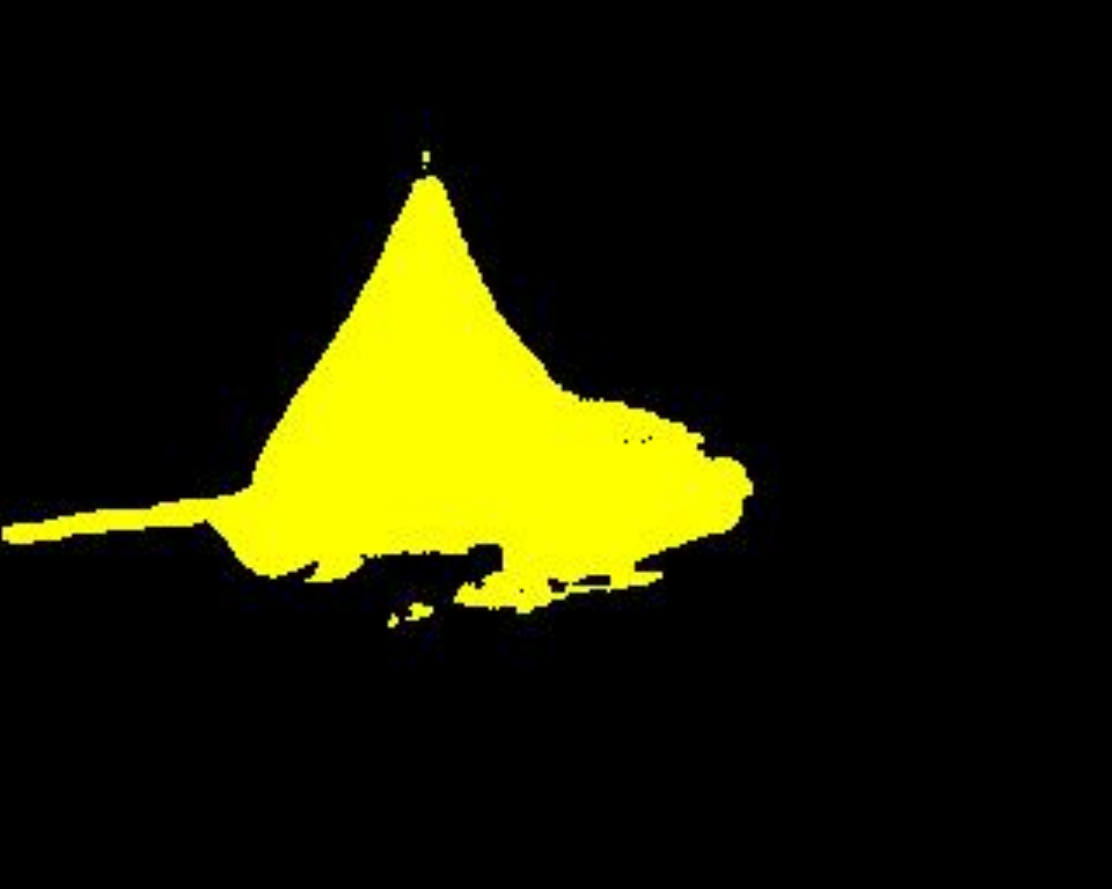}\\

\includegraphics[width=3cm, height=2cm]{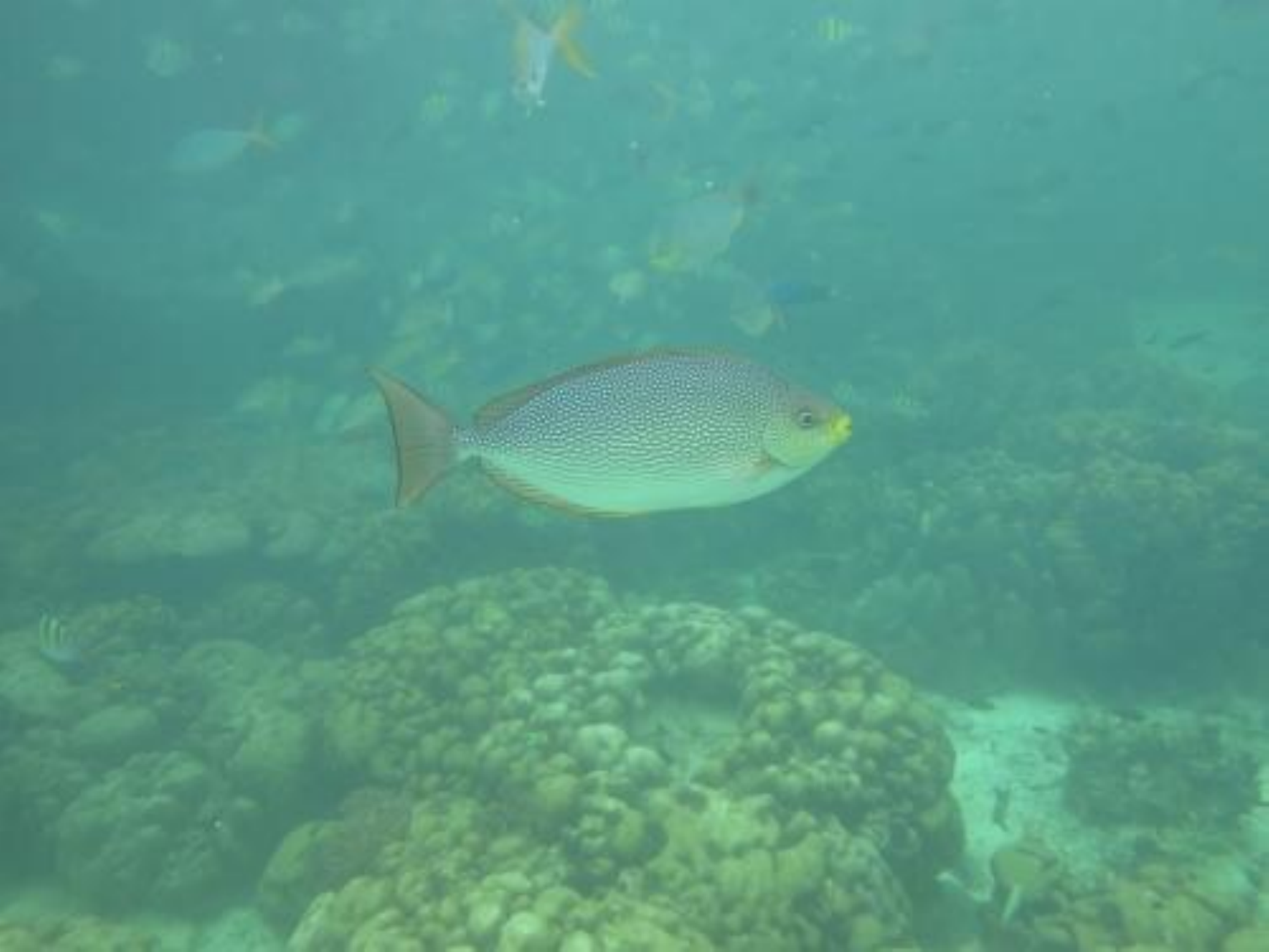}&
\includegraphics[width=3cm, height=2cm]{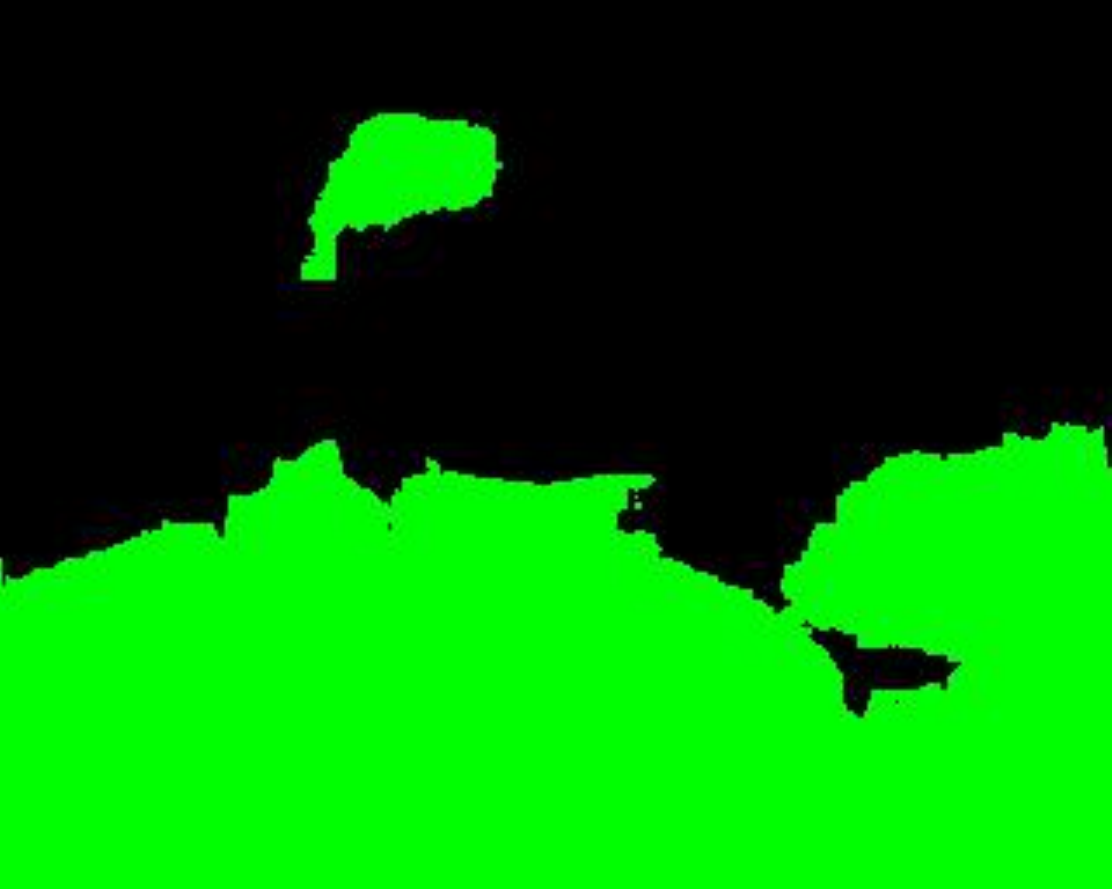}&
\includegraphics[width=3cm, height=2cm]{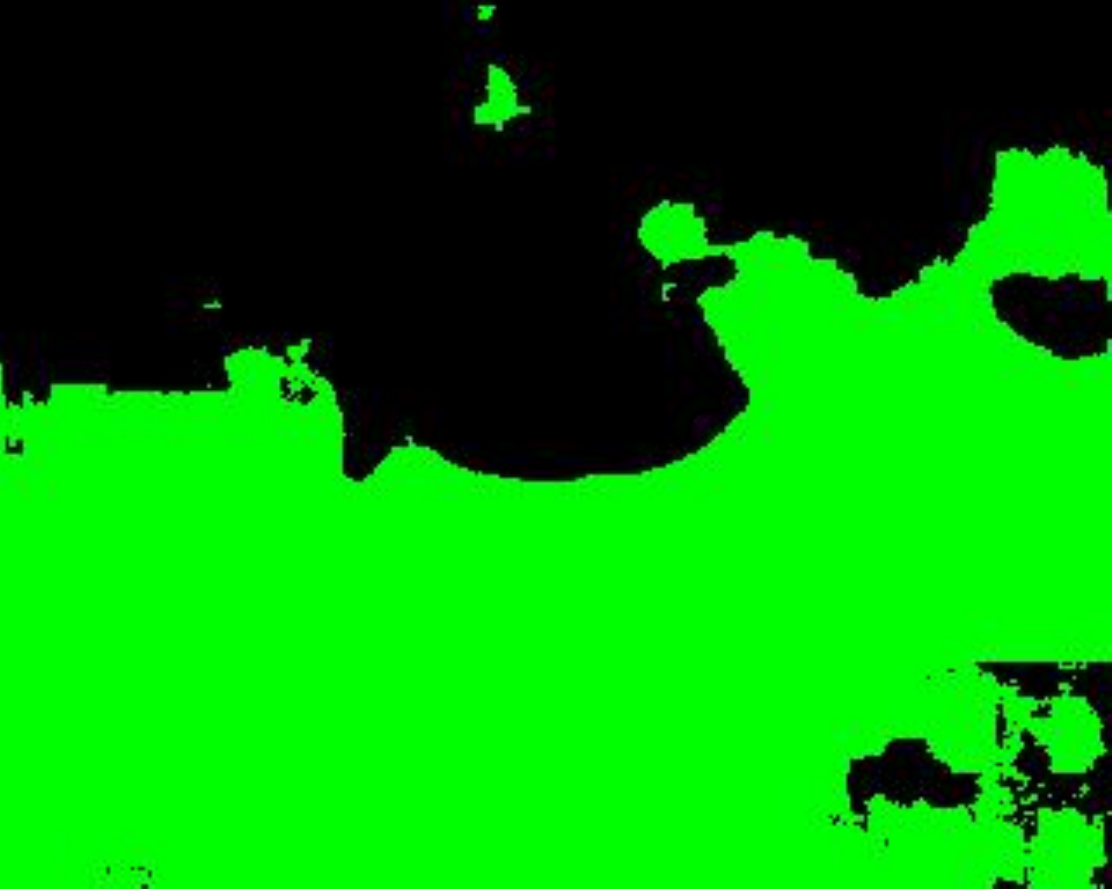}&
\includegraphics[width=3cm, height=2cm]{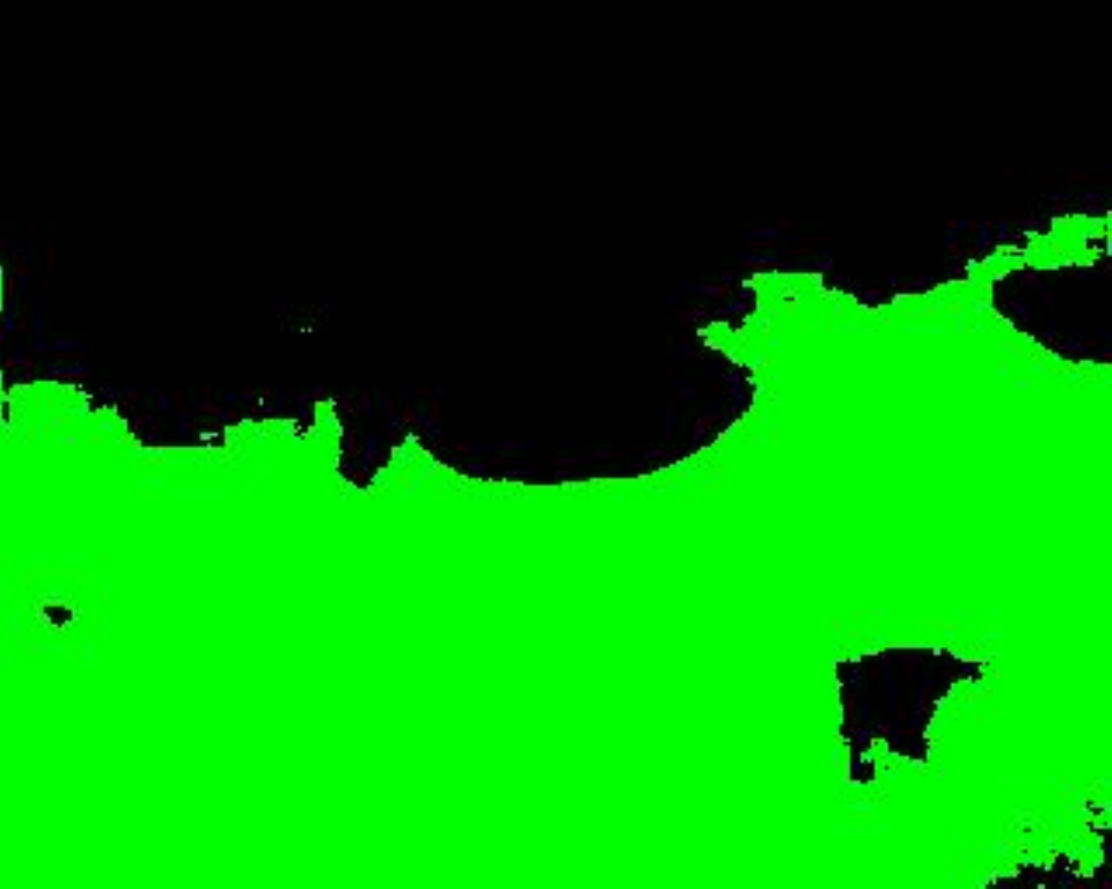}&
\includegraphics[width=3cm, height=2cm]{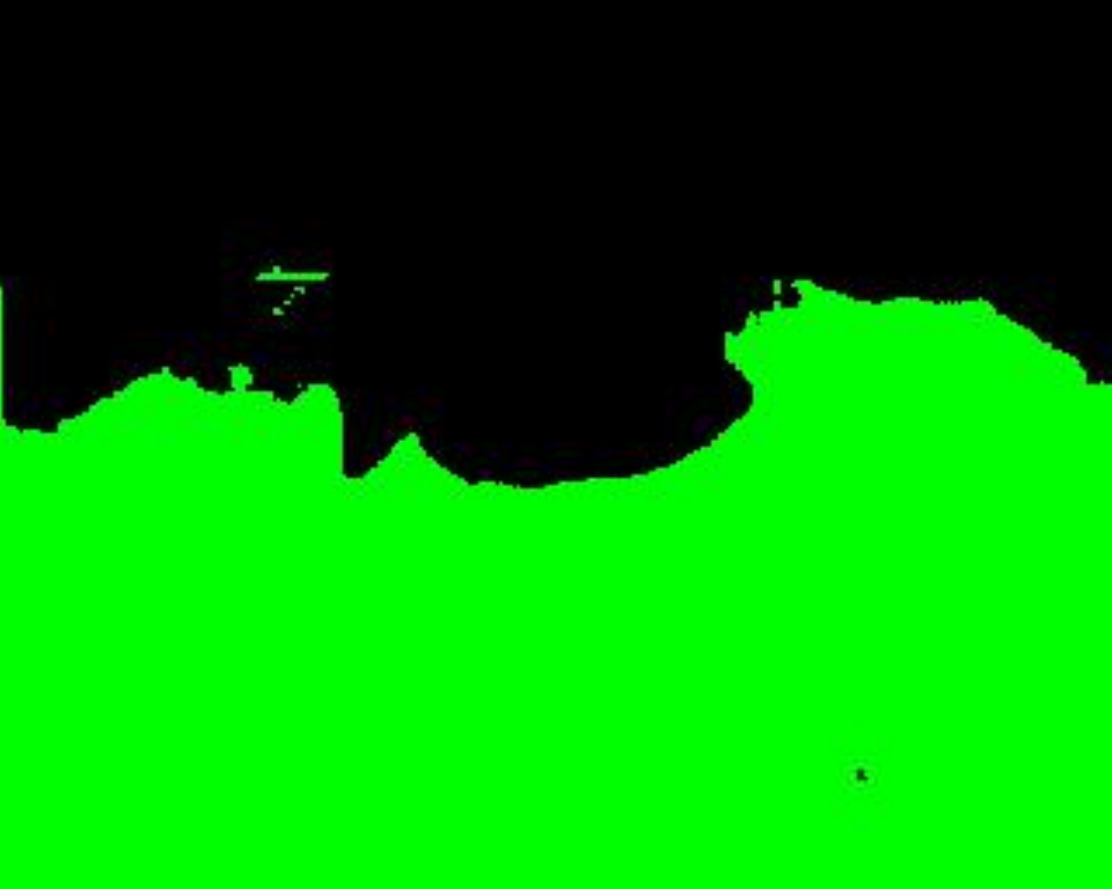}&
\includegraphics[width=3cm, height=2cm]{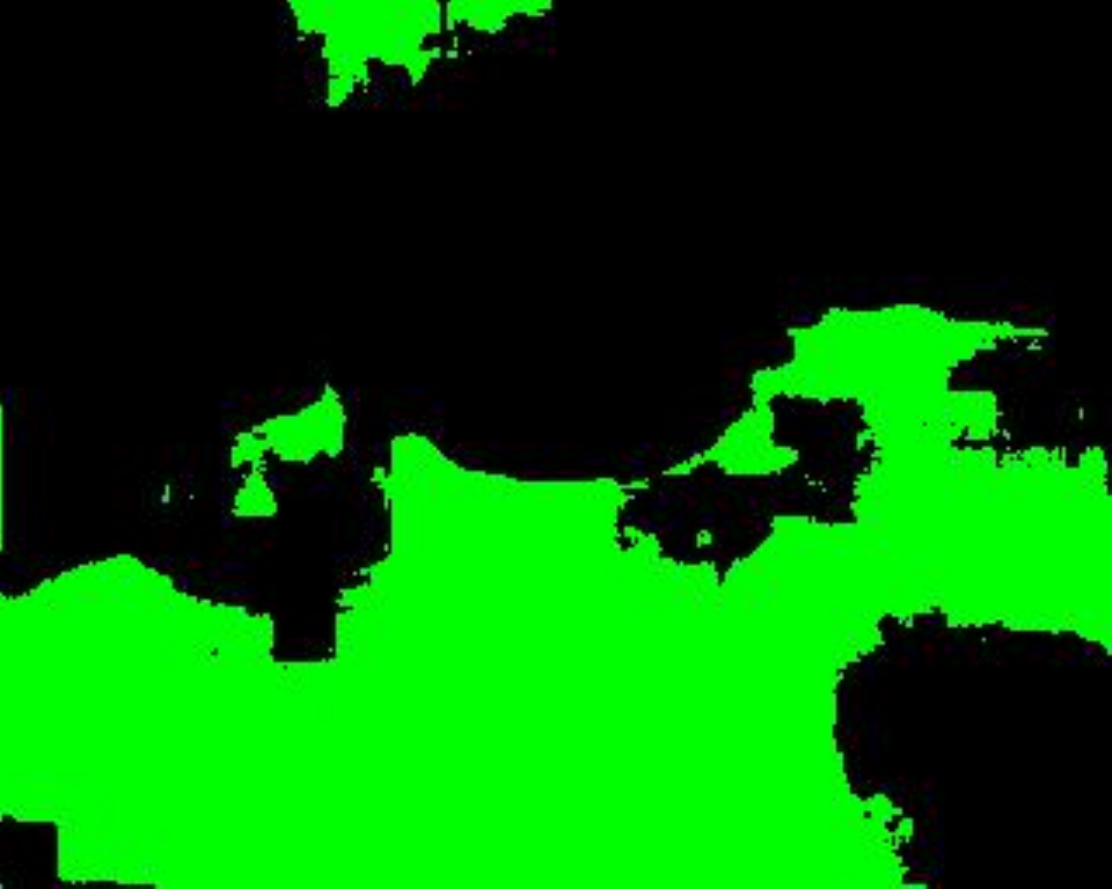}&
\includegraphics[width=3cm, height=2cm]{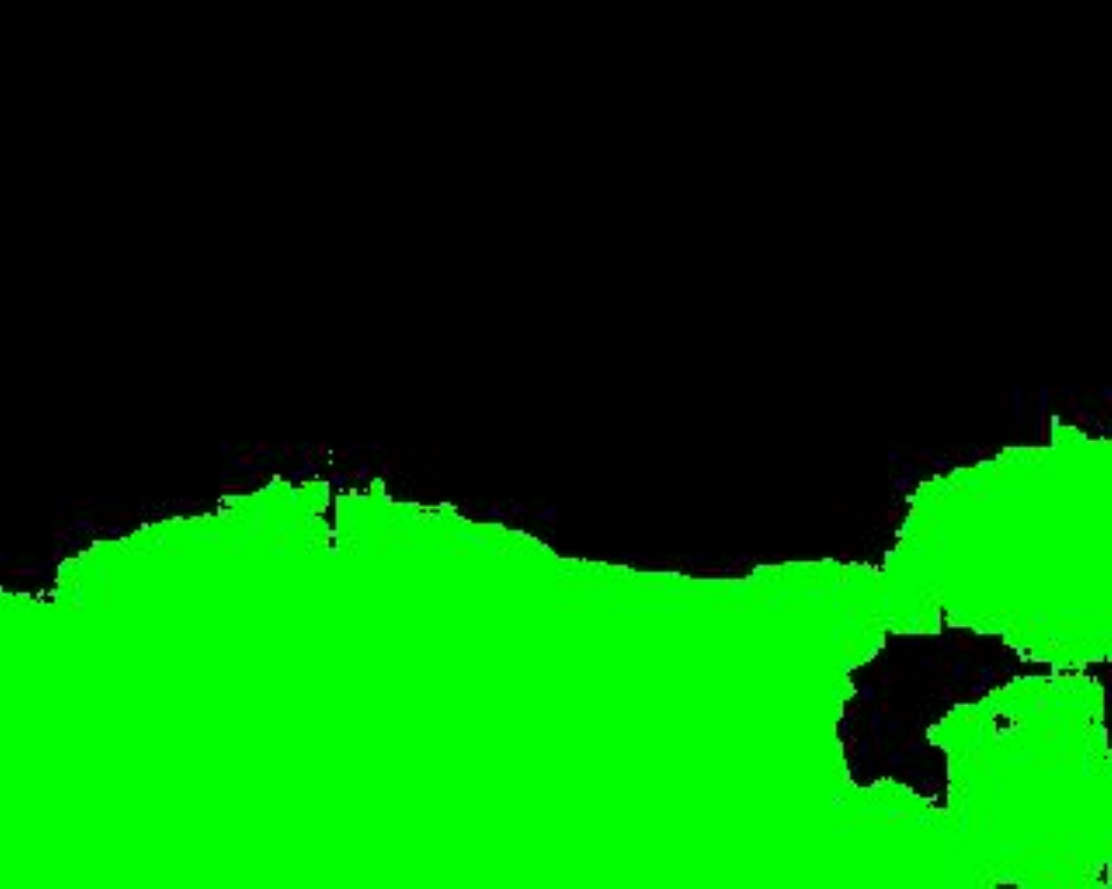}&
\includegraphics[width=3cm, height=2cm]{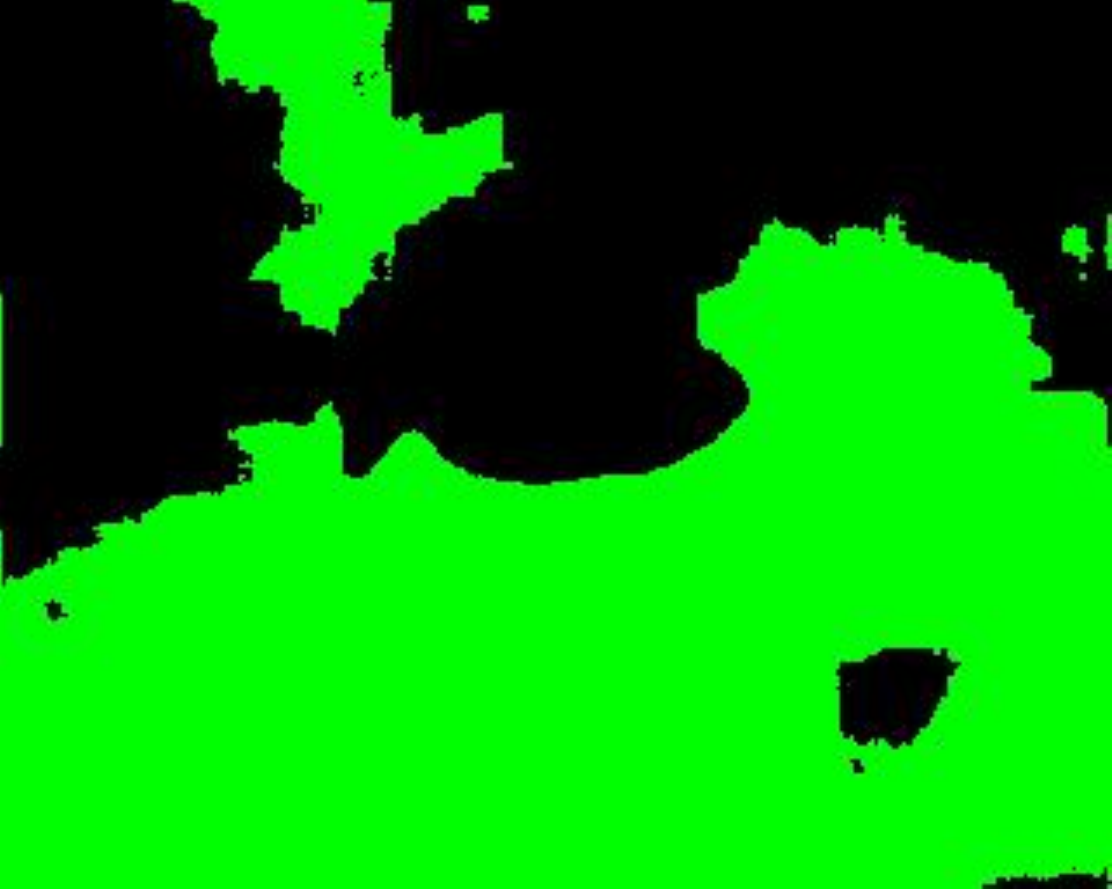}&
\includegraphics[width=3cm, height=2cm]{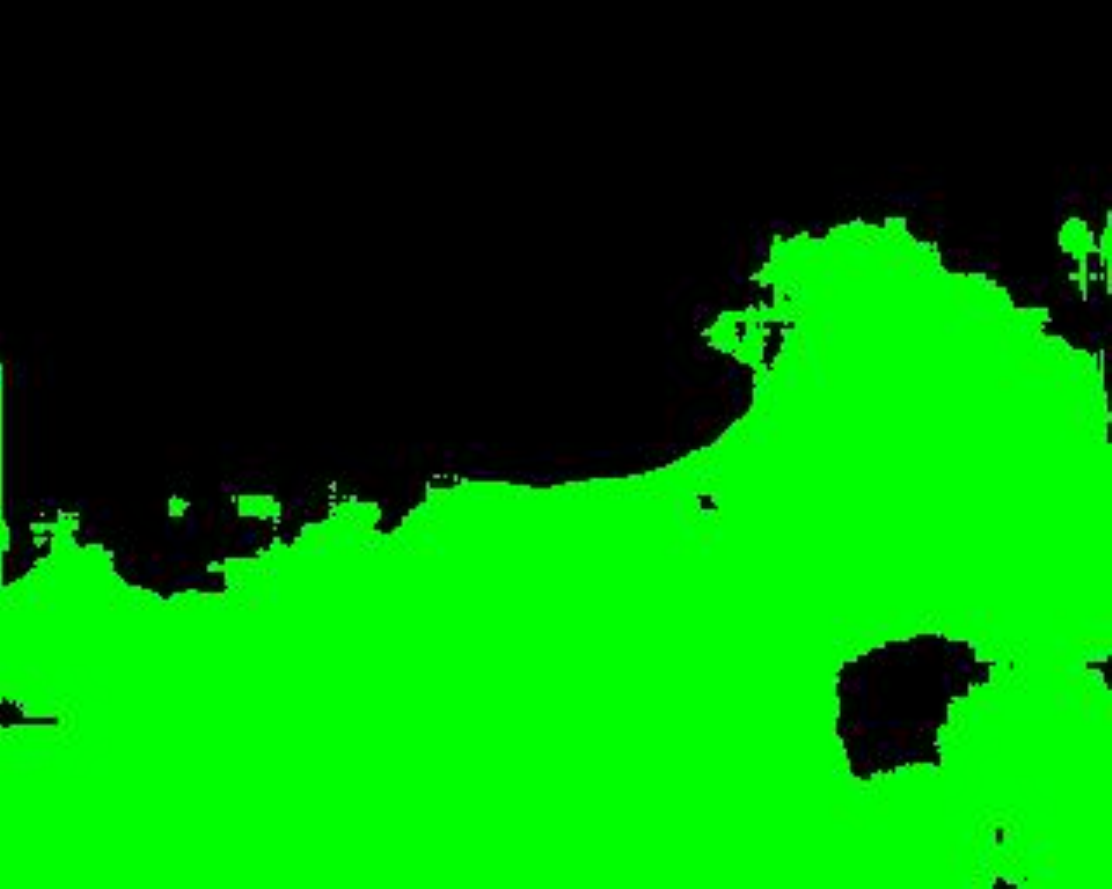}\\

{ Scene} & { Degraded} & { Retinex-based \cite{retinex-based} } & { Fusion-based \cite{ancuti1} } & { GDCP \cite{gdcp} } & { Haze Lines \cite{berman_pami_20} } & { Deep SESR \cite{sesr} } & { Deep WaveNet} & { Ground Truth}\\

%
%

%

\includegraphics[width=3cm, height=2cm]{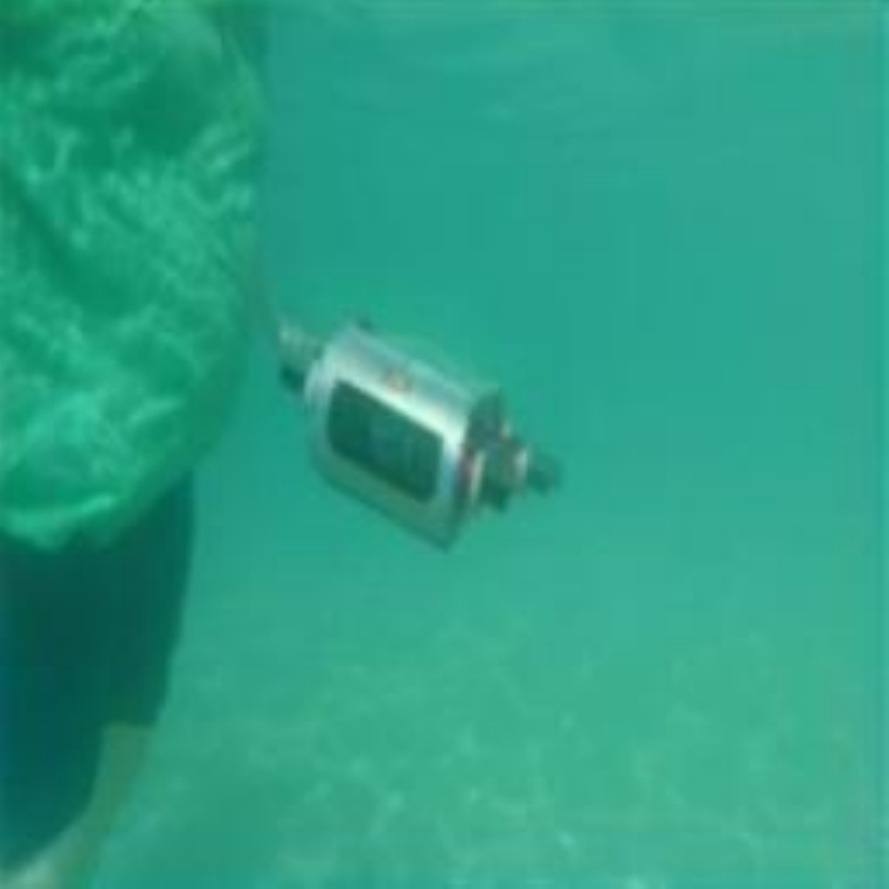}&
\includegraphics[width=3cm, height=2cm]{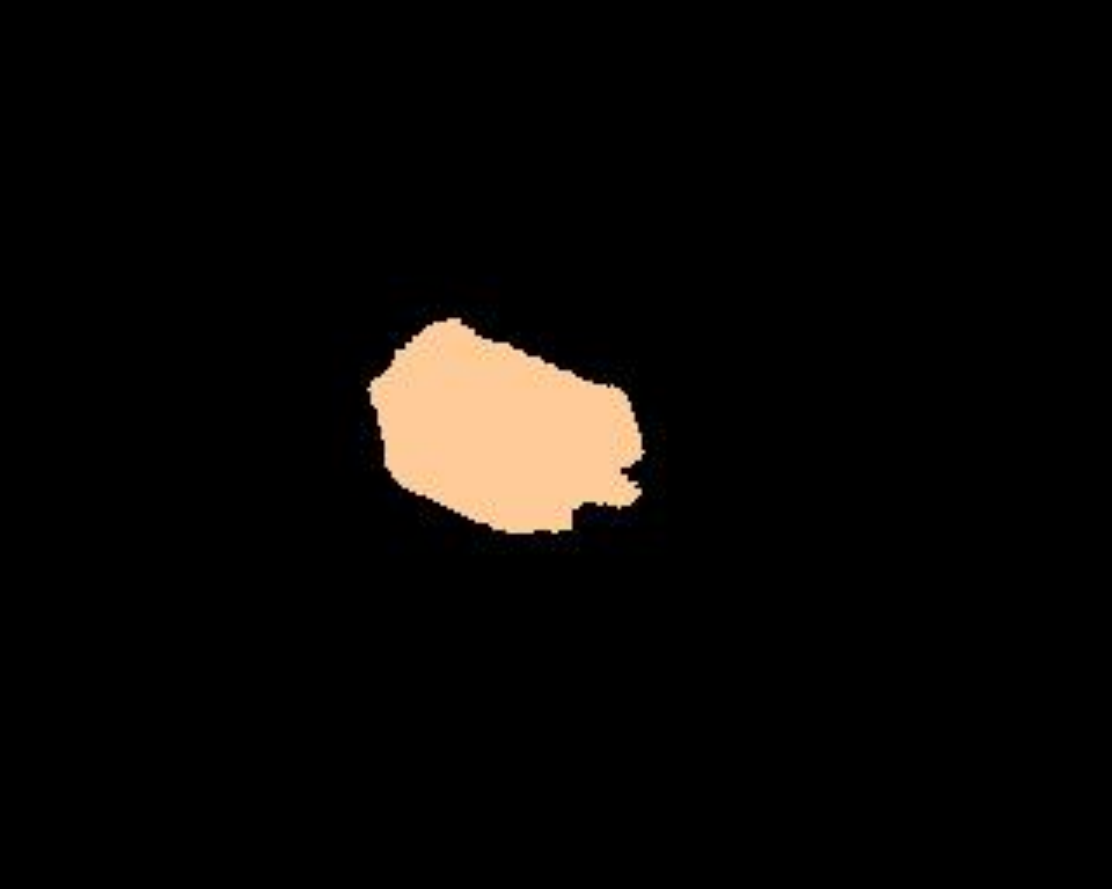}&
\includegraphics[width=3cm, height=2cm]{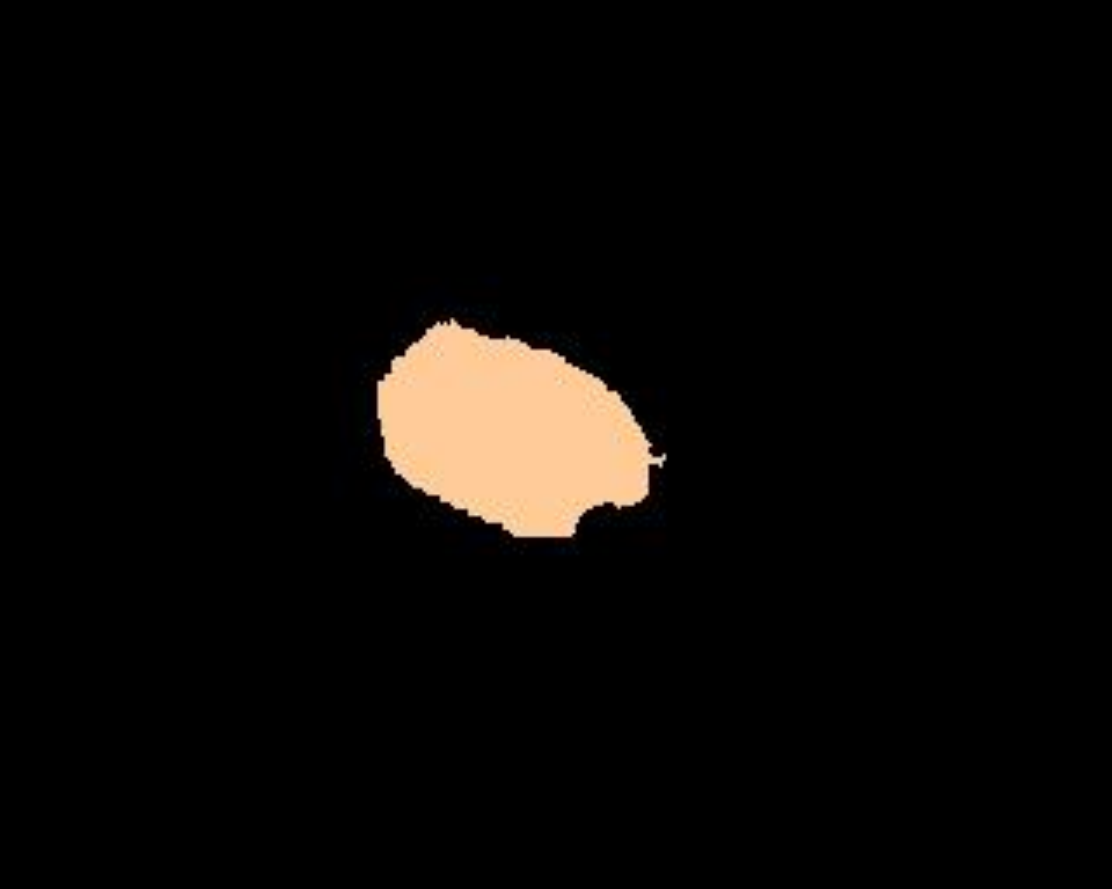}&
\includegraphics[width=3cm, height=2cm]{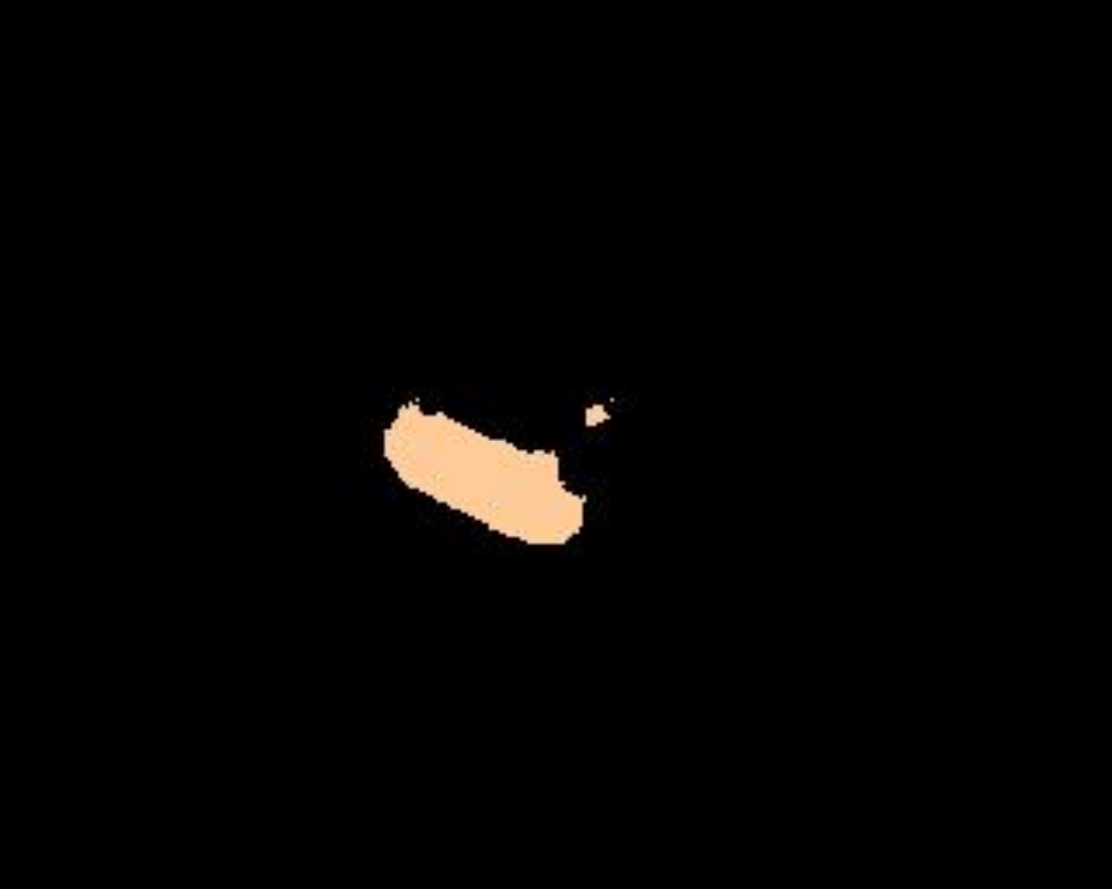}&
\includegraphics[width=3cm, height=2cm]{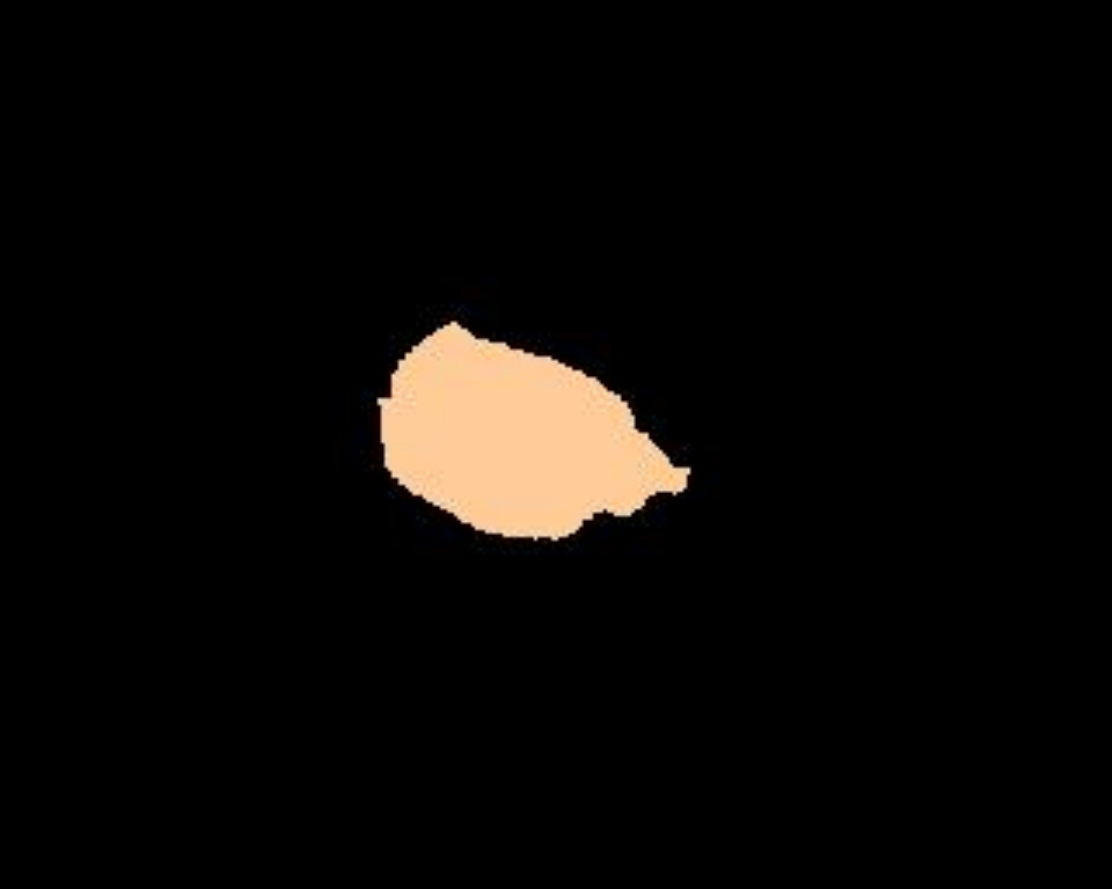}&
\includegraphics[width=3cm, height=2cm]{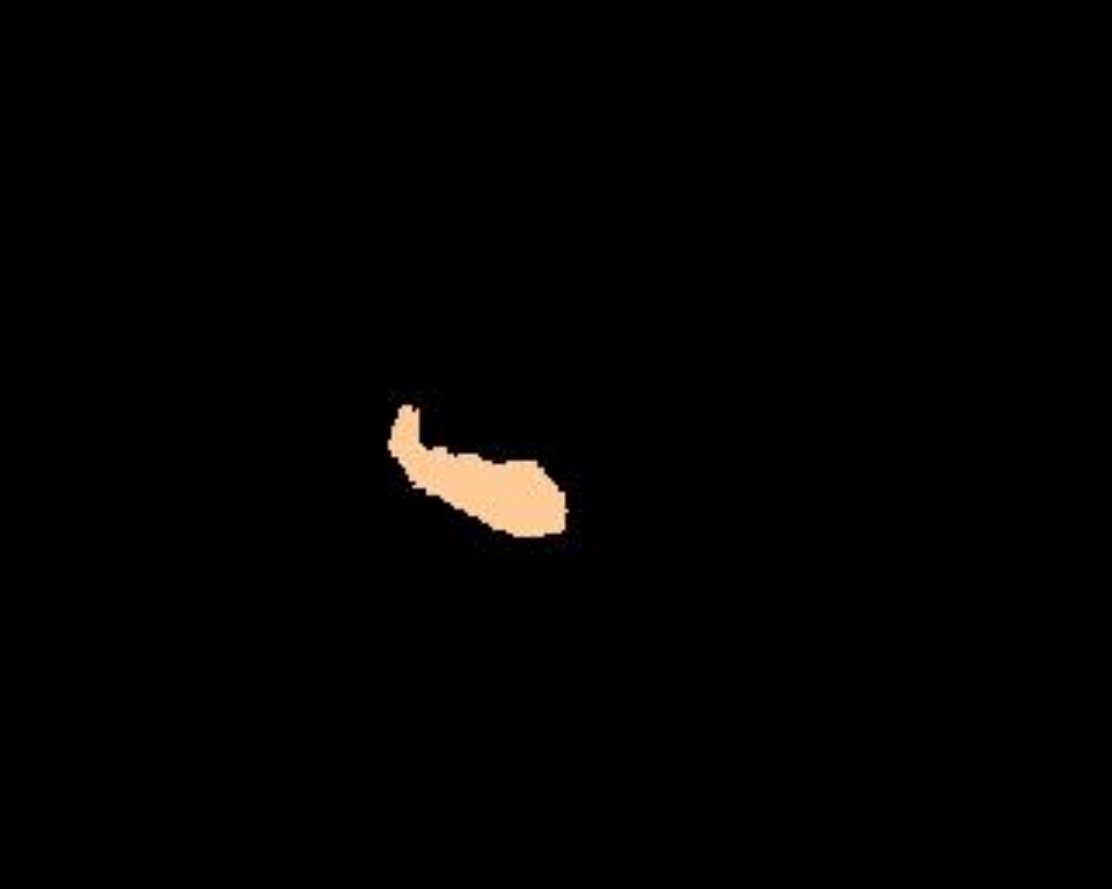}&
\includegraphics[width=3cm, height=2cm]{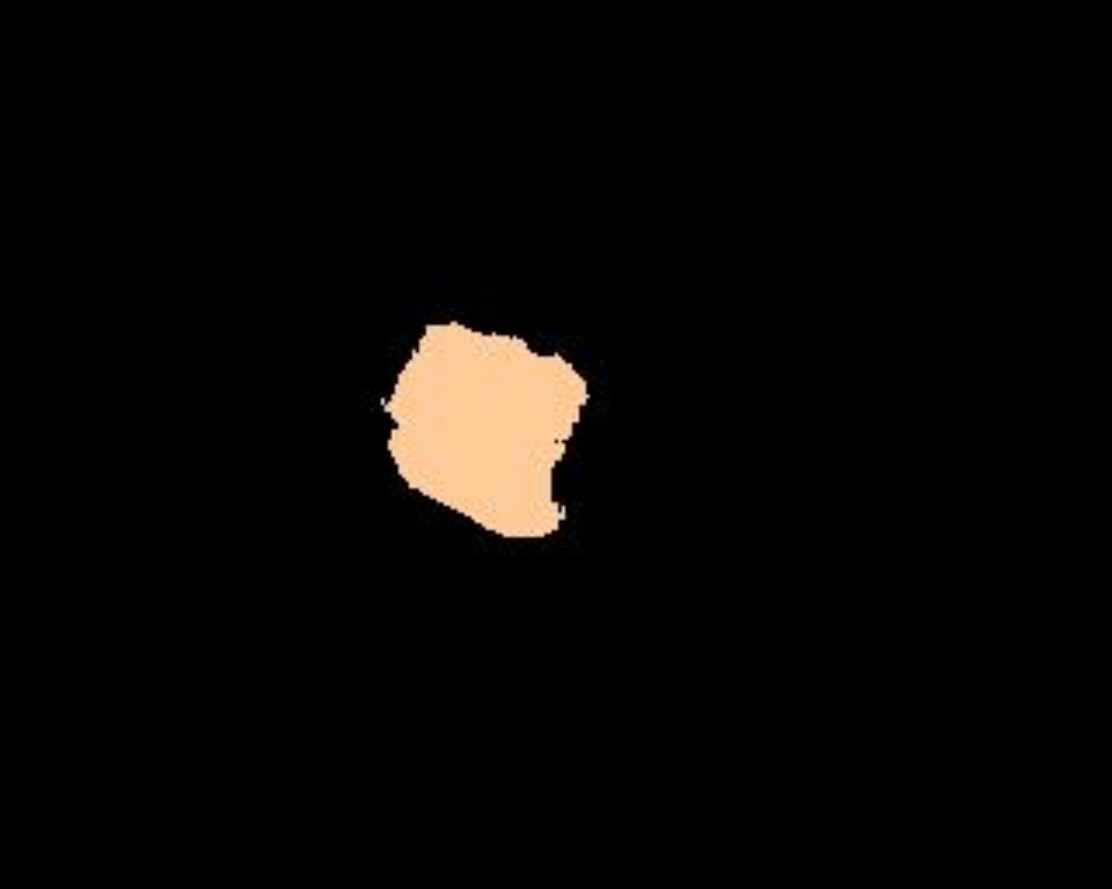}&
\includegraphics[width=3cm, height=2cm]{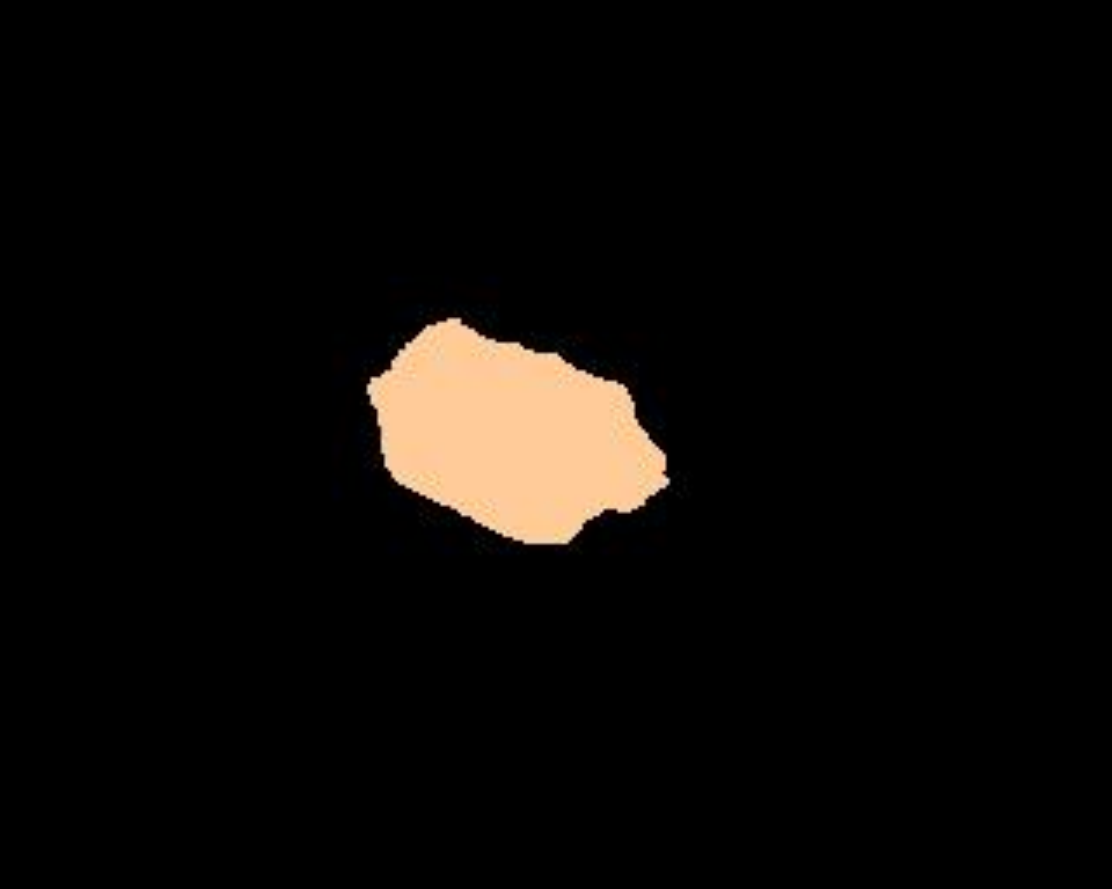}&
\includegraphics[width=3cm, height=2cm]{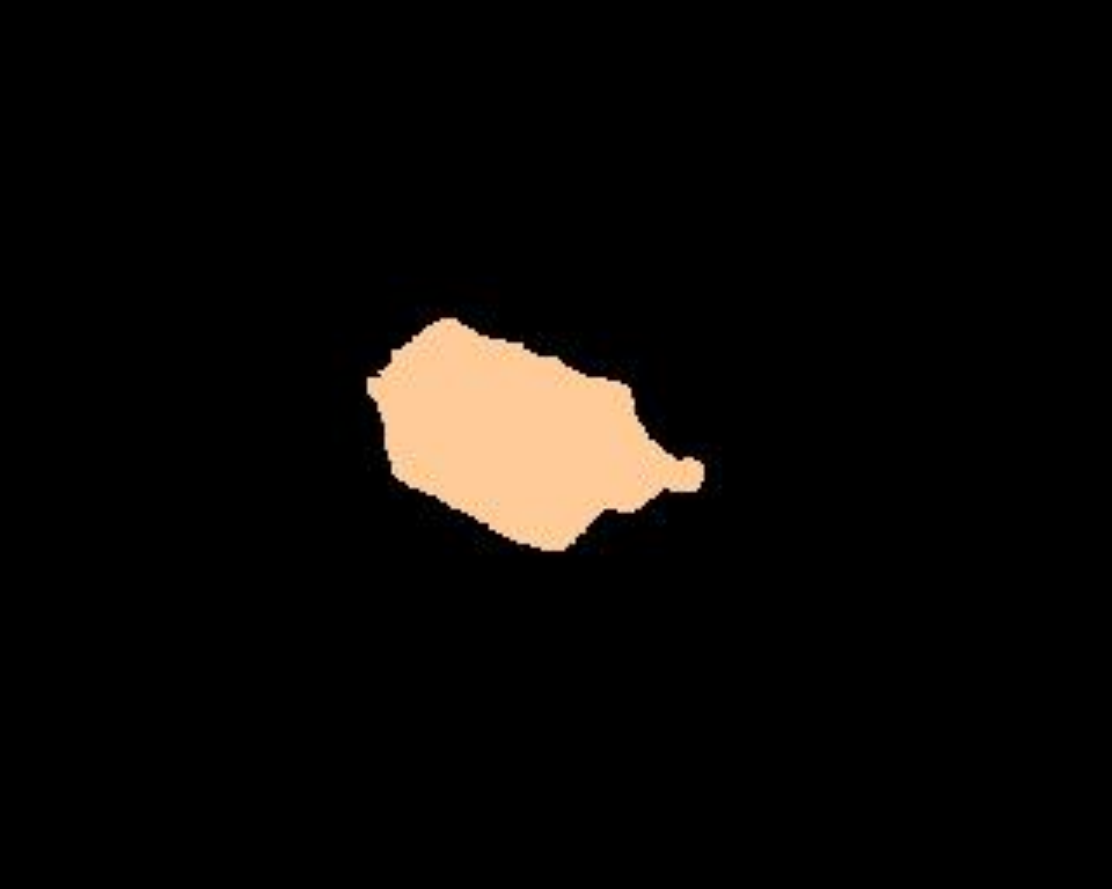}\\

\includegraphics[width=3cm, height=2cm]{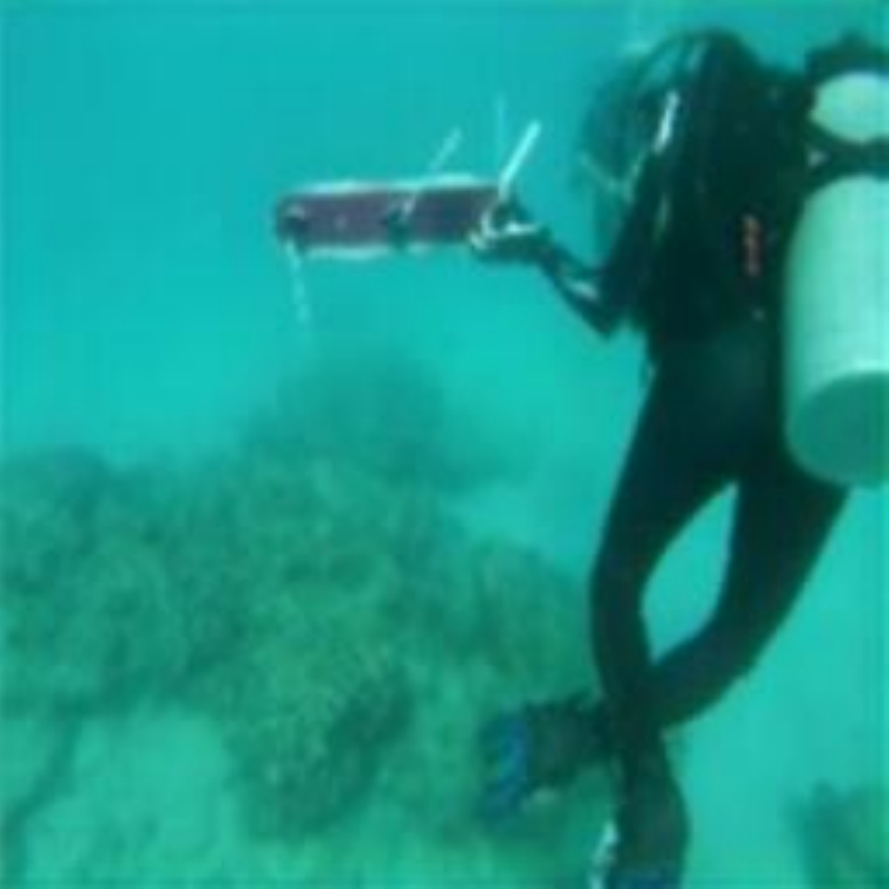}&
\includegraphics[width=3cm, height=2cm]{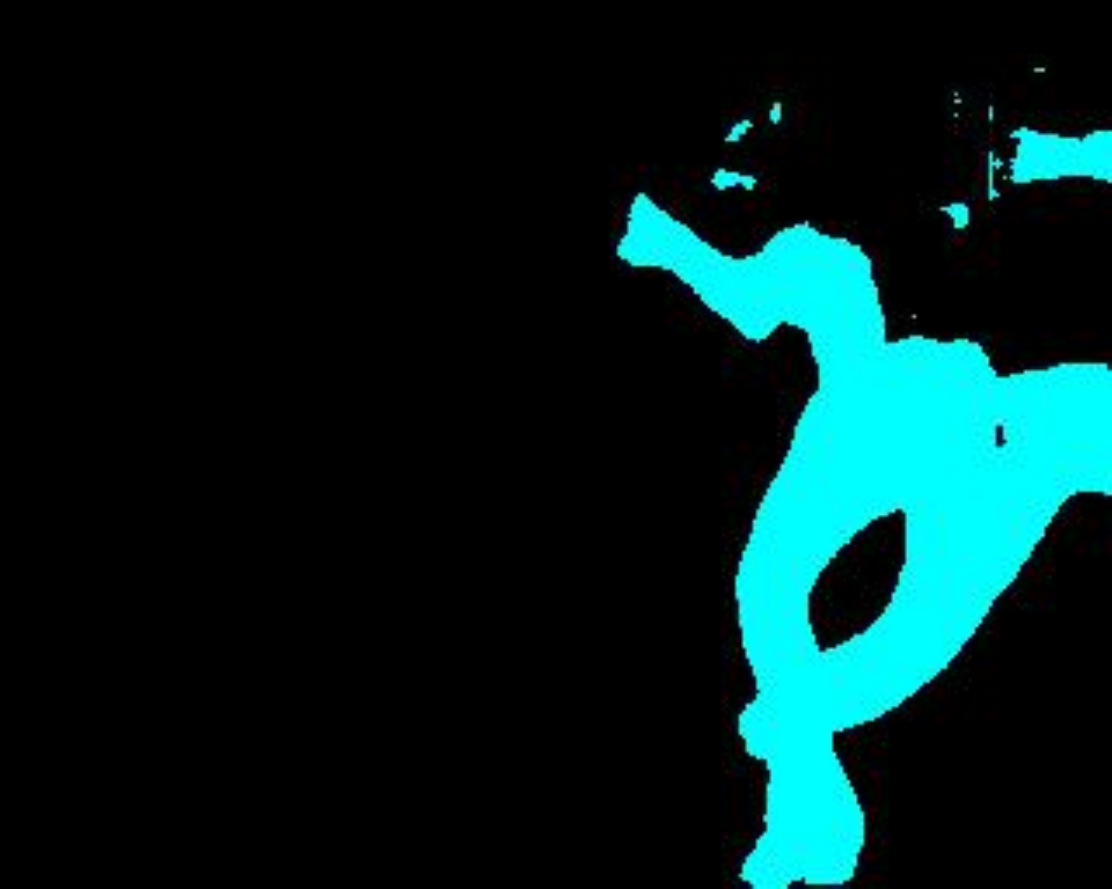}&
\includegraphics[width=3cm, height=2cm]{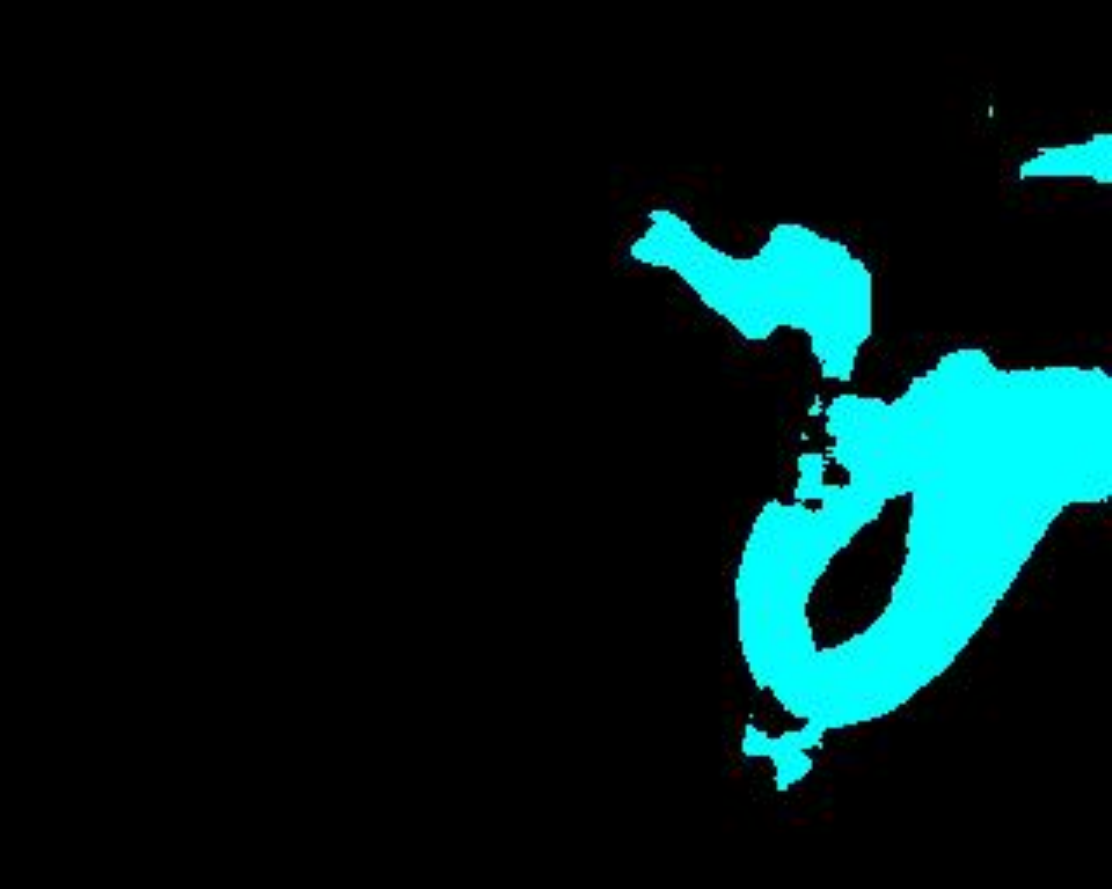}&
\includegraphics[width=3cm, height=2cm]{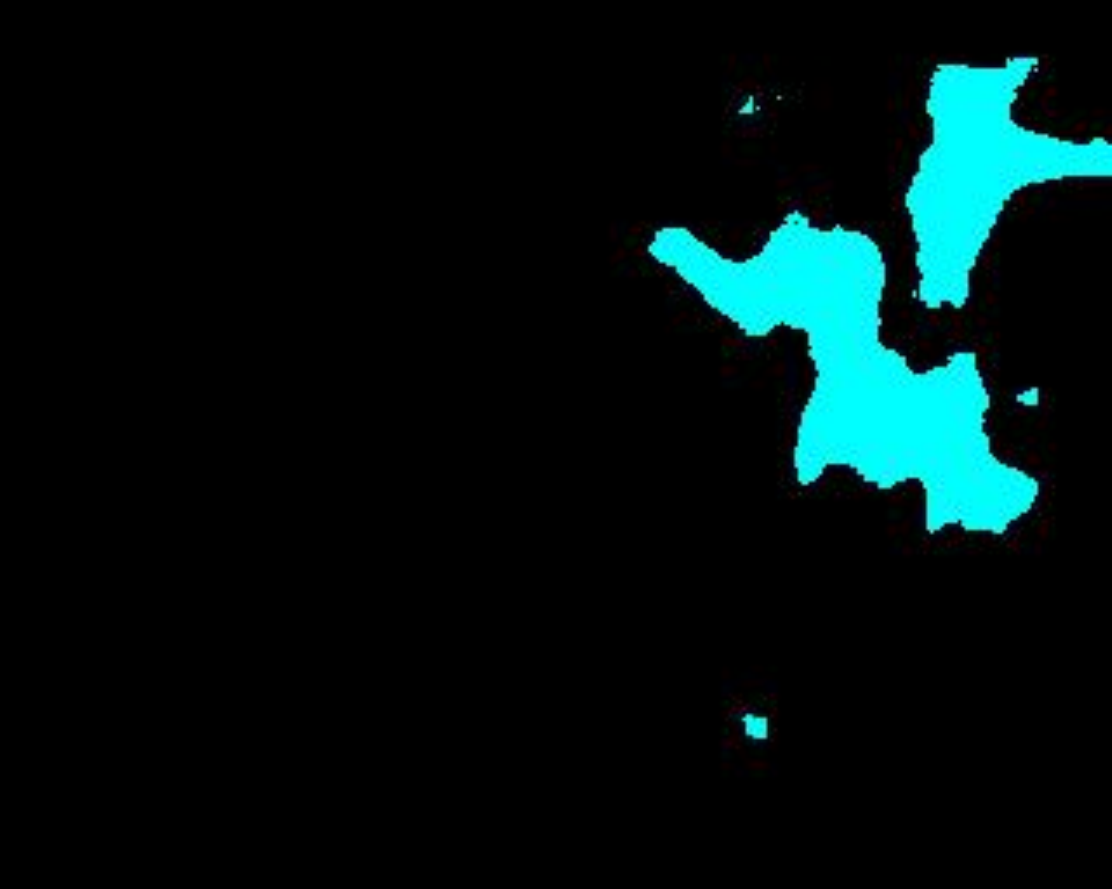}&
\includegraphics[width=3cm, height=2cm]{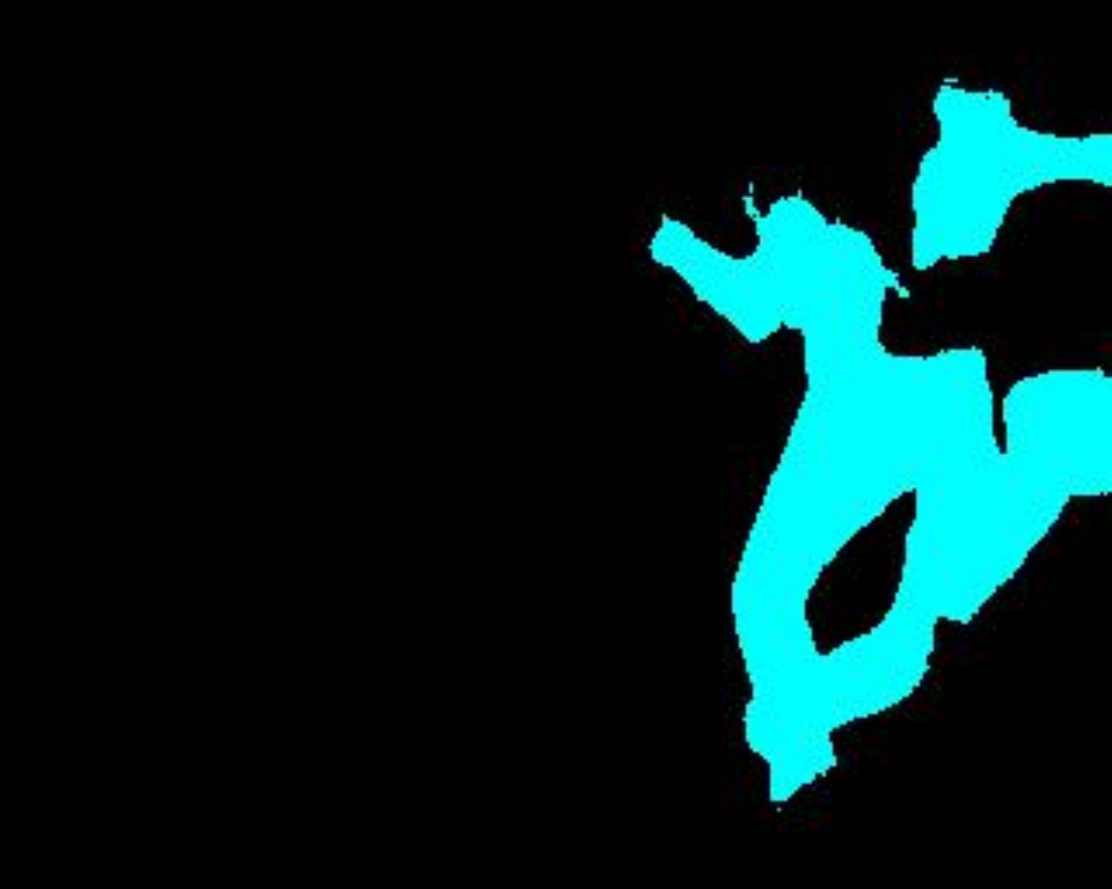}&
\includegraphics[width=3cm, height=2cm]{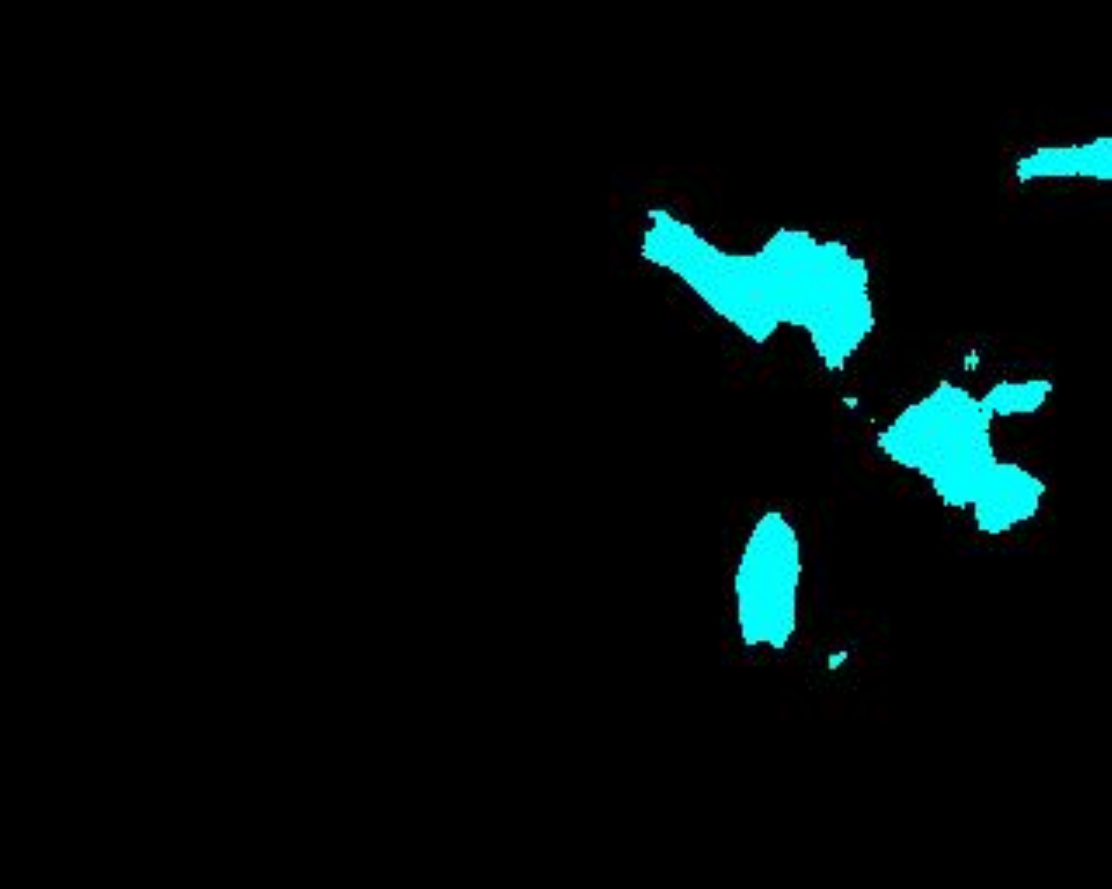}&
\includegraphics[width=3cm, height=2cm]{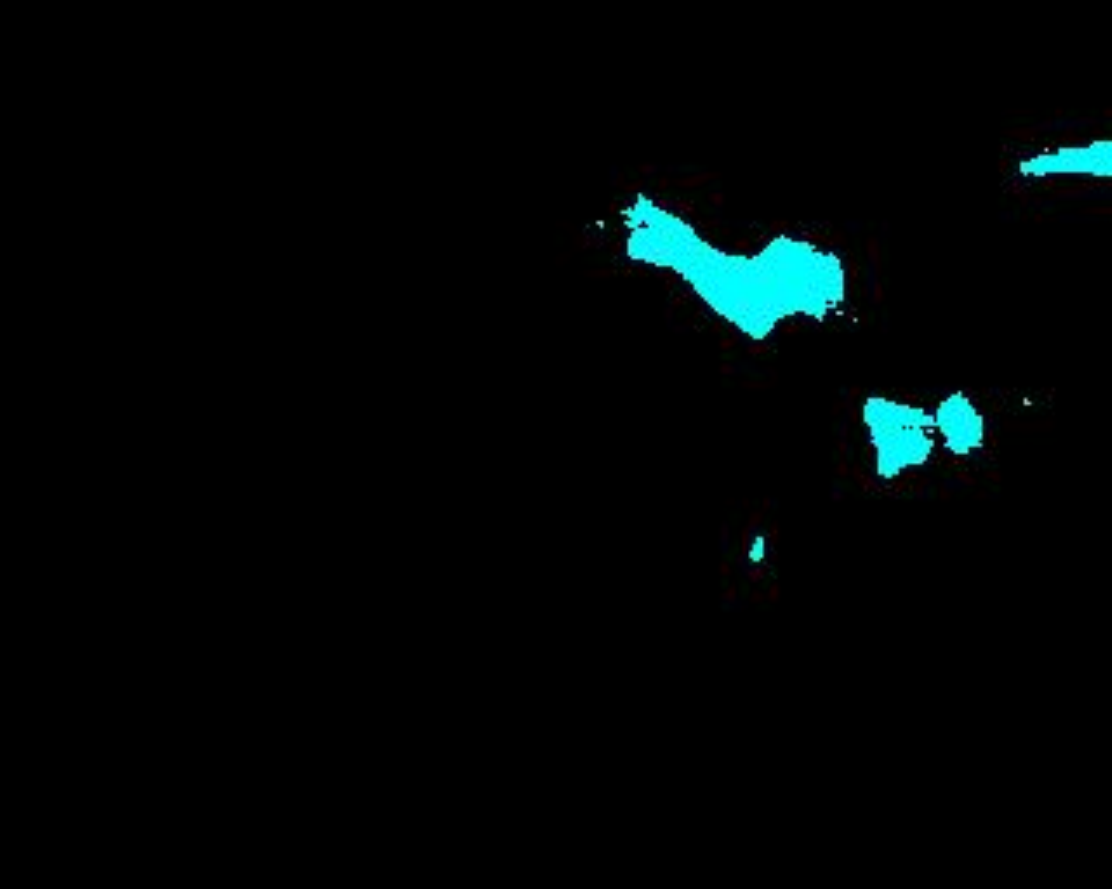}&
\includegraphics[width=3cm, height=2cm]{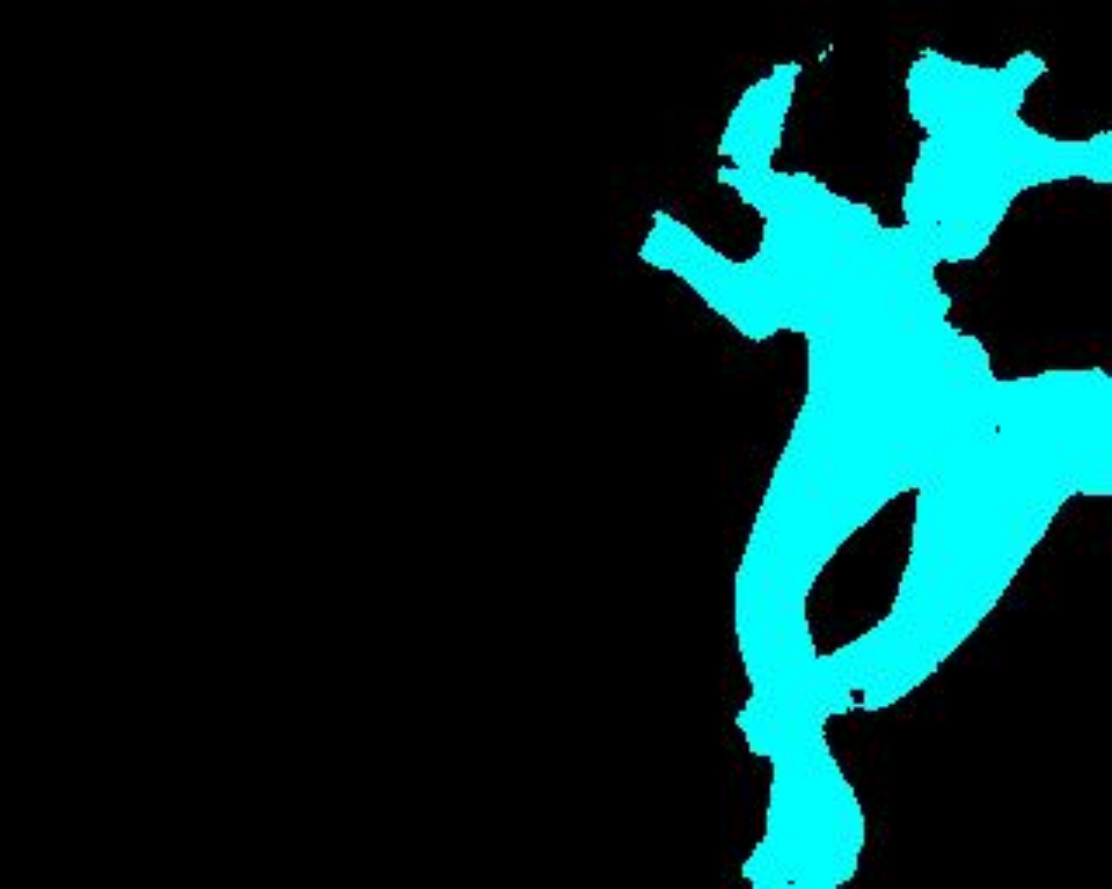}&
\includegraphics[width=3cm, height=2cm]{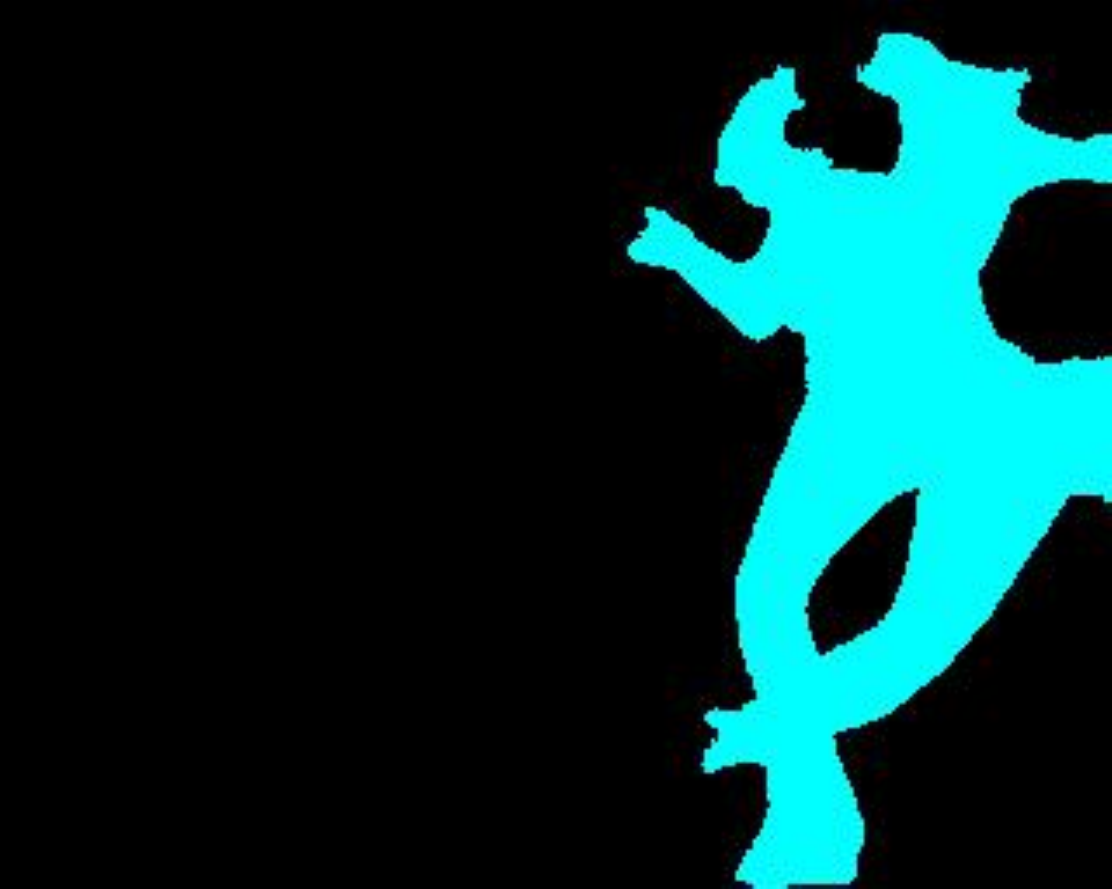}\\

\includegraphics[width=3cm, height=2cm]{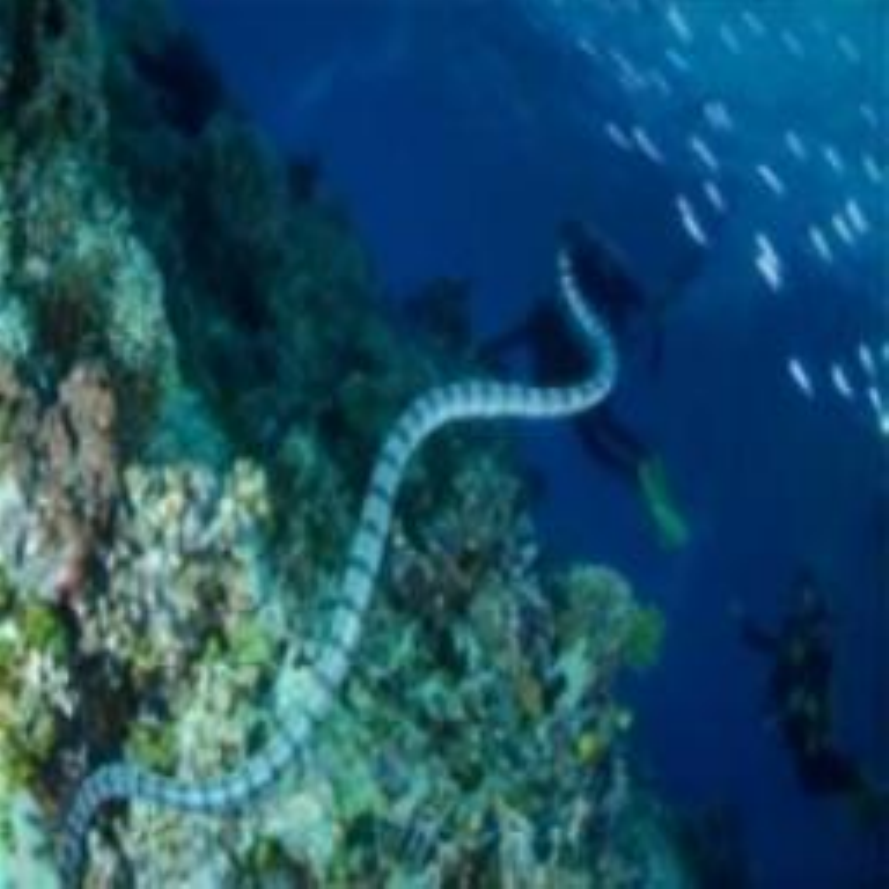}&
\includegraphics[width=3cm, height=2cm]{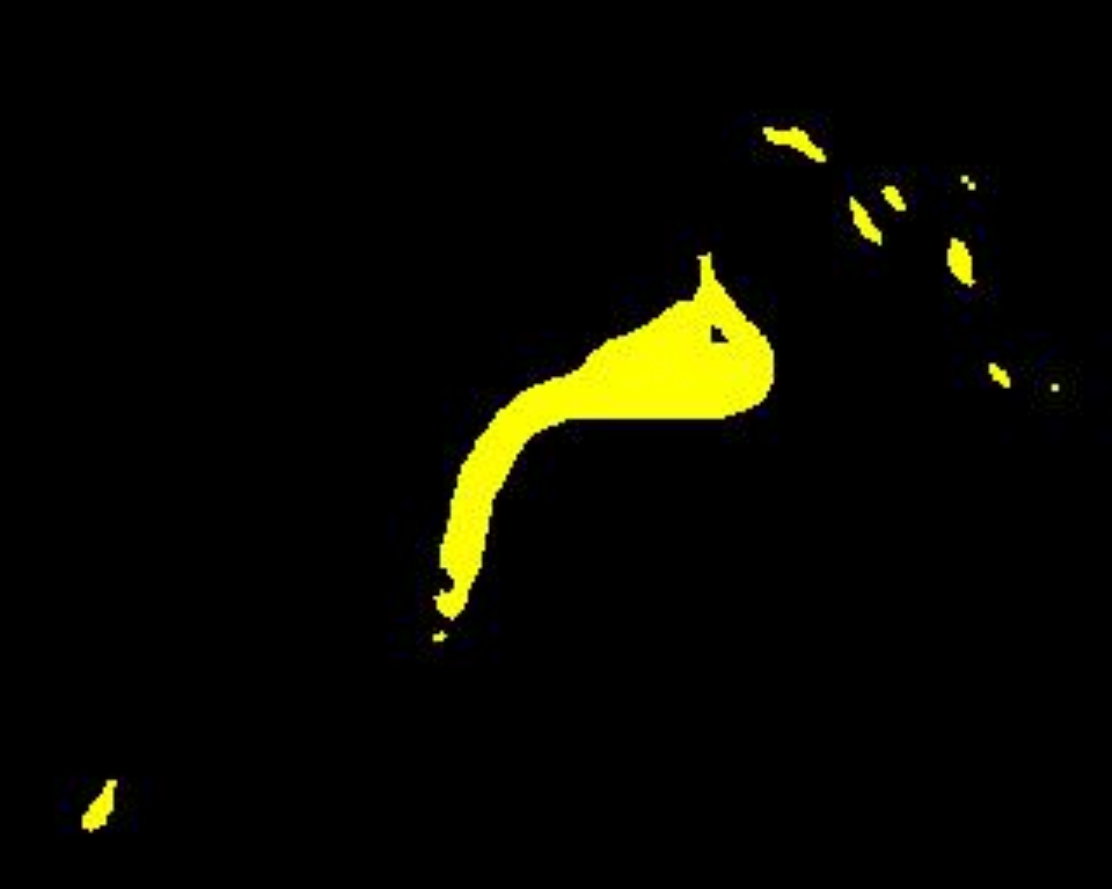}&
\includegraphics[width=3cm, height=2cm]{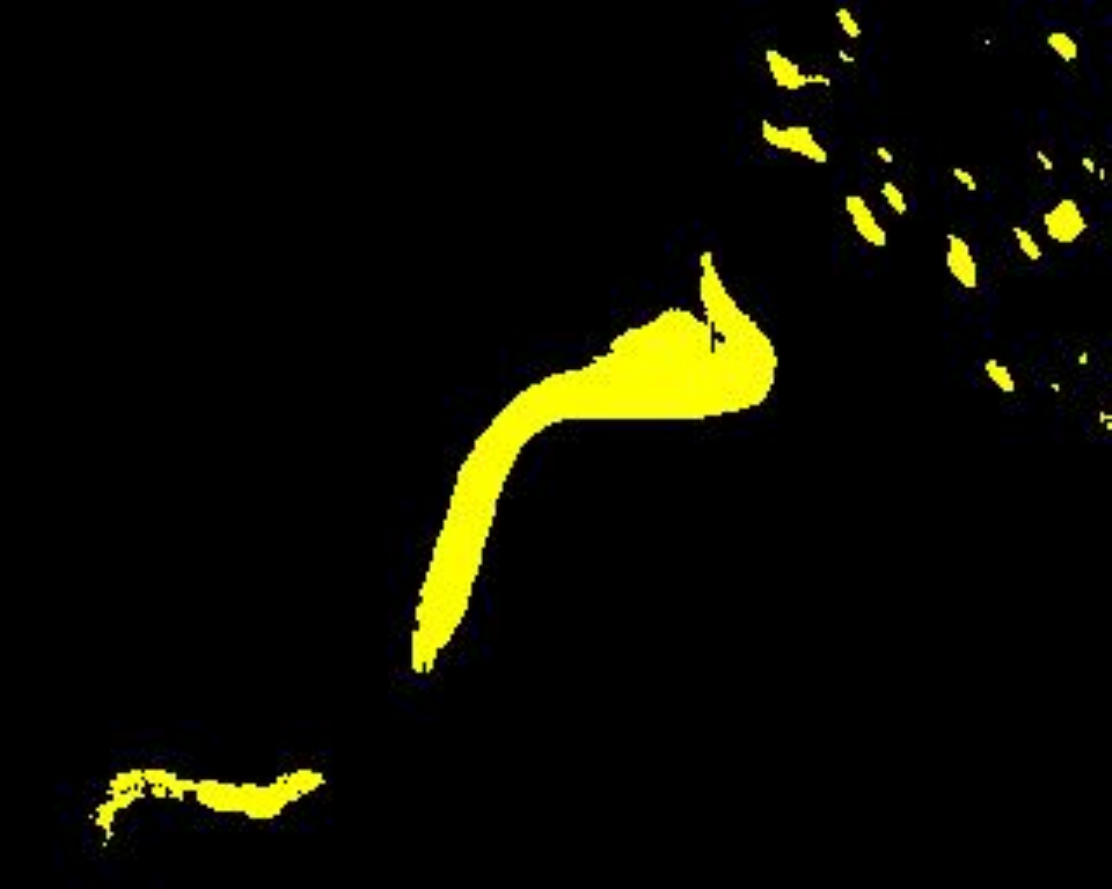}&
\includegraphics[width=3cm, height=2cm]{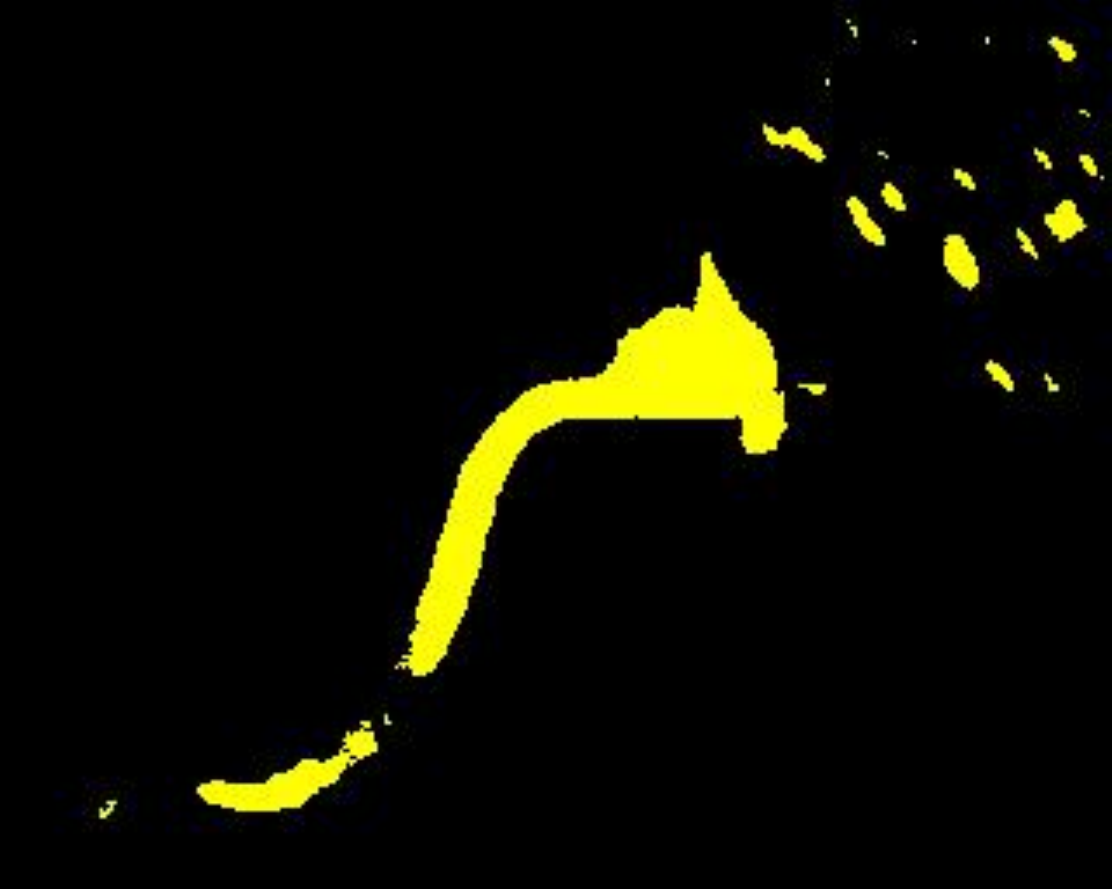}&
\includegraphics[width=3cm, height=2cm]{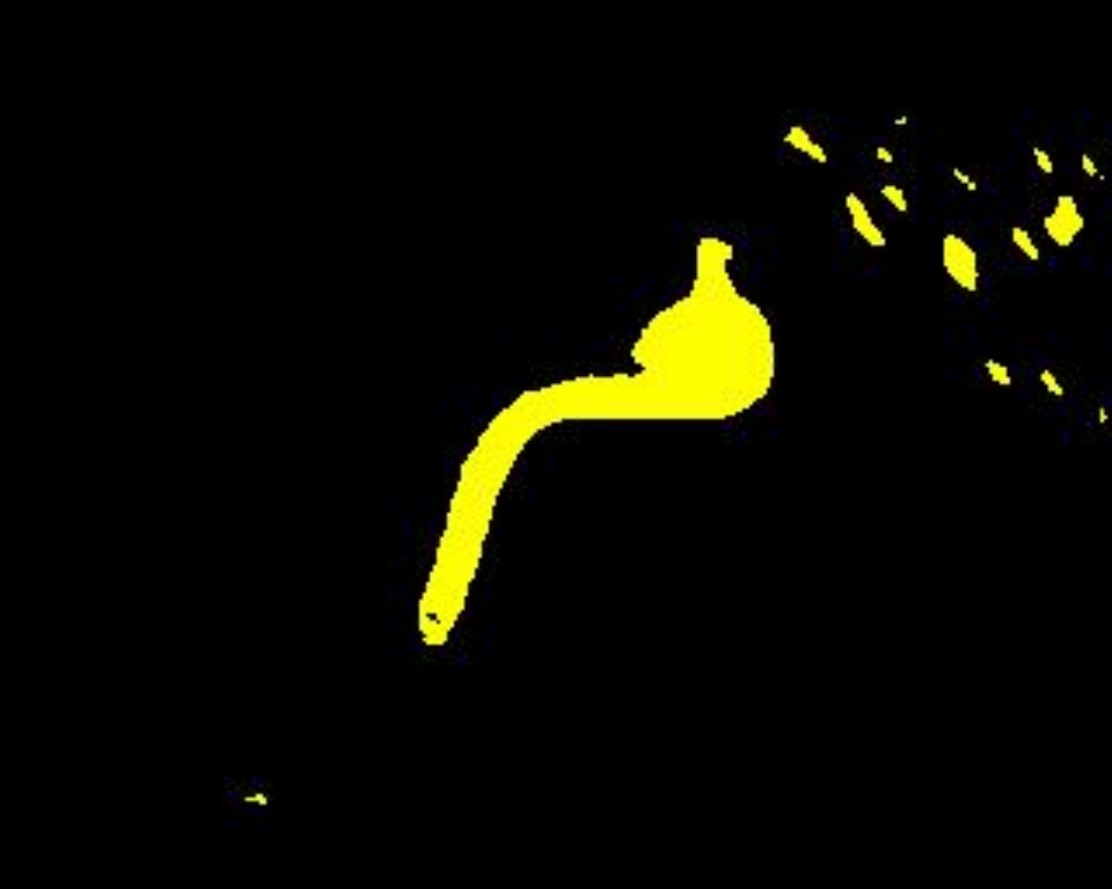}&
\includegraphics[width=3cm, height=2cm]{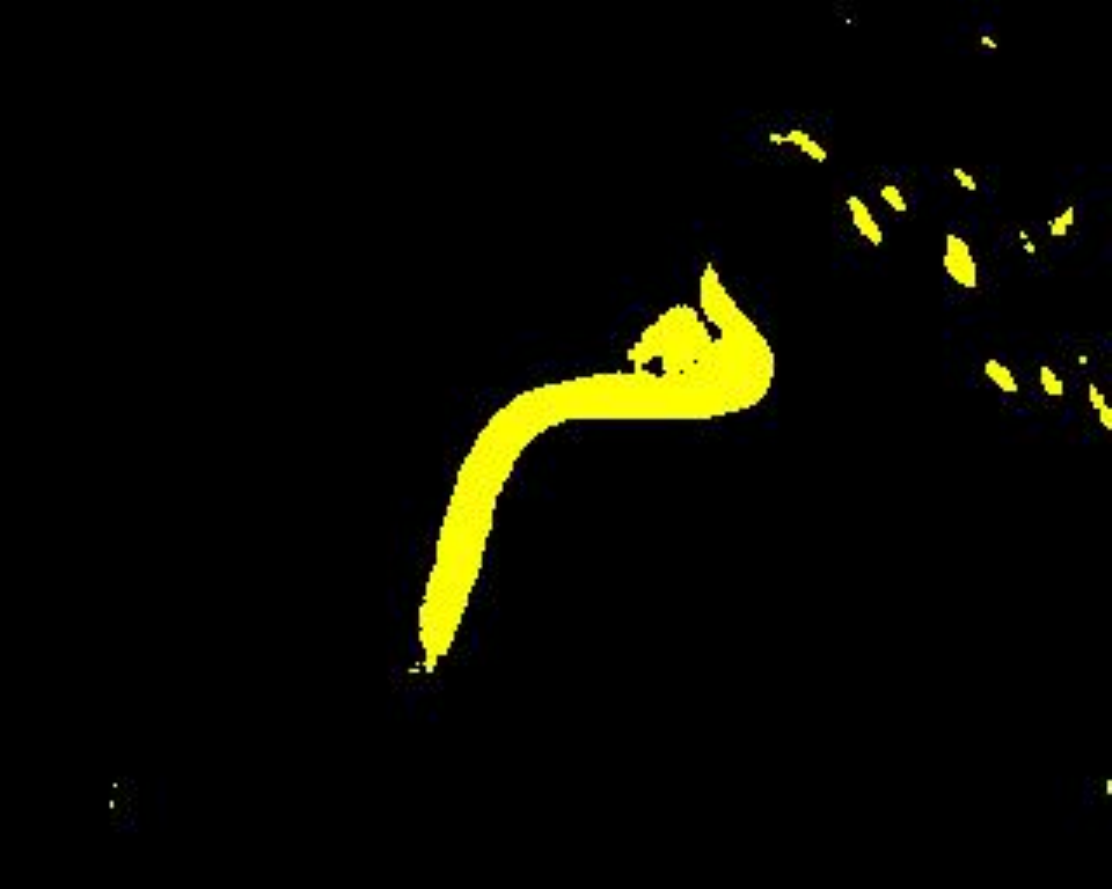}&
\includegraphics[width=3cm, height=2cm]{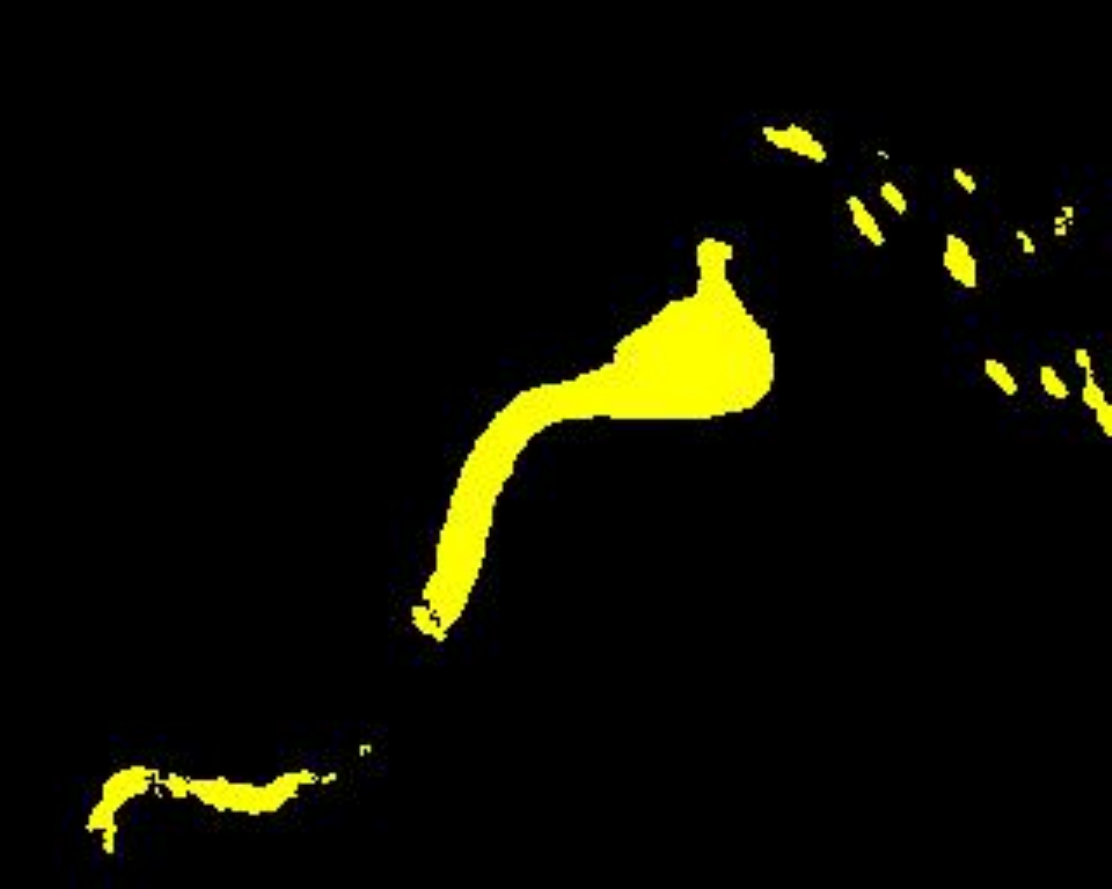}&
\includegraphics[width=3cm, height=2cm]{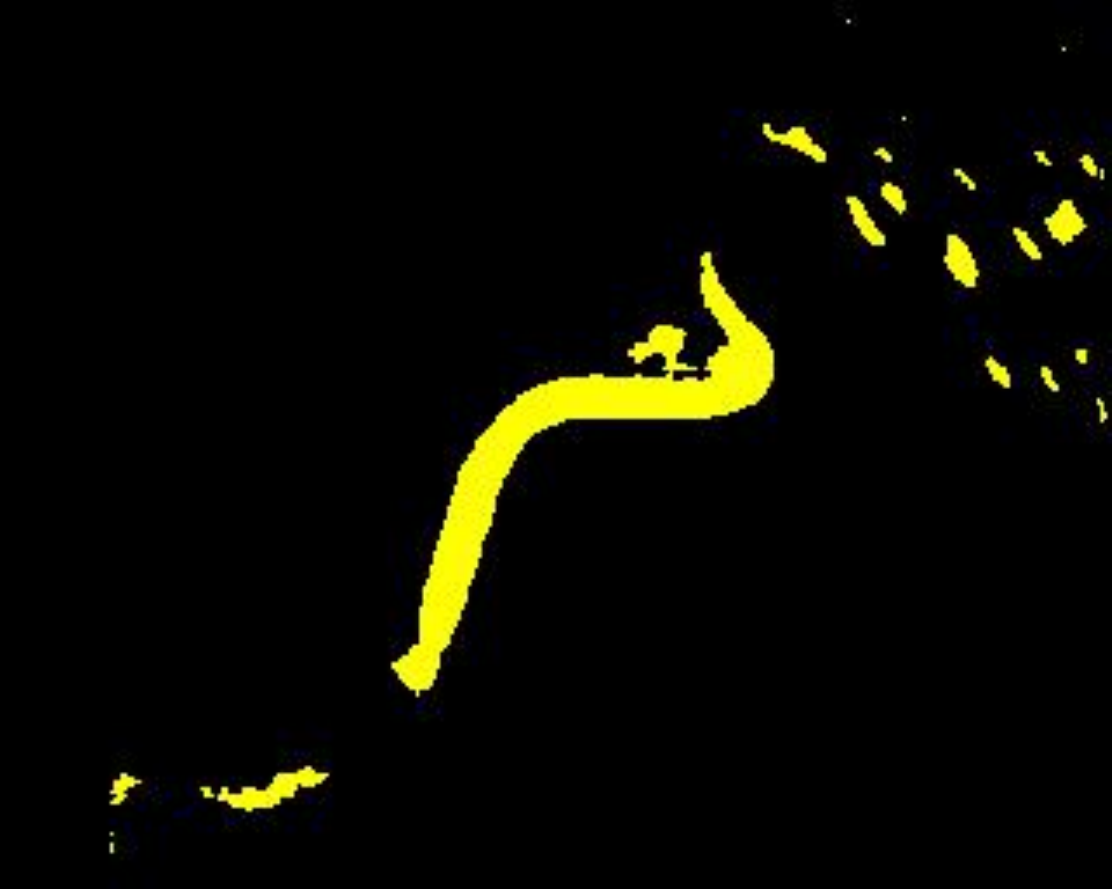}&
\includegraphics[width=3cm, height=2cm]{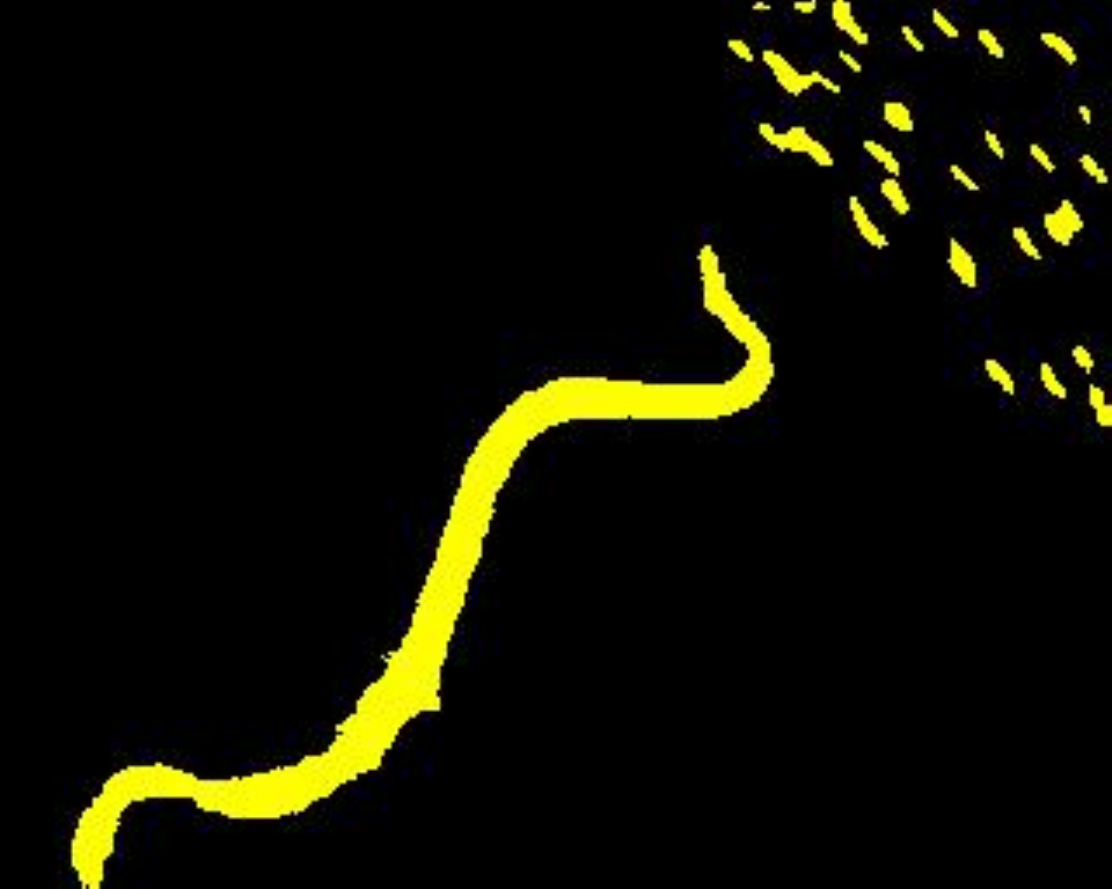}\\

\includegraphics[width=3cm, height=2cm]{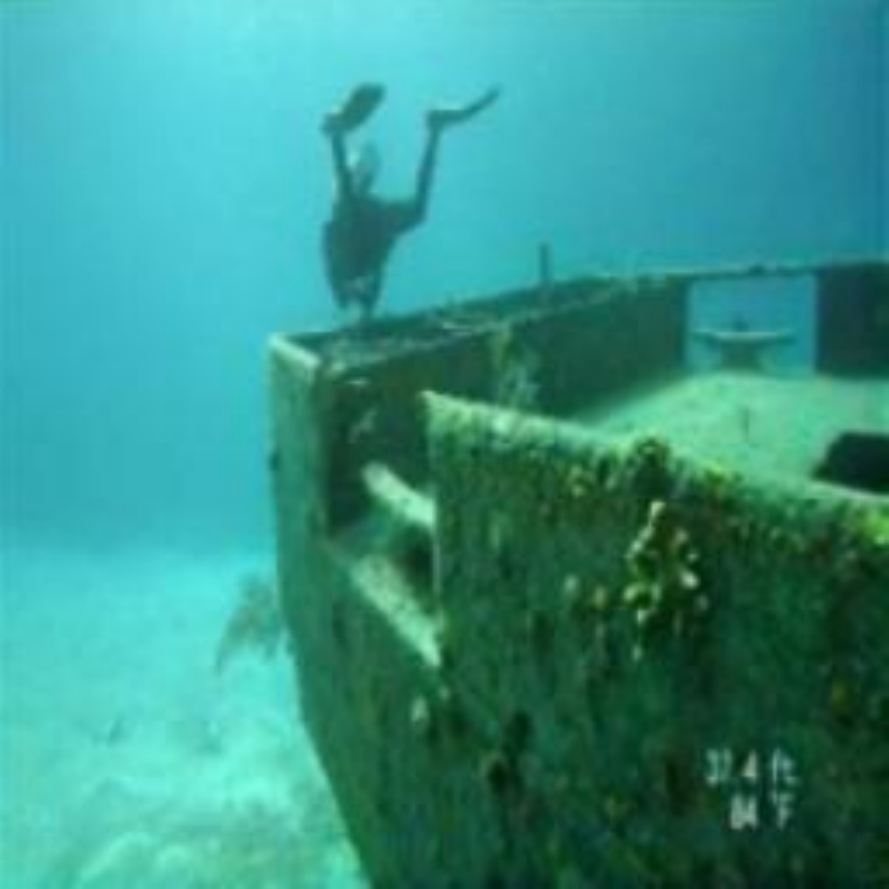}&
\includegraphics[width=3cm, height=2cm]{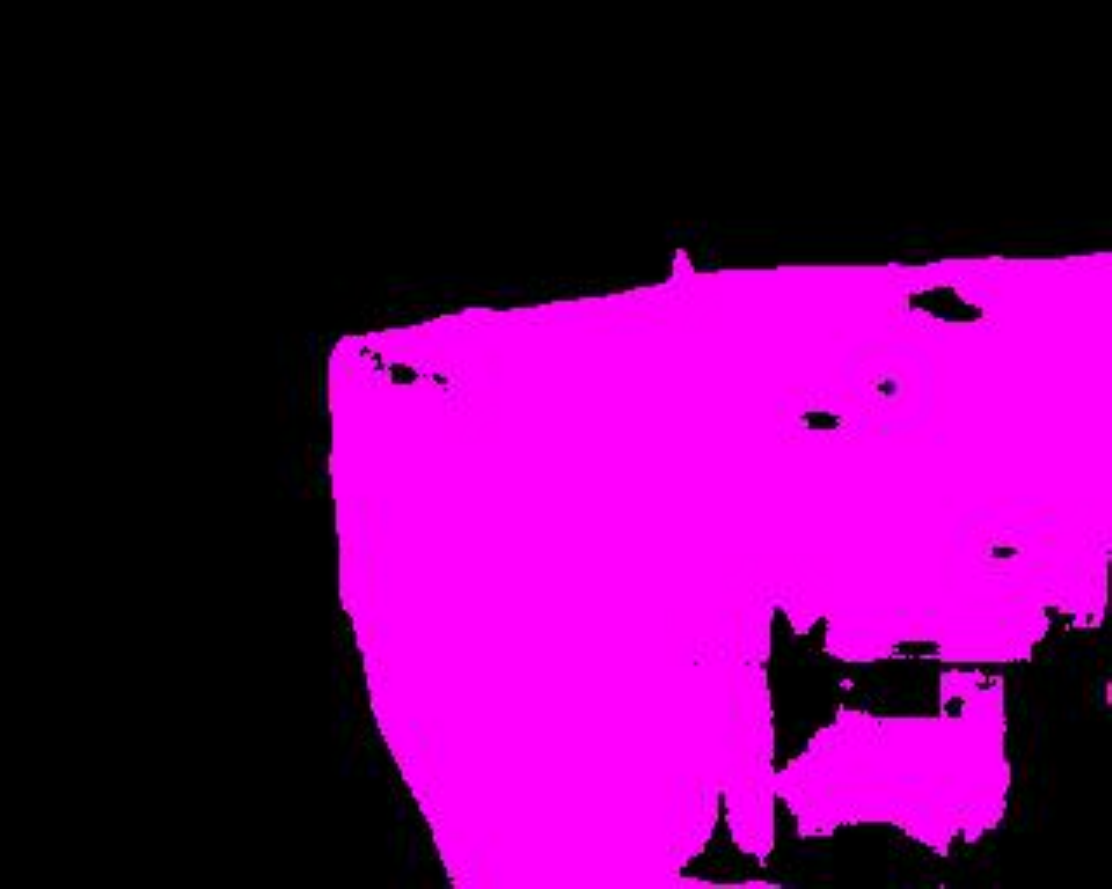}&
\includegraphics[width=3cm, height=2cm]{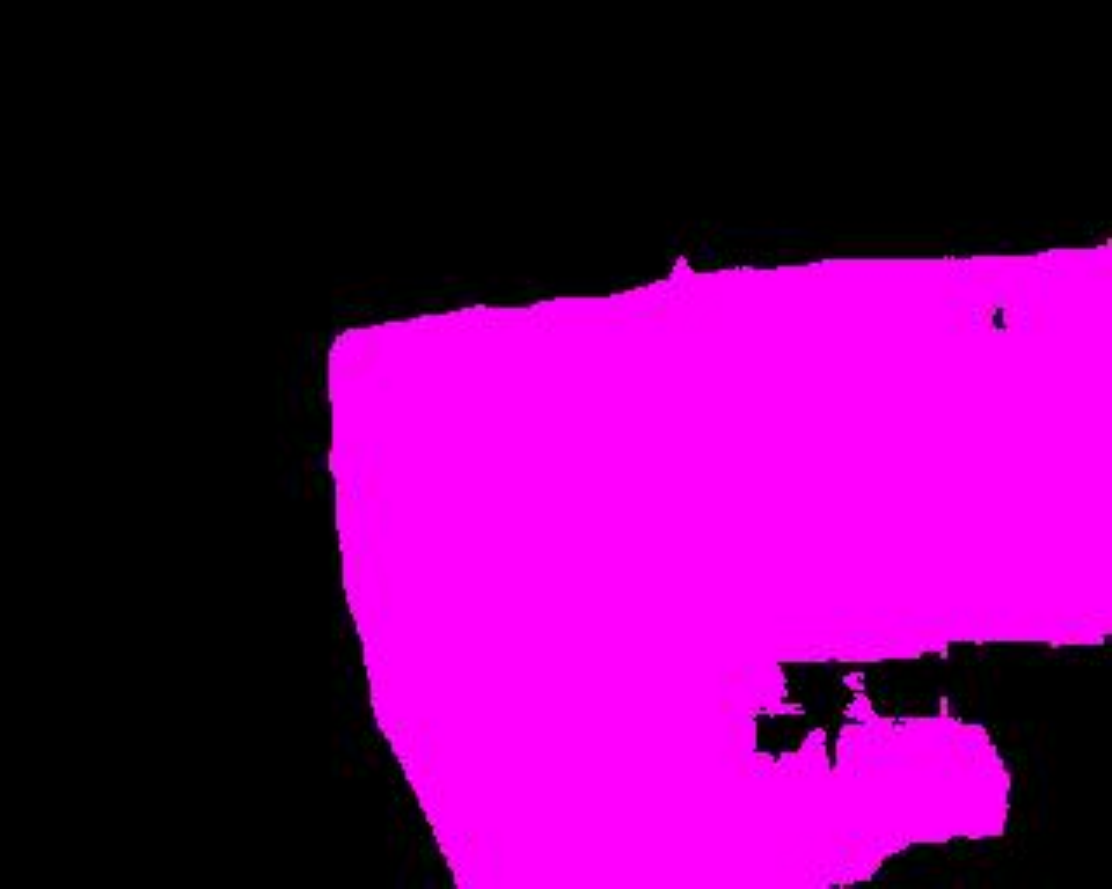}&
\includegraphics[width=3cm, height=2cm]{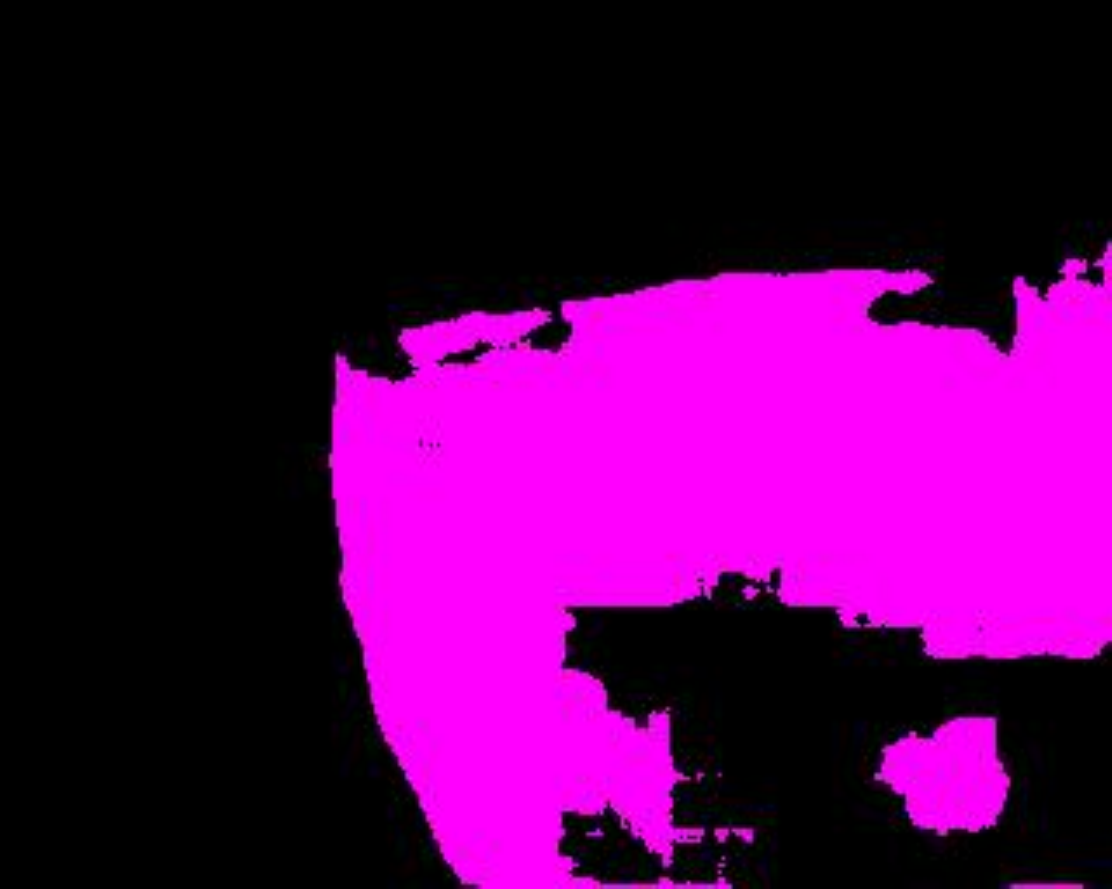}&
\includegraphics[width=3cm, height=2cm]{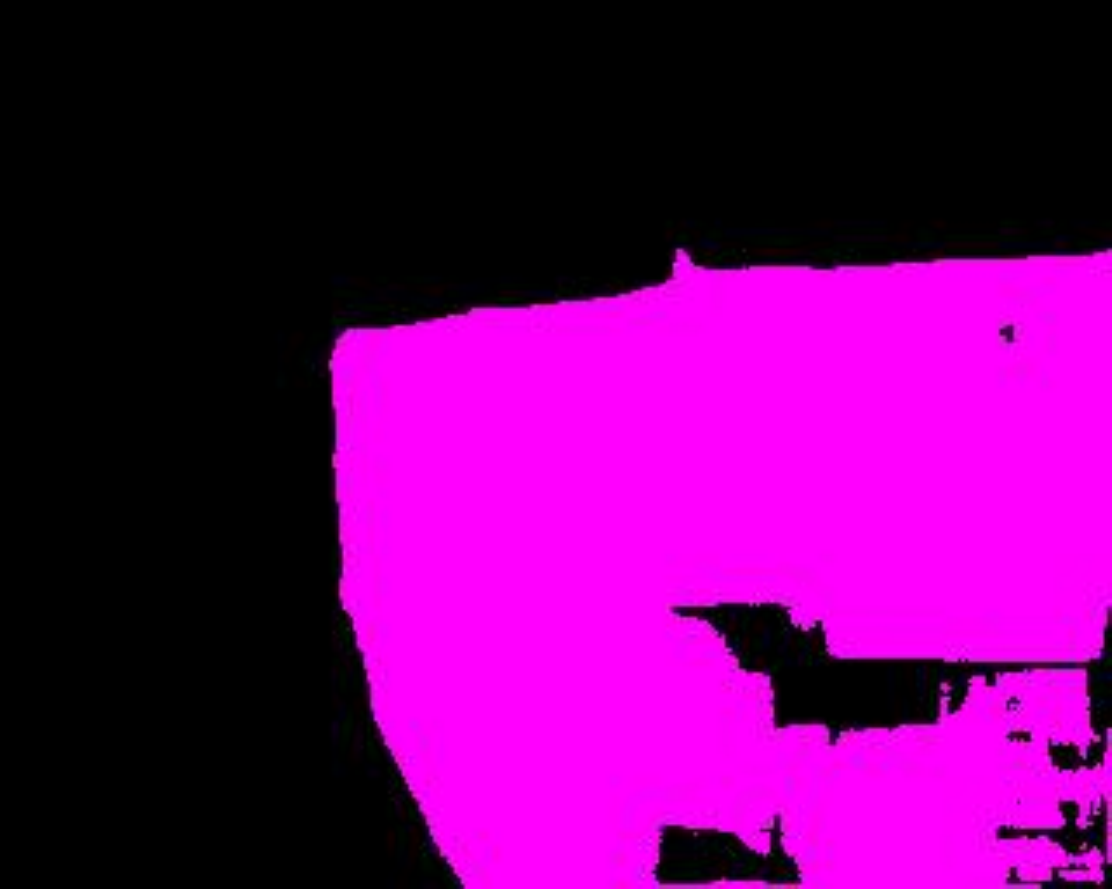}&
\includegraphics[width=3cm, height=2cm]{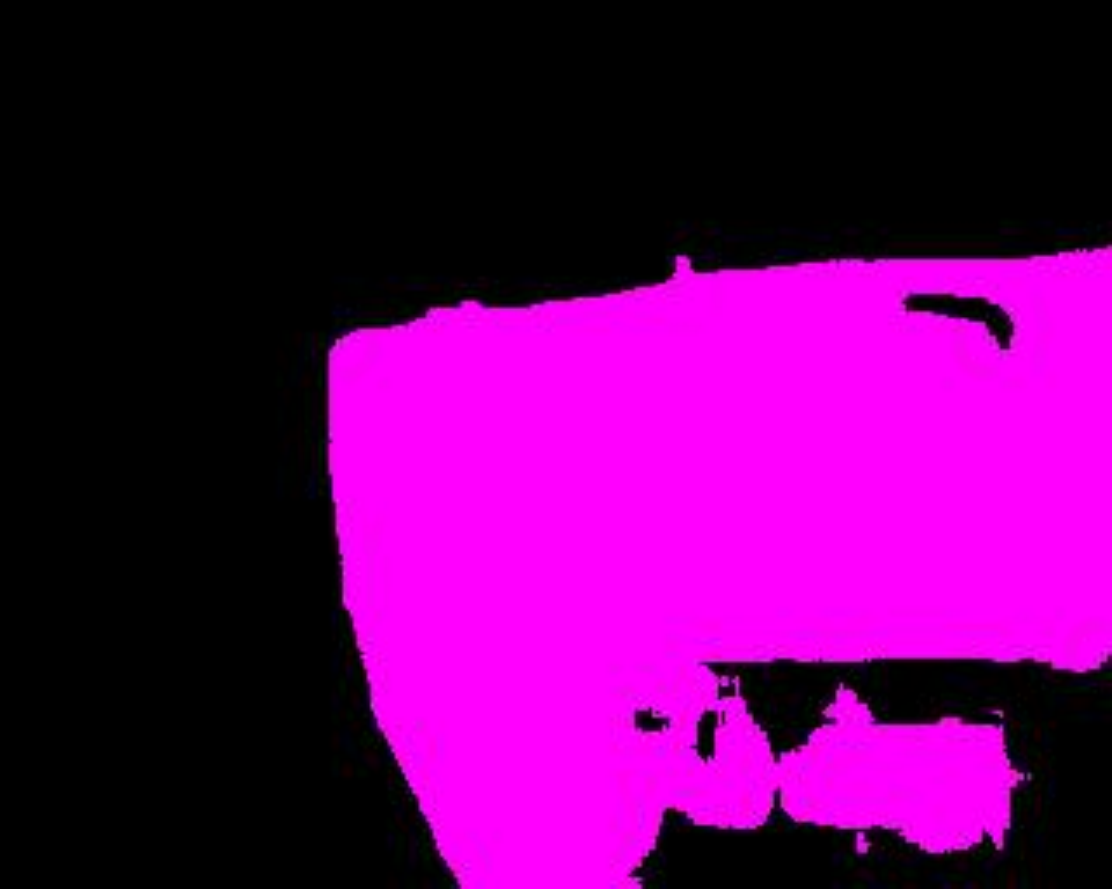}&
\includegraphics[width=3cm, height=2cm]{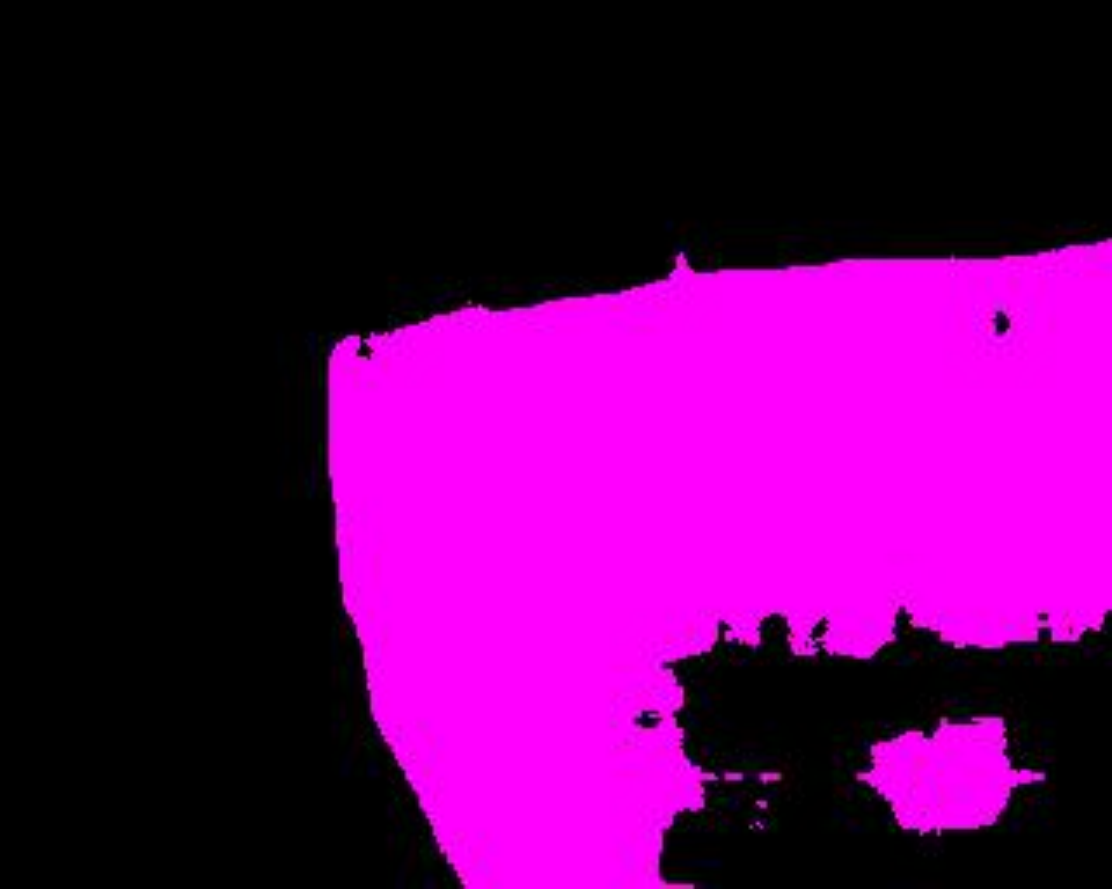}&
\includegraphics[width=3cm, height=2cm]{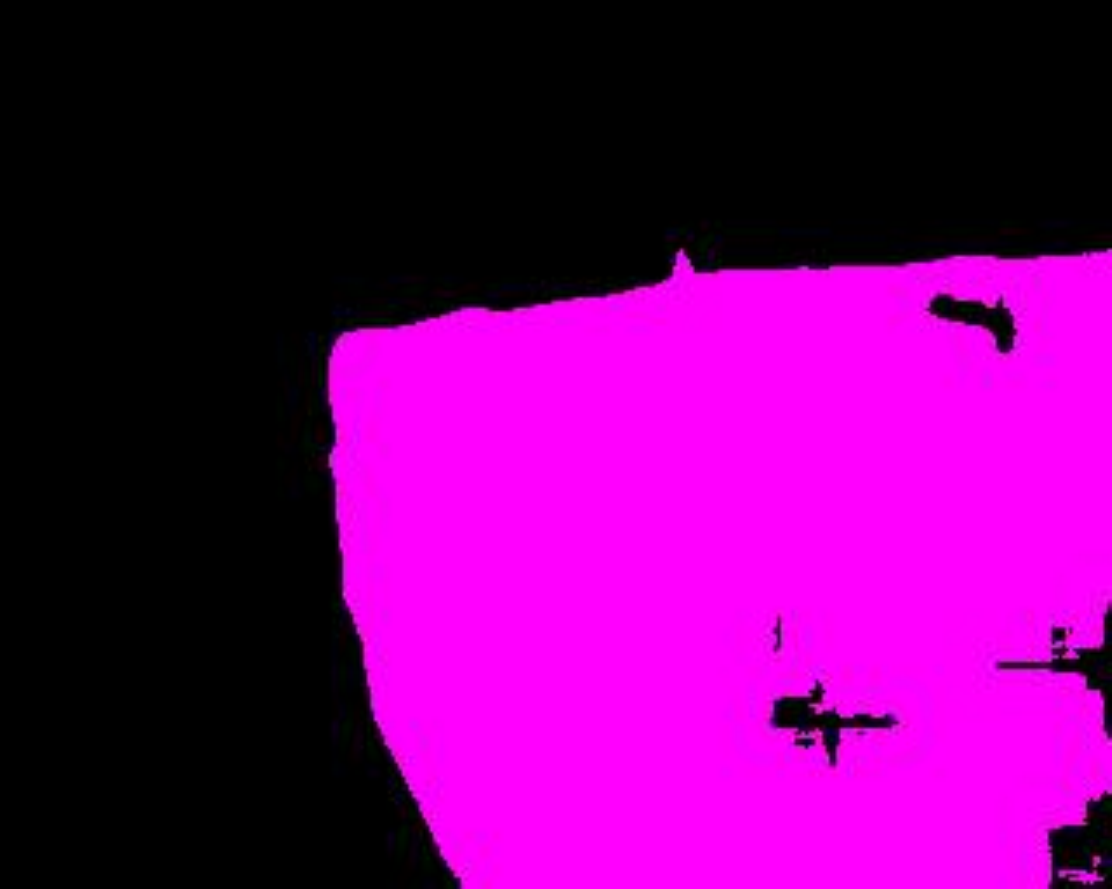}&
\includegraphics[width=3cm, height=2cm]{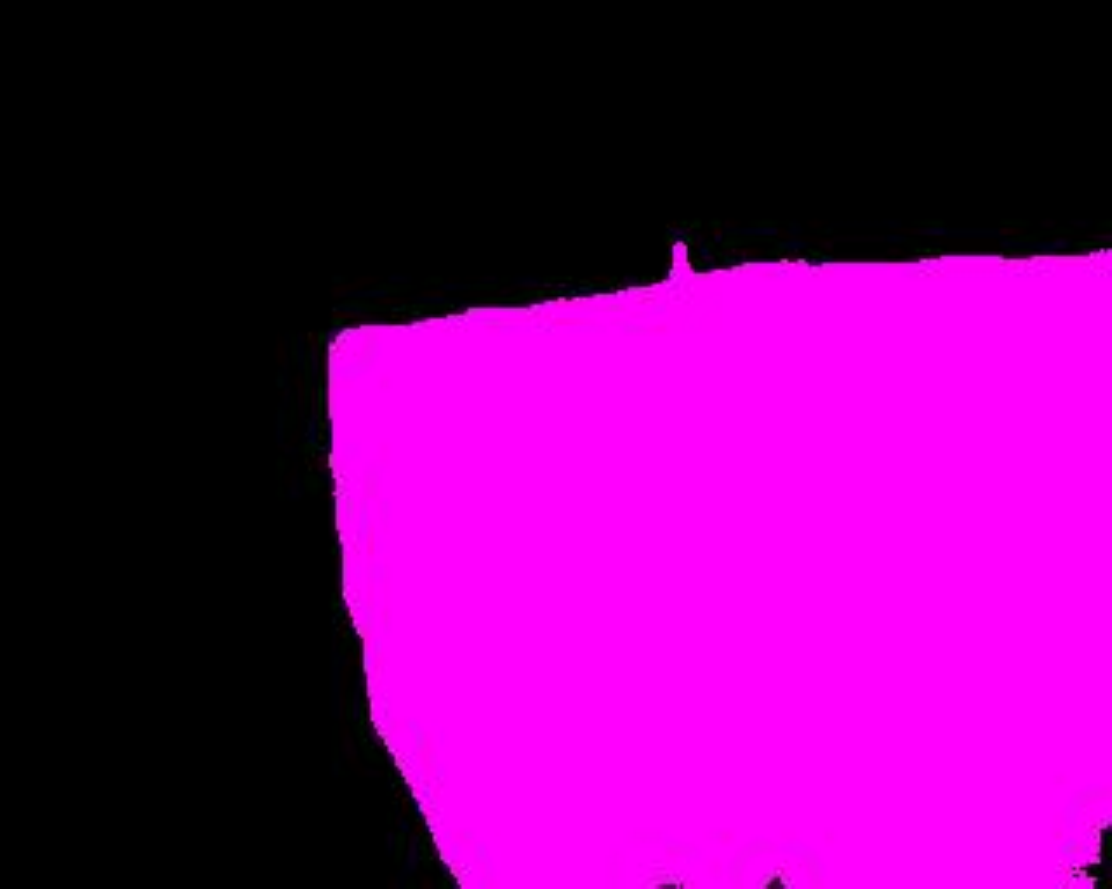}\\

{ Scene} & { Degraded} & { Funie-GAN \cite{funiegan} } & { Funie-GAN-UP \cite{funiegan} } & { Deep-SESR \cite{sesr} } & { UGAN \cite{ugan} } & { UGAN-P \cite{ugan} } & { Deep WaveNet} & { Ground Truth}\\

\end{tabular}}
\caption{Qualitative demonstration of the segmentation maps obtained by using the enhanced images from various UIR existing works. \mycircle{cyan} Human divers, \mycircle{pink} Wrecks and ruins, \mycircle{brown} Robots and instruments, \mycircle{yellow} Fish and vertebrates, \mycircle{green} Reefs and invertebrates.}
\label{fig:seg_maps}
\end{figure*}

\begin{figure*}
\resizebox{\textwidth}{!}{
\setlength{\tabcolsep}{1pt}
\begin{tabular}{ccccccc}

\includegraphics[width=3cm, height=2cm]{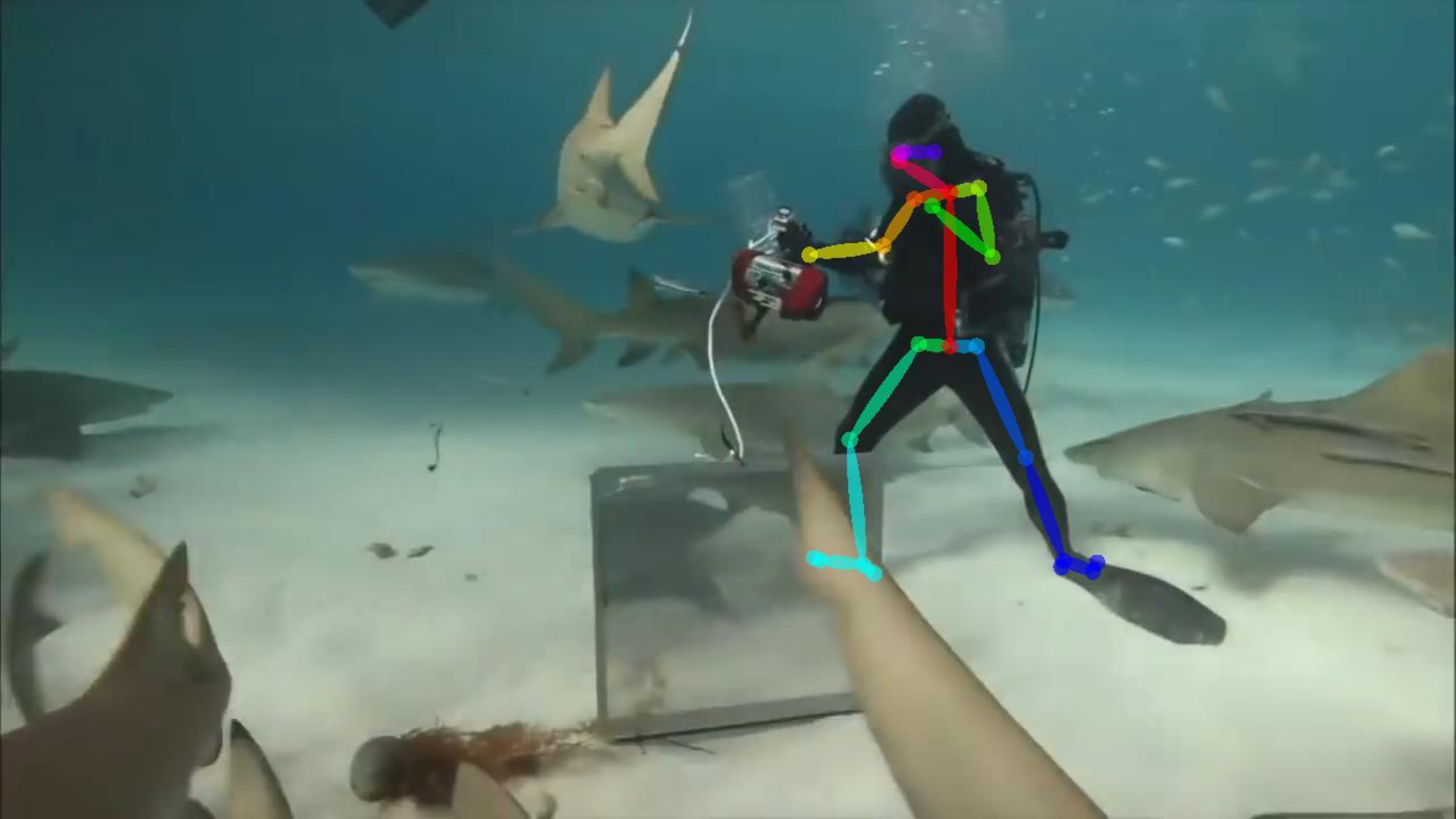}&
\includegraphics[width=3cm, height=2cm]{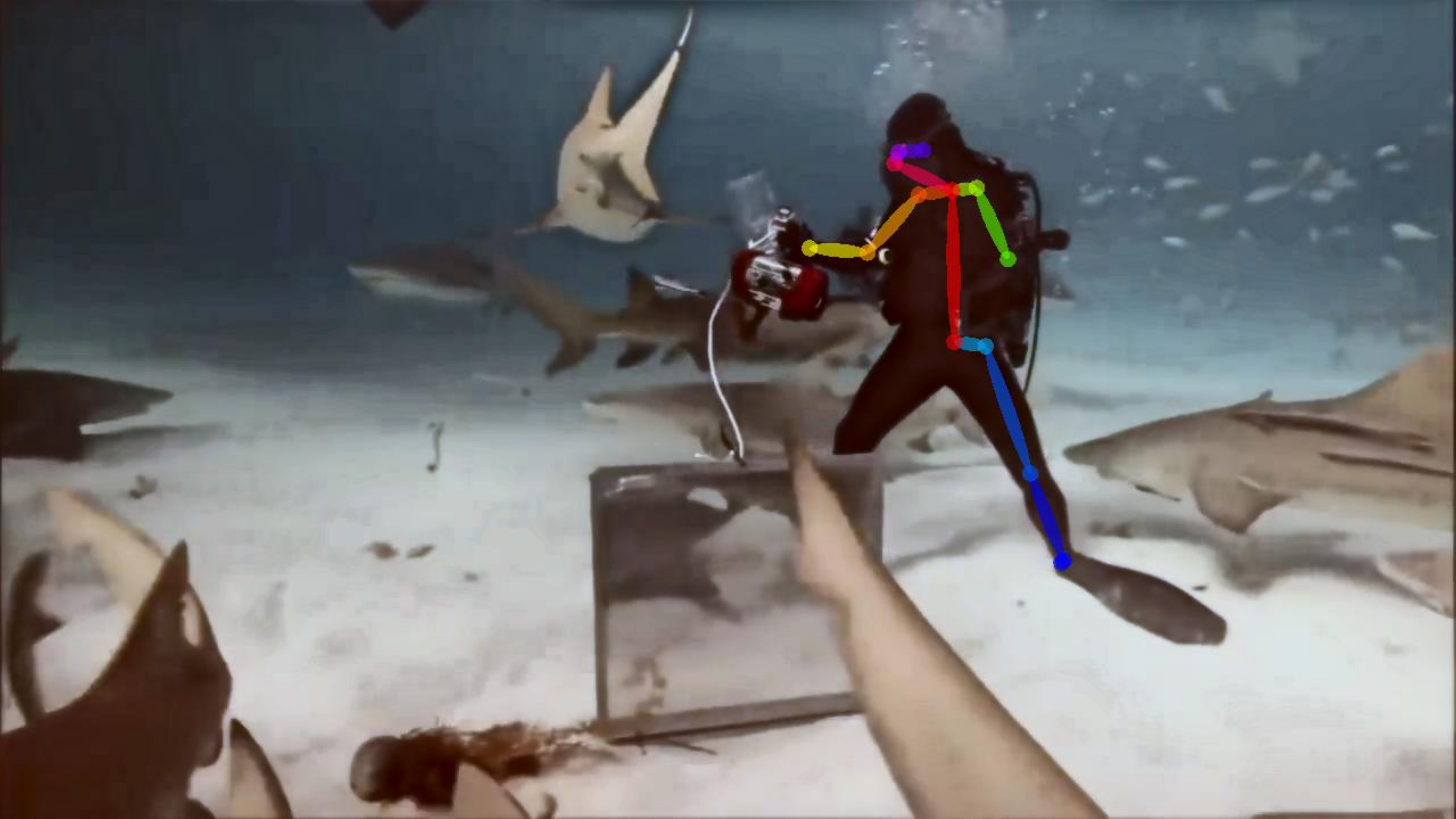}&
\includegraphics[width=3cm, height=2cm]{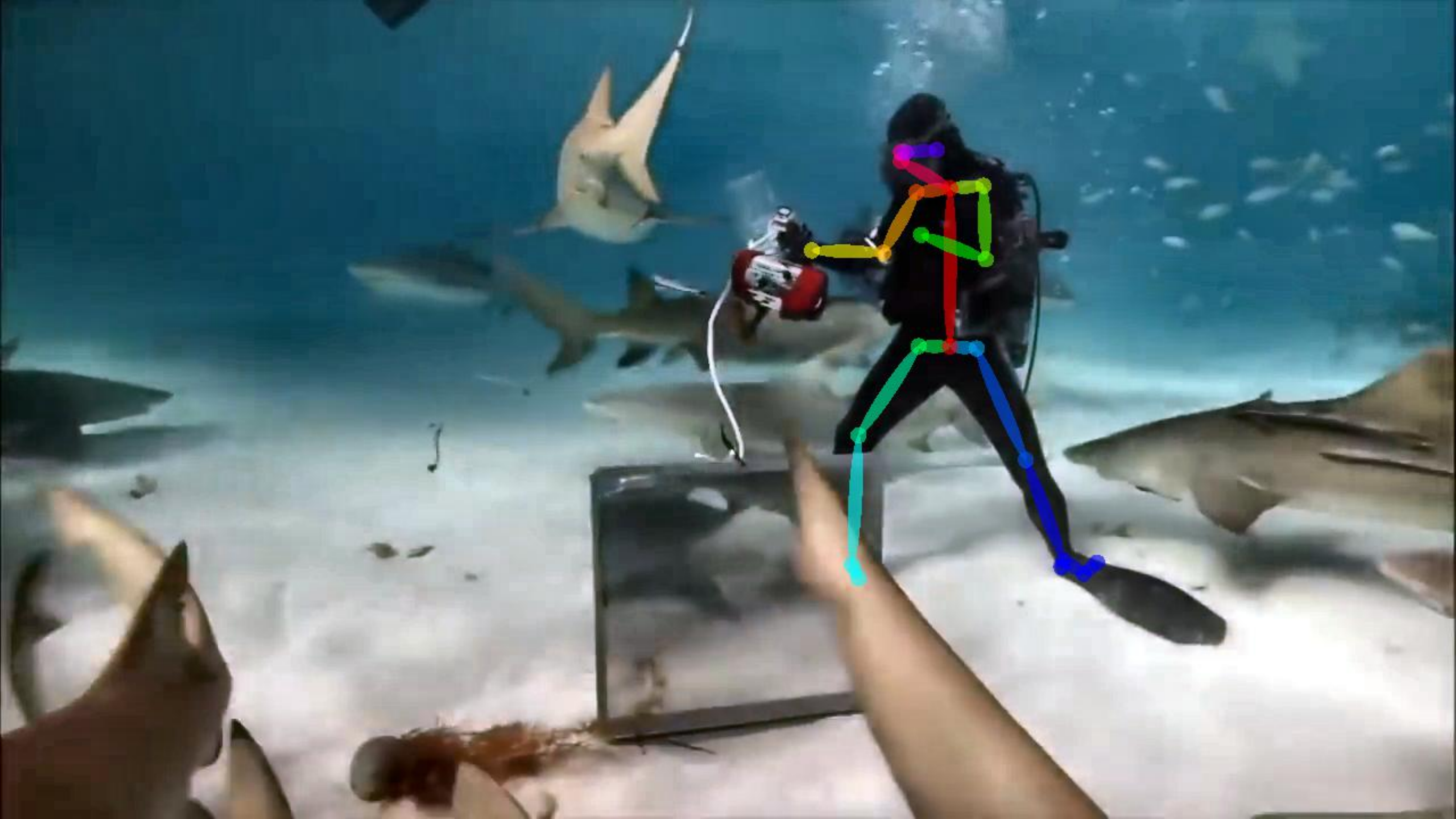}&
\includegraphics[width=3cm, height=2cm]{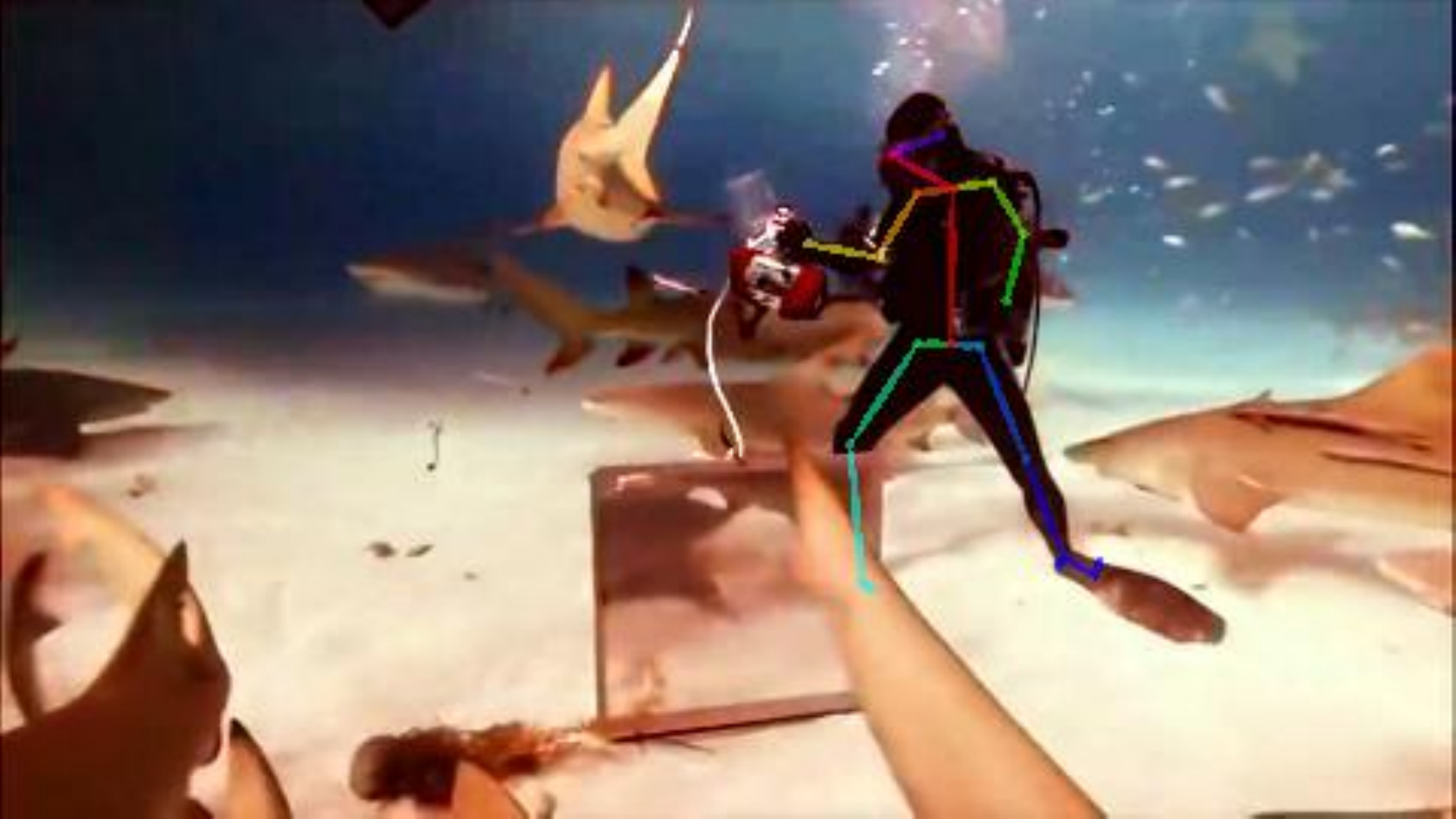}&
\includegraphics[width=3cm, height=2cm]{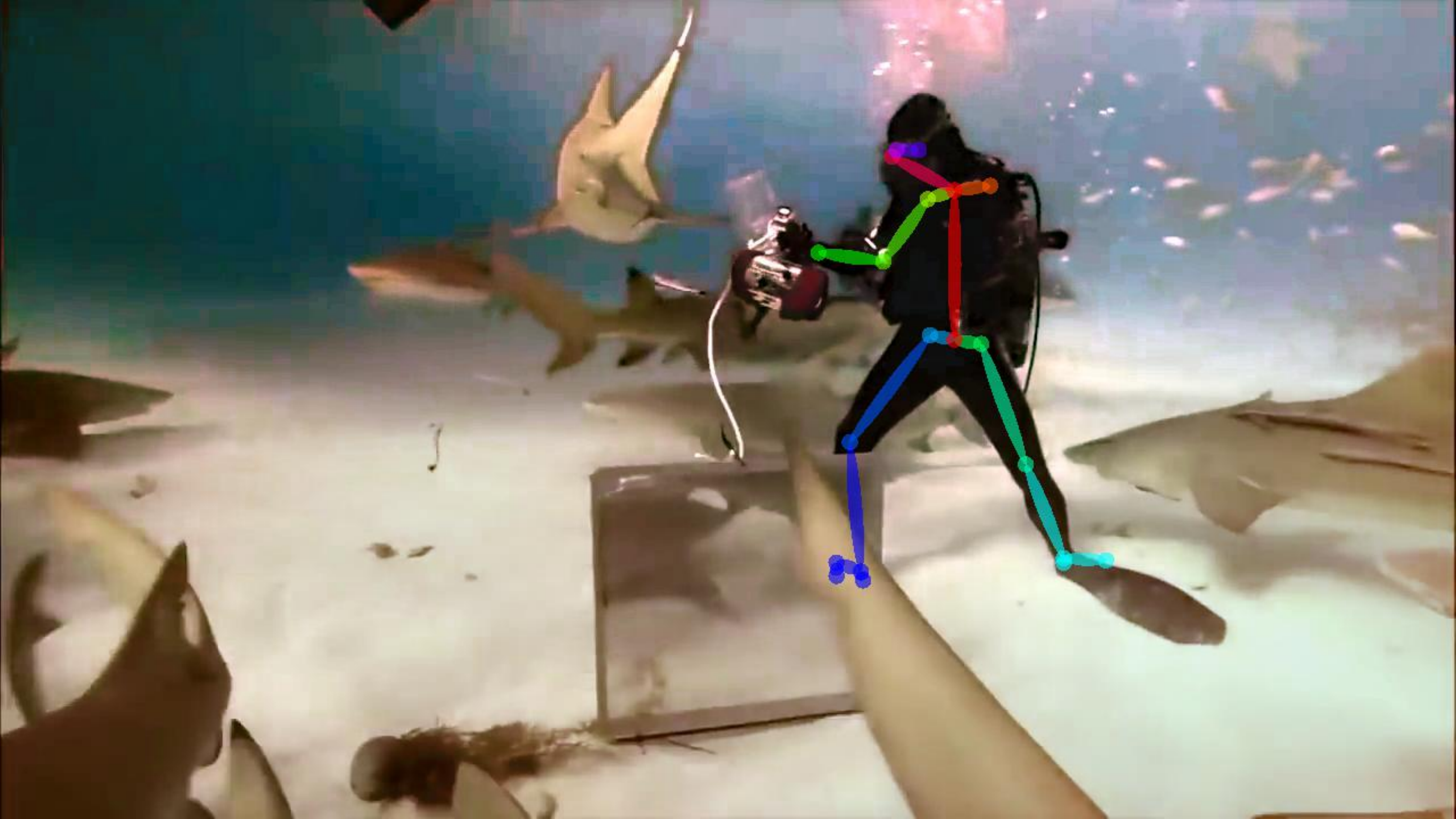}&
\includegraphics[width=3cm, height=2cm]{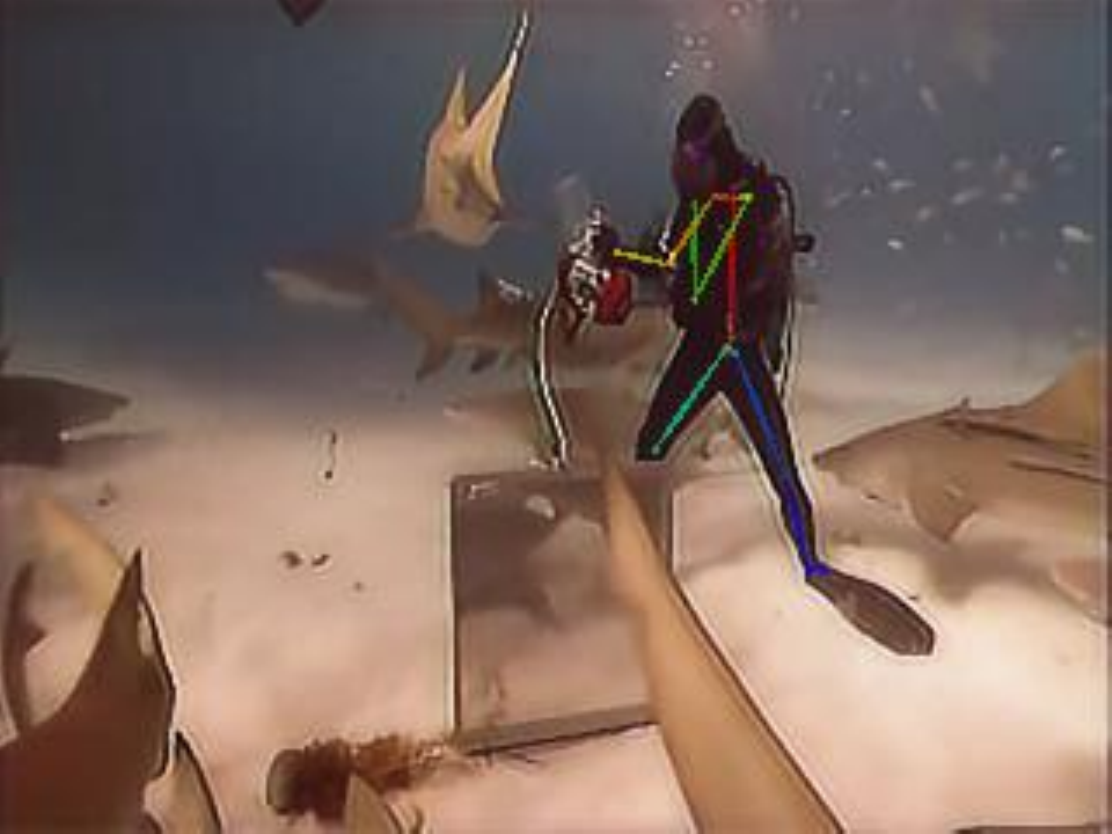}&
\includegraphics[width=3cm, height=2cm]{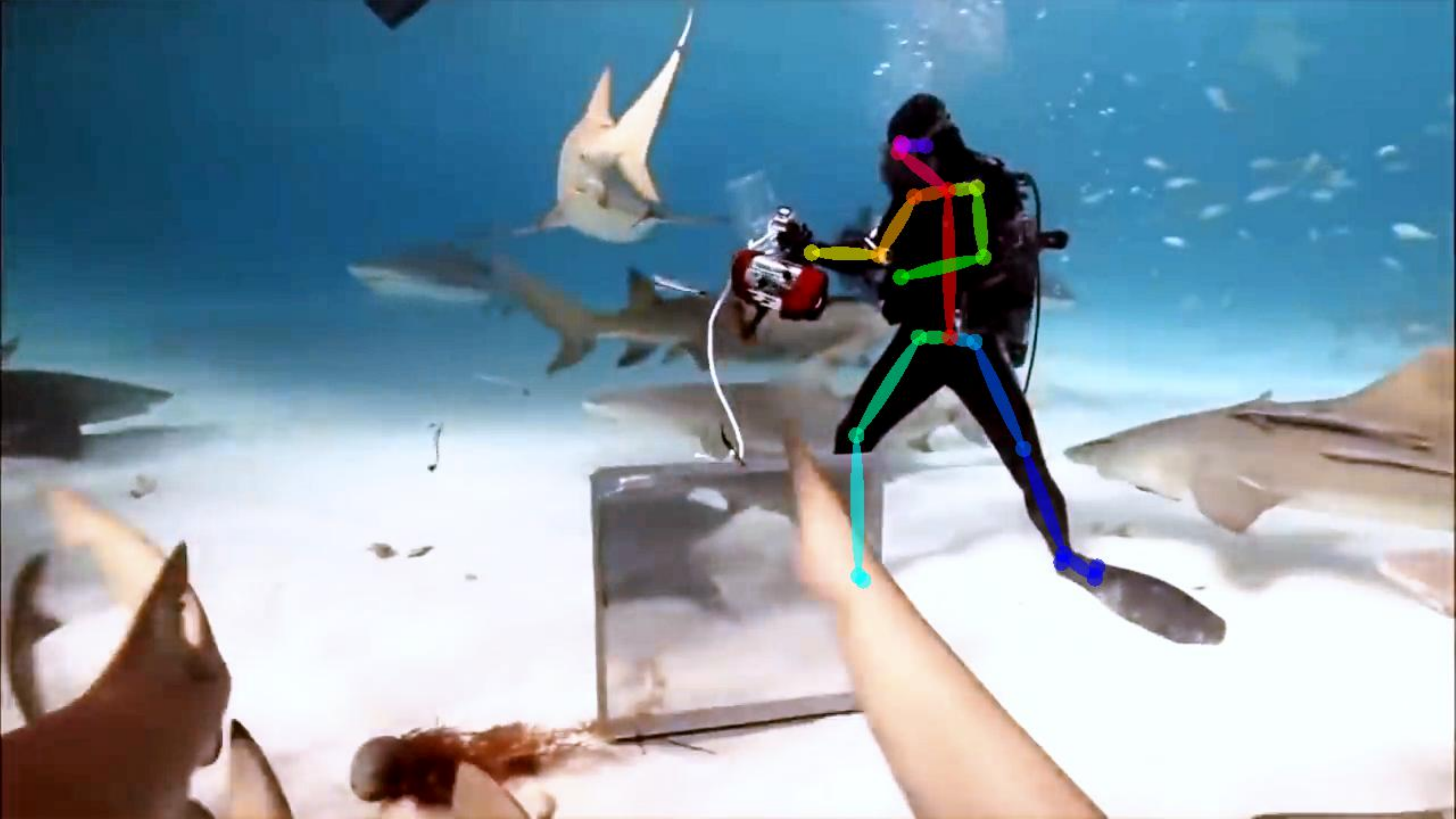}\\

\includegraphics[width=3cm, height=2cm]{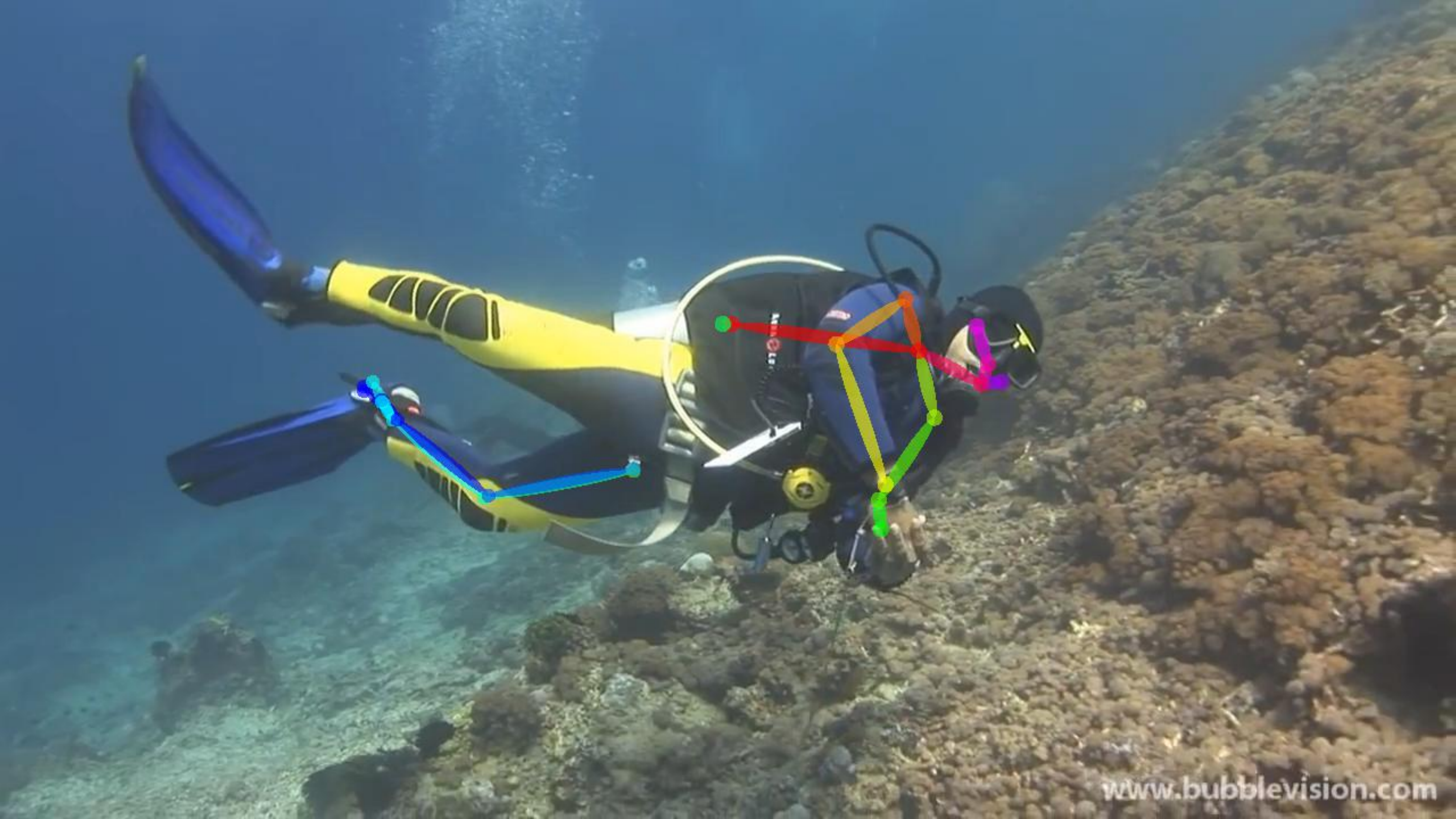}&
\includegraphics[width=3cm, height=2cm]{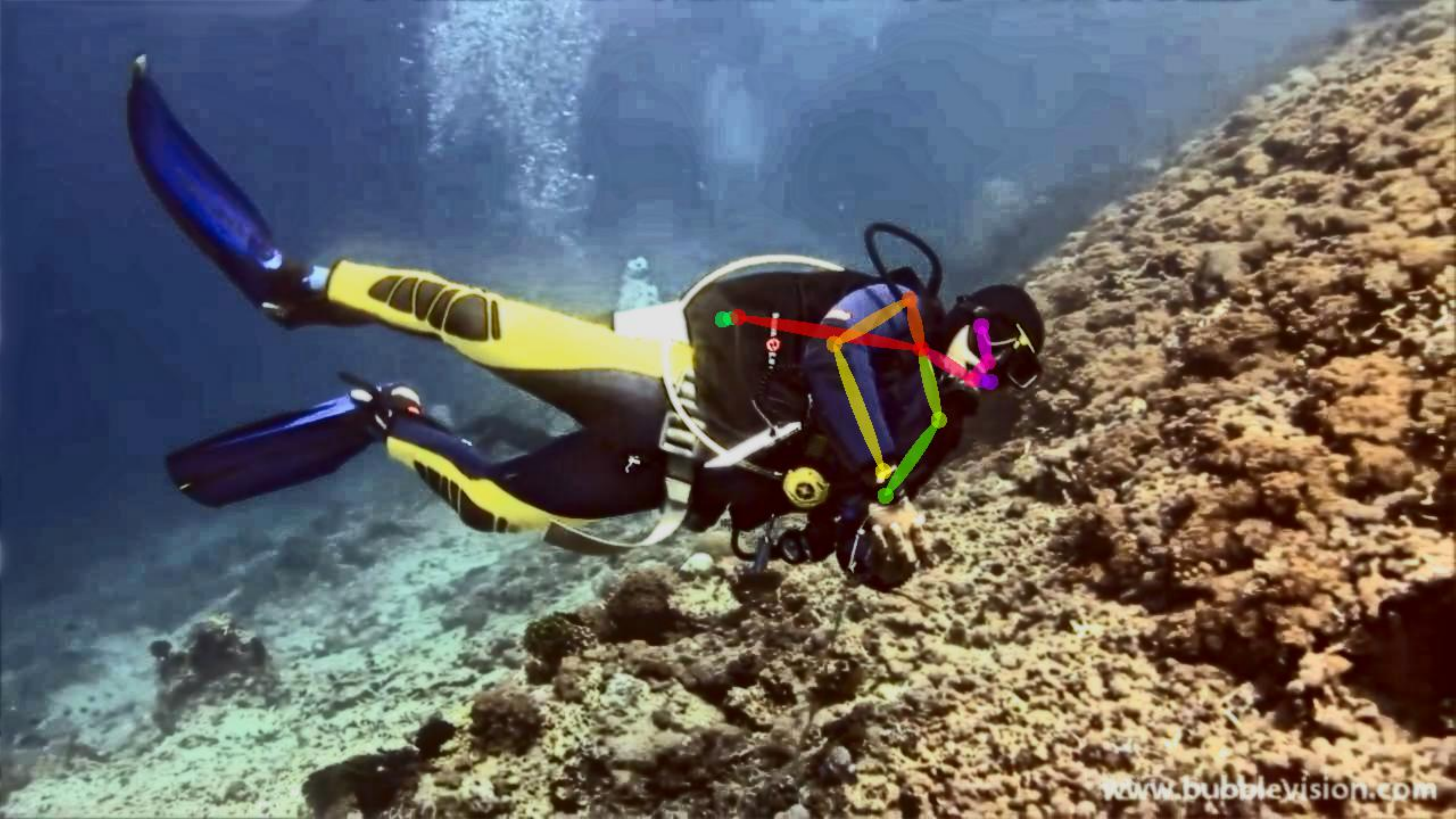}&
\includegraphics[width=3cm, height=2cm]{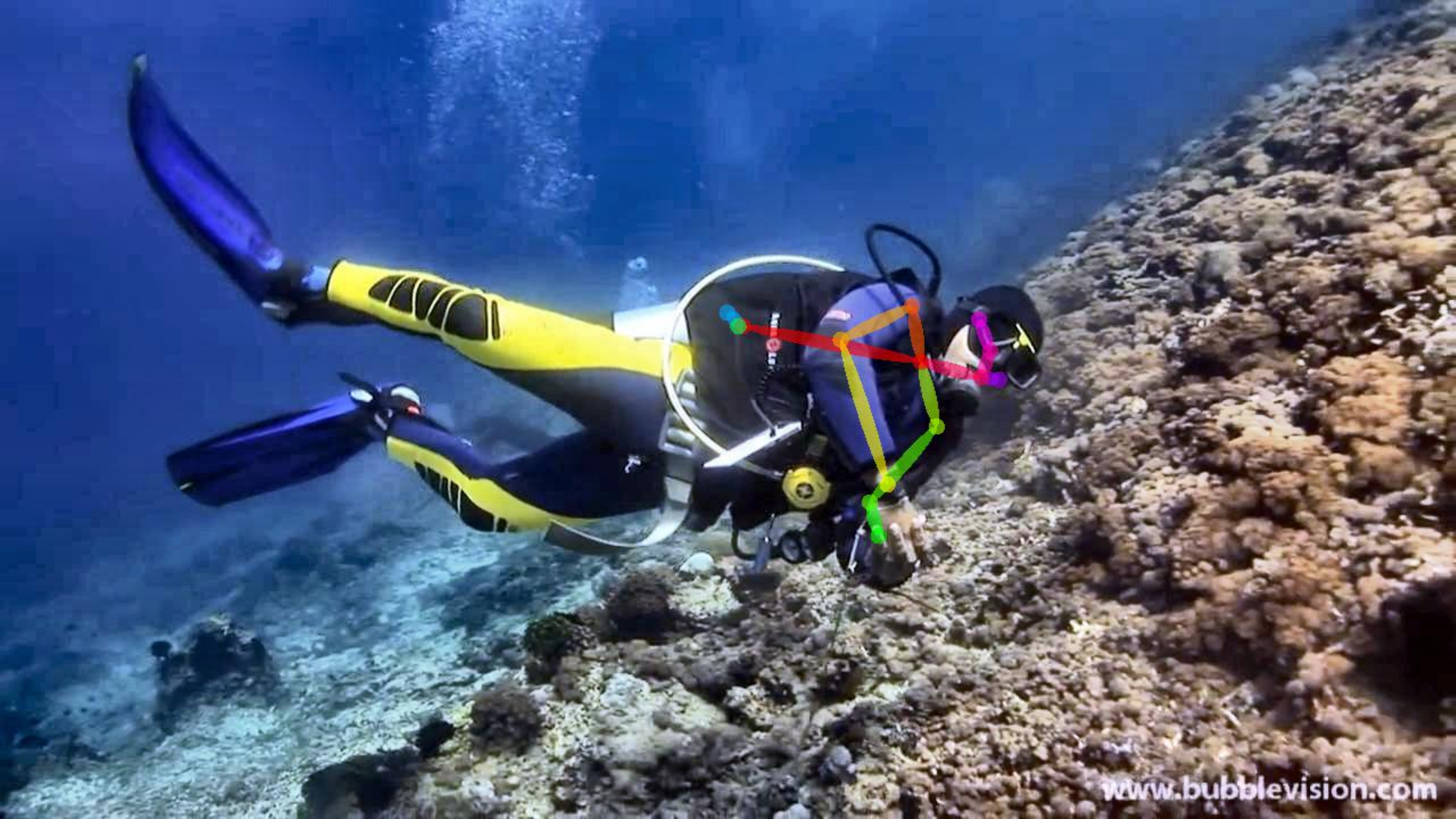}&
\includegraphics[width=3cm, height=2cm]{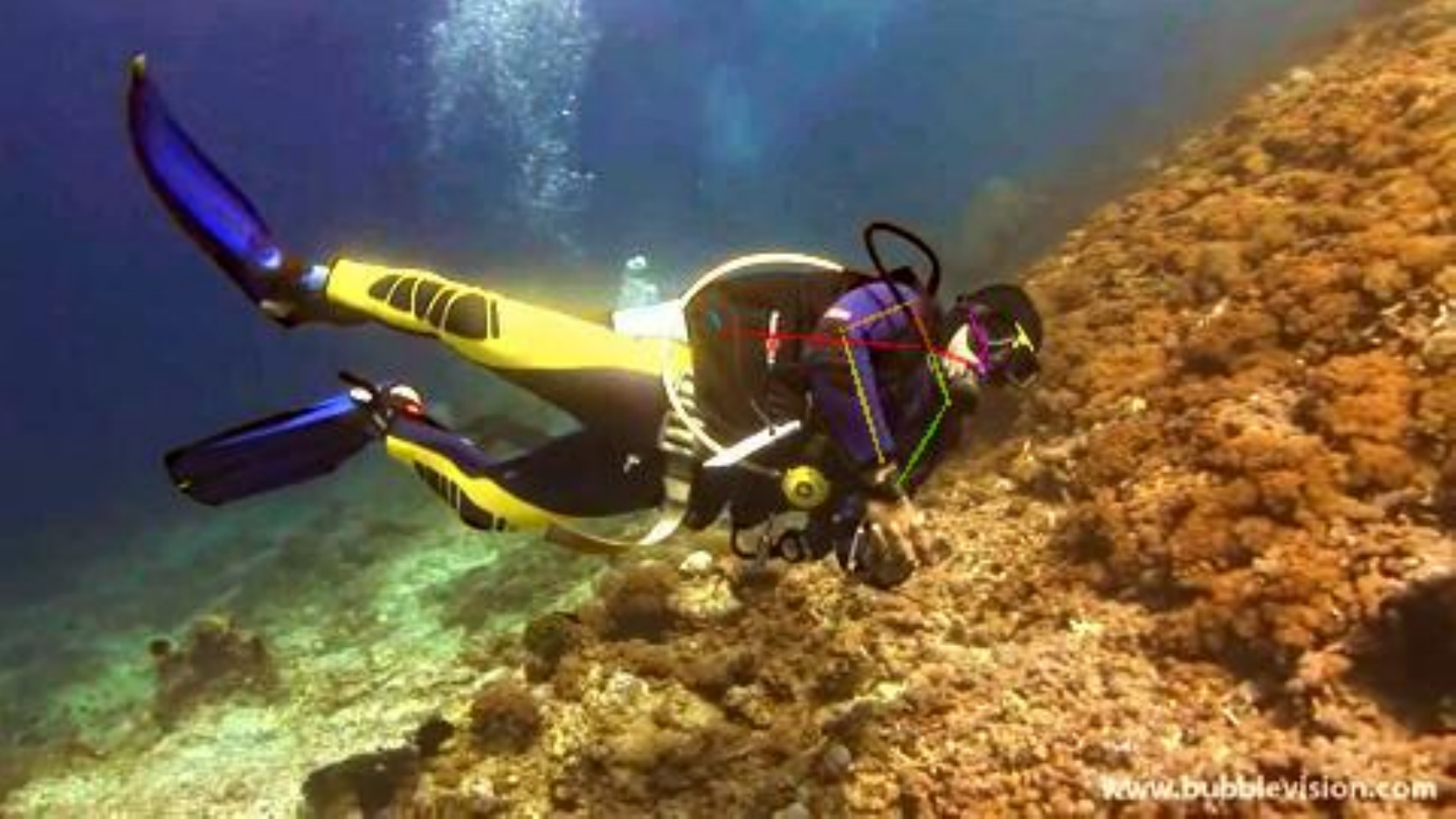}&
\includegraphics[width=3cm, height=2cm]{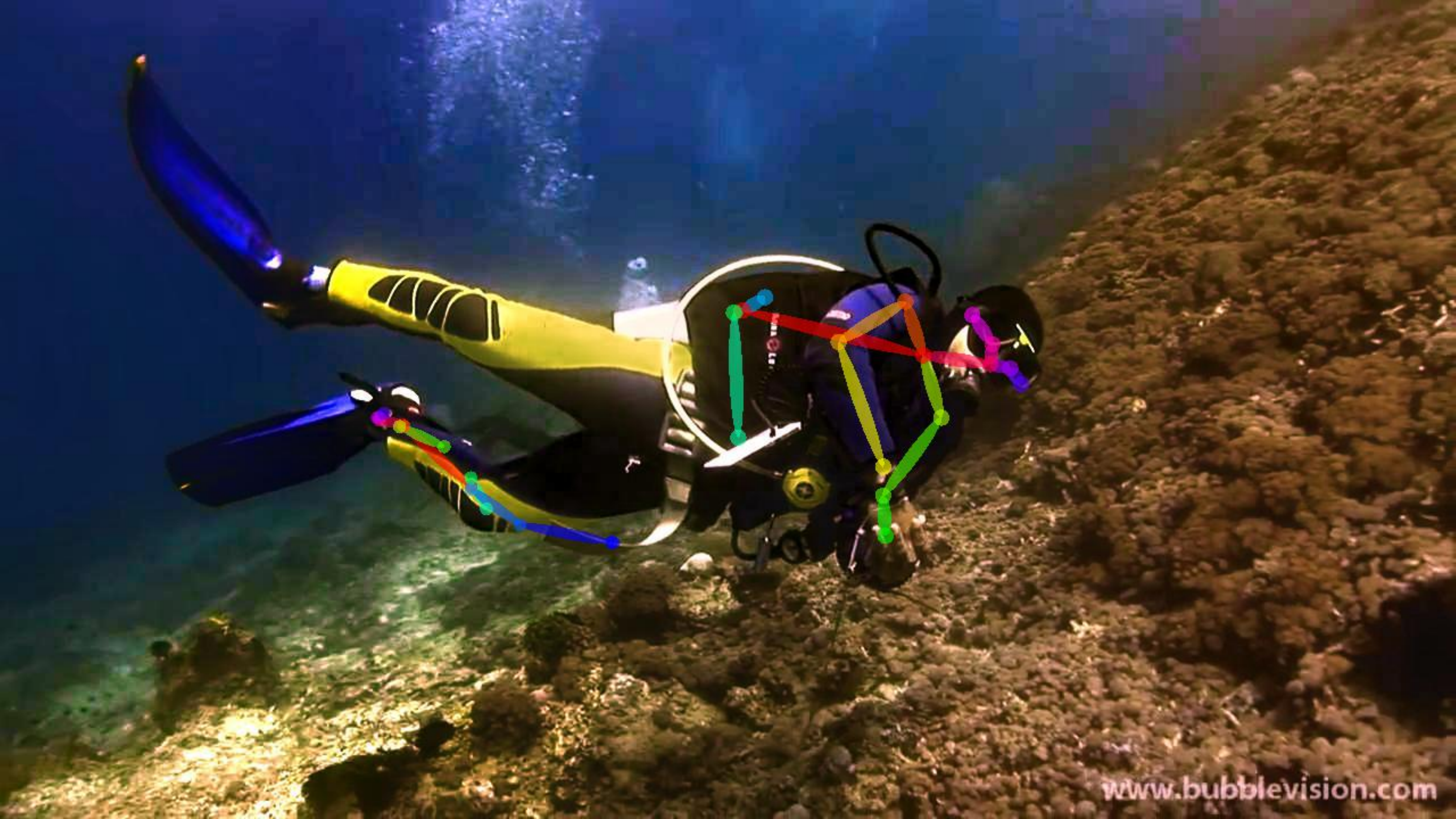}&
\includegraphics[width=3cm, height=2cm]{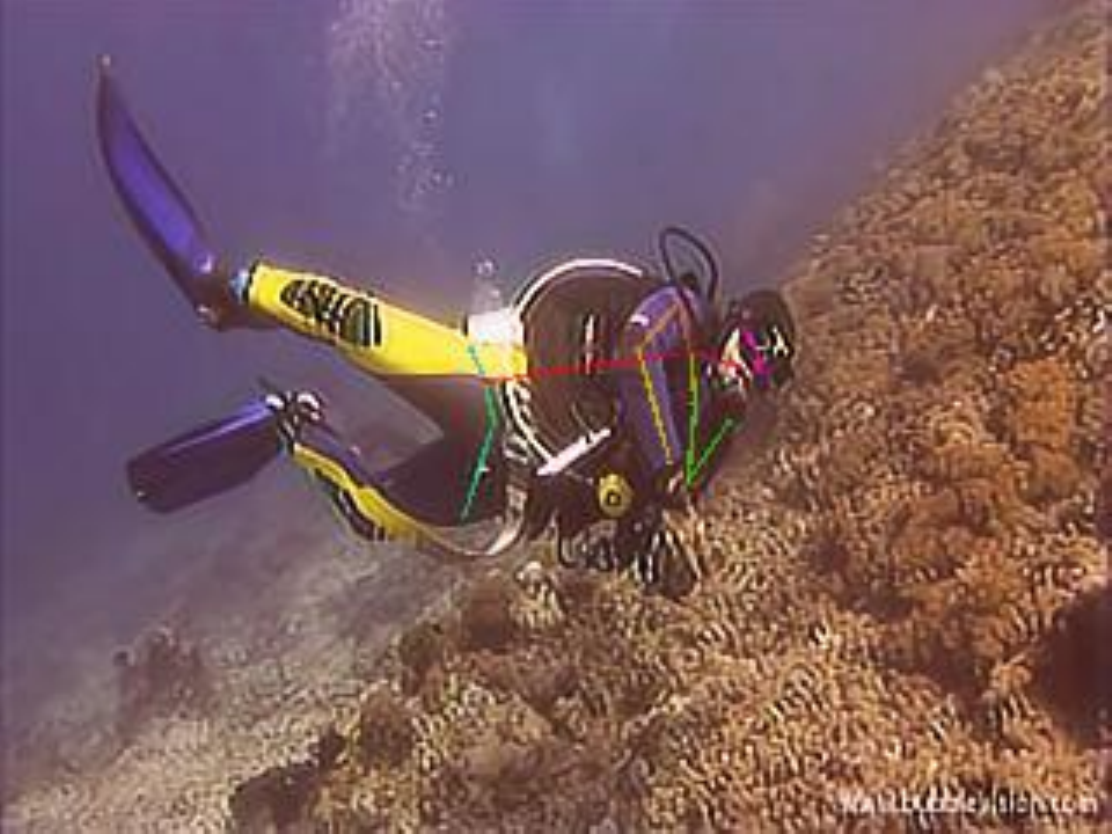}&
\includegraphics[width=3cm, height=2cm]{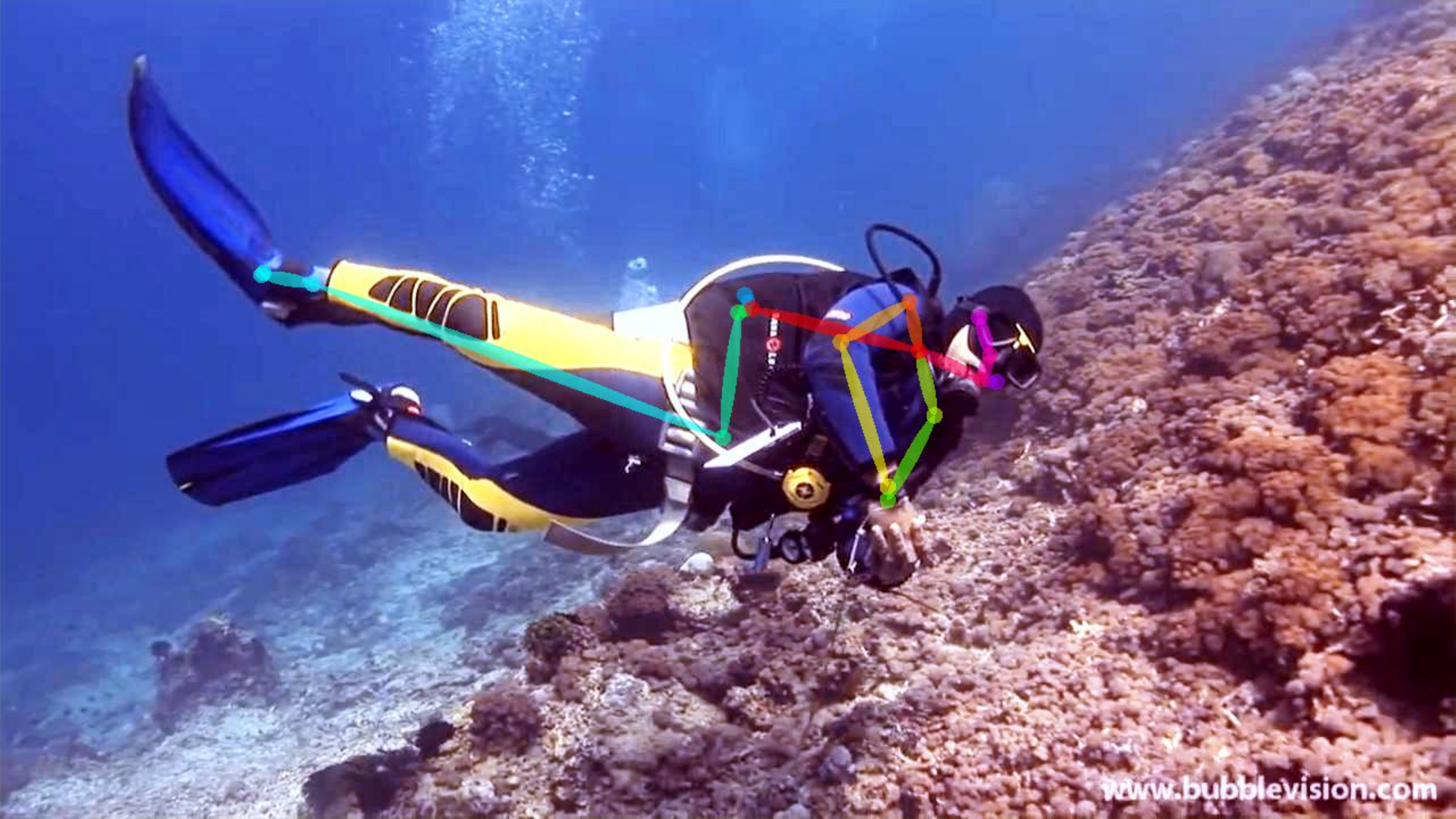}\\

\includegraphics[width=3cm, height=2cm]{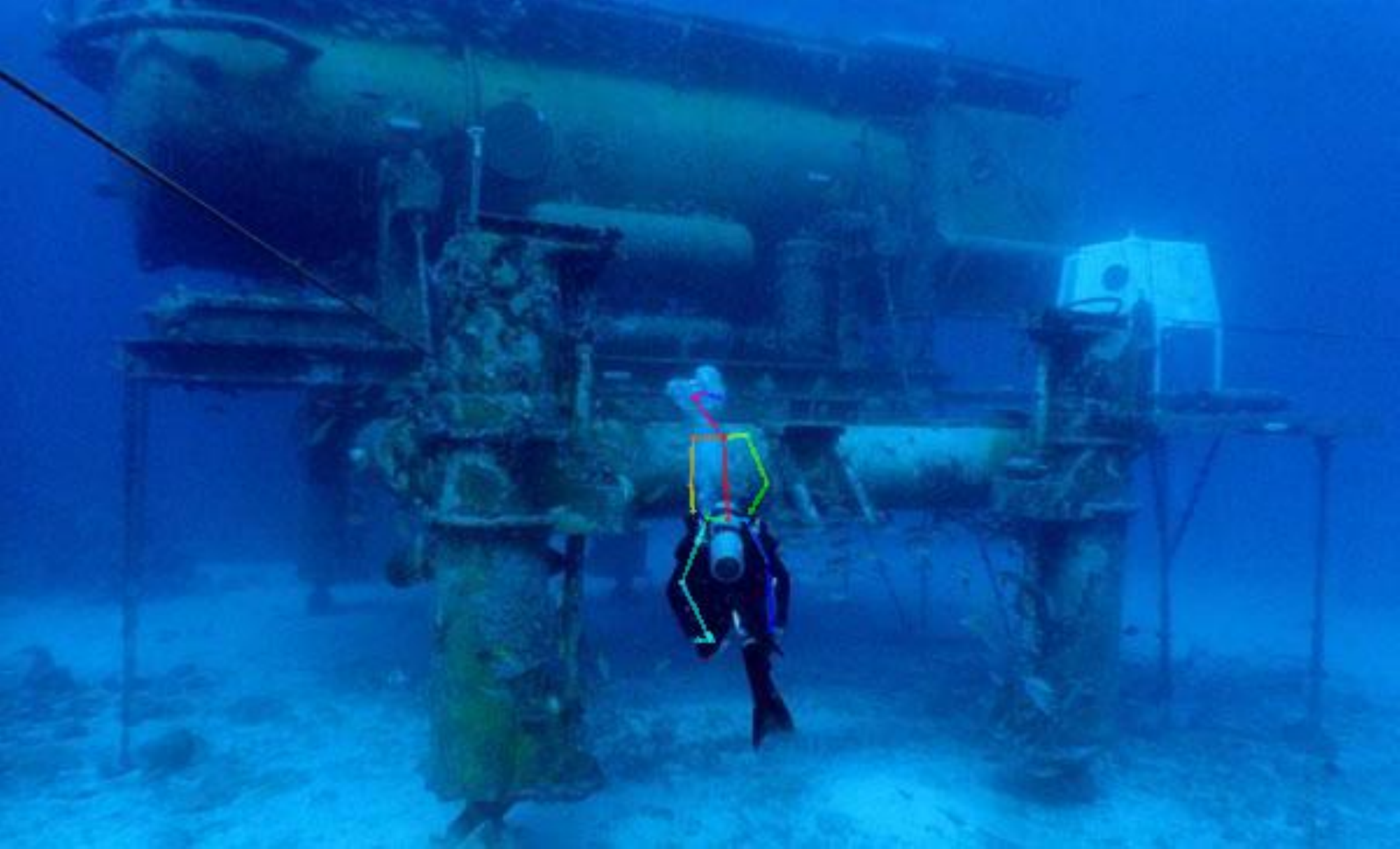}&
\includegraphics[width=3cm, height=2cm]{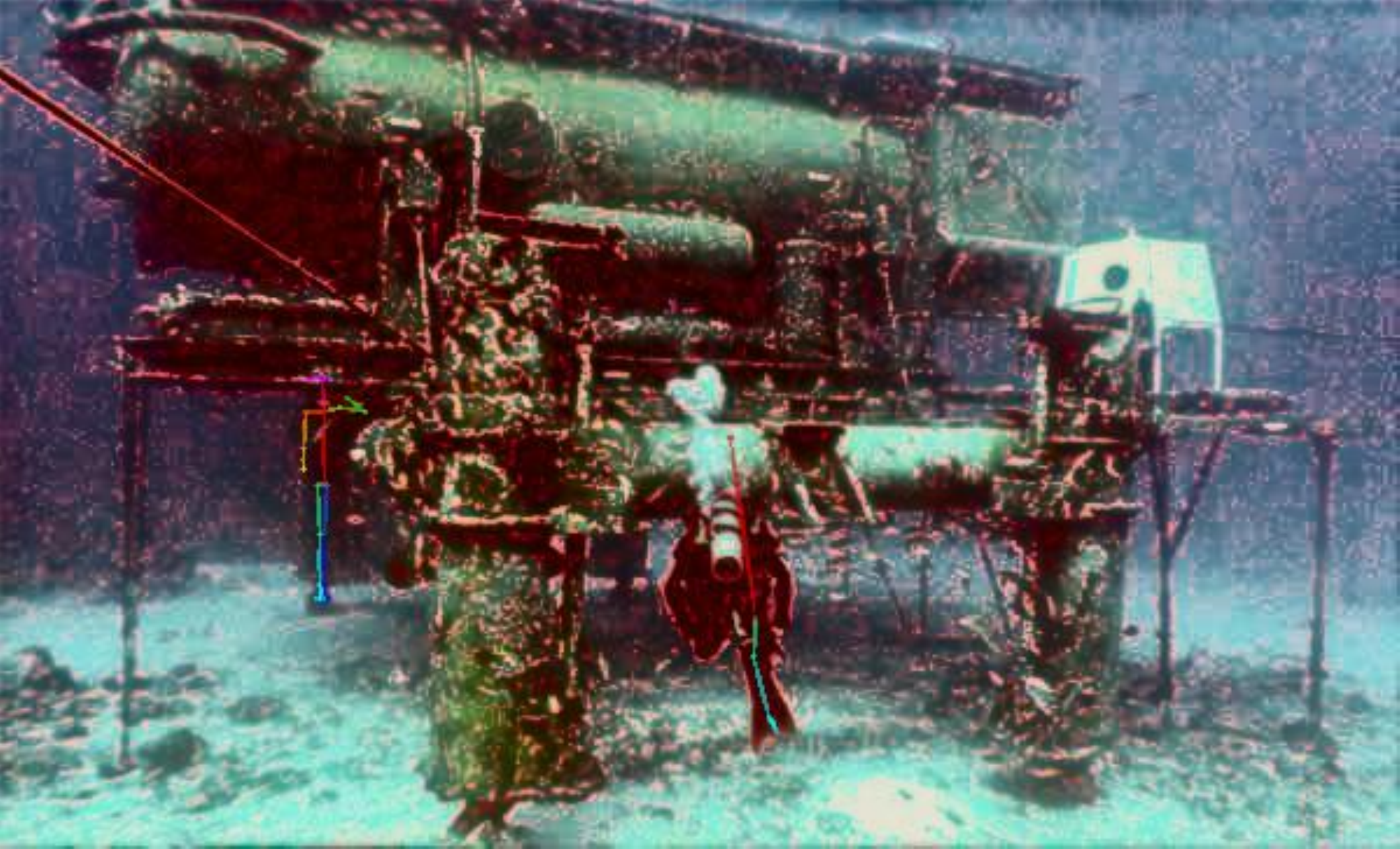}&
\includegraphics[width=3cm, height=2cm]{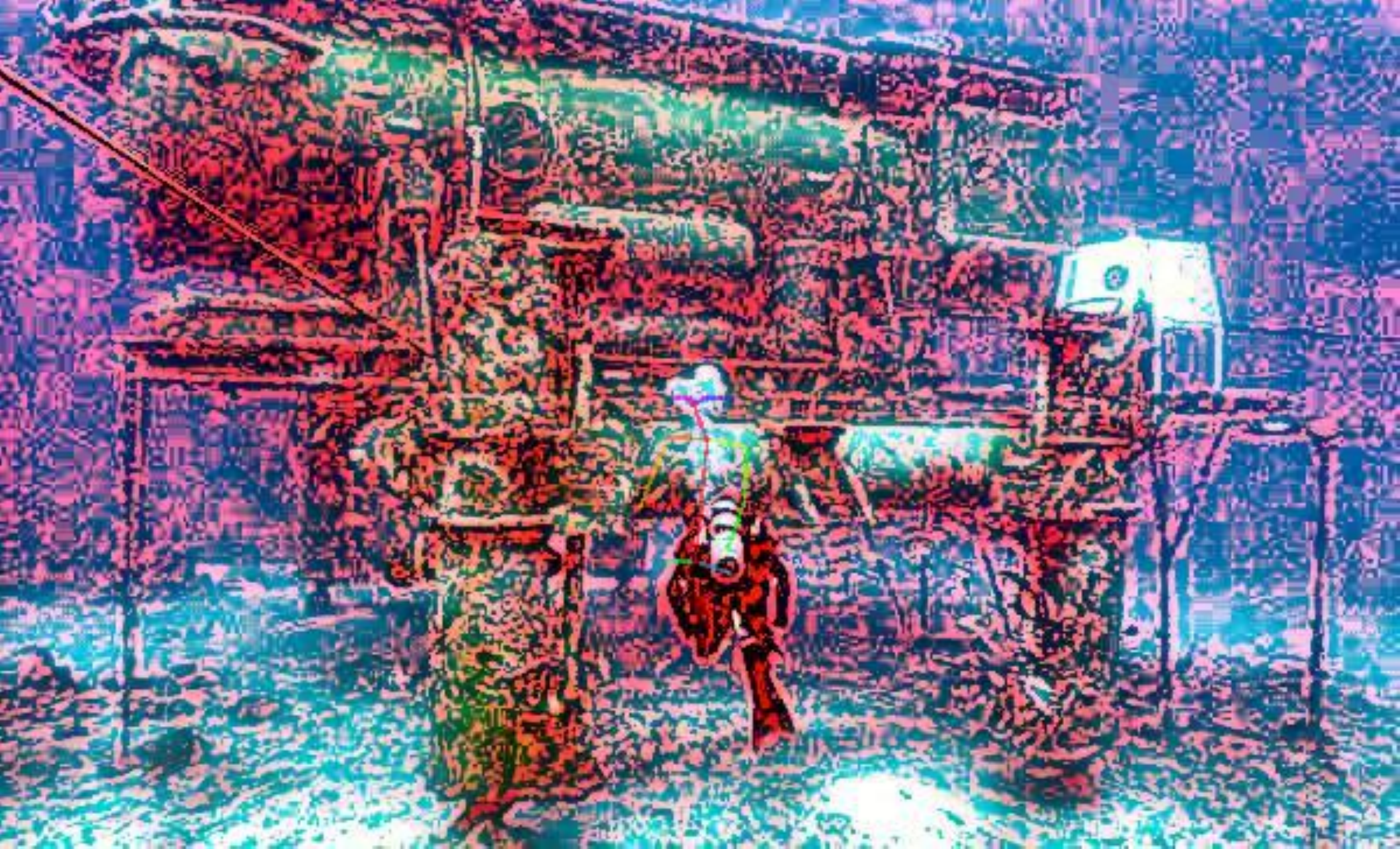}&
\includegraphics[width=3cm, height=2cm]{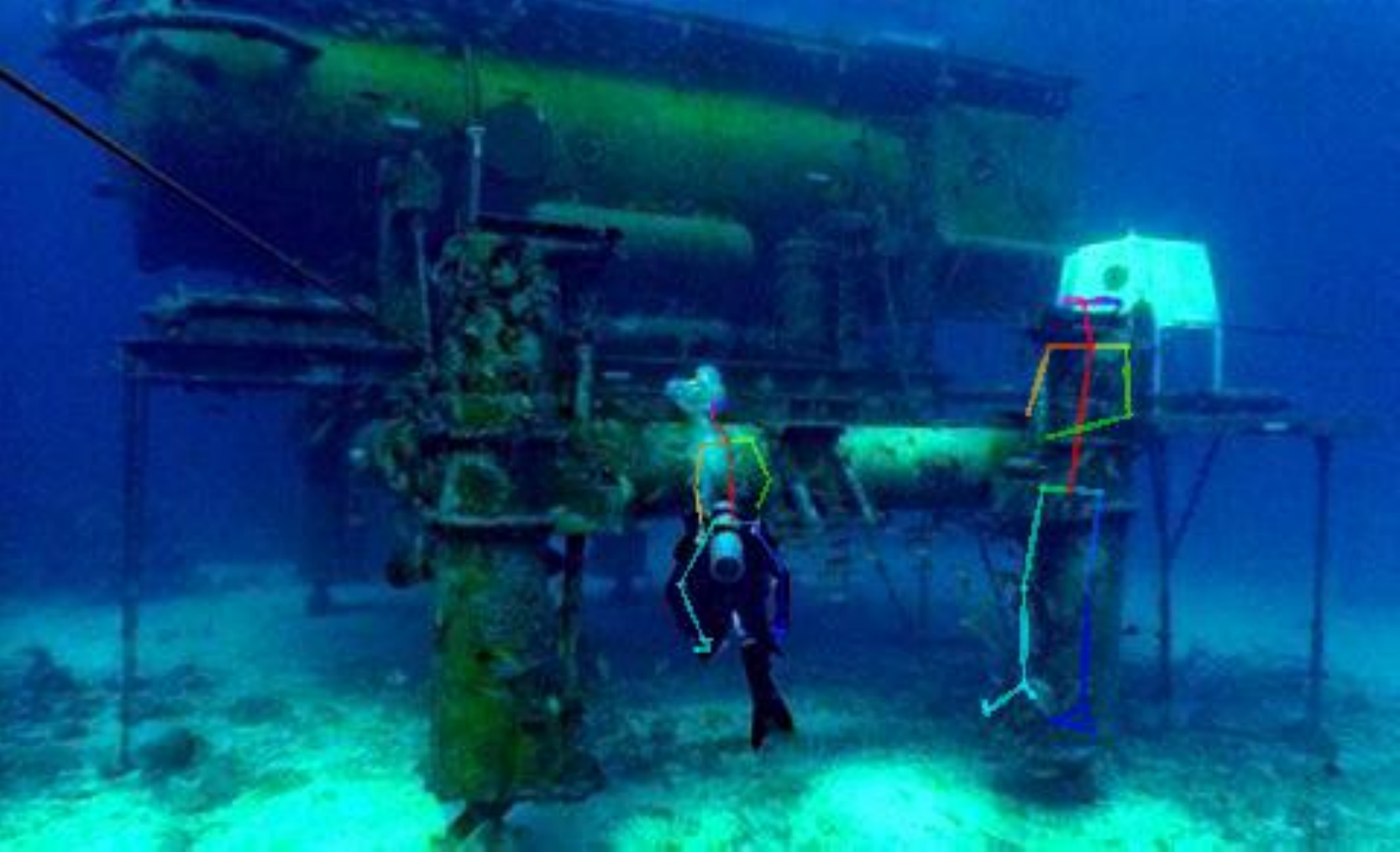}&
\includegraphics[width=3cm, height=2cm]{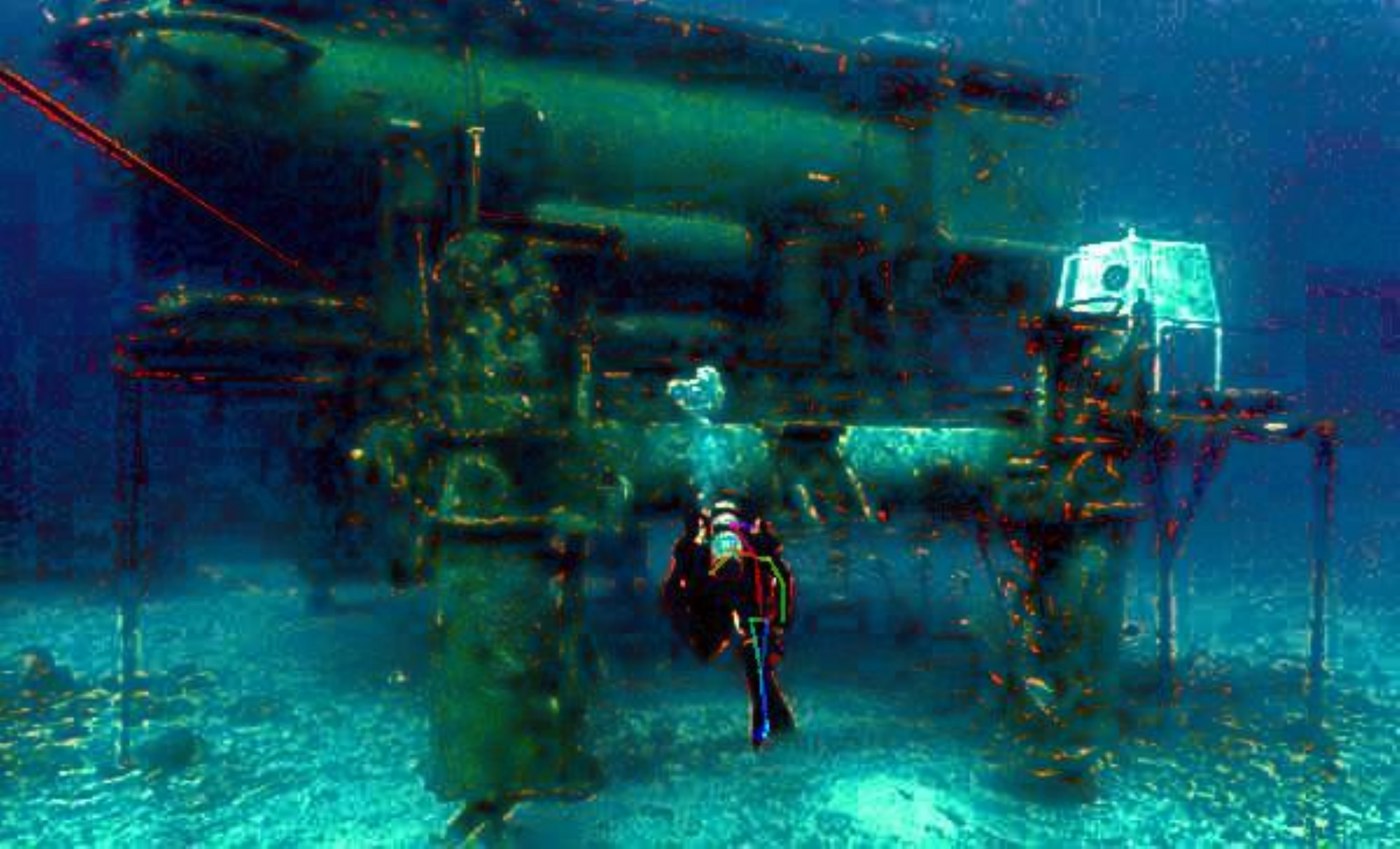}&
\includegraphics[width=3cm, height=2cm]{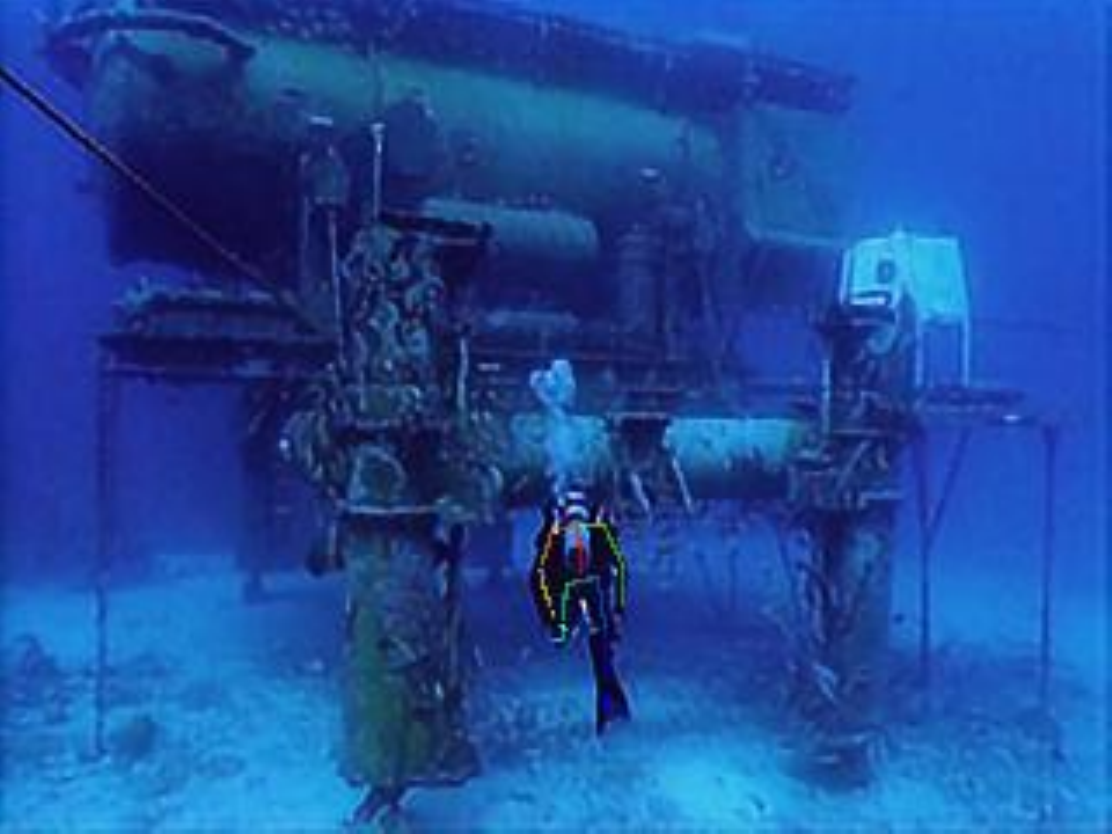}&
\includegraphics[width=3cm, height=2cm]{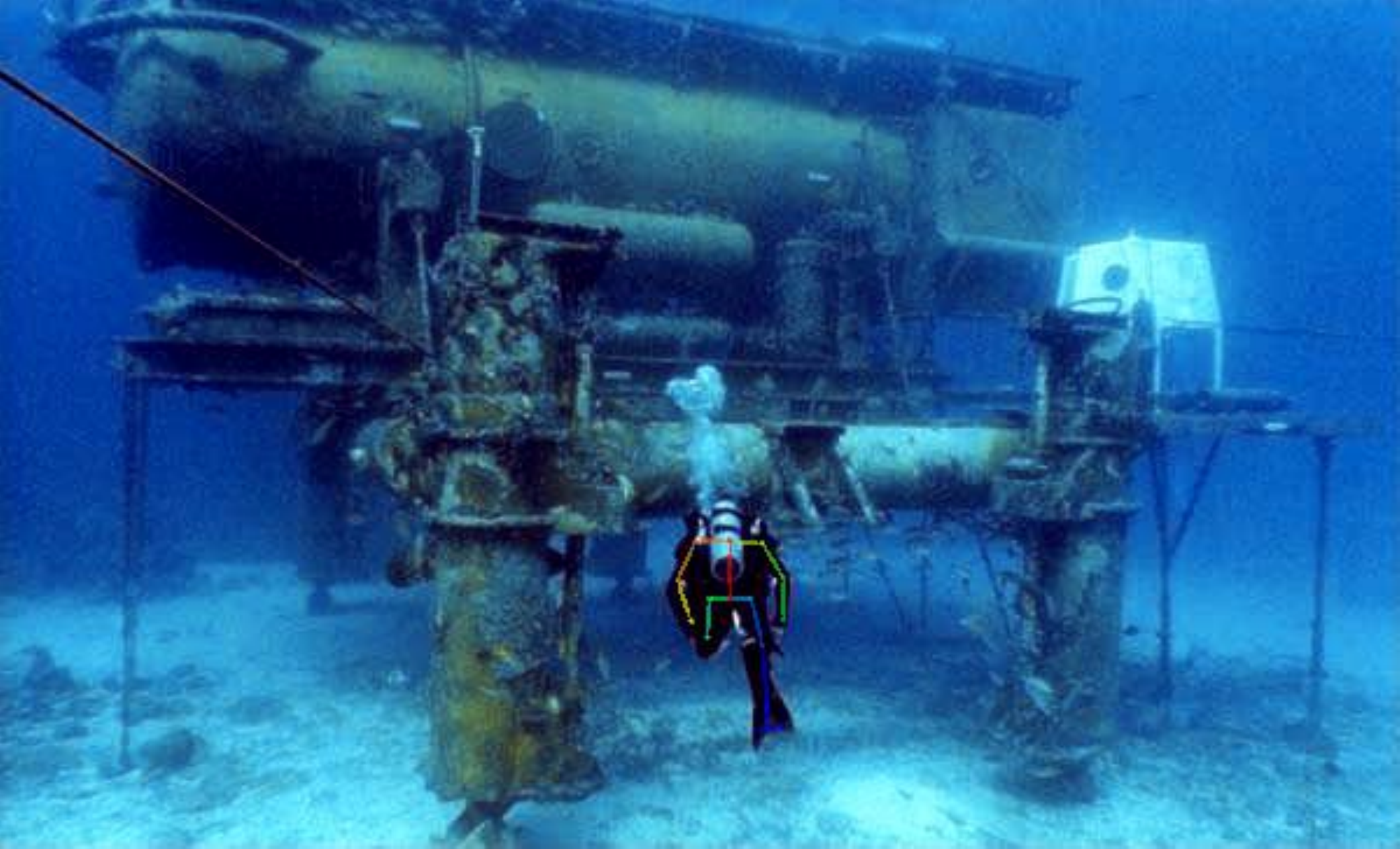}\\

\includegraphics[width=3cm, height=2cm]{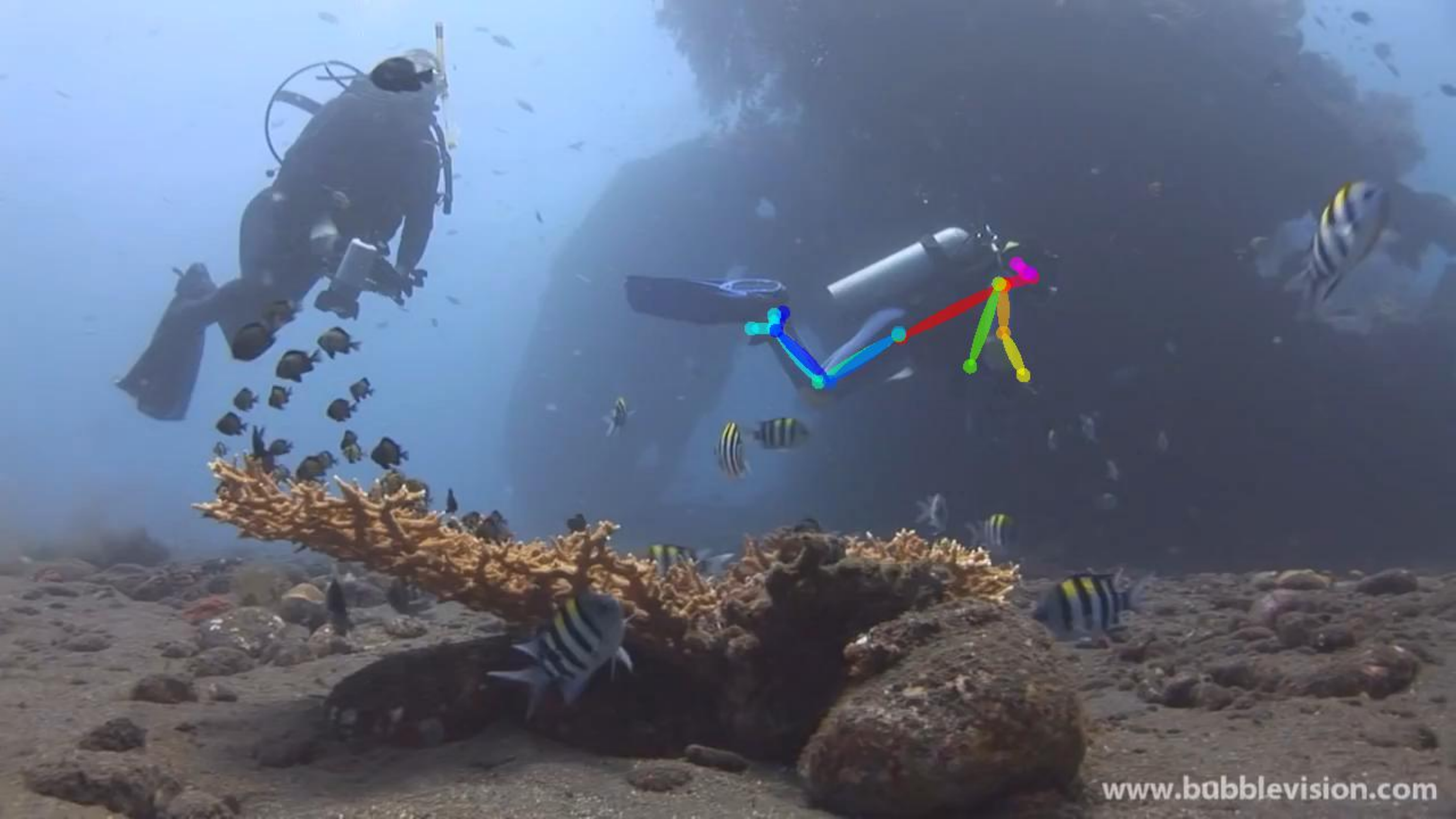}&
\includegraphics[width=3cm, height=2cm]{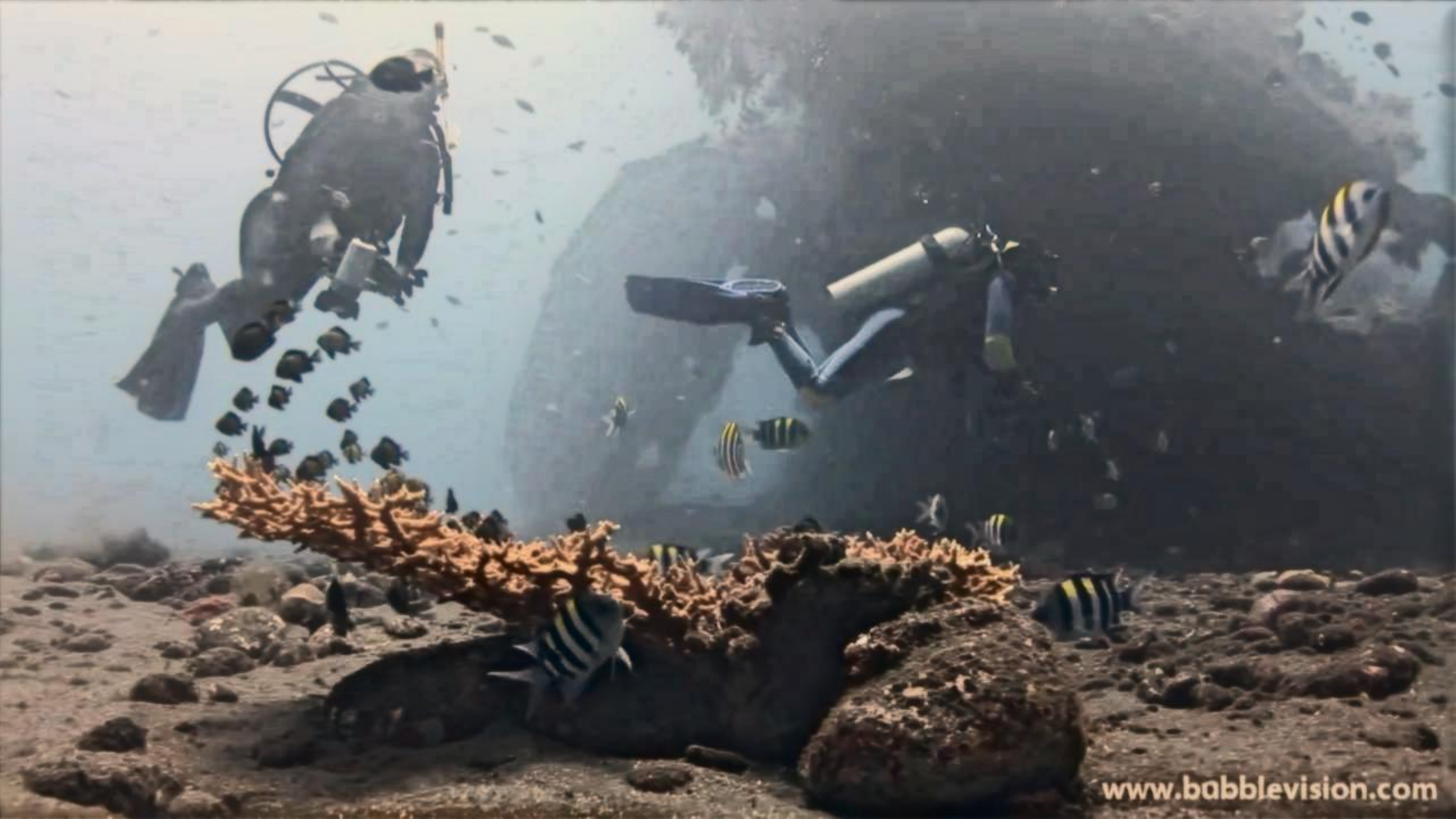}&
\includegraphics[width=3cm, height=2cm]{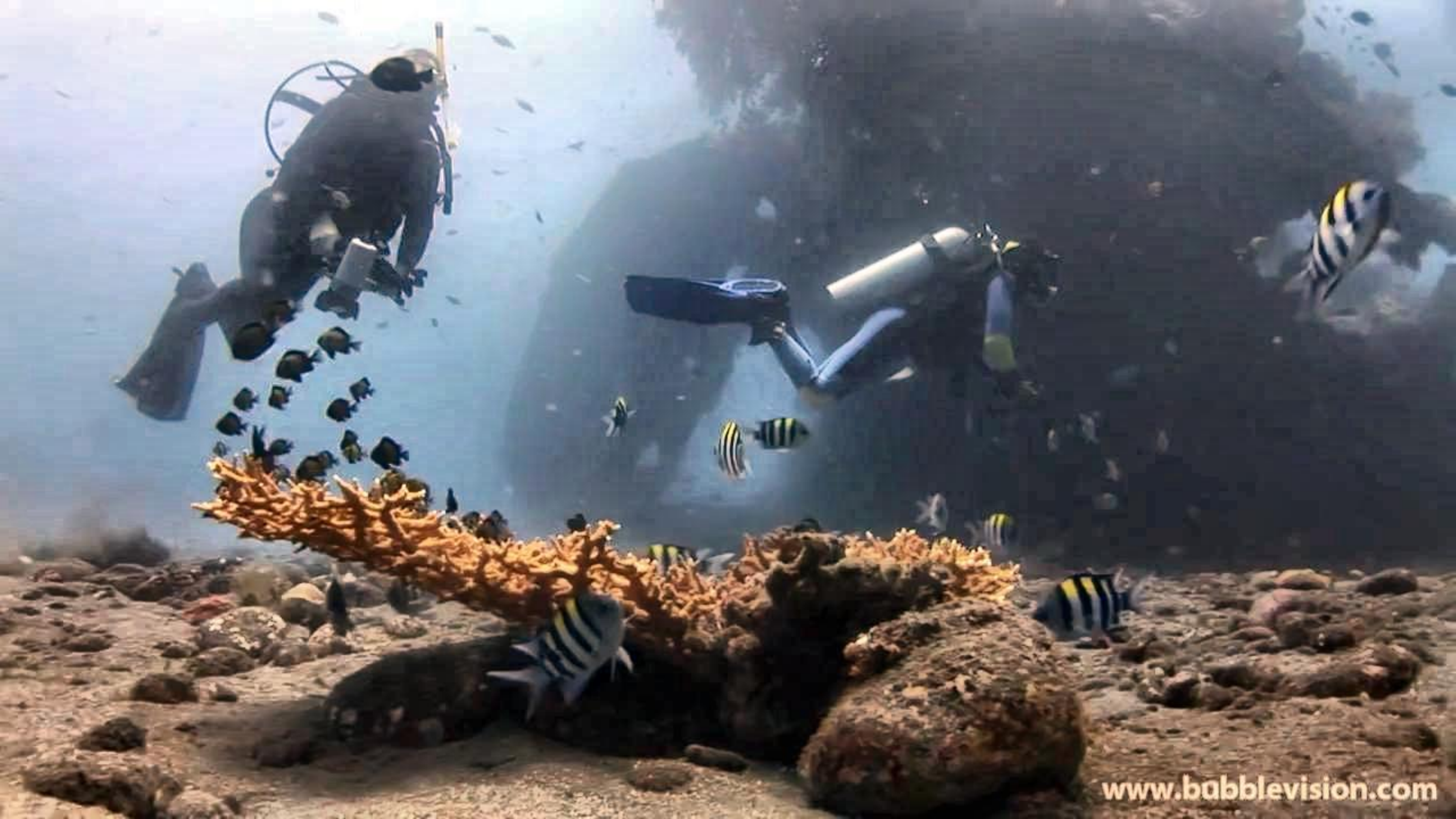}&
\includegraphics[width=3cm, height=2cm]{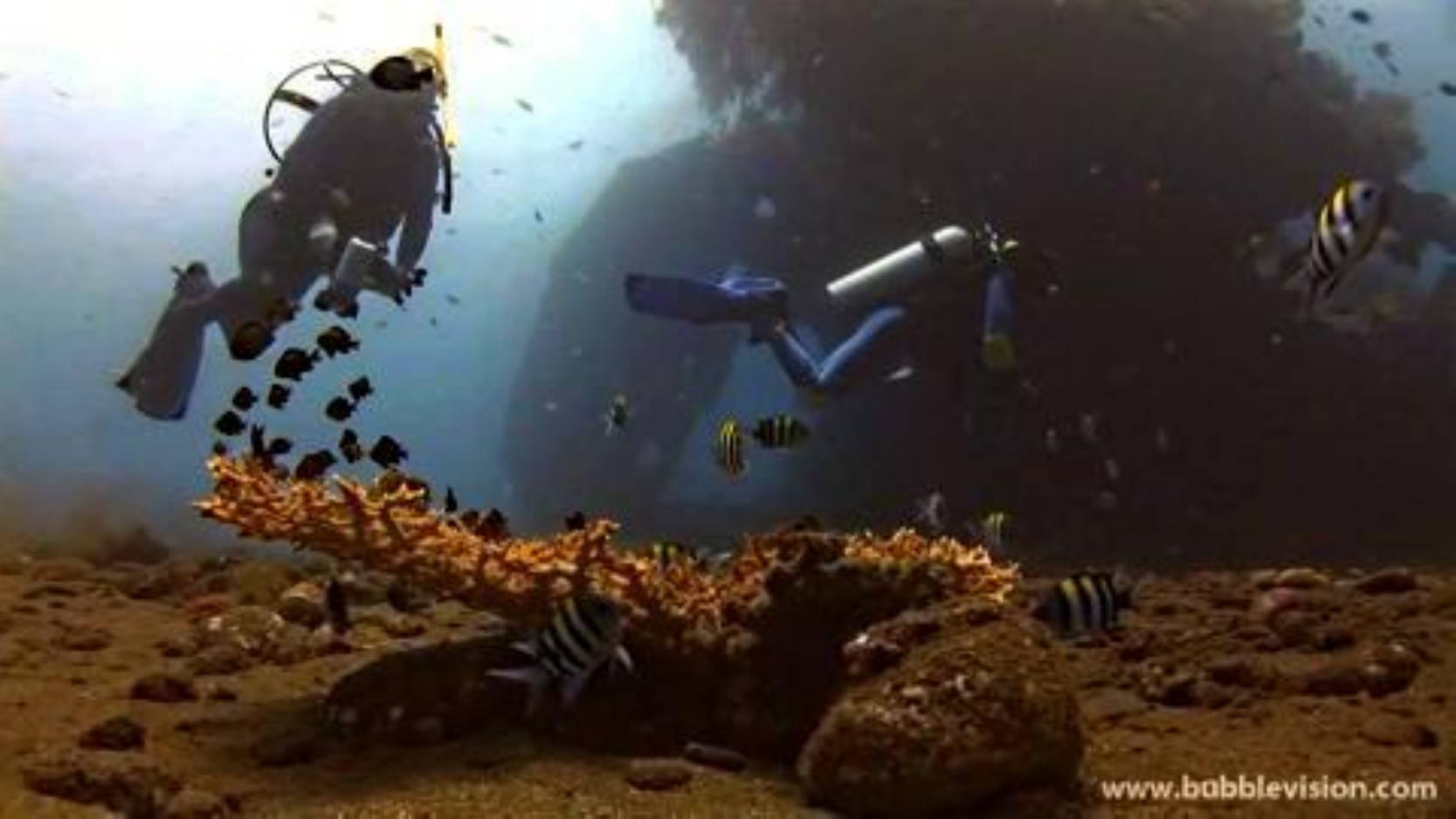}&
\includegraphics[width=3cm, height=2cm]{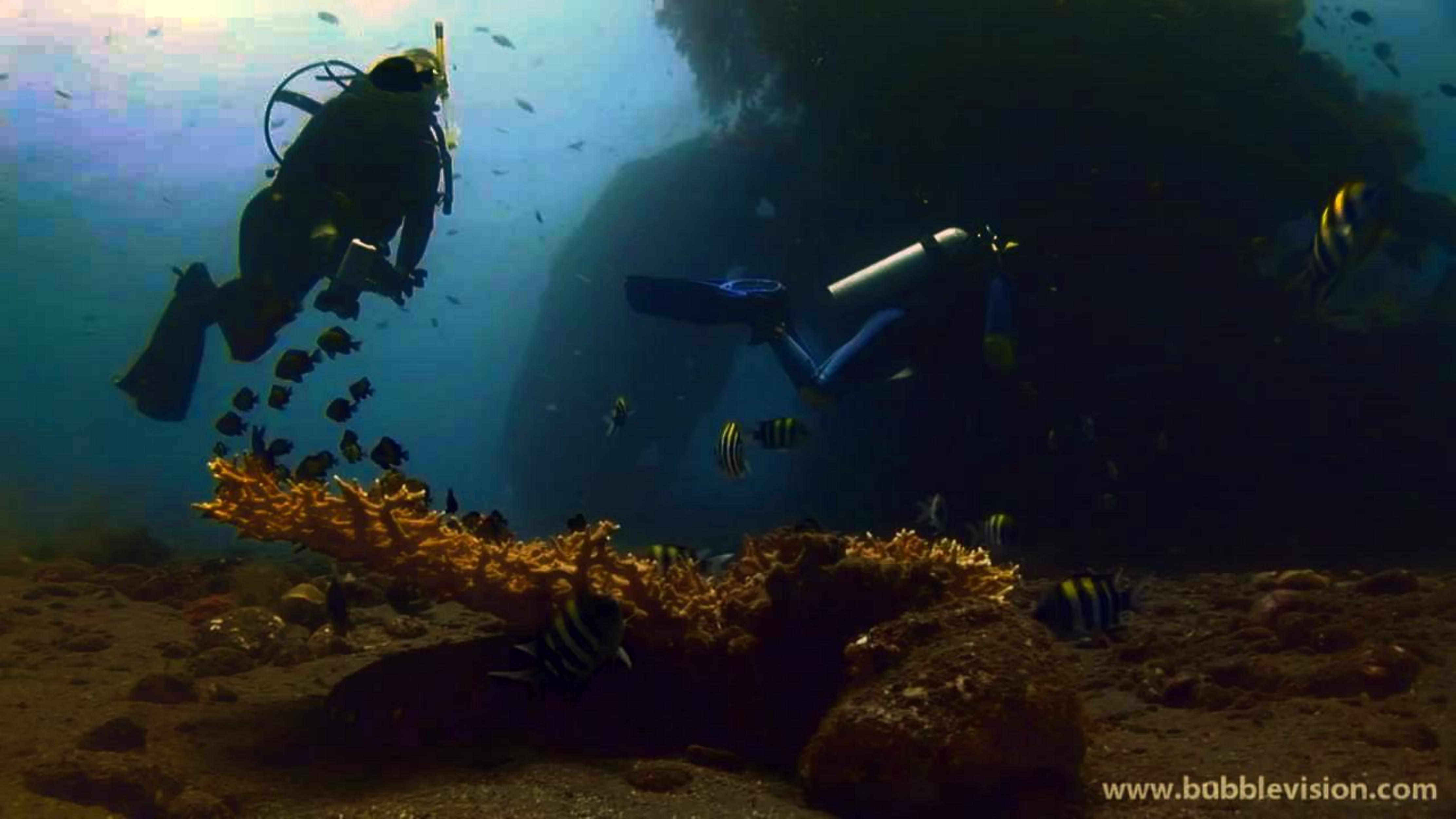}&
\includegraphics[width=3cm, height=2cm]{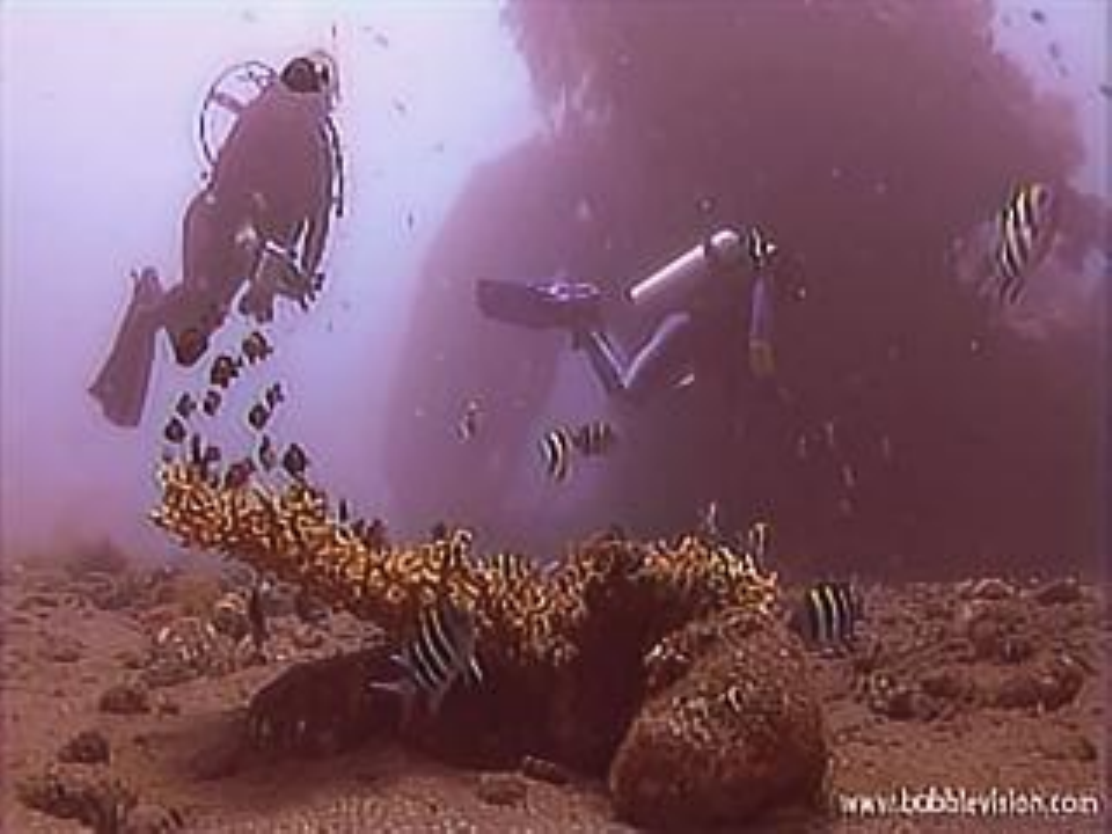}&
\includegraphics[width=3cm, height=2cm]{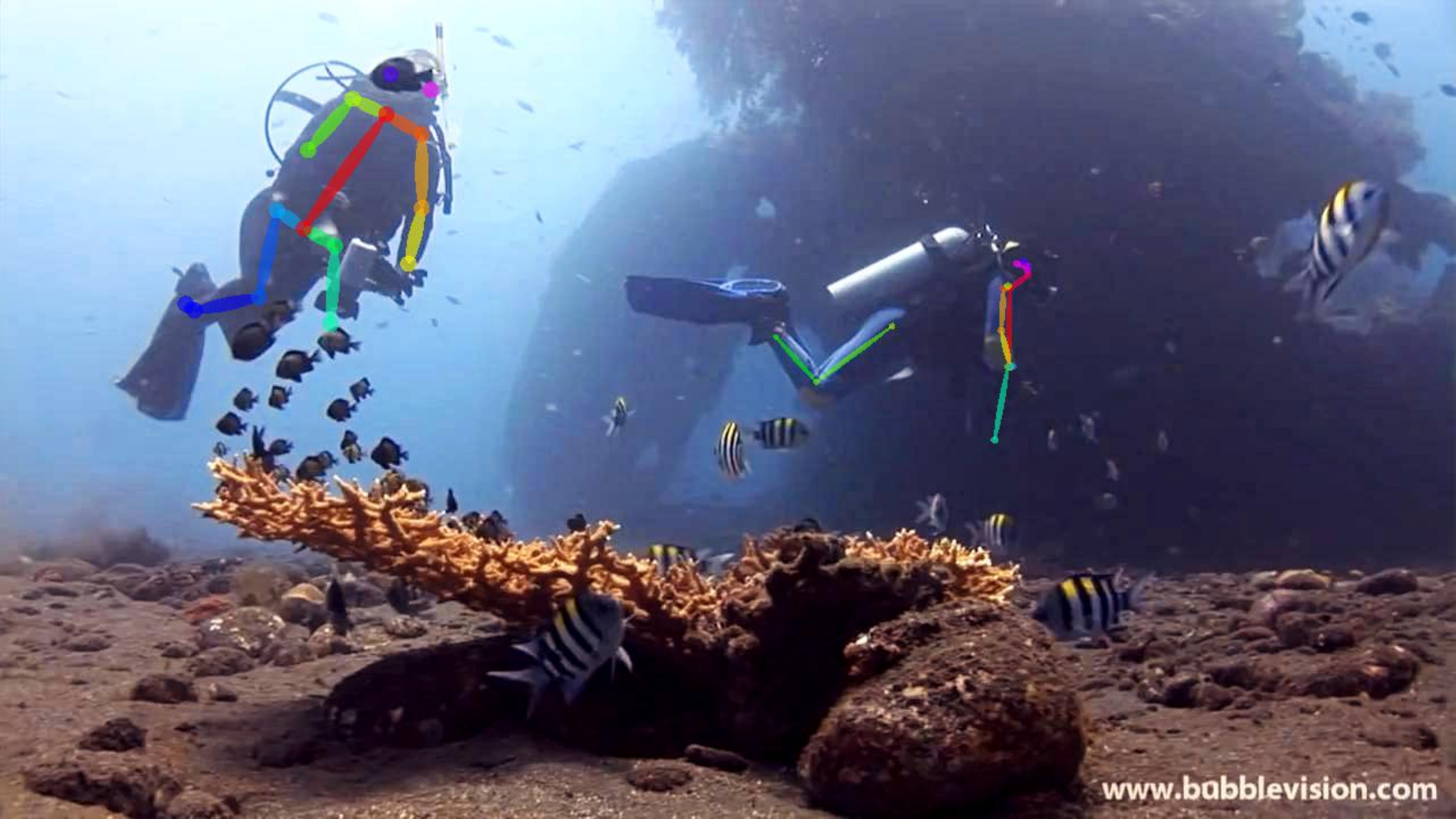}\\

 { Degraded} & { Retinex-based \cite{retinex-based} } & { Fusion-based \cite{ancuti1} } & { GDCP \cite{gdcp} } & { Haze Lines \cite{berman_pami_20} } & { Deep SESR \cite{sesr} } & { Deep WaveNet} \\

\end{tabular}}
\caption{Qualitative demonstration of the 2D pose estimation task by using the enhanced images from various UIR existing works.}
\label{fig:2d_pose}
\end{figure*}

Quantitative results have been presented in Tables \ref{tab:euvp}, \ref{tab:uieb} and \ref{tab:sr}. A comprehensive comparison against best-published works for underwater image enhancement has been presented in Tables \ref{tab:euvp} and \ref{tab:uieb}. Whereas the results on underwater single image super-resolution have been given in Table \ref{tab:sr}. It can be observed from Table \ref{tab:euvp} that the proposed model has outperformed the existing works on almost all adopted image quality metrics. While on PSNR and SSIM, the proposed work has shown a significant improvement of $5.68$\%, $3.75$\%, respectively, a notable increment of $\sim .7$ and $\sim 11$ has been observed in terms of VIF and NIQE over one of the most recent works in UIR, namely Deep SESR \cite{sesr}. It can also be observed that the proposed model has procured supremacy on $7$ out of $12$ adopted image quality metrics in the case of underwater image enhancement. The quantitative comparison on \texttt{UIEB} test set has been shown in Table \ref{tab:uieb}. It can be observed from Table \ref{tab:uieb} that the proposed scheme has significantly outperformed the existing best-published works on the \texttt{UIEB} test set in terms of all mentioned metrics. Especially in the case of PSNR, on which the proposed model has shown a notable improvement of $13$\% over WaterNet \cite{uieb}. 

We have also presented the visual comparison of the proposed work against existing methods in Figure \ref{fig:uieb2}. It can be observed from Figure \ref{fig:uieb2} that the existing methods Retinex \cite{retinex-based} and Haze Lines \cite{berman_pami_20} suffer from under and over-saturation of colors in enhanced images. While the methods like GDCP \cite{gdcp} and Fusion-based \cite{ancuti1} fail to recover the degraded images, the proposed method outputs the most visually pleasant enhanced images. We have shown the qualitative results of \texttt{UIEB Challenge} test-set as well. It can be observed that the enhanced results obtained by utilizing the proposed work retain the bluish essence of the water better than the existing works such as Ucolor \cite{tip21}. \textcolor{black}{The quantitative result on \texttt{Challenge} set has been shown in Table \ref{tab:tc60}. It can be observed that the proposed Deep WaveNet model has outperformed the existing best-published works on underwater image enhancement with a significant improvement in terms of UIQM and NIQE.} It may be due to the wavelength-specific receptive field sizes of the kernels dedicated to each color channel and adaptive usage in residual learning. 

Table \ref{tab:sr} presents the quantitative results on the underwater single image super-resolution task (SISR). Note that all the enlisted best-published methods for traditional SISR in Table \ref{tab:sr} have been trained and tested on the underwater dataset. However, due to the unavailability of the executable codes, we have directly presented the values mentioned in Deep SESR \cite{sesr} for SRCNN, SRResNet, and SRGAN. It can be observed from Table \ref{tab:sr} that the proposed wavelength-specific multi-contextual deep CNN has outperformed the existing best-published works for the task of underwater SISR.  Even though a slight underperformance has been observed in PSNR, a significant performance gain has been observed in terms of SSIM and UIQM. Particularly, in the case of $4\times$, where the proposed work has outperformed the Deep SESR \cite{sesr} by an improvement of $\sim 12$\% in SSIM.
  
A qualitative comparison has also been given in Fig. \ref{fig:results_sr} for the task of underwater SISR. It can be observed that the results produced by SRDRM \cite{srdrm_srdrmgan} and SRDRM-GAN \cite{srdrm_srdrmgan} do not entirely remove the color distortions in the enhanced images. Further, the results obtained by using Deep SESR \cite{sesr} still comprise of original noise traces. Whereas the enhanced images generated by using Deep WaveNet are artifacts-free and perceptually similar to ground truth images. \textbf{{More quantitative and qualitative results on underwater image enhancement and SISR are given in the provided supplementary material. We encourage the readers to refer it.}}

\section{Effect on High-Level Vision Tasks}
\label{sec:more_results}
We have also verified the robustness of the enhanced images produced by the Deep WaveNet against existing UIR methods on a few underwater high-level vision tasks. For this, in particular, we have considered underwater single image semantic segmentation and diver's 2D pose estimation tasks. 
\subsection{Underwater Image Semantic Segmentation}
Semantic segmentation deals with scene understanding by assigning the dense labels to all pixels in an image \cite{segmentation_intro}. Its applications span from ground-level autonomous driving to satellite remote sensing. For the task of underwater image semantic segmentation, we have considered the recent work \texttt{SUIM}\footnote{\url{https://github.com/xahidbuffon/SUIM}} \cite{suim} that assigns each pixel in one of the five groups, namely: (a) \texttt{Human divers}, (b) \texttt{Wrecks and ruins}, (c) \texttt{Robots and instruments}, (d) \texttt{Fish and vertebrates}, and (e) \texttt{Reefs and invertebrates}. It can be observed from Fig. \ref{fig:seg_maps} that the segmentation maps generated by utilizing the enhanced images of the proposed method are more refined than those of other existing best-published works. It may be due to attributing a larger contextual size to the blue channel that reflects the global coherence, resulting in quantitatively enhanced underwater images.

\subsection{Underwater Divers 2D Pose Estimation}
In addition, we have also considered one of the most challenging high-level vision tasks, \textit{i.e.}, diver's 2D pose estimation, in underwater imaging. For this, following \cite{funiegan}, we have utilized \texttt{OpenPose}\footnote{\url{https://github.com/CMU-Perceptual-Computing-Lab/openpose}} \cite{openpose1, openpose2, openpose3, openpose4} method to estimate the 2D pose of the human body in an image. Particularly, we have utilized enhanced images generated by the proposed model and existing works for this task, as shown in Fig. \ref{fig:2d_pose}. It can be observed that the pose key points reflected in the enhanced images of the proposed model are more accurate than those of the existing works \cite{retinex-based, ancuti1, gdcp,berman_pami_20, sesr} and original degraded images. Further, it can be concluded that an efficient model for a low-level vision task may also help in improving the performance of the relevant high-level vision tasks.

\subsection{Failure Case}

\begin{figure}[t]
\setlength{\tabcolsep}{0em}
\resizebox{\linewidth}{!}{
\begin{tabular}{ccccccc}
\includegraphics[width=3cm, height=3cm]{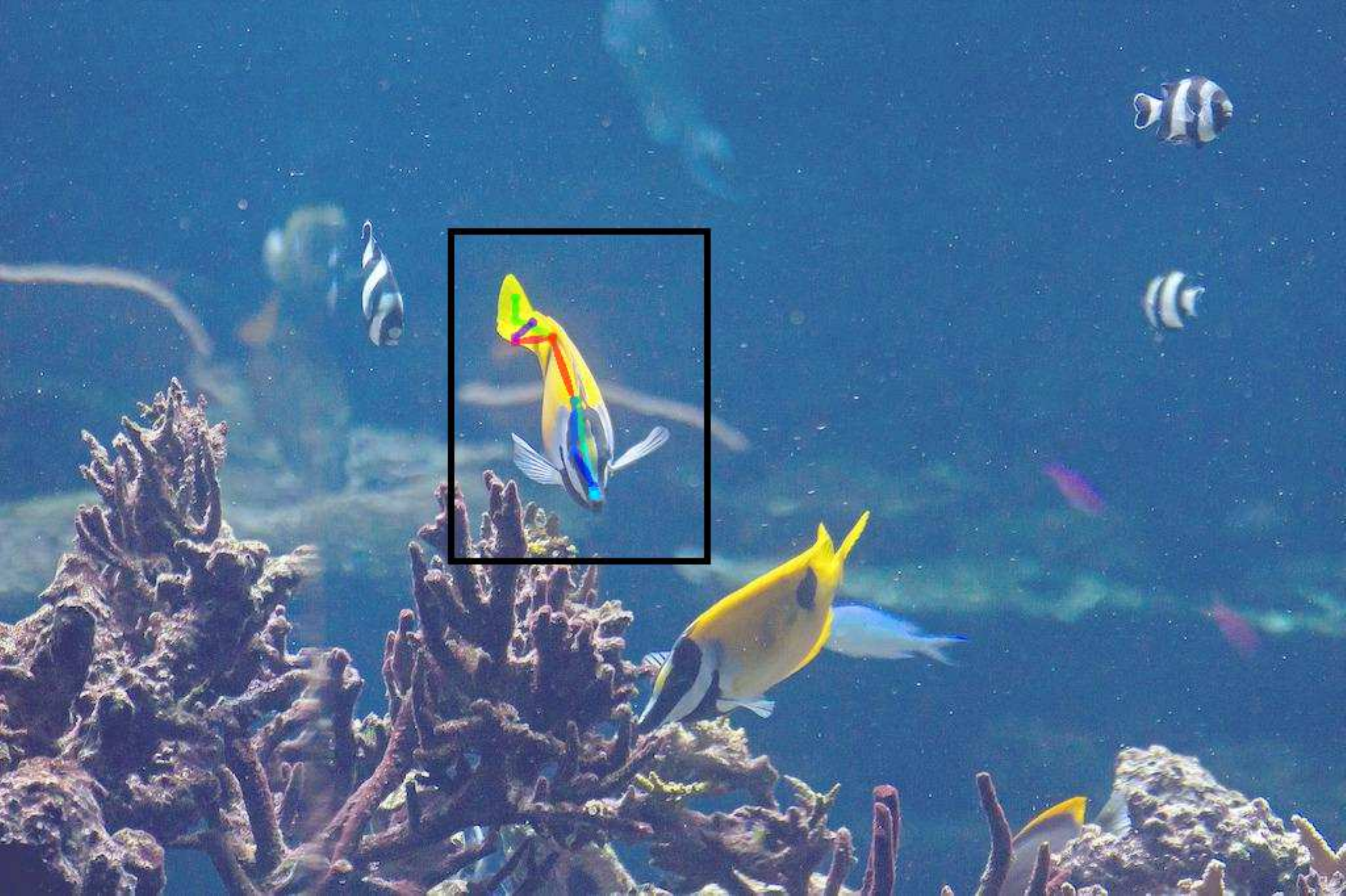} &
\includegraphics[width=3cm, height=3cm]{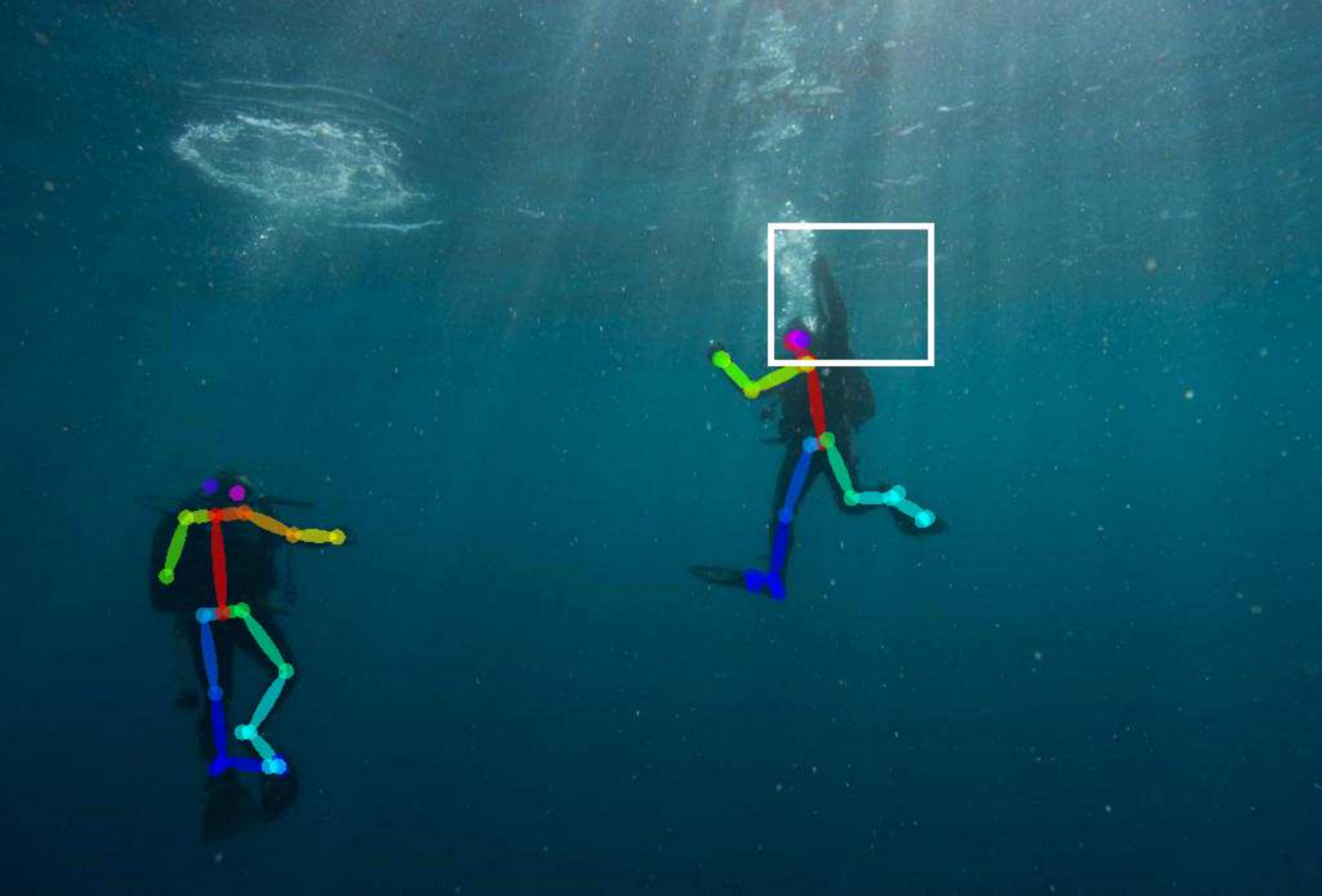} &
\includegraphics[width=3cm, height=3cm]{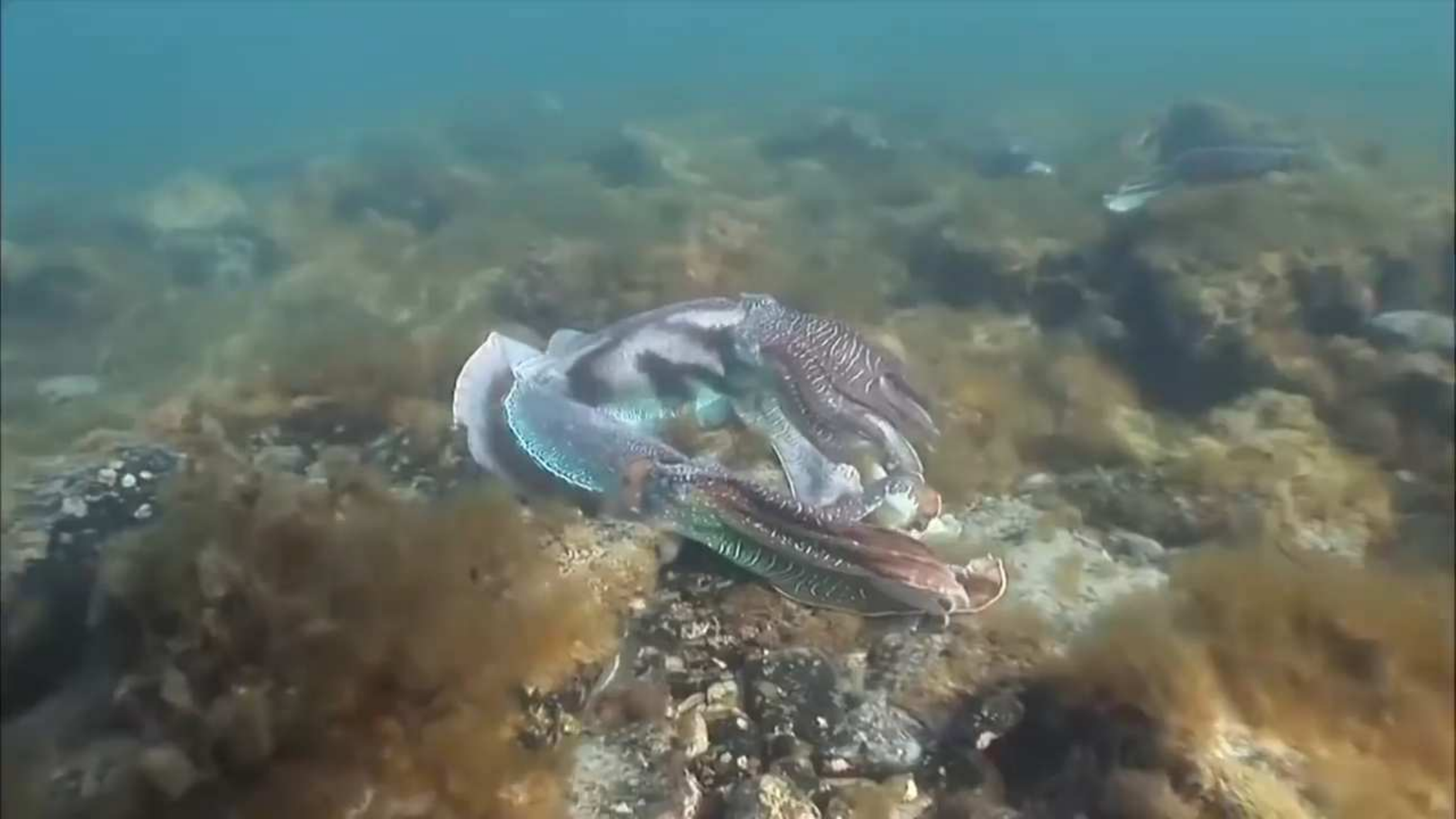} &
\includegraphics[width=3cm, height=3cm]{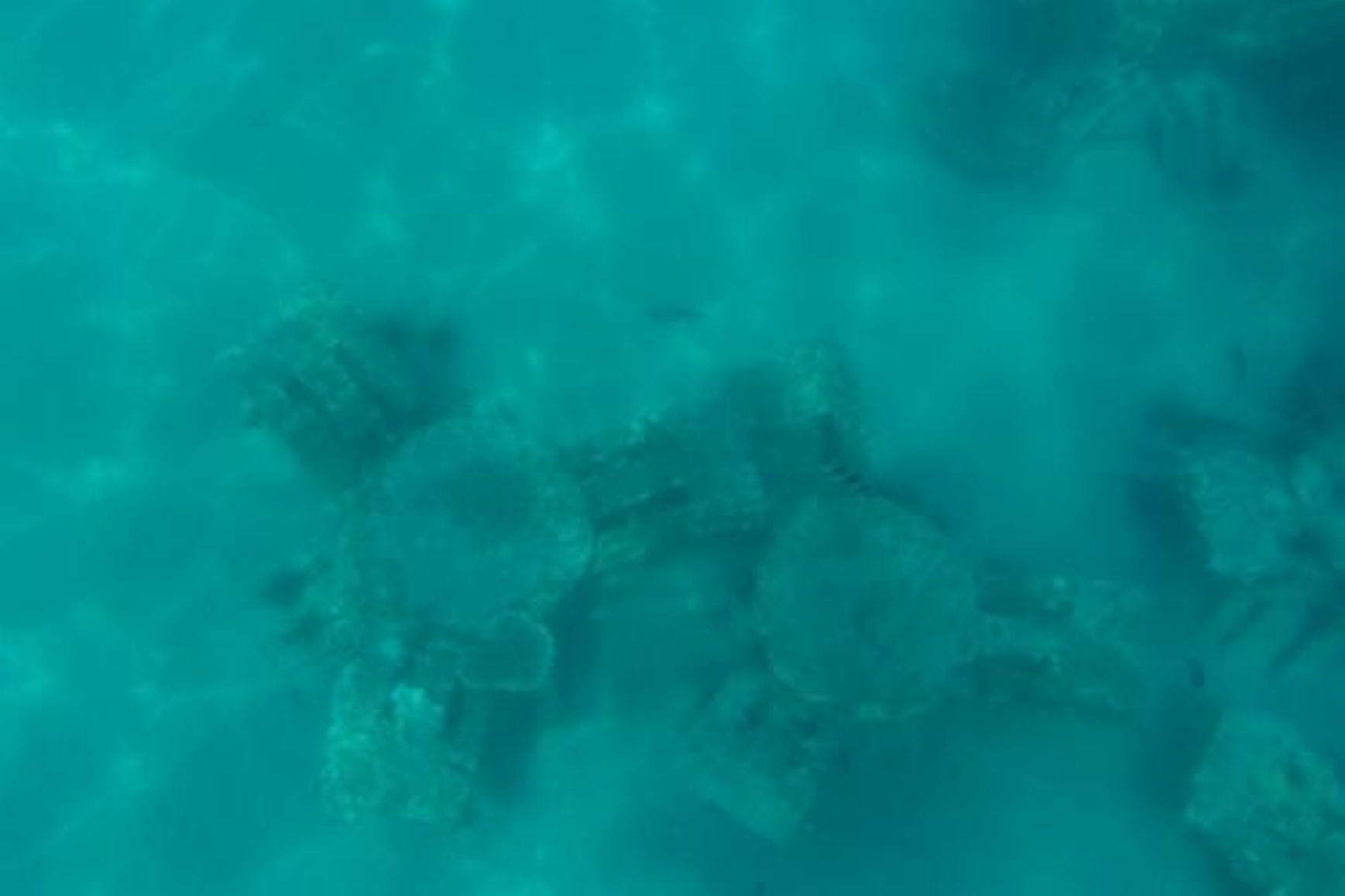} &
\includegraphics[width=3cm, height=3cm]{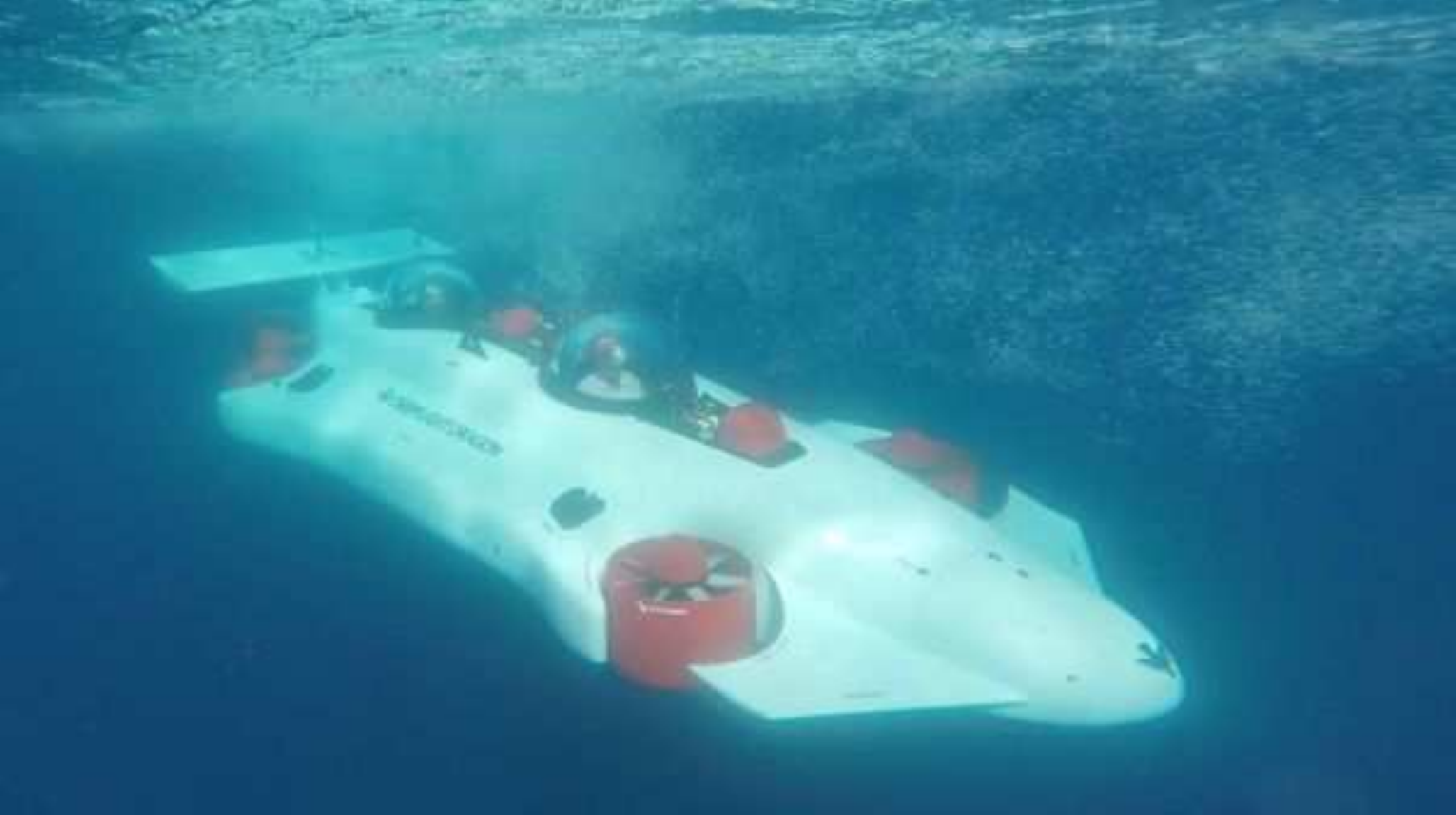} &
\includegraphics[width=3cm, height=3cm]{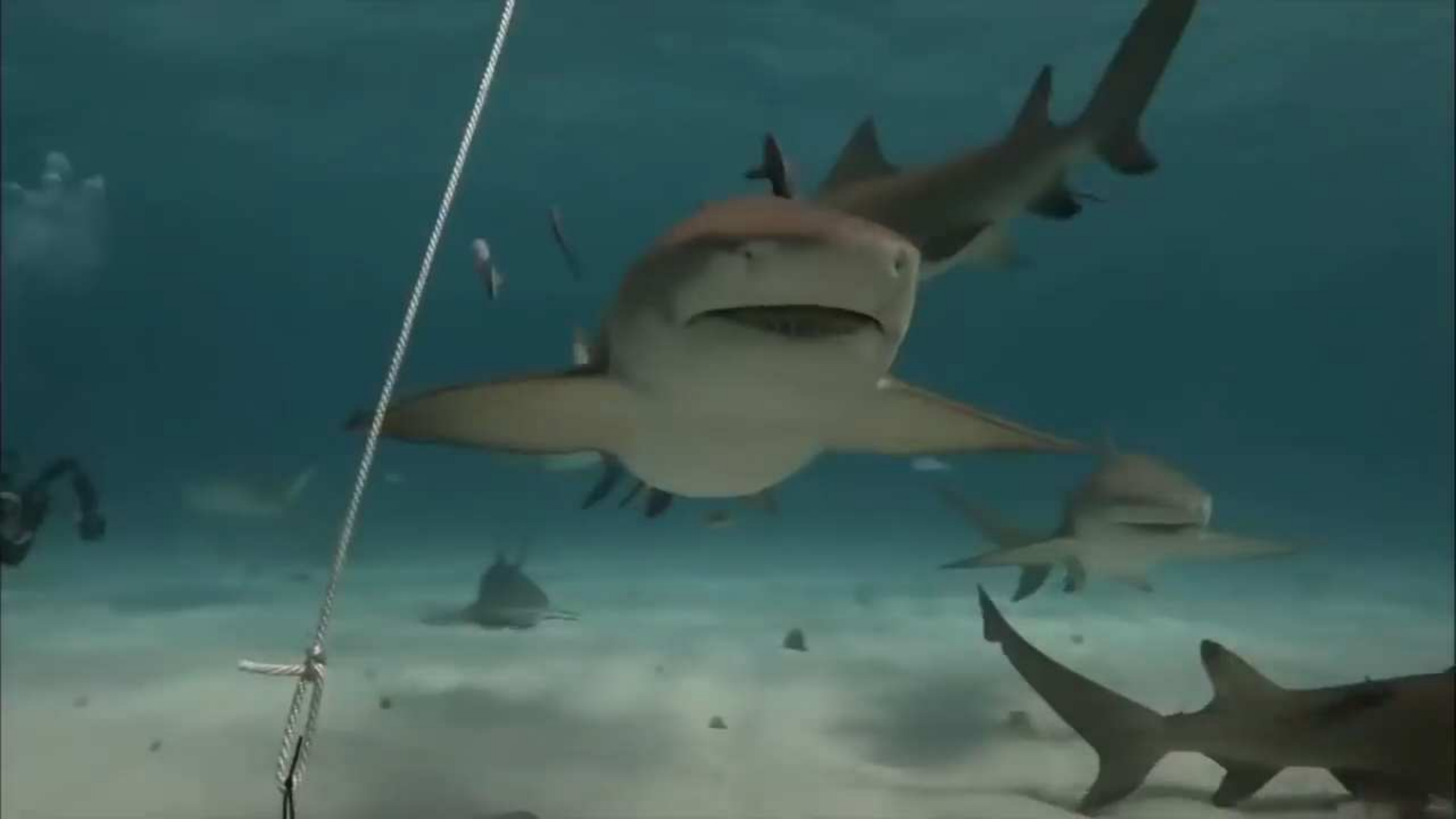} &
\includegraphics[width=3cm, height=3cm]{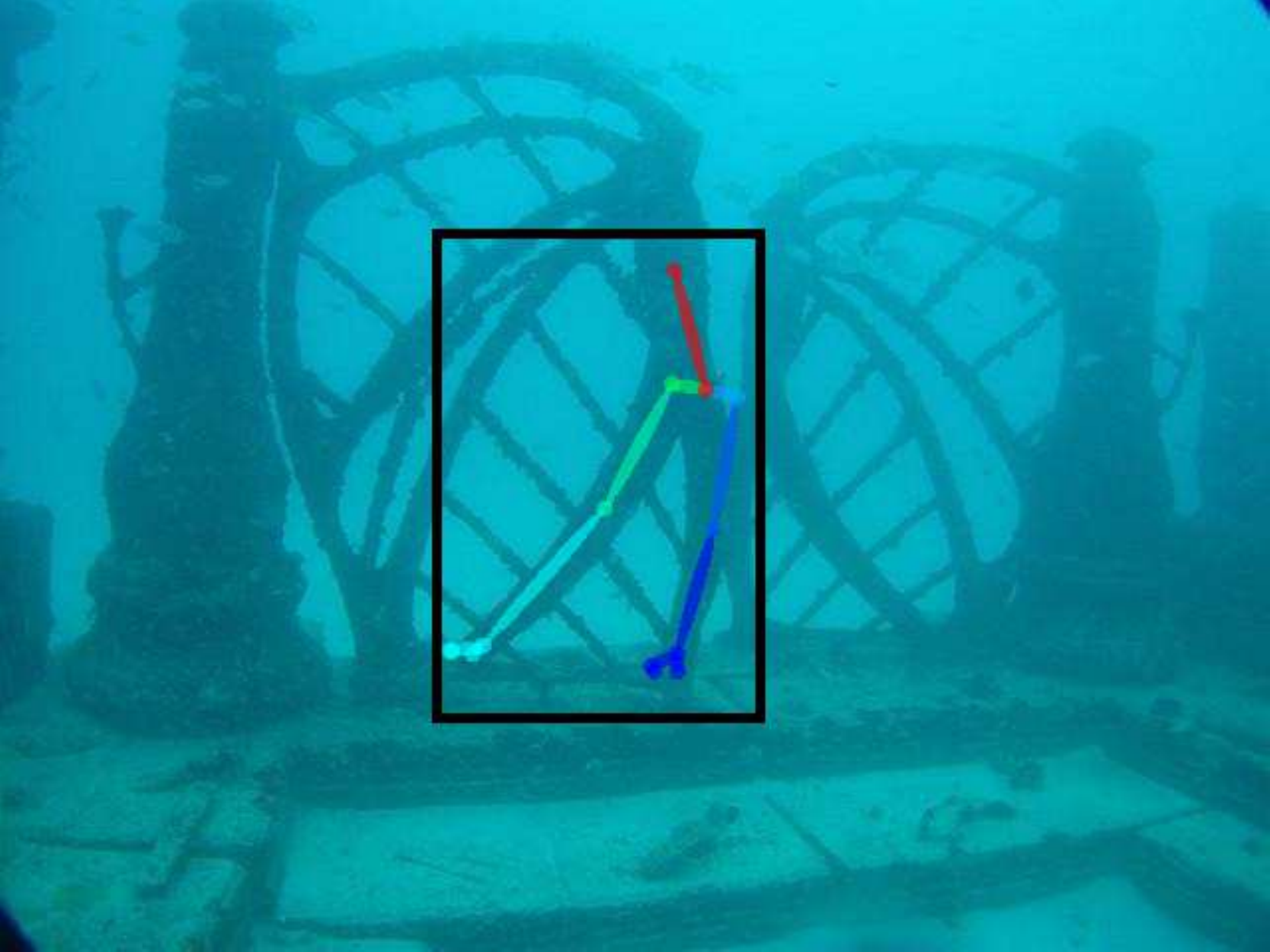} \\

\includegraphics[width=3cm, height=3cm]{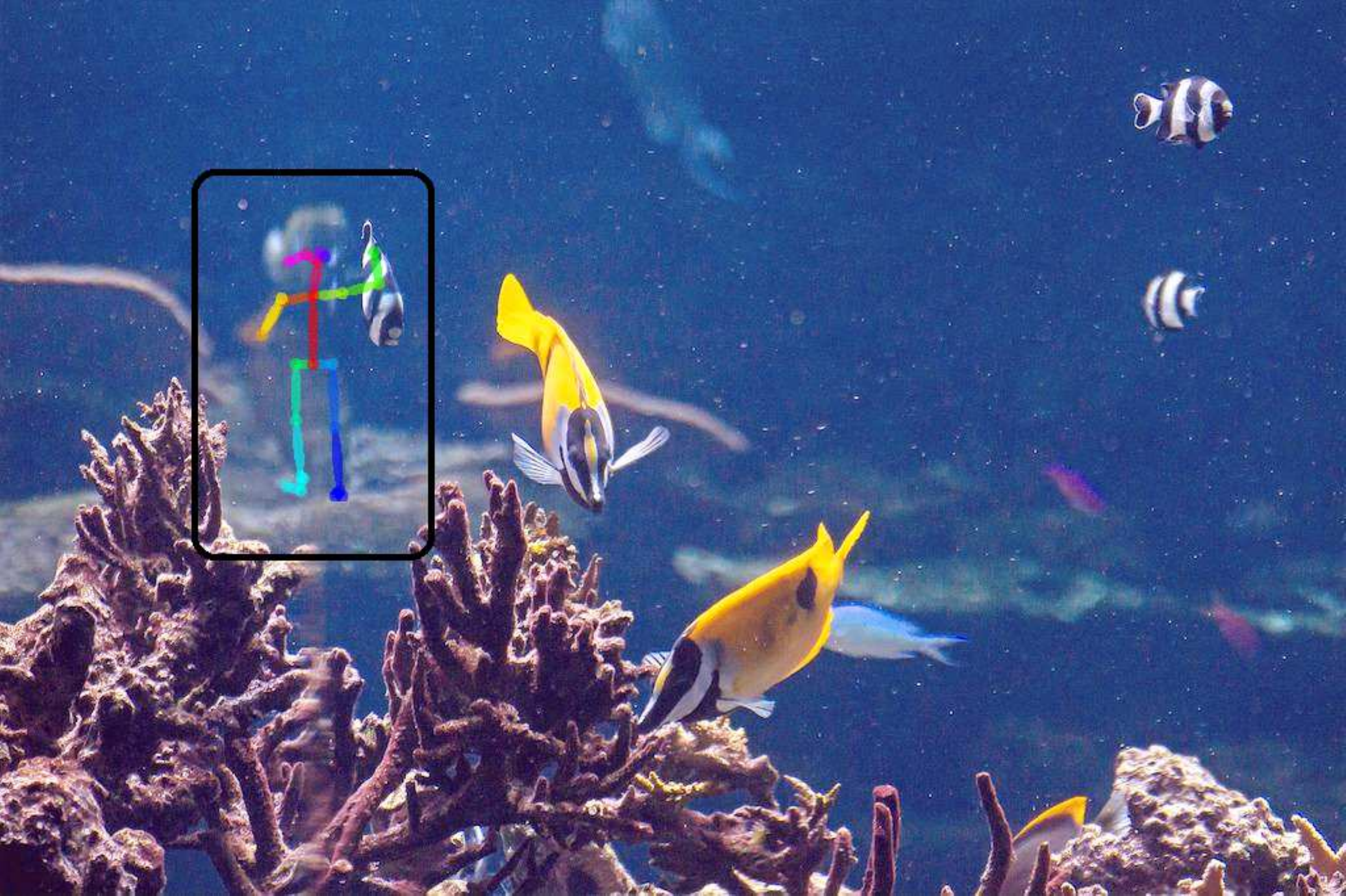} &
\includegraphics[width=3cm, height=3cm]{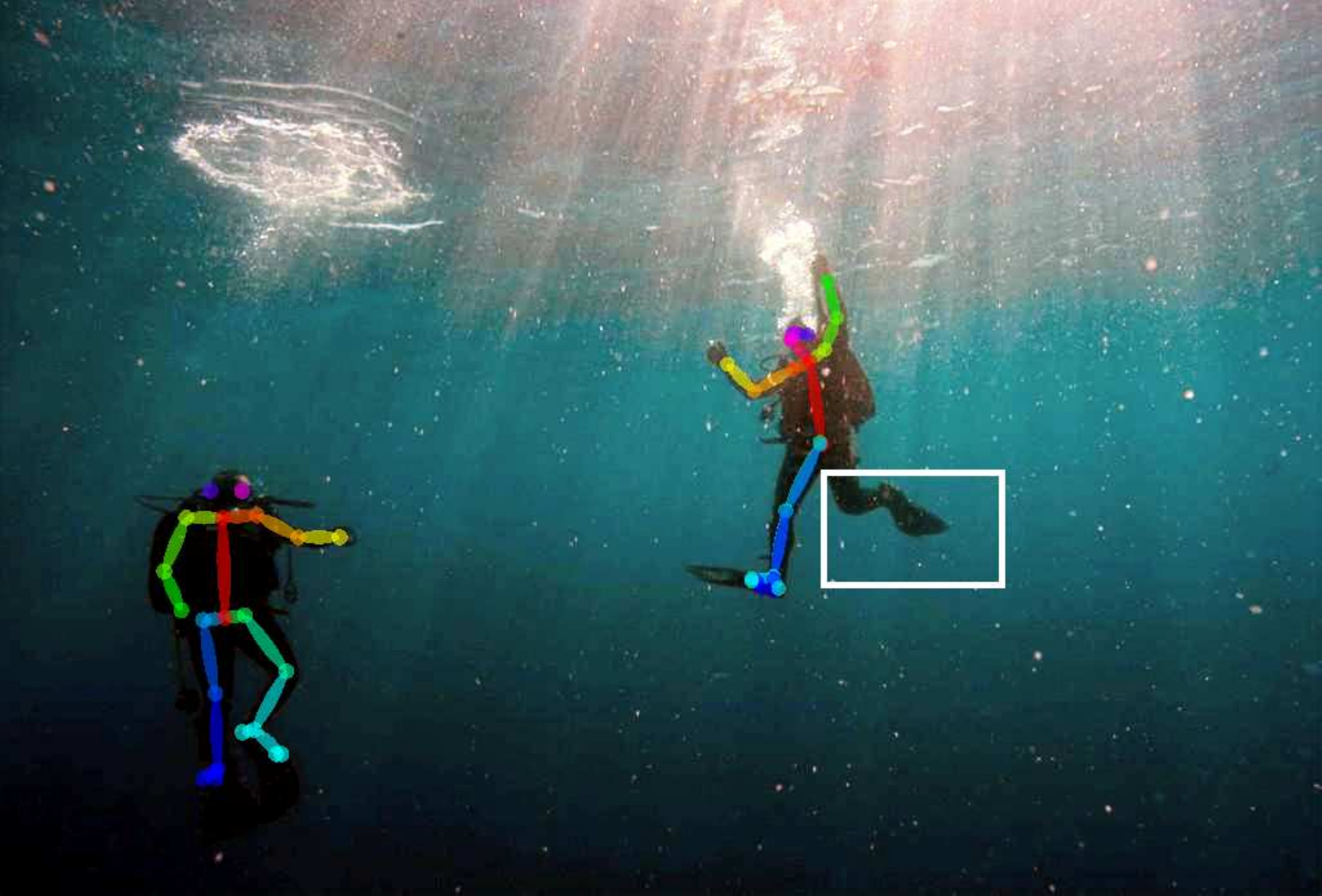} &
\includegraphics[width=3cm, height=3cm]{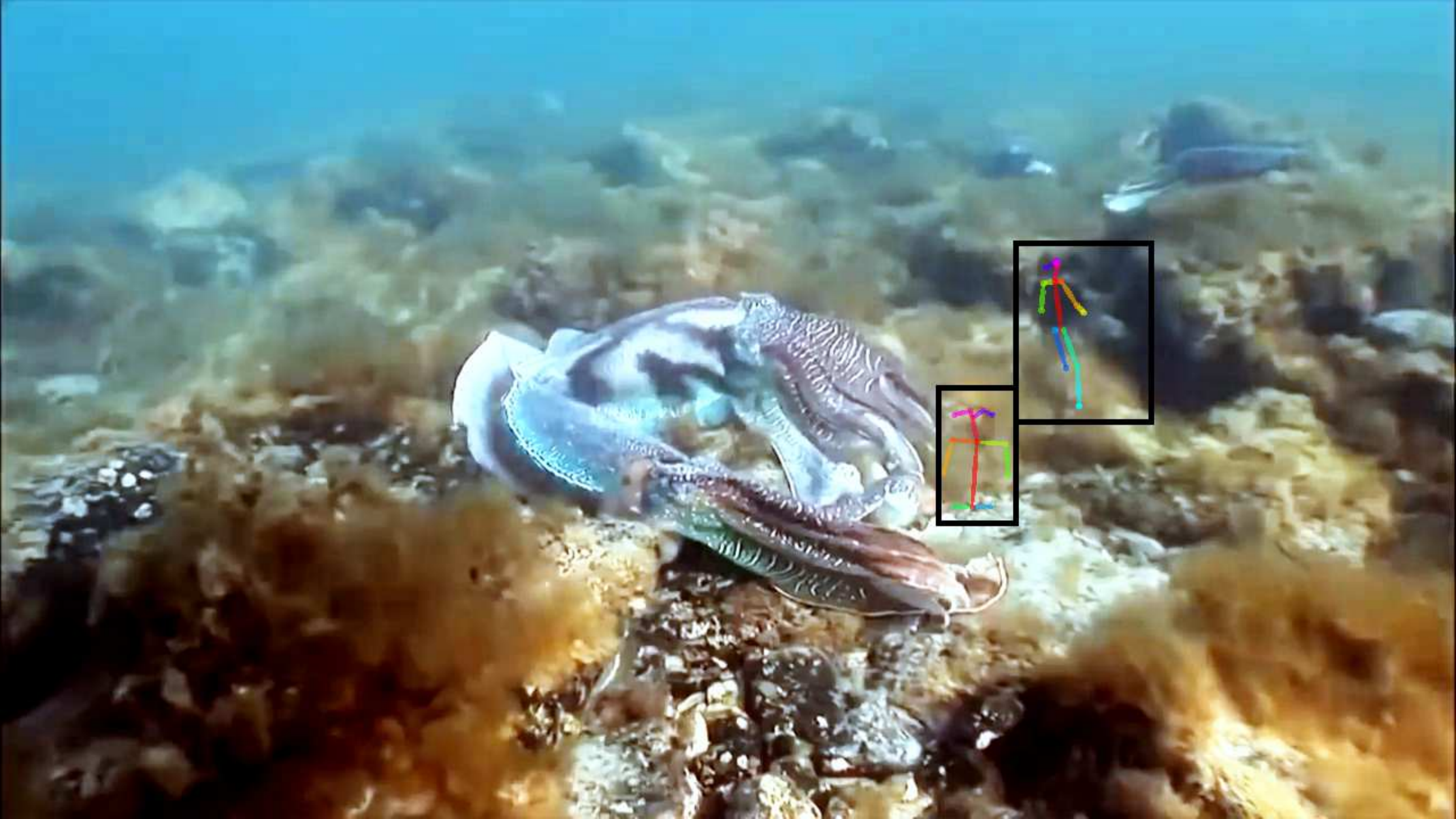} &
\includegraphics[width=3cm, height=3cm]{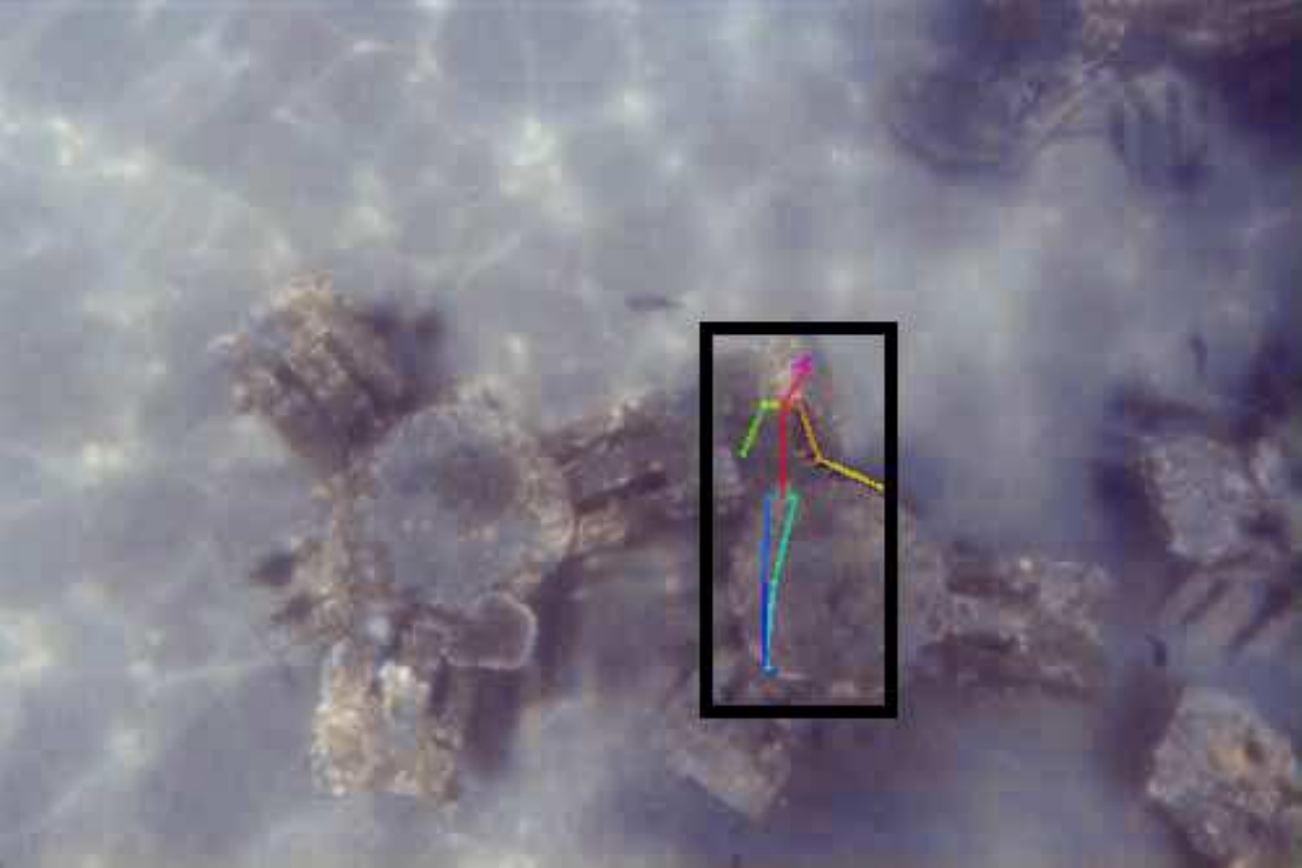} &
\includegraphics[width=3cm, height=3cm]{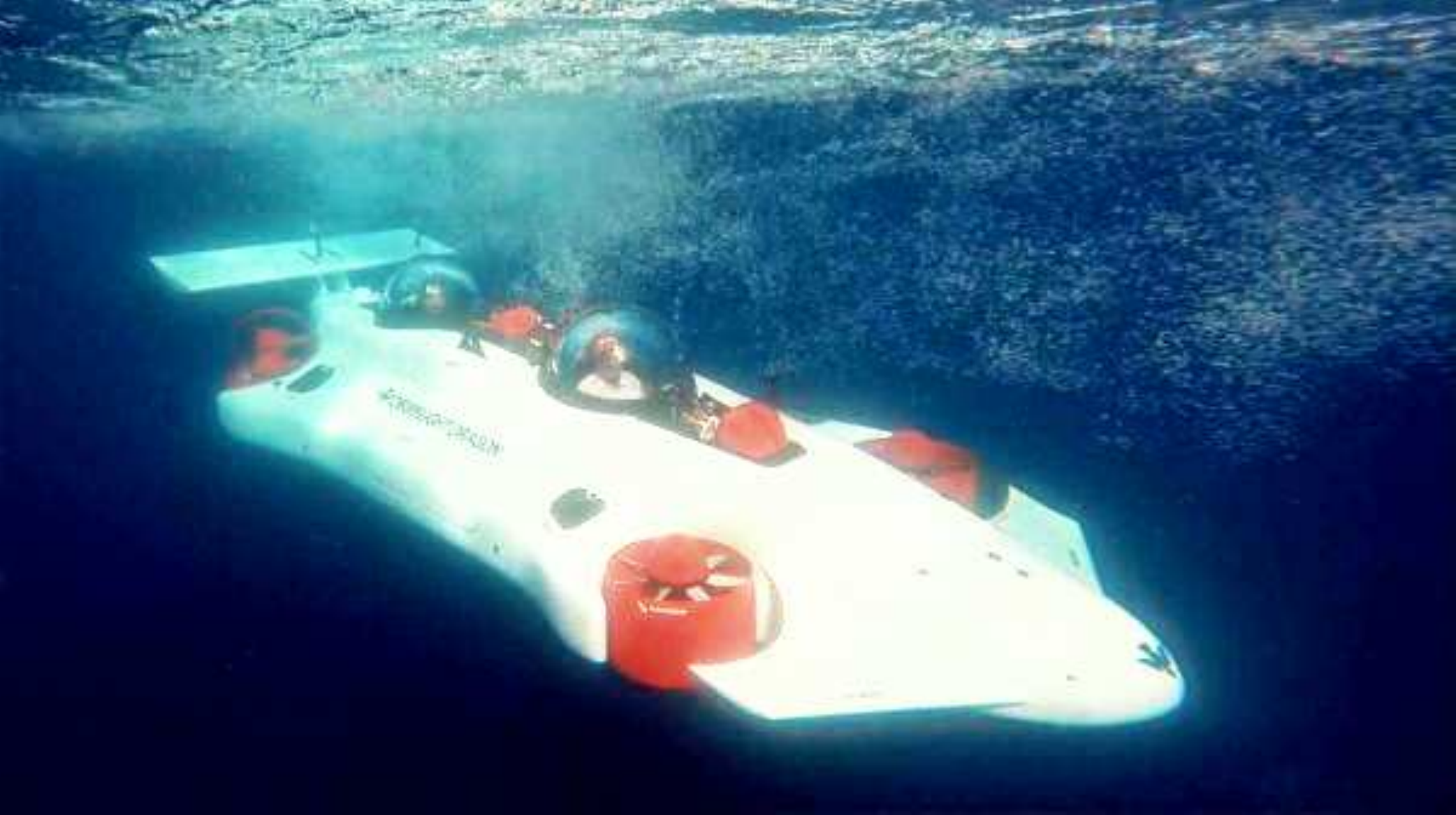} &
\includegraphics[width=3cm, height=3cm]{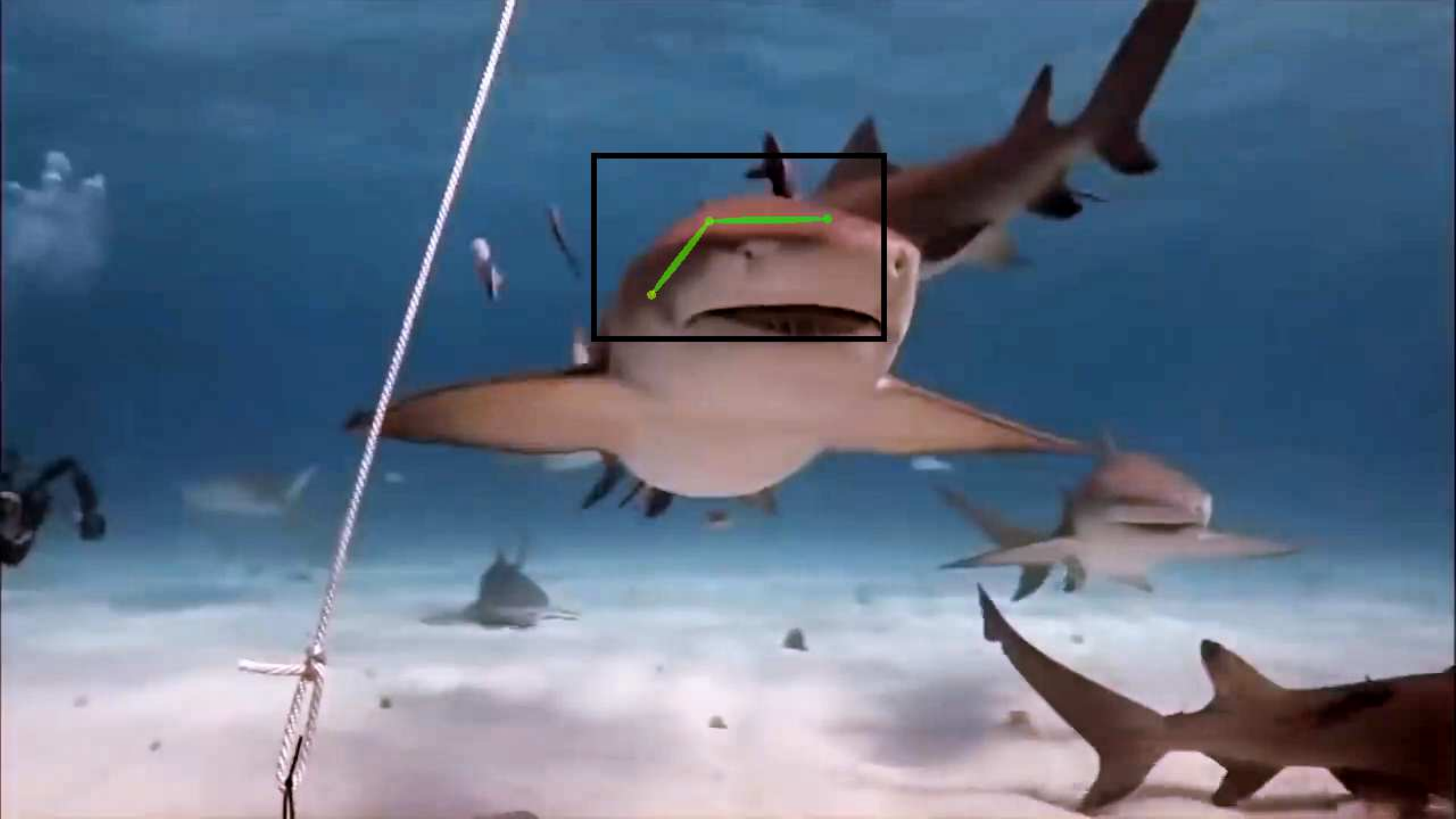} &
\includegraphics[width=3cm, height=3cm]{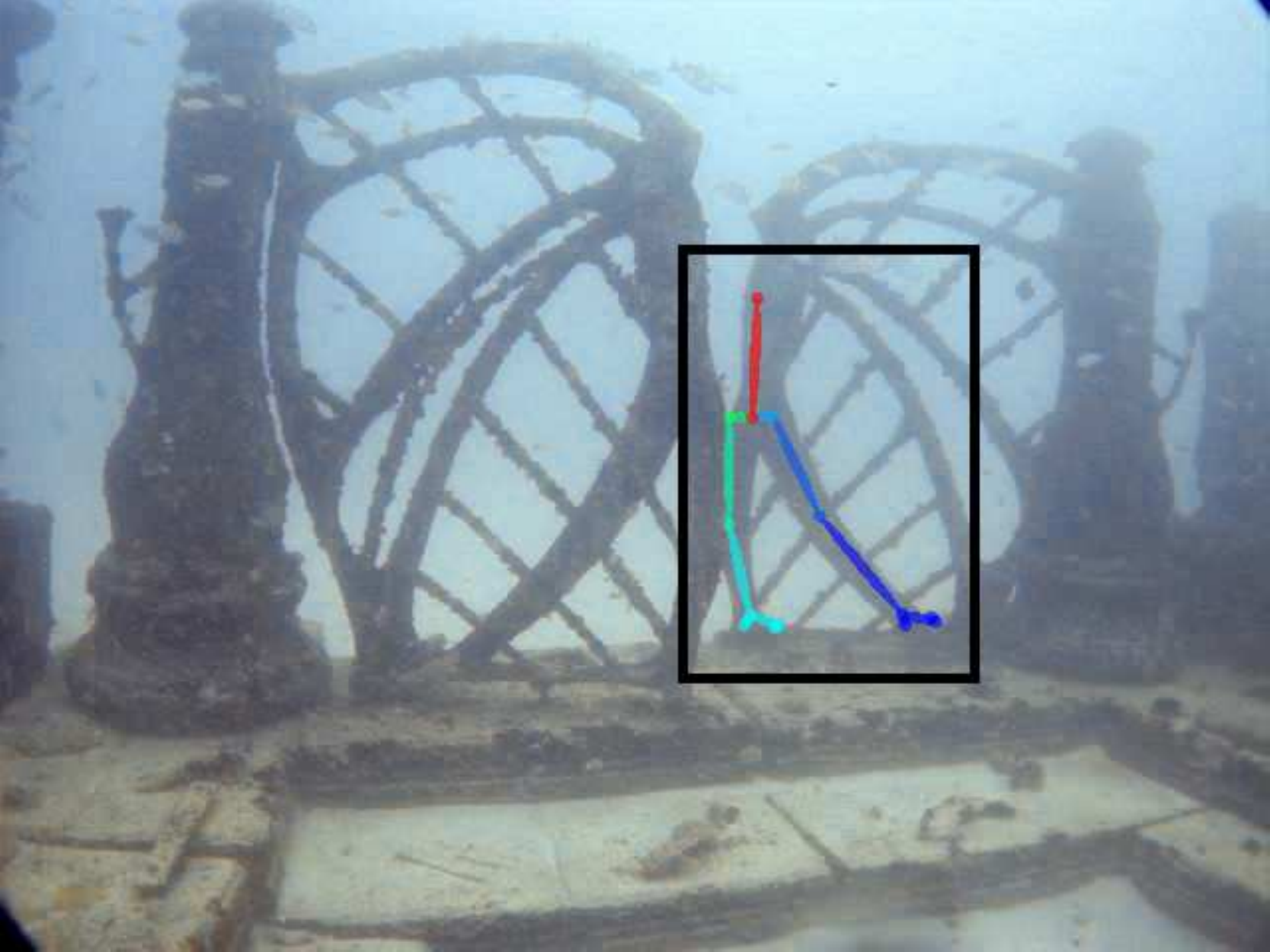} \\

\includegraphics[width=3cm, height=3cm]{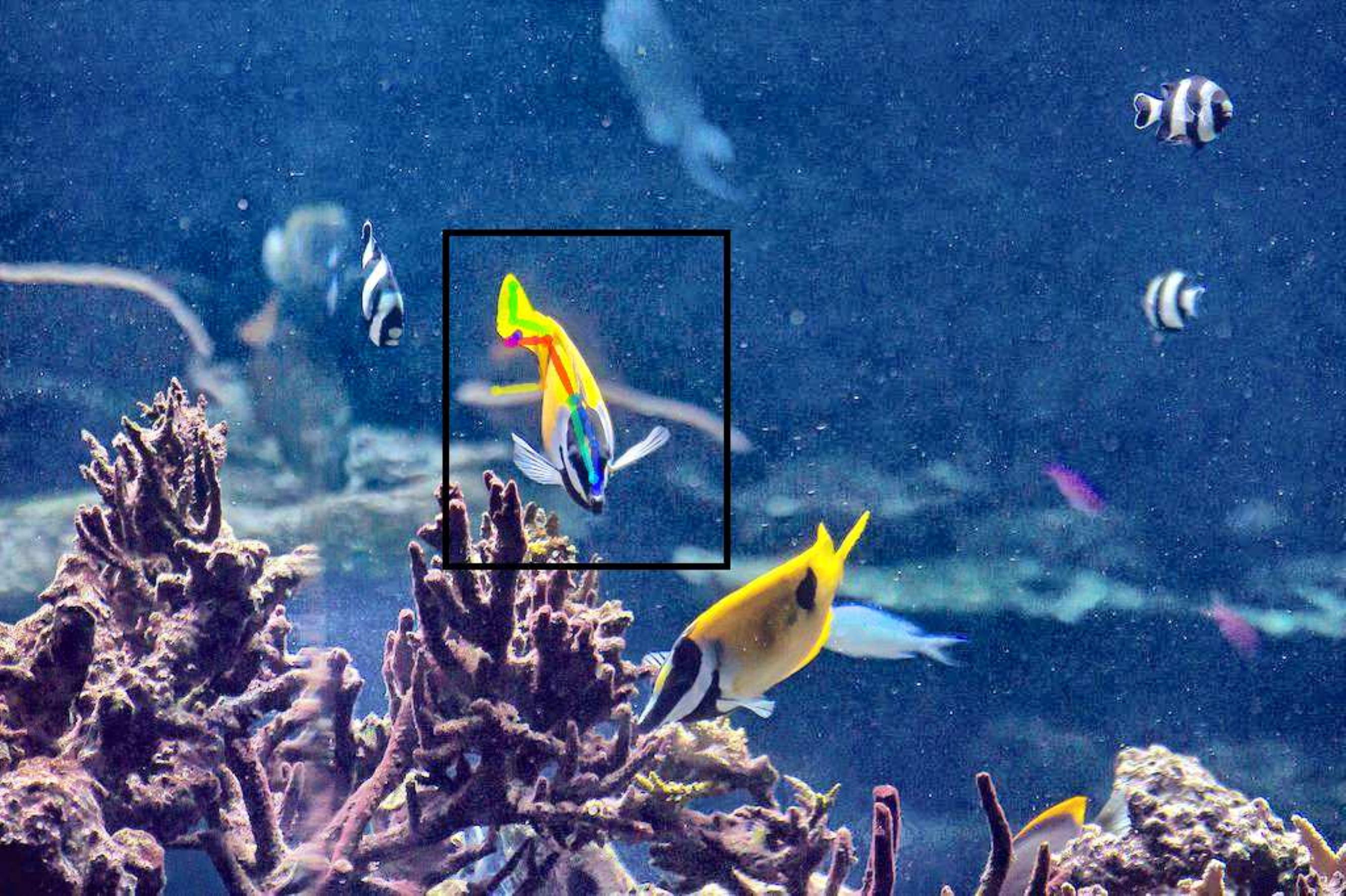} &
\includegraphics[width=3cm, height=3cm]{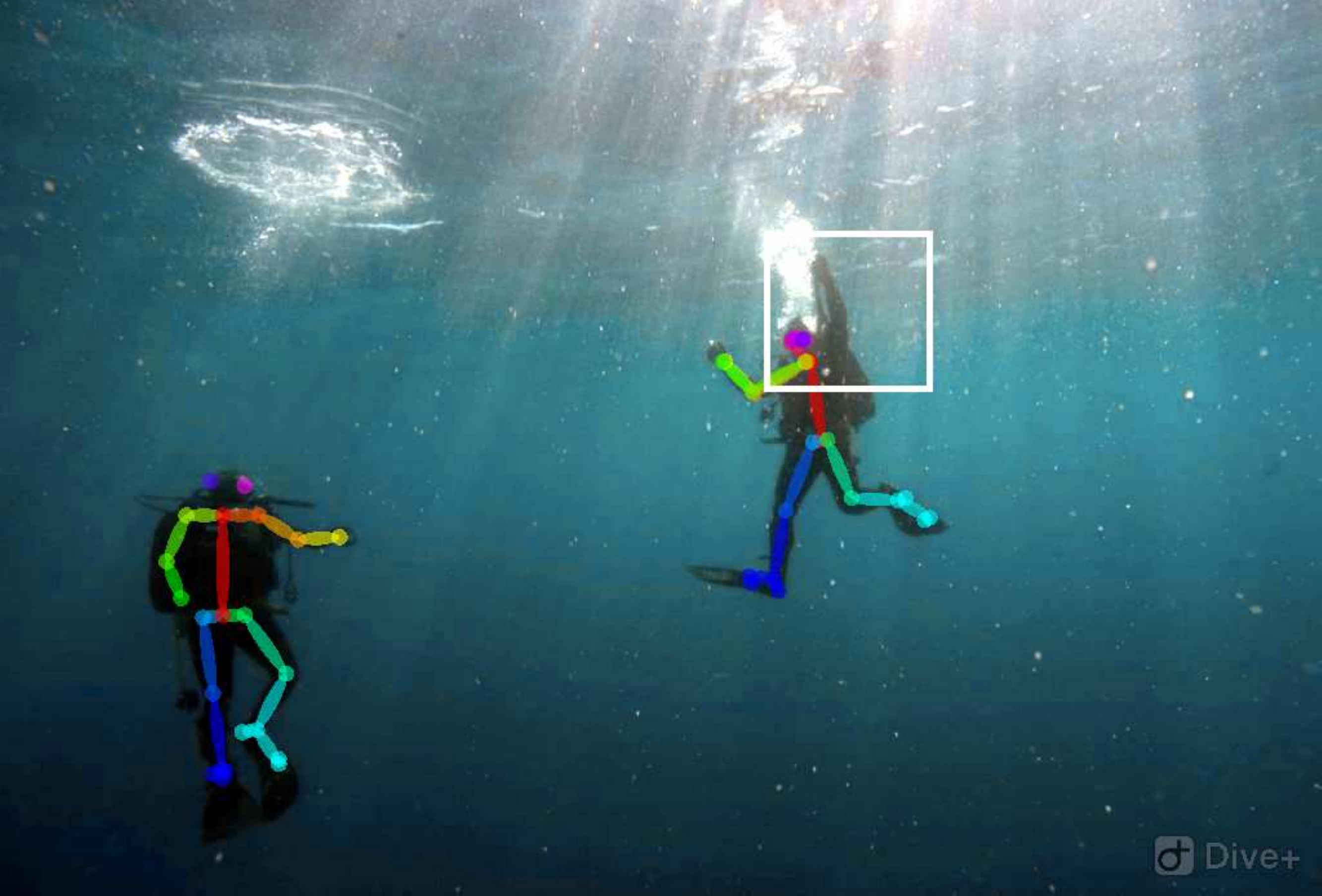} &
\includegraphics[width=3cm, height=3cm]{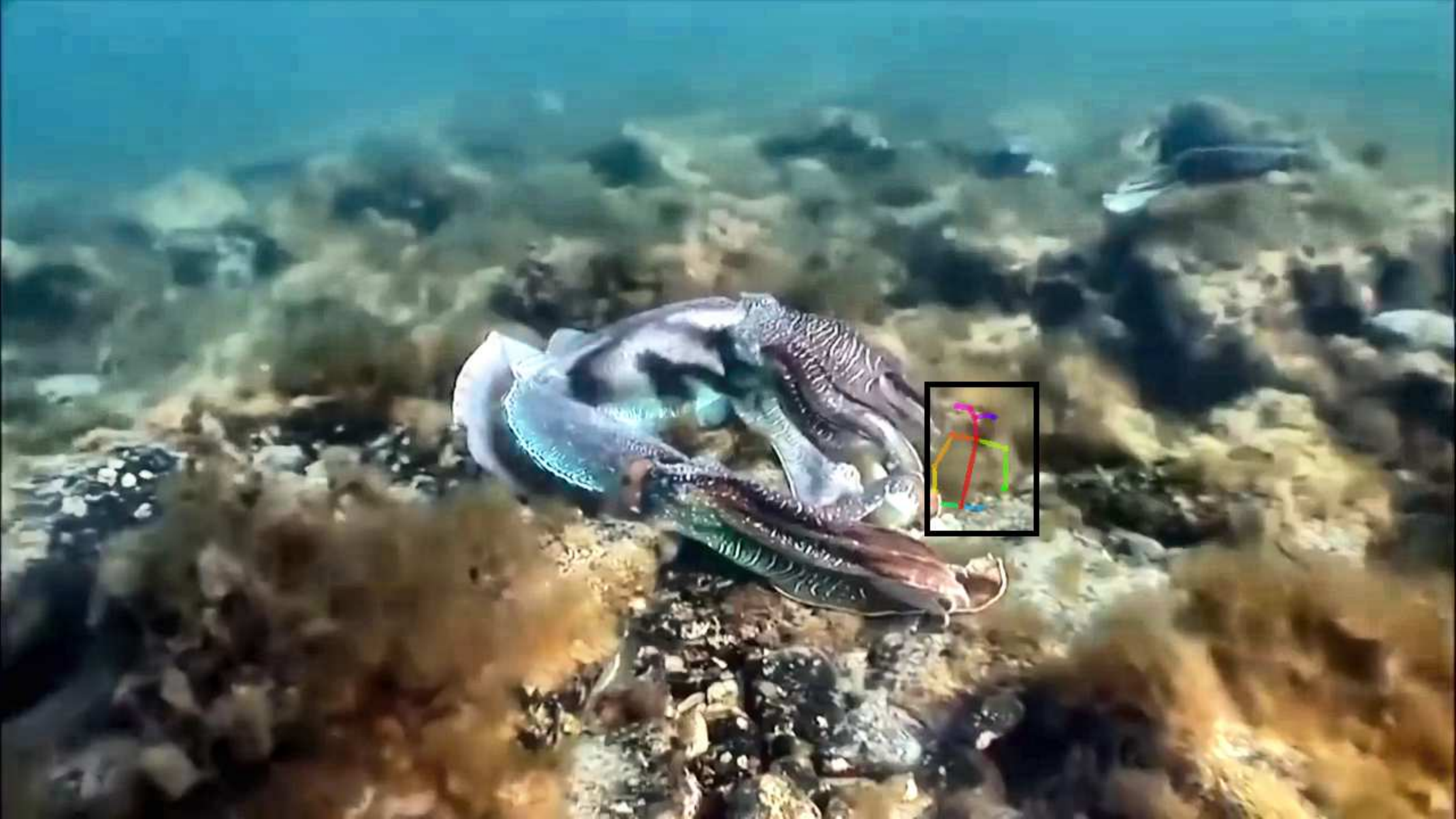} &
\includegraphics[width=3cm, height=3cm]{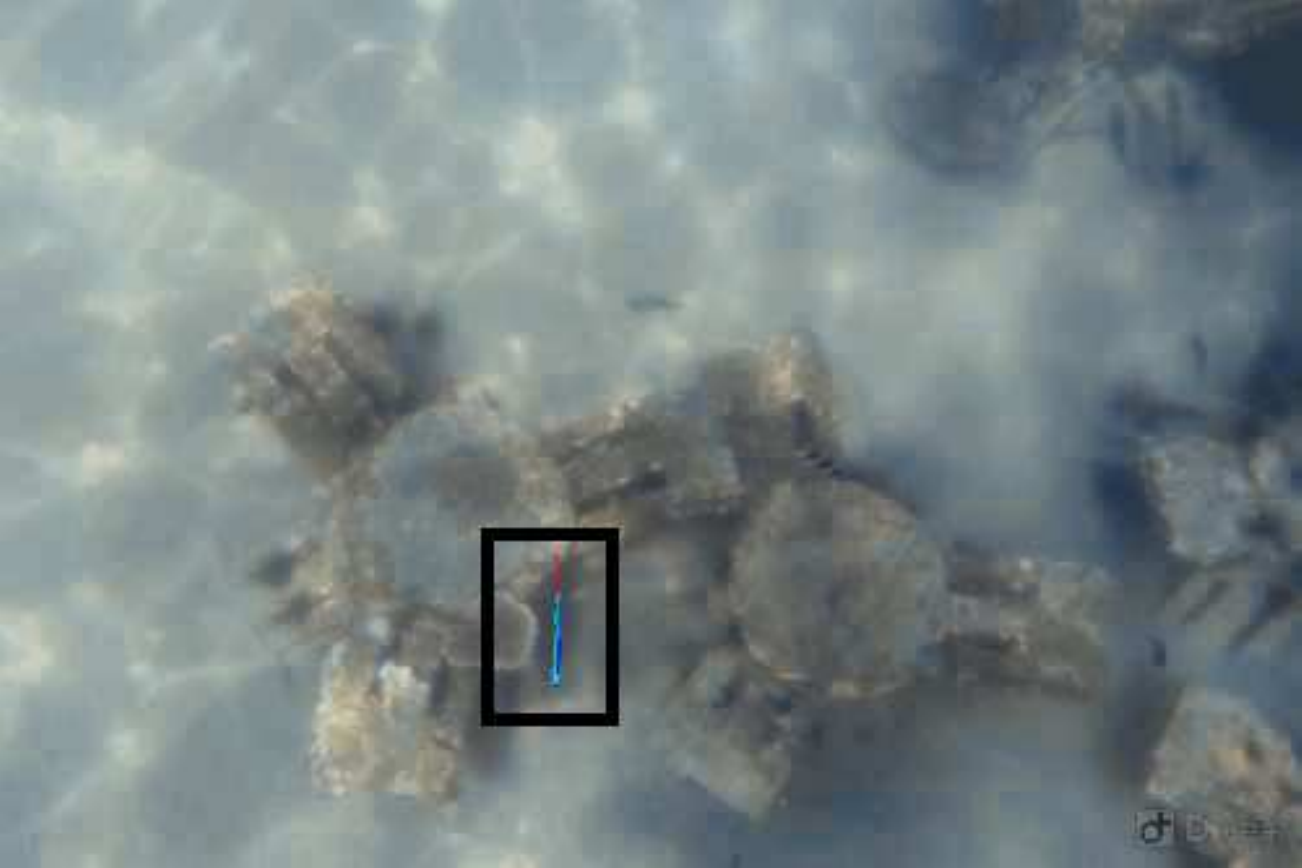} &
\includegraphics[width=3cm, height=3cm]{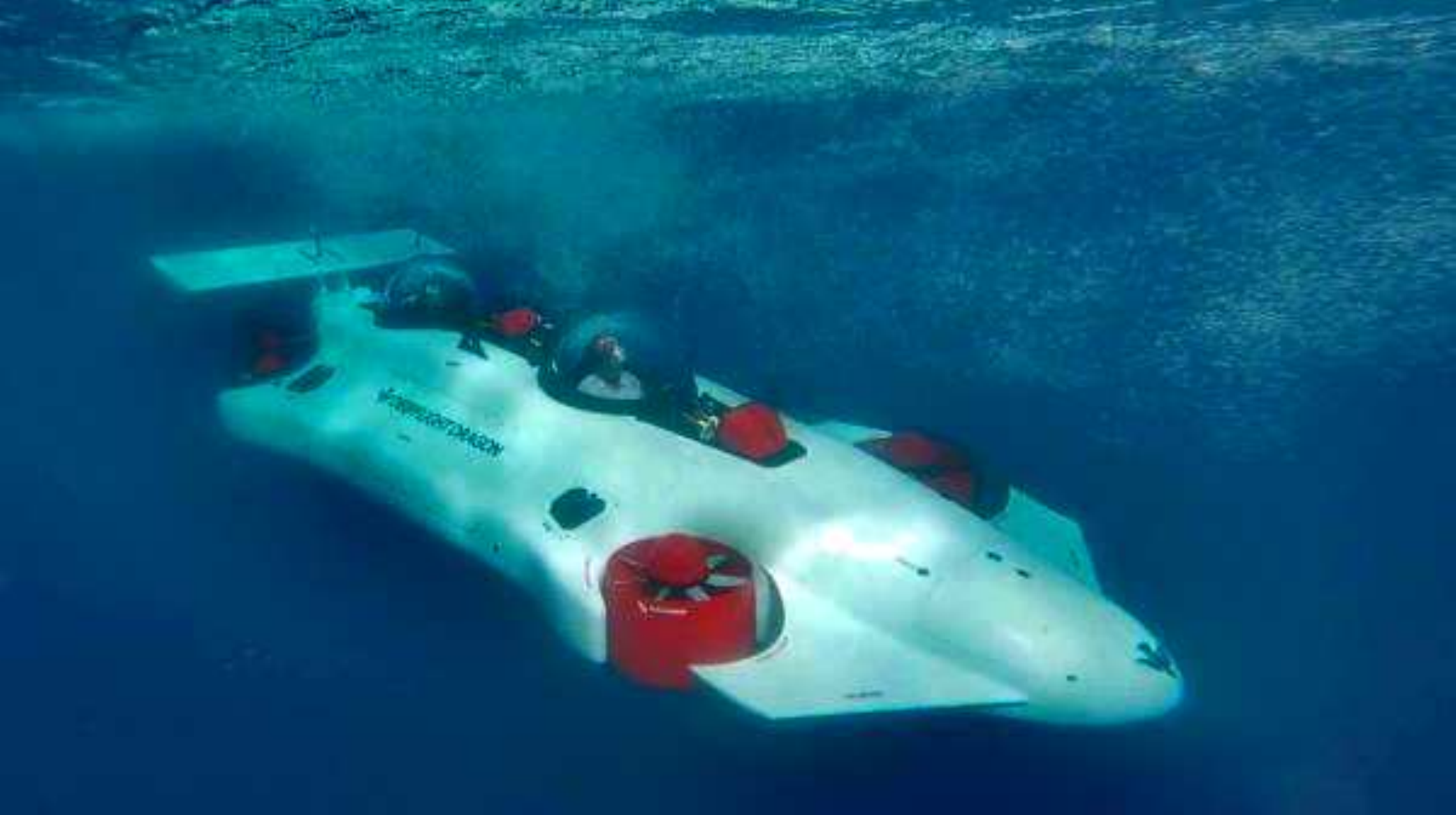} &
\includegraphics[width=3cm, height=3cm]{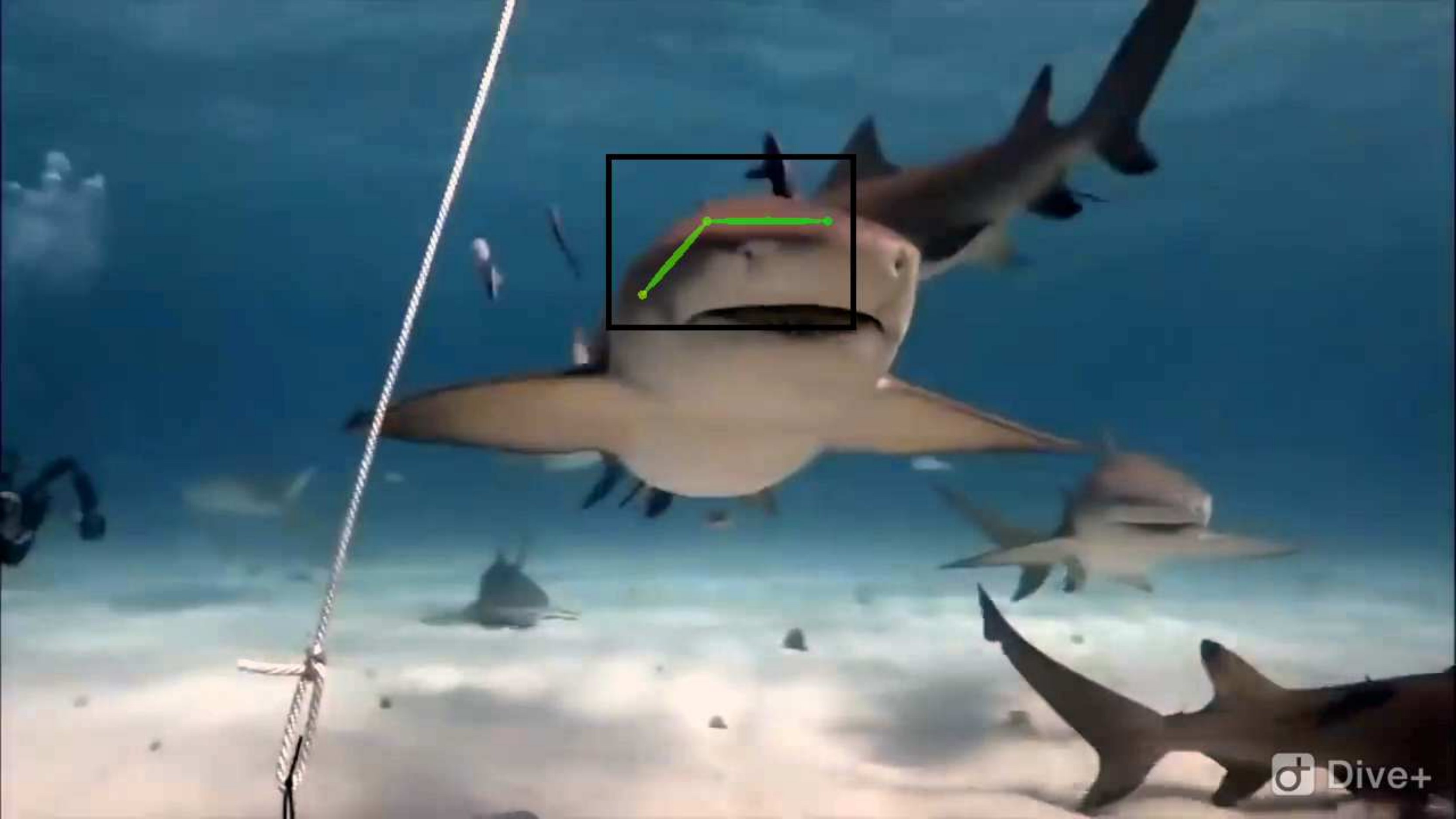} &
\includegraphics[width=3cm, height=3cm]{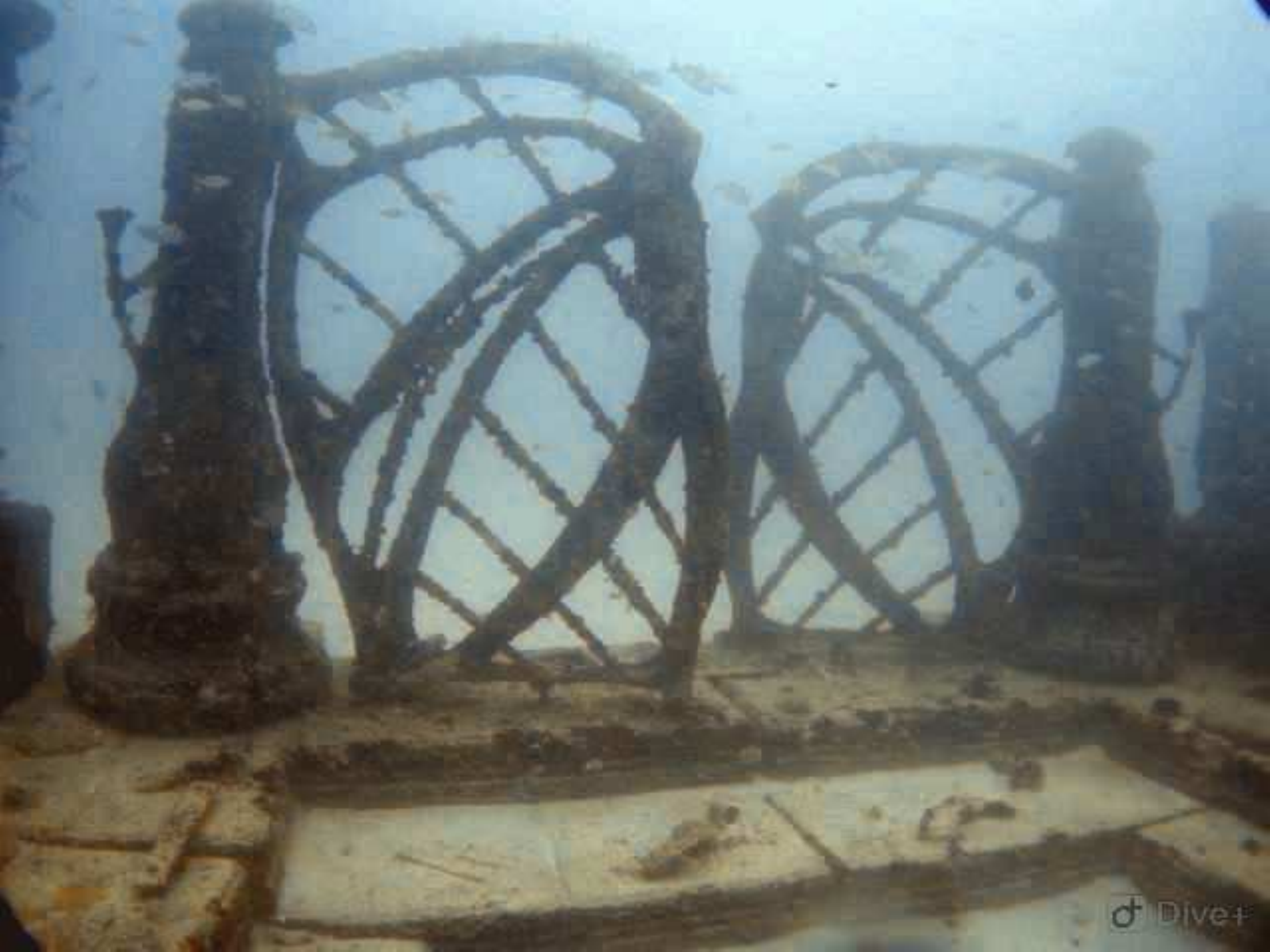} \\

(a) & (b) & (c) & (d) & (e) & (f) & (g) \\

\end{tabular}}
\caption{\small{\textbf{Failure case:} Visual demonstration of the cases where enhancement declines the high-level vision performance. \textit{Top:} Degraded underwater images, \textit{middle:} enhanced images using Deep WaveNet, and \textit{bottom:} clean underwater images.}}
\label{fig:high_issue}
\end{figure}
\textcolor{black}{We have found some inconsistent samples where enhancement declines the high-level vision performance. To elaborate, consider Figure \ref{fig:high_issue}, where \textit{top-row} denotes the degraded underwater images, \textit{middle-row} shows the enhanced underwater images by using Deep WaveNet and \textit{bottom-row} presents the clean underwater images. Each image presents the detected 2D pose of the human body using \texttt{OpenPose} framework. \texttt{OpenPose} is a multi-person system to jointly detect the human body, hand, facial, and foot key points (in total 135 key points) on single images.}

\textcolor{black}{It can be observed from Figure \ref{fig:high_issue} (a) that \texttt{OpenPose} detected the 2D pose of a \textit{fish (a human alike diver)} in the degraded image (\textit{top-row; shown using a black box}). Whereas the same framework detected a random background object as human in the enhanced image (\textit{middle-row}) instead of \textit{fish}.  While the enhancement helps in alleviating the false positive of 2D pose detected for \textit{fish}, it also induces another false positive of 2D pose detected for random-illusion being human-alike. Such a phenomenon can also be observed in Figures \ref{fig:high_issue} (c, d, f, \textit{and} g). It can also be observed from Figure \ref{fig:high_issue} (b) that although \texttt{OpenPose} detected the \textit{hand} part of the diver in enhanced image, it missed out the \textit{leg} part which has been earlier detected in degraded underwater image, as shown in \textit{white-box}. It may be because \texttt{OpenPose} has not seen the human-alike objects, such as \textit{fish}, during its learning. Therefore, it may have generated those false positives in the degraded and enhanced images. However, for the cases like Figure \ref{fig:high_issue} (b), the enhancement declined the high-level vision performance. It may be because of several reasons, including (i) local/global change in the distribution of color/texture of the underwater image after enhancement or (ii) a completely new environment (underwater) for \texttt{OpenPose}, which in general expects the outdoor images as inputs. As it can be observed from Figure \ref{fig:high_issue} (e), even \texttt{OpenPose} could not identify the humans in any version of the underwater image. The more robust framework for 2D pose estimation of human divers in the underwater scenario will be our future goal towards this direction.}

\begin{table}[t]
\centering
\caption{Quantitative results achieved using different cost functions on \texttt{EUVP} and \texttt{UIEB} datasets for the task of underwater image enhancement.}
\label{tab:ablation_loss_enhancement}
\resizebox{0.4\textwidth}{!}{
\begin{tabular}{lcccc}
\toprule
\midrule
Metrics & \multicolumn{2}{c}{EUVP} & \multicolumn{2}{c}{UIEB}\\
\cmidrule{2-5}
 & $\mathcal{L}_2$ & $\mathcal{L}_2 + \lambda_P \mathcal{L}_P$ & $\mathcal{L}_2$ & $\mathcal{L}_2 + \lambda_P \mathcal{L}_P$ \\
 \midrule
 \midrule
 
 MSE & $.59$ & \textcolor{red}{$\mathbf{.29}$} & $.68$& \textcolor{red}{$\mathbf{.60}$}\\
 PSNR & $27.51$ & \textcolor{red}{$\mathbf{28.62}$} & $21.29$ & \textcolor{red}{$\mathbf{21.57}$} \\
 SSIM & $.79$ & \textcolor{red}{$\mathbf{.83}$} & $.79$ & \textcolor{red}{$\mathbf{.80}$} \\
 \bottomrule
 \bottomrule
\end{tabular}}
\end{table}

\section{Ablation Study}
\label{sec:ablation}

\begin{table*}[t]
\centering
\caption{Quantitative results achieved using different cost functions on \texttt{UFO-120} dataset for underwater image super-resolution.}
\label{tab:ablation_loss_sr}
\resizebox{\textwidth}{!}{
\begin{tabular}{lccccccccc}
  \toprule
 \midrule
Methods & \multicolumn{3}{c}{\text{PSNR}} & \multicolumn{3}{c}{\text{SSIM}} & \multicolumn{3}{c}{\text{UIQM}}\\  
  \cmidrule{2-10}

& $2\times$ & $3\times$ & $4\times$ & $2\times$ & $3\times$ & $4\times$ & $2\times$ & $3\times$ & $4\times$\\

\midrule  
\midrule
$\mathcal{L}_2$ & $22.95\pm2.9$ & {\color{blue}$\mathbf{24.78\pm2.4}$} & $24.32\pm2.6$ & {\color{blue}$\mathbf{.73\pm.06}$} & $.71\pm.07$ & {\color{blue}$\mathbf{.70\pm.08}$} & $2.72\pm0.60$ & $2.95\pm0.48$ & $2.73\pm0.60$\\ 

 $\mathcal{L}_2 + \lambda_P.\mathcal{L}_P$ & {\color{blue}$\mathbf{24.38\pm2.3}$} & $24.62\pm2.5$ & {\color{blue}$\mathbf{24.60\pm2.6}$} & $.73\pm.09$ & {\color{blue}$\mathbf{.72\pm.09}$} & $.69\pm.08$ & {\color{blue}$\mathbf{2.82\pm0.63}$} & \textcolor{red}{$\mathbf{2.97\pm0.55}$} & {\color{blue}$\mathbf{2.86\pm0.61}$} \\ 
 
  $\mathcal{L}_2 + \lambda_P.\mathcal{L}_P + \lambda_S.\mathcal{L}_{SSIM}$ & \textcolor{red}{$\mathbf{25.71 \pm 3.0}$}& {\color{red}$\mathbf{25.23 \pm 2.7}$} & {\color{red}$\mathbf{25.08 \pm 2.9}$}& \textcolor{red}{$\mathbf{.77 \pm .07}$}& {\color{red}$\mathbf{.76 \pm .07}$} &{\color{red}$\mathbf{.74 \pm .07}$}& \textcolor{red}{$\mathbf{2.99\pm0.57}$}& {\color{blue}$\mathbf{2.96\pm0.60}$}& \textcolor{red}{$\mathbf{2.97\pm0.59}$} \\
  
  \bottomrule
	\bottomrule
\end{tabular}}
\end{table*}

\subsection{Effect of cost functions}
To demonstrate the effect of various loss functions incorporated in the proposed model, we have performed the following set of baselines:
	\begin{itemize}
	\item $\mathcal{L}_2$: The proposed model is trained for the task of underwater image enhancement on \texttt{EUVP} and \texttt{UIEB} datasets, as shown in Table. \ref{tab:ablation_loss_enhancement}, using $\mathcal{L}_2$ loss only.
	
	\item $\mathcal{L}_2$: The proposed model is trained for the task of underwater image super-resolution on \texttt{UFO-120} dataset, as shown in Table. \ref{tab:ablation_loss_sr}, using $\mathcal{L}_2$ loss only.
	
	\item $\mathcal{L}_2 + \lambda_P.\mathcal{L}_P$: The proposed model is trained for the task of underwater image super-resolution on \texttt{UFO-120} dataset, as shown in Table. \ref{tab:ablation_loss_sr}, using $\mathcal{L}_2 + \lambda_P.\mathcal{L}_P$ loss only.
	\end{itemize}	
A significant improvement can be observed due to the addition of $\mathcal{L}_P$ over $\mathcal{L}_2$ for the task of underwater image enhancement. The perceptual loss might have helped in retaining the low-level features in the enhanced images. Coming to underwater SISR, it has been observed (see Table \ref{tab:ablation_loss_sr}) that the $\mathcal{L}_2$ yields minimal performance due to its widely known behavior of inducing blurriness in the denoised images. Although the inclusion of perceptual loss $\mathcal{L}_P$ over $\mathcal{L}_2$ significantly improved the PSNR in all resolution configurations, there has not been much increment in SSIM values. Interestingly, the addition of SSIM-based loss function $\mathcal{L}_{SSIM}$ has remarkably improved both the SSIM and PSNR values over earlier cost functions.

\subsection{Effect of Wavelength-driven Contexual Sizes and CBAM}
\label{sec:ablation_wave_cbam}
\begin{figure*}[t]
\centering
\resizebox{\textwidth}{!}{
\setlength{\tabcolsep}{1pt}
\begin{tabular}{ccccccccccc}

\includegraphics[width=0.095\textwidth]{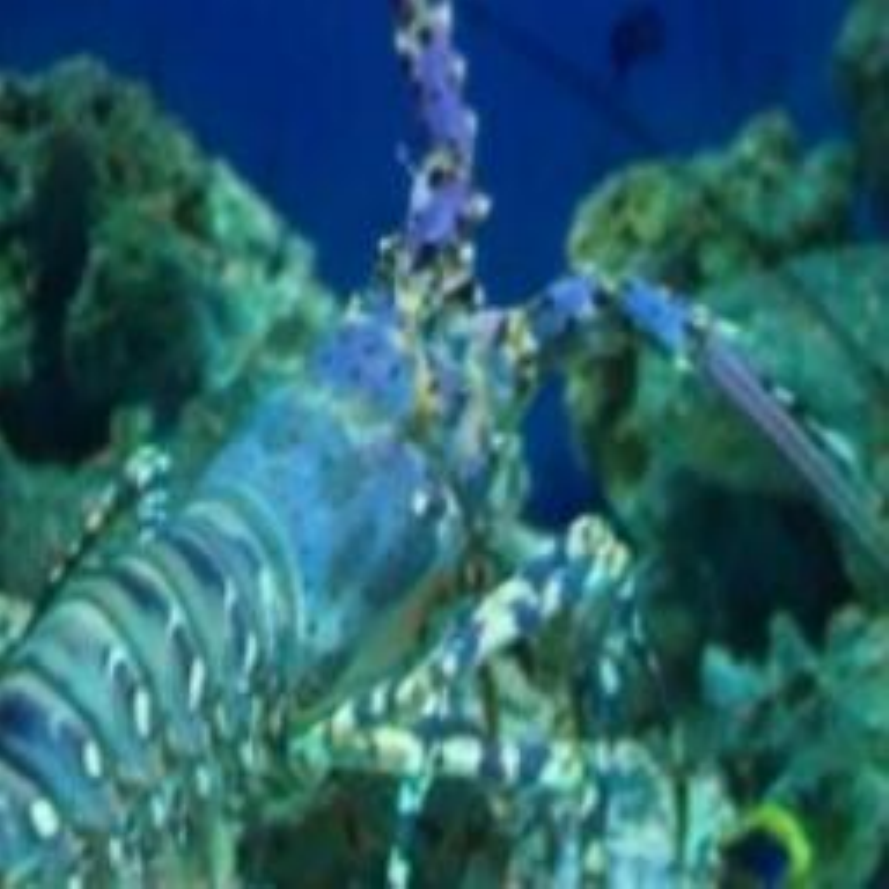}
& 
\includegraphics[width=0.095\textwidth]{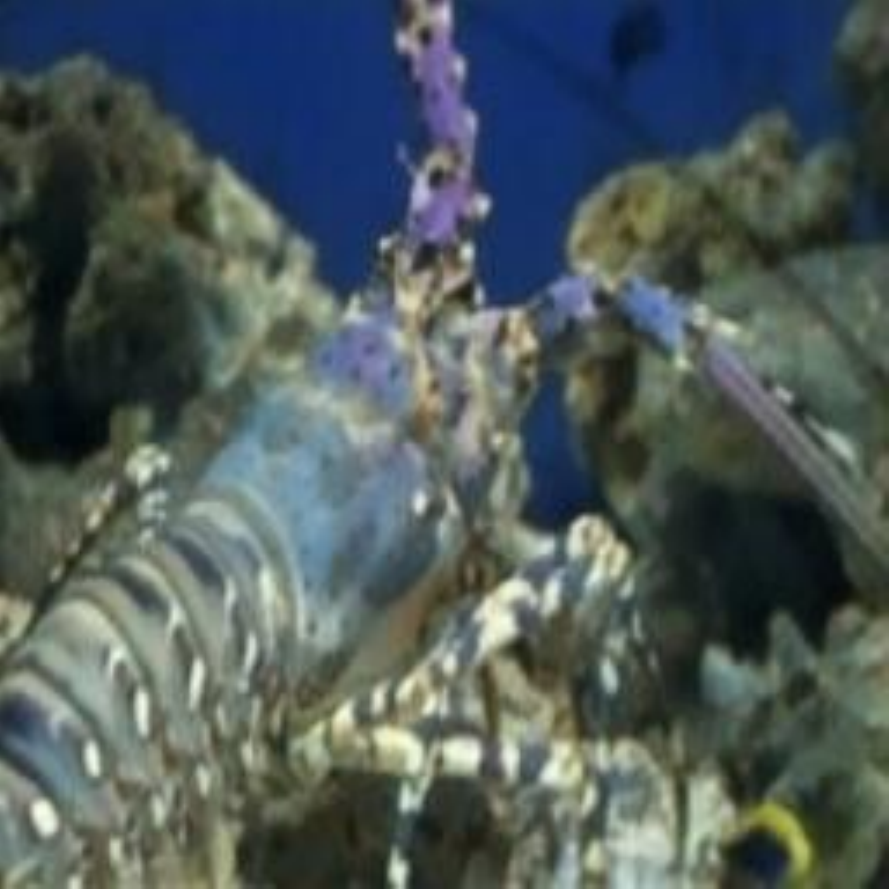}
&
\includegraphics[width=0.095\textwidth]{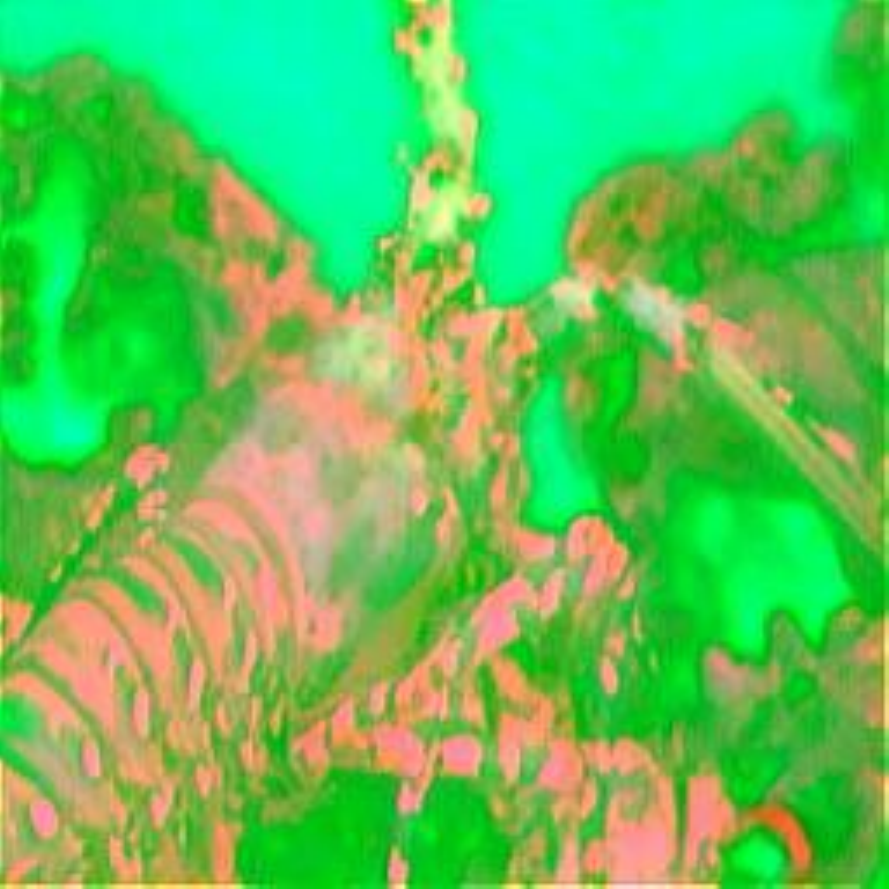}
&
\includegraphics[width=0.095\textwidth]{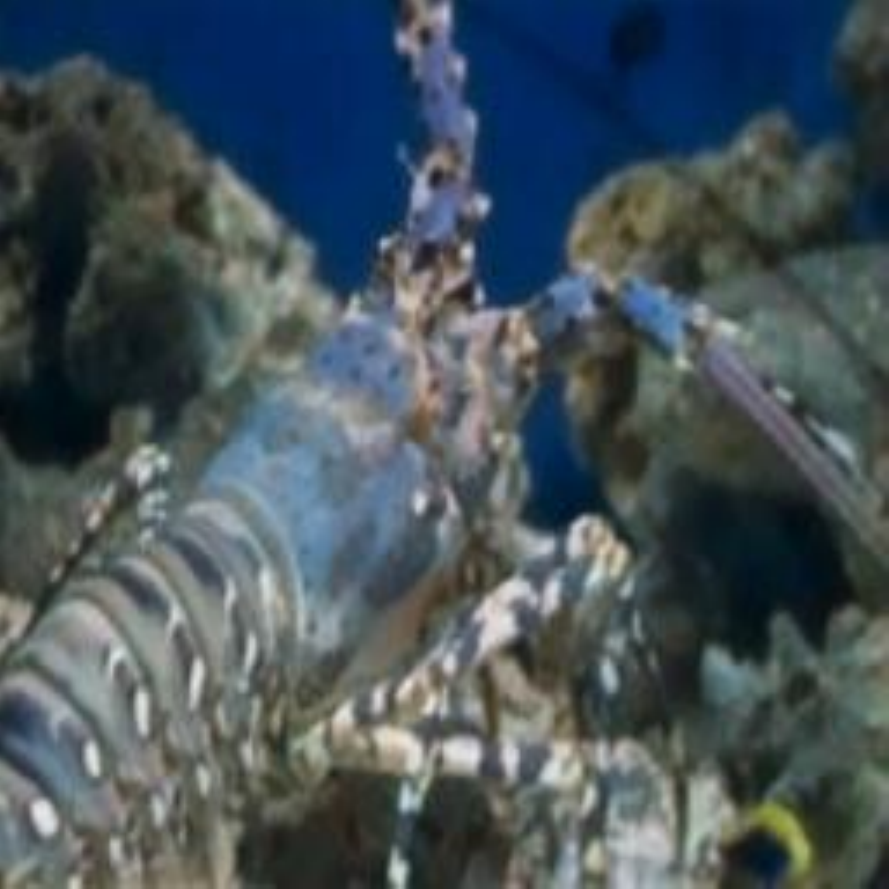}
&
\includegraphics[width=0.095\textwidth]{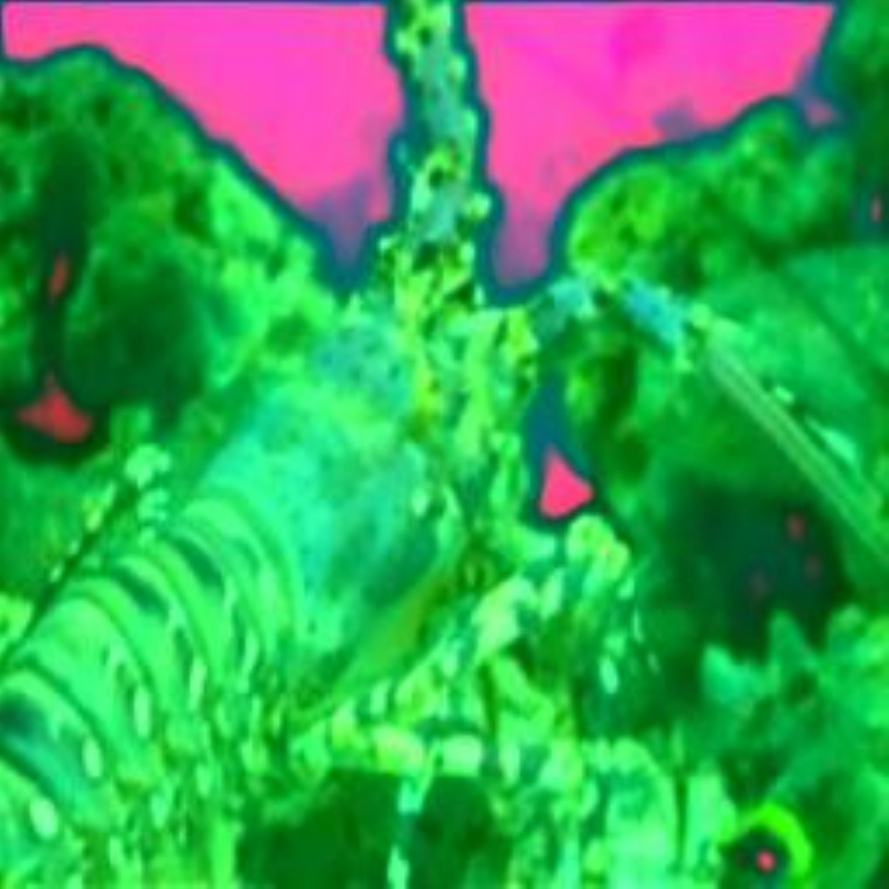}
&
\includegraphics[width=0.095\textwidth]{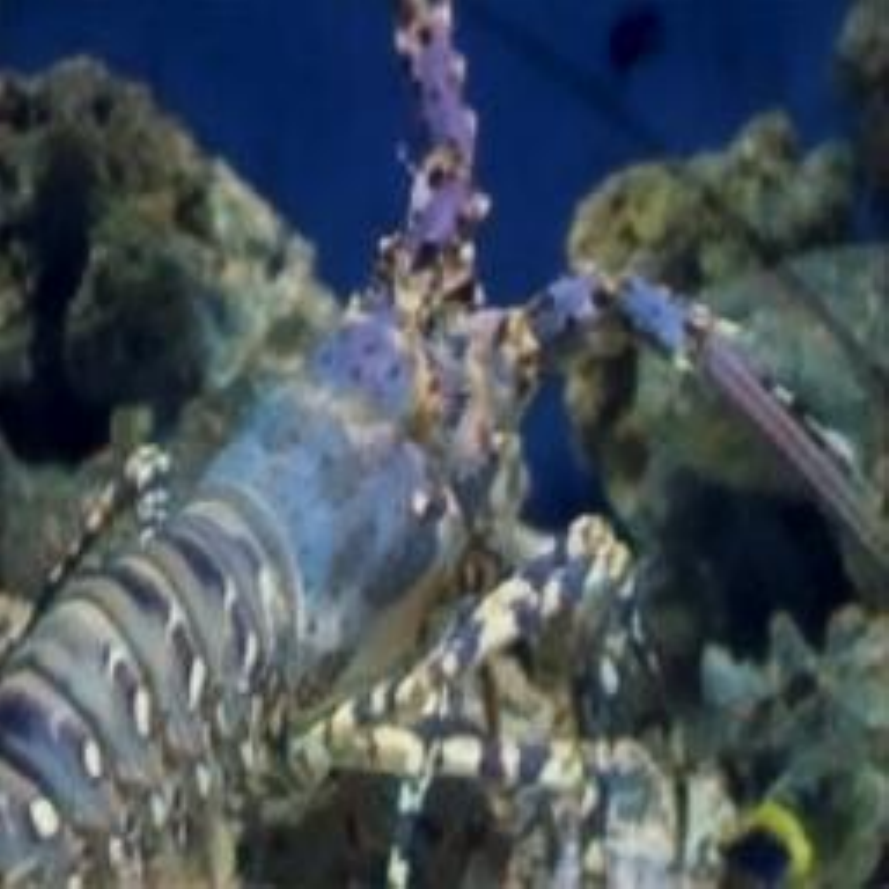}
&
\includegraphics[width=0.095\textwidth]{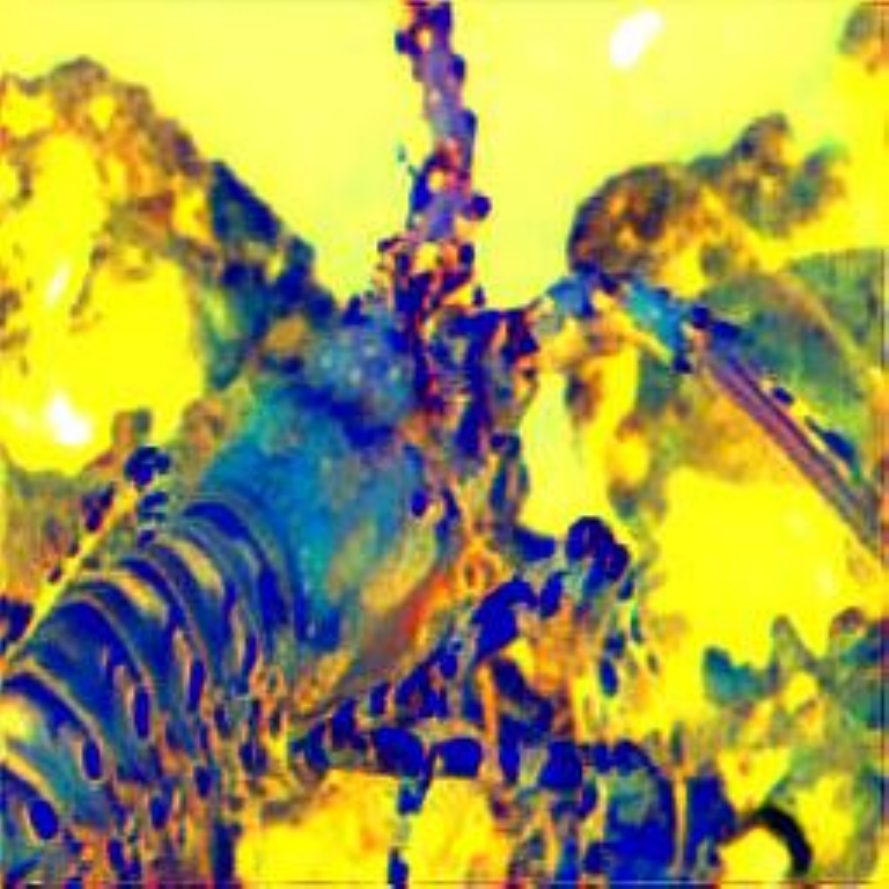}
&
\includegraphics[width=0.095\textwidth]{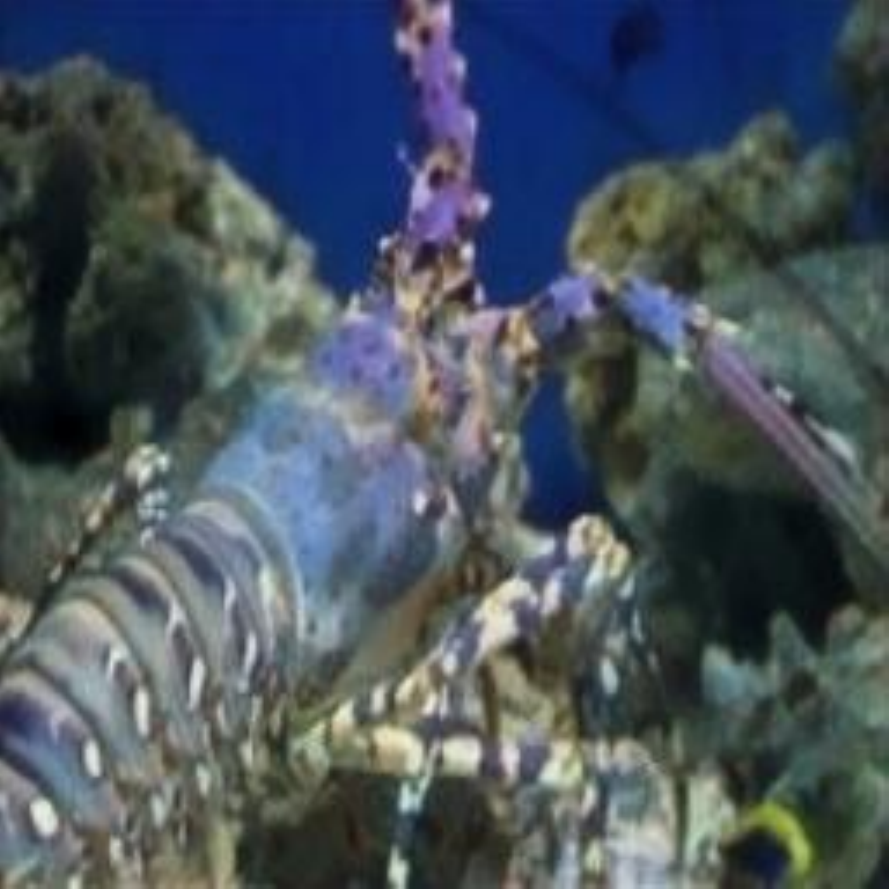}
&
\includegraphics[width=0.095\textwidth]{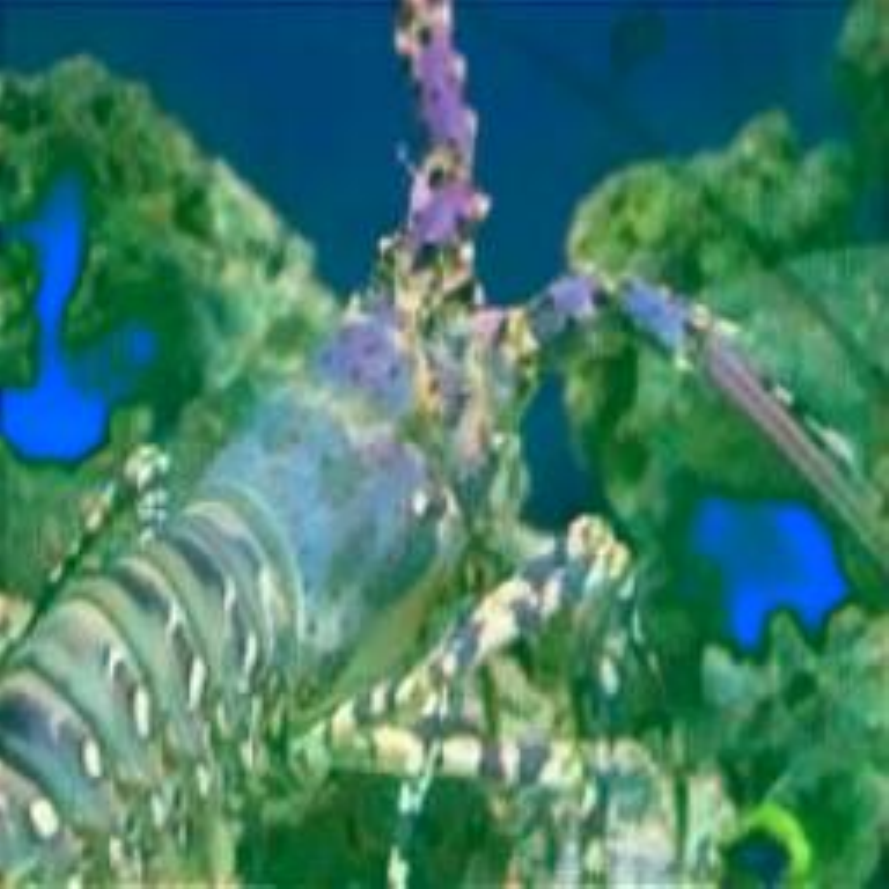}
&
\includegraphics[width=0.095\textwidth]{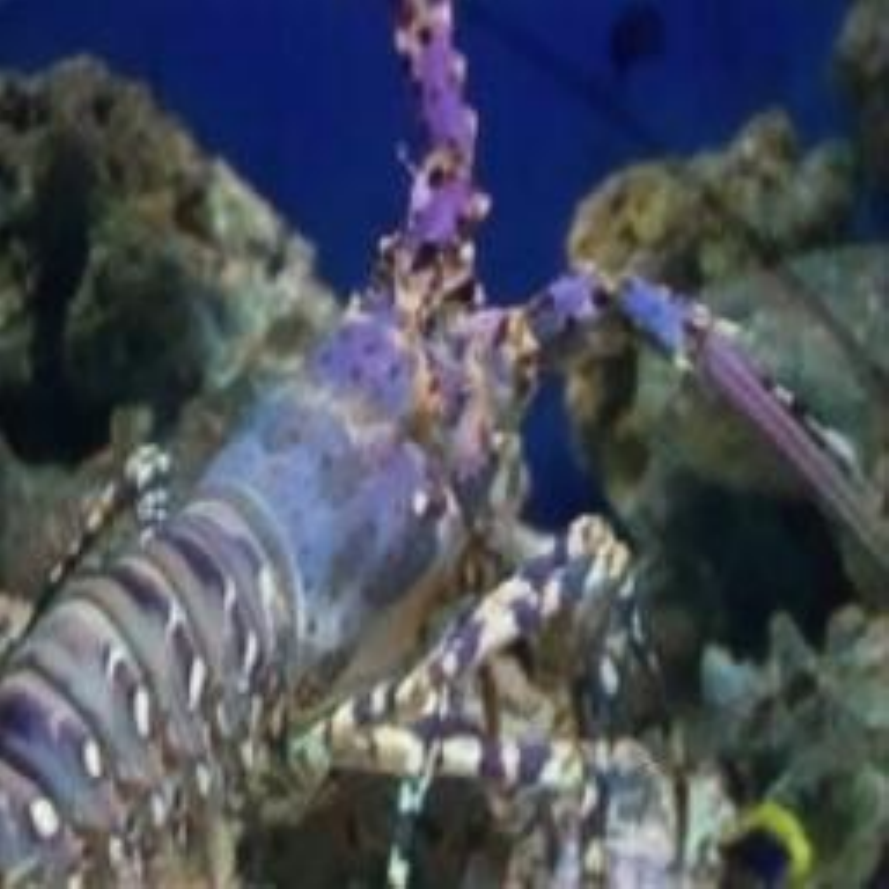}
&
\includegraphics[width=0.095\textwidth]{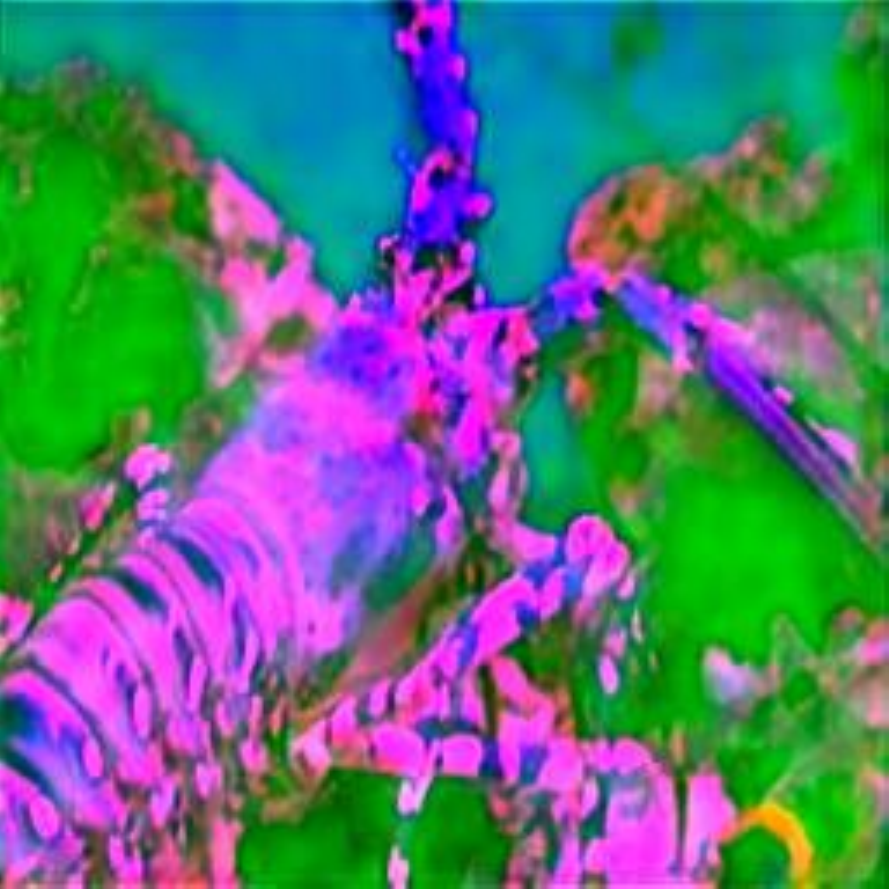}\\

\includegraphics[width=0.095\textwidth]{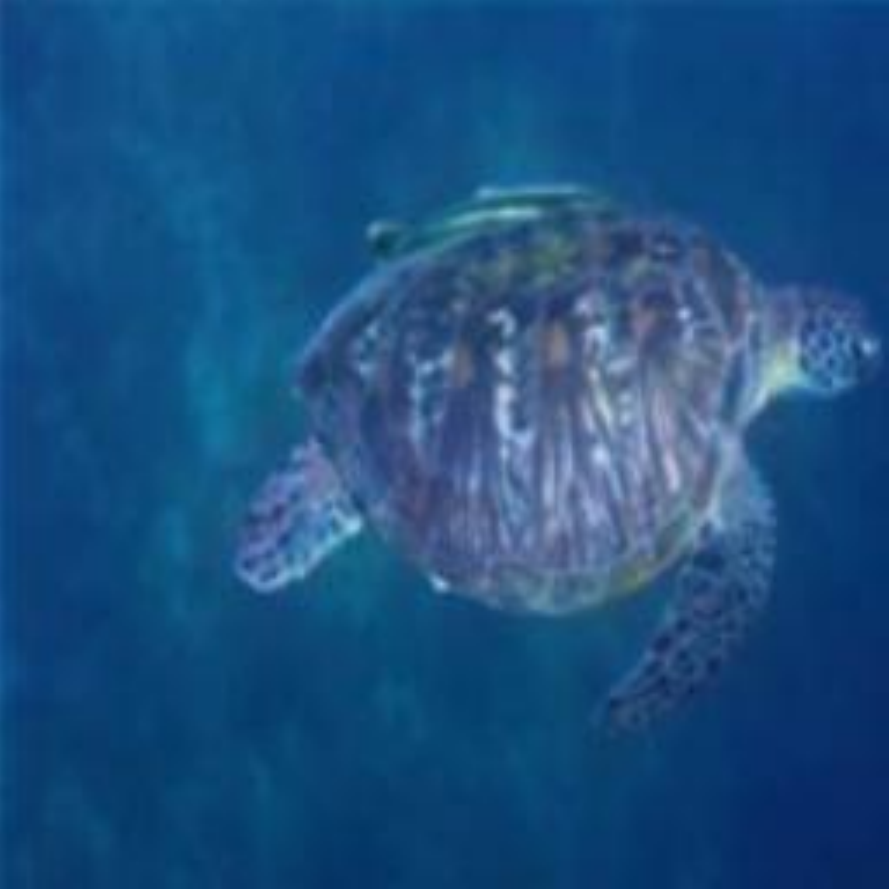}
& 
\includegraphics[width=0.095\textwidth]{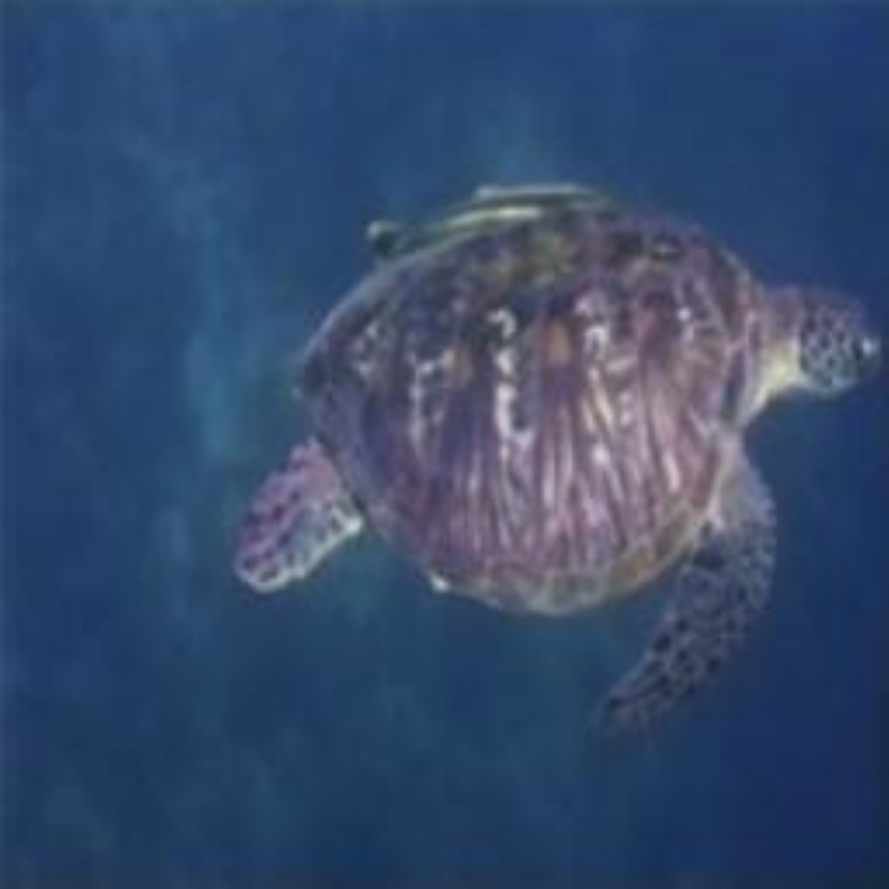}
&
\includegraphics[width=0.095\textwidth]{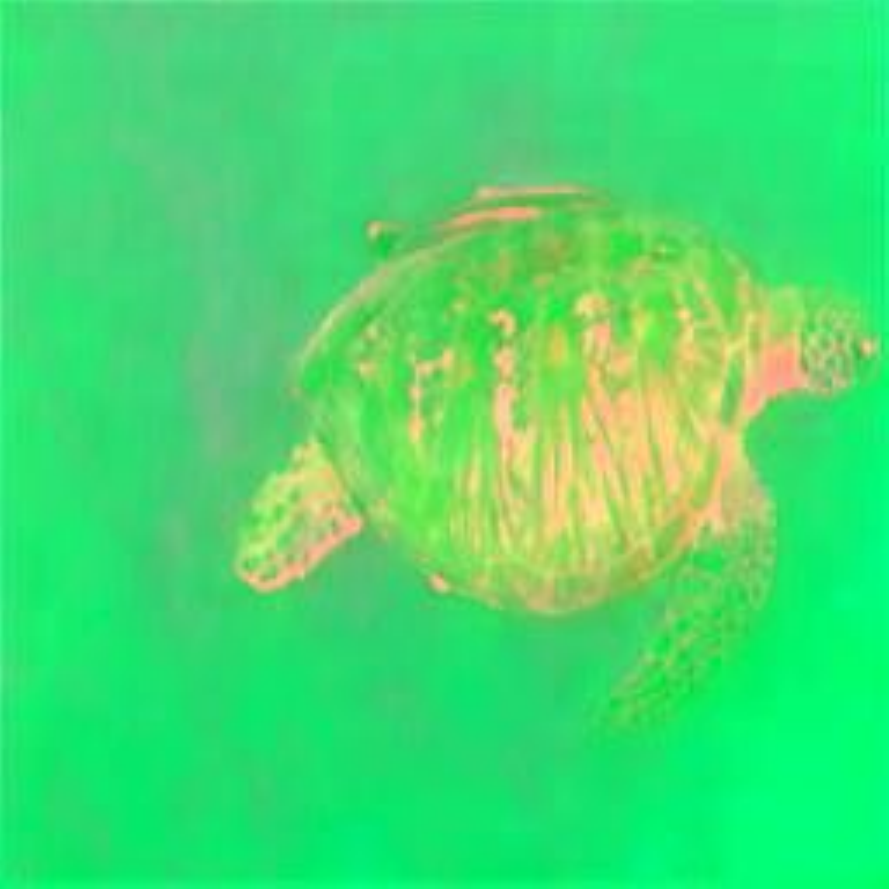}
&
\includegraphics[width=0.095\textwidth]{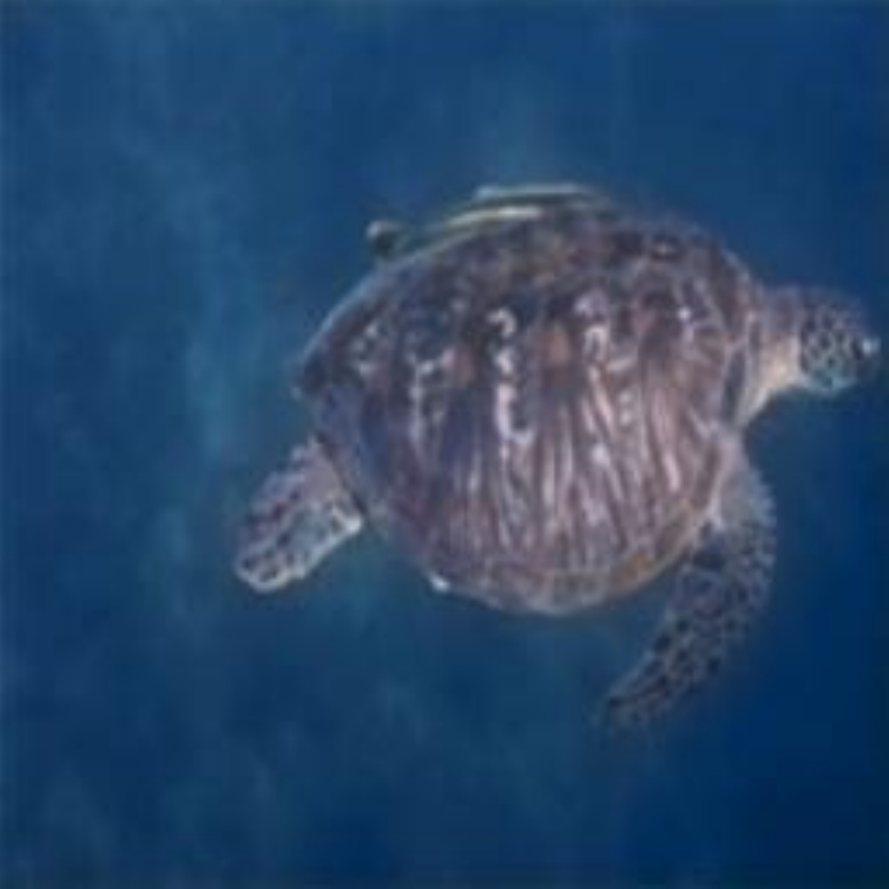}
&
\includegraphics[width=0.095\textwidth]{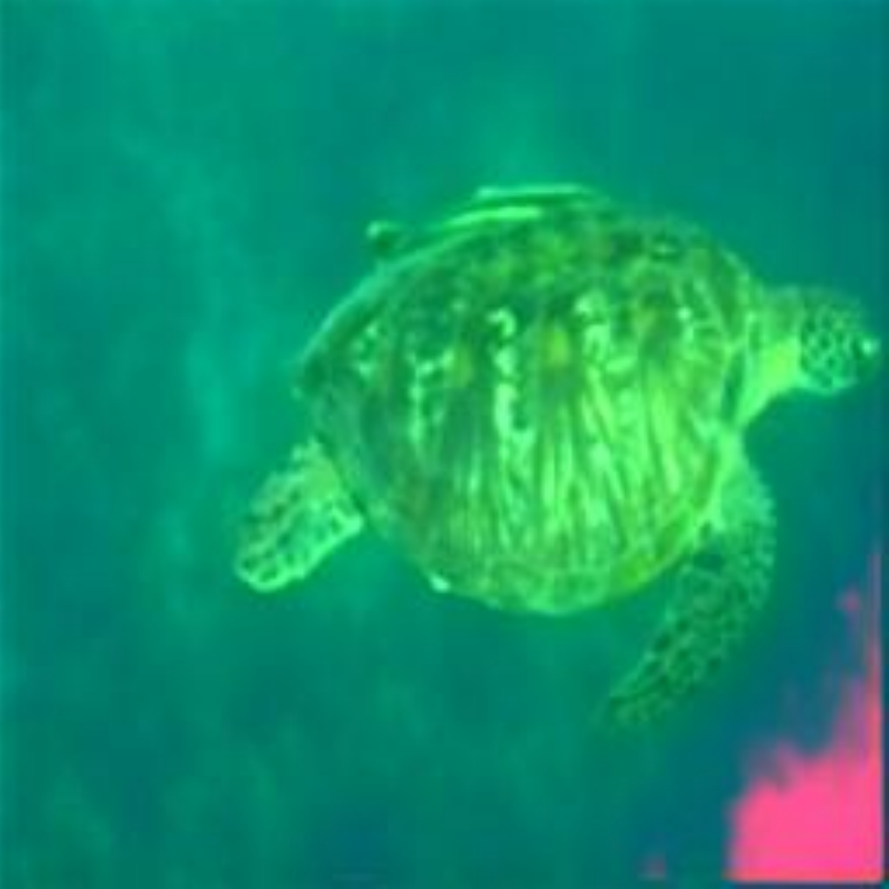}
&
\includegraphics[width=0.095\textwidth]{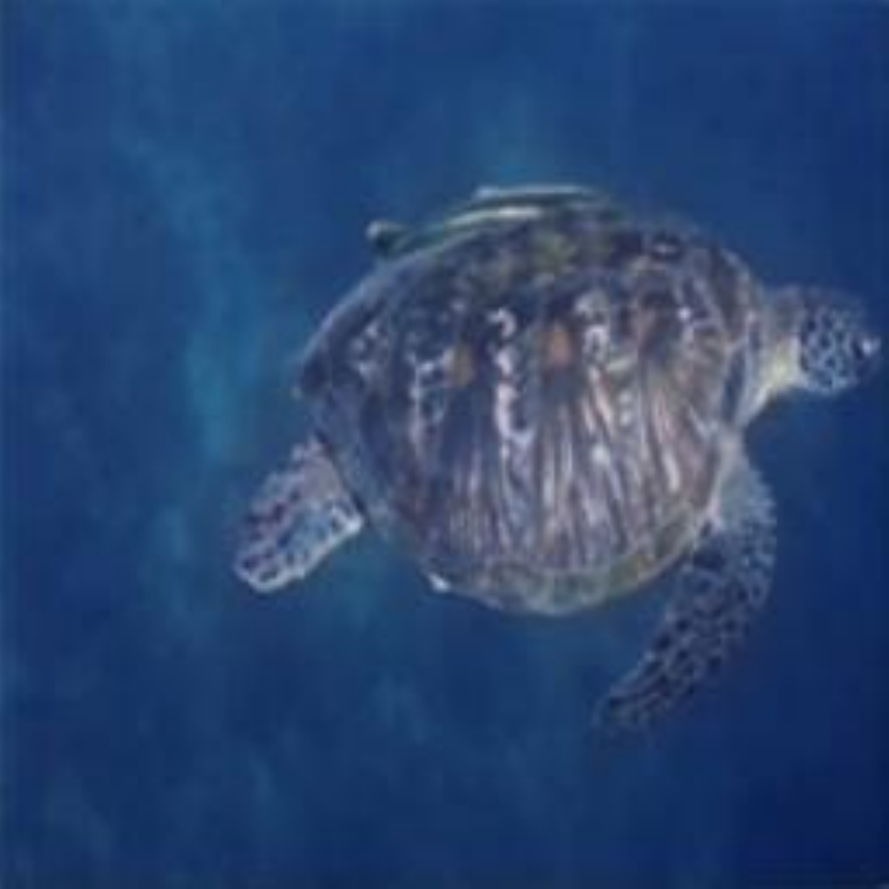}
&
\includegraphics[width=0.095\textwidth]{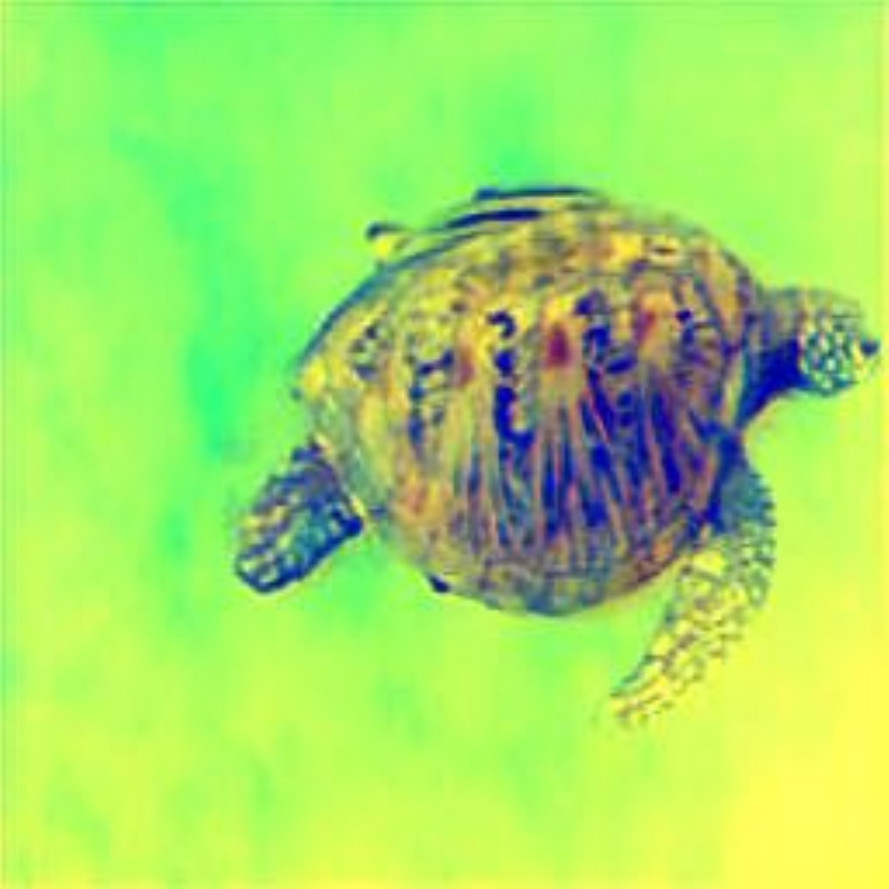}
&
\includegraphics[width=0.095\textwidth]{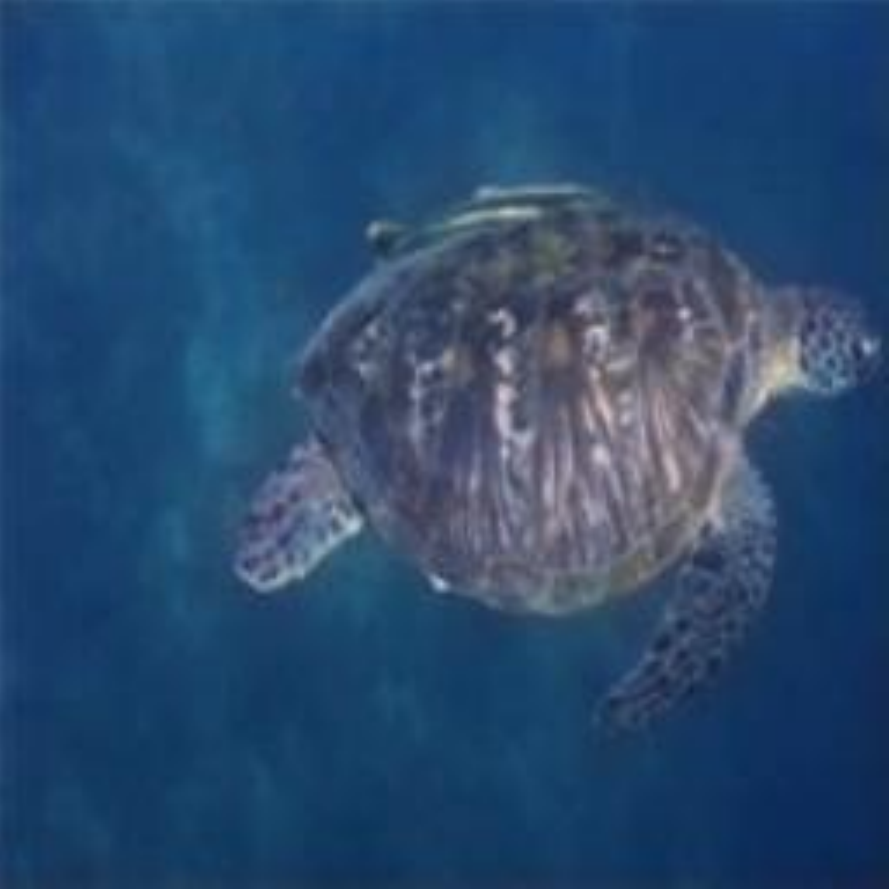}
&
\includegraphics[width=0.095\textwidth]{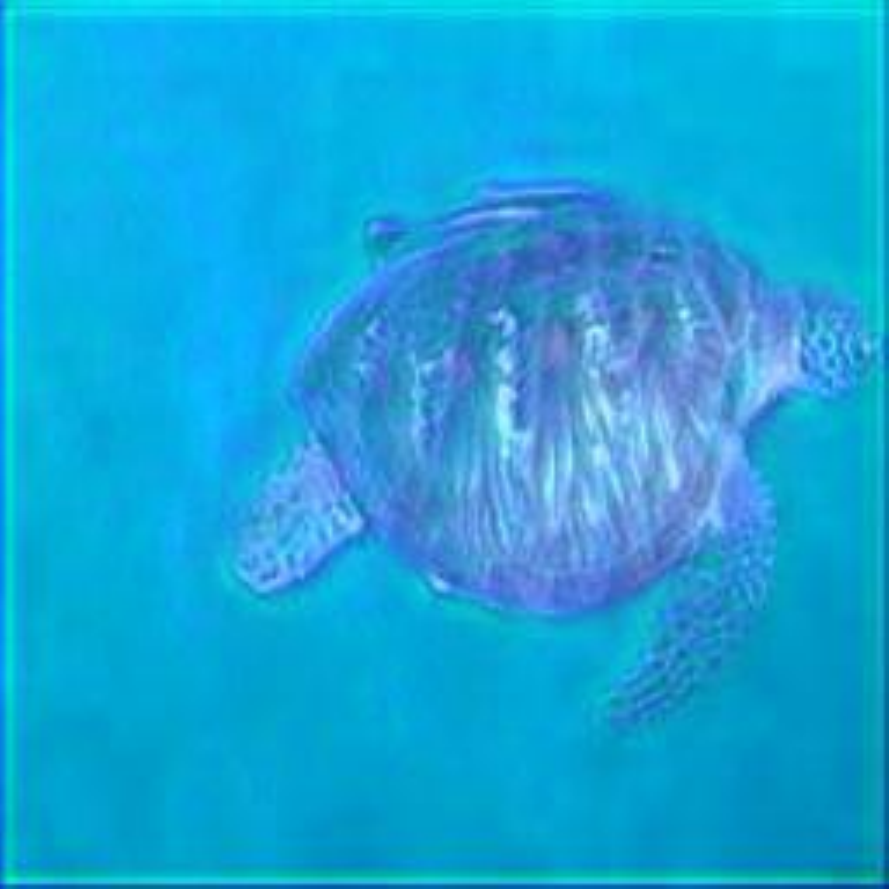}
&
\includegraphics[width=0.095\textwidth]{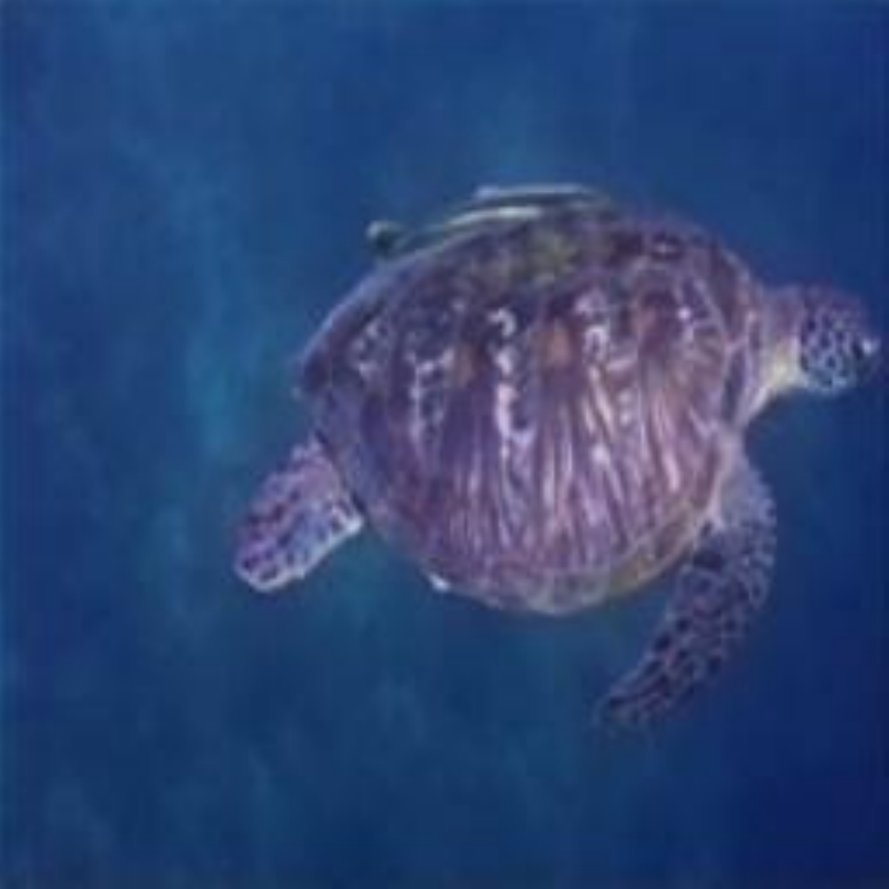}
&
\includegraphics[width=0.095\textwidth]{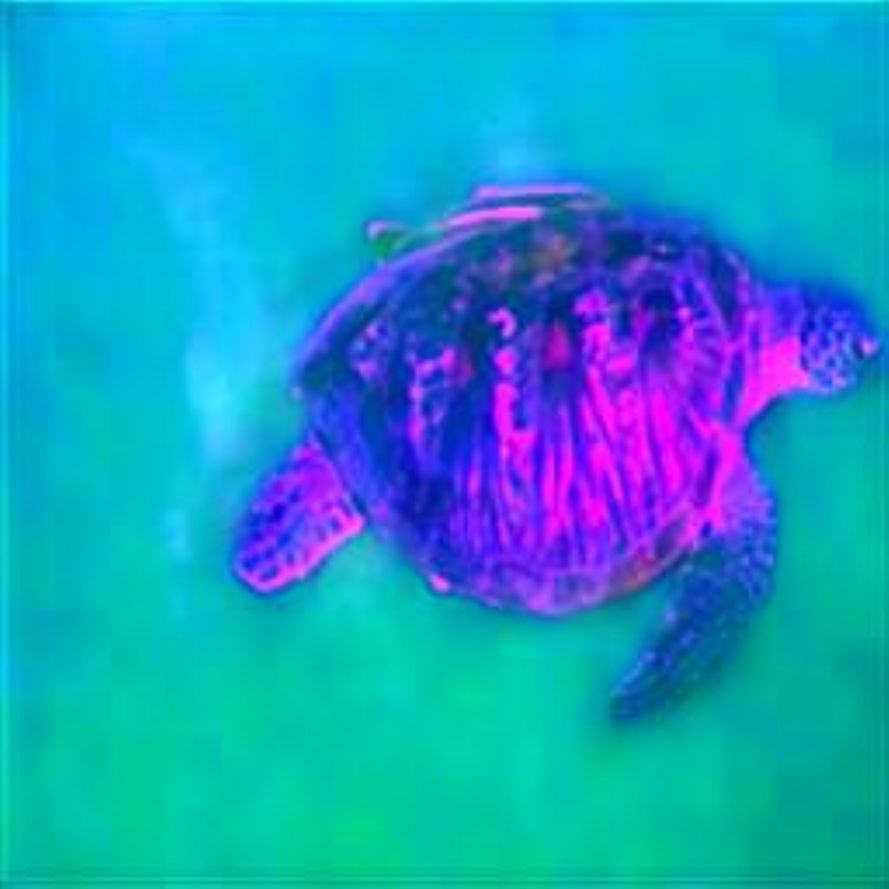}\\

\includegraphics[width=0.095\textwidth]{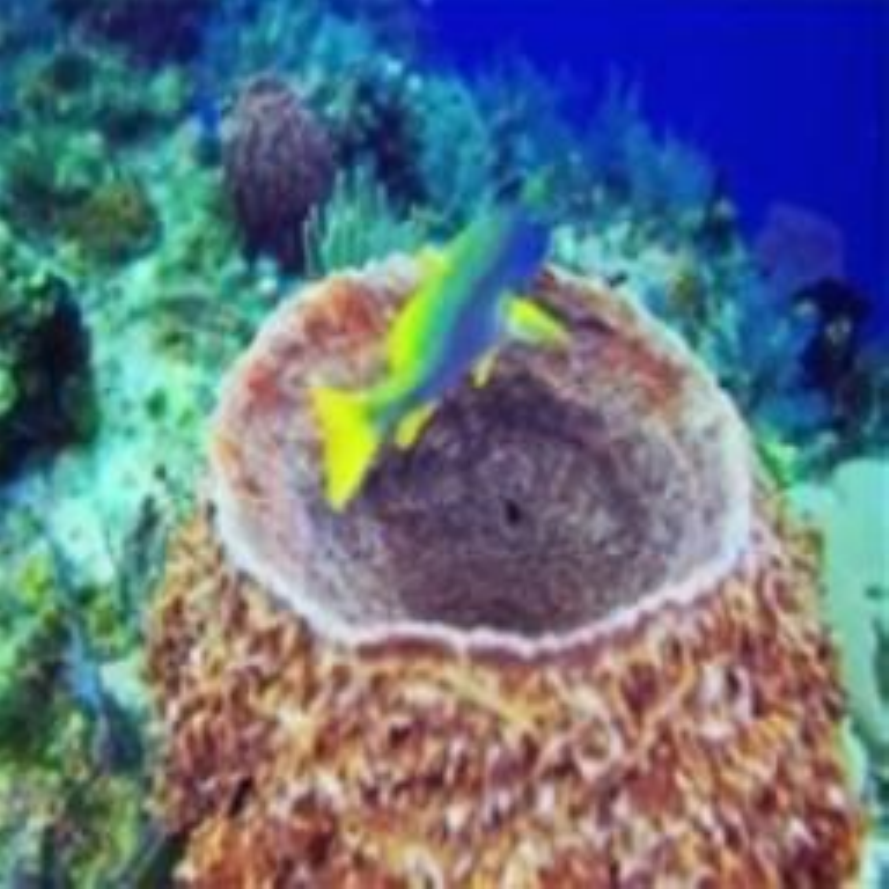}
& 
\includegraphics[width=0.095\textwidth]{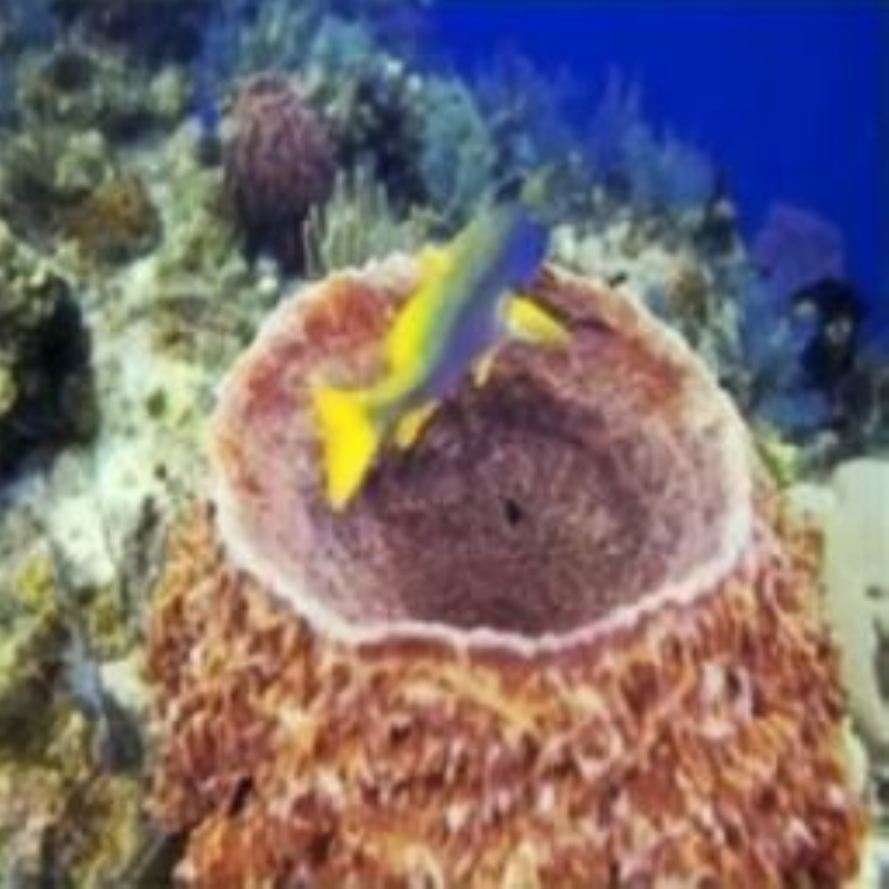}
&
\includegraphics[width=0.095\textwidth]{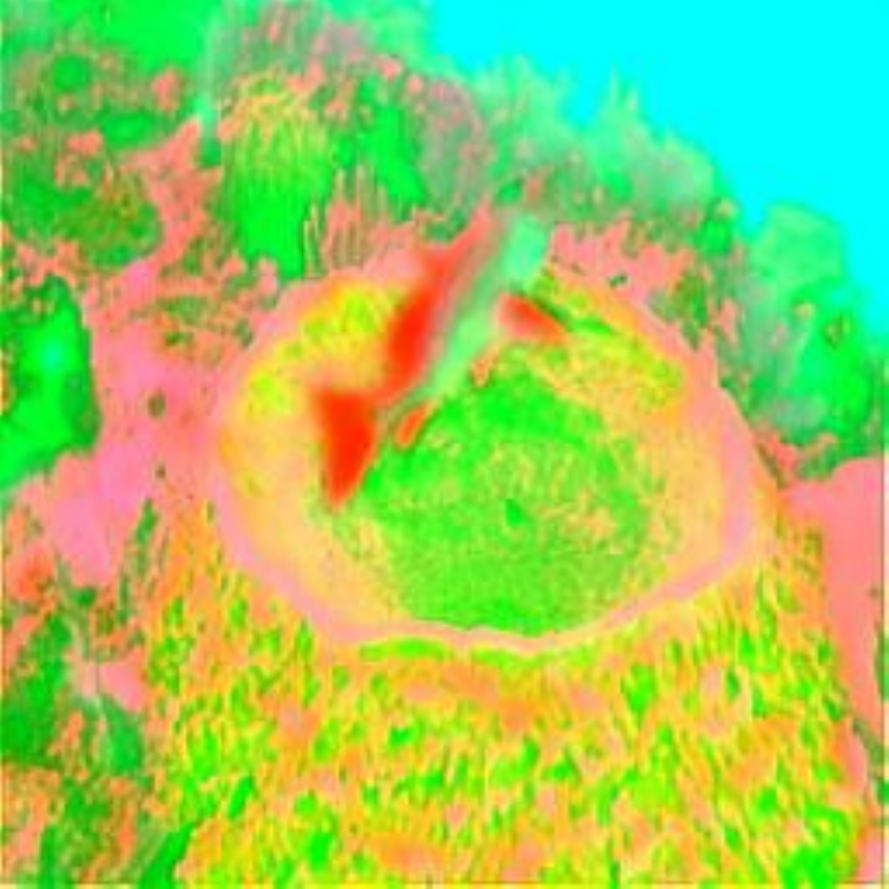}
&
\includegraphics[width=0.095\textwidth]{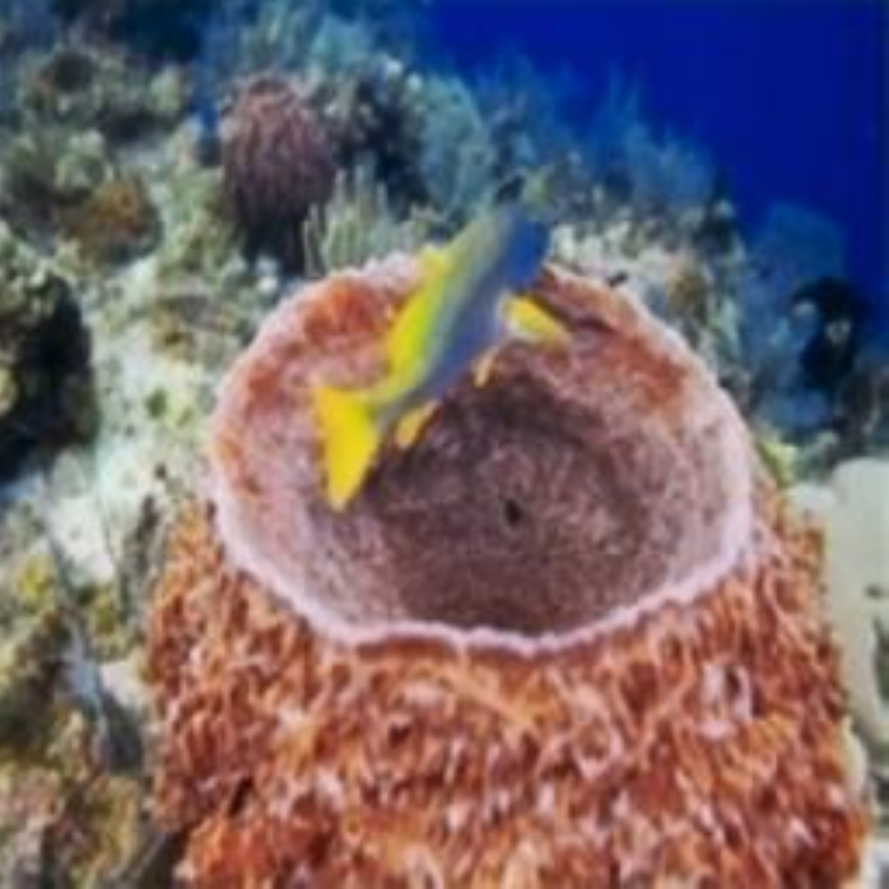}
&
\includegraphics[width=0.095\textwidth]{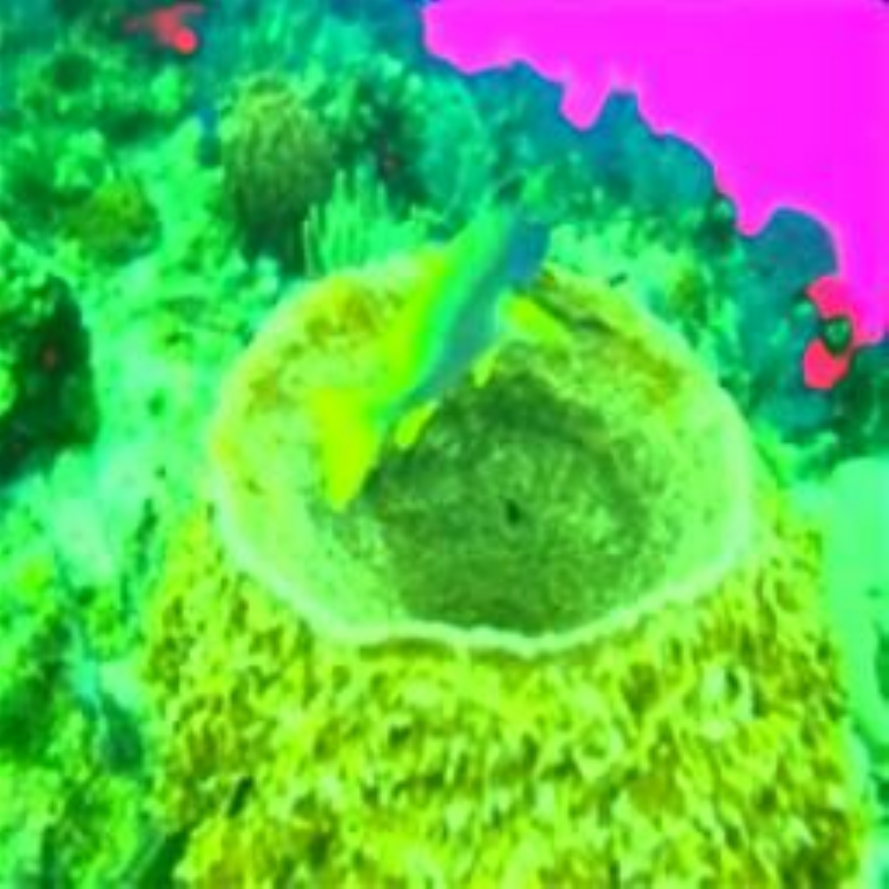}
&
\includegraphics[width=0.095\textwidth]{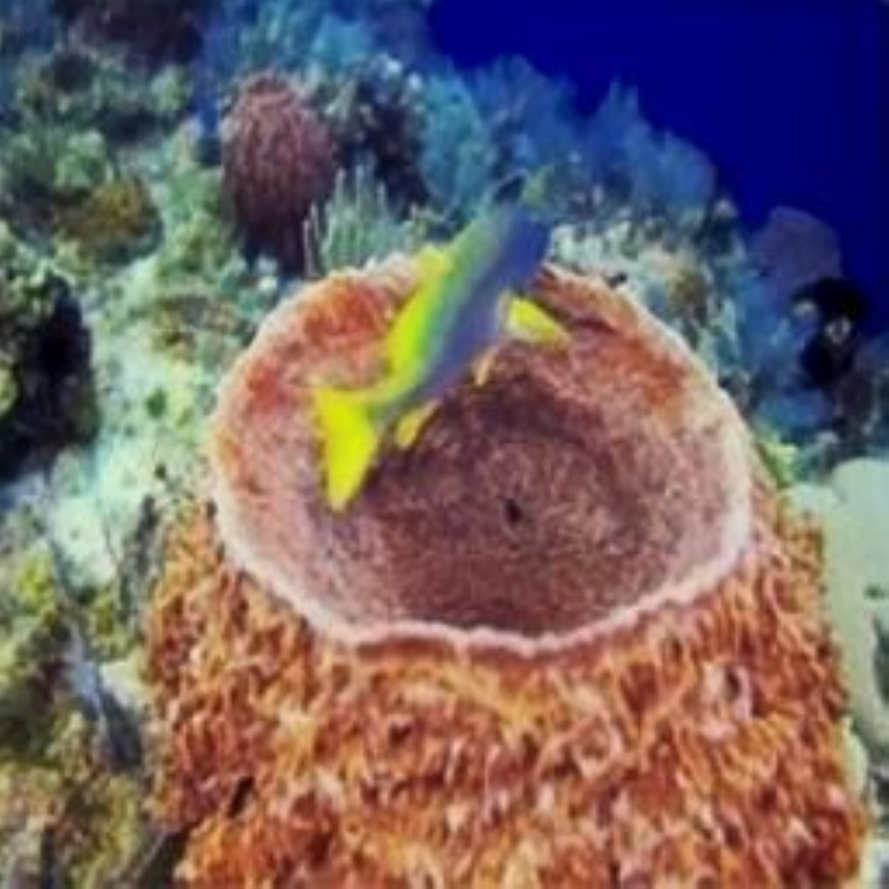}
&
\includegraphics[width=0.095\textwidth]{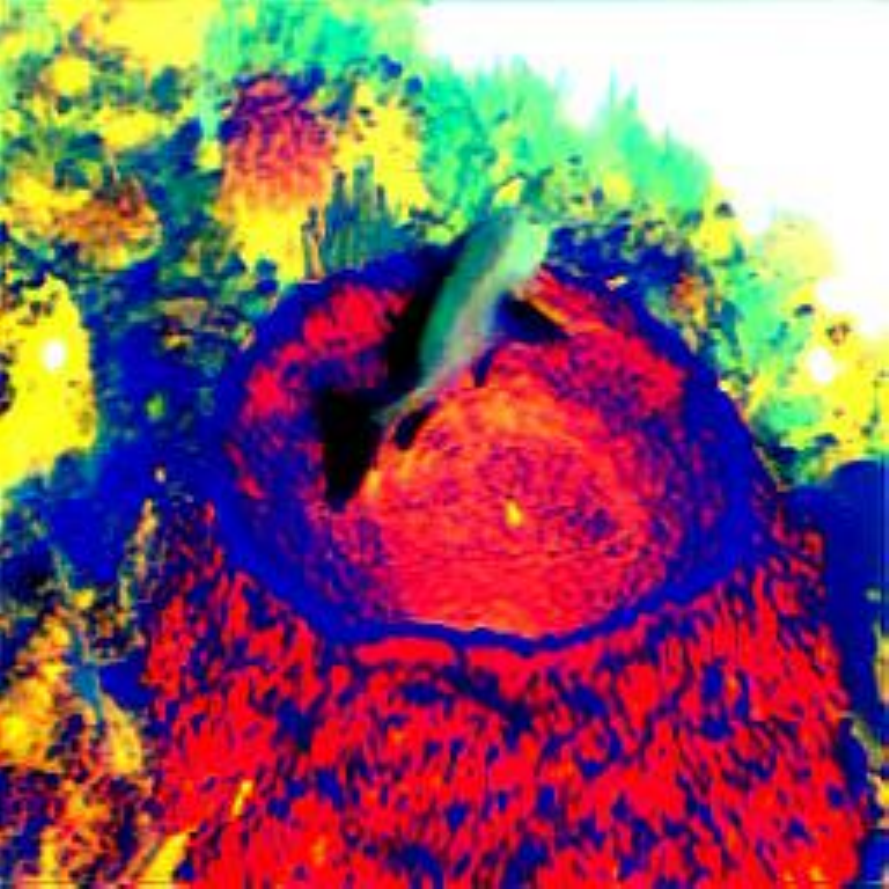}
&
\includegraphics[width=0.095\textwidth]{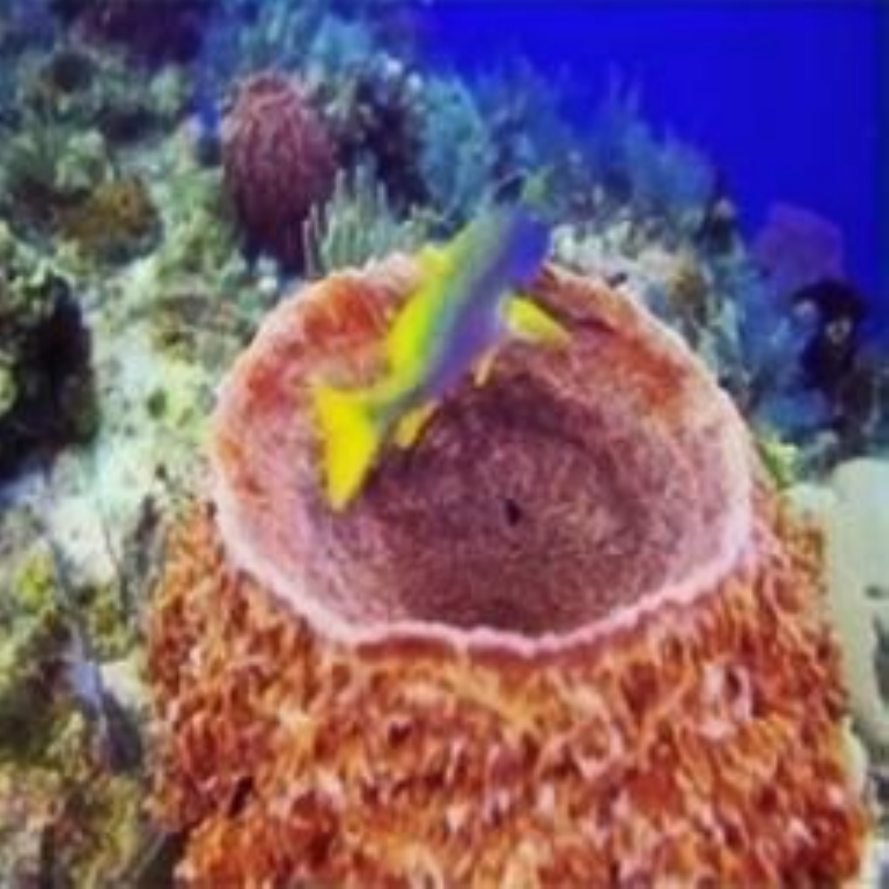}
&
\includegraphics[width=0.095\textwidth]{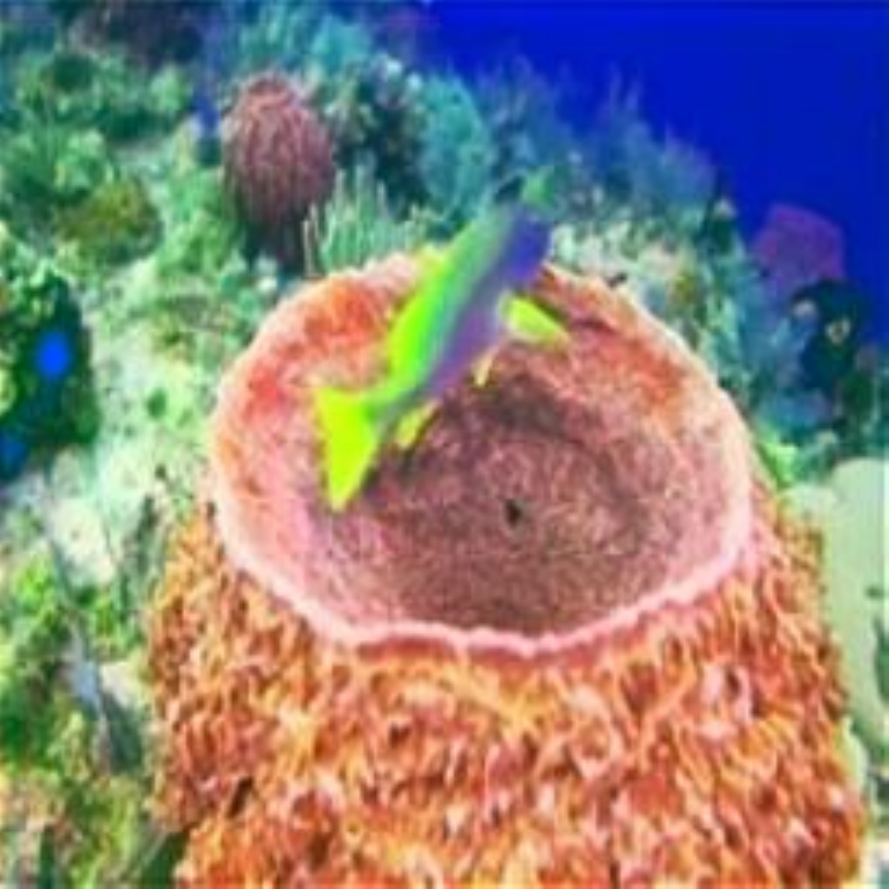}
&
\includegraphics[width=0.095\textwidth]{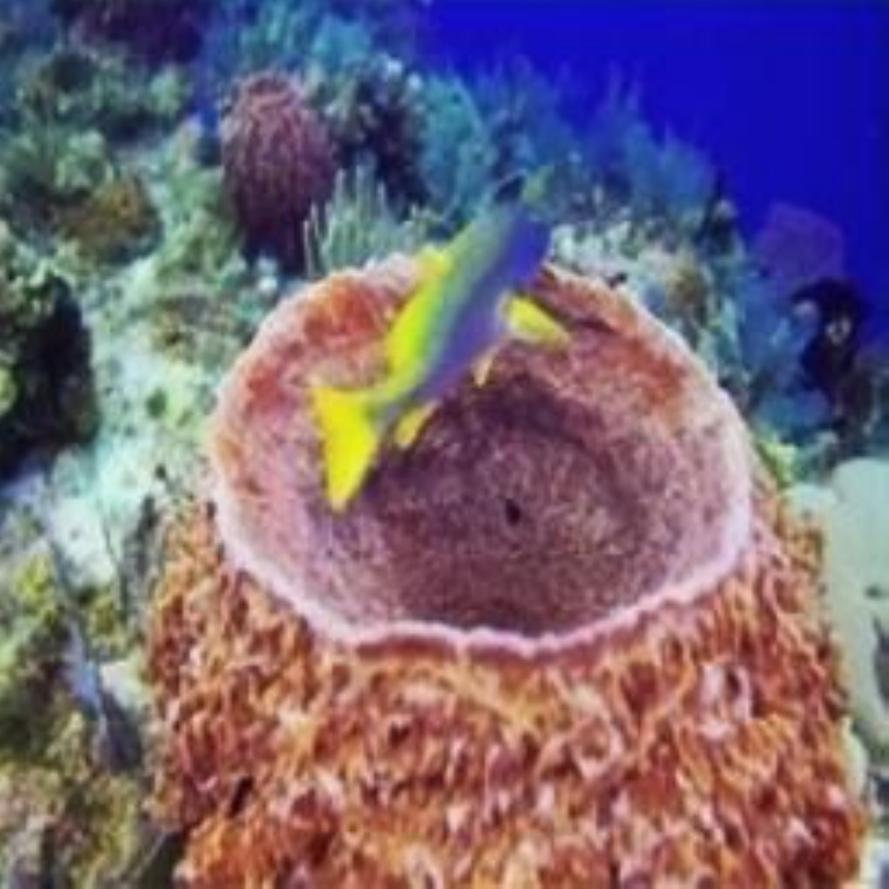}
&
\includegraphics[width=0.095\textwidth]{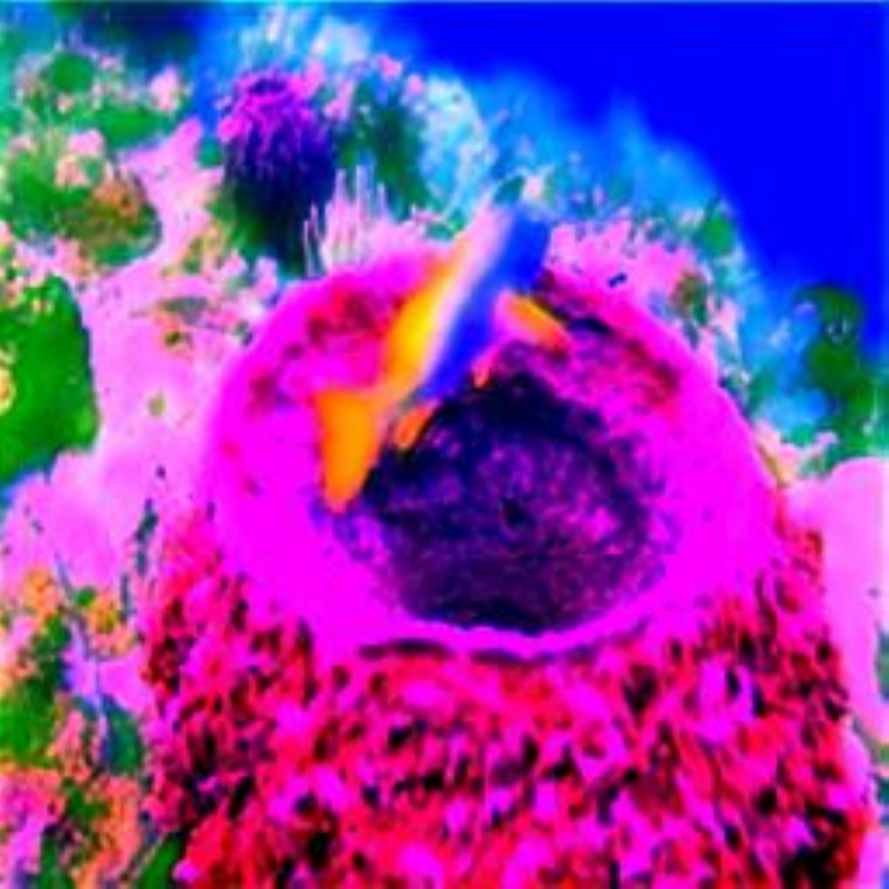}\\

\includegraphics[width=0.095\textwidth]{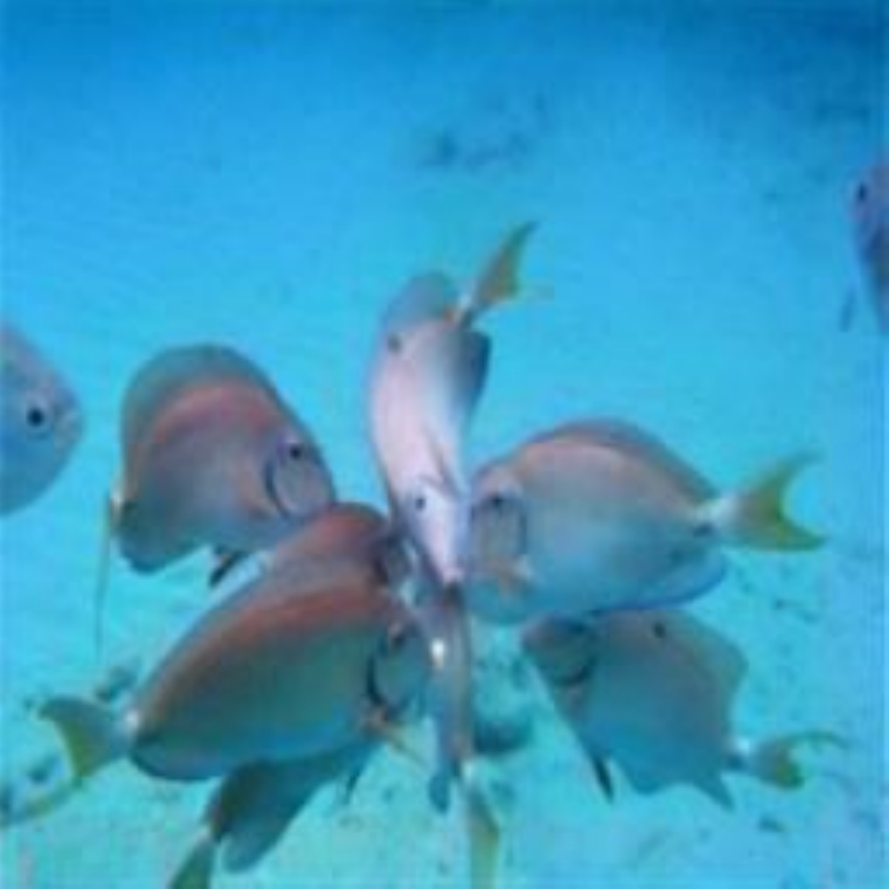}
& 
\includegraphics[width=0.095\textwidth]{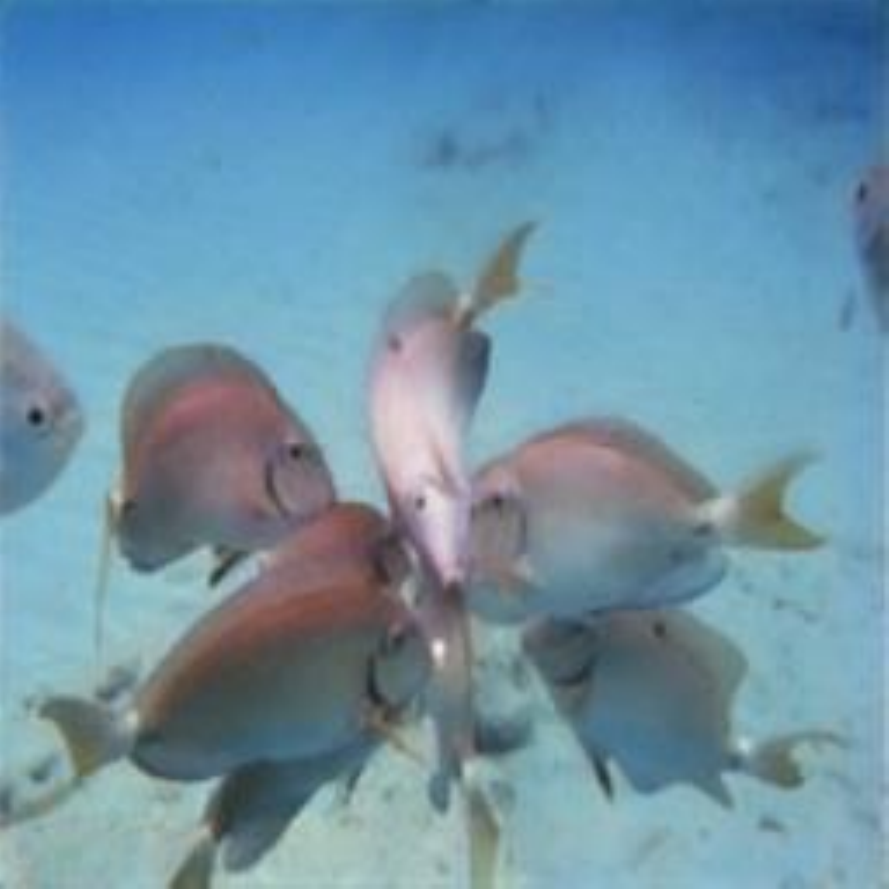}
&
\includegraphics[width=0.095\textwidth]{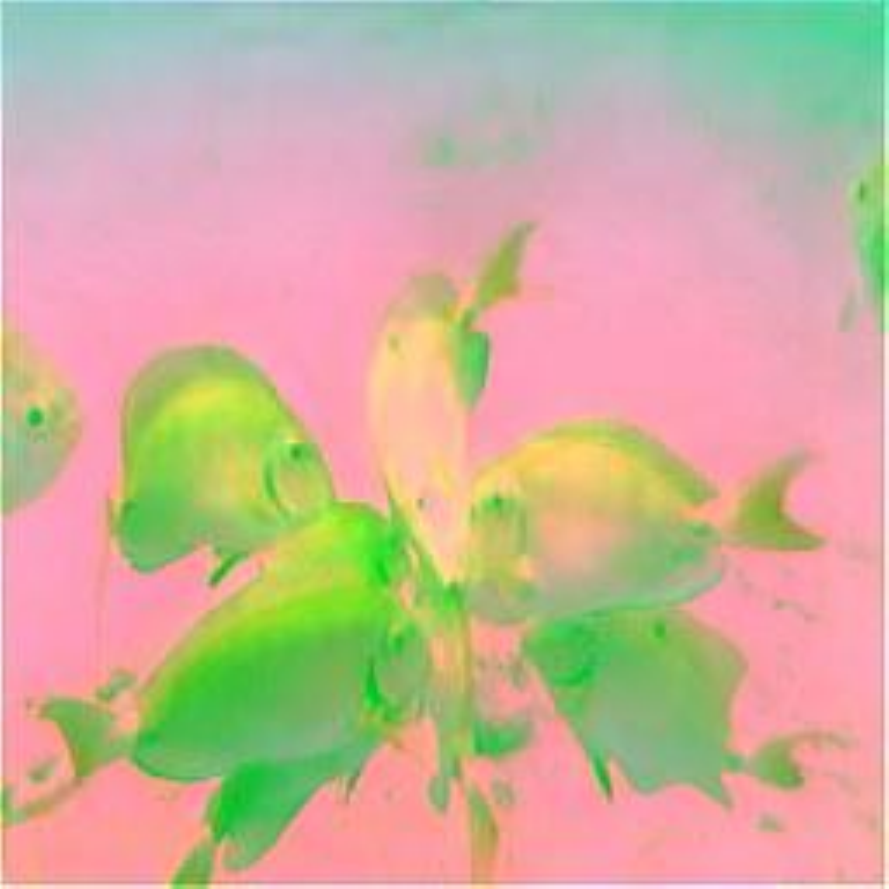}
&
\includegraphics[width=0.095\textwidth]{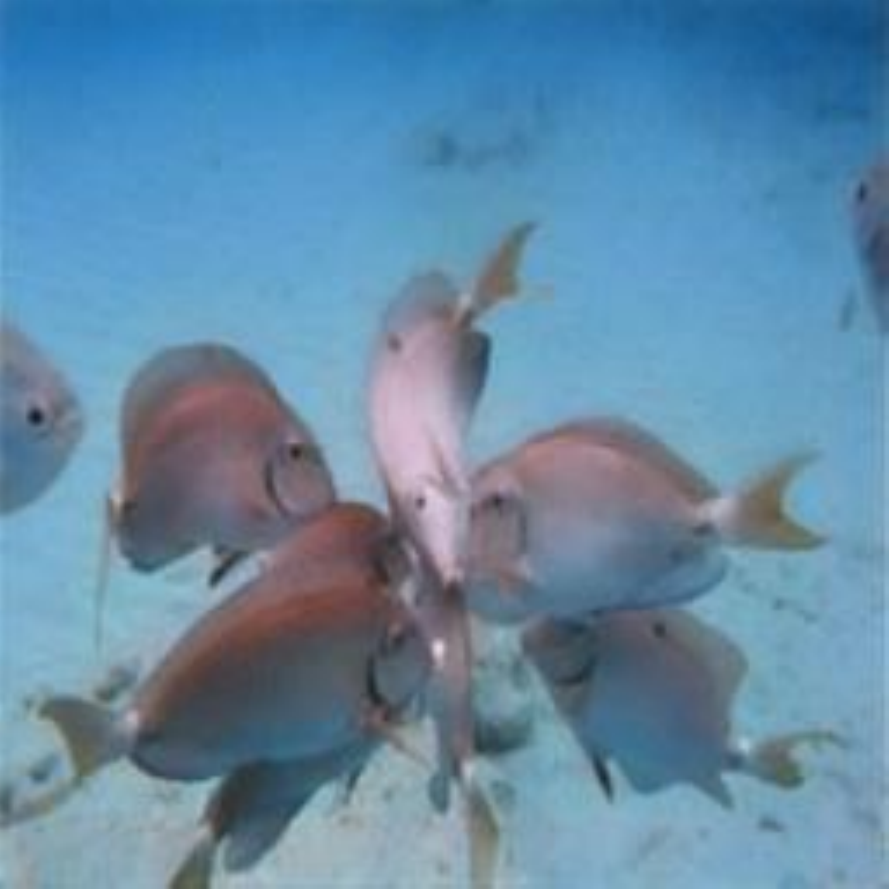}
&
\includegraphics[width=0.095\textwidth]{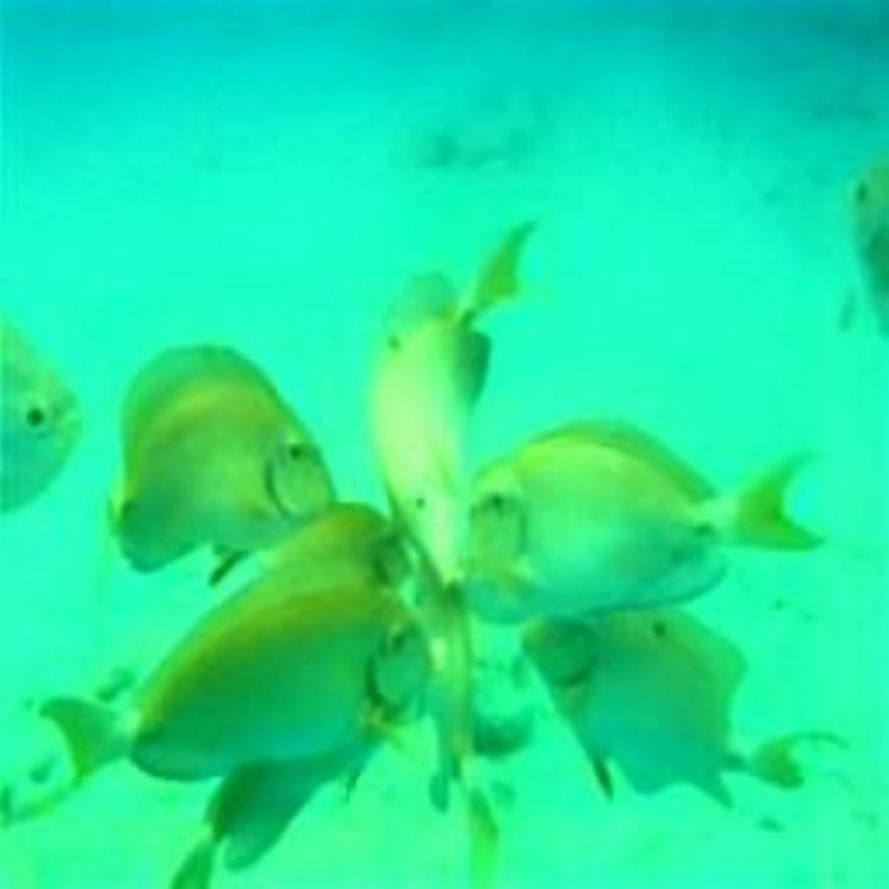}
&
\includegraphics[width=0.095\textwidth]{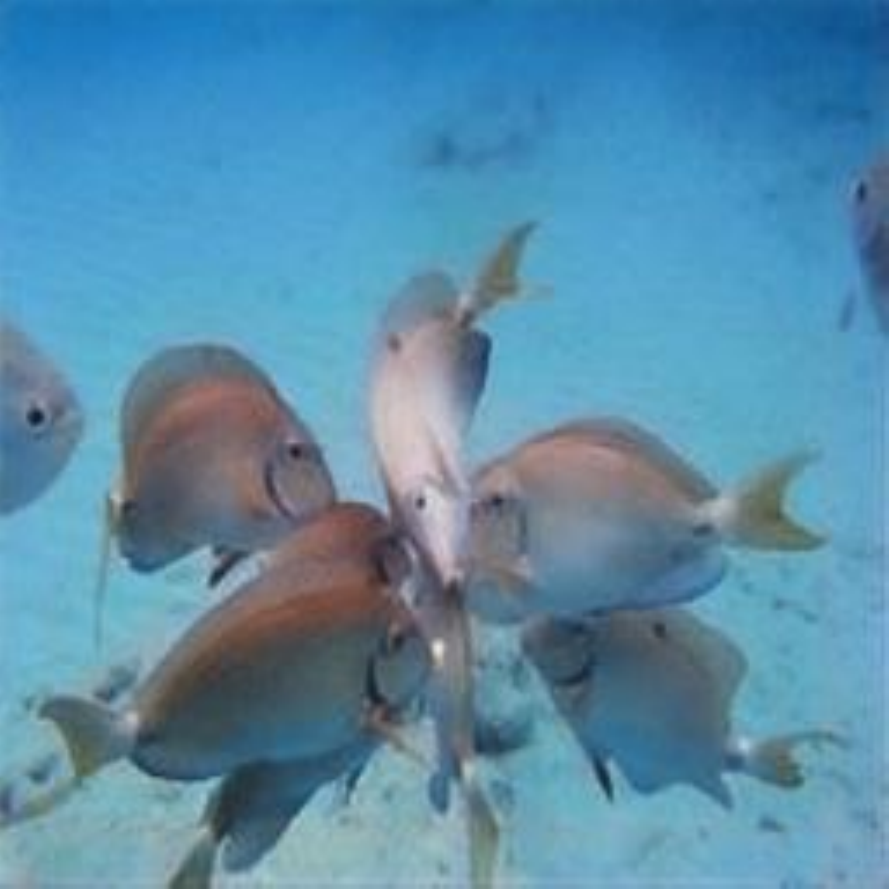}
&
\includegraphics[width=0.095\textwidth]{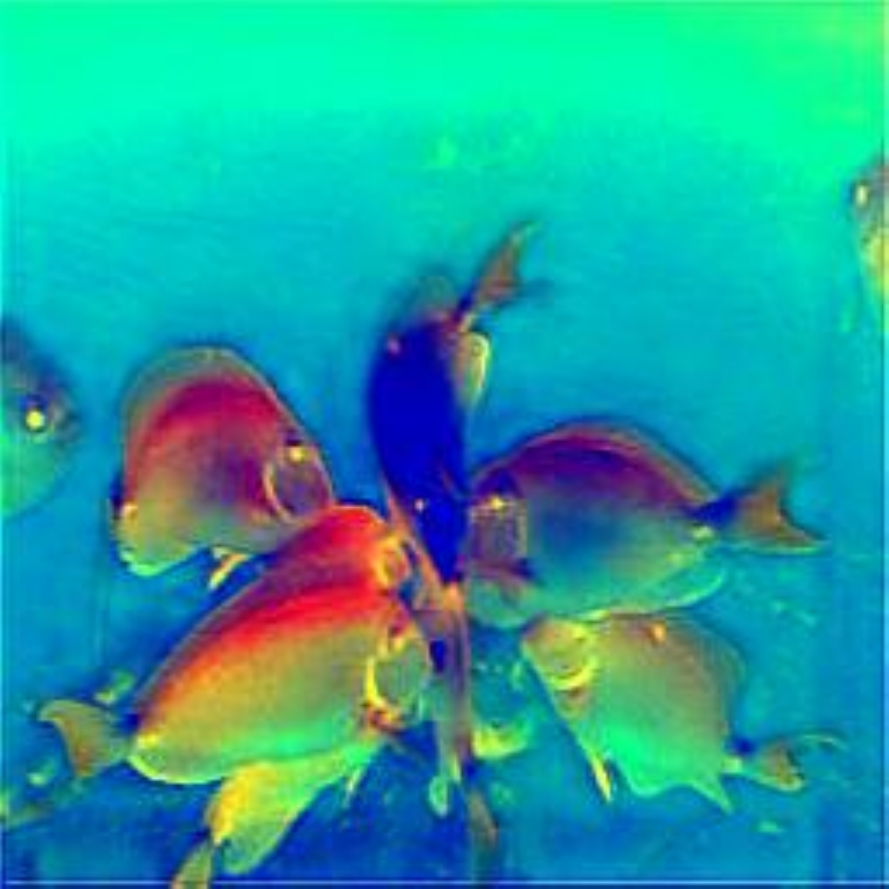}
&
\includegraphics[width=0.095\textwidth]{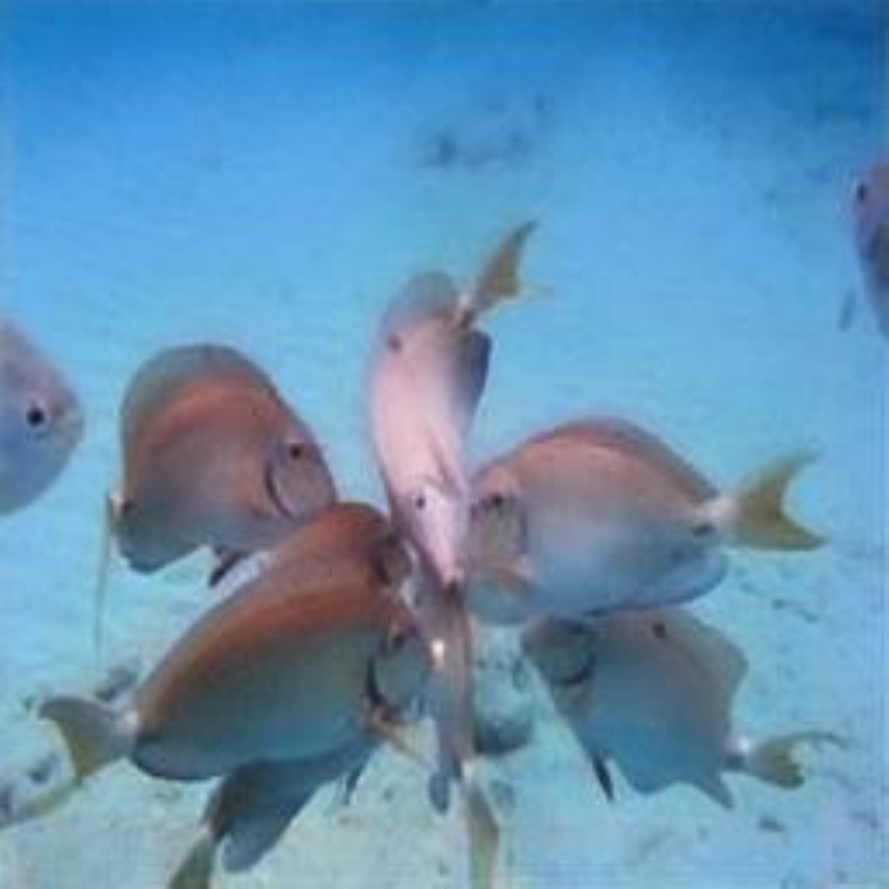}
&
\includegraphics[width=0.095\textwidth]{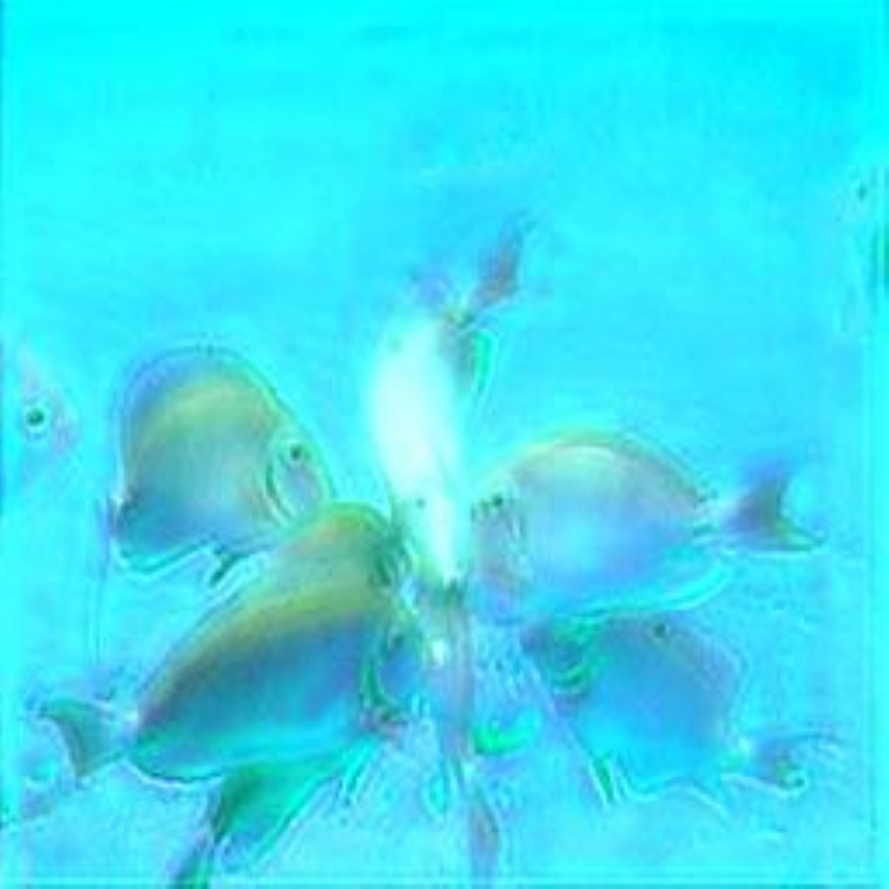}
&
\includegraphics[width=0.095\textwidth]{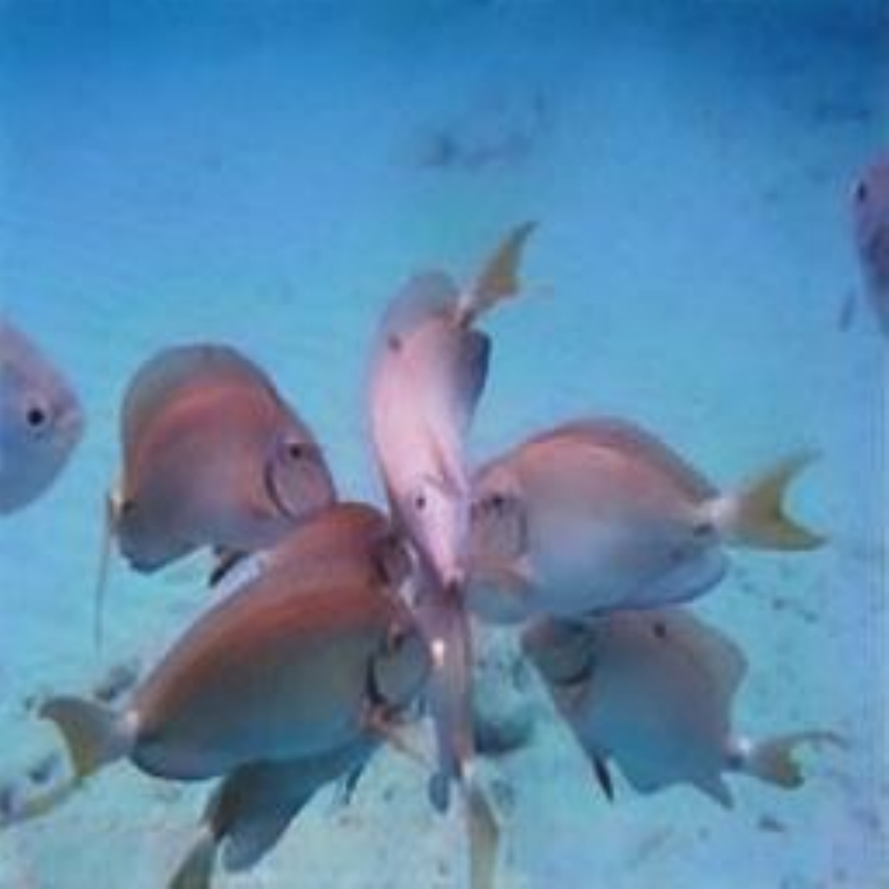}
&
\includegraphics[width=0.095\textwidth]{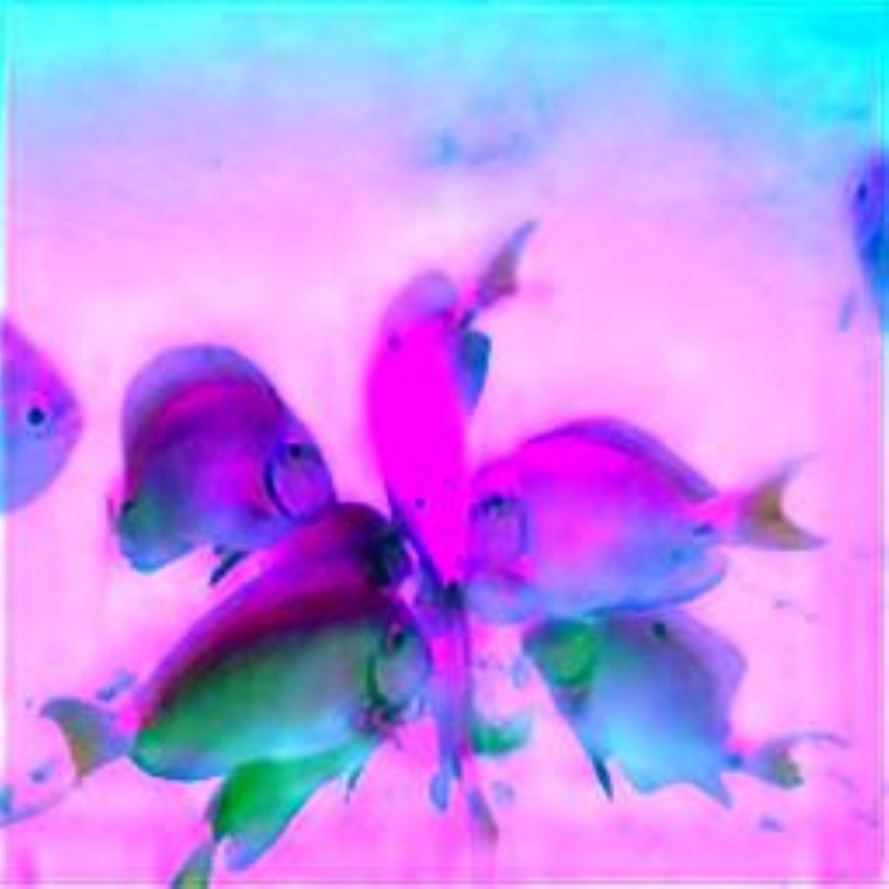}\\

\includegraphics[width=0.095\textwidth]{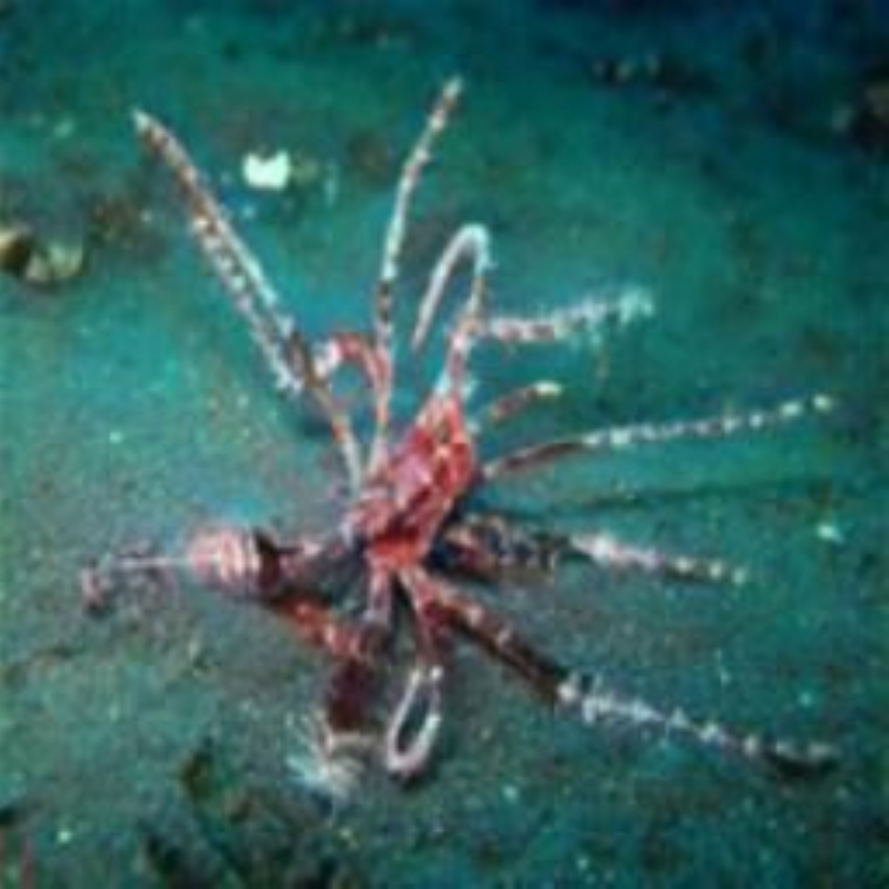}
& 
\includegraphics[width=0.095\textwidth]{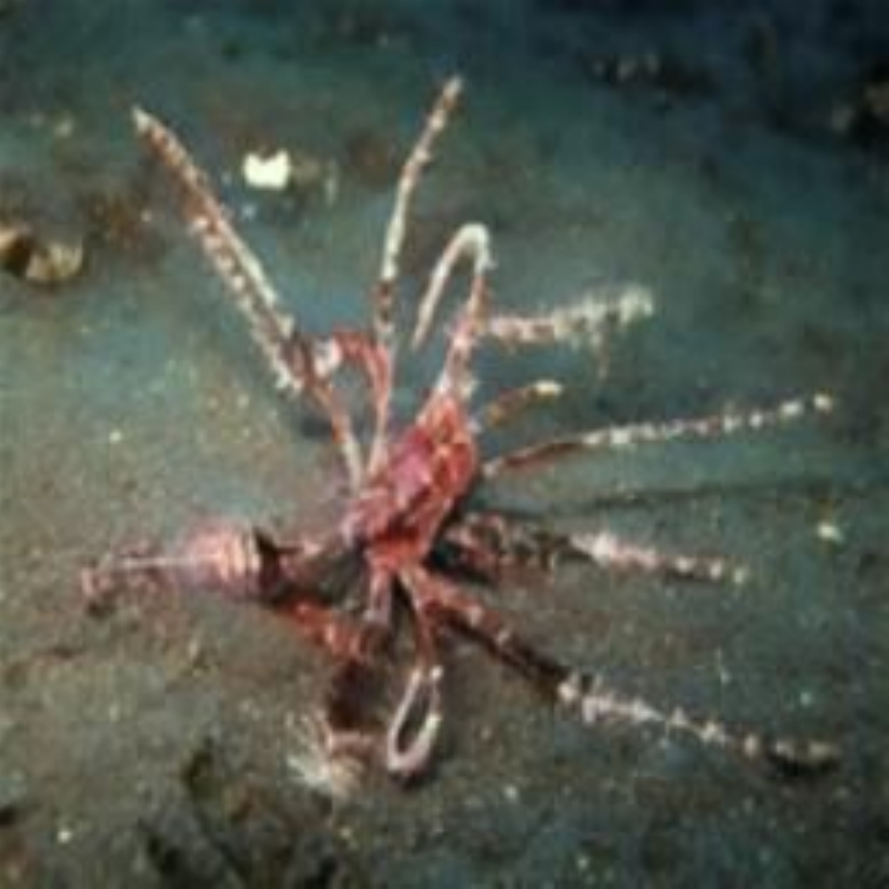}
&
\includegraphics[width=0.095\textwidth]{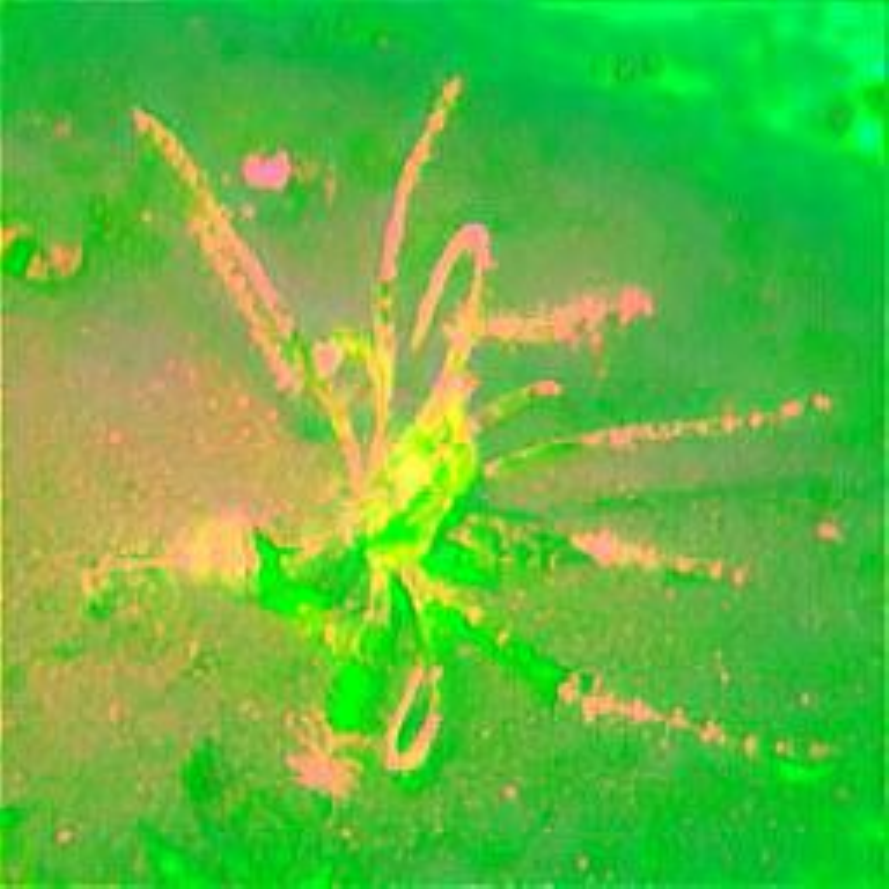}
&
\includegraphics[width=0.095\textwidth]{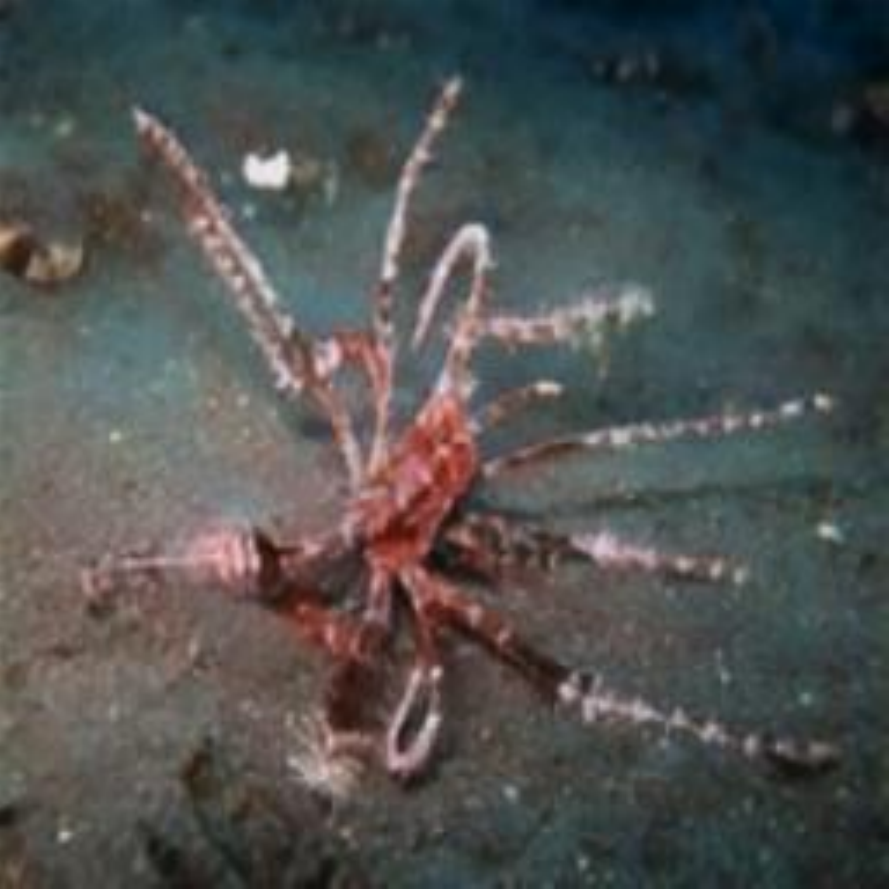}
&
\includegraphics[width=0.095\textwidth]{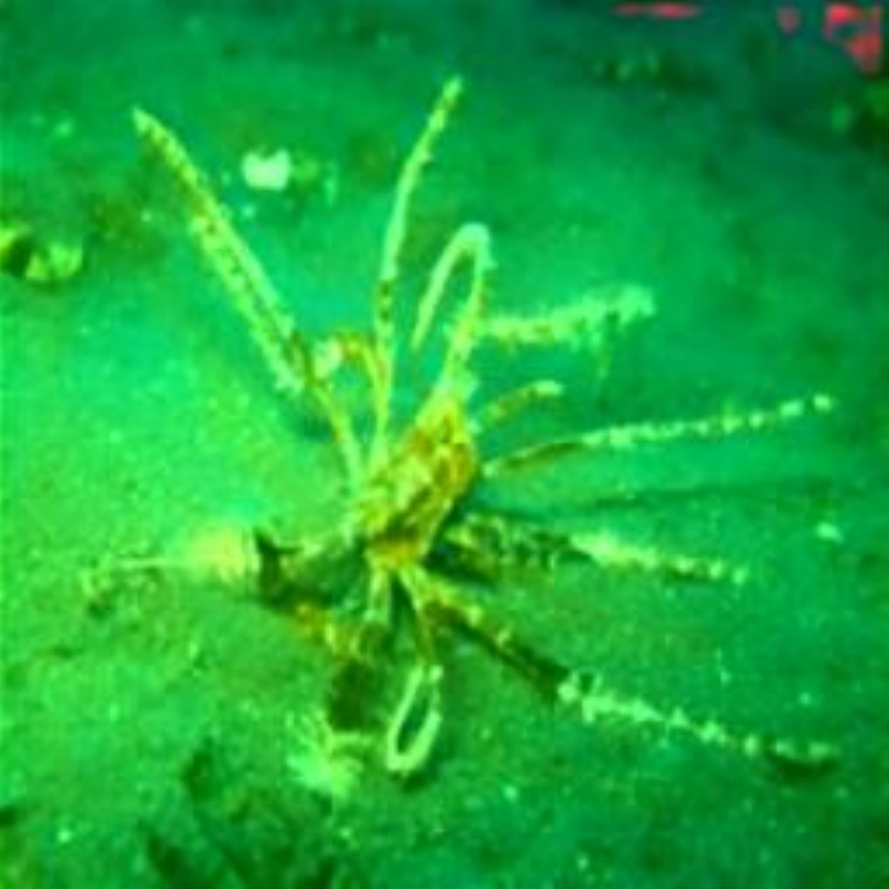}
&
\includegraphics[width=0.095\textwidth]{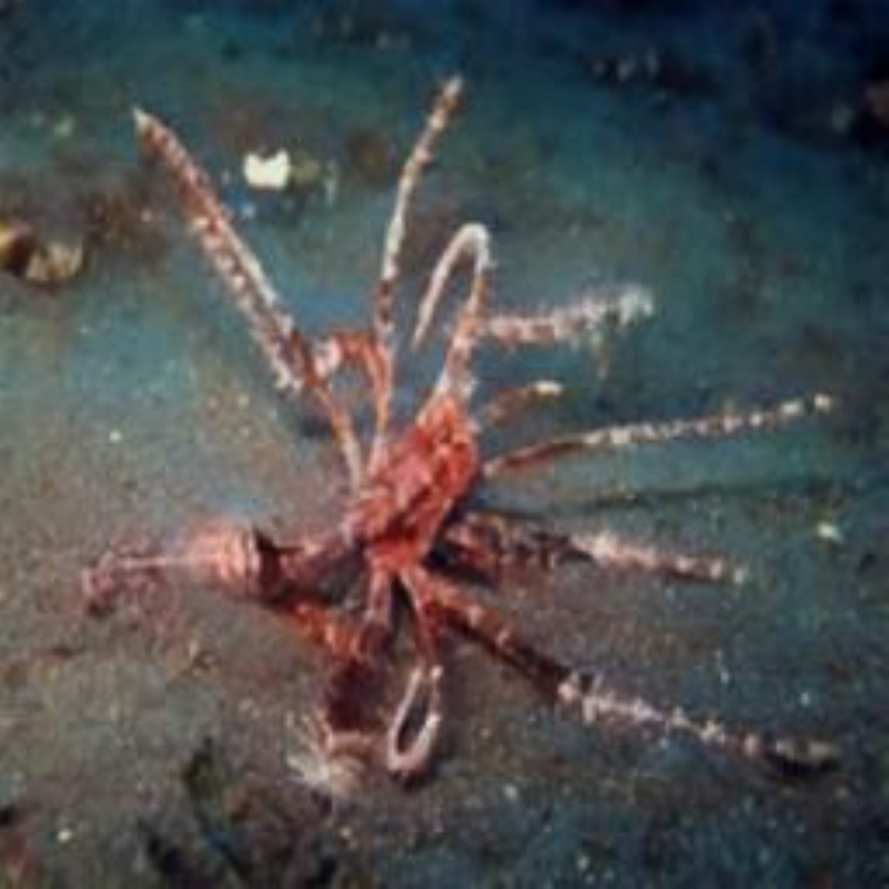}
&
\includegraphics[width=0.095\textwidth]{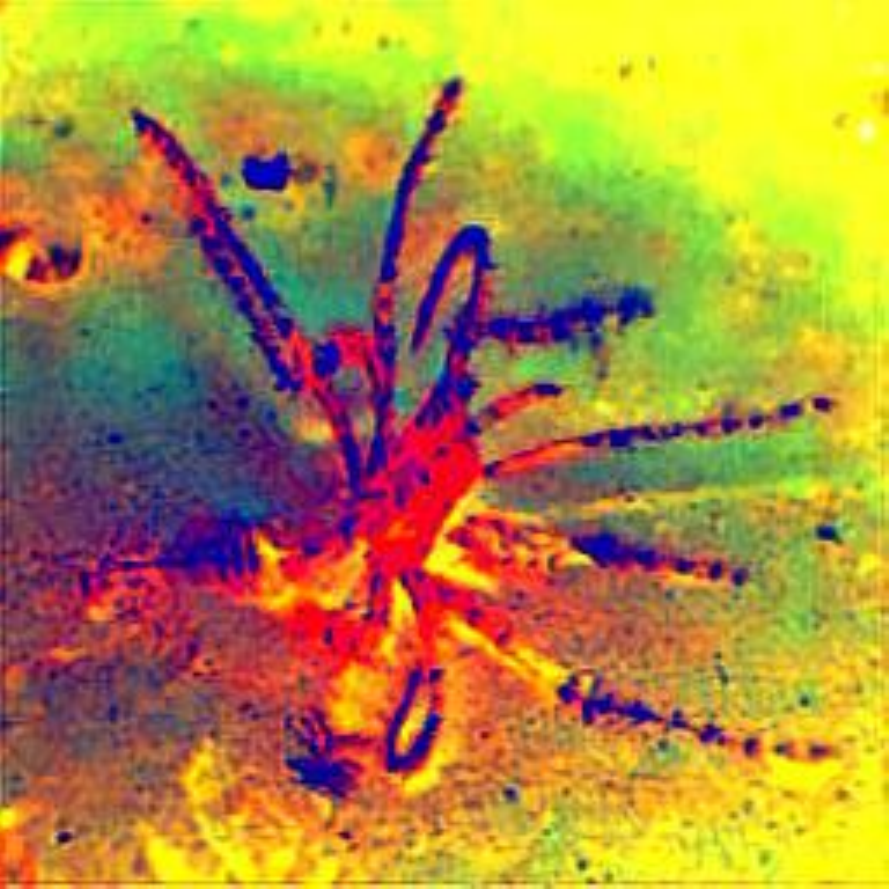}
&
\includegraphics[width=0.095\textwidth]{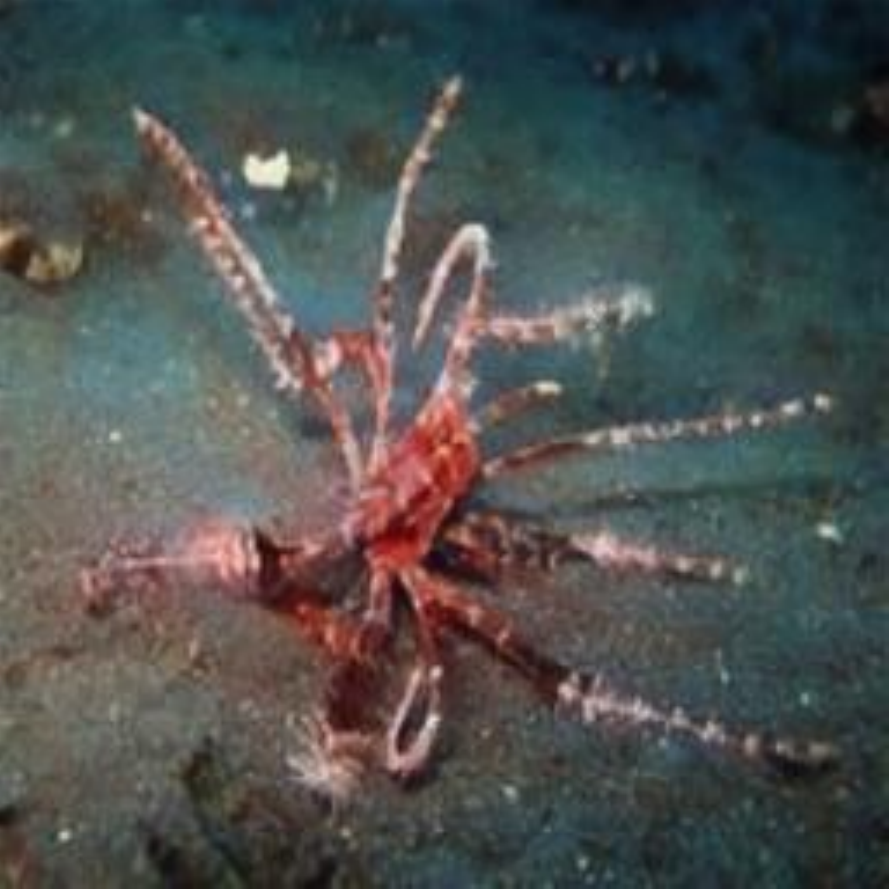}
&
\includegraphics[width=0.095\textwidth]{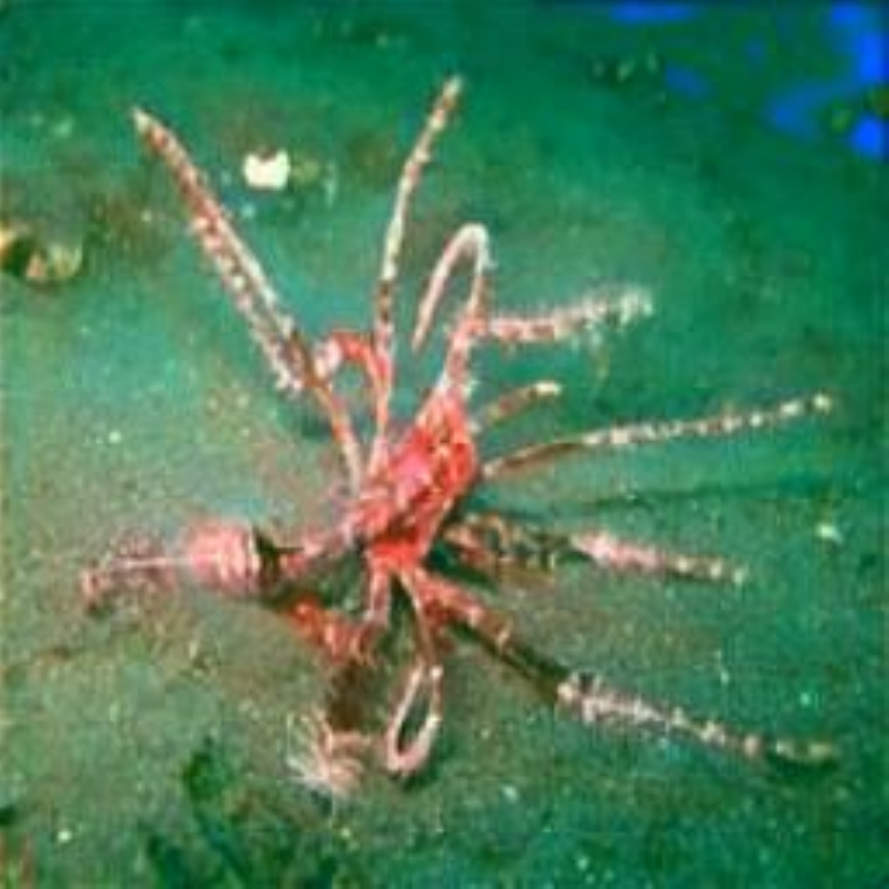}
&
\includegraphics[width=0.095\textwidth]{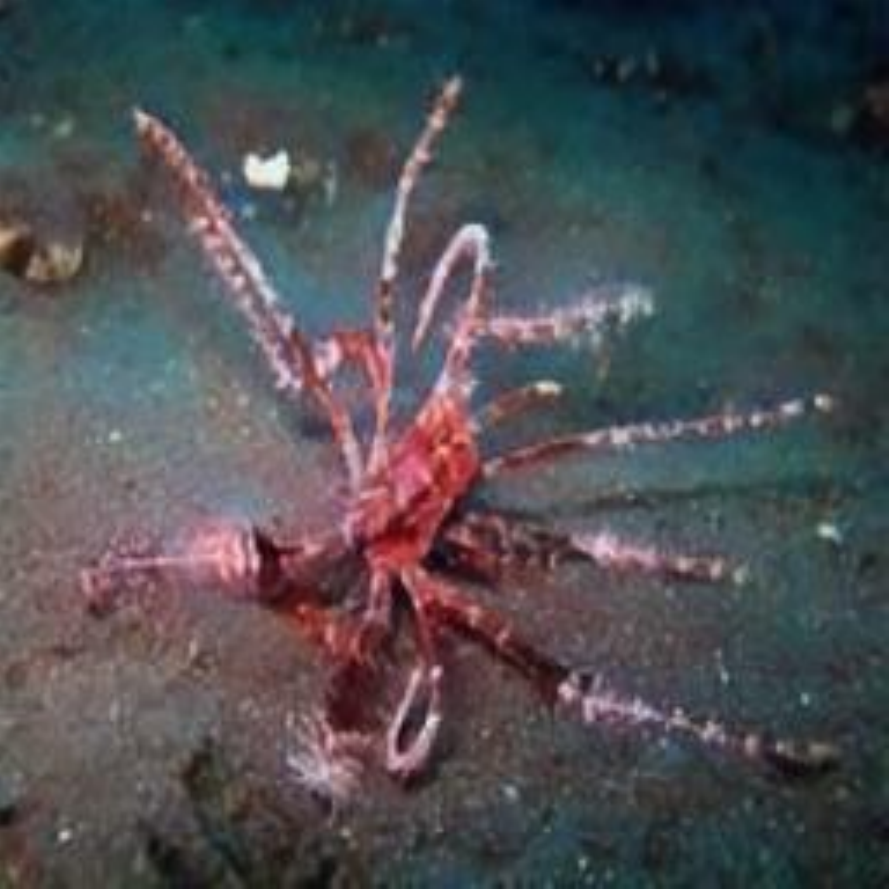}
&
\includegraphics[width=0.095\textwidth]{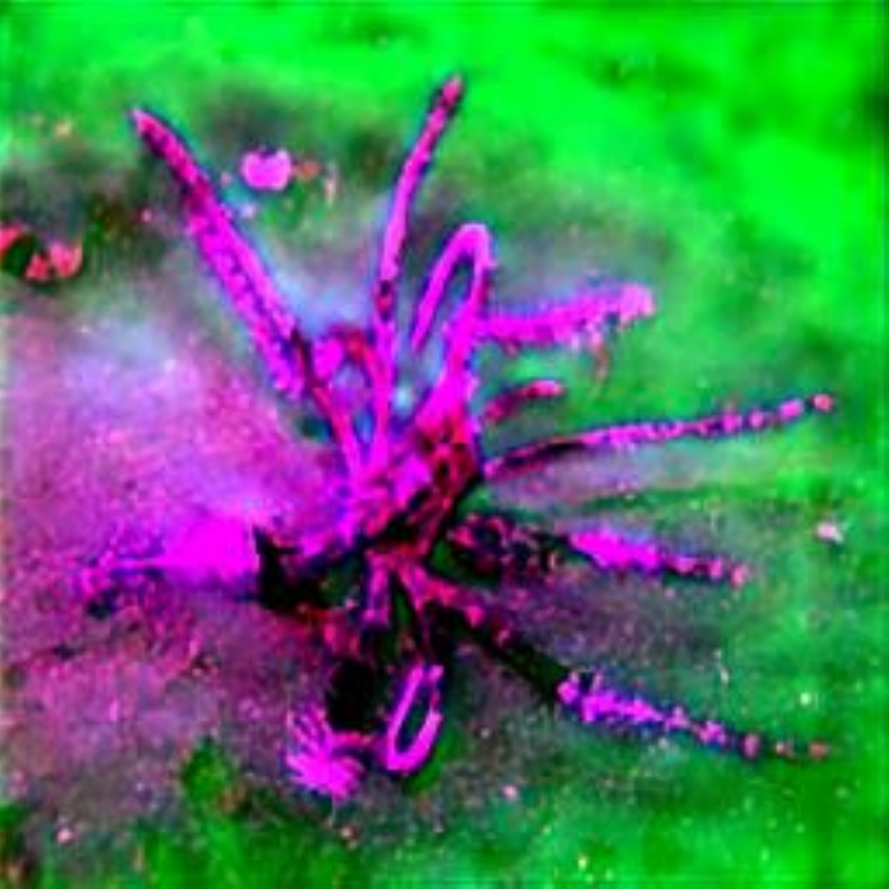}\\

(a) & (b) & (c) & (d) & (e) & (f) & (g) & (h) & (i) & (j) & (k) \\

\end{tabular}}
\caption{Qualitative demonstration of results obtained by utilizing different baseline modules as mentioned in Section \ref{sec:ablation_wave_cbam}. Deep WaveNet (DW)-* Res. denotes learned global color correction residual $M^3_{<3,5,7>}$ (see Section \ref{proposed_approach}) using the respective baseline. (a) Degraded, (b) Deep WaveNet-1, (c) DW-1 Res., (d) Deep WaveNet-2, (e) DW-2 Res., (f) Deep WaveNet-3, (g) DW-3 Res., (h) Deep WaveNet-M, (i) DW-M Res., (j) Deep WaveNet, (k) DW Res. Please zoom the figures to observe the difference.}
\label{fig:ablation_modules}
\end{figure*}
\begin{table}[t]
\begin{center}
\caption{Comparison against different baselines on \texttt{EUVP} dataset for the task of image enhancement. $\bigtriangledown$ denotes lower is better.}
\label{tab:euvp_model}
\vspace*{-\baselineskip}
\resizebox{0.6\textwidth}{!}{
\begin{tabular}{lccccc}
\toprule
\midrule
\text{Methods} & Wavelength-driven & CBAM & \text{MSE}$\bigtriangledown$ & \text{PSNR} & \text{SSIM}\\
\midrule
\midrule

Deep WaveNet-1 & \textcolor{red}{\ding{55}} & \textcolor{red}{\ding{55}} & $.37$ & $27.45$ & $.80$ \\

Deep WaveNet-2 & \textcolor{green}{\ding{51}} & \textcolor{red}{\ding{55}} & $.37$ & $27.62$ & $.81$ \\

Deep WaveNet-3 & \textcolor{red}{\ding{55}} & \textcolor{green}{\ding{51}} & {\color{blue}$\mathbf{.36}$} & {\color{black}$27.99$} & {\color{blue}$\mathbf{.82}$} \\

Deep WaveNet-M & \textcolor{red}{\ding{55}} & \textcolor{green}{\ding{51}} & $.57$ & {\color{blue}$\mathbf{28.02}$} & $.81$ \\

Deep WaveNet & \textcolor{green}{\ding{51}} & \textcolor{green}{\ding{51}} & {\textcolor{red}{$\mathbf{.29}$}}&{\color{red}$\mathbf{28.62}$}&{\color{red}$\mathbf{.83}$}\\
\bottomrule
\bottomrule
\end{tabular}}

\end{center}
\end{table}

To show the effect of wavelength-driven contextual size formulation and CBAM \cite{cbam}, we have performed the following baselines on \texttt{EUVP} dataset 
\begin{enumerate}
\item Deep WaveNet-1: The proposed model (see Fig. \ref{fig:main_model}, Section \ref{proposed_approach}), where a contextual size of $3\times 3$ has been set for every layer. Further, the CBAM \cite{cbam} modules have been removed throughout.

\item Deep WaveNet-2: The proposed model (see Fig. \ref{fig:main_model}, Section \ref{proposed_approach}), where the wavelength-driven contextual setting has been used without any CBAM \cite{cbam} module.

\item Deep WaveNet-3: The proposed model (see Fig. \ref{fig:main_model}, Section \ref{proposed_approach}), where a contextual size of $3\times 3$ has been set for every layer without removing the CBAM \cite{cbam} modules.

\item Deep WaveNet-M: Similar to the proposed model in Fig. \ref{fig:main_model}, except in stage 1, where instead of processing individual channels, each convolution layer ($3\times 3$, $5\times 5$, $7\times 7$) process the whole RGB input image. This can be another approach of utilizing multi-contextual features. 
\end{enumerate} 

For simplicity, we have given a categorical description of each baseline in Table \ref{tab:euvp_model}. To summarize, it has been observed that the model (Deep WaveNet-1) with the homogeneous contextual size of $3 \times 3$ without CBAM \cite{cbam} has a comparatively minimum performance. Over Deep WaveNet-1, there has been a significant improvement due to the addition of wavelength-driven multi-contextual formulation (Deep WaveNet-2). However, the performance further notably improved when incorporated with CBAM \cite{cbam} module (Deep WaveNet). To justify the role of wavelength-driven multi-contextual formulation and CBAM \cite{cbam} module, consider Fig. \ref{fig:ablation_modules}, where we have shown the global color-correction residuals $M^3_{<3,5,7>}$ (see Section \ref{proposed_approach}) learned by the proposed baselines. It can be clearly observed that the residuals (Deep WaveNet-1 Res.) learned by the proposed baseline Deep WaveNet-1 do not comprise of much useful details that can help in improving the end performance. Similarly, adding wavelength-driven multi-contextual setting (Deep WaveNet-2 Res.) or CBAM \cite{cbam} (Deep WaveNet-3 Res.) may not convey much about what the network is trying to learn for the end enhancement. 

Deep WaveNet-M is architecturally similar to Deep WaveNet (see Fig. \ref{proposed_approach}). However, in stage 1, it processes the whole RGB image with different contexts instead of channel-wise processing. As a result, although the multi-contextual formulation may learn local and global coherence spatially. But it may fail to address the same channel-wise. As each channel of the image is being processed by a homogeneous contextual size. This formulation may be reasonable in natural outdoor images where light attenuation is assumed to be uniform \cite{berman_pami_20}. Whereas it is not the case with underwater images (see Fig. \ref{fig:intro_wavelength}). As evident from Fig. \ref{fig:ablation_modules} and Table \ref{tab:euvp_model}, the residuals learned by the Deep WaveNet-M are unrefined compared to Deep WaveNet, hence delivered a slight under-performance. When the Deep WaveNet-1 is reinforced with wavelength-driven multi-contextual design and CBAM \cite{cbam} module, it can be said that the proposed model is trying to retain the blueish and greenish essence of water and marine life (see Deep WaveNet Res.).  For \textit{e.g.}, the blueness around the \texttt{turtle} in Fig. \ref{fig:ablation_modules} (Row 2, DW Res.), or segmenting water and marine life in (Row 3, DW Res.). Therefore, it may be concluded that the addition of both wavelength-driven multi-contextual design and CBAM \cite{cbam} module may have helped the proposed model in learning essential features for an efficient UIR.

\subsubsection{Run-time and Memory Usages of Different Baselines:}
\textcolor{black}{We have already presented the qualitative and quantitative results of different baselines of Deep WaveNet in above section. We now present the memory usage and computational overhead of each baseline during training and inference in Table \ref{tab:comem}.  }
\begin{table}[h]
\caption{{Computational overhead of different baselines in terms of run-time and memory usage.}}
\label{tab:comem}
\vspace*{-\baselineskip}
\centering
\resizebox{\linewidth}{!}{
\begin{tabular}{c|c|c|c|c}
\hline
\hline
\multirow{ 2}{*}{Baseline} & \multicolumn{2}{|c}{Training} & \multicolumn{2}{|c}{Inference}\\
\cline{2-5}
 & Time (\textit{per epoch in minutes}) & Memory (\textit{per batch in GB}) & Time (\textit{per image in seconds}) & Memory (\textit{per image in GB})\\
\hline
Deep WaveNet-1 & 5.71 & 3.43 & 0.01 & 1.51 \\
Deep WaveNet-2 & 17.83 & 3.67 & 0.05 & 1.75 \\
Deep WaveNet-3 & 11.87 & 4.02 & 0.03 & 2.21 \\
Deep WaveNet-M & 40.17 & 4.74 & 0.13 & 2.73 \\
Deep WaveNet & 36.74 & 4.50 & 0.10 & 2.64 \\
\hline
\hline
\end{tabular}}
\end{table}
\textcolor{black}{The experiments have been conducted using Nvidia P100 16 GB GPU. Note that images of size $256 \times 256$ have been used with batch size of 5 and 1 during training and testing, respectively.}

\section{Conclusions}
\label{sec:conclusions}
In this paper, we proposed a novel deep learning-based model for simultaneous underwater image enhancement and super-resolution. We proposed to utilize the distinctive receptive field size specific to each channel of the image, driven by its wavelength. We showed that such a type of multi-contextual formulation helps in learning the diverse local and global features of each channel of the underwater images.  Further, the learned features are adaptively refined using a block attention mechanism that remarkably enhanced the performance of the proposed scheme. Our proposed model is architecturally extensible to support spatial super-resolution of the enhanced underwater images. We have shown the supremacy of the proposed scheme across various benchmark datasets over existing best-published works. We also provided extensive ablation experiments to justify the contributions of different modules. In the future, we would like to extend this approach for underwater video enhancement and super-resolution.

%

\bibliographystyle{ACM-Reference-Format}
\bibliography{sample-base}


\begin{thebibliography}{90}


\ifx \showCODEN    \undefined \def \showCODEN     #1{\unskip}     \fi
\ifx \showDOI      \undefined \def \showDOI       #1{#1}\fi
\ifx \showISBNx    \undefined \def \showISBNx     #1{\unskip}     \fi
\ifx \showISBNxiii \undefined \def \showISBNxiii  #1{\unskip}     \fi
\ifx \showISSN     \undefined \def \showISSN      #1{\unskip}     \fi
\ifx \showLCCN     \undefined \def \showLCCN      #1{\unskip}     \fi
\ifx \shownote     \undefined \def \shownote      #1{#1}          \fi
\ifx \showarticletitle \undefined \def \showarticletitle #1{#1}   \fi
\ifx \showURL      \undefined \def \showURL       {\relax}        \fi
\providecommand\bibfield[2]{#2}
\providecommand\bibinfo[2]{#2}
\providecommand\natexlab[1]{#1}
\providecommand\showeprint[2][]{arXiv:#2}

\bibitem[\protect\citeauthoryear{Alexandridis and Zapranis}{Alexandridis and
  Zapranis}{2013}]%
        {WCNN}
\bibfield{author}{\bibinfo{person}{Antonios~K. Alexandridis} {and}
  \bibinfo{person}{Achilleas~D. Zapranis}.} \bibinfo{year}{2013}\natexlab{}.
\newblock \showarticletitle{Wavelet neural networks: A practical guide}.
\newblock \bibinfo{journal}{\emph{Neural Networks}}  \bibinfo{volume}{42}
  (\bibinfo{year}{2013}), \bibinfo{pages}{1--27}.
\newblock
\showISSN{0893-6080}
\urldef\tempurl%
\url{https://doi.org/10.1016/j.neunet.2013.01.008}
\showDOI{\tempurl}


\bibitem[\protect\citeauthoryear{{Ancuti}, {Ancuti}, {Haber}, and
  {Bekaert}}{{Ancuti} et~al\mbox{.}}{2012}]%
        {ancuti1}
\bibfield{author}{\bibinfo{person}{C. {Ancuti}}, \bibinfo{person}{C.~O.
  {Ancuti}}, \bibinfo{person}{T. {Haber}}, {and} \bibinfo{person}{P.
  {Bekaert}}.} \bibinfo{year}{2012}\natexlab{}.
\newblock \showarticletitle{Enhancing underwater images and videos by fusion}.
  In \bibinfo{booktitle}{\emph{2012 IEEE Conference on Computer Vision and
  Pattern Recognition}}. \bibinfo{pages}{81--88}.
\newblock
\urldef\tempurl%
\url{https://doi.org/10.1109/CVPR.2012.6247661}
\showDOI{\tempurl}


\bibitem[\protect\citeauthoryear{{Ancuti}, {Ancuti}, {De Vleeschouwer}, and
  {Bekaert}}{{Ancuti} et~al\mbox{.}}{2018}]%
        {ancuti_tip_18}
\bibfield{author}{\bibinfo{person}{C.~O. {Ancuti}}, \bibinfo{person}{C.
  {Ancuti}}, \bibinfo{person}{C. {De Vleeschouwer}}, {and} \bibinfo{person}{P.
  {Bekaert}}.} \bibinfo{year}{2018}\natexlab{}.
\newblock \showarticletitle{Color Balance and Fusion for Underwater Image
  Enhancement}.
\newblock \bibinfo{journal}{\emph{IEEE Transactions on Image Processing}}
  \bibinfo{volume}{27}, \bibinfo{number}{1} (\bibinfo{year}{2018}),
  \bibinfo{pages}{379--393}.
\newblock
\urldef\tempurl%
\url{https://doi.org/10.1109/TIP.2017.2759252}
\showDOI{\tempurl}


\bibitem[\protect\citeauthoryear{Anonymous}{Anonymous}{2022}]%
        {anonymous2022patches}
\bibfield{author}{\bibinfo{person}{Anonymous}.}
  \bibinfo{year}{2022}\natexlab{}.
\newblock \showarticletitle{Patches Are All You Need?}. In
  \bibinfo{booktitle}{\emph{Submitted to The Tenth International Conference on
  Learning Representations}}.
\newblock
\urldef\tempurl%
\url{https://openreview.net/forum?id=TVHS5Y4dNvM}
\showURL{%
\tempurl}
\newblock
\shownote{under review.}


\bibitem[\protect\citeauthoryear{{Bai}, {Zhang}, {Pan}, and {Zhao}}{{Bai}
  et~al\mbox{.}}{2020}]%
        {bai}
\bibfield{author}{\bibinfo{person}{L. {Bai}}, \bibinfo{person}{W. {Zhang}},
  \bibinfo{person}{X. {Pan}}, {and} \bibinfo{person}{C. {Zhao}}.}
  \bibinfo{year}{2020}\natexlab{}.
\newblock \showarticletitle{Underwater Image Enhancement Based on Global and
  Local Equalization of Histogram and Dual-Image Multi-Scale Fusion}.
\newblock \bibinfo{journal}{\emph{IEEE Access}}  \bibinfo{volume}{8}
  (\bibinfo{year}{2020}), \bibinfo{pages}{128973--128990}.
\newblock
\urldef\tempurl%
\url{https://doi.org/10.1109/ACCESS.2020.3009161}
\showDOI{\tempurl}


\bibitem[\protect\citeauthoryear{{Berman}, {Levy}, {Avidan}, and
  {Treibitz}}{{Berman} et~al\mbox{.}}{2020}]%
        {berman_pami_20}
\bibfield{author}{\bibinfo{person}{D. {Berman}}, \bibinfo{person}{D. {Levy}},
  \bibinfo{person}{S. {Avidan}}, {and} \bibinfo{person}{T. {Treibitz}}.}
  \bibinfo{year}{2020}\natexlab{}.
\newblock \showarticletitle{Underwater Single Image Color Restoration Using
  Haze-Lines and a New Quantitative Dataset}.
\newblock \bibinfo{journal}{\emph{IEEE Transactions on Pattern Analysis and
  Machine Intelligence}} (\bibinfo{year}{2020}), \bibinfo{pages}{1--1}.
\newblock
\urldef\tempurl%
\url{https://doi.org/10.1109/TPAMI.2020.2977624}
\showDOI{\tempurl}


\bibitem[\protect\citeauthoryear{{Cao}, {Hidalgo Martinez}, {Simon}, {Wei}, and
  {Sheikh}}{{Cao} et~al\mbox{.}}{2019}]%
        {openpose1}
\bibfield{author}{\bibinfo{person}{Z. {Cao}}, \bibinfo{person}{G. {Hidalgo
  Martinez}}, \bibinfo{person}{T. {Simon}}, \bibinfo{person}{S. {Wei}}, {and}
  \bibinfo{person}{Y.~A. {Sheikh}}.} \bibinfo{year}{2019}\natexlab{}.
\newblock \showarticletitle{OpenPose: Realtime Multi-Person 2D Pose Estimation
  using Part Affinity Fields}.
\newblock \bibinfo{journal}{\emph{IEEE Transactions on Pattern Analysis and
  Machine Intelligence}} (\bibinfo{year}{2019}).
\newblock


\bibitem[\protect\citeauthoryear{Cao, Simon, Wei, and Sheikh}{Cao
  et~al\mbox{.}}{2017}]%
        {openpose3}
\bibfield{author}{\bibinfo{person}{Zhe Cao}, \bibinfo{person}{Tomas Simon},
  \bibinfo{person}{Shih-En Wei}, {and} \bibinfo{person}{Yaser Sheikh}.}
  \bibinfo{year}{2017}\natexlab{}.
\newblock \showarticletitle{Realtime Multi-Person 2D Pose Estimation using Part
  Affinity Fields}. In \bibinfo{booktitle}{\emph{CVPR}}.
\newblock


\bibitem[\protect\citeauthoryear{{Chen}, {Jiang}, {Tong}, {Liu}, {Zhao},
  {Zhang}, {Dong}, and {Zhou}}{{Chen} et~al\mbox{.}}{2020a}]%
        {chen_csvt}
\bibfield{author}{\bibinfo{person}{L. {Chen}}, \bibinfo{person}{Z. {Jiang}},
  \bibinfo{person}{L. {Tong}}, \bibinfo{person}{Z. {Liu}}, \bibinfo{person}{A.
  {Zhao}}, \bibinfo{person}{Q. {Zhang}}, \bibinfo{person}{J. {Dong}}, {and}
  \bibinfo{person}{H. {Zhou}}.} \bibinfo{year}{2020}\natexlab{a}.
\newblock \showarticletitle{Perceptual underwater image enhancement with deep
  learning and physical priors}.
\newblock \bibinfo{journal}{\emph{IEEE Transactions on Circuits and Systems for
  Video Technology}} (\bibinfo{year}{2020}), \bibinfo{pages}{1--1}.
\newblock
\urldef\tempurl%
\url{https://doi.org/10.1109/TCSVT.2020.3035108}
\showDOI{\tempurl}


\bibitem[\protect\citeauthoryear{{Chen}, {Niu}, {Zeng}, and {Pan}}{{Chen}
  et~al\mbox{.}}{2020b}]%
        {chen_access_2020}
\bibfield{author}{\bibinfo{person}{Y. {Chen}}, \bibinfo{person}{K. {Niu}},
  \bibinfo{person}{Z. {Zeng}}, {and} \bibinfo{person}{Y. {Pan}}.}
  \bibinfo{year}{2020}\natexlab{b}.
\newblock \showarticletitle{A Wavelet Based Deep Learning Method for Underwater
  Image Super Resolution Reconstruction}.
\newblock \bibinfo{journal}{\emph{IEEE Access}}  \bibinfo{volume}{8}
  (\bibinfo{year}{2020}), \bibinfo{pages}{117759--117769}.
\newblock
\urldef\tempurl%
\url{https://doi.org/10.1109/ACCESS.2020.3004141}
\showDOI{\tempurl}


\bibitem[\protect\citeauthoryear{{Chiang} and {Chen}}{{Chiang} and
  {Chen}}{2012}]%
        {chiang}
\bibfield{author}{\bibinfo{person}{J.~Y. {Chiang}} {and} \bibinfo{person}{Y.
  {Chen}}.} \bibinfo{year}{2012}\natexlab{}.
\newblock \showarticletitle{Underwater Image Enhancement by Wavelength
  Compensation and Dehazing}.
\newblock \bibinfo{journal}{\emph{IEEE Transactions on Image Processing}}
  \bibinfo{volume}{21}, \bibinfo{number}{4} (\bibinfo{year}{2012}),
  \bibinfo{pages}{1756--1769}.
\newblock
\urldef\tempurl%
\url{https://doi.org/10.1109/TIP.2011.2179666}
\showDOI{\tempurl}


\bibitem[\protect\citeauthoryear{Deng, Dong, Socher, Li, Li, and Fei-Fei}{Deng
  et~al\mbox{.}}{2009}]%
        {imagenet}
\bibfield{author}{\bibinfo{person}{Jia Deng}, \bibinfo{person}{Wei Dong},
  \bibinfo{person}{Richard Socher}, \bibinfo{person}{Li-Jia Li},
  \bibinfo{person}{Kai Li}, {and} \bibinfo{person}{Li Fei-Fei}.}
  \bibinfo{year}{2009}\natexlab{}.
\newblock \showarticletitle{ImageNet: A large-scale hierarchical image
  database}. In \bibinfo{booktitle}{\emph{2009 IEEE Conference on Computer
  Vision and Pattern Recognition}}. \bibinfo{pages}{248--255}.
\newblock
\urldef\tempurl%
\url{https://doi.org/10.1109/CVPR.2009.5206848}
\showDOI{\tempurl}


\bibitem[\protect\citeauthoryear{Divakar and Venkatesh~Babu}{Divakar and
  Venkatesh~Babu}{2017}]%
        {divakar_cvpr}
\bibfield{author}{\bibinfo{person}{Nithish Divakar} {and} \bibinfo{person}{R.
  Venkatesh~Babu}.} \bibinfo{year}{2017}\natexlab{}.
\newblock \showarticletitle{Image Denoising via CNNs: An Adversarial Approach}.
  In \bibinfo{booktitle}{\emph{Proceedings of the IEEE Conference on Computer
  Vision and Pattern Recognition (CVPR) Workshops}}.
\newblock


\bibitem[\protect\citeauthoryear{Dong, Deng, Loy, and Tang}{Dong
  et~al\mbox{.}}{2015}]%
        {deconv1}
\bibfield{author}{\bibinfo{person}{Chao Dong}, \bibinfo{person}{Yubin Deng},
  \bibinfo{person}{Chen~Change Loy}, {and} \bibinfo{person}{Xiaoou Tang}.}
  \bibinfo{year}{2015}\natexlab{}.
\newblock \showarticletitle{Compression Artifacts Reduction by a Deep
  Convolutional Network}. In \bibinfo{booktitle}{\emph{2015 IEEE International
  Conference on Computer Vision (ICCV)}}. \bibinfo{pages}{576--584}.
\newblock
\urldef\tempurl%
\url{https://doi.org/10.1109/ICCV.2015.73}
\showDOI{\tempurl}


\bibitem[\protect\citeauthoryear{Dong, Loy, He, and Tang}{Dong
  et~al\mbox{.}}{2016}]%
        {srcnn}
\bibfield{author}{\bibinfo{person}{Chao Dong}, \bibinfo{person}{Chen~Change
  Loy}, \bibinfo{person}{Kaiming He}, {and} \bibinfo{person}{Xiaoou Tang}.}
  \bibinfo{year}{2016}\natexlab{}.
\newblock \showarticletitle{Image Super-Resolution Using Deep Convolutional
  Networks}.
\newblock \bibinfo{journal}{\emph{IEEE Transactions on Pattern Analysis and
  Machine Intelligence}} \bibinfo{volume}{38}, \bibinfo{number}{2}
  (\bibinfo{year}{2016}), \bibinfo{pages}{295--307}.
\newblock
\urldef\tempurl%
\url{https://doi.org/10.1109/TPAMI.2015.2439281}
\showDOI{\tempurl}


\bibitem[\protect\citeauthoryear{{Dudhane}, {Hambarde}, {Patil}, and
  {Murala}}{{Dudhane} et~al\mbox{.}}{2020}]%
        {dudhane}
\bibfield{author}{\bibinfo{person}{A. {Dudhane}}, \bibinfo{person}{P.
  {Hambarde}}, \bibinfo{person}{P. {Patil}}, {and} \bibinfo{person}{S.
  {Murala}}.} \bibinfo{year}{2020}\natexlab{}.
\newblock \showarticletitle{Deep Underwater Image Restoration and Beyond}.
\newblock \bibinfo{journal}{\emph{IEEE Signal Processing Letters}}
  \bibinfo{volume}{27} (\bibinfo{year}{2020}), \bibinfo{pages}{675--679}.
\newblock
\urldef\tempurl%
\url{https://doi.org/10.1109/LSP.2020.2988590}
\showDOI{\tempurl}


\bibitem[\protect\citeauthoryear{Fabbri, Islam, and Sattar}{Fabbri
  et~al\mbox{.}}{2018}]%
        {ugan}
\bibfield{author}{\bibinfo{person}{Cameron Fabbri}, \bibinfo{person}{Md~Jahidul
  Islam}, {and} \bibinfo{person}{Junaed Sattar}.}
  \bibinfo{year}{2018}\natexlab{}.
\newblock \showarticletitle{Enhancing Underwater Imagery Using Generative
  Adversarial Networks}. In \bibinfo{booktitle}{\emph{2018 IEEE International
  Conference on Robotics and Automation (ICRA)}}.
\newblock
\urldef\tempurl%
\url{https://doi.org/10.1109/ICRA.2018.8460552}
\showDOI{\tempurl}


\bibitem[\protect\citeauthoryear{Fu, Zhuang, Huang, Liao, Zhang, and Ding}{Fu
  et~al\mbox{.}}{2014}]%
        {retinex-based}
\bibfield{author}{\bibinfo{person}{Xueyang Fu}, \bibinfo{person}{Peixian
  Zhuang}, \bibinfo{person}{Yue Huang}, \bibinfo{person}{Yinghao Liao},
  \bibinfo{person}{Xiao-Ping Zhang}, {and} \bibinfo{person}{Xinghao Ding}.}
  \bibinfo{year}{2014}\natexlab{}.
\newblock \showarticletitle{A retinex-based enhancing approach for single
  underwater image}. In \bibinfo{booktitle}{\emph{2014 IEEE International
  Conference on Image Processing (ICIP)}}.
\newblock
\urldef\tempurl%
\url{https://doi.org/10.1109/ICIP.2014.7025927}
\showDOI{\tempurl}


\bibitem[\protect\citeauthoryear{Goodfellow, Pouget-Abadie, Mirza, Xu,
  Warde-Farley, Ozair, Courville, and Bengio}{Goodfellow
  et~al\mbox{.}}{[n.d.]}]%
        {gan}
\bibfield{author}{\bibinfo{person}{Ian Goodfellow}, \bibinfo{person}{Jean
  Pouget-Abadie}, \bibinfo{person}{Mehdi Mirza}, \bibinfo{person}{Bing Xu},
  \bibinfo{person}{David Warde-Farley}, \bibinfo{person}{Sherjil Ozair},
  \bibinfo{person}{Aaron Courville}, {and} \bibinfo{person}{Yoshua Bengio}.}
  \bibinfo{year}{[n.d.]}\natexlab{}.
\newblock \showarticletitle{Generative Adversarial Nets}. In
  \bibinfo{booktitle}{\emph{Advances in Neural Information Processing
  Systems}}, \bibfield{editor}{\bibinfo{person}{Z.~Ghahramani},
  \bibinfo{person}{M.~Welling}, \bibinfo{person}{C.~Cortes},
  \bibinfo{person}{N.~Lawrence}, {and} \bibinfo{person}{K.~Q. Weinberger}}
  (Eds.). \bibinfo{publisher}{Curran Associates, Inc.}
\newblock


\bibitem[\protect\citeauthoryear{{Guo}, {Li}, and {Zhuang}}{{Guo}
  et~al\mbox{.}}{2020}]%
        {guo_oe}
\bibfield{author}{\bibinfo{person}{Y. {Guo}}, \bibinfo{person}{H. {Li}}, {and}
  \bibinfo{person}{P. {Zhuang}}.} \bibinfo{year}{2020}\natexlab{}.
\newblock \showarticletitle{Underwater Image Enhancement Using a Multiscale
  Dense Generative Adversarial Network}.
\newblock \bibinfo{journal}{\emph{IEEE Journal of Oceanic Engineering}}
  \bibinfo{volume}{45}, \bibinfo{number}{3} (\bibinfo{year}{2020}),
  \bibinfo{pages}{862--870}.
\newblock
\urldef\tempurl%
\url{https://doi.org/10.1109/JOE.2019.2911447}
\showDOI{\tempurl}


\bibitem[\protect\citeauthoryear{{Han}, {Guan}, {Yu}, {Liu}, and {Zheng}}{{Han}
  et~al\mbox{.}}{2020}]%
        {han_access_2020}
\bibfield{author}{\bibinfo{person}{R. {Han}}, \bibinfo{person}{Y. {Guan}},
  \bibinfo{person}{Z. {Yu}}, \bibinfo{person}{P. {Liu}}, {and}
  \bibinfo{person}{H. {Zheng}}.} \bibinfo{year}{2020}\natexlab{}.
\newblock \showarticletitle{Underwater Image Enhancement Based on a Spiral
  Generative Adversarial Framework}.
\newblock \bibinfo{journal}{\emph{IEEE Access}}  \bibinfo{volume}{8}
  (\bibinfo{year}{2020}), \bibinfo{pages}{218838--218852}.
\newblock
\urldef\tempurl%
\url{https://doi.org/10.1109/ACCESS.2020.3041280}
\showDOI{\tempurl}


\bibitem[\protect\citeauthoryear{{He}, {Sun}, and {Tang}}{{He}
  et~al\mbox{.}}{2011}]%
        {dcp}
\bibfield{author}{\bibinfo{person}{K. {He}}, \bibinfo{person}{J. {Sun}}, {and}
  \bibinfo{person}{X. {Tang}}.} \bibinfo{year}{2011}\natexlab{}.
\newblock \showarticletitle{Single Image Haze Removal Using Dark Channel
  Prior}.
\newblock \bibinfo{journal}{\emph{IEEE Transactions on Pattern Analysis and
  Machine Intelligence}} \bibinfo{volume}{33}, \bibinfo{number}{12}
  (\bibinfo{year}{2011}), \bibinfo{pages}{2341--2353}.
\newblock
\urldef\tempurl%
\url{https://doi.org/10.1109/TPAMI.2010.168}
\showDOI{\tempurl}


\bibitem[\protect\citeauthoryear{{Hou}, {Zhao}, {Pan}, {Yang}, {Tan}, and
  {Li}}{{Hou} et~al\mbox{.}}{2020}]%
        {huo_access_20}
\bibfield{author}{\bibinfo{person}{G. {Hou}}, \bibinfo{person}{X. {Zhao}},
  \bibinfo{person}{Z. {Pan}}, \bibinfo{person}{H. {Yang}}, \bibinfo{person}{L.
  {Tan}}, {and} \bibinfo{person}{J. {Li}}.} \bibinfo{year}{2020}\natexlab{}.
\newblock \showarticletitle{Benchmarking Underwater Image Enhancement and
  Restoration, and Beyond}.
\newblock \bibinfo{journal}{\emph{IEEE Access}}  \bibinfo{volume}{8}
  (\bibinfo{year}{2020}), \bibinfo{pages}{122078--122091}.
\newblock
\urldef\tempurl%
\url{https://doi.org/10.1109/ACCESS.2020.3006359}
\showDOI{\tempurl}


\bibitem[\protect\citeauthoryear{Ioffe and Szegedy}{Ioffe and Szegedy}{2015}]%
        {bn}
\bibfield{author}{\bibinfo{person}{Sergey Ioffe} {and}
  \bibinfo{person}{Christian Szegedy}.} \bibinfo{year}{2015}\natexlab{}.
\newblock \showarticletitle{Batch Normalization: Accelerating Deep Network
  Training by Reducing Internal Covariate Shift}. In
  \bibinfo{booktitle}{\emph{Proceedings of the 32nd International Conference on
  International Conference on Machine Learning - Volume 37}} (Lille, France)
  \emph{(\bibinfo{series}{ICML'15})}. \bibinfo{publisher}{JMLR.org},
  \bibinfo{numpages}{9}~pages.
\newblock


\bibitem[\protect\citeauthoryear{Islam, Edge, Xiao, Luo, Mehtaz, Morse, Enan,
  and Sattar}{Islam et~al\mbox{.}}{2020a}]%
        {suim}
\bibfield{author}{\bibinfo{person}{Md~Jahidul Islam}, \bibinfo{person}{Chelsey
  Edge}, \bibinfo{person}{Yuyang Xiao}, \bibinfo{person}{Peigen Luo},
  \bibinfo{person}{Muntaqim Mehtaz}, \bibinfo{person}{Christopher Morse},
  \bibinfo{person}{Sadman~Sakib Enan}, {and} \bibinfo{person}{Junaed Sattar}.}
  \bibinfo{year}{2020}\natexlab{a}.
\newblock \showarticletitle{{Semantic Segmentation of Underwater Imagery:
  Dataset and Benchmark}}. In \bibinfo{booktitle}{\emph{IEEE/RSJ International
  Conference on Intelligent Robots and Systems (IROS)}}. IEEE/RSJ.
\newblock


\bibitem[\protect\citeauthoryear{Islam, Luo, and Sattar}{Islam
  et~al\mbox{.}}{2020b}]%
        {sesr}
\bibfield{author}{\bibinfo{person}{Md~Jahidul Islam}, \bibinfo{person}{Peigen
  Luo}, {and} \bibinfo{person}{Junaed Sattar}.}
  \bibinfo{year}{2020}\natexlab{b}.
\newblock \showarticletitle{{Simultaneous Enhancement and Super-Resolution of
  Underwater Imagery for Improved Visual Perception}}. In
  \bibinfo{booktitle}{\emph{Robotics: Science and Systems (RSS)}}.
  \bibinfo{address}{Corvalis, Oregon, USA}.
\newblock
\urldef\tempurl%
\url{https://doi.org/{10.15607/RSS.2020.XVI.018}}
\showDOI{\tempurl}


\bibitem[\protect\citeauthoryear{Islam, Sakib~Enan, Luo, and Sattar}{Islam
  et~al\mbox{.}}{2020c}]%
        {srdrm_srdrmgan}
\bibfield{author}{\bibinfo{person}{Md~Jahidul Islam}, \bibinfo{person}{Sadman
  Sakib~Enan}, \bibinfo{person}{Peigen Luo}, {and} \bibinfo{person}{Junaed
  Sattar}.} \bibinfo{year}{2020}\natexlab{c}.
\newblock \showarticletitle{Underwater Image Super-Resolution using Deep
  Residual Multipliers}. In \bibinfo{booktitle}{\emph{2020 IEEE International
  Conference on Robotics and Automation (ICRA)}}.
\newblock
\urldef\tempurl%
\url{https://doi.org/10.1109/ICRA40945.2020.9197213}
\showDOI{\tempurl}


\bibitem[\protect\citeauthoryear{Islam, Xia, and Sattar}{Islam
  et~al\mbox{.}}{2020d}]%
        {euvp}
\bibfield{author}{\bibinfo{person}{Md~Jahidul Islam}, \bibinfo{person}{Youya
  Xia}, {and} \bibinfo{person}{Junaed Sattar}.}
  \bibinfo{year}{2020}\natexlab{d}.
\newblock \showarticletitle{Fast Underwater Image Enhancement for Improved
  Visual Perception}.
\newblock \bibinfo{journal}{\emph{IEEE Robotics and Automation Letters}}
  \bibinfo{volume}{5}, \bibinfo{number}{2} (\bibinfo{year}{2020}),
  \bibinfo{pages}{3227--3234}.
\newblock
\urldef\tempurl%
\url{https://doi.org/10.1109/LRA.2020.2974710}
\showDOI{\tempurl}


\bibitem[\protect\citeauthoryear{Islam, Xia, and Sattar}{Islam
  et~al\mbox{.}}{2020e}]%
        {funiegan}
\bibfield{author}{\bibinfo{person}{Md~Jahidul Islam}, \bibinfo{person}{Youya
  Xia}, {and} \bibinfo{person}{Junaed Sattar}.}
  \bibinfo{year}{2020}\natexlab{e}.
\newblock \showarticletitle{Fast Underwater Image Enhancement for Improved
  Visual Perception}.
\newblock \bibinfo{journal}{\emph{IEEE Robotics and Automation Letters}}
  \bibinfo{volume}{5}, \bibinfo{number}{2} (\bibinfo{year}{2020}).
\newblock
\urldef\tempurl%
\url{https://doi.org/10.1109/LRA.2020.2974710}
\showDOI{\tempurl}


\bibitem[\protect\citeauthoryear{Jin, Iqbal, Bobkov, Zou, Li, and
  Steinbach}{Jin et~al\mbox{.}}{2020}]%
        {sisr6}
\bibfield{author}{\bibinfo{person}{Zhi Jin}, \bibinfo{person}{Muhammad~Zafar
  Iqbal}, \bibinfo{person}{Dmytro Bobkov}, \bibinfo{person}{Wenbin Zou},
  \bibinfo{person}{Xia Li}, {and} \bibinfo{person}{Eckehard Steinbach}.}
  \bibinfo{year}{2020}\natexlab{}.
\newblock \showarticletitle{A Flexible Deep CNN Framework for Image
  Restoration}.
\newblock \bibinfo{journal}{\emph{IEEE Transactions on Multimedia}}
  \bibinfo{volume}{22}, \bibinfo{number}{4} (\bibinfo{year}{2020}),
  \bibinfo{pages}{1055--1068}.
\newblock
\urldef\tempurl%
\url{https://doi.org/10.1109/TMM.2019.2938340}
\showDOI{\tempurl}


\bibitem[\protect\citeauthoryear{Johnson, Alahi, and Fei-Fei}{Johnson
  et~al\mbox{.}}{2016}]%
        {perceptual}
\bibfield{author}{\bibinfo{person}{Justin Johnson}, \bibinfo{person}{Alexandre
  Alahi}, {and} \bibinfo{person}{Li Fei-Fei}.} \bibinfo{year}{2016}\natexlab{}.
\newblock \showarticletitle{Perceptual losses for real-time style transfer and
  super-resolution}. In \bibinfo{booktitle}{\emph{European Conference on
  Computer Vision}}.
\newblock


\bibitem[\protect\citeauthoryear{Kim, Lee, and Lee}{Kim et~al\mbox{.}}{2016}]%
        {VDSR}
\bibfield{author}{\bibinfo{person}{Jiwon Kim}, \bibinfo{person}{Jung~Kwon Lee},
  {and} \bibinfo{person}{Kyoung~Mu Lee}.} \bibinfo{year}{2016}\natexlab{}.
\newblock \showarticletitle{Accurate Image Super-Resolution Using Very Deep
  Convolutional Networks}. In \bibinfo{booktitle}{\emph{The IEEE Conference on
  Computer Vision and Pattern Recognition (CVPR Oral)}}.
\newblock


\bibitem[\protect\citeauthoryear{Kingma and Ba}{Kingma and Ba}{2015}]%
        {adam}
\bibfield{author}{\bibinfo{person}{Diederik~P. Kingma} {and}
  \bibinfo{person}{Jimmy Ba}.} \bibinfo{year}{2015}\natexlab{}.
\newblock \showarticletitle{Adam: {A} Method for Stochastic Optimization}. In
  \bibinfo{booktitle}{\emph{3rd International Conference on Learning
  Representations, {ICLR} 2015, San Diego, CA, USA, May 7-9, 2015, Conference
  Track Proceedings}}, \bibfield{editor}{\bibinfo{person}{Yoshua Bengio} {and}
  \bibinfo{person}{Yann LeCun}} (Eds.).
\newblock


\bibitem[\protect\citeauthoryear{Krizhevsky, Sutskever, and Hinton}{Krizhevsky
  et~al\mbox{.}}{2012}]%
        {ilsvrc}
\bibfield{author}{\bibinfo{person}{Alex Krizhevsky}, \bibinfo{person}{Ilya
  Sutskever}, {and} \bibinfo{person}{Geoffrey~E. Hinton}.}
  \bibinfo{year}{2012}\natexlab{}.
\newblock \showarticletitle{ImageNet Classification with Deep Convolutional
  Neural Networks}. In \bibinfo{booktitle}{\emph{Proceedings of the 25th
  International Conference on Neural Information Processing Systems - Volume
  1}} \emph{(\bibinfo{series}{NIPS'12})}. \bibinfo{pages}{1097–1105}.
\newblock


\bibitem[\protect\citeauthoryear{Ledig, Theis, Huszar, Caballero, Cunningham,
  Acosta, Aitken, Tejani, Totz, Wang, and Shi}{Ledig et~al\mbox{.}}{2017}]%
        {srresnet_srgan}
\bibfield{author}{\bibinfo{person}{Christian Ledig}, \bibinfo{person}{Lucas
  Theis}, \bibinfo{person}{Ferenc Huszar}, \bibinfo{person}{Jose Caballero},
  \bibinfo{person}{Andrew Cunningham}, \bibinfo{person}{Alejandro Acosta},
  \bibinfo{person}{Andrew Aitken}, \bibinfo{person}{Alykhan Tejani},
  \bibinfo{person}{Johannes Totz}, \bibinfo{person}{Zehan Wang}, {and}
  \bibinfo{person}{Wenzhe Shi}.} \bibinfo{year}{2017}\natexlab{}.
\newblock \showarticletitle{Photo-Realistic Single Image Super-Resolution Using
  a Generative Adversarial Network}. In \bibinfo{booktitle}{\emph{Proceedings
  of the IEEE Conference on Computer Vision and Pattern Recognition (CVPR)}}.
\newblock


\bibitem[\protect\citeauthoryear{Li, Anwar, Hou, Cong, Guo, and Ren}{Li
  et~al\mbox{.}}{2021}]%
        {tip21}
\bibfield{author}{\bibinfo{person}{Chongyi Li}, \bibinfo{person}{Saeed Anwar},
  \bibinfo{person}{Junhui Hou}, \bibinfo{person}{Runmin Cong},
  \bibinfo{person}{Chunle Guo}, {and} \bibinfo{person}{Wenqi Ren}.}
  \bibinfo{year}{2021}\natexlab{}.
\newblock \showarticletitle{Underwater Image Enhancement via Medium
  Transmission-Guided Multi-Color Space Embedding}.
\newblock \bibinfo{journal}{\emph{IEEE Transactions on Image Processing}}
  \bibinfo{volume}{30} (\bibinfo{year}{2021}).
\newblock
\urldef\tempurl%
\url{https://doi.org/10.1109/TIP.2021.3076367}
\showDOI{\tempurl}


\bibitem[\protect\citeauthoryear{Li, Guo, Ren, Cong, Hou, Kwong, and Tao}{Li
  et~al\mbox{.}}{2020}]%
        {uieb}
\bibfield{author}{\bibinfo{person}{Chongyi Li}, \bibinfo{person}{Chunle Guo},
  \bibinfo{person}{Wenqi Ren}, \bibinfo{person}{Runmin Cong},
  \bibinfo{person}{Junhui Hou}, \bibinfo{person}{Sam Kwong}, {and}
  \bibinfo{person}{Dacheng Tao}.} \bibinfo{year}{2020}\natexlab{}.
\newblock \showarticletitle{An Underwater Image Enhancement Benchmark Dataset
  and Beyond}.
\newblock \bibinfo{journal}{\emph{IEEE Transactions on Image Processing}}
  \bibinfo{volume}{29} (\bibinfo{year}{2020}), \bibinfo{pages}{4376--4389}.
\newblock
\urldef\tempurl%
\url{https://doi.org/10.1109/TIP.2019.2955241}
\showDOI{\tempurl}


\bibitem[\protect\citeauthoryear{{Li}, {Guo}, {Cong}, {Pang}, and {Wang}}{{Li}
  et~al\mbox{.}}{2016a}]%
        {li_tip}
\bibfield{author}{\bibinfo{person}{C. {Li}}, \bibinfo{person}{J. {Guo}},
  \bibinfo{person}{R. {Cong}}, \bibinfo{person}{Y. {Pang}}, {and}
  \bibinfo{person}{B. {Wang}}.} \bibinfo{year}{2016}\natexlab{a}.
\newblock \showarticletitle{Underwater Image Enhancement by Dehazing With
  Minimum Information Loss and Histogram Distribution Prior}.
\newblock \bibinfo{journal}{\emph{IEEE Transactions on Image Processing}}
  \bibinfo{volume}{25}, \bibinfo{number}{12} (\bibinfo{year}{2016}),
  \bibinfo{pages}{5664--5677}.
\newblock
\urldef\tempurl%
\url{https://doi.org/10.1109/TIP.2016.2612882}
\showDOI{\tempurl}


\bibitem[\protect\citeauthoryear{{Li}, {Guo}, and {Guo}}{{Li}
  et~al\mbox{.}}{2018a}]%
        {li_spl}
\bibfield{author}{\bibinfo{person}{C. {Li}}, \bibinfo{person}{J. {Guo}}, {and}
  \bibinfo{person}{C. {Guo}}.} \bibinfo{year}{2018}\natexlab{a}.
\newblock \showarticletitle{Emerging From Water: Underwater Image Color
  Correction Based on Weakly Supervised Color Transfer}.
\newblock \bibinfo{journal}{\emph{IEEE Signal Processing Letters}}
  \bibinfo{volume}{25}, \bibinfo{number}{3} (\bibinfo{year}{2018}),
  \bibinfo{pages}{323--327}.
\newblock
\urldef\tempurl%
\url{https://doi.org/10.1109/LSP.2018.2792050}
\showDOI{\tempurl}


\bibitem[\protect\citeauthoryear{{Li}, {Quo}, {Pang}, {Chen}, and {Wang}}{{Li}
  et~al\mbox{.}}{2016b}]%
        {li}
\bibfield{author}{\bibinfo{person}{C. {Li}}, \bibinfo{person}{J. {Quo}},
  \bibinfo{person}{Y. {Pang}}, \bibinfo{person}{S. {Chen}}, {and}
  \bibinfo{person}{J. {Wang}}.} \bibinfo{year}{2016}\natexlab{b}.
\newblock \showarticletitle{Single underwater image restoration by blue-green
  channels dehazing and red channel correction}. In
  \bibinfo{booktitle}{\emph{2016 IEEE International Conference on Acoustics,
  Speech and Signal Processing (ICASSP)}}. \bibinfo{pages}{1731--1735}.
\newblock
\urldef\tempurl%
\url{https://doi.org/10.1109/ICASSP.2016.7471973}
\showDOI{\tempurl}


\bibitem[\protect\citeauthoryear{{Li}, {Liu}, {Yang}, {Sun}, and {Guo}}{{Li}
  et~al\mbox{.}}{2018b}]%
        {li_tip_retinex}
\bibfield{author}{\bibinfo{person}{M. {Li}}, \bibinfo{person}{J. {Liu}},
  \bibinfo{person}{W. {Yang}}, \bibinfo{person}{X. {Sun}}, {and}
  \bibinfo{person}{Z. {Guo}}.} \bibinfo{year}{2018}\natexlab{b}.
\newblock \showarticletitle{Structure-Revealing Low-Light Image Enhancement Via
  Robust Retinex Model}.
\newblock \bibinfo{journal}{\emph{IEEE Transactions on Image Processing}}
  \bibinfo{volume}{27}, \bibinfo{number}{6} (\bibinfo{year}{2018}),
  \bibinfo{pages}{2828--2841}.
\newblock
\urldef\tempurl%
\url{https://doi.org/10.1109/TIP.2018.2810539}
\showDOI{\tempurl}


\bibitem[\protect\citeauthoryear{{Li}, {Ma}, {Zhang}, {Li}, {Ge}, {Li}, and
  {Serikawa}}{{Li} et~al\mbox{.}}{2019}]%
        {mlp_access}
\bibfield{author}{\bibinfo{person}{Y. {Li}}, \bibinfo{person}{C. {Ma}},
  \bibinfo{person}{T. {Zhang}}, \bibinfo{person}{J. {Li}}, \bibinfo{person}{Z.
  {Ge}}, \bibinfo{person}{Y. {Li}}, {and} \bibinfo{person}{S. {Serikawa}}.}
  \bibinfo{year}{2019}\natexlab{}.
\newblock \showarticletitle{Underwater Image High Definition Display Using the
  Multilayer Perceptron and Color Feature-Based SRCNN}.
\newblock \bibinfo{journal}{\emph{IEEE Access}}  \bibinfo{volume}{7}
  (\bibinfo{year}{2019}), \bibinfo{pages}{83721--83728}.
\newblock
\urldef\tempurl%
\url{https://doi.org/10.1109/ACCESS.2019.2925209}
\showDOI{\tempurl}


\bibitem[\protect\citeauthoryear{Liu, Cheng, Wang, Zhang, and Huang}{Liu
  et~al\mbox{.}}{2019}]%
        {low-level-1}
\bibfield{author}{\bibinfo{person}{Ding Liu}, \bibinfo{person}{Bowen Cheng},
  \bibinfo{person}{Zhangyang Wang}, \bibinfo{person}{Haichao Zhang}, {and}
  \bibinfo{person}{Thomas~S. Huang}.} \bibinfo{year}{2019}\natexlab{}.
\newblock \showarticletitle{Enhance Visual Recognition Under Adverse Conditions
  via Deep Networks}.
\newblock \bibinfo{journal}{\emph{IEEE Transactions on Image Processing}}
  \bibinfo{volume}{28}, \bibinfo{number}{9} (\bibinfo{year}{2019}),
  \bibinfo{pages}{4401--4412}.
\newblock
\urldef\tempurl%
\url{https://doi.org/10.1109/TIP.2019.2908802}
\showDOI{\tempurl}


\bibitem[\protect\citeauthoryear{Liu, Wen, Liu, Wang, and Huang}{Liu
  et~al\mbox{.}}{2018a}]%
        {ijcai18}
\bibfield{author}{\bibinfo{person}{Ding Liu}, \bibinfo{person}{Bihan Wen},
  \bibinfo{person}{Xianming Liu}, \bibinfo{person}{Zhangyang Wang}, {and}
  \bibinfo{person}{Thomas~S. Huang}.} \bibinfo{year}{2018}\natexlab{a}.
\newblock \showarticletitle{When Image Denoising Meets High-Level Vision Tasks:
  A Deep Learning Approach}. In \bibinfo{booktitle}{\emph{Proceedings of the
  27th International Joint Conference on Artificial Intelligence}} (Stockholm,
  Sweden) \emph{(\bibinfo{series}{IJCAI'18})}. \bibinfo{publisher}{AAAI Press},
  \bibinfo{pages}{842–848}.
\newblock
\showISBNx{9780999241127}


\bibitem[\protect\citeauthoryear{{Liu}, {Wang}, {Qi}, {Zhang}, {Zheng}, and
  {Yu}}{{Liu} et~al\mbox{.}}{2019}]%
        {residual_access}
\bibfield{author}{\bibinfo{person}{P. {Liu}}, \bibinfo{person}{G. {Wang}},
  \bibinfo{person}{H. {Qi}}, \bibinfo{person}{C. {Zhang}}, \bibinfo{person}{H.
  {Zheng}}, {and} \bibinfo{person}{Z. {Yu}}.} \bibinfo{year}{2019}\natexlab{}.
\newblock \showarticletitle{Underwater Image Enhancement With a Deep Residual
  Framework}.
\newblock \bibinfo{journal}{\emph{IEEE Access}}  \bibinfo{volume}{7}
  (\bibinfo{year}{2019}), \bibinfo{pages}{94614--94629}.
\newblock
\urldef\tempurl%
\url{https://doi.org/10.1109/ACCESS.2019.2928976}
\showDOI{\tempurl}


\bibitem[\protect\citeauthoryear{Liu, Zhang, Zhang, Lin, and Zuo}{Liu
  et~al\mbox{.}}{2018b}]%
        {large_kernel_issue_intro}
\bibfield{author}{\bibinfo{person}{Pengju Liu}, \bibinfo{person}{Hongzhi
  Zhang}, \bibinfo{person}{Kai Zhang}, \bibinfo{person}{Liang Lin}, {and}
  \bibinfo{person}{Wangmeng Zuo}.} \bibinfo{year}{2018}\natexlab{b}.
\newblock \showarticletitle{Multi-level Wavelet-CNN for Image Restoration}. In
  \bibinfo{booktitle}{\emph{2018 IEEE/CVF Conference on Computer Vision and
  Pattern Recognition Workshops (CVPRW)}}.
\newblock
\urldef\tempurl%
\url{https://doi.org/10.1109/CVPRW.2018.00121}
\showDOI{\tempurl}


\bibitem[\protect\citeauthoryear{{Liu}, {Gao}, and {Chen}}{{Liu}
  et~al\mbox{.}}{2020}]%
        {liu_gan}
\bibfield{author}{\bibinfo{person}{X. {Liu}}, \bibinfo{person}{Z. {Gao}}, {and}
  \bibinfo{person}{B.~M. {Chen}}.} \bibinfo{year}{2020}\natexlab{}.
\newblock \showarticletitle{MLFcGAN: Multilevel Feature Fusion-Based
  Conditional GAN for Underwater Image Color Correction}.
\newblock \bibinfo{journal}{\emph{IEEE Geoscience and Remote Sensing Letters}}
  \bibinfo{volume}{17}, \bibinfo{number}{9} (\bibinfo{year}{2020}).
\newblock
\urldef\tempurl%
\url{https://doi.org/10.1109/LGRS.2019.2950056}
\showDOI{\tempurl}


\bibitem[\protect\citeauthoryear{Long, Shelhamer, and Darrell}{Long
  et~al\mbox{.}}{2015}]%
        {back_conv}
\bibfield{author}{\bibinfo{person}{Jonathan Long}, \bibinfo{person}{Evan
  Shelhamer}, {and} \bibinfo{person}{Trevor Darrell}.}
  \bibinfo{year}{2015}\natexlab{}.
\newblock \showarticletitle{Fully convolutional networks for semantic
  segmentation}. In \bibinfo{booktitle}{\emph{2015 IEEE Conference on Computer
  Vision and Pattern Recognition (CVPR)}}. \bibinfo{pages}{3431--3440}.
\newblock
\urldef\tempurl%
\url{https://doi.org/10.1109/CVPR.2015.7298965}
\showDOI{\tempurl}


\bibitem[\protect\citeauthoryear{Lu, Li, Nakashima, Kim, and Serikawa}{Lu
  et~al\mbox{.}}{2017}]%
        {uwsisr}
\bibfield{author}{\bibinfo{person}{Huimin Lu}, \bibinfo{person}{Yujie Li},
  \bibinfo{person}{Shota Nakashima}, \bibinfo{person}{Hyongseop Kim}, {and}
  \bibinfo{person}{Seiichi Serikawa}.} \bibinfo{year}{2017}\natexlab{}.
\newblock \showarticletitle{Underwater Image Super-Resolution by Descattering
  and Fusion}.
\newblock \bibinfo{journal}{\emph{IEEE Access}}  \bibinfo{volume}{5}
  (\bibinfo{year}{2017}), \bibinfo{pages}{670--679}.
\newblock
\urldef\tempurl%
\url{https://doi.org/10.1109/ACCESS.2017.2648845}
\showDOI{\tempurl}


\bibitem[\protect\citeauthoryear{Ma, Fan, Yang, Zhang, and Zhu}{Ma
  et~al\mbox{.}}{2018}]%
        {jma_model}
\bibfield{author}{\bibinfo{person}{Jinxiang Ma}, \bibinfo{person}{Xinnan Fan},
  \bibinfo{person}{Simon~X. Yang}, \bibinfo{person}{Xuewu Zhang}, {and}
  \bibinfo{person}{Xifang Zhu}.} \bibinfo{year}{2018}\natexlab{}.
\newblock \showarticletitle{Contrast Limited Adaptive Histogram
  Equalization-Based Fusion in YIQ and HSI Color Spaces for Underwater Image
  Enhancement}.
\newblock \bibinfo{journal}{\emph{International Journal of Pattern Recognition
  and Artificial Intelligence}} \bibinfo{volume}{32}, \bibinfo{number}{07}
  (\bibinfo{year}{2018}), \bibinfo{pages}{1854018}.
\newblock
\urldef\tempurl%
\url{https://doi.org/10.1142/S0218001418540186}
\showDOI{\tempurl}
\showeprint{https://doi.org/10.1142/S0218001418540186}


\bibitem[\protect\citeauthoryear{{Mittal}, {Soundararajan}, and
  {Bovik}}{{Mittal} et~al\mbox{.}}{2013}]%
        {niqe}
\bibfield{author}{\bibinfo{person}{A. {Mittal}}, \bibinfo{person}{R.
  {Soundararajan}}, {and} \bibinfo{person}{A.~C. {Bovik}}.}
  \bibinfo{year}{2013}\natexlab{}.
\newblock \showarticletitle{Making a “Completely Blind” Image Quality
  Analyzer}.
\newblock \bibinfo{journal}{\emph{IEEE Signal Processing Letters}}
  \bibinfo{volume}{20}, \bibinfo{number}{3} (\bibinfo{year}{2013}),
  \bibinfo{pages}{209--212}.
\newblock
\urldef\tempurl%
\url{https://doi.org/10.1109/LSP.2012.2227726}
\showDOI{\tempurl}


\bibitem[\protect\citeauthoryear{{Panetta}, {Gao}, and {Agaian}}{{Panetta}
  et~al\mbox{.}}{2016}]%
        {uiqm}
\bibfield{author}{\bibinfo{person}{K. {Panetta}}, \bibinfo{person}{C. {Gao}},
  {and} \bibinfo{person}{S. {Agaian}}.} \bibinfo{year}{2016}\natexlab{}.
\newblock \showarticletitle{Human-Visual-System-Inspired Underwater Image
  Quality Measures}.
\newblock \bibinfo{journal}{\emph{IEEE Journal of Oceanic Engineering}}
  \bibinfo{volume}{41}, \bibinfo{number}{3} (\bibinfo{year}{2016}),
  \bibinfo{pages}{541--551}.
\newblock
\urldef\tempurl%
\url{https://doi.org/10.1109/JOE.2015.2469915}
\showDOI{\tempurl}


\bibitem[\protect\citeauthoryear{Paszke, Gross, Massa, Lerer, Bradbury, Chanan,
  Killeen, Lin, Gimelshein, Antiga, Desmaison, Kopf, Yang, DeVito, Raison,
  Tejani, Chilamkurthy, Steiner, Fang, Bai, and Chintala}{Paszke
  et~al\mbox{.}}{2019}]%
        {pytorch}
\bibfield{author}{\bibinfo{person}{Adam Paszke}, \bibinfo{person}{Sam Gross},
  \bibinfo{person}{Francisco Massa}, \bibinfo{person}{Adam Lerer},
  \bibinfo{person}{James Bradbury}, \bibinfo{person}{Gregory Chanan},
  \bibinfo{person}{Trevor Killeen}, \bibinfo{person}{Zeming Lin},
  \bibinfo{person}{Natalia Gimelshein}, \bibinfo{person}{Luca Antiga},
  \bibinfo{person}{Alban Desmaison}, \bibinfo{person}{Andreas Kopf},
  \bibinfo{person}{Edward Yang}, \bibinfo{person}{Zachary DeVito},
  \bibinfo{person}{Martin Raison}, \bibinfo{person}{Alykhan Tejani},
  \bibinfo{person}{Sasank Chilamkurthy}, \bibinfo{person}{Benoit Steiner},
  \bibinfo{person}{Lu Fang}, \bibinfo{person}{Junjie Bai}, {and}
  \bibinfo{person}{Soumith Chintala}.} \bibinfo{year}{2019}\natexlab{}.
\newblock \showarticletitle{PyTorch: An Imperative Style, High-Performance Deep
  Learning Library}.
\newblock In \bibinfo{booktitle}{\emph{Advances in Neural Information
  Processing Systems 32}}. \bibinfo{publisher}{Curran Associates, Inc.},
  \bibinfo{pages}{8024--8035}.
\newblock


\bibitem[\protect\citeauthoryear{Peng, Zhang, Yu, Luo, and Sun}{Peng
  et~al\mbox{.}}{2017}]%
        {large_kernel}
\bibfield{author}{\bibinfo{person}{Chao Peng}, \bibinfo{person}{Xiangyu Zhang},
  \bibinfo{person}{Gang Yu}, \bibinfo{person}{Guiming Luo}, {and}
  \bibinfo{person}{Jian Sun}.} \bibinfo{year}{2017}\natexlab{}.
\newblock \showarticletitle{Large Kernel Matters — Improve Semantic
  Segmentation by Global Convolutional Network}. In
  \bibinfo{booktitle}{\emph{2017 IEEE Conference on Computer Vision and Pattern
  Recognition (CVPR)}}. \bibinfo{pages}{1743--1751}.
\newblock
\urldef\tempurl%
\url{https://doi.org/10.1109/CVPR.2017.189}
\showDOI{\tempurl}


\bibitem[\protect\citeauthoryear{{Peng} and {Cosman}}{{Peng} and
  {Cosman}}{2017}]%
        {peng_tip}
\bibfield{author}{\bibinfo{person}{Y. {Peng}} {and} \bibinfo{person}{P.~C.
  {Cosman}}.} \bibinfo{year}{2017}\natexlab{}.
\newblock \showarticletitle{Underwater Image Restoration Based on Image
  Blurriness and Light Absorption}.
\newblock \bibinfo{journal}{\emph{IEEE Transactions on Image Processing}}
  \bibinfo{volume}{26}, \bibinfo{number}{4} (\bibinfo{year}{2017}),
  \bibinfo{pages}{1579--1594}.
\newblock
\urldef\tempurl%
\url{https://doi.org/10.1109/TIP.2017.2663846}
\showDOI{\tempurl}


\bibitem[\protect\citeauthoryear{{Peng}, {Zhao}, and {Cosman}}{{Peng}
  et~al\mbox{.}}{2015}]%
        {peng}
\bibfield{author}{\bibinfo{person}{Y. {Peng}}, \bibinfo{person}{X. {Zhao}},
  {and} \bibinfo{person}{P.~C. {Cosman}}.} \bibinfo{year}{2015}\natexlab{}.
\newblock \showarticletitle{Single underwater image enhancement using depth
  estimation based on blurriness}. In \bibinfo{booktitle}{\emph{2015 IEEE
  International Conference on Image Processing (ICIP)}}.
\newblock
\urldef\tempurl%
\url{https://doi.org/10.1109/ICIP.2015.7351749}
\showDOI{\tempurl}


\bibitem[\protect\citeauthoryear{Peng, Cao, and Cosman}{Peng
  et~al\mbox{.}}{2018}]%
        {gdcp}
\bibfield{author}{\bibinfo{person}{Yan-Tsung Peng}, \bibinfo{person}{Keming
  Cao}, {and} \bibinfo{person}{Pamela~C. Cosman}.}
  \bibinfo{year}{2018}\natexlab{}.
\newblock \showarticletitle{Generalization of the Dark Channel Prior for Single
  Image Restoration}.
\newblock \bibinfo{journal}{\emph{IEEE Transactions on Image Processing}}
  \bibinfo{volume}{27}, \bibinfo{number}{6} (\bibinfo{year}{2018}),
  \bibinfo{pages}{2856--2868}.
\newblock
\urldef\tempurl%
\url{https://doi.org/10.1109/TIP.2018.2813092}
\showDOI{\tempurl}


\bibitem[\protect\citeauthoryear{Sharma, Jain, and Sur}{Sharma
  et~al\mbox{.}}{2020}]%
        {ours_wacv}
\bibfield{author}{\bibinfo{person}{Prasen Sharma}, \bibinfo{person}{Priyankar
  Jain}, {and} \bibinfo{person}{Arijit Sur}.} \bibinfo{year}{2020}\natexlab{}.
\newblock \showarticletitle{Scale-aware Conditional Generative Adversarial
  Network for Image Dehazing}. In \bibinfo{booktitle}{\emph{Proceedings of the
  IEEE/CVF Winter Conference on Applications of Computer Vision (WACV)}}.
\newblock


\bibitem[\protect\citeauthoryear{Sharma, Jain, and Sur}{Sharma
  et~al\mbox{.}}{2019}]%
        {ours_icip}
\bibfield{author}{\bibinfo{person}{Prasen~Kumar Sharma},
  \bibinfo{person}{Priyankar Jain}, {and} \bibinfo{person}{Arijit Sur}.}
  \bibinfo{year}{2019}\natexlab{}.
\newblock \showarticletitle{Dual-Domain Single Image De-Raining Using
  Conditional Generative Adversarial Network}. In
  \bibinfo{booktitle}{\emph{2019 IEEE International Conference on Image
  Processing (ICIP)}}.
\newblock
\urldef\tempurl%
\url{https://doi.org/10.1109/ICIP.2019.8803353}
\showDOI{\tempurl}


\bibitem[\protect\citeauthoryear{{Sheikh} and {Bovik}}{{Sheikh} and
  {Bovik}}{2006}]%
        {vif}
\bibfield{author}{\bibinfo{person}{H.~R. {Sheikh}} {and} \bibinfo{person}{A.~C.
  {Bovik}}.} \bibinfo{year}{2006}\natexlab{}.
\newblock \showarticletitle{Image information and visual quality}.
\newblock \bibinfo{journal}{\emph{IEEE Transactions on Image Processing}}
  \bibinfo{volume}{15}, \bibinfo{number}{2} (\bibinfo{year}{2006}).
\newblock
\urldef\tempurl%
\url{https://doi.org/10.1109/TIP.2005.859378}
\showDOI{\tempurl}


\bibitem[\protect\citeauthoryear{Shi, Caballero, Huszar, Totz, Aitken, Bishop,
  Rueckert, and Wang}{Shi et~al\mbox{.}}{2016}]%
        {ps}
\bibfield{author}{\bibinfo{person}{Wenzhe Shi}, \bibinfo{person}{Jose
  Caballero}, \bibinfo{person}{Ferenc Huszar}, \bibinfo{person}{Johannes Totz},
  \bibinfo{person}{Andrew~P. Aitken}, \bibinfo{person}{Rob Bishop},
  \bibinfo{person}{Daniel Rueckert}, {and} \bibinfo{person}{Zehan Wang}.}
  \bibinfo{year}{2016}\natexlab{}.
\newblock \showarticletitle{Real-Time Single Image and Video Super-Resolution
  Using an Efficient Sub-Pixel Convolutional Neural Network}. In
  \bibinfo{booktitle}{\emph{Proceedings of the IEEE Conference on Computer
  Vision and Pattern Recognition (CVPR)}}.
\newblock


\bibitem[\protect\citeauthoryear{Simon, Joo, Matthews, and Sheikh}{Simon
  et~al\mbox{.}}{2017}]%
        {openpose2}
\bibfield{author}{\bibinfo{person}{Tomas Simon}, \bibinfo{person}{Hanbyul Joo},
  \bibinfo{person}{Iain Matthews}, {and} \bibinfo{person}{Yaser Sheikh}.}
  \bibinfo{year}{2017}\natexlab{}.
\newblock \showarticletitle{Hand Keypoint Detection in Single Images using
  Multiview Bootstrapping}. In \bibinfo{booktitle}{\emph{CVPR}}.
\newblock


\bibitem[\protect\citeauthoryear{Simonyan and Zisserman}{Simonyan and
  Zisserman}{[n.d.]}]%
        {vgg}
\bibfield{author}{\bibinfo{person}{Karen Simonyan} {and}
  \bibinfo{person}{Andrew Zisserman}.} \bibinfo{year}{[n.d.]}\natexlab{}.
\newblock \showarticletitle{Very Deep Convolutional Networks for Large-Scale
  Image Recognition}.
\newblock \bibinfo{journal}{\emph{CoRR}} (\bibinfo{year}{[n.\,d.]}).
\newblock


\bibitem[\protect\citeauthoryear{Szegedy, Zaremba, Sutskever, Bruna, Erhan,
  Goodfellow, and Fergus}{Szegedy et~al\mbox{.}}{2014}]%
        {perturb1}
\bibfield{author}{\bibinfo{person}{Christian Szegedy},
  \bibinfo{person}{Wojciech Zaremba}, \bibinfo{person}{Ilya Sutskever},
  \bibinfo{person}{Joan Bruna}, \bibinfo{person}{Dumitru Erhan},
  \bibinfo{person}{Ian Goodfellow}, {and} \bibinfo{person}{Rob Fergus}.}
  \bibinfo{year}{2014}\natexlab{}.
\newblock \showarticletitle{Intriguing properties of neural networks}. In
  \bibinfo{booktitle}{\emph{International Conference on Learning
  Representations}}.
\newblock


\bibitem[\protect\citeauthoryear{Tian, Xu, Zuo, Zhang, Fei, and Lin}{Tian
  et~al\mbox{.}}{2021}]%
        {sisr4}
\bibfield{author}{\bibinfo{person}{Chunwei Tian}, \bibinfo{person}{Yong Xu},
  \bibinfo{person}{Wangmeng Zuo}, \bibinfo{person}{Bob Zhang},
  \bibinfo{person}{Lunke Fei}, {and} \bibinfo{person}{Chia-Wen Lin}.}
  \bibinfo{year}{2021}\natexlab{}.
\newblock \showarticletitle{Coarse-to-Fine CNN for Image Super-Resolution}.
\newblock \bibinfo{journal}{\emph{IEEE Transactions on Multimedia}}
  \bibinfo{volume}{23} (\bibinfo{year}{2021}), \bibinfo{pages}{1489--1502}.
\newblock
\urldef\tempurl%
\url{https://doi.org/10.1109/TMM.2020.2999182}
\showDOI{\tempurl}


\bibitem[\protect\citeauthoryear{Vaswani, Shazeer, Parmar, Uszkoreit, Jones,
  Gomez, Kaiser, and Polosukhin}{Vaswani et~al\mbox{.}}{[n.d.]}]%
        {attention}
\bibfield{author}{\bibinfo{person}{Ashish Vaswani}, \bibinfo{person}{Noam
  Shazeer}, \bibinfo{person}{Niki Parmar}, \bibinfo{person}{Jakob Uszkoreit},
  \bibinfo{person}{Llion Jones}, \bibinfo{person}{Aidan~N Gomez},
  \bibinfo{person}{\L~ukasz Kaiser}, {and} \bibinfo{person}{Illia Polosukhin}.}
  \bibinfo{year}{[n.d.]}\natexlab{}.
\newblock \showarticletitle{Attention is All you Need}. In
  \bibinfo{booktitle}{\emph{Advances in Neural Information Processing
  Systems}}, \bibfield{editor}{\bibinfo{person}{I.~Guyon},
  \bibinfo{person}{U.~V. Luxburg}, \bibinfo{person}{S.~Bengio},
  \bibinfo{person}{H.~Wallach}, \bibinfo{person}{R.~Fergus},
  \bibinfo{person}{S.~Vishwanathan}, {and} \bibinfo{person}{R.~Garnett}}
  (Eds.). \bibinfo{publisher}{Curran Associates, Inc.}
\newblock


\bibitem[\protect\citeauthoryear{{Wang}, {Li}, {Deng}, {Du}, {Zhuang}, {Liang},
  and {Liu}}{{Wang} et~al\mbox{.}}{2020}]%
        {wang_ca_gan}
\bibfield{author}{\bibinfo{person}{J. {Wang}}, \bibinfo{person}{P. {Li}},
  \bibinfo{person}{J. {Deng}}, \bibinfo{person}{Y. {Du}}, \bibinfo{person}{J.
  {Zhuang}}, \bibinfo{person}{P. {Liang}}, {and} \bibinfo{person}{P. {Liu}}.}
  \bibinfo{year}{2020}\natexlab{}.
\newblock \showarticletitle{CA-GAN: Class-Condition Attention GAN for
  Underwater Image Enhancement}.
\newblock \bibinfo{journal}{\emph{IEEE Access}}  \bibinfo{volume}{8}
  (\bibinfo{year}{2020}), \bibinfo{pages}{130719--130728}.
\newblock
\urldef\tempurl%
\url{https://doi.org/10.1109/ACCESS.2020.3003351}
\showDOI{\tempurl}


\bibitem[\protect\citeauthoryear{{Wang}, {Wang}, {Gao}, {Lu}, and
  {Zhang}}{{Wang} et~al\mbox{.}}{2019}]%
        {wang_access_2019}
\bibfield{author}{\bibinfo{person}{J. {Wang}}, \bibinfo{person}{H. {Wang}},
  \bibinfo{person}{G. {Gao}}, \bibinfo{person}{H. {Lu}}, {and}
  \bibinfo{person}{Z. {Zhang}}.} \bibinfo{year}{2019}\natexlab{}.
\newblock \showarticletitle{Single Underwater Image Enhancement Based on
  $L_{P}$ -Norm Decomposition}.
\newblock \bibinfo{journal}{\emph{IEEE Access}}  \bibinfo{volume}{7}
  (\bibinfo{year}{2019}), \bibinfo{pages}{145199--145213}.
\newblock
\urldef\tempurl%
\url{https://doi.org/10.1109/ACCESS.2019.2945576}
\showDOI{\tempurl}


\bibitem[\protect\citeauthoryear{Wang, Li, Zhu, Tian, and Shan}{Wang
  et~al\mbox{.}}{2020}]%
        {segmentation_intro}
\bibfield{author}{\bibinfo{person}{Li Wang}, \bibinfo{person}{Dong Li},
  \bibinfo{person}{Yousong Zhu}, \bibinfo{person}{Lu Tian}, {and}
  \bibinfo{person}{Yi Shan}.} \bibinfo{year}{2020}\natexlab{}.
\newblock \showarticletitle{Dual Super-Resolution Learning for Semantic
  Segmentation}. In \bibinfo{booktitle}{\emph{IEEE/CVF Conference on Computer
  Vision and Pattern Recognition (CVPR)}}.
\newblock


\bibitem[\protect\citeauthoryear{{Wang}, {Zheng}, and {Zheng}}{{Wang}
  et~al\mbox{.}}{2017c}]%
        {wang_access}
\bibfield{author}{\bibinfo{person}{N. {Wang}}, \bibinfo{person}{H. {Zheng}},
  {and} \bibinfo{person}{B. {Zheng}}.} \bibinfo{year}{2017}\natexlab{c}.
\newblock \showarticletitle{Underwater Image Restoration via Maximum
  Attenuation Identification}.
\newblock \bibinfo{journal}{\emph{IEEE Access}}  \bibinfo{volume}{5}
  (\bibinfo{year}{2017}).
\newblock
\urldef\tempurl%
\url{https://doi.org/10.1109/ACCESS.2017.2753796}
\showDOI{\tempurl}


\bibitem[\protect\citeauthoryear{Wang, Chen, Yuan, Liu, Huang, Hou, and
  Cottrell}{Wang et~al\mbox{.}}{2018}]%
        {gridding_effect}
\bibfield{author}{\bibinfo{person}{P. Wang}, \bibinfo{person}{P. Chen},
  \bibinfo{person}{Y. Yuan}, \bibinfo{person}{D. Liu}, \bibinfo{person}{Z.
  Huang}, \bibinfo{person}{X. Hou}, {and} \bibinfo{person}{G. Cottrell}.}
  \bibinfo{year}{2018}\natexlab{}.
\newblock \showarticletitle{Understanding Convolution for Semantic
  Segmentation}. In \bibinfo{booktitle}{\emph{2018 IEEE Winter Conference on
  Applications of Computer Vision (WACV)}}. \bibinfo{publisher}{IEEE Computer
  Society}, \bibinfo{address}{Los Alamitos, CA, USA},
  \bibinfo{pages}{1451--1460}.
\newblock
\urldef\tempurl%
\url{https://doi.org/10.1109/WACV.2018.00163}
\showDOI{\tempurl}


\bibitem[\protect\citeauthoryear{{Wang}, {Ma}, {Yeganeh}, {Wang}, and
  {Lin}}{{Wang} et~al\mbox{.}}{2015}]%
        {pcqi}
\bibfield{author}{\bibinfo{person}{S. {Wang}}, \bibinfo{person}{K. {Ma}},
  \bibinfo{person}{H. {Yeganeh}}, \bibinfo{person}{Z. {Wang}}, {and}
  \bibinfo{person}{W. {Lin}}.} \bibinfo{year}{2015}\natexlab{}.
\newblock \showarticletitle{A Patch-Structure Representation Method for Quality
  Assessment of Contrast Changed Images}.
\newblock \bibinfo{journal}{\emph{IEEE Signal Processing Letters}}
  \bibinfo{volume}{22}, \bibinfo{number}{12} (\bibinfo{year}{2015}).
\newblock
\urldef\tempurl%
\url{https://doi.org/10.1109/LSP.2015.2487369}
\showDOI{\tempurl}


\bibitem[\protect\citeauthoryear{{Wang}, {Liu}, and {Chau}}{{Wang}
  et~al\mbox{.}}{2017a}]%
        {wang_iscas}
\bibfield{author}{\bibinfo{person}{Y. {Wang}}, \bibinfo{person}{H. {Liu}},
  {and} \bibinfo{person}{L. {Chau}}.} \bibinfo{year}{2017}\natexlab{a}.
\newblock \showarticletitle{Single underwater image restoration using
  attenuation-curve prior}. In \bibinfo{booktitle}{\emph{2017 IEEE
  International Symposium on Circuits and Systems (ISCAS)}}.
  \bibinfo{pages}{1--4}.
\newblock
\urldef\tempurl%
\url{https://doi.org/10.1109/ISCAS.2017.8050994}
\showDOI{\tempurl}


\bibitem[\protect\citeauthoryear{{Wang}, {Zhang}, {Cao}, and {Wang}}{{Wang}
  et~al\mbox{.}}{2017b}]%
        {wang_cnn_icip}
\bibfield{author}{\bibinfo{person}{Y. {Wang}}, \bibinfo{person}{J. {Zhang}},
  \bibinfo{person}{Y. {Cao}}, {and} \bibinfo{person}{Z. {Wang}}.}
  \bibinfo{year}{2017}\natexlab{b}.
\newblock \showarticletitle{A deep CNN method for underwater image
  enhancement}. In \bibinfo{booktitle}{\emph{2017 IEEE International Conference
  on Image Processing (ICIP)}}. \bibinfo{pages}{1382--1386}.
\newblock
\urldef\tempurl%
\url{https://doi.org/10.1109/ICIP.2017.8296508}
\showDOI{\tempurl}


\bibitem[\protect\citeauthoryear{Wang, Bovik, Sheikh, and Simoncelli}{Wang
  et~al\mbox{.}}{2004}]%
        {ssim}
\bibfield{author}{\bibinfo{person}{Zhou Wang}, \bibinfo{person}{A.~C. Bovik},
  \bibinfo{person}{H.~R. Sheikh}, {and} \bibinfo{person}{E.~P. Simoncelli}.}
  \bibinfo{year}{2004}\natexlab{}.
\newblock \showarticletitle{Image Quality Assessment: From Error Visibility to
  Structural Similarity}.
\newblock  \bibinfo{volume}{13}, \bibinfo{number}{4} (\bibinfo{year}{2004}).
\newblock
\showISSN{1057-7149}


\bibitem[\protect\citeauthoryear{Wang, Chang, Yang, Liu, and Huang}{Wang
  et~al\mbox{.}}{2016}]%
        {low-level-2}
\bibfield{author}{\bibinfo{person}{Zhangyang Wang}, \bibinfo{person}{Shiyu
  Chang}, \bibinfo{person}{Yingzhen Yang}, \bibinfo{person}{Ding Liu}, {and}
  \bibinfo{person}{Thomas~S. Huang}.} \bibinfo{year}{2016}\natexlab{}.
\newblock \showarticletitle{Studying Very Low Resolution Recognition Using Deep
  Networks}. In \bibinfo{booktitle}{\emph{2016 IEEE Conference on Computer
  Vision and Pattern Recognition (CVPR)}}.
\newblock
\urldef\tempurl%
\url{https://doi.org/10.1109/CVPR.2016.518}
\showDOI{\tempurl}


\bibitem[\protect\citeauthoryear{Wei, Ramakrishna, Kanade, and Sheikh}{Wei
  et~al\mbox{.}}{2016}]%
        {openpose4}
\bibfield{author}{\bibinfo{person}{Shih-En Wei}, \bibinfo{person}{Varun
  Ramakrishna}, \bibinfo{person}{Takeo Kanade}, {and} \bibinfo{person}{Yaser
  Sheikh}.} \bibinfo{year}{2016}\natexlab{}.
\newblock \showarticletitle{Convolutional pose machines}. In
  \bibinfo{booktitle}{\emph{CVPR}}.
\newblock


\bibitem[\protect\citeauthoryear{{Wen}, {Tian}, {Huang}, and {Gao}}{{Wen}
  et~al\mbox{.}}{2013}]%
        {wen}
\bibfield{author}{\bibinfo{person}{H. {Wen}}, \bibinfo{person}{Y. {Tian}},
  \bibinfo{person}{T. {Huang}}, {and} \bibinfo{person}{W. {Gao}}.}
  \bibinfo{year}{2013}\natexlab{}.
\newblock \showarticletitle{Single underwater image enhancement with a new
  optical model}. In \bibinfo{booktitle}{\emph{2013 IEEE International
  Symposium on Circuits and Systems (ISCAS)}}. \bibinfo{pages}{753--756}.
\newblock
\urldef\tempurl%
\url{https://doi.org/10.1109/ISCAS.2013.6571956}
\showDOI{\tempurl}


\bibitem[\protect\citeauthoryear{Woo, Park, Lee, and Kweon}{Woo
  et~al\mbox{.}}{2018}]%
        {cbam}
\bibfield{author}{\bibinfo{person}{Sanghyun Woo}, \bibinfo{person}{Jongchan
  Park}, \bibinfo{person}{Joon-Young Lee}, {and} \bibinfo{person}{In~So
  Kweon}.} \bibinfo{year}{2018}\natexlab{}.
\newblock \showarticletitle{CBAM: Convolutional Block Attention Module}. In
  \bibinfo{booktitle}{\emph{Proceedings of the European Conference on Computer
  Vision (ECCV)}}.
\newblock


\bibitem[\protect\citeauthoryear{{Xu}, {Sun}, and {Zhang}}{{Xu}
  et~al\mbox{.}}{2006}]%
        {wnn}
\bibfield{author}{\bibinfo{person}{J. {Xu}}, \bibinfo{person}{J. {Sun}}, {and}
  \bibinfo{person}{C. {Zhang}}.} \bibinfo{year}{2006}\natexlab{}.
\newblock \showarticletitle{Non-linear Algorithm for Contrast Enhancement for
  Image Using Wavelet Neural Network}. In \bibinfo{booktitle}{\emph{2006 9th
  International Conference on Control, Automation, Robotics and Vision}}.
  \bibinfo{pages}{1--6}.
\newblock
\urldef\tempurl%
\url{https://doi.org/10.1109/ICARCV.2006.345382}
\showDOI{\tempurl}


\bibitem[\protect\citeauthoryear{{Yang}, {Wang}, {Yue}, {Fu}, and {Hou}}{{Yang}
  et~al\mbox{.}}{2017}]%
        {yang_icip}
\bibfield{author}{\bibinfo{person}{J. {Yang}}, \bibinfo{person}{X. {Wang}},
  \bibinfo{person}{H. {Yue}}, \bibinfo{person}{X. {Fu}}, {and}
  \bibinfo{person}{C. {Hou}}.} \bibinfo{year}{2017}\natexlab{}.
\newblock \showarticletitle{Underwater image enhancement based on
  structure-texture decomposition}. In \bibinfo{booktitle}{\emph{2017 IEEE
  International Conference on Image Processing (ICIP)}}.
\newblock
\urldef\tempurl%
\url{https://doi.org/10.1109/ICIP.2017.8296473}
\showDOI{\tempurl}


\bibitem[\protect\citeauthoryear{{Yang} and {Sowmya}}{{Yang} and
  {Sowmya}}{2015}]%
        {uciqe}
\bibfield{author}{\bibinfo{person}{M. {Yang}} {and} \bibinfo{person}{A.
  {Sowmya}}.} \bibinfo{year}{2015}\natexlab{}.
\newblock \showarticletitle{An Underwater Color Image Quality Evaluation
  Metric}.
\newblock \bibinfo{journal}{\emph{IEEE Transactions on Image Processing}}
  \bibinfo{volume}{24}, \bibinfo{number}{12} (\bibinfo{year}{2015}),
  \bibinfo{pages}{6062--6071}.
\newblock
\urldef\tempurl%
\url{https://doi.org/10.1109/TIP.2015.2491020}
\showDOI{\tempurl}


\bibitem[\protect\citeauthoryear{{Yang}, {Chen}, {Feng}, and {Ma}}{{Yang}
  et~al\mbox{.}}{2019}]%
        {yang_unified}
\bibfield{author}{\bibinfo{person}{S. {Yang}}, \bibinfo{person}{Z. {Chen}},
  \bibinfo{person}{Z. {Feng}}, {and} \bibinfo{person}{X. {Ma}}.}
  \bibinfo{year}{2019}\natexlab{}.
\newblock \showarticletitle{Underwater Image Enhancement Using Scene
  Depth-Based Adaptive Background Light Estimation and Dark Channel Prior
  Algorithms}.
\newblock \bibinfo{journal}{\emph{IEEE Access}}  \bibinfo{volume}{7}
  (\bibinfo{year}{2019}).
\newblock
\urldef\tempurl%
\url{https://doi.org/10.1109/ACCESS.2019.2953463}
\showDOI{\tempurl}


\bibitem[\protect\citeauthoryear{Yang, Zhang, Tian, Wang, Xue, and Liao}{Yang
  et~al\mbox{.}}{2019}]%
        {sisr5}
\bibfield{author}{\bibinfo{person}{Wenming Yang}, \bibinfo{person}{Xuechen
  Zhang}, \bibinfo{person}{Yapeng Tian}, \bibinfo{person}{Wei Wang},
  \bibinfo{person}{Jing-Hao Xue}, {and} \bibinfo{person}{Qingmin Liao}.}
  \bibinfo{year}{2019}\natexlab{}.
\newblock \showarticletitle{Deep Learning for Single Image Super-Resolution: A
  Brief Review}.
\newblock \bibinfo{journal}{\emph{Trans. Multi.}} \bibinfo{volume}{21},
  \bibinfo{number}{12} (\bibinfo{date}{Dec.} \bibinfo{year}{2019}),
  \bibinfo{pages}{3106–3121}.
\newblock
\showISSN{1520-9210}
\urldef\tempurl%
\url{https://doi.org/10.1109/TMM.2019.2919431}
\showDOI{\tempurl}


\bibitem[\protect\citeauthoryear{Yu and Koltun}{Yu and Koltun}{2016}]%
        {dilated_filtering}
\bibfield{author}{\bibinfo{person}{Fisher Yu} {and} \bibinfo{person}{Vladlen
  Koltun}.} \bibinfo{year}{2016}\natexlab{}.
\newblock \showarticletitle{Multi-Scale Context Aggregation by Dilated
  Convolutions}. In \bibinfo{booktitle}{\emph{International Conference on
  Learning Representations (ICLR)}}.
\newblock


\bibitem[\protect\citeauthoryear{Yu, Dong, Loy, and Tang}{Yu
  et~al\mbox{.}}{2016}]%
        {deconv2}
\bibfield{author}{\bibinfo{person}{Ke Yu}, \bibinfo{person}{Chao Dong},
  \bibinfo{person}{Chen~Change Loy}, {and} \bibinfo{person}{Xiaoou Tang}.}
  \bibinfo{year}{2016}\natexlab{}.
\newblock \showarticletitle{Deep Convolution Networks for Compression Artifacts
  Reduction}.
\newblock \bibinfo{journal}{\emph{CoRR}}  \bibinfo{volume}{abs/1608.02778}
  (\bibinfo{year}{2016}).
\newblock
\showeprint[arxiv]{1608.02778}


\bibitem[\protect\citeauthoryear{Yu, Zhang, Zhang, Wang, Lin, Xu, Bai, and
  Yuille}{Yu et~al\mbox{.}}{2021}]%
        {mgmatting}
\bibfield{author}{\bibinfo{person}{Qihang Yu}, \bibinfo{person}{Jianming
  Zhang}, \bibinfo{person}{He Zhang}, \bibinfo{person}{Yilin Wang},
  \bibinfo{person}{Zhe Lin}, \bibinfo{person}{Ning Xu}, \bibinfo{person}{Yutong
  Bai}, {and} \bibinfo{person}{Alan Yuille}.} \bibinfo{year}{2021}\natexlab{}.
\newblock \showarticletitle{Mask Guided Matting via Progressive Refinement
  Network}. In \bibinfo{booktitle}{\emph{Proceedings of the IEEE/CVF Conference
  on Computer Vision and Pattern Recognition (CVPR)}}.
  \bibinfo{pages}{1154--1163}.
\newblock


\bibitem[\protect\citeauthoryear{Zeiler, Taylor, and Fergus}{Zeiler
  et~al\mbox{.}}{2011}]%
        {deconv}
\bibfield{author}{\bibinfo{person}{Matthew~D. Zeiler},
  \bibinfo{person}{Graham~W. Taylor}, {and} \bibinfo{person}{Rob Fergus}.}
  \bibinfo{year}{2011}\natexlab{}.
\newblock \showarticletitle{Adaptive deconvolutional networks for mid and high
  level feature learning}. In \bibinfo{booktitle}{\emph{2011 International
  Conference on Computer Vision}}. \bibinfo{pages}{2018--2025}.
\newblock
\urldef\tempurl%
\url{https://doi.org/10.1109/ICCV.2011.6126474}
\showDOI{\tempurl}


\bibitem[\protect\citeauthoryear{Zhang, Tian, Kong, Zhong, and Fu}{Zhang
  et~al\mbox{.}}{2018}]%
        {sisr3}
\bibfield{author}{\bibinfo{person}{Yulun Zhang}, \bibinfo{person}{Yapeng Tian},
  \bibinfo{person}{Yu Kong}, \bibinfo{person}{Bineng Zhong}, {and}
  \bibinfo{person}{Yun Fu}.} \bibinfo{year}{2018}\natexlab{}.
\newblock \showarticletitle{Residual Dense Network for Image Super-Resolution}.
  In \bibinfo{booktitle}{\emph{Proceedings of the IEEE Conference on Computer
  Vision and Pattern Recognition (CVPR)}}.
\newblock


\bibitem[\protect\citeauthoryear{Zhu, Park, Isola, and Efros}{Zhu
  et~al\mbox{.}}{2017}]%
        {cyclegan}
\bibfield{author}{\bibinfo{person}{Jun-Yan Zhu}, \bibinfo{person}{Taesung
  Park}, \bibinfo{person}{Phillip Isola}, {and} \bibinfo{person}{Alexei~A
  Efros}.} \bibinfo{year}{2017}\natexlab{}.
\newblock \showarticletitle{Unpaired Image-to-Image Translation using
  Cycle-Consistent Adversarial Networks}. In \bibinfo{booktitle}{\emph{Computer
  Vision (ICCV), 2017 IEEE International Conference on}}.
\newblock


\end{thebibliography}

\end{document}